\documentclass[prb,aps,showpacs,twocolumn,preprintnumbers,amsmath,amssymb,superscriptaddress,longbibliography,nofootinbib]{revtex4-2}
\usepackage[english]{babel}
\usepackage{amsmath,amssymb,bbm,mathrsfs,mathtools,multirow,diagbox,xcolor,%
graphicx,tabularx,comment,amsfonts,dsfont,lettrine,units,comment}

\usepackage{graphicx}
\usepackage[colorlinks=True,linkcolor=red,citecolor=blue,urlcolor=blue]{hyperref}
\usepackage{bookmark}
\usepackage{xcolor}
\usepackage{chngcntr}
\usepackage{braket}
\usepackage{nicefrac}
\usepackage{bm}
\usepackage{bbm}
\usepackage{bbold}
\usepackage{braket}
\usepackage{relsize}
\usepackage{microtype}
\usepackage{lineno}
\usepackage{array}
\newcolumntype{C}{>{$}c<{$}}
\usepackage{newtxtext} 
\usepackage{orcidlink}
\usepackage[normalem]{ulem} 
\usepackage{wasysym} 
\usepackage{cancel} 
\usepackage{float}

\renewcommand{\dag}{\dagger}

\def\infinity{\infty}

\newcommand{\bs}{\boldsymbol}

\usepackage{graphicx}
\usepackage{bm}

\DeclareMathOperator{\Tr}{Tr}

\DeclareMathOperator{\sgn}{sgn}

\usepackage{hyperref}
\hypersetup{
    colorlinks=true,
    linkcolor=blue,
    filecolor=blue,      
    urlcolor=blue,
    citecolor=blue
}
 
\makeatletter\renewcommand*\env@matrix[1][*\c@MaxMatrixCols c]{%
  \hskip -\arraycolsep
  \let\@ifnextchar\new@ifnextchar
  \array{#1}}
\makeatother

\begin{document}

\title{Geometry-Enforced Topological Chiral Fermions in Amorphous Chiral Metals}

\author{Justin Schirmann\orcidlink{0009-0007-7030-0155}}
\thanks{Corresponding author:~\href{mailto:jschirm@sandia.gov}{jschirm@sandia.gov}}
\affiliation{Univ. Grenoble Alpes, CNRS, Grenoble INP, Institut Néel, 38000 Grenoble, France}

\author{Adolfo G. Grushin\orcidlink{0000-0001-7678-7100}}
\thanks{Corresponding author:~\href{mailto:grushin@dipc.org}{grushin@dipc.org}.~This work was jointly supervised by A.~G.~G. and B.~J.~W.}
\affiliation{Donostia International Physics Center (DIPC),
Paseo Manuel de Lardiz\'{a}bal 4, 20018, Donostia-San Sebasti\'{a}n, Spain}
\affiliation{IKERBASQUE, Basque Foundation for Science, Maria Diaz de Haro 3, 48013 Bilbao, Spain}
\affiliation{Univ. Grenoble Alpes, CNRS, Grenoble INP, Institut Néel, 38000 Grenoble, France}

\author{Benjamin J.~Wieder\orcidlink{0000-0003-2540-6202}}
\thanks{Primary address for B. J. W.: Institut de Physique Th\'eorique, Universit\'{e} Paris-Saclay, CEA, CNRS, F-91191 Gif-sur-Yvette, France.  Corresponding author:~\href{mailto:benjamin.wieder@ipht.fr}{benjamin.wieder@ipht.fr}.~This work was jointly supervised by A.~G.~G. and B.~J.~W.}
\affiliation{Institut de Physique Th\'eorique, Universit\'{e} Paris-Saclay, CEA, CNRS, F-91191 Gif-sur-Yvette, France}
\affiliation{Department of Physics, Massachusetts Institute of Technology, Cambridge, MA 02139, USA}

\begin{abstract}
Since the prediction and observation of topological Weyl semimetals (chiral TSMs), there have been enormous efforts to characterize further condensed matter realizations of chiral fermions.
These efforts were dramatically accelerated by the subsequent discovery of a profound link between low-energy topological and lattice chirality in structurally chiral crystals.  
Though TSMs are well understood in the limit of perfect translation symmetry, real solid-state materials host defects and disorder, and may even be rendered amorphous down to all but the smallest system length scales.
Previous theoretical studies have concluded that chiral TSMs transition into trivial diffusive metals at moderate disorder scales, raising concerns that chiral TSM states may only be accessible in highly crystalline samples.
In this work, we identify large families of chiral TSMs that instead persist under strong structural disorder -- even into the amorphous regime.
We show that amorphous chiral TSM phases can be stabilized in particular by the presence of long-range order in the \emph{local structural chirality}.  
We present extensive analytic and numerical calculations demonstrating the existence of both Weyl and higher-charge chiral fermions in amorphous metals whose topology and spin and orbital angular momentum textures are tunable via the interplay of average symmetry and geometry. 
To distinguish and generate new realizations of strongly disordered chiral fermions, we introduce an analytic approach grounded in symmetry group theory.
We then introduce an amorphous Wilson loop numerical method to identify chiral fermions in metals with 3D structural disorder.
Our findings bridge the crystalline and strongly disordered regimes of chiral TSMs, and indicate a clear route toward engineering geometry-enforced topology in non-crystalline materials and metamaterials.
\end{abstract}

\maketitle

{
\vspace{0.2in}
\centerline{\bf I. INTRODUCTION}
\vspace{0.1in}
}

The study of electronic structure and response in solid-state materials has been greatly transformed through the recognition of the intertwined roles of symmetry, geometry, and topology in crystalline solids~\cite{HasanKaneReview,WenZooReview,Wieder22}.
Though the ability of crystal symmetry to both constrain and enrich electronic topological phases was uncovered relatively soon after the discovery of the first topological insulators [TIs] and topological semimetals [TSMs]~\cite{HasanKaneReview,WenZooReview,Wieder22,AshvinWeyl,Armitage2018}, the role of geometry in topological phases was only more recently recognized~\cite{KramersWeyl,JiabinQuantumGeometryReview}.

One of the most fundamental geometric properties in nature is \emph{chirality}.  
Chirality plays a central role in characterizing a diverse range of systems including elementary particles~\cite{PeskinSchroederQFT}, biomolecules like DNA~\cite{WatsonCrickDNA}, and mesoscopic samples of solid-state materials~\cite{FlackChiral}.
Crucially, whereas many material properties arise from the presence of symmetries, chirality conversely emerges from the \emph{absence} of symmetry~\cite{KamienLubenskyChiralParameter}.
For example in high-energy theory, a \emph{chiral fermion} is a particle that can only be present in the absence of parity inversion symmetry, because parity transforms a chiral fermion [\emph{e.g.} a right-handed Weyl fermion] into a distinct partner particle [\emph{e.g.} a left-handed Weyl fermion]~\cite{WeylOriginal1929}.

It has previously been established that in gapless, semimetallic, solid-state systems with lattice translation symmetry, the low-energy, ${\bf k}\cdot {\bf p}$ Hamiltonian of a nodal degeneracy can take the same form as the Hamiltonian of a particle in high-energy theory~\cite{MeleDirac,Haldane1988}.
For example, in a 3D Weyl semimetal, the valence and conduction bands meet in twofold nodal degeneracies with linear dispersion in all three directions in crystal momentum [${\bf k}$] space, and hence closely resemble high-energy Weyl fermions~\cite{AshvinWeyl}.
The analogy between high-energy chiral fermions and solid-state Weyl points is further bolstered by recognizing that conventional Weyl fermions are monopoles of Berry curvature with a quantized chiral charge~\cite{VolovikWeylMonopole}:
\begin{equation}
C = \frac{1}{2\pi}\oint_{S}\Tr[{\bf F}({\bf k})]\cdot d{\bf A},
\label{eq:MainChiralCharge}
\end{equation}
where $S$ is a sphere surrounding the nodal point and ${\bf F}({\bf k})$ is the matrix Berry curvature, such that $C$ is odd under the action of spatial inversion $\mathcal{I}$.
Because the chiral charge $C$ of a nodal degeneracy is integer-quantized and cannot change without closing an energy gap or changing the system symmetry, then $C$ is a topological invariant. 
Semimetals in which nodal degeneracies carry nonzero $C$ are therefore further termed \emph{chiral TSMs}~\cite{Wieder22,Armitage2018}.
Chiral TSM phases may exhibit exotic bulk electromagnetic response effects and arc-like topological surface states.
However, these effects can only be distinguished from trivial responses and surface states in chiral TSMs for which the bulk chiral fermions exhibit a large separation in ${\bf k}$-space, and in which the bulk Fermi pockets carry nontrivial $C$ over large energy windows.

Soon after their prediction in magnetic pyrochlore iridates~\cite{AshvinWeyl}, chiral TSM phases were theoretically predicted and experimentally observed in the TaAs family~\cite{AndreiWeyl,SuyangWeyl,YulinWeylExp}.
However, the Weyl points in TaAs arise from spin-orbit-coupling- [SOC-] driven band inversion at generic ${\bf k}$ points, and are therefore only weakly separated in energy and momentum space. 
Specifically, in real materials with dispersive [non-flat] bands, the band dispersion is typically dominated by orbital interactions, with SOC acting on much weaker energy scales~\cite{Wieder22}.
More generally, band-inversion-driven chiral TSMs like TaAs carry a relatively limited energy range for probing responses attributable to bulk chiral fermions, and can exhibit Weyl-point configurations that are highly sensitive to numerical simulation and experimental sample details~\cite{AlexeyType2,TMDHOTI,IlyaIdealMagneticWeyl,CDWWeyl}.

Condensed-matter chiral fermions are sometimes presented as electronic-structure analogs of magnetic monopoles.
Like the magnetic charge $q_{m}$ of a magnetic monopole, the chiral charge $C$ of a chiral fermion quasiparticle is odd under $\mathcal{I}$.
Importantly, however, whereas $q_{m}$ is odd under the action of time-reversal $\mathcal{T}$, $C$ in Eq.~(\ref{eq:MainChiralCharge}) is \emph{even} under $\mathcal{T}$. 
Hence in crystals without $\mathcal{I}$ and with $\mathcal{T}$, it is possible for chiral fermions to appear pinned to $\mathcal{T}$-invariant ${\bf k}$ [TRIM] points for which ${\bf k}=-{\bf k}\text{ mod }{\bf K}_{\mu}$ where ${\bf K}_{\mu}$ is a reciprocal lattice vector [see Appendix~\ref{app:corepDefs} and Refs.~\cite{ManesNewFermion,KramersWeyl}].
In Ref.~\cite{KramersWeyl}, the authors more generally recognized that $C$ is odd under the action of any rotoinversion [improper rotation] $g$ of the form:
\begin{equation}
g = \{\mathcal{I}\times C_{ni}|{\bf v}\},
\label{eq:MainTextRotoinversion}
\end{equation}
where $C_{ni}$ is a proper rotation by $360 / n$ degrees about the $i$-axis and ${\bf v}$ is a discrete [possibly fractional] lattice translation.
This implies that chiral fermions may only lie at TRIM points in 3D crystals if and only if the system space group does not contain rotoinversion symmetries.
The absence in a 3D crystal of not just $\mathcal{I}$ symmetry, but also other rotoinversion symmetries like mirror reflection $M_{i}=\mathcal{I}\times C_{2i}$ crucially implies that the crystal is \emph{structurally chiral} and respects the symmetries of a chiral space group [see Appendix~\ref{app:symDefs} for the definition of chiral symmetry groups employed in this work].

Building on this recognition, structurally chiral TSM phases were predicted~\cite{chang2017large,tang2017CoSi} and subsequently confirmed through extensive experimental investigations in B20 chiral cubic materials including CoSi, RhSi, and AlPt~\cite{CoSiObserveJapan,CoSiObserveHasan,CoSiObserveChina,AlPtObserve}.
Crucially, unlike in TaAs, the bulk Fermi surface in the B20 family almost entirely originates from maximally separated chiral fermions pinned to TRIM points at opposite corners of the 3D Brillouin zone [BZ].
Further unlike in TaAs, the chiral fermion dispersion in the B20 family primarily arises from large, purely orbital interactions -- rather than SOC -- leading to topologically nontrivial energy windows in the eV range~\cite{chang2017large}.
Importantly, it was theoretically predicted~\cite{KramersWeyl} and subsequently experimentally demonstrated~\cite{AlPtObserve,PdGaObserve} that in structurally chiral TSMs, there exists a profound link between disparate notions of chirality across system length scales.
Specifically, it was shown in Refs.~\cite{KramersWeyl,AlPtObserve,PdGaObserve} that the signs of the low-energy topological chiral charges $C$ of chiral fermions in structurally chiral crystals are governed by the lattice-scale structural chirality [handedness], implying that chiral fermions at the same TRIM point in oppositely handed crystal enantiomers carry oppositely signed $C$.
Owing to the above properties, structurally chiral TSMs have emerged as a rich platform for exploring advantageous, industry-relevant effects including enantioselective photoresponse~\cite{deJuanAdolfoCPGE,ReesCPGEObserve}, tunable spin and orbital angular momentum [OAM] textures~\cite{OAMmultifold2,OAMmultifold3}, and enantioselective and high-efficiency catalysis~\cite{ChiralCatalysisPdGa,ClaudiaChiralCatalysis,ClaudiaNatureEnergyCatalysis}.

The above discussion of chiral TSM states is highly reliant on the existence of perfect lattice translation symmetry, which permits chiral fermions with compensating $C$ to lie at distinct well-defined crystal momenta ${\bf k}$.  
However, real-material samples necessarily contain defects and disorder, or may even be prepared as alloys or in quasicrystalline or amorphous structural regimes. 
For TI states, it is theoretically well-established that the bulk topology is stable to disorder that does not close the mobility gap, and can survive in amorphous structural phases~\cite{agarwala_topological_2017,Mitchell2018,marsal_topological_2020,Zilberberg:21,Corbae_2023}.
This can be understood by recognizing that TI states are solely protected by on-site [internal] symmetries like $\mathcal{T}$ and U(1) charge conservation that do not act on spatial coordinates~\cite{HasanKaneReview,WenZooReview}, and hence remain exactly preserved in the presence of structural disorder [see Appendix~\ref{app:DiffTypesDisorder}].
Theoretical predictions of amorphous TIs have been further bolstered by the observation of mesoscopic response and spectroscopic signatures consistent with TI states in strongly disordered and amorphous samples of bismuth-based materials~\cite{corbae_evidence_2020,Ciocys2023,Banerjee:2017jd,DC2018} and in non-crystalline metamaterials~\cite{Mitchell2018,Liu2020,ZhangDelplace2023scattering}.

Conversely for TSMs with strong disorder, the theoretical predictions are less promising.
Employing a combination of analytic and numerical methods to investigate the survival of bulk chiral TSM phases and their associated surface Fermi arcs under a large range of disorder implementations and non-crystalline settings, previous theoretical works largely concluded that while chiral TSMs are perturbatively stable to weak disorder, stronger disorder appears to trivialize chiral TSM states~\cite{Chen2015,Altland2015,Gorbar2016,Slager2017,Buchhold2018b,yang_topological_2019,Pixley2021,Franca2024}.
The earlier works specifically observed that idealized models of chiral TSMs subject to strong disorder transition into topologically trivial diffusive metals or insulators, with the surface and bulk phase transitions potentially occurring at distinct disorder scales.

However, recent experiments on alloyed, strongly disordered, and amorphous gapless materials suggest the tantalizing possibility that under certain conditions, TSM states can survive up to much stronger disorder scales.  
First, bulk and surface signatures of a chiral TSM state were observed in angle-resolved photoemission spectroscopy [ARPES] experiments on the structurally chiral substitutional alloy Rh$_{1-x}$Ni$_{x}$Si, despite the manifest bulk translation symmetry breaking induced by the non-stoichiometric sample composition~\cite{ZahidLadderMultigap}.
More strikingly, highly non-crystalline samples of chiral TSM materials have been experimentally shown to exhibit mesoscopic responses -- such as anomalous Hall conductivities -- that are comparable to crystalline samples of the same material, where the crystalline-limit responses have been attributed to bulk chiral fermions~\cite{Li2021aWTex,Fujiwara2023kagome,KarelAmorphousBerryCMG,Asir2025,Molinari2023,Rocchino2024}.
These studies provide evidence that TSM materials may retain topological information via local structural order and bonding environments, which can persist in non-crystalline samples owing to local chemical constraints [\emph{e.g.} preferred bonding configurations and oxidation states].
Furthermore, spin-ARPES experiments on amorphous Bi$_2$Se$_3$ have revealed the presence of gapless spectral features on 2D surfaces that closely resemble and carry the same spin textures as the surface Dirac cones in crystalline Bi$_2$Se$_3$, the archetypal 3D TI~\cite{corbae_evidence_2020,Ciocys2023}.
Because the gapless 2D surfaces of 3D TIs are equivalent to anomalous 2D Dirac semimetals -- as they carry one quarter of the degrees of freedom of the simplest free-standing [lattice-regularized] spinful $\mathcal{T}$-invariant 2D Dirac semimetal~\cite{DiracInsulator} -- then in this sense, amorphous TSM states have \emph{already} been directly observed in experiment.

Inspired by the above experiments, we revisit in this work the theoretical question of whether a chiral TSM state may be stabilized on an amorphous lattice.
By more carefully treating the larger order parameter space admitted in systems with local [\emph{e.g.} spin, orbital, or sublattice] degrees of freedom, we discover that contrary to the previous theoretical studies~\cite{Chen2015,Altland2015,Gorbar2016,Slager2017,Buchhold2018b,yang_topological_2019,Pixley2021,Franca2024}, there exist broad families of chiral TSMs that remain stable in the presence of strong structural disorder.
As detailed below and in Appendices~\ref{app:pseudoK},~\ref{app:corepAmorphous}, and~\ref{app:DiffTypesDisorder}, the amorphous chiral TSMs identified in this work are specifically stabilized by long-range order in the \emph{local structural chirality}, such that the finite-range spin and orbital interactions in each local region [patch] carry the same handedness over large domains with respect to the system size.

To take advantage of local chirality order in the amorphous regime, we focus in this work on structurally chiral TSMs that, like the CoSi family~\cite{chang2017large,tang2017CoSi,CoSiObserveJapan,CoSiObserveHasan,CoSiObserveChina,AlPtObserve}, host chiral fermions close to the Fermi level at the TRIM point $\Gamma$ [${\bf k}={\bf 0}$] in the crystalline limit.
As discussed in Refs.~\cite{marsal_topological_2020,corbae_evidence_2020,Ciocys2023} and Appendices~\ref{app:pseudoK} and~\ref{app:PhysicalObservables}, strongly disordered and amorphous systems also exhibit a Fourier-transformed energy spectrum as a function of the plane-wave pseudo-momentum ${\bf p}$, albeit a spectrum that is isotropic due to the lack of preferred system directions, and with a ${\bf p}$-dependent sharpness because ${\bf p}$ is only an approximate quantum number.
Nevertheless, the spectral functions of strongly disordered systems are consistently observed to be sharpest at the longest system wavelengths, and hence smallest pseudo-momenta ${\bf p} \approx {\bf 0}$ [see Appendix~\ref{app:PhysicalObservables} and Refs.~\cite{marsal_topological_2020,corbae_evidence_2020,Ciocys2023}].
Building on this result, we discover that structurally chiral TSMs with crystalline-limit $\Gamma$-point chiral fermions remain gapless and topological up to strong disorder scales in the presence of long-range local chirality order, and \emph{continue} to host chiral fermions at ${\bf p}={\bf 0}$ with relatively sharp dispersion relations in the amorphous regime.

Our ability to surpass the disorder bounds on chiral TSM states in previous theoretical and numerical studies stems directly from the above requirements of crystalline-limit $\Gamma$-point chiral fermions and long-range local chirality order in the structurally [positionally] disordered regime.
Specifically, the majority of previous analyses of strongly disordered chiral TSMs focused on band-inversion-driven Weyl semimetal states like TaAs, in which oppositely charged Weyl points lie at $\pm {\bf k}$~\cite{Chen2015,Altland2015,Gorbar2016,Slager2017,Buchhold2018b,yang_topological_2019}.
Because structural disorder renders systems spectrally isotropic, bulk Fermi pockets with compensating chiral charges $\pm C$ in band-inversion chiral TSMs are easily mixed by disorder, and are expected to merge into trivial bulk Fermi rings [spheres] as the system approaches the spectrally-isotropic limit [see Appendices~\ref{app:pseudoK} and~\ref{app:models} and Refs.~\cite{Ciocys2023,Corbae_2023}].
Conversely in structurally chiral TSMs, $\Gamma$-point chiral fermions and nodal degeneracies with compensating $C$ lie at unequal $|{\bf k}|$~\cite{KramersWeyl}, and are hence less easily merged by structural disorder.
Instead, we observe that $\Gamma$-point chiral fermions are most straightforwardly trivialized by disorder that drives the system towards average structural \emph{achirality} [\emph{i.e.} towards a regime in which the average system symmetry group contains a rotoinversion symmetry of the form of Eq.~(\ref{eq:MainTextRotoinversion})]. 
While a handful of earlier theoretical works also analyzed the disorder stability of structurally chiral TSM models with TRIM-point chiral fermions~\cite{Pixley2021,Franca2024}, the earlier works employed disorder realizations that were structurally achiral on the average, or overlooked the role of average or exact structural chirality in their analyses, and hence did not observe the amorphous chiral fermions identified in this work.

To confirm the presence of non-crystalline bulk chiral fermions in strongly disordered systems with local chirality order, we perform extensive analytic and numerical calculations, whose complete details are provided in the Supplementary Appendices.
First, using Fourier-transformed Green's functions in amorphous semimetal models [Appendices~\ref{app:PhysicalObservables} and~\ref{app:surfaceGreens}], we compute the bulk and surface spectra and spin and OAM textures, which can approximately indicate the chiral charges of nodal degeneracies~\cite{KramersWeyl,OAMmultifold2,OAMmultifold3}.
As we will show below in the main text and in Appendices~\ref{app:amorphousKramers},~\ref{app:amorphousCharge2}, and~\ref{app:amorphousMultifold}, though the surface Fermi arcs of amorphous chiral TSMs are generically obscured by isotropic spectral features away from ${\bf p}={\bf 0}$~\cite{Ciocys2023}, amorphous chiral TSMs can continue to exhibit experimentally detectable monopole-like spin and OAM textures like those of their crystalline counterparts~\cite{KramersWeyl,OAMmultifold2,OAMmultifold3}.

Next, the relatively sharp spectral functions of amorphous systems near ${\bf p}\approx{\bf 0}$ suggests that a mean-field  [effective single-particle] Hamiltonian can be used to approximately capture the exact [many-particle, Fourier-transformed] system properties at small ${\bf p}$.
Motivated by this observation, we use Fourier-transformed Green's functions [Appendix~\ref{app:PhysicalObservables}] to construct pseudo-momentum-space effective Hamiltonians near ${\bf p}={\bf 0}$ in amorphous chiral TSM models.
We further perform numerical benchmarking to confirm that the ${\bf p}={\bf 0}$ nodal degeneracies and Berry phases of the effective Hamiltonian are numerically stable and independent of our computational implementation details [Appendix~\ref{app:EffectiveHamiltonian}].
Using the eigenstates of the effective Hamiltonian, we then introduce in Appendix~\ref{sec:WilsonBerry} a generalization of the non-Abelian Wilson loop method~\cite{AndreiXiZ2,DiracInsulator,TMDHOTI,Z2Pack,Wieder22} to demonstrate that ${\bf p}={\bf 0}$ nodal degeneracies in amorphous semimetals can precisely be identified as non-crystalline TRIM-point chiral fermions with nontrivial topological chiral charges $C$.
Building on this result, we then show that, analogously to their crystalline counterparts~\cite{KramersWeyl,AlPtObserve,PdGaObserve}, non-crystalline chiral fermions in amorphous chiral TSMs carry chiral charges that are tunable via the \emph{average} [net] local structural chirality [Appendix~\ref{app:models}].

While previous works have employed ${\bf k}$-dependent topological markers to infer the existence of non-crystalline chiral fermions in 3D systems composed of identical, regularly stacked 2D quasicrystals [see Refs.~\cite{Grossi2023b,WeylQuasicrystalSCBott} and Appendix~\ref{sec:WilsonBerry}], the amorphous Wilson loop calculations performed in this work represent the first demonstration of quantized topological invariants in 3D metals with \emph{complete} structural disorder.
To place our findings within the context of previous studies of enforced nodal degeneracies in crystalline systems [see Appendices~\ref{app:symDefs} and~\ref{app:corepDefs} and Refs.~\cite{ManesNewFermion,DDP,NewFermions}], we also introduce a rigorous analytic framework grounded in symmetry group theory in which non-crystalline chiral fermions are classified via the irreducible corepresentations [coreps] of \emph{chiral average symmetry groups} [Appendices~\ref{app:pseudoK} and~\ref{app:corepAmorphous}].
We conclude by discussing potential real-material venues and physically motivated sample conditions for observing signatures of amorphous chiral fermions.

In the main text, we focus our analysis on non-crystalline generalizations of the relatively simple structurally chiral Kramers-Weyl model [Ref.~\cite{KramersWeyl} and Appendix~\ref{app:Kramers}], as well as a more realistic model that in the crystalline limit exhibits the same low-energy chiral multifold fermions [threefold nodal degeneracies] as B20 CoSi [Refs.~\cite{chang2017large,tang2017CoSi} and Appendix~\ref{sec:Multifold}].
In Appendix~\ref{sec:Charge2}, we also introduce and analyze a structurally chiral model with a quadratically dispersing, $|C|=2$, non-crystalline chiral [double-Weyl~\cite{AndreiMultiWeyl}] fermion at ${\bf p}={\bf 0}$.
For each of the models analyzed in this work, we employ multiple tight-binding schemes to approach and approximate the amorphous structural regime [Appendix~\ref{app:lattices}].
Importantly, we have numerically confirmed that when each model is simulated with the highest degree of structural disorder [\emph{e.g.} on random lattices], the topological features of each model remain stable under the subsequent addition of weak Anderson [on-site chemical potential] disorder.
Taken together, our calculations reveal that non-crystalline chiral fermions represent a strikingly general feature of 3D amorphous systems with local structural chirality order.

{
\vspace{0.2in}
\centerline{\bf II. NUMERICAL WORKFLOW}
\vspace{0.1in}
}

\begin{figure*}[t]
\centering
\includegraphics[width=\linewidth]{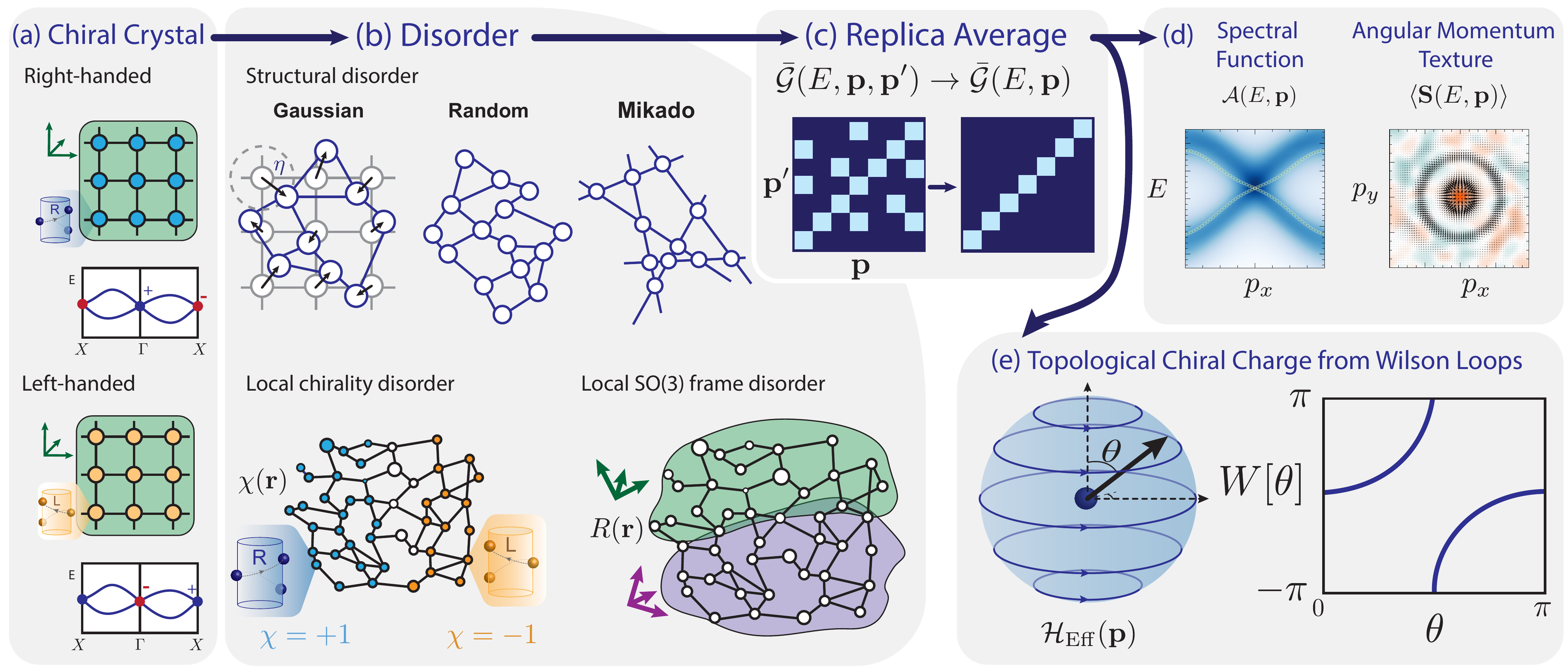}
\caption{{\bf Schematic numerical workflow.}
(a) To construct a non-crystalline chiral topological semimetal [TSM] state, we begin by introducing a tight-binding Hamiltonian $\mathcal{H}$ [Eq.~(\ref{eq:Genrealspacemodel})] that is structurally chiral when placed on a crystalline lattice and can be realized in well-defined right-handed [$R$, blue helix and lattice] and left-handed [$L$, yellow helix and lattice] enantiomers [see Appendix~\ref{app:symDefs}].
The lattice-scale structural chirality enforces that chiral [\emph{e.g.} Kramers-Weyl] fermions at time-reversal-invariant crystal momenta ${\bf k}$ carry oppositely signed chiral charges $C$ [Eq.~(\ref{eq:MainChiralCharge}), blue and red $\pm$ signs at $\Gamma$ and $X$ in (a), see Refs.~\cite{KramersWeyl,AlPtObserve,PdGaObserve} and Appendix~\ref{app:models}].
(b) Each position-space $\mathcal{H}$ in this work is defined such that $\mathcal{H}$ may equally straightforwardly be implemented on both crystalline and structurally disordered lattices.
As detailed in Appendix~\ref{app:lattices} and the text preceding Eq.~(\ref{eq:deltaComponentsMain}), we employ three distinct forms of structural disorder -- Gaussian disorder, random lattices, and Mikado lattices -- to approximate the amorphous regime.
The models in this work also contain internal spin and orbital degrees of freedom on each lattice site, such that the \emph{hopping reference frame orientation} relative to the global coordinate axes may also be disordered independently from the lattice regularity [see Appendix~\ref{app:DiffTypesDisorder} and the text following Eq.~(\ref{eq:TmatrixRotFrameMain})].
The orientational disorder in each model specifically admits a decomposition into local chirality disorder characterized by the $\mathbb{Z}_{2}$ scalar order parameter $\chi({\bf r})=\pm 1$ [blue and yellow lattice sites in (b)], and local frame orientation disorder characterized by the SO(3) matrix order parameter $R({\bf r})$ [green and purple axes and regions in (b)].
To identify a TSM state, it is necessary to establish a momentum-space system description in which nodal degeneracies can be assigned topological invariants~\cite{Armitage2018,Wieder22}.
Though structurally disordered systems cannot be characterized in terms of an exact crystal momentum ${\bf k}$, they may still be characterized in terms of an approximate plane-wave pseudo-momentum ${\bf p}$ [Eq.~(\ref{eq:planewavesMain})].
(c) Using ${\bf p}$, we Fourier-transform the real-space system Green's function [$\mathcal{G}(E)$ in Eq.~(\ref{eq:RealSpaceGreenMain})] to obtain the momentum-resolved matrix Green's function $\mathcal{G}(E,{\bf p},{\bf p}')$ [Eq.~(\ref{eq:MomGreenFuncMain}) and Appendix~\ref{app:PhysicalObservables}].
Though $\mathcal{G}(E,{\bf p},{\bf p}')$ is an exact quantity, it is generically a function of two pseudo-momenta ${\bf p}$ and ${\bf p}'$, obscuring a direct connection to crystalline quantities computed from Green's functions that only depend on a single ${\bf k}$.
However, we find that after averaging over an ensemble of replicas with independently generated structural and internal degree-of-freedom disorder, $\mathcal{G}(E,{\bf p},{\bf p}')$ in practice reduces on the average to a one-momentum [average] matrix Green's function $\bar{\mathcal{G}}(E,{\bf p})$ [Eq.~(\ref{eq:averageOneMomentumGreenMain}) and Appendix~\ref{app:PhysicalObservables}].
(d) Using $\bar{\mathcal{G}}(E,{\bf p})$, we then compute the spectral function $\bar{A}(E,\mathbf{p})$ [Eq.~(\ref{eq:SpecFuncMain}) and Appendix~\ref{app:PhysicalObservables}] and spin and orbital angular momentum textures [Eqs.~(\ref{eq:SpinDOSMain}) and (\ref{eq:OAMDOSMain}) and Appendices~\ref{app:amorphousKramers} and~\ref{app:amorphousMultifold}].
We also use $\bar{\mathcal{G}}(E,{\bf p})$ to construct a single-particle effective Hamiltonian $\mathcal{H}_\text{Eff}({\bf p})$, which we find to exhibit numerically stable dispersion relations and Berry phases near ${\bf p}={\bf 0}$ over a wide range of reference energy cuts $E_{C}$ in Eq.~(\ref{eq:AvgHEffMain}) [Appendix~\ref{app:EffectiveHamiltonian}].
(e) Motivated by this observation, we lastly introduce an amorphous Wilson loop method to quantitatively evaluate the topology of ${\bf p}={\bf 0}$ nodal degeneracies in strongly disordered systems [Eq.~(\ref{eq:WilsonElementsMain}) and Appendix~\ref{sec:WilsonBerry}].
Across all three structural disorder implementation methods in (b), we find that each of the models in this work exhibits the same spectral and topologically nontrivial features in the presence of long-range local chirality order in $\chi({\bf r})$, providing strong evidence that each model realizes an amorphous chiral TSM state.}
\label{fig:Fig1flow}
\end{figure*}

{
\centerline{\bf A. Model construction}
\vspace{0.1in}
}
In Fig.~\ref{fig:Fig1flow}, we schematically depict our numerical approach to generate and identify non-crystalline chiral fermions.
To begin, for each of the models in this work, we first introduce a set of lattice sites indexed by $\alpha$, where each site $\alpha$ lies at a position ${\bf r}_{\alpha}$.
At each site $\alpha$, we may then place internal [on-site] spin and orbital degrees of freedom that lie at ${\bf r}_{\alpha}$, or we may also introduce sublattice degrees of freedom that lie on additional sublattice-indexed sites away from ${\bf r}_{\alpha}$.
Because it is more challenging to numerically disorder lattice models with sublattice degrees of freedom while implementing hopping interactions that independently preserve the relative positions, orientations, and handedness of local structural motifs~\cite{Franca2024}, we throughout this work limit our focus to models with only on-site spin and orbital degrees of freedom, and require that each site carries the same $N_{\text{orb}}$ internal degrees of freedom.
Nevertheless, as we will shortly show, models with only on-site degrees of freedom still retain a rich order parameter space that includes orientational and chirality order parameters akin to those in models with enforced local sublattice structures [see Fig.~\ref{fig:Fig1flow}(b) and Appendix~\ref{app:lattices}].

In this work, we consider models that can be defined both on regular [crystalline] lattices, as well as on positionally disordered [\emph{e.g.} random] lattices.
To that end, for each model in this work, we must define the hopping interactions in position space, which we accomplish using a second-quantized Hamiltonian $\mathcal{H}$ of the form:
\begin{equation}
\mathcal{H} = \sum_{\langle \alpha\beta \rangle}\sum_{l,l'}c_{\alpha,l}^{\dagger}T^{ll'}_{\alpha\beta}c^{\phantom{}}_{\beta,l'} + \sum_{\alpha}\sum_{l,l'}c_{\alpha,l}^{\dagger}M^{ll'}_{\alpha}c^{\phantom{}}_{\alpha,l'},
\label{eq:Genrealspacemodel}
\end{equation}
where the operator $c^{\dagger}_{\alpha,l}$ creates a particle at the site $\alpha$ with an internal degree of freedom $l$, $l$ and $l'$ run from $1$ to $N_{\text{orb}}$, and where the $\langle$ and $\rangle$ symbols denote summation over pairs of sites $\alpha$ and $\beta$ [see Appendices~\ref{app:PhysicalObservables} and~\ref{app:models} for further details].
In Eq.~(\ref{eq:Genrealspacemodel}), the elements of the $N_{\text{orb}}\times N_{\text{orb}}$ intersite hopping matrix $T_{\alpha\beta}$ for each pair of sites $\alpha$ and $\beta$:
\begin{equation}
T^{ll'}_{\alpha\beta} = \langle {\bf r}_\alpha,l | \mathcal{H} | {\bf r}_\beta,l'\rangle,
\label{eq:MainThopDef}
\end{equation}
and the $N_{\text{orb}}\times N_{\text{orb}}$ on-site mass matrix $M_{\alpha}$ for each site $\alpha$:
\begin{equation}
M^{ll'}_{\alpha} = \langle {\bf r}_\alpha,l | \mathcal{H} | {\bf r}_\alpha,l'\rangle,
\label{eq:MainMDef}
\end{equation}
are defined through inner products with respect to the single-particle tight-binding basis states $|{\bf r}_\alpha, l\rangle$ [further detailed in Appendix~\ref{app:PhysicalObservables}].

Importantly, when $\mathcal{H}$ is placed on a regular lattice with 3D lattice translation symmetry, we require that $\mathcal{H}$ respects the symmetries of a chiral space group [Appendix~\ref{app:symDefs}].
We further require the chiral space group of $\mathcal{H}$ to specifically contain at least one high-symmetry rotation axis [Appendix~\ref{app:symDefs} and Ref.~\cite{FlackChiral}], such that $\mathcal{H}$ characterizes the electronic spectrum of a structurally chiral crystal with well-defined right-handed [$R$] and left-handed [$L$] enantiomers [Fig.~\ref{fig:Fig1flow}(a)].
Together, these requirements ensure that $\Gamma$-point chiral fermions are admitted in the Fourier-transformed band structure of $\mathcal{H}$, and that the chiral charge at $\Gamma$ is controlled by the lattice-scale structural chirality $C_{\mathcal{H}}=\pm 1$ [handedness, see Fig.~\ref{fig:Fig1flow}(a), Appendices~\ref{app:symDefs} and~\ref{app:corepDefs}, and Refs.~\cite{KramersWeyl,AlPtObserve,PdGaObserve}].
In the crystalline limit of $\mathcal{H}$, $C_{\mathcal{H}}$ is a spatially homogeneous quantity, which physically corresponds to the case of a homochiral single-crystal sample without defects or disorder.
However, we will shortly show that by allowing the signs of hopping parameters in $T_{\alpha\beta}$ [Eqs.~(\ref{eq:Genrealspacemodel}) and~(\ref{eq:MainThopDef})] to vary for bonds originating from each site $\alpha$, we may also promote $C_{\mathcal{H}}$ to a spatially varying local chirality $\chi_{\alpha}$ whose spatial ordering acts as a knob for generating and controlling the properties of non-crystalline chiral fermions [Appendices~\ref{app:DiffTypesDisorder},~\ref{app:amorphousKramers},~\ref{app:amorphousCharge2}, and~\ref{app:amorphousMultifold}].

{
\vspace{0.2in}
\centerline{\bf B. Disorder implementation}
\vspace{0.1in}
}
Central to the present work, the position-space Hamiltonian $\mathcal{H}$ in Eq.~(\ref{eq:Genrealspacemodel}) can also characterize non-crystalline systems.
We may for example implement $\mathcal{H}$ on a lattice with irregularly spaced sites $\alpha$ [structural disorder], with unequal masses or chemical potentials in $M_{\alpha}$ [Eq.~(\ref{eq:MainMDef})] on each site [Anderson disorder], or in a manner in which the intersite hopping matrix $T_{\alpha\beta}$ [Eq.~(\ref{eq:MainThopDef})] is unequal in its magnitude or matrix elements for pairs of sites $\alpha$ and $\beta$ [respectively bond or orientational disorder].
Despite the many available forms of disorder, previous theoretical studies of topology and electronic structure in amorphous solids have primarily focused on relatively simple models with only structural or Anderson disorder, and have largely overlooked the role played by internal or local degrees of freedom and their relative [physical or hopping-frame] orientations.
While more recent studies of non-crystalline topological phases have employed more elaborate tight-binding models with internal spin and orbital degrees of freedom~\cite{agarwala_topological_2017,yang_topological_2019,corbae_evidence_2020,Ciocys2023}, the possibility of hopping frame orientation disorder was still not considered in these works.

Though not previously emphasized in earlier studies of amorphous topological phases, this contrast between orientational and structural order is well-established in other condensed-matter settings.
For example, liquid crystals composed of rod-like molecules can experimentally exhibit nematic phases in which the molecular rods lie at random positions, but on the average point in the same direction [see Appendix~\ref{app:SmecticNematicDisorder} and Ref.~\cite{KamienLubenskyChiralParameter}].
Furthermore, amorphous materials have been experimentally shown to host states of matter that can \emph{only} be captured in non-crystalline tight-binding models with specific local or internal degrees of freedom like electronic spins or magnetic moments, whose relative alignments give rise to orientational order via the spatial uniformity of intersite interactions.
Amorphous Bi$_2$Se$_3$, for example, has been shown to host a non-crystalline 3D TI state~\cite{corbae_evidence_2020,Ciocys2023}, which can only be captured in tight-binding models with internal spin-1/2 degrees of freedom~\cite{HasanKaneReview,WenZooReview}.
Overall, we find that a more careful treatment of the relative roles of structural and orientational order is in fact \emph{essential} for identifying an order parameter regime that stabilizes amorphous chiral TSM states.

In this work, we seek to model amorphous solid-state systems in which the local units largely carry chemically and electronically similar environments [up to the orientations and handedness of the local units].
This focus is supported by previous experimental studies of non-crystalline samples of established TSM materials like WTe$_2$~\cite{Li2021aWTex}, Fe$_x$Sn$_{1-x}$~\cite{Fujiwara2023kagome}, Co$_2$MnGa~\cite{KarelAmorphousBerryCMG}, and NbP~\cite{Asir2025}, in which signatures of topological mesoscopic response were observed and attributed primarily to the presence of local structural order.
Therefore, we focus primarily on models subject to the following three independent forms of disorder [see Fig.~\ref{fig:Fig1flow}(b) and Appendix~\ref{app:DiffTypesDisorder}]: 
\begin{enumerate}
    \item Structural disorder.
    \item Local hopping frame chirality disorder.
    \item Local hopping frame orientation disorder.
\end{enumerate}
Importantly, as detailed below and demonstrated through extensive numerical calculations documented in Appendix~\ref{app:models}, we find that these three forms of disorder are sufficient to construct amorphous chiral TSM states that exhibit tunable chiral charges and are perturbatively stable to the subsequent addition of Anderson disorder.

To approximate the amorphous regime, we employ three distinct forms of structural disorder: Gaussian disorder, random lattices, and Mikado lattices [Fig.~\ref{fig:Fig1flow}(b)].
After implementing each form of structural disorder, we next apply a relaxation procedure to narrow the distribution of intersite bond distances and recover a qualitatively sharp average nearest-neighbor site spacing $\bar{a}$, as is observed in real solid-state amorphous materials [see Appendix~\ref{app:DiffTypesDisorder} and Refs.~\cite{Ciocys2023,Corbae_2023}].
We then carefully compare system observables obtained using each disorder scheme.

First, to implement \emph{Gaussian structural disorder}, we begin with a 3D crystalline lattice with translation symmetry in three linearly independent directions.
We then displace each site $\alpha$ by a random vector ${\bs \delta}_{\alpha}$:
\begin{equation}
{\bs \delta}_{\alpha} = \begin{pmatrix}
\delta_{x,a}\\
\delta_{y,a}\\
\delta_{z,a}
\end{pmatrix},
\label{eq:deltaComponentsMain}
\end{equation}
where each component $\delta_{i,\alpha}$ is drawn from a Gaussian [normal] distribution $\mathcal{N}(0,\eta)$ in which $0$ denotes the mean value and $\eta$ indicates the standard deviation [see Fig.~\ref{fig:Fig1flow}(b), Appendix~\ref{app:DiffTypesDisorder}, and Ref.~\cite{Franca2024}].
Consequently, after displacing each site $\alpha$, the displacement components $\delta_{i,\alpha}$ in Eq.~(\ref{eq:deltaComponentsMain}) each exhibit a distribution governed by a Gaussian probability density function:
\begin{equation}
P(\delta_{i,\alpha}) = \frac{1}{\sqrt{2\pi\eta^{2}}}\exp\left(-\frac{\delta_{i,\alpha}^2}{2\eta^2}\right).
\label{eq:GaussianDisorderMain}
\end{equation}

From a physical perspective, Gaussian disorder can be considered akin to melting and subsequently quenching a crystalline material close to its melting temperature~\cite{Franca2024}.
In this picture, the Gaussian-disordered site displacements are distributed with a standard deviation of $\eta^2 \propto k_{B}T$ in Eq.~(\ref{eq:GaussianDisorderMain}), where $T$ is the maximum temperature of the melt just prior to the system quench.  
From a modeling perspective, Gaussian disorder is advantageous because it allows one to continuously track system properties from a clearly defined reference crystal [$\eta=0$] into the regime of strong structural disorder [quantitatively defined in Appendix~\ref{app:PhysicalObservables}].
However, for the same reason -- namely that Gaussian disorder retains some ``memory'' of the $\eta=0$ crystal lattice -- numerical quantities obtained from Gaussian disorder may not always truly approximate the behavior of maximally structurally disordered [\emph{i.e.} amorphous] systems.

To avoid this issue, we also analyze our non-crystalline semimetal models on lattices without any ``memory'' of reference crystalline limits.
These include \emph{random lattices}~\cite{agarwala_topological_2017,yang_topological_2019,Mitchell2018}, in which each site $\alpha$ is assigned a random position in 3D space [Fig.~\ref{fig:Fig1flow}(b)].
Unlike with Gaussian disorder, random lattices do not carry unique reference crystalline limits, and are therefore inherently non-crystalline.

We additionally implement non-crystalline chiral TSM states on \emph{Mikado lattices} [see Fig.~\ref{fig:Fig1flow}(b), Appendix~\ref{app:DiffTypesDisorder}, and Ref.~\cite{marsal_topological_2020}].
To generate 3D Mikado lattices, we first randomly place 2D planes throughout 3D space, and assign each plane a randomly oriented normal vector.
The random 2D planes in turn intersect to form a network of 1D lines.
We then place lattice sites at each 0D intersection point between lines in the network.
Like the random lattices detailed above, and unlike in the case of Gaussian structural disorder, Mikado lattices do not carry unique reference crystalline limits, and are therefore also inherently non-crystalline.

In each of the structural disorder implementation schemes in Fig.~\ref{fig:Fig1flow}(b), the spatial coordinates of pairs of sites $\alpha$ and $\beta$ are related by the intersite separation vector ${\bf d}_{\alpha\beta}$, given by:
\begin{equation}
    \mathbf{d}_{\alpha\beta}
= \begin{pmatrix}
x_{\alpha}-x_{\beta}\\y_{\alpha}-y_{\beta}\\z_{\alpha}-z_{\beta}
\end{pmatrix}.
\label{eq:dVectorDefMain}
\end{equation} 
Using ${\bf d}_{\alpha\beta}$, we then define for each hopping interaction in each model in this work a hopping rangedness function $f(|\mathbf{d}_{\alpha\beta}|)$. 
For each model and disorder implementation, $f(|\mathbf{d}_{\alpha\beta}|)$ contains an exponential decay contribution $\exp(a-|\mathbf{d}_{\alpha\beta}|)$, as well as a Heaviside function piece that ensures that $f(|\mathbf{d}_{\alpha\beta}|)$ is only nonvanishing for sites defined to be ``connected'' [see Appendices~\ref{app:DiffTypesDisorder} and~\ref{app:models}].
When employing Gaussian disorder and random lattices, we allow all sites within a radius $R_{0}$ of each other to be connected, leading to models without a fixed bond coordination.
Conversely on Mikado lattices, each site carries the same fixed bond coordination [albeit with bond strengths that are modulated by the exponential piece of $f(|\mathbf{d}_{\alpha\beta}|)$].
In this sense, Mikado lattices most closely emulate real solid-state amorphous materials, which largely consist of local units with the same bond coordination, owing to their chemically similar local environments
~\cite{Corbae_2023,Fujiwara2023kagome,KarelAmorphousBerryCMG,weaire_electronic_1971,AmorphousTeDFT,AmorphousChalcogenideNatCommH}.
As we will explicitly demonstrate below and in Appendices~\ref{app:amorphousKramers},~\ref{app:amorphousCharge2}, and~\ref{app:amorphousMultifold}, each of our non-crystalline tight-binding models exhibits the same spectral and topological features across all three structural disorder implementation methods. 
This provides strong evidence that our structural disorder methods converge to the same amorphous limit, and that in this limit, the non-crystalline tight-binding models realize numerically stable amorphous chiral TSM states.

More subtly, even in non-crystalline tight-binding models consisting of point-like sites with internal spin and orbital degrees of freedom, the intersite hopping matrix elements of $T_{\alpha\beta}$ in Eq.~(\ref{eq:MainThopDef}) are also free to carry a spatially varying relationship to the global coordinate axes, giving rise to \emph{internal degree-of-freedom disorder}.
As an illustrative example, consider a non-crystalline tight-binding model for which the only terms in the Hamiltonian $\mathcal{H}$ [Eq.~(\ref{eq:Genrealspacemodel})] arise from an intersite hopping matrix:
\begin{equation}
T^{\sigma}_{\alpha\beta} = iv_{x}[{\bf d}_{\alpha\beta}\cdot\hat{x}]\sigma^{x} + iv_{y}[{\bf d}_{\alpha\beta}\cdot\hat{y}]\sigma^{y} + iv_{z}[{\bf d}_{\alpha\beta}\cdot\hat{z}]\sigma^{z}, 
\label{eq:TmatrixRotFrameMain}
\end{equation}
where ${\bf d}_{\alpha\beta}$ is defined in Eq.~(\ref{eq:dVectorDefMain}) and where each site $\alpha$ carries an internal spin-1/2 degree of freedom parameterized by the Pauli matrices $\sigma^{i}$. 
In the crystalline limit, Eq.~(\ref{eq:TmatrixRotFrameMain}) characterizes the Dresselhaus SOC hopping term in the right-handed enantiomer of the Kramers-Weyl tight-binding model [see Ref.~\cite{KramersWeyl} and Appendix~\ref{app:Kramers}], which we will shortly study below in further detail.

Though not necessarily obvious from the form of Eq.~(\ref{eq:TmatrixRotFrameMain}), even when $T^{\sigma}_{\alpha\beta}$ is implemented on a structurally disordered lattice, the resulting system in fact exhibits perfect \emph{orientational order}.  
For example, $T^{\sigma}_{\alpha\beta}$ in a structurally disordered system remains proportional to $\sigma^{x}$ for all pairs of sites $\alpha$ and $\beta$ related by ${\bf d}_{\alpha\beta}\propto \hat{x}$ in Eq.~(\ref{eq:dVectorDefMain}), despite the absence of well-defined crystallographic axes that can be aligned to the global coordinate- [\emph{i.e.} lab-] frame $x$-direction. 
The existence of an internal hopping frame in $T^{\sigma}_{\alpha\beta}$ can be seen by re-expressing Eq.~(\ref{eq:TmatrixRotFrameMain}) in the form:
\begin{equation}
T^{\sigma}_{\alpha\beta} = i{\bf d}_{\alpha\beta}^\mathsf{T}Q_{\sigma}{\bs \sigma},
\label{eq:frameExplicitMain}
\end{equation}
where:
\begin{equation}
{\bs \sigma} =  \begin{pmatrix}
\sigma^{x} \\ \sigma^{y} \\ \sigma^{z}\end{pmatrix},
\label{eq:tempSigmaVecMAIN}
\end{equation}
and
\begin{equation}
Q_{\sigma} = \begin{pmatrix}
              {\bf f}_{x} & {\bf f}_{y} & {\bf f}_{z}
          \end{pmatrix}.
\label{eq:QframeDefMain}
\end{equation}
In Eqs.~(\ref{eq:frameExplicitMain}) and~(\ref{eq:QframeDefMain}), $Q_{\sigma}$ is a row vector of three hopping frame vectors ${\bf f}_{i}$:
\begin{equation}
{\bf f}_{x} = \begin{pmatrix}
v_{x} \\ 
0 \\ 
0\end{pmatrix},\ {\bf f}_{y} = \begin{pmatrix}
0 \\ 
v_{y} \\ 
0\end{pmatrix},\ {\bf f}_{z} = \begin{pmatrix}
0 \\ 
0 \\ 
v_{z}\end{pmatrix},
\end{equation}
where each ${\bf f}_{i}$ is aligned with a global Cartesian coordinate axis.

In a realistic model of a strongly disordered solid-state material, however, no particular local region in position space should have physical properties that depend on the global coordinate frame, such as the $x$-direction in an arbitrary coordinate system.
Hence, to model real solid-state amorphous materials, we must also disorder the local internal frame orientation so that the non-crystalline model does not implicitly contain orbital hopping or SOC terms that are locked in any local region to the global coordinate frame.

In amorphous materials, the local sites predominantly exhibit bonds at the same relative angles, owing to their similar local chemical environments~\cite{Corbae_2023,Fujiwara2023kagome,KarelAmorphousBerryCMG,weaire_electronic_1971,AmorphousTeDFT,AmorphousChalcogenideNatCommH}.
To incorporate this observation into our theoretical analysis, we consider for each model and hopping interaction in this work the most general length-conserving [orthogonal] transformation that preserves the magnitude of the inner products $|{\bf f}_{i}\cdot {\bf f}_{j}|$ while also preserving the relative angles between each ${\bf f}_{i}$ [\emph{i.e.} the local ``chemistry'', see Appendix~\ref{app:DiffTypesDisorder}].
For the case of $T^{\sigma}_{\alpha\beta}$ in Eq.~(\ref{eq:frameExplicitMain}), the most general transformation that satisfies these constraints can be implemented by augmenting Eq.~(\ref{eq:frameExplicitMain}) with a spatially varying matrix $M_{\alpha\beta}$ that is a representation of the group O(3):
\begin{equation}
T^{\sigma}_{\alpha\beta} = i{\bf d}_{\alpha\beta}^\mathsf{T}M_{\alpha\beta}Q_{\sigma}{\bs \sigma}.
\label{eq:frameModificationMAIN}
\end{equation}

Tellingly, O(3) can be re-expressed as the direct product of the proper rotation group SO(3) and the group $\mathbb{Z}_{2\chi}=\left\{E,\mathcal{I}_{\chi}\right\}$ where $E$ is the identity element and $\mathcal{I}_{\chi}$ is spatial inversion in the internal coordinate space of the frame vectors ${\bf f}_{x,y,z}$ [Appendix~\ref{app:DiffTypesDisorder}].
Importantly, the elements in SO(3) rotate the frame vectors while preserving their relative \emph{handedness}: 
\begin{equation}
C_{\bf f} = \sgn[({\bf f}_{x}\times{\bf f}_{y})\cdot{\bf f}_{z}],
\label{eq:mainFrameChiralitySign}
\end{equation}
Conversely, the internal rotoinversion elements in the set $\mathcal{I}_{\chi}\text{SO}(3)$ also rotate the frame vectors, but reverse the frame handedness [$C_{\bf f}\rightarrow -C_{\bf f}$ in Eq.~(\ref{eq:mainFrameChiralitySign})].

The decomposition of O(3) into $\mathbb{Z}_{2\chi}\times\text{SO}(3)$ indicates that $M_{\alpha\beta}$ in Eq.~(\ref{eq:frameModificationMAIN}) may be represented as the product of a $\mathbb{Z}_{2}$ scalar $\chi_{\alpha\beta}=\pm 1$ and an SO(3) proper rotation matrix $R_{\alpha\beta}$:
\begin{equation}
M_{\alpha\beta} = \chi_{\alpha\beta}R_{\alpha\beta}.
\label{eq:tempMmatMain}
\end{equation}
Following the implementation of analogous local reference frame transformations in lattice gauge theory~\cite{KogutLatticeGaugeRMP}, we next choose the specific forms of $\chi_{\alpha\beta}$ and $R_{\alpha\beta}$ to be:
\begin{equation}
\chi_{\alpha\beta} = \frac{1}{2}\left(\chi_{\alpha}+\chi_{\beta}\right),\ R_{\alpha\beta} = R_{\beta}R_{\alpha}^\mathsf{T},
\label{eq:temp2siteFrameBreakdownMain}
\end{equation}
where $\chi_{\alpha}$ and $\chi_{\beta}$ are scalars that for now take the values $\pm 1$, and $R_{\alpha}$ and $R_{\beta}$ are free SO(3) matrices.
Inserting Eqs.~(\ref{eq:tempMmatMain}) and~(\ref{eq:temp2siteFrameBreakdownMain}) into Eq.~(\ref{eq:frameModificationMAIN}) then gives:
\begin{equation}
T^{\sigma}_{\alpha\beta} = i\left(\frac{\chi_{\alpha}+\chi_{\beta}}{2}\right)\left[R_{\alpha}R^\mathsf{T}_{\beta}{\bf d}_{\alpha\beta}\right]^\mathsf{T}Q_{\sigma}{\bs \sigma},
\label{eq:frameModification2MAIN}
\end{equation}
in which $\chi_{\alpha}$, $\chi_{\beta}$, $R_{\alpha}$, and $R_{\beta}$ individually vary for each site $\alpha$ and $\beta$.

The forms of $\chi_{\alpha\beta}$ and $R_{\alpha\beta}$ in Eq.~(\ref{eq:temp2siteFrameBreakdownMain}) were specifically chosen so that properties of each bond [here respectively the chirality and reference frame] are refactored into properties of each \emph{site} with equal weights given to the two sites $\alpha$ and $\beta$ participating in each bond.
For example, by defining the bond chirality $\chi_{\alpha\beta}$ as the equal-weighted average of the $\alpha$ and $\beta$ site chiralities, we can then interpret sites as ``locally chiral'' objects [such as chiral molecules], while still preserving the overall bond sign $\chi_{\alpha\beta}=\pm 1$ that distinguishes the left- and right-handed enantiomers of $T^{\sigma}_{\alpha\beta}$ in Eq.~(\ref{eq:TmatrixRotFrameMain}).

The assignment of local site chirality in Eq.~(\ref{eq:temp2siteFrameBreakdownMain}) also carries the added benefit, which we will explore later in the main text and in Appendices~\ref{app:amorphousKramers} and~\ref{app:amorphousMultifold}, that we may now also extend consideration to systems in which some -- or even most -- sites carry a vanishing local structural chirality [$\chi_{\alpha}=0$], leading to the appearance of both chiral and \emph{achiral} domains.
As extensively discussed in Appendices~\ref{app:pseudoK},~\ref{app:corepAmorphous}, and~\ref{app:models}, structurally disordered systems with $\chi_{\alpha}=0$ respect achiral average symmetry groups, and hence do not admit ${\bf p}={\bf 0}$ chiral fermions.
Therefore, by extending the analysis in Eqs.~(\ref{eq:TmatrixRotFrameMain}) through~(\ref{eq:frameModification2MAIN}) to include amorphous systems with both chiral [$\chi_{\alpha}=\pm 1$] and achiral [$\chi_{\alpha}=0$] sites, we can analyze physically motivated cases in which the structural chirality has been partially or nearly completely suppressed.
We can then examine these more realistic amorphous systems for surviving signatures of topological chiral fermions.

In the specific case of $T^{\sigma}_{\alpha\beta}$ in Eq.~(\ref{eq:frameModification2MAIN}), the orientational order previously encoded in the spatially varying O(3) matrix $M_{\alpha\beta}$ [Eq.~(\ref{eq:frameModificationMAIN})] now originates from two orientational parameters on each site $\alpha$: a continuous SO(3) \emph{frame orientation} parameterized by the matrix $R_{\alpha}$, and a $\mathbb{Z}_{2}$ Ising-spin-like discrete scalar orientational parameter $\chi_{\alpha}=\pm 1$ that corresponds to the \emph{local frame chirality} [Appendix~\ref{app:DiffTypesDisorder}]. 
A physical understanding for $R_{\alpha}$ and $\chi_{\alpha}$ may be obtained by briefly returning to the crystalline limit.
Specifically, when Eq.~(\ref{eq:frameModification2MAIN}) is realized on a spatially homogeneous crystalline lattice, $R_{\alpha}$ controls the relative orientations of the spin texture and the crystallographic axes, and $\chi_{\alpha}$ controls the handedness of hopping [Dresselhaus SOC] that breaks rotoinversion symmetries and gives rise to structural chirality [see Ref.~\cite{KramersWeyl} and Appendix~\ref{app:Kramers}].
While a relatively established concept in chemistry~\cite{LocalChiralityMoleculeReviewNatChem}, liquid crystals~\cite{KamienLubenskyChiralParameter,LocalChiralityDomainLiquidCrystal}, electromagnetic response~\cite{lindell1994electromagneticBook}, and quantum spin liquids~\cite{LocalChiralityVillain,LocalChiralityWenZee}, the notion of local chirality has received comparably less attention in the study of amorphous solids.
In the context of $T^{\sigma}_{\alpha\beta}$ for the Kramers-Weyl model [Eq.~(\ref{eq:frameModification2MAIN})], $\chi_{\alpha}$ can specifically be viewed as a $\mathcal{T}$-invariant generalization of the local spin chirality order parameter introduced in Refs.~\cite{LocalChiralityVillain,LocalChiralityWenZee} to distinguish spin-liquid ground states.

In the tight-binding models introduced and analyzed in this work [Appendix~\ref{app:models}], the hopping matrix elements in general have more complicated relationships with the local coordinate frame than $T^{\sigma}_{\alpha\beta}$ in Eq.~(\ref{eq:frameExplicitMain}). 
However, each hopping matrix element for each bond between two sites $\alpha$ and $\beta$ still enters the Hamiltonian via tensor contractions with the intersite hopping vector ${\bf d}_{\alpha\beta}$ [Eq.~(\ref{eq:dVectorDefMain})].
Hence rather than identify for each model the specific functional dependence of each hopping interaction on the local frame vectors ${\bf f}_{i}$, we instead divide the O(3) frame transformations in Eqs.~(\ref{eq:tempMmatMain}) and~(\ref{eq:temp2siteFrameBreakdownMain}) into two families of hopping matrix modifications, which independently give rise to \emph{local frame orientation disorder} and \emph{local chirality disorder} [Fig.~\ref{fig:Fig1flow}(b)].

To implement local frame disorder, we first recognize that in each model in this work [Appendix~\ref{app:models}], intersite hopping interactions enter the Hamiltonian $\mathcal{H}$ via contractions of vectors and tensors of Pauli matrices with the intersite displacement vectors ${\bf d}_{\alpha\beta}$ [Eq.~(\ref{eq:dVectorDefMain})].
In each case, the Pauli matrices that enter $\mathcal{H}$ for each bond specifically depend on the \emph{orientation} of ${\bf d}_{\alpha\beta}$.
Hence, to implement local frame disorder, we apply randomly generated local rotations using the matrix product $R_{\beta}R_{\alpha}^\mathsf{T}$ in Eqs.~(\ref{eq:temp2siteFrameBreakdownMain}) and~(\ref{eq:frameModification2MAIN}) to transform ${\bf d}_{\alpha\beta}$ into the rotated bond vector $\tilde{\bf d}_{\alpha\beta}$ in each term in $\mathcal{H}$:
\begin{equation}
{\bf d}_{\alpha\beta} \rightarrow R_{\alpha}R_{\beta}^{\mathsf{T}}{\bf d}_{\alpha\beta} \underset{def}{\equiv} \tilde{{\bf d}}_{\alpha\beta}.
\label{appeq:rotaMain}
\end{equation}

Next, to implement local chirality disorder, we recall that each of the tight-binding models studied in this work respects the symmetries of a chiral space group [defined in Appendix~\ref{app:symDefs}] when placed on a crystalline lattice.
Using this, we identify the terms in each model that in the crystalline limit break rotoinversion symmetries of the form of Eq.~(\ref{eq:MainTextRotoinversion}), and hence generate structural chirality.
Generalizing the $\mathbb{Z}_{2}$ local frame chirality introduced for $T^{\sigma}_{\alpha\beta}$ in the text preceding Eq.~(\ref{eq:frameModification2MAIN}), we then multiply a chirality- [handedness-] controlling [sub]set of the rotoinversion-symmetry-breaking hopping terms in each model by a scalar prefactor proportional to $\chi_{\alpha}+\chi_{\beta}$, in which $\chi_{\alpha}$ represents the local chirality [handedness] of each lattice site $\alpha$.

The decomposition of the O(3) local frame transformations into $\chi_{\alpha}$ and $R_{\alpha}$ in Eqs.~(\ref{eq:tempMmatMain}) and~(\ref{eq:temp2siteFrameBreakdownMain}) allows us to study systems in which the local frame orientation $R_{\alpha}$ and chirality $\chi_{\alpha}$ are separately ordered or disordered.
To implement orientational disorder in the non-crystalline tight-binding models in this work, we specifically construct \emph{frame orientation domains} in which $R_{\alpha}$ is a spatially constant random matrix drawn from a Gaussian distribution over SO(3) [or SO(2) for the double-Weyl model in Appendix~\ref{app:amorphousCharge2}], and \emph{chirality domains} in which $\chi_{\alpha}$ takes the constant values $\chi_{\alpha}=\pm 1$ [respectively corresponding to right- and left-handed sites] or $\chi_{\alpha}=0$ [corresponding to achiral sites].
Across all of the models in this work, we consistently observe non-crystalline chiral TSM states even in the absence of long-range structural and frame orientational order [\emph{i.e.} when the frame orientation domains are much smaller than the system size], provided that $\chi_{\alpha}$ exhibits long-range order [\emph{i.e.} when the majority chirality domain is comparable in scale to the system size, see Appendices~\ref{app:DiffTypesDisorder},~\ref{app:amorphousKramers},~\ref{app:amorphousCharge2}, and~\ref{app:amorphousMultifold}].
To draw connection with potential real-material realizations, we lastly note that in the language of Ref.~\cite{FlackChiral}, a physical sample lying in this topologically chiral regime of structural disorder, orientational disorder, and local chirality order is termed a disordered non-racemic [mixed enantiomeric] solid.

{
\vspace{0.2in}
\centerline{\bf C. Computing observables and topology}
\vspace{0.1in}
}
To classify a gapless system as a TSM, and to compute observables associated with the bulk nodal degeneracies, it is necessary to establish a momentum-space system description in which spectrally isolated nodal degeneracies near the Fermi energy can be assigned topological invariants [see Refs.~\cite{Armitage2018,Wieder22} and Appendix~\ref{sec:WilsonBerry}].
Unlike crystals, amorphous materials lack lattice translation symmetry, preventing the definition of an exact crystal momentum ${\bf k}$ that can be used to construct a Fourier-transformed Bloch Hamiltonian $\mathcal{H}({\bf k})$.
However, previous works have shown that amorphous systems can still exhibit well-resolved topological spectral features when characterized in terms of the plane-wave pseudo-momentum ${\bf p}$, which takes unbounded values from ${\bf 0}$ to ${\bs \infty}$, and that these features are physically observable in angle-resolved photoemission experiments~\cite{marsal_topological_2020,corbae_evidence_2020,Ciocys2023}.
Formally, ${\bf p}$ enters our calculations via the Fourier-transformed plane-wave states:
\begin{equation} 
\lvert \mathbf{p},l\rangle = \frac{1}{\sqrt{N_{\text{sites}}}}\sum_{\alpha=1}^{N_{\text{sites}}} \exp(i \mathbf{p}\cdot \mathbf{r}_\alpha)\lvert \mathbf{r}_\alpha,l\rangle,
\label{eq:planewavesMain}
\end{equation}
where $N_{\text{sites}}$ is the number of lattice sites and $\lvert \mathbf{r}_\alpha,l\rangle$ is a position-space tight-binding basis state at the site $\alpha$ with the internal degree-of-freedom index $l$ [see Appendix~\ref{app:PhysicalObservables} for complete details].
Like the crystal momentum ${\bf k}$ in a pristine system, the pseudo-momentum ${\bf p}$ in a disordered system also carries units and meaningful values that similarly derive from the units and relative values of the site positions ${\bf r}_{\alpha}$ via Eq.~(\ref{eq:planewavesMain}).

Though the plane-wave states $\lvert \mathbf{p},l\rangle$ are not exact energy eigenstates in an amorphous system, we may still define exact and physically measurable quantities using expectation values with respect to the set of $\lvert \mathbf{p},l\rangle$.
To construct a quantitative, Fourier-transformed description of the non-crystalline TSM models in this work, we begin by using the system Hamiltonian $\mathcal{H}$ [Eq.~(\ref{eq:Genrealspacemodel})] to define the real-space Green's function at a given energy $E$:
\begin{equation}
\mathcal{G}\left(E\right) = \left(\left(E + i \varepsilon\right)\mathbb{1}_{N_{\text{sites}}N_{\text{orb}}}-\mathcal{H}\right)^{-1},
\label{eq:RealSpaceGreenMain}
\end{equation} 
where $\varepsilon$ is a spectral broadening parameter whose value is set by the details of our computational methods [see Appendix~\ref{app:PhysicalObservables}].
We next use $\lvert \mathbf{p},l\rangle$ in Eq.~(\ref{eq:planewavesMain}) to Fourier-transform the position-space system Green's function $\mathcal{G}(E)$ in Eq.~(\ref{eq:RealSpaceGreenMain}):
\begin{equation}
\begin{split}
    \left[\mathcal{G}(E,\mathbf{p},\mathbf{p}')\right]^{ll'} &= \left<\mathbf{p},l\right|\mathcal{G}(E)|\mathbf{p}',l'\rangle,
\end{split}
\label{eq:MomGreenFuncMain}
\end{equation}
where $\mathcal{G}(E,{\bf p},{\bf p}')$ is the exact, [pseudo-] momentum-resolved $N_{\text{orb}}\times N_{\text{orb}}$ matrix Green's function [the reduction from the $N_{\text{sites}}N_{\text{orb}}\times N_{\text{sites}}N_{\text{orb}}$ matrix $\mathcal{G}(E)$ to the $N_{\text{orb}}\times N_{\text{orb}}$ matrix $\mathcal{G}(E,{\bf p},{\bf p}')$ in Eq.~(\ref{eq:MomGreenFuncMain}) is explicitly detailed in Appendix~\ref{app:PhysicalObservables}].
Unlike in a crystal, $\mathcal{G}(E,{\bf p},{\bf p}')$ in Eq.~(\ref{eq:MomGreenFuncMain}) generically carries both diagonal-in-momentum [${\bf p}={\bf p}'$] and off-diagonal-in-momentum [${\bf p}\neq{\bf p}'$] matrix elements in a structurally disordered system.

In this work, we seek to model solid-state amorphous materials that have thermodynamically large numbers of atoms.
However, our computational resources limit calculations to relatively small systems [$\sim 20^{3}$ sites].
Hence, to generate a momentum-space Green's function that is more representative of a real solid-state amorphous material, we construct a two-momentum \emph{average Green's function} $\bar{\mathcal{G}}(E,\mathbf{p},\mathbf{p'})$ by averaging the Fourier-transformed matrix Green's function $\mathcal{G}_{i}(E,{\bf p},{\bf p}')$ in Eq.~(\ref{eq:MomGreenFuncMain}) over an ensemble of $20-50$ \emph{replicas}~\cite{ParisiReplicaCourse} that each contain independently generated structural and internal degree-of-freedom disorder and are indexed below by $i$ [Fig.~\ref{fig:Fig1flow}(c)]:
\begin{equation}
\bar{\mathcal{G}}(E,\mathbf{p},\mathbf{p'}) = \frac{1}{N_{\text{rep}}}\sum_{i=1}^{N_{\text{rep}}} \mathcal{G}_{i}(E,\mathbf{p},\mathbf{p'}),
\label{eq:averageTWoMomentumGreenMain}
\end{equation}
where $N_{\text{rep}}$ is the number of replicas, and where the same structural disorder implementation scheme is used for each replica in the ensemble [Fig.~\ref{fig:Fig1flow}(b) and Appendix~\ref{app:DiffTypesDisorder}].
Physical observables and topological invariants extracted from $\bar{\mathcal{G}}(E,\mathbf{p},\mathbf{p'})$ in Eq.~(\ref{eq:averageTWoMomentumGreenMain}) can in turn provide improved estimates of self-averaging spectral and topological properties in amorphous systems, such as the spectral function~\cite{corbae_evidence_2020,Ciocys2023}.

Though the exact, replica-averaged, momentum-resolved Green's function $\bar{\mathcal{G}}(E,\mathbf{p},\mathbf{p'})$ in Eq.~(\ref{eq:averageTWoMomentumGreenMain}) strictly remains a function of two pseudo-momenta ${\bf p}$ and ${\bf p}'$, we find by direct computation that for the tight-binding models studied in this work, the off-diagonal-in-momentum elements of $\bar{\mathcal{G}}(E,\mathbf{p},\mathbf{p'})$ become much smaller than the diagonal-in-momentum elements for large-$\eta$ Gaussian disorder and random lattices when $N_{\text{rep}}\geq 20$ in Eq.~(\ref{eq:averageTWoMomentumGreenMain}) [see Appendix~\ref{app:PhysicalObservables}].
Using this observation, we then quantitatively define \emph{strong structural disorder} as the model regime [\emph{e.g} the value of $\eta$ in Eq.~(\ref{eq:GaussianDisorderMain})] in which the off-diagonal-in-momentum matrix elements of $\bar{\mathcal{G}}(E,\mathbf{p},\mathbf{p'})$ vanish.
In the case of Gaussian structural disorder, we specifically find that for all of the models in this work, the onset of strong structural disorder occurs in the vicinity of $\eta\approx 0.5$.
For ensembles of structurally disordered models lying in the regime of strong structural disorder, we may therefore approximately restrict consideration to a one-momentum, Fourier-transformed average matrix Green's function $\bar{\mathcal{G}}(E,{\bf p})$ constructed from the diagonal-in-momentum elements of $\bar{\mathcal{G}}(E,\mathbf{p},\mathbf{p'})$ [Fig.~\ref{fig:Fig1flow}(c)]:
\begin{equation}
\bar{\mathcal{G}}(E,\mathbf{p}) = \frac{1}{N_{\text{rep}}}\sum_{i=1}^{N_{\text{rep}}} \mathcal{G}_{i}(E,\mathbf{p},\mathbf{p}).
\label{eq:averageOneMomentumGreenMain}
\end{equation}

Using $\bar{\mathcal{G}}(E,{\bf p})$ in Eq.~(\ref{eq:averageOneMomentumGreenMain}), we may then compute experimental observables in amorphous systems -- such as the spectral function and spin and orbital angular momentum textures -- that can be contrasted with analogous quantities in crystals [Fig.~\ref{fig:Fig1flow}(d)].
To begin, previous works, have shown that the disorder-averaged, one-momentum Green's function $\bar{\mathcal{G}}(E,\mathbf{p})$ in Eq.~(\ref{eq:averageOneMomentumGreenMain}) can be used to define a disorder-averaged, momentum-resolved spectral function $\bar{A}(E,\mathbf{p})$ in a non-crystalline system [Fig.~\ref{fig:Fig1flow}(d), see Refs.~\cite{marsal_topological_2020,corbae_evidence_2020,Ciocys2023} and Appendix~\ref{app:PhysicalObservables}]:
\begin{equation}
\bar{A}(E,\mathbf{p}) = -\frac{1}{\pi}\text{Im}\left\{\Tr\left[\bar{\mathcal{G}}\left(E,\mathbf{p}\right)\right]\right\}.
\label{eq:SpecFuncMain}
\end{equation}

Unlike in a crystal, however, $\bar{A}(E,\mathbf{p})$ in a structurally disordered system exhibits a ${\bf p}$-dependent sharpness, because the pseudo-momentum ${\bf p}$ is only an approximate system quantum number because of the absence of lattice translation symmetry.
In a non-crystalline system with strong structural disorder, $\bar{A}(E,\mathbf{p})$ is specifically sharpest at the smallest values of ${\bf p}$.
This can be understood by recognizing that the ${\bf p}\approx {\bf 0}$ spectrum characterizes the system at its smallest momenta and hence longest wavelengths, and that strongly disordered [\emph{e.g.} amorphous] systems can exhibit approximate continuous translation symmetry when viewed on length scales much longer than the average-nearest-neighbor intersite spacing [Appendix~\ref{app:pseudoK}].
Together, this implies the possibility of a well-defined dispersion relation in an amorphous system near ${\bf p}={\bf 0}$ that is independent of short-wavelength model details.

However at larger $|{\bf p}|$, and hence shorter wavelengths, the absence of continuous and exact lattice translation symmetry has a stronger effect on the energy spectrum, leading to significant disorder broadening in $\bar{A}(E,\mathbf{p})$ [see Appendices~\ref{app:pseudoK} and~\ref{app:PhysicalObservables}].
Large-$|{\bf p}|$ spectral features in disordered tight-binding models are also in general model-dependent, because they can depend on the details of the spatially decaying tails of the underlying tight-binding basis states and the choice of model disorder implementation scheme.
For these reasons, we will throughout this work largely limit our analysis to spectral features at small-to-moderate values of $|{\bf p}|$.

Next, in addition to their spectral functions [dispersion relations], condensed-matter chiral fermions are also frequently characterized through a combination of their spin~\cite{KramersWeyl,chang2017large} and OAM~\cite{OAMmultifold2,OAMmultifold3} textures, where here and throughout this work, we use OAM to refer to the ``local'' contribution to the OAM from the atomic orbital basis states [see Appendices~\ref{app:amorphousCharge2} and~\ref{app:amorphousMultifold}].
To define the angular momentum texture in a disordered system, we first note that in real materials and more complicated tight-binding model with large numbers of bands, neither the total angular momentum ${\bf J}$, nor the spin ${\bf S}$, nor the OAM ${\bf L}$ of a momentum-space energy eigenstate is generically quantized, due to the combined effects of SOC and OAM terms and interband matrix elements~\cite{KramersWeyl,chang2017large}.
For example, just as $S^z$ symmetry is generically broken in real materials by SOC~\cite{HasanKaneReview,WenZooReview}, the atomic OAM component $L^z$ is similarly non-conserved in realistic models with interband matrix elements.
Nevertheless, for a region in energy and momentum space with high spectral weight [\emph{i.e.} a large spectral function $\bar{A}(E,{\bf p})$ corresponding to Bloch eigenstates in the crystalline limit], we may still determine the \emph{degree} of ${\bf J}$, ${\bf S}$, or ${\bf L}$ polarization -- as well as the angular momentum orientations [textures] of highly polarized states -- by computing angular-momentum-dependent spectral functions [Fig.~\ref{fig:Fig1flow}(d)].  
To compute the spin texture in models with internal spin-1/2 degrees of freedom, we specifically construct the spin-dependent spectral function vector $\langle\mathbf{S}(E,\mathbf{p})\rangle$, whose components are given by:
\begin{equation}
\left\langle S^{i} \left(E,\mathbf{p}\right)\right\rangle = -\frac{1}{\pi} \text{Im}\left\{\text{Tr}\left[\hat{S}^{i} \bar{\mathcal{G}}(E,\mathbf{p})\right]\right\},
\label{eq:SpinDOSMain}
\end{equation}
where $\bar{\mathcal{G}}(E,{\bf p})$ is the disorder- [replica-] averaged momentum-resolved matrix Green's function [Eq.~(\ref{eq:averageOneMomentumGreenMain})] and $\hat{S}^{i}$ is the $i$-th component of the spin operator vector $\hat{\mathbf{S}}=(\hat{S}^x,\hat{S}^y,\hat{S}^z)$ [see Ref.~\cite{chang2017large} and Appendix~\ref{app:amorphousKramers}].
Similarly, to compute the OAM texture in models with internal orbital degrees of freedom, we construct an OAM-dependent spectral function vector $\langle\mathbf{L}(E,\mathbf{p})\rangle$, whose components are given by:
\begin{equation}
\left\langle L^i \left(E,\mathbf{p}\right)\right\rangle = -\frac{1}{\pi}\text{Im}\left\{\text{Tr}\left[\hat{L}^i \bar{\mathcal{G}}(E,\mathbf{p})\right]\right\},
\label{eq:OAMDOSMain}
\end{equation}
where $\bar{\mathcal{G}}(E,{\bf p})$ is defined in Eq.~(\ref{eq:averageOneMomentumGreenMain}) and $\hat{L}^i$ is the $i$-th component of the OAM operator vector $\hat{\bf L}=(\hat{L}^x,\hat{L}^y,\hat{L}^z)$ [see Appendices~\ref{app:amorphousCharge2} and~\ref{app:amorphousMultifold}].

Lastly, to precisely determine if energetically isolated spectral features in $\bar{A}(E,\mathbf{p})$ [Eq.~(\ref{eq:SpecFuncMain})] are non-crystalline chiral fermions, it is necessary to introduce pseudo-momentum-space topological invariants.
Previous works have shown that the sharpest features in $\bar{A}(E,{\bf p})$ and the global bulk topology of gapped amorphous systems can be accurately approximated by constructing a single-particle \emph{effective Hamiltonian} $\mathcal{H}_\text{Eff}({\bf p})$ using the [average] one-momentum Green's function matrix $\bar{\mathcal{G}}(E,\mathbf{p})$~\cite{varjas_topological_2019,marsal_topological_2020,corbae_evidence_2020}. 
The effective Hamiltonian $\mathcal{H}_\text{Eff}({\bf p})$ was specifically first introduced in Ref.~\cite{varjas_topological_2019} to diagnose topology in 2D quasicrystals, and was then adapted to amorphous systems in Refs.~\cite{marsal_topological_2020,corbae_evidence_2020}.
For non-crystalline tight-binding models that consist only of sites with the same $N_{\text{orb}}$ internal spin and orbital degrees of freedom -- like those analyzed in this work [see Appendices~\ref{app:lattices} and~\ref{app:models} and the text surrounding Eq.~(\ref{eq:Genrealspacemodel})] -- we define the $N_{\text{orb}}\times N_{\text{orb}}$ Hermitian effective Hamiltonian matrix $\mathcal{H}_\text{Eff}({\bf p})$ by restricting $\bar{\mathcal{G}}(E,\mathbf{p})$ in Eq.~(\ref{eq:averageOneMomentumGreenMain}) to a reference energy cut $E_{C}$ and computing:
\begin{equation}
\mathcal{H}_{\text{Eff}}(\mathbf{p}) = -\frac{1}{2}\left(\bar{\mathcal{G}}^{-1}(E_C,\mathbf{p})  + \left[\bar{\mathcal{G}}^{-1}(E_C,\mathbf{p}) \right]^{\dagger}\right) - E_{\text{C}}.
\label{eq:AvgHEffMain}
\end{equation}

Though it is often notationally suppressed, $\mathcal{H}_\text{Eff}({\bf p})$ is in general highly dependent on the choice of $E_{C}$ in Eq.~(\ref{eq:AvgHEffMain}).
However for the non-crystalline models studied in this work, we find that $\mathcal{H}_\text{Eff}({\bf p})$ exhibits the same low-energy, small-$|{\bf p}|$ [${\bf k}\cdot {\bf p}$] polynomial dispersion relation and [non-Abelian] Berry phases over a wide range of $E_{C}$ [see Appendix~\ref{app:EffectiveHamiltonian} for explicit $\mathcal{H}_\text{Eff}({\bf p})$ benchmarking calculation details].
Together, this suggests that mean-field topological invariants computed from the eigenstates of $\mathcal{H}_\text{Eff}({\bf p})$ in the local vicinity of non-crystalline ${\bf p}={\bf 0}$ nodal degeneracies are numerically stable and can accurately reproduce the full many-particle system topology at small $|{\bf p}|$. 
Conversely at larger $|{\bf p}|$, the diffuse spread in energy of the matrix elements of $\bar{\mathcal{G}}(E,\mathbf{p})$, combined with the possible presence of zero eigenvalues, suggests that caution is warranted in attributing $\mathcal{H}_\text{Eff}({\bf p})$ in Eq.~(\ref{eq:AvgHEffMain}) to physical topological features.
The large-$|{\bf p}|$ behavior of $\bar{\mathcal{G}}(E,\mathbf{p})$ and the potential failure of $\mathcal{H}_\text{Eff}({\bf p})$ to diagnose physical topological invariants at larger $|{\bf p}|$ is in this sense reminiscent of similar limitations of the single-particle Green's function in characterizing strongly correlated systems~\cite{MottFailureBradlynPhilipps,MottFailureGoldman}.

Having found that ${\bf p}={\bf 0}$ spectral features in non-crystalline systems can be well-captured by mean-field effective Hamiltonians [Eq.~(\ref{eq:AvgHEffMain})], we next establish a numerical method for evaluating non-crystalline nodal topology.
We begin by recalling that in crystalline nodal semimetals, the most general tool for identifying TSM states is the non-Abelian Wilson loop [holonomy] matrix~\cite{AndreiXiZ2,DiracInsulator,TMDHOTI,Z2Pack,Wieder22} evaluated on closed contours surrounding each bulk nodal degeneracy.
In particular, for point-like nodal degeneracies in 3D crystals, the topological chiral charge $C$ [Eq.~(\ref{eq:MainChiralCharge})] is nontrivial if the Wilson loop eigenvalues on a sphere surrounding the nodal point exhibit a nontrivial winding number~\cite{Z2Pack}.

Motivated by the numerical stability of the dispersion relation and Berry phases of $\mathcal{H}_\text{Eff}({\bf p})$ near ${\bf p}= {\bf 0}$ [see Appendix~\ref{app:EffectiveHamiltonian} and the text surrounding Eq.~(\ref{eq:AvgHEffMain})], we introduce an \emph{amorphous Wilson loop method} in which the eigenstates of $\mathcal{H}_{\text{Eff}}(\mathbf{p})$ are used to infer the chiral charges of many-particle nodal degeneracies at ${\bf p}={\bf 0}$ in strongly disordered 3D TSMs [Fig.~\ref{fig:Fig1flow}(e)].
To begin, we first construct the effective Hamiltonian matrix $\mathcal{H}_{\text{Eff}}(\mathbf{p})$ using Eq.~(\ref{eq:AvgHEffMain}) with $E_{C}$ set to lie at the largest spectral weight at ${\bf p}={\bf 0}$, which we find to maximize the qualitative spectral accuracy of $\mathcal{H}_\text{Eff}({\bf p})$ [see Appendices~\ref{app:EffectiveHamiltonian},~\ref{app:amorphousKramers},~\ref{app:amorphousCharge2}, and~\ref{app:amorphousMultifold}].
This choice of $E_{C}$ in $\mathcal{H}_{\text{Eff}}(\mathbf{p})$ also allows sensitivity to rare regions that could give rise to a finite density of states exactly at the nodal point of a disordered chiral fermion, potentially affecting the quantization of its topological chiral charge~\cite{Pixley2021} [though in our calculations, we do not observe variation in topological invariants computed from $\mathcal{H}_{\text{Eff}}(\mathbf{p})$ as $E_{C}$ is varied].
We then diagonalize $\mathcal{H}_{\text{Eff}}(\mathbf{p})$ at each pseudo-momentum point ${\bf p}[\theta]$ at the polar angle $\theta$ on a sphere surrounding ${\bf p}={\bf 0}$ [blue transparent sphere in Fig.~\ref{fig:Fig1flow}(e)].
Next, we construct the matrix projector onto the $N_{\text{occ}}$ occupied eigenstates $|u^{j}_{{\bf p}[\theta]}\rangle$ of $\mathcal{H}_{\text{Eff}}(\mathbf{p})$:
\begin{equation}
P_{\bf p}[\theta] = \sum_{j=1}^{N_{\text{occ}}}|u^{j}_{{\bf p}[\theta]}\rangle\langle u^{j}_{{\bf p}[\theta]}|.
\label{eq:OccupiedProjectorWilsonTEMP}
\end{equation}

The topological chiral charge of a ${\bf p}={\bf 0}$ nodal degeneracy can be determined using the eigenvalues of the Wilson loop matrix $W[\theta]$ [non-Abelian Berry phases] computed on a series of counterclockwise parallel curves indexed by the polar angle $\theta$ of a pseudo-momentum sphere enclosing ${\bf p}={\bf 0}$~\cite{Z2Pack,Armitage2018}.
To obtain the $N_{\text{occ}}\times N_{\text{occ}}$ unitary Wilson loop matrix $W[\theta]$ for each parallel closed contour on the sphere, we compute the path-ordered product of projectors $P_{\bf p}[\theta]$ [Eq.~(\ref{eq:OccupiedProjectorWilsonTEMP})] at each polar angle $\theta$ restricted to the image of the $N_{\text{occ}}$ occupied eigenstates of $\mathcal{H}_{\text{Eff}}(\mathbf{p})$, such that the matrix elements of $W[\theta]$ are given by:
\begin{equation}
\bigg[W[\theta]\bigg]^{jk} = \langle u^{j}_{{\bf p}_{0}[\theta]}|\left(\mathcal{P}_{\theta}\prod_{\bf p[\theta]}P_{\bf p}[\theta]\right)|u^{k}_{{\bf p}_{0}[\theta]}\rangle,
\label{eq:WilsonElementsMain}
\end{equation}
where ${\bf p}_{0}[\theta]$ is the Wilson loop base point and $\mathcal{P}_{\theta}$ denotes the path-ordered product along the counterclockwise blue contour at $\theta$ in Fig.~\ref{fig:Fig1flow}(e) [see Appendix~\ref{sec:WilsonBerry} for complete details].

The eigenvalues of the unitary Wilson loop matrix $W[\theta]$ in Eq.~(\ref{eq:WilsonElementsMain}) take the form of $N_{\text{occ}}$ gauge-invariant complex phases $\exp(i\gamma_{n}[\theta])$ that are smooth and continuous functions of $\theta$ for energetically isolated nodal degeneracies at ${\bf p}={\bf 0}$. 
For the set of Wilson loop eigenvalues $\{\exp(i\gamma_{n}[\theta])\}$, the phase angles $\{\gamma_{n}[\theta]\}$ are well-defined modulo $2\pi$ and are termed the non-Abelian Berry phases.
Finally, the winding number of the set of non-Abelian Berry phases $\{\gamma_{n}[\theta]\}$ on a sphere surrounding a nodal point at ${\bf p}={\bf 0}$ corresponds to the surface integral of the Berry curvature flux [Eq.~(\ref{eq:MainChiralCharge})], and hence indicates the quantized topological chiral charge $C\in\mathbb{Z}$ of the nodal degeneracy~\cite{Z2Pack}.
We therefore in this work identify ${\bf p}={\bf 0}$ nodal spectral features with $C\neq 0$ in strongly disordered tight-binding models as condensed-matter realizations of 3D amorphous chiral fermions.

\begin{figure*}[t]
\centering
\includegraphics[width=\linewidth]{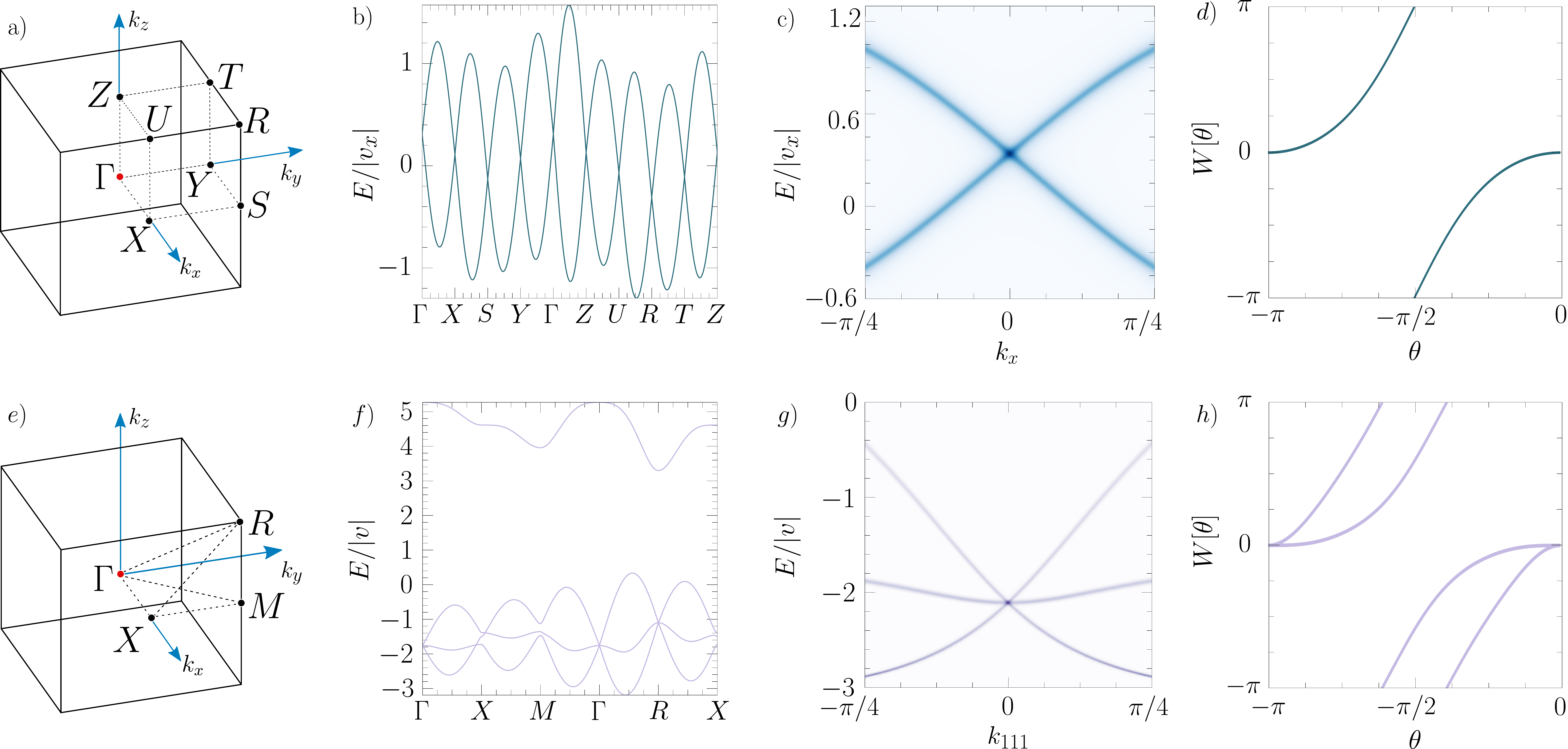}
\caption{{\bf Kramers-Weyl and multifold fermions in the crystalline limit.}
(a) The Brillouin zone [BZ] of chiral space group 16 ($P222$), the orthorhombic space group of the Kramers-Weyl model in the crystalline limit [Eq.~(\ref{eq:HKWMaink}), see Appendices~\ref{app:symDefs} and~\ref{app:PristineKramers} and Ref.~\cite{KramersWeyl}].
(b) Bulk band structure of the right-handed [$R$] enantiomer of the crystalline Kramers-Weyl model.
The spectrum at each time-reversal-invariant momentum [TRIM] point in (b) consists of a twofold nodal degeneracy with linear dispersion [Eq.~(\ref{eq:mainHKP})], and corresponds to a Kramers-Weyl chiral fermion with chiral charge $|C|=1$ [Eq.~(\ref{eq:MainChiralCharge})].
(c) The spectral function $A(E,\mathbf{p})$ [Appendix~\ref{app:PhysicalObservables}] of the Kramers-Weyl fermion at the $\Gamma$ point [${\bf k}={\bf 0}$] in (b).
(d) The Abelian Wilson loop spectrum [Eq.~(\ref{eq:WilsonElementsMain}) and Appendix~\ref{sec:WilsonBerry}] of the lower band, computed on a sphere surrounding the nodal degeneracy at the $\Gamma$ point in (b,c).
The Wilson loop spectrum in (d) winds once in the positive direction, indicating that the Kramers-Weyl fermion at $\Gamma$ carries a chiral charge of $C=1$ for the $R$ model enantiomer.  
In the $L$ model enantiomer, the chiral charge of the Kramers-Weyl fermion at $\Gamma$ is instead $C=-1$, and is therefore directly determined by the lattice-scale structural chirality $C^{\mathrm{KW}}_{\mathcal{H}}$ [Eq.~(\ref{eq:mainChiralCharge}), see Appendix~\ref{app:PristineKramers} and Ref.~\cite{KramersWeyl}].
(e) The BZ of chiral space group 195 ($P23$), the cubic space group of the symmorphic multifold fermion model introduced in this work tuned to the crystalline limit [Eq.~(\ref{eq:HamBloch3FMain}), see Appendices~\ref{app:symDefs} and~\ref{app:PristineMulifold}]. 
(f) Bulk band structure of the $R$ enantiomer of the multifold fermion model.
In (f), the lower three bands meet in symmetry-enforced threefold nodal degeneracies at the TRIM points $\Gamma$ and $R$ that correspond to spin-1 chiral multifold fermions~\cite{ManesNewFermion,NewFermions,chang2017large,tang2017CoSi,KramersWeyl}.
(g) The spectral function of the multifold fermion at the $\Gamma$ point in (f), plotted along $k_{111} = (1/\sqrt{3})(k_{x}+k_{y}+k_{z})$.
(h) The non-Abelian Wilson loop spectrum of the lower two bands, computed on a sphere surrounding the threefold degeneracy at the $\Gamma$ point.
In (h), the Wilson loop eigenvalues as a set wind twice in the positive direction, indicating that the lower two bands of the multifold fermion at $\Gamma$ together carry a chiral charge of $C=2$.
In the $L$ model enantiomer, the lower two bands of the multifold fermion at $\Gamma$ instead carry a chiral charge of $C=-2$ [Appendix~\ref{app:PristineMulifold}].
Like the Kramers-Weyl fermions in (b,c) and the chiral multifold fermions experimentally studied in Refs.~\cite{AlPtObserve,PdGaObserve}, the spin-1 chiral fermion at $\Gamma$ in (f,g) therefore similarly carries a low-energy topological chirality that is inherited from and controlled by the lattice-scale structural chirality $C^{\mathrm{3F}}_{\mathcal{H}}$ [Eq.~(\ref{eq:multifoldStructuralChiMain})].
Numerical details for panels (b-d) and (f-h) are provided in Appendices~\ref{app:PristineKramers} and~\ref{app:PristineMulifold}, respectively.}
\label{fig:MainBands}
\end{figure*}

{
\vspace{0.2in}
\centerline{\bf III. AMORPHOUS CHIRAL FERMIONS}
\vspace{0.1in}
}

{
\centerline{\bf A. Kramers-Weyl and multifold fermion models}
\vspace{0.1in}
}
Having established our numerical workflow, we next introduce chiral TSM tight-binding models that can be realized both in the crystalline limit and with tunable structural and orientational disorder.
We begin by considering the intuitive example of the Kramers-Weyl model, which consists of a single Kramers pair of spinful [spin-1/2] $s$ orbitals on each lattice site coupled by nearest-neighbor spinless [orbital] hopping and Dresselhaus SOC [see Appendix~\ref{app:Kramers} and Ref.~\cite{KramersWeyl}].
When allowing for structural, local frame orientation, and local chirality disorder [Fig.~\ref{fig:Fig1flow}(b) and Appendix~\ref{app:DiffTypesDisorder}], the crystalline and non-crystalline Kramers-Weyl models lack on-site mass matrices [$M_{\alpha}=\mathbb{0}$ in the absence of Anderson disorder, see Eq.~(\ref{eq:MainMDef}) and the following text], and are instead fully specified by the intersite hopping matrix:
\begin{align}
T^{\mathrm{KW}}_{\alpha\beta} &= \frac{1}{2}f(|\mathbf{d}_{\alpha\beta}|)\ \times \nonumber \\
&\bigg(i\left(\frac{\chi_{\alpha}+\chi_{\beta}}{2}\right)\left(\tilde{\mathbf{d}}_{\alpha\beta}\right)^\mathsf{T}V\bm{\sigma} + \left[\left(\tilde{\mathbf{d}}_{\alpha\beta}\right)^\mathsf{T}Q_{t}\tilde{\mathbf{d}}_{\alpha\beta}\right]\sigma^0\bigg), \nonumber \\
\label{eq:amorphousKWTmatrixFinalChiralityMain}
\end{align}
where ${\bs \sigma}$ is a vector of $2\times 2$ Pauli matrices that act on the internal spin-1/2 degrees of freedom [Eq.~(\ref{eq:tempSigmaVecMAIN})], $\sigma^{0}$ is the $2\times 2$ identity matrix in spin space, $\chi_{\alpha}$ [$\chi_{\beta}$] is the local chirality of the site $\alpha$ [$\beta$] detailed in Eq.~(\ref{eq:temp2siteFrameBreakdownMain}) and the surrounding text, and $\tilde{\mathbf{d}}_{\alpha\beta}$ is the SO(3)-rotated bond vector defined in Eq.~(\ref{appeq:rotaMain}).
The scalar prefactor $f(|\mathbf{d}_{\alpha\beta}|)$ in Eq.~(\ref{eq:amorphousKWTmatrixFinalChiralityMain}) is a hopping rangedness function that rescales the hopping magnitude -- but not its sign or the relative values of its matrix elements -- based on the intersite distance $|\tilde{\bf d}_{\alpha\beta}|=|{\bf d}_{\alpha\beta}|$:
\begin{equation}
f(|{\bf d}_{\alpha\beta}|) = \Theta(R_0-|{\bf d}_{\alpha\beta}|)\exp(a-|{\bf d}_{\alpha\beta}|),
\label{eq:KWHeavisideMain}
 \end{equation}
where $\Theta$ denotes the Heaviside function, $a=1$ is a length that becomes the lattice spacing in each of the $x,y,z$ directions in the crystalline limit, and $R_0$ is the maximum allowed hopping distance [see Appendix~\ref{app:PristineKramers} and Ref.~\cite{agarwala_topological_2017}].
Lastly, in Eq.~(\ref{eq:KWHeavisideMain}), $V$ and $Q_{t}$ are $3\times 3$ matrices of hopping coefficients that respectively encode the nearest-neighbor Dresselhaus SOC and $s$-orbital-like hopping:
\begin{equation}
V = \begin{pmatrix}
              v_x&0&0\\
              0&v_y&0\\
              0&0&v_z
    \end{pmatrix},\ Q_{t} = \begin{pmatrix}
              t_x&0&0\\
              0&t_y&0\\
              0&0&t_z          
    \end{pmatrix}.
\label{eq:KWmainMats}
\end{equation}

In the crystalline limit in which the sites $\alpha$ form a regular orthorhombic lattice with a spatially homogeneous hopping reference frame [$R_{\alpha}=\mathds{1}$ for all $\alpha$ and $\chi_{\alpha}=\chi_{\beta}$ for all pairs of sites $\alpha$ and $\beta$], the Kramers-Weyl Hamiltonian [Eqs.~(\ref{eq:amorphousKWTmatrixFinalChiralityMain}),~(\ref{eq:KWHeavisideMain}), and~(\ref{eq:KWmainMats})] takes the form of a $2\times 2$ Bloch Hamiltonian in ${\bf k}$-space:
\begin{equation}
\mathcal{H}^{\mathrm{KW}}({\bf k})= \sum_{i=x,y,z}t_{i}\cos(k_i)\sigma^0 + \sum_{i=x,y,z}v_{i}\sin(k_i)\sigma^i,
\label{eq:HKWMaink}
\end{equation}
where $\chi_{\alpha}$ has been absorbed into the definitions of the Dresselhaus SOC hopping parameters $v_{i}$ to match the form of the Kramers-Weyl Bloch Hamiltonian in Ref.~\cite{KramersWeyl}.
The lattice-scale structural chirality [handedness] $C^{\mathrm{KW}}_{\mathcal{H}}$ of $\mathcal{H}^{\mathrm{KW}}({\bf k})$ in Eq.~(\ref{eq:HKWMaink}) is hence governed by the signs of the $v_{i}$ parameters~\cite{KramersWeyl}:
\begin{equation}
C^{\mathrm{KW}}_{\mathcal{H}} = \text{sgn}\left(\prod_{i=x,y,z} v_{i}\right),
\label{eq:appStructuralChiralityMain}
\end{equation}
such that $C^{\mathrm{KW}}_{\mathcal{H}}=1$ corresponds to the right-handed [$R$] model enantiomer and $C^{\mathrm{KW}}_{\mathcal{H}}=-1$ corresponds to the left-handed [$L$] enantiomer.
When the $t_{i}$ or $v_{i}$ parameters are chosen to be different in the $i=x,y,z$ directions, and when $v_{i}\neq 0$ for all $i$ [\emph{i.e.} SOC is nonvanishing in all directions], both enantiomers of $\mathcal{H}^{\mathrm{KW}}({\bf k})$ respect the symmetries of nonmagnetic chiral space group 16 ($P222$), which is generated by twofold rotations around the $i=x,y,z$ axes $C_{2i}$, $\mathcal{T}$ symmetry, and 3D orthorhombic lattice translations [see Appendices~\ref{app:symDefs} and~\ref{app:PristineKramers}].

In Fig.~\ref{fig:MainBands}(a,b), we plot the band structure of the right-handed enantiomer of $\mathcal{H}^{\mathrm{KW}}({\bf k})$, and in Fig.~\ref{fig:MainBands}(c) we plot the spectral function $A(E,\mathbf{p})$ at the $\Gamma$ point [see Appendix~\ref{app:PristineKramers} for numerical model parameters].
When $v_{i}\neq 0$ for all $i=x,y,z$, $\mathcal{H}^{\mathrm{KW}}({\bf k})$ at half filling realizes a chiral TSM state with eight conventional [$|C|=1$] Weyl points that are pinned to the eight TRIM points.
The twofold nodal degeneracies at the TRIM points in Fig.~\ref{fig:MainBands}(b,c) can therefore also be interpreted as Kramers pairs whose gaplessness is protected by spinful $\mathcal{T}$, and were hence termed Kramers-Weyl fermions in Ref.~\cite{KramersWeyl}.
For this reason, unlike conventional Weyl fermions at generic crystal momenta, the Kramers-Weyl points in Fig.~\ref{fig:MainBands}(b,c) can further be labeled by little group small coreps, and specifically transform in two-dimensional, double-valued [spinful] small coreps of the TRIM-point little groups [see Refs.~\cite{KramersWeyl,chang2017large} and Appendices~\ref{app:corepDefs} and~\ref{app:PristineKramers}].

At each TRIM point ${\bf k}_{\mathcal{T}}$ in the Kramers-Weyl Bloch Hamiltonian [Eq.~(\ref{eq:HKWMaink})], the low-energy ${\bf k}\cdot {\bf p}$ Hamiltonian $\mathcal{H}_{{\bf k}_{\mathcal{T}}}({\bf q})$ takes the canonical form of a $|C|=1$ Weyl fermion~\cite{AshvinWeyl,KramersWeyl}:
\begin{equation}
\mathcal{H}_{{\bf k}_{\mathcal{T}}}({\bf q}) = \sum_{i=x,y,z} e^{i\left({\bf k}_{\mathcal{T}}\cdot \hat{\bf r}_{i}\right)}v_{i}q_{i}\sigma^{i},
\label{eq:mainHKP}
\end{equation}
where ${\bf q} \approx {\bf k} - {\bf k}_{\mathcal{T}}$ and $\hat{\bf r}_{i}$ is a unit vector in the $i=x,y,z$ direction.
As detailed in Ref.~\cite{KramersWeyl} and Appendix~\ref{app:PristineKramers}, the chiral charge $C_{{\bf k}_{\mathcal{T}}}$ of the Kramers-Weyl fermion at each TRIM point [Eq.~(\ref{eq:MainChiralCharge})] is hence given by:
\begin{equation}
C_{{\bf k}_{\mathcal{T}}} = \text{sgn}\left(\prod_{i=x,y,z} e^{i\left({\bf k}_{\mathcal{T}}\cdot \hat{\bf r}_{i}\right)}v_{i}\right).
\label{eq:imageProofKWChargeMain}
\end{equation}

Most importantly, combining Eq.~(\ref{eq:imageProofKWChargeMain}) with the lattice-scale structural chirality $C^{\mathrm{KW}}_{\mathcal{H}}$ in Eq.~(\ref{eq:appStructuralChiralityMain}) produces the following relation:
\begin{equation}
C_{{\bf k}_{\mathcal{T}}} = \left(\prod_{i=x,y,z} e^{i\left({\bf k}_{\mathcal{T}}\cdot \hat{\bf r}_{i}\right)}\right)C^{\mathrm{KW}}_{\mathcal{H}}.
\label{eq:mainChiralCharge}
\end{equation}
As theoretically established in Ref.~\cite{KramersWeyl} and subsequently experimentally demonstrated in Refs.~\cite{AlPtObserve,PdGaObserve}, Eq.~(\ref{eq:mainChiralCharge}) represents the simplest example of the statement that the low-energy topological chirality of nodal degeneracies in structurally chiral metals is directly inherited from and controlled by the lattice-scale geometry [structural chirality].
In Eq.~(\ref{eq:mainChiralCharge}), the chiral charges $C_{{\bf k}_{\mathcal{T}}}$ of the eight Kramers-Weyl fermions are specifically controlled by the lattice-scale structural chirality $C^{\mathrm{KW}}_{\mathcal{H}}$ through the signs of the $v_{i}$ SOC parameters in Eqs.~(\ref{eq:amorphousKWTmatrixFinalChiralityMain}),~(\ref{eq:KWmainMats}), and~(\ref{eq:HKWMaink}).

To numerically confirm Eq.~(\ref{eq:mainChiralCharge}), we have computed in Fig.~\ref{fig:MainBands}(d) the $\Gamma$-point sphere Wilson loop spectrum of the $R$ model enantiomer [Eq.~(\ref{eq:WilsonElementsMain}), see Appendix~\ref{sec:WilsonBerry} and Refs.~\cite{AndreiXiZ2,Z2Pack,Wieder22,DiracInsulator,TMDHOTI}].
The Wilson loop spectrum in Fig.~\ref{fig:MainBands}(d) winds once in the positive direction, indicating that the nodal degeneracy at $\Gamma$ carries a chiral charge of $C=1$.
We further find in Appendix~\ref{app:PristineKramers} that the Wilson loop winding at $\Gamma$ is reversed in direction when $C^{\mathrm{KW}}_{\mathcal{H}}=-1$, such that the chiral charge is instead $C=-1$ for the $L$ enantiomer, consistent with Eq.~(\ref{eq:mainChiralCharge}).
As we will shortly show below and detailed at length in Appendices~\ref{app:amorphousKramers},~\ref{app:amorphousCharge2}, and~\ref{app:amorphousMultifold}, we find that the relationship between structural and topological chirality established in Ref.~\cite{KramersWeyl} and exemplified by Eq.~(\ref{eq:mainChiralCharge}) can remarkably survive in highly disordered -- and even amorphous -- metals.

Though the Kramers-Weyl model represents the theoretically simplest manifestation of a structurally chiral TSM state, the most ideal real-material structurally chiral TSMs occur in B20 chiral cubic materials like CoSi, RhSi, and AlPt~\cite{chang2017large,tang2017CoSi,CoSiObserveJapan,CoSiObserveHasan,CoSiObserveChina,AlPtObserve}, which instead exhibit weak-SOC $\Gamma$-point \emph{multifold} fermions, rather than energetically isolated strong-SOC Kramers-Weyl fermions.
B20 CoSi, RhSi, and AlPt 
specifically exhibit threefold degeneracies [per spin] close to the Fermi level that consist of linearly dispersing upper and lower topological bands and a central trivial band that is flat to leading order [Fig.~\ref{fig:MainBands}(g)].
For this reason, the weak-SOC [spinless] threefold degeneracies have been termed condensed-matter realizations of spin-1 chiral fermions~\cite{ManesNewFermion,NewFermions,chang2017large,tang2017CoSi,KramersWeyl}.

Interestingly, CoSi has recently been grown amorphous, and exhibits unconventional magnetoresistance~\cite{Molinari2023} and resistivity scaling~\cite{Rocchino2024} in amorphous thin films.
This further hints that topological aspects of chiral TSMs may survive into the amorphous regime, and has motivated theoretical searches for signatures of strongly disordered TSM states in the B20 family~\cite{Franca2024}.
However, the B20 crystal structure is nonsymmorphic, and is specifically generated by screw symmetries that give rise to sublattice degrees of freedom.
For this reason, multifold fermion tight-binding models based on B20 materials are difficult to disorder while maintaining a sense of local [average] chirality compared to models with only internal spin and orbital degrees of freedom, limiting their ability to exhibit sharp signatures of amorphous chiral TSM states under standard numerical disorder implementation schemes [see Ref.~\cite{Franca2024} and Appendix~\ref{app:DiffTypesDisorder}]. 
Inspired by the above experimental and theoretical studies, we introduce a symmorphic model with a $\Gamma$-point chiral multifold fermion that only arises from on-site [internal] orbital degrees of freedom, and can hence more straightforwardly be disordered and analyzed in the same manner as the Kramers-Weyl model [Fig.~\ref{fig:Fig1flow}(b) and Appendix~\ref{app:amorphousMultifold}].

\begin{figure*}
\centering
\includegraphics[width=\linewidth]{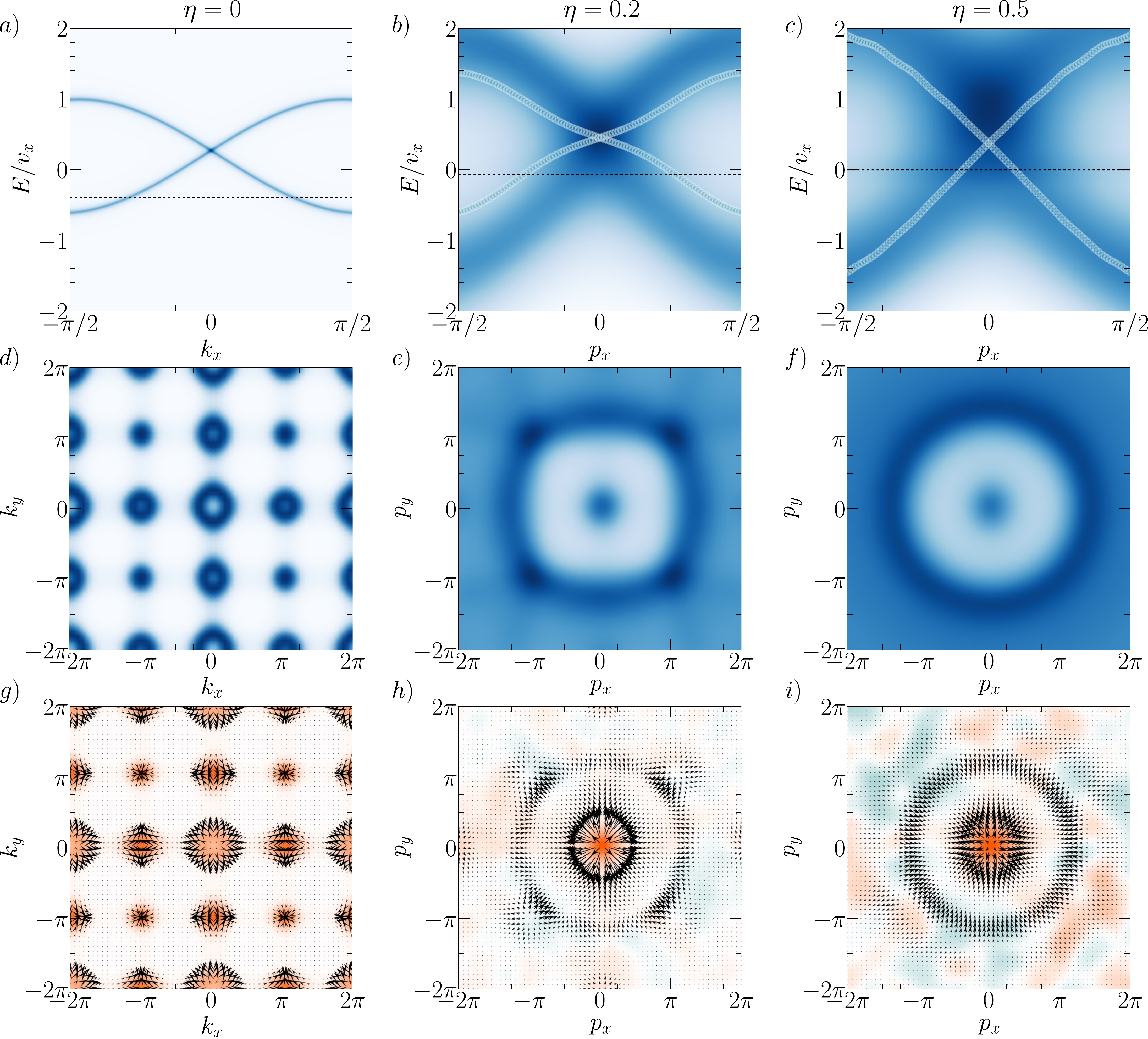}    
\caption{ 
\textbf{Energy spectrum and spin texture of the disordered Kramers-Weyl model.}
(a-f) The spectral function $\bar{A}(E,{\bf p})$ [Eq.~(\ref{eq:SpecFuncMain})] averaged over 50 replicas of the Kramers-Weyl model [Eqs.~(\ref{eq:amorphousKWTmatrixFinalChiralityMain}),~(\ref{eq:KWHeavisideMain}), and~(\ref{eq:KWmainMats}) and Ref.~\cite{KramersWeyl}] with increasing Gaussian structural disorder and local frame disorder implemented with the same standard deviation $\eta$ [Eq.~(\ref{eq:GaussianDisorderMain})], and with local chirality domains of unequal volume [70\% right-handed, 30\% left-handed] within each disorder replica [see Fig.~\ref{fig:Fig1flow}(b) and Appendix~\ref{app:DiffTypesDisorder}].
Panels (a-c) show $\bar{A}(E,{\bf p})$ as a function of $E$ and $p_{x}$ [$k_{x}$ in the $\eta=0$ crystalline limit in (a)] and panels (d-f) show $\bar{A}(E,{\bf p})$ as a function of $p_{x,y}$ [$k_{x,y}$ in (d)] at $p_{z}=0.1$ and fixed $E$ [the dashed lines in (a-c), respectively].
In both the moderate [$\eta=0.2$ in (b)] and strong [$\eta=0.5$ in (c)] structural disorder regimes [see Appendix~\ref{app:PhysicalObservables}], $\bar{A}(E,{\bf p})$ exhibits a linearly dispersing feature centered around ${\bf p}={\bf 0}$ that becomes increasingly diffuse as $\eta$ is increased and upwardly shifted in energy by a disorder-renormalized chemical potential, but nevertheless continues to strongly resemble the crystalline Kramers-Weyl fermion in (a).
In (b,c) we also plot in light blue circles the band structure of the mean-field effective Hamiltonian $\mathcal{H}_{\mathrm{Eff}}({\bf p})$ [Eq.~(\ref{eq:AvgHEffMain}) and Appendix~\ref{app:EffectiveHamiltonian}], which for all $\eta$ exhibits a linear nodal degeneracy at ${\bf p}={\bf 0}$ that precisely corresponds to a $|C|=1$ Kramers-Weyl fermion [see Figs.~\ref{fig:Fig3randomKW3F}(d) and~\ref{fig:Fig4_KW_main_chiralities}(a-e)].
In addition to the linear spectral feature at ${\bf p}={\bf 0}$, the disordered systems in (e,f) exhibit broadened, ring- [sphere-] like spectral features in the vicinity of $|{\bf p}|=\pi/\bar{a}$ and $|{\bf p}|=\pi\sqrt{2}/\bar{a}$ where $\bar{a}=1$ is the average nearest-neighbor spacing, similar to the ring-like, higher-Brillouin-zone Dirac surface-state repetitions recently observed in amorphous Bi$_{2}$Se$_{3}$~\cite{corbae_evidence_2020,Ciocys2023}.
(g-i) The spin texture [Eq.~(\ref{eq:SpinDOSMain})] of the non-crystalline Kramers-Weyl systems in (d-f), respectively.
In panels (g-i), the in-plane components of the spin texture $\langle S^{x,y}(E,\mathbf{p})\rangle$ are represented as arrows with log-scale lengths, while the out-of-plane component $\langle S^{z}(E,\mathbf{p})\rangle$ is represented through a log-scale color map in which orange is positive and teal is negative.
(g) In the crystalline limit, the Kramers-Weyl fermions at each TRIM point exhibit perfect monopole-like spin textures that are locked to their chiral charges~\cite{KramersWeyl}.
(h,i) As the disorder scale $\eta$ is increased, the Kramers-Weyl fermions away from ${\bf p}={\bf 0}$ become merged into ring-like spectral features with largely isotropic spin textures that are inherited from their crystalline-limit [$\eta=0$] spin textures, and hence chiral charges.
This suggests the possibility that the ring-like feature in (e,f,h,i) at $|{\bf p}|=\pi$ [$|{\bf p}|=\pi\sqrt{2}$] is a many-particle disordered Kramers-Weyl fermion with the opposite [same] chiral charge as the linear nodal degeneracy at ${\bf p}={\bf 0}$.
Complete calculation details for all panels are provided in Appendix~\ref{app:amorphousKramers}.}
\label{fig:Fig2KWmainspec}
\end{figure*}

To construct our symmorphic multifold fermion model, we begin by placing four spinless orbitals at each site of a lattice.  
In order, the four basis states specifically consist of spinless $s$, $p_{x}$, $p_{y}$, and $p_{z}$ orbitals.
The four basis states can also alternatively be understood as the \emph{molecular} orbitals of tightly bound tetrahedral molecules with four atoms [or four states on $sp^{3}$ bonds], such as those appearing in the decoupled tetrahedron limit of the Weaire-Thorpe model of amorphous silicon~\cite{weaire_electronic_1971}.
In real space, the four atomic orbitals are then coupled via a combination of on-site crystal field splitting, achiral spinless orbital hopping, and hopping proportional to $\chi_{\alpha}$ that gives rise to structural chirality in the crystalline limit [see Appendix~\ref{app:DiffTypesDisorder} and the text following Eq.~(\ref{appeq:rotaMain})].
Complete details of the real-space tight-binding implementation of our symmorphic crystalline and non-crystalline multifold fermion models are provided in Appendix~\ref{sec:Multifold}.

When placed in a crystalline limit in which the sites $\alpha$ form a regular cubic lattice with a spatially homogeneous hopping reference frame [Fig.~\ref{fig:Fig1flow}(a,b)], the multifold fermion model introduced in Appendix~\ref{sec:Multifold} takes the form of a $4\times 4$ Bloch Hamiltonian in ${\bf k}$-space:
\begin{eqnarray}
\mathcal{H}^{\mathrm{3F}}({\bf k}) &=& t_1\left[\cos(k_x)\mu^z + \cos(k_y)\tau^z + \cos(k_z)\tau^z\mu^z\right]\nonumber \\
&+& t_2\left[\cos(k_x)\tau^z  + \cos(k_y)\tau^z\mu^z + \cos(k_z)\mu^z\right]\nonumber \\
&+& t_3\left[\cos(k_x)\tau^z\mu^z + \cos(k_y)\mu^z + \cos(k_z)\tau^z\right]\nonumber \\
&+& v\left[\sin(k_x)\tau^y + \sin(k_y)\tau^z\mu^y + \sin(k_z)\tau^x\mu^y\right] \nonumber \\
&+& m \left[\mu^z+\tau^z\mu^z + \tau^z\right],
\label{eq:HamBloch3FMain}
\end{eqnarray}
where $\tau^{i}$ and $\mu^{i}$ are $2\times 2$ Pauli matrices that together act within the $4\times 4$ space of the internal spinless $s$ and $p_{x,y,z}$ orbital degrees of freedom, and where the local chirality $\chi_{\alpha}$ from the position-space non-crystalline model has been absorbed into the definition of $v$ [see Appendix~\ref{app:amorphousMultifold}].
The lattice-scale structural chirality $C^{\mathrm{3F}}_{\mathcal{H}}$ of $\mathcal{H}^{\mathrm{3F}}({\bf k})$ in Eq.~(\ref{eq:HamBloch3FMain}) is consequently governed by the sign of $v$:
\begin{equation}
C^{\mathrm{3F}}_{\mathcal{H}} = \text{sgn}(v),
\label{eq:multifoldStructuralChiMain}
\end{equation}
such that, similarly to the Kramers-Weyl model in Eq.~(\ref{eq:appStructuralChiralityMain}), $C^{\mathrm{3F}}_{\mathcal{H}}=1$ corresponds to the right-handed [$R$] model enantiomer and $C^{\mathrm{3F}}_{\mathcal{H}}=-1$ corresponds to the left-handed [$L$] enantiomer.
When $v\neq 0$, both enantiomers of $\mathcal{H}^{\mathrm{3F}}({\bf k})$ respect the symmetries of nonmagnetic symmorphic chiral cubic space group 195 ($P23$), which contains the same twofold rotation symmetries $C_{2i}$ as the space group of the Kramers-Weyl model [albeit now as single-group symmetries due to the spinless orbital basis states], as well as the additional threefold cubic rotation symmetry $C_{3,111}$ [see Appendices~\ref{app:symDefs} and~\ref{app:PristineMulifold}].

In the multifold fermion model in Eq.~(\ref{eq:HamBloch3FMain}), the intersite $t_{i}$ terms correspond to positively and negatively signed, nearest-neighbor, same-orbital [$s-s$ and $p_{x,y,z}-p_{x,y,z}$] achiral spinless hoppings that are equivalent to linear combinations of inter-tetrahedron hopping terms in the Weaire-Thorpe model of amorphous silicon taken close to the limit of decoupled tetrahedra~\cite{weaire_electronic_1971}.
Conversely, the intersite $v$ term in Eq.~(\ref{eq:HamBloch3FMain}) corresponds to a new ``inter-tetrahedron'' OAM coupling that breaks rotoinversion symmetries and hence renders the system structurally chiral [Appendices~\ref{app:symDefs} and~\ref{app:PristineMulifold}].
Notably, unlike in more simplified multifold fermion models, the $v$ term in Eq.~(\ref{eq:HamBloch3FMain}) does not induce a pure monopole-like OAM texture [Eq.~(\ref{eq:OAMDOSMain})] for the point-like degeneracies in the spectrum of $\mathcal{H}^{\mathrm{3F}}({\bf k})$.
The intersite $v$ term instead implements the combined effects of OAM-crystal-momentum-locking and interband matrix elements that remove OAM quantization, which are generically permitted in many-band models and present in real materials~\cite{KramersWeyl,OAMmultifold2,OAMmultifold3}.
Lastly, unlike the crystalline Kramers-Weyl model [Eq.~(\ref{eq:HKWMaink})], the multifold fermion model in Eq.~(\ref{eq:HamBloch3FMain}) additionally contains an on-site mass [crystal field] term proportional to $m$ that splits the four Bloch eigenstates at ${\bf k}={\bf 0}$ into a single and a threefold degeneracy.

In Fig.~\ref{fig:MainBands}(e,f), we plot the band structure of the right-handed enantiomer of $\mathcal{H}^{\mathrm{3F}}({\bf k})$, and in Fig.~\ref{fig:MainBands}(g) we plot the spectral function $A(E,\mathbf{p})$ at the $\Gamma$ point [see Appendix~\ref{app:PristineMulifold} for numerical model parameters].
For all nonzero values of the hopping parameters in Eq.~(\ref{eq:HamBloch3FMain}), the energy spectrum exhibits symmetry-enforced, threefold-degenerate spin-1 chiral [multifold] fermions at $\Gamma$ and $R$ that transform in three-dimensional, single-valued [spinless] little group small coreps [Fig.~\ref{fig:MainBands}(f,g), see Appendices~\ref{app:corepDefs} and~\ref{app:PristineMulifold}].
For the choice of parameters in Fig.~\ref{fig:MainBands}(f), the energy spectrum also exhibits a singly degenerate trivial band at much higher energies [$E/|v| \sim 4$] that is disconnected from the other three bands in the spectrum at all ${\bf k}$ points.
Lastly, though there are no enforced chiral fermions exactly at $X$ and $M$ in Fig.~\ref{fig:MainBands}(f), the energy spectrum additionally contains clusters of conventional Weyl fermions with nonvanishing net chiral charges that link both the first and second bands and the second and third bands within a close vicinity of $X$ and $M$ [see Appendix~\ref{app:PristineMulifold} for further details].

Previous works~\cite{NewFermions,chang2017large,tang2017CoSi,KramersWeyl} have shown that spin-1 multifold fermions with the same ${\bf k}\cdot{\bf p}$ Hamiltonian as the threefold degeneracy at $\Gamma$ in Fig.~\ref{fig:MainBands}(f,g) carry chiral charges of $|C|=2$ for their lower two bands as a set.
To numerically confirm this result for the multifold model in Eq.~(\ref{eq:HamBloch3FMain}), we have computed in Fig.~\ref{fig:MainBands}(h) the non-Abelian Wilson loop spectrum of the lower two bands of the $R$ model enantiomer on a sphere surrounding the $\Gamma$ point [Eq.~(\ref{eq:WilsonElementsMain}), see Appendix~\ref{sec:WilsonBerry} and Refs.~\cite{AndreiXiZ2,Z2Pack,Wieder22,DiracInsulator,TMDHOTI}].
The two Wilson loop eigenvalues in Fig.~\ref{fig:MainBands}(h) wind twice as a set in the positive direction, indicating that the lower two bands of the multifold fermion at $\Gamma$ together carry a chiral charge of $C=2$.
We further find in Appendix~\ref{app:PristineMulifold} that the two-band sphere Wilson loop spectrum at $\Gamma$ instead winds twice in the negative direction for the $L$ model enantiomer, indicating a chiral charge of $C=-2$.
We additionally confirmed by direct computation in Appendix~\ref{app:PristineMulifold} that the lower two bands of the multifold fermion at the $R$ point in Fig.~\ref{fig:MainBands}(f) carry a chiral charge of $C=-2$ for the $R$ model enantiomer, and $C=2$ for the $L$ model enantiomer.
Together, this implies that like the Kramers-Weyl fermions in Eqs.~(\ref{eq:HKWMaink}),~(\ref{eq:mainHKP}), and~(\ref{eq:mainChiralCharge}) [Appendix~\ref{app:PristineKramers} and Ref.~\cite{KramersWeyl}], and like the chiral multifold fermions experimentally studied in Refs.~\cite{AlPtObserve,PdGaObserve}, the spin-1 chiral fermions in the symmorphic multifold model introduced in this work [Eq.~(\ref{eq:HamBloch3FMain}) and Appendix~\ref{app:PristineMulifold}] carry a low-energy topological chirality that is directly inherited from and controlled by the lattice-scale structural chirality [$C^{\mathrm{3F}}_{\mathcal{H}}$ in Eq.~(\ref{eq:multifoldStructuralChiMain})].

{
\vspace{0.2in}
\centerline{\bf B. $\Gamma$-point chiral fermions with strong structural disorder}
\vspace{0.1in}
}
We will next demonstrate that the $\Gamma$-point chiral fermions in the crystalline Kramers-Weyl and multifold fermion models [Fig.~\ref{fig:MainBands}] can survive in the amorphous structural regime, and that in this regime their topology is tunable via the local chirality order.
We first motivate our numerical calculations by briefly returning to the $\Gamma$-point [${\bf k}_{\mathcal{T}}={\bf 0}$] crystalline Kramers-Weyl ${\bf k}\cdot{\bf p}$ Hamiltonian [Eq.~(\ref{eq:mainHKP})].
As first noted in Ref.~\cite{KramersWeyl}, Eq.~(\ref{eq:mainHKP}) with ${\bf k}_{\mathcal{T}}={\bf 0}$ can be tuned to an isotropic form that satisfies the symmetries of any of the chiral crystallographic $\Gamma$-point little groups by setting the $v_{i}$ coefficients to be equal $v_{i}=\tilde{v}$ for all $i=x,y,z$.
However, by now relaxing the constraint of lattice translation symmetry and substituting the pseudo-momentum ${\bf p}$ for ${\bf q}$, we further recognize that Eq.~(\ref{eq:mainHKP}) in the isotropic limit can also be viewed as the ${\bf k}\cdot{\bf p}$ \emph{effective} Hamiltonian [Eq.~(\ref{eq:AvgHEffMain})] of an amorphous system near ${\bf p}={\bf 0}$:
\begin{equation}
\mathcal{H}_{\text{Eff}}({\bf p}) = \tilde{v}\left(p_{x}\sigma^{x} + p_{y}\sigma^{y} + p_{z}\sigma^{z}\right).
\label{eq:mainKWdisorderIsotropic}
\end{equation}
Crucially, because Eq.~(\ref{eq:mainKWdisorderIsotropic}) takes the form of an isotropic $|C|=1$ Weyl fermion~\cite{KramersWeyl,AshvinWeyl}, Eq.~(\ref{eq:mainKWdisorderIsotropic}) can specifically only be the ${\bf p}\approx {\bf 0}$ effective Hamiltonian of an amorphous system that lacks average inversion or other rotoinversion symmetries like mirror, and is hence structurally chiral on the average [\emph{i.e.} has a chiral average symmetry group, see Appendix~\ref{app:pseudoK}].
Therefore, to identify a subset of all possible amorphous chiral fermions [those with crystalline counterparts], we may employ a strategy in which we deform the ${\bf k}\cdot{\bf p}$ Hamiltonians of $\Gamma$-point crystalline chiral fermions to a limit with continuous rotation symmetries, and then ask whether there exist strongly disordered systems governed by the same ${\bf k}\cdot{\bf p}$ effective Hamiltonian near ${\bf p}={\bf 0}$ [Appendix~\ref{app:corepAmorphous}].

The most natural candidate for a strongly disordered system with Eq.~(\ref{eq:mainKWdisorderIsotropic}) as its ${\bf p}\approx{\bf 0}$ effective Hamiltonian is the Kramers-Weyl model subject to strong structural disorder and long-range local chirality order [Fig.~\ref{fig:Fig1flow}(b)].
To numerically investigate this scenario, we plot in Fig.~\ref{fig:Fig2KWmainspec}(a-c) the spectral function $\bar{A}(E,{\bf p})$ [Eq.~(\ref{eq:SpecFuncMain})] as a function of $E$ and $p_{x}$ averaged over 50 replicas of the Kramers-Weyl model [Eqs.~(\ref{eq:amorphousKWTmatrixFinalChiralityMain}),~(\ref{eq:KWHeavisideMain}), and~(\ref{eq:KWmainMats})] with increasing Gaussian structural disorder [Eq.~(\ref{eq:GaussianDisorderMain})], local frame disorder [Eq.~(\ref{appeq:rotaMain})] implemented with the same standard deviation $\eta$ as the lattice disorder, and chirality domains of unequal volume [70\% right-handed, 30\% left-handed] within each disorder replica [see Fig.~\ref{fig:Fig1flow}(b) and Appendix~\ref{app:DiffTypesDisorder}, complete calculation details for Fig.~\ref{fig:Fig2KWmainspec} are provided in Appendix~\ref{app:amorphousKramers}].

For the Gaussian-disordered systems with $\eta=0.2$ and $\eta=0.5$ in Fig.~\ref{fig:Fig2KWmainspec}(b,c), we observe a linearly dispersing feature centered around ${\bf p}={\bf 0}$ that exhibits increased spectral broadening for increasing disorder, but nevertheless strongly resembles the crystalline $\Gamma$-point Kramers-Weyl fermion in Figs.~\ref{fig:MainBands}(b,c) and~\ref{fig:Fig2KWmainspec}(a). 
Notably, the nodal spectral features in Fig.~\ref{fig:Fig2KWmainspec}(a-c) at ${\bf p}={\bf 0}$ shift upwards in energy as $\eta$ is increased, reflecting the presence of an increasingly strong disorder-renormalized chemical potential.
Like the Dirac-cone surface state of the amorphous 3D TI Bi$_2$Se$_3$ observed in recent ARPES experiments~\cite{corbae_evidence_2020,Ciocys2023}, the Fermi velocity of the linear spectral feature in Fig.~\ref{fig:Fig2KWmainspec}(a-c) is also renormalized to larger values as $\eta$ is increased.
Previous theoretical works~\cite{Ziman1971} have importantly predicted that in amorphous systems with local structural order [\emph{e.g.} fixed bond coordination~\cite{weaire_electronic_1971}] -- like the models analyzed in this work [Appendix~\ref{app:models}] -- ${\bf p}\approx {\bf 0}$ spectral features can correspond to delocalized [extended] states that exhibit a large overlap with plane-wave states and therefore carry well-defined ${\bf p}$.
Such extended states have correspondingly been observed near ${\bf p}={\bf 0}$ in both tight-binding and experimental studies of the amorphous 3D TI Bi$_2$Se$_3$~\cite{corbae_evidence_2020,Ciocys2023}, and even in liquid lead~\cite{Kim2011}.

In Fig.~\ref{fig:Fig2KWmainspec}(b,c), we also plot the bands of the effective Hamiltonian $\mathcal{H}_{\mathrm{Eff}}({\bf p})$ [Eq.~(\ref{eq:AvgHEffMain})] with light blue circles. 
In both Fig.~\ref{fig:Fig2KWmainspec}(b,c), we constructed $\mathcal{H}_{\mathrm{Eff}}({\bf p})$ using a reference energy cut $E_{C}$ centered at the maximum spectral function at ${\bf p}={\bf 0}$ to maximize the qualitative spectral accuracy of $\mathcal{H}_{\mathrm{Eff}}({\bf p})$ [see Appendix~\ref{app:EffectiveHamiltonian}].
Like $\bar{A}(E,{\bf p})$, the effective Hamiltonian bands in Fig.~\ref{fig:Fig2KWmainspec}(b,c) also continue to exhibit linear nodal degeneracies at ${\bf p}={\bf 0}$ in the presence of increasingly strong structural disorder.

We next construct constant-energy spectral cuts of the disordered Kramers-Weyl model to explore spectral features at higher momenta.
In Fig.~\ref{fig:Fig2KWmainspec}(d-f) we respectively plot the average spectral function $\bar{A}(E,{\bf p})$ of the crystalline and non-crystalline Kramers-Weyl systems in Fig.~\ref{fig:Fig2KWmainspec}(a-c), but now at $p_{z}=0.1$ and fixed $E$, rather than as functions of $E$ and $p_{x}$. 
Interestingly, as $\eta$ is increased, $\bar{A}(E,{\bf p})$ develops 3D sphere-like spectral features that appear as 2D rings in the vicinity of $|{\bf p}|=\pi/\bar{a}$ and $|{\bf p}|=\pi\sqrt{2}/\bar{a}$ in the constant-energy and fixed-$p_{z}$ spectra in Fig.~\ref{fig:Fig2KWmainspec}(e,f), where $\bar{a}=1$ is the average nearest-neighbor spacing.
The ring-like features in Fig.~\ref{fig:Fig2KWmainspec}(e,f) can be understood as originating from disorder-broadening and averaging the crystalline Kramers-Weyl fermions at $|{\bf k}|=\pi$ and $|{\bf k}|=\pi\sqrt{2}$ in Fig.~\ref{fig:Fig2KWmainspec}(d) over random orientations and lattice spacings, giving rise to the characteristic isotropic spectral features of an amorphous system~\cite{corbae_evidence_2020,Ciocys2023}.

The appearance of 3D sphere-like Kramers-Weyl features in Fig.~\ref{fig:Fig2KWmainspec}(e,f) is also analogous to the 2D surface states of amorphous Bi$_{2}$Se$_{3}$.
Specifically, in crystalline Bi$_{2}$Se$_{3}$, topological Dirac fermions lie at $\bar{\Gamma}$ [${\bf k}={\bf 0}$] in the surface BZ, and at all surface ${\bf k}$ points related to $\bar{\Gamma}$ by reciprocal lattice vectors with $|{\bf K}|=2\pi/a$ where $a$ is the surface lattice spacing.  
In the experimentally-obtained surface spectrum of amorphous Bi$_{2}$Se$_{3}$, one Dirac-cone surface state continues to lie at ${\bf p}={\bf 0}$, but the other surface Dirac cones in higher surface BZs merge into ring-like spectral features, with the first Dirac ring specifically lying at $|{\bf p}|=2\pi/\bar{a}$ where $\bar{a}$ is the average in-plane [surface] nearest-neighbor atomic spacing~\cite{corbae_evidence_2020,Ciocys2023}.

\begin{figure*}[t]
\centering
\includegraphics[width=\linewidth]{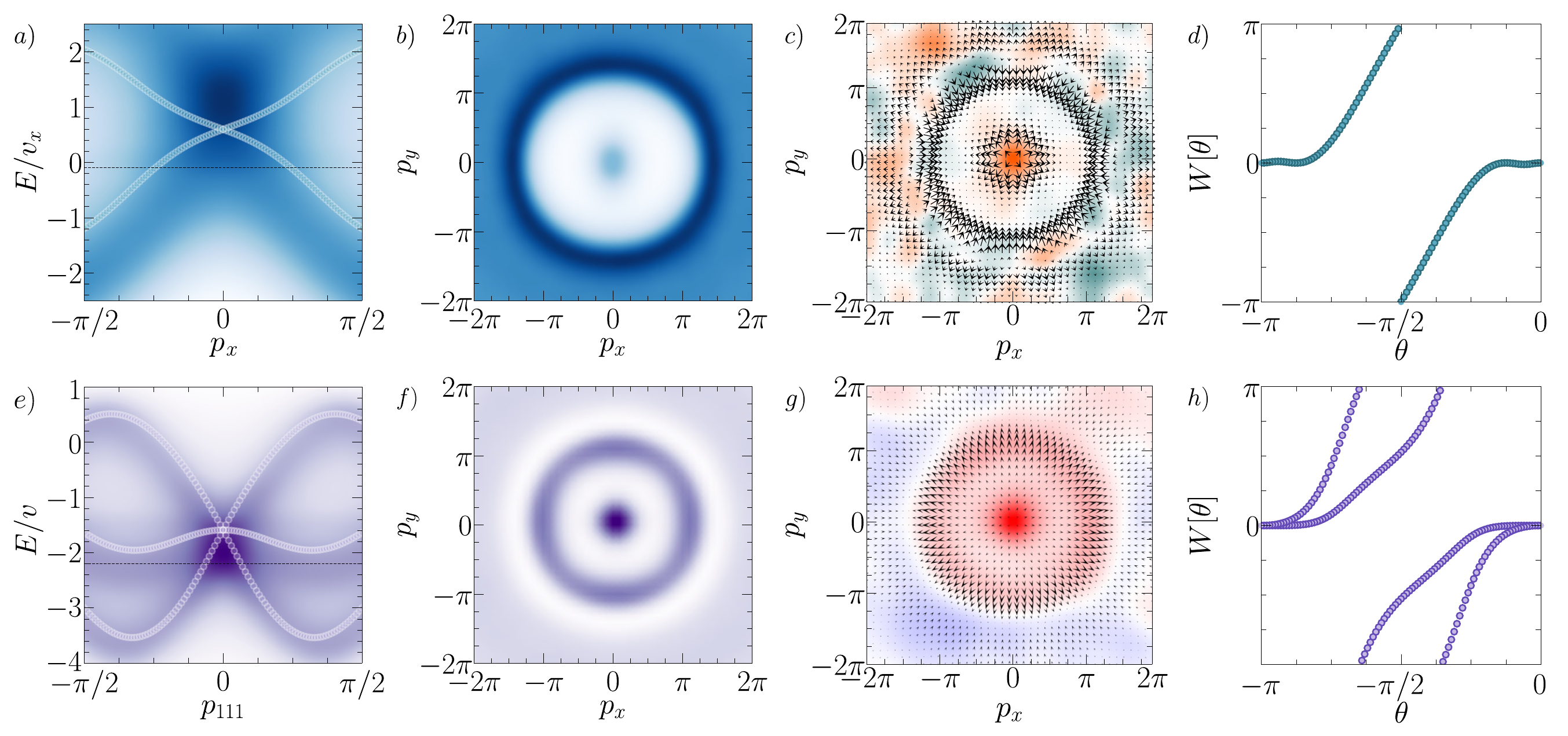}    
\caption{\textbf{Random-lattice Kramers-Weyl and chiral multifold fermions.}
(a-c) Energy spectrum $\bar{A}(E,\mathbf{p})$ [Eq.~(\ref{eq:SpecFuncMain})] and spin texture $\langle\mathbf{S}(E,\mathbf{p})\rangle$ [Eq.~(\ref{eq:SpinDOSMain})] averaged over 50 replicas of the non-crystalline Kramers-Weyl model [Eqs.~(\ref{eq:amorphousKWTmatrixFinalChiralityMain}),~(\ref{eq:KWHeavisideMain}), and~(\ref{eq:KWmainMats}) and Ref.~\cite{KramersWeyl}] on 3D random lattices with strong random frame disorder parameterized by $\eta=0.5$ and contiguous chirality domains of unequal volume such that $2/3$ of sites are right-handed [$n_R=2/3$] and $1/3$ of sites are left-handed [$n_L=1-n_{R}=1/3$] within each disorder replica [see Fig.~\ref{fig:Fig1flow}(b) and Appendix~\ref{app:DiffTypesDisorder}].
Panel (a) shows $\bar{A}(E,{\bf p})$ as a function of $E$ and $p_{x}$, and panels (b,c) respectively show $\bar{A}(E,{\bf p})$ and $\langle\mathbf{S}(E,\mathbf{p})\rangle$ as functions of $p_{x,y}$ at $p_{z}=0.1$ and fixed $E$ [the dashed line in (a)].
In (c), $\langle S^{x,y}(E,\mathbf{p})\rangle$ are represented as arrows with log-scale lengths and $\langle S^{z}(E,\mathbf{p})\rangle$ is represented using a log-scale color map in which orange is positive and teal is negative.
Like in the Kramers-Weyl model with strong Gaussian disorder [Fig.~\ref{fig:Fig2KWmainspec}(c,f)], the random-lattice Kramers-Weyl model exhibits (a) a linear spectral feature at ${\bf p}={\bf 0}$ and (b) broadened, ring- [sphere-] like features at $|{\bf p}|=\pi/\bar{a}$ and $|{\bf p}|=\pi\sqrt{2}/\bar{a}$, where $\bar{a}=1$ is the average nearest-neighbor spacing.
Also like in the Gaussian-disordered Kramers-Weyl model [Fig.~\ref{fig:Fig2KWmainspec}(i)], the ${\bf p}={\bf 0}$ feature exhibits (c) a nearly perfect outward-pointing, monopole-like spin texture, and the ring-like features at $|{\bf p}|=\pi$ and $|{\bf p}|=\pi\sqrt{2}$ respectively exhibit inward- and outward-pointing spin textures.
(d) The Abelian amorphous Wilson loop spectrum  [Eq.~(\ref{eq:WilsonElementsMain}) and Appendix~\ref{sec:WilsonBerry}] of the linear spectral feature at ${\bf p}={\bf 0}$ in (a), computed over the lower band of the effective Hamiltonian $\mathcal{H}_{\text{Eff}}(\mathbf{p})$ [light blue circles in (a), see Eq.~(\ref{eq:AvgHEffMain}) and Appendix~\ref{app:EffectiveHamiltonian}].
The Wilson spectrum in (d) winds once in the positive direction as a function of the sphere polar angle $\theta$, quantitatively identifying the nodal spectral feature at ${\bf p}={\bf 0}$ in (a) as an amorphous Kramers-Weyl fermion with a chiral charge of $C=1$ [Eq.~(\ref{eq:MainChiralCharge})].
(e-g) $\bar{A}(E,{\bf p})$ and orbital angular momentum [OAM] texture $\langle\mathbf{L}(E,\mathbf{p})\rangle$ [Eq.~(\ref{eq:OAMDOSMain})] averaged over 50 replicas of the symmorphic multifold fermion model [see Eq.~(\ref{eq:HamBloch3FMain}) and Appendix~\ref{sec:Multifold}] on 3D random lattices with $\eta=0.5$ random frame disorder and contiguous chirality domains with $n_R=2/3$ and $n_L=1/3$.
Panel (e) shows $\bar{A}(E,{\bf p})$ as a function of $E$ and $p_{111} = (1/\sqrt{3})(p_{x}+p_{y}+p_{z})$, and panels (f,g) respectively show $\bar{A}(E,{\bf p})$ and $\langle\mathbf{L}(E,\mathbf{p})\rangle$ as functions of $p_{x,y}$ at $p_{z}=0.1$ and fixed $E$ [the dashed line in (e)].
In (g), $\langle L^{x,y}(E,\mathbf{p})\rangle$ are represented as arrows with log-scale lengths and $\langle L^{z}(E,\mathbf{p})\rangle$ is represented using a log-scale color map in which red is positive and blue is negative.
The random-lattice multifold model in (e) exhibits a disorder-broadened threefold nodal degeneracy at ${\bf p}={\bf 0}$ that to leading order consists of two linearly dispersing bands and a central nondispersing band, closely resembling the crystalline $\Gamma$-point chiral multifold fermion in Fig.~\ref{fig:MainBands}(g).
$\bar{A}(E,{\bf p})$ in (f) also exhibits ring-like spectral features at larger $|{\bf p}|$ whose origin and properties are further detailed in Appendix~\ref{app:amorphousMultifold}.
The multifold spectral feature at ${\bf p}={\bf 0}$ in (e,f) exhibits (g) a nearly perfect outward-pointing, monopole-like OAM texture like that observed in recent experimental studies of crystalline chiral multifold fermions~\cite{OAMmultifold2,OAMmultifold3}.
(h) The non-Abelian amorphous Wilson loop spectrum of the threefold spectral feature at ${\bf p}={\bf 0}$ in (e), computed over the lower two bands of $\mathcal{H}_{\text{Eff}}({\bf p})$ [light purple circles in (e)].
The two Wilson loop eigenvalues in (h) wind twice as a set in the positive direction, confirming that the threefold spectral feature in (e) is an amorphous chiral multifold fermion whose lower two bands together carry $C=2$.
Complete calculation details for panels (a-d) and (e-h) are provided in Appendices~\ref{app:amorphousKramers} and~\ref{app:amorphousMultifold}, respectively.}
\label{fig:Fig3randomKW3F}
\end{figure*}

To gain further insight on the possibility of nontrivial topology in the disordered Kramers-Weyl model, we next compute the spin texture of the spectral features in Fig.~\ref{fig:Fig2KWmainspec}(d-f).
In Ref.~\cite{KramersWeyl}, it was shown that energetically isolated Kramers-Weyl fermions can exhibit monopole-like spin textures that approximately indicate the sign of their topological chiral charge.  
In Fig.~\ref{fig:Fig2KWmainspec}(g-i), we plot the spin texture $\langle\mathbf{S}(E,\mathbf{p})\rangle$ [Eq.~(\ref{eq:SpinDOSMain})] of the disordered Kramers-Weyl model with the same parameters previously employed to generate the energy spectra in Fig.~\ref{fig:Fig2KWmainspec}(a-f).
In the crystalline limit [$\eta=0$ in Fig.~\ref{fig:Fig2KWmainspec}(g)], the $C=+1$ Kramers-Weyl fermions at $k_{x}=k_{y}=0,\pi$ exhibit outward-pointing spin textures along odd numbers of principal axes and inward-pointing spin textures among the remaining [even numbers of] principal axes, and the $C=-1$ Kramers-Weyl fermions at $(k_{x},k_{y})=(\pi,0)$ and $(k_{x},k_{y})=(0,\pi)$ exhibit spin textures that respectively point inward in the $k_{x}$- and $k_{y}$-directions and outward along the other two principal axes. 
In the presence of increasingly strong Gaussian structural disorder and fixed local chirality order, the central spectral feature -- which corresponds to the nodal degeneracy at ${\bf p}={\bf 0}$ in Fig.~\ref{fig:Fig2KWmainspec}(a-c) -- retains its nearly perfect outward-pointing, monopole-like spin texture, suggesting a picture [which we will shortly use Wilson loops to make precise] in which the linear nodal degeneracy at ${\bf p}={\bf 0}$ in Fig.~\ref{fig:Fig2KWmainspec}(c,f,i) is a strongly disordered Kramers-Weyl fermion with a positive chiral charge owing to the average system right-handedness.

Interestingly, as $\eta$ is increased without varying the degree of local chirality order, we observe that the spin texture [anti]monopoles at $|{\bf p}|=\pi$ and $|{\bf p}|=\pi\sqrt{2}$ merge into 3D sphere-like spectral features with largely isotropic spin textures.
Specifically, the crystalline Kramers-Weyl fermions at $|{\bf k}|=\pi$ in Fig.~\ref{fig:Fig2KWmainspec}(g) merge into a ring-like feature at $|{\bf p}|=\pi$ with an inward-pointing spin texture and high spectral weight, and the crystalline Kramers-Weyl fermions at $|{\bf k}|=\pi\sqrt{2}$ merge into a ring-like feature at $|{\bf p}|=\pi\sqrt{2}$ with an outward-pointing spin texture and relatively weaker spectral weight.
Importantly, the varying $\langle S^{x,y}(E,\mathbf{p})\rangle$ components at different pseudo-momenta ${\bf p}$ along the monopole- and ring-like features in Fig.~\ref{fig:Fig2KWmainspec}(i) highlight that even though the energy spectra of strongly disordered systems are isotropic [Fig.~\ref{fig:Fig2KWmainspec}(f)], this \emph{does not imply} that disordered systems are identical at each ${\bf p}$ with the same magnitude $|{\bf p}|$.
In the limit of strong structural disorder, system properties like the spin texture $\langle\mathbf{S}(E,\mathbf{p})\rangle$ at two ${\bf p}$ with the same $|{\bf p}|$ instead become \emph{related} by the action of emergent continuous proper rotation symmetries in the average system symmetry group [Appendix~\ref{app:pseudoK}].

Because the energy spectrum is more diffuse at larger momenta, and because we have only shown that the effective Hamiltonian method is numerically stable near ${\bf p}={\bf 0}$ [Appendix~\ref{app:EffectiveHamiltonian}], the analytic and numerical tools in this work cannot be used to diagnose the topology of the disorder-broadened, ring-like spectral features in Fig.~\ref{fig:Fig2KWmainspec}(e,f,h,i).
Nevertheless, the alternating inward- and outward-pointing spin textures of the spectral rings in Fig.~\ref{fig:Fig2KWmainspec}(h,i) suggest the possibility that the ring-like feature at $|{\bf p}|=\pi$ [$|{\bf p}|=\pi\sqrt{2}$] is a many-particle disordered Kramers-Weyl fermion with the opposite [same] chiral charge as the linear nodal degeneracy at ${\bf p}={\bf 0}$.

However, one might be concerned that the energy spectra and spin textures in Fig.~\ref{fig:Fig2KWmainspec} are specific to our use of Gaussian structural disorder, which admits a well-defined crystalline limit.
To show that the spectral features and spin texture in Fig.~\ref{fig:Fig2KWmainspec}(c,f,i) are instead more generic features of the Kramers-Weyl model with strong structural disorder and local chirality order, we next analyze the Kramers-Weyl model on randomly generated lattices without well-defined [unique] crystalline limits.
In Fig.~\ref{fig:Fig3randomKW3F}(a,b), we plot the spectral function $\bar{A}(E,{\bf p})$ averaged over 50 replicas of the Kramers-Weyl model [Eqs.~(\ref{eq:amorphousKWTmatrixFinalChiralityMain}),~(\ref{eq:KWHeavisideMain}), and~(\ref{eq:KWmainMats})] on lattices with randomly located sites in three dimensions [Fig.~\ref{fig:Fig1flow}(b) and Appendix~\ref{app:DiffTypesDisorder}], strong random frame disorder parameterized by the standard deviation $\eta=0.5$ [Eq.~(\ref{appeq:rotaMain})], and contiguous chirality domains of unequal volume such that $2/3$ of sites are right-handed and $1/3$ of sites are left-handed within each disorder replica [complete calculation details for Fig.~\ref{fig:Fig3randomKW3F}(a-d) are provided in Appendix~\ref{app:amorphousKramers}].
Like in the large-$\eta$ Gaussian-disordered Kramers-Weyl model [Fig.~\ref{fig:Fig2KWmainspec}(c,f)], $\bar{A}(E,{\bf p})$ in the random-lattice Kramers-Weyl model exhibits a linearly dispersing spectral feature at ${\bf p}={\bf 0}$ [Fig.~\ref{fig:Fig3randomKW3F}(a)], as well as broadened, ring- [sphere-] like features at $|{\bf p}|=\pi/\bar{a}$ and $|{\bf p}|=\pi\sqrt{2}/\bar{a}$, where $\bar{a}=1$ is the average nearest-neighbor spacing [Fig.~\ref{fig:Fig3randomKW3F}(b)].

In Fig.~\ref{fig:Fig3randomKW3F}(c), we next plot the spin texture of the random-lattice Kramers-Weyl model.
As in the Gaussian-disordered system in Fig.~\ref{fig:Fig2KWmainspec}(i), the ${\bf p}={\bf 0}$ linear spectral feature in Fig.~\ref{fig:Fig3randomKW3F}(c) exhibits a nearly perfect outward-pointing, monopole-like spin texture, consistent with the average system right-handedness [see the text surrounding Eq.~(\ref{eq:mainChiralCharge})], and further hinting that the linear feature at ${\bf p}={\bf 0}$ is a non-crystalline topological chiral fermion.
Also like in the Gaussian-disordered model in Fig.~\ref{fig:Fig2KWmainspec}(i), the ring-like spectral feature at $|{\bf p}|=\pi$ in Fig.~\ref{fig:Fig3randomKW3F}(c) exhibits an inward-pointing spin texture, and the ring-like feature at $|{\bf p}|=\pi\sqrt{2}$ exhibits an outward-pointing spin texture.
In Appendix~\ref{app:amorphousKramers}, we also analyze the non-crystalline Kramers-Weyl model on Mikado lattices, for which we observe the same spectral features and spin texture as the random-lattice Kramers-Weyl model in Fig.~\ref{fig:Fig3randomKW3F}(a-c).
Along with the large-$\eta$ Gaussian-disorder calculations in Fig.~\ref{fig:Fig2KWmainspec}(c,f,i), this strongly implies that our different structural disorder implementation schemes [Fig.~\ref{fig:Fig1flow}(b) and Appendix~\ref{app:DiffTypesDisorder}] converge to the same amorphous limit.

Having shown that the Kramers-Weyl model with strong structural disorder and local chirality order exhibits a linear nodal degeneracy with a monopole-like spin texture at ${\bf p}={\bf 0}$, we next employ an amorphous-system generalization of the Wilson loop method~\cite{AndreiXiZ2,DiracInsulator,TMDHOTI,Z2Pack,Wieder22} to precisely show that the ${\bf p}={\bf 0}$ nodal degeneracy in Figs.~\ref{fig:Fig2KWmainspec} and~\ref{fig:Fig3randomKW3F}(a-c) is a non-crystalline [amorphous] Kramers-Weyl fermion with a nontrivial topological chiral charge.
We begin by constructing an effective Hamiltonian $\mathcal{H}_{\text{Eff}}(\mathbf{p})$ for the random-lattice Kramers-Weyl model [light blue circles in Fig.~\ref{fig:Fig3randomKW3F}(a)] with the reference energy cut [$E_{C}$ in Eq.~(\ref{eq:AvgHEffMain})] set to the energy where $\bar{A}(E,{\bf p})$ is largest at ${\bf p}={\bf 0}$.
We then compute in Fig.~\ref{fig:Fig3randomKW3F}(d) the Wilson loop spectrum of the occupied [lower] band of $\mathcal{H}_{\mathrm{Eff}}({\bf p})$ on a sphere surrounding the nodal degeneracy at ${\bf p}={\bf 0}$ [see Fig.~\ref{fig:Fig1flow}(e), Appendix~\ref{sec:WilsonBerry}, and the text surrounding Eq.~(\ref{eq:WilsonElementsMain})].
The amorphous Wilson spectrum in Fig.~\ref{fig:Fig3randomKW3F}(d) winds once in the positive direction, indicating that the nodal spectral feature at ${\bf p}={\bf 0}$ is a $C=1$ amorphous Kramers-Weyl fermion.
We have also performed the analogous sphere Wilson loop calculation for the Gaussian-disordered and Mikado-lattice Kramers-Weyl models in Fig.~\ref{fig:Fig2KWmainspec} and Appendix~\ref{app:amorphousKramers}, respectively, and in both cases observe the same $C=1$ Wilson loop winding at ${\bf p}={\bf 0}$ in systems that are right-handed on the average. 
Importantly, we also continue to observe $C=1$ amorphous Wilson loop winding as the reference energy cut $E_{C}$ used to generate $\mathcal{H}_{\text{Eff}}(\mathbf{p})$ is varied over a wide energy range [Appendix~\ref{app:EffectiveHamiltonian}].
This provides strong evidence that single-particle topological invariants computed from $\mathcal{H}_{\text{Eff}}(\mathbf{p})$ near ${\bf p}={\bf 0}$ are numerically stable, and may accurately capture the exact [many-particle] topology of ${\bf p}={\bf 0}$ nodal degeneracies in amorphous TSMs.

To show that the disorder-robustness of $\Gamma$-point chiral fermions is not limited to the Kramers-Weyl model, we next consider the more material-relevant case~\cite{chang2017large,tang2017CoSi,CoSiObserveJapan,CoSiObserveHasan,CoSiObserveChina,AlPtObserve} of $\Gamma$-point multifold fermions.
In Fig.~\ref{fig:Fig3randomKW3F}(e,f), we plot the spectral function $\bar{A}(E,{\bf p})$ averaged over 50 replicas of the symmorphic multifold fermion model [see Fig.~\ref{fig:MainBands}(e-h), Eq.~(\ref{eq:HamBloch3FMain}), and Appendix~\ref{sec:Multifold}] on 3D random lattices with strong random frame disorder parameterized by $\eta=0.5$ and contiguous chirality domains of unequal volume such that $2/3$ of sites are right-handed and $1/3$ of sites are left-handed within each disorder replica [complete calculation details for Fig.~\ref{fig:Fig3randomKW3F}(e-h) are provided in Appendix~\ref{app:amorphousMultifold}].
The energy spectrum in Fig.~\ref{fig:Fig3randomKW3F}(e) continues to exhibit a threefold [multifold] nodal degeneracy at ${\bf p}={\bf 0}$ that to leading order consists of two linearly dispersing spectral features and a central nondispersing feature with significant disorder broadening.
Like in the Gaussian-disordered and random-lattice Kramers-Weyl models [Figs.~\ref{fig:Fig2KWmainspec}(f) and~\ref{fig:Fig3randomKW3F}(b), respectively], the random-lattice multifold model in Fig.~\ref{fig:Fig3randomKW3F}(f) also exhibits ring-like features away from ${\bf p}={\bf 0}$.
The larger-$|{\bf p}|$ spectral features in Fig.~\ref{fig:Fig3randomKW3F}(f) arise from more complicated disorder-driven processes than the analogous spectral rings in the Kramers-Weyl model, and are further detailed in Appendix~\ref{app:amorphousMultifold}.

Because the multifold fermion model is constructed from spinless orbital basis states, we cannot compute its spin texture.
However, we may instead compute the \emph{OAM texture} [Eq.~(\ref{eq:OAMDOSMain})].
Previous circular dichroism ARPES [CD-ARPES] experiments~\cite{OAMmultifold2,OAMmultifold3} have shown that spin-1 chiral multifold fermions with weak SOC can exhibit monopole-like OAM textures analogous to the monopole-like spin textures of idealized Kramers-Weyl fermions with strong SOC [see Fig.~\ref{fig:Fig2KWmainspec}(g), Appendix~\ref{app:amorphousKramers}, and Ref.~\cite{KramersWeyl}].
In the random-lattice multifold model with local chirality order, we correspondingly observe a nearly perfect outward-pointing, monopole-like OAM texture for the multifold spectral feature at ${\bf p}={\bf 0}$ [Fig.~\ref{fig:Fig3randomKW3F}(g)], suggesting that the multifold feature is a strongly disordered spin-1 chiral fermion. 
In Appendix~\ref{app:amorphousMultifold}, we also analyze the non-crystalline multifold model with Gaussian disorder, and observe the same spectral features and OAM texture as the random-lattice multifold model in Fig.~\ref{fig:Fig3randomKW3F}(e-g).
This indicates that the spectral features and OAM texture in Fig.~\ref{fig:Fig3randomKW3F}(e-g) are inherent to the multifold model in the [structurally chiral] amorphous limit, as opposed to artifacts of a particular choice of structural disorder implementation scheme.

Lastly, to precisely characterize the topology of the ${\bf p}={\bf 0}$ spectral feature in the random-lattice multifold model, we compute the amorphous sphere Wilson loop spectrum at ${\bf p}={\bf 0}$.
To obtain the Wilson spectrum, we first construct an effective Hamiltonian $\mathcal{H}_{\mathrm{Eff}}({\bf p})$ [light purple circles in Fig.~\ref{fig:Fig3randomKW3F}(e)] with the reference energy cut $E_{C}$ in Eq.~(\ref{eq:AvgHEffMain}) set to the energy of the largest spectral weight $\bar{A}(E,{\bf p})$ at ${\bf p}={\bf 0}$ in Fig.~\ref{fig:Fig3randomKW3F}(e) [excluding the higher-energy states that arise from disordering the singly-degenerate trivial band at $E/|v| \approx 4.5$ in Fig.~\ref{fig:MainBands}(f)].
We then use the lower two bands of $\mathcal{H}_{\mathrm{Eff}}({\bf p})$ in energy to compute the non-Abelian [matrix] Wilson loop spectrum on a sphere surrounding the threefold degeneracy at ${\bf p}={\bf 0}$ [Fig.~\ref{fig:Fig3randomKW3F}(h), see Eq.~(\ref{eq:WilsonElementsMain}) and Appendix~\ref{sec:WilsonBerry}].
The two Wilson loop eigenvalues in Fig.~\ref{fig:Fig3randomKW3F}(h) wind twice as a set in the positive direction, confirming that the nodal spectral feature at ${\bf p}={\bf 0}$ is an amorphous chiral multifold fermion whose lower two bands together carry a chiral charge of $C=2$.
We have also performed the analogous sphere Wilson loop calculation for the Gaussian-disordered multifold model in Appendix~\ref{app:amorphousMultifold}, and again observe $C=2$ sphere Wilson winding at ${\bf p}={\bf 0}$ for systems that are right-handed on the average.
Crucially, as previously for the non-crystalline Kramers-Weyl model, we continue to observe $C=2$ amorphous Wilson loop winding in the non-crystalline multifold model as the reference energy $E_{C}$ in $\mathcal{H}_{\mathrm{Eff}}({\bf p})$ is varied over a wide range, confirming the numerical stability of the ${\bf p}={\bf 0}$ chiral charge [Appendix~\ref{app:EffectiveHamiltonian}].

\begin{figure*}[t]
\centering
\includegraphics[width=\linewidth]{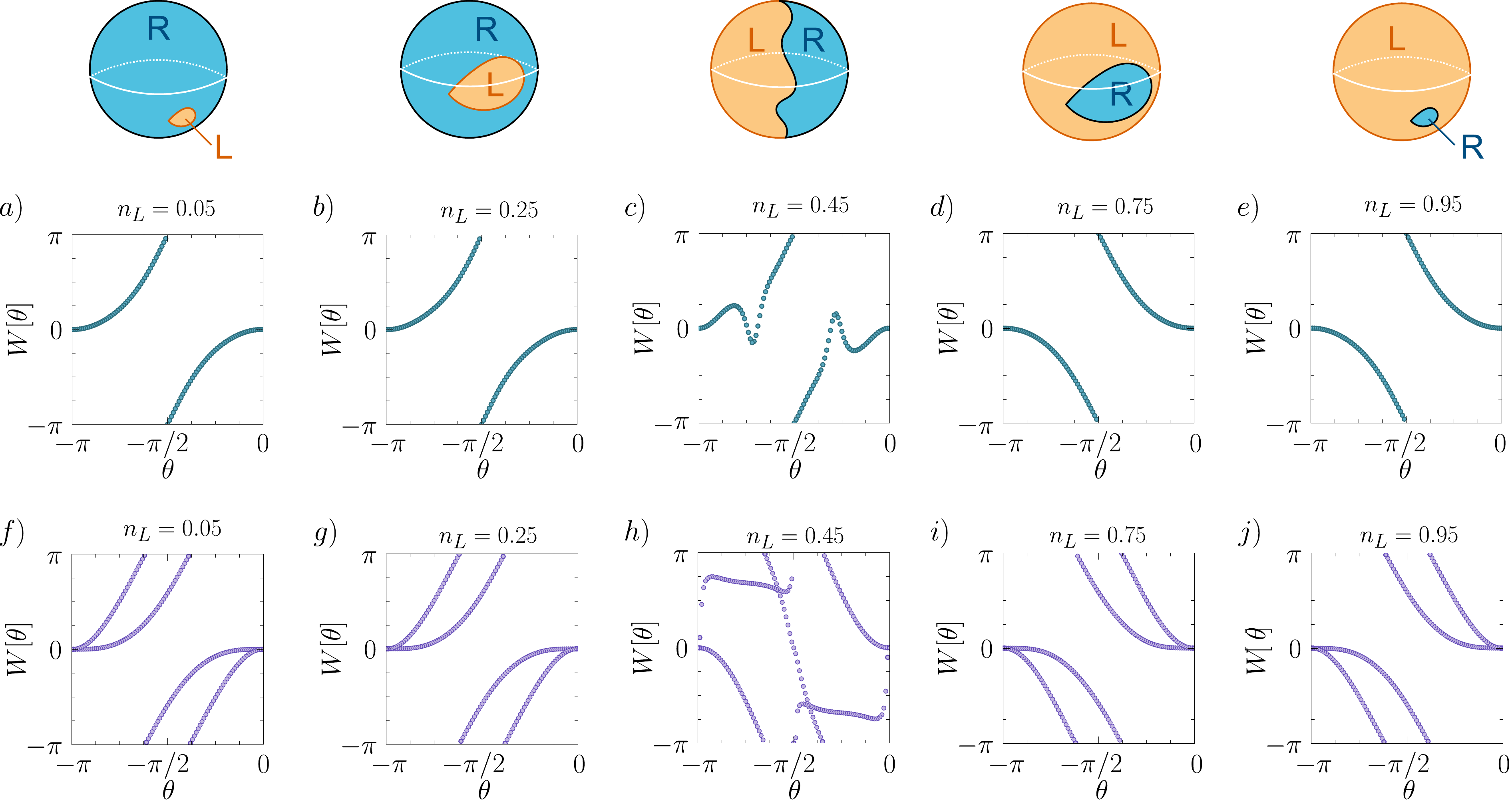}
\caption{\textbf{Controlling topology with average structural chirality in disordered chiral semimetals.}
To generate this figure, we first construct five ensembles of (a-e) the non-crystalline Kramers-Weyl model [Eqs.~(\ref{eq:amorphousKWTmatrixFinalChiralityMain}),~(\ref{eq:KWHeavisideMain}), and~(\ref{eq:KWmainMats}) and Ref.~\cite{KramersWeyl}] and (f-j) five ensembles of the symmorphic multifold model [Eq.~(\ref{eq:HamBloch3FMain}) and Appendix~\ref{sec:Multifold}] with 50 replicas each, where each replica is subject to Gaussian lattice and local frame disorder with $\eta=0.2$ in Eq.~(\ref{eq:GaussianDisorderMain}) and contains contiguous right- and left-handed chirality domains with varying concentrations respectively given by $n_R=N_R/N_{\mathrm{sites}}$ and $n_{L}=1-n_{R}$ [respectively schematically depicted with blue $R$ and yellow $L$ regions in the top row, see Fig.~\ref{fig:Fig1flow}(a,b) and Appendix~\ref{app:DiffTypesDisorder}].
For each ensemble, we construct an effective Hamiltonian $\mathcal{H}_{\mathrm{Eff}}({\bf p})$ using the replica-averaged Green's function $\bar{\mathcal{G}}(E,\mathbf{p})$ [Eqs.~(\ref{eq:averageOneMomentumGreenMain}) and~(\ref{eq:AvgHEffMain})], and then use the eigenstates of $\mathcal{H}_{\mathrm{Eff}}({\bf p})$ to compute the amorphous [disordered] Wilson loop spectrum on a sphere surrounding the nodal degeneracy at ${\bf p}={\bf 0}$ [Eqs.~(\ref{eq:MainChiralCharge}) and~(\ref{eq:WilsonElementsMain}) and Appendices~\ref{app:EffectiveHamiltonian} and~\ref{sec:WilsonBerry}].
For the Abelian [scalar] Wilson loop of the lower band of $\mathcal{H}_{\mathrm{Eff}}({\bf p})$ in the Kramers-Weyl model, we observe a quantized winding number as a function of the sphere polar angle $\theta$ of (a,b) $C=1$ for $n_{L}<0.5$, (d,e) $C=-1$ for $n_{L}>0.5$, and (c) a region in the vicinity of $n_{L}\approx 0.5$ with a non-smooth Wilson spectrum.
Similarly, for the non-Abelian Wilson loop of the lowest two bands of $\mathcal{H}_{\mathrm{Eff}}({\bf p})$ in the multifold model, the two Wilson loop eigenvalues as a set exhibit quantized winding numbers of (f,g) $C=2$ for $n_{L}<0.5$ and (i,j) $C=-2$ for $n_{L}>0.5$, outside of again (h) a region in the vicinity of $n_{L}\approx 0.5$ with a non-smooth Wilson spectrum.
The Wilson spectra in (a-j) therefore indicate that both the linear spectral feature at ${\bf p}={\bf 0}$ in Figs.~\ref{fig:Fig2KWmainspec}(b,c) and~\ref{fig:Fig3randomKW3F}(a) and the threefold spectral feature at ${\bf p}={\bf 0}$ in Fig.~\ref{fig:Fig3randomKW3F}(e) are disordered chiral fermions whose topological chirality is tunable via the average system structural chirality, generalizing the results of Refs.~\cite{KramersWeyl,AlPtObserve,PdGaObserve} to the structurally disordered regime.
Complete calculation details for panels (a-e) and (f-j) are provided in Appendices~\ref{app:amorphousKramers} and~\ref{app:amorphousMultifold}, respectively.}
\label{fig:Fig4_KW_main_chiralities}
\end{figure*}

The data in Fig.~\ref{fig:Fig3randomKW3F} raise an important question.
Namely, both the random-lattice Kramers-Weyl model in Fig.~\ref{fig:Fig3randomKW3F}(a-d) and the random-lattice multifold model in Fig.~\ref{fig:Fig3randomKW3F}(e-h) respect the same average symmetry group of $\mathcal{T}$ and real-space SO(3) proper rotations.
Nevertheless, the two models exhibit spectrally and topologically distinct features at ${\bf p}={\bf 0}$.
As shown in Appendices~\ref{app:pseudoK},~\ref{app:corepAmorphous},~\ref{app:amorphousKramers}, and~\ref{app:amorphousMultifold}, the amorphous Kramers-Weyl and multifold fermions in Fig.~\ref{fig:Fig3randomKW3F} can in fact be distinguished via the \emph{representations} of the average system symmetry group, most precisely the ${\bf p}={\bf 0}$ average little group $\tilde{G}_{\Gamma,3}$.
Specifically, the amorphous Kramers-Weyl fermion in Fig.~\ref{fig:Fig3randomKW3F}(a-d) transforms in a two-dimensional, double-valued [spinful] small corep of the double group $\tilde{G}_{\Gamma,3}$, whereas the amorphous chiral multifold fermion in Fig.~\ref{fig:Fig3randomKW3F}(e-h) transforms in a three-dimensional, single-valued [spinless] small corep of the single group $\tilde{G}_{\Gamma,3}$.
This demonstrates that even in amorphous systems with the same average symmetry group, different local degrees of freedom and microscopic interactions can give rise to spectral features that transform in different \emph{irreducible coreps} of the average symmetry group, leading to the appearance of topologically distinct states~\cite{marsal_topological_2020}.

Finally, in crystalline chiral TSMs, it has been theoretically and experimentally shown that the low-energy topology of chiral fermions is controlled by the discrete lattice-scale structural chirality [see Refs.~\cite{KramersWeyl,AlPtObserve,PdGaObserve} and the text surrounding Eq.~(\ref{eq:mainChiralCharge})].
To determine whether an analogous relationship holds in amorphous chiral TSMs, we next analyze the topology of the Kramers-Weyl and multifold models with strong structural disorder and varying local chirality order.
Unlike in single crystals, the structural chirality in a structurally disordered system composed of locally chiral patches [\emph{e.g.} sites with $\chi_{\alpha}=\pm 1$ in Eq.~(\ref{eq:temp2siteFrameBreakdownMain})] is a \emph{continuous} parameter, and can be tuned by varying the relative volumes of right-handed [$\chi_{\alpha}=1$] and left-handed [$\chi_{\alpha}=-1$] chirality domains.

\begin{figure*}[t]
\centering
\includegraphics[width=\linewidth]{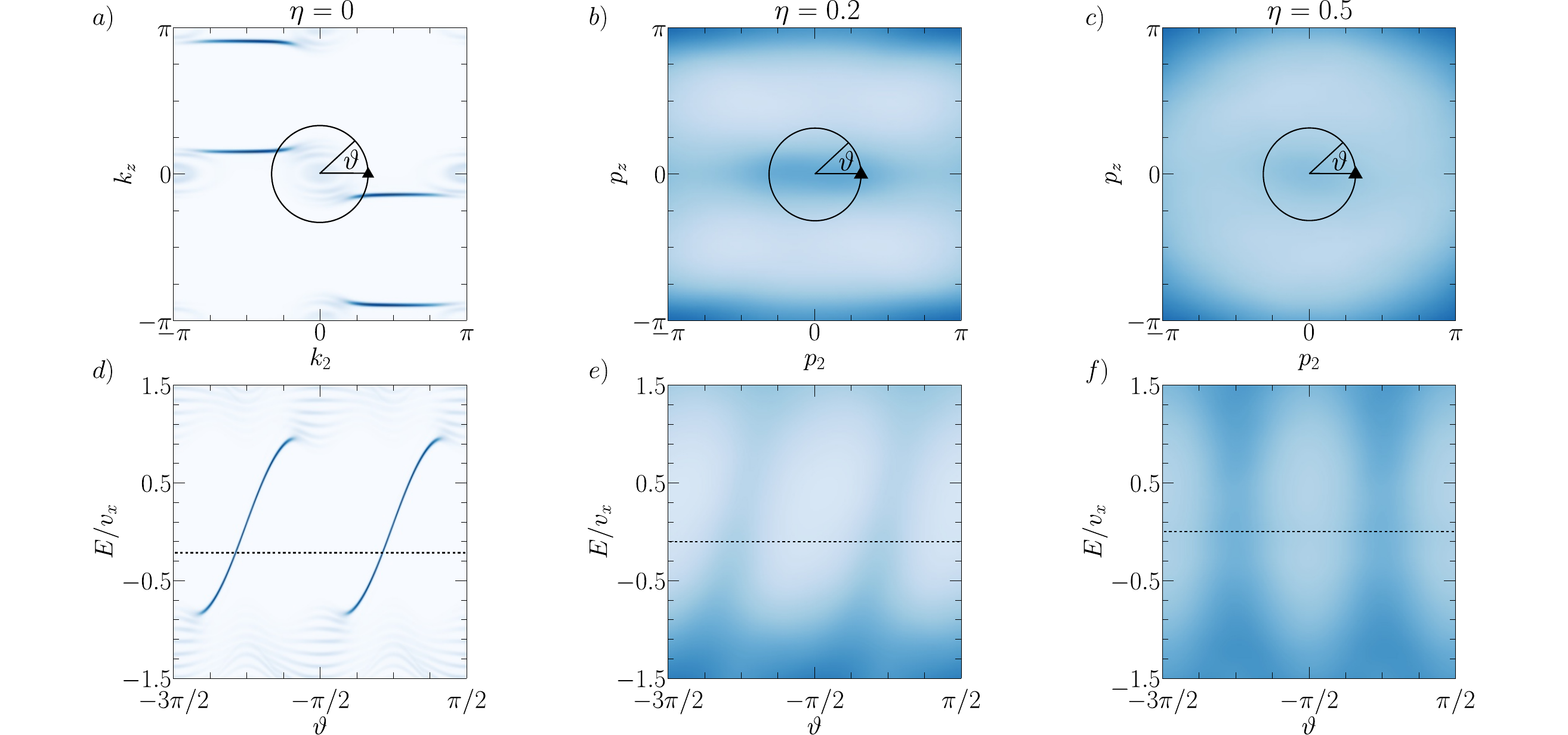}
\caption{\textbf{Absence of topological surface Fermi arcs in amorphous Kramers-Weyl semimetals.}
(a-f) The $(\hat{x}+\hat{y})$-normal surface spectral function $\bar{A}_{\text{surf}}(E,{\bf p})$ [Appendix~\ref{app:surfaceGreens}] averaged over 50 replicas of the non-crystalline Kramers-Weyl model with increasing Gaussian structural disorder parameterized by $\eta$ in Eq.~(\ref{eq:GaussianDisorderMain}) and the same local frame disorder implementation, structural chirality imbalance percentages, and numerical parameters as the bulk calculations in Fig.~\ref{fig:Fig2KWmainspec}.
In (a-c), we plot $\bar{A}_{\text{surf}}(E,{\bf p})$ as a function of $p_{2}=(1/\sqrt{2})(p_{x}-p_{y})$ and $p_{z}$ for increasing $\eta$ at a fixed relative energy $E/v_{x}$ [the dashed line in (d-f)]. 
We then in (d-f) plot $\bar{A}_{\text{surf}}(E,{\bf p})$ as a function of energy on counterclockwise circular paths, parameterized by $\vartheta$, surrounding $p_{2}=p_{z}=0$ for the systems in (a-c), respectively.
For $\eta=0.2$ in (b,e), $\bar{A}_{\text{surf}}(E,{\bf p})$ continues to exhibits two clear -- but diffuse -- Fermi-arc surface states with the same connectivity and topological chirality [positive slopes] as the right-handed enantiomer of the crystalline [$\eta=0$] Kramers-Weyl model in (a,d) [see Appendix~\ref{app:PristineKramers} and Ref.~\cite{KramersWeyl}].
In the strong-disorder regime [$\eta=0.5$, see Appendix~\ref{app:PhysicalObservables}], the Fermi arcs are nearly indistinguishable in (c) the constant-energy spectral function, having largely merged with the bulk Kramers-Weyl Fermi pocket at ${\bf p}={\bf 0}$, as well as the isotropic [sphere-like] bulk spectral features at $|{\bf p}|={\bf \pi}$ and larger momenta in Fig.~\ref{fig:Fig2KWmainspec}(f,i).
The triviality of the strong-disorder surface states can also be seen through their dispersion in (f). 
Specifically, the surface-state Fermi velocities in (f) have become renormalized by disorder to nearly infinite [singular] values, such that the surface spectral features no longer exhibit well-defined 1D topological chirality [positive, non-singular slopes].
Complete calculation details for all panels are provided in Appendix~\ref{app:amorphousKramers}.}
\label{fig:Fig5_KW_arcs}
\end{figure*}

In Fig.~\ref{fig:Fig4_KW_main_chiralities}(a-e), we begin by plotting the ${\bf p}={\bf 0}$ sphere Wilson loop spectrum [see Fig.~\ref{fig:Fig3randomKW3F}(d) and Appendices~\ref{app:EffectiveHamiltonian} and~\ref{sec:WilsonBerry}] of five disorder ensembles of the non-crystalline Kramers-Weyl model with 50 replicas each, where each replica is subject to Gaussian lattice and local frame disorder with $\eta=0.2$ in Eq.~(\ref{eq:GaussianDisorderMain}) and contains contiguous right- and left-handed chirality domains with varying concentrations respectively given by $n_R=N_R/N_{\mathrm{sites}}$ and $n_{L}=1-n_{R}$ [complete calculation details for Fig.~\ref{fig:Fig4_KW_main_chiralities}(a-e) are provided in Appendix~\ref{app:amorphousKramers}].
Beginning in Fig.~\ref{fig:Fig4_KW_main_chiralities}(a) with a disordered Kramers-Weyl system containing almost entirely right-handed sites [$n_{L}=0.05$] and continuing in increasing $n_{L}$ to the system in Fig.~\ref{fig:Fig4_KW_main_chiralities}(e) with almost entirely left-handed sites [$n_{L}=0.95$], we observe a quantized Wilson loop winding of $C=1$ for $n_{L}<0.5$ [Fig.~\ref{fig:Fig4_KW_main_chiralities}(a,b)] and $C=-1$ for $n_{L}>0.5$ [Fig.~\ref{fig:Fig4_KW_main_chiralities}(d,e)].
In the vicinity of $n_{L}\approx 0.5$,
the Wilson loop eigenvalues become non-smooth, indicating that the sphere Wilson loop is within the close vicinity of a topological quantum critical point [energy gap closure]; the onset of this behavior can be seen in Fig.~\ref{fig:Fig4_KW_main_chiralities}(c).
The Wilson loop spectra in Fig.~\ref{fig:Fig4_KW_main_chiralities}(a-e) indicate that the nodal degeneracy at ${\bf p}={\bf 0}$ is a disordered chiral [Kramers-Weyl] fermion that, analogous to its crystalline counterpart [Eq.~(\ref{eq:mainChiralCharge})], carries a topological chirality that is tunable via the average system structural chirality.

To show the generality of this result, in Fig.~\ref{fig:Fig4_KW_main_chiralities}(f-j), we plot the two-band ${\bf p}={\bf 0}$ sphere Wilson spectrum [see Fig.~\ref{fig:Fig3randomKW3F}(h) and Appendices~\ref{app:EffectiveHamiltonian} and~\ref{sec:WilsonBerry}] of five ensembles of the symmorphic multifold model with 50 replicas each, where each replica is subject to Gaussian lattice and local frame disorder with $\eta=0.2$ and again contains contiguous right- and left-handed chirality domains with the respective concentrations $n_R=N_R/N_{\mathrm{sites}}$ and $n_{L}=1-n_{R}$ [complete calculation details for Fig.~\ref{fig:Fig4_KW_main_chiralities}(f-j) are provided in Appendix~\ref{app:amorphousMultifold}].
Beginning in Fig.~\ref{fig:Fig4_KW_main_chiralities}(f) with a multifold system composed almost entirely of right-handed sites [$n_{L}=0.05$] and continuing in increasing $n_{L}$ to the system in Fig.~\ref{fig:Fig4_KW_main_chiralities}(j) with almost entirely left-handed sites [$n_{L}=0.95$], we observe that the two Wilson loop eigenvalues as a set exhibit quantized winding numbers of $C=2$ for $n_{L}<0.5$ [Fig.~\ref{fig:Fig4_KW_main_chiralities}(f,g)] and $C=-2$ for $n_{L}>0.5$ [Fig.~\ref{fig:Fig4_KW_main_chiralities}(i,j)].
In the vicinity of $n_{L}\approx 0.5$, the Wilson loop eigenvalues again become non-smooth [for example at $n_{L}=0.45$ in Fig.~\ref{fig:Fig4_KW_main_chiralities}(h)], implying that the system lies within the close vicinity of an energy gap closure.
The non-Abelian Wilson spectra in Fig.~\ref{fig:Fig4_KW_main_chiralities}(f-j) indicate that the threefold spectral feature at ${\bf p}={\bf 0}$ in the non-crystalline multifold model [Fig.~\ref{fig:Fig3randomKW3F}(e)] is a disordered spin-1 chiral multifold fermion that, like its crystalline counterpart [Appendix~\ref{app:PristineMulifold}] and like the disordered Kramers-Weyl fermion in Fig.~\ref{fig:Fig4_KW_main_chiralities}(a-e), carries a topological chirality that is inherently linked to the average system structural chirality.

Overall, the Wilson loop spectra in Figs.~\ref{fig:Fig3randomKW3F}(d,h) and~\ref{fig:Fig4_KW_main_chiralities} and Appendices~\ref{app:amorphousKramers},~\ref{app:amorphousCharge2}, and~\ref{app:amorphousMultifold} represent a central result of the present work, as they for the first time provide precise \emph{quantized} indicators of nodal [gapless] topology in fully structurally disordered 3D metals, and represent the first calculation of non-Abelian Berry phases [multiband Wilson loop eigenvalues] in 3D amorphous systems. For each of the non-crystalline TSM models in this work [Appendices~\ref{app:amorphousKramers},~\ref{app:amorphousCharge2}, and~\ref{app:amorphousMultifold}], we have further numerically confirmed that the topological chiral charge at ${\bf p}={\bf 0}$ remains quantized and controlled by the ratio $n_{R}/n_{L}$ independent of whether we employ Gaussian disorder or place the model on random and Mikado lattices. 
We have hence more generally shown that the link between low-energy topology and local structural chirality order illustrated in Fig.~\ref{fig:Fig4_KW_main_chiralities} is a reproducible, model-independent property of amorphous chiral TSMs.

{
\vspace{0.2in}
\centerline{\bf IV. DISCUSSION}
\vspace{0.1in}
}

{
\centerline{\bf A. Experimental signatures}
\vspace{0.1in}
}
We conclude by discussing potential experimental observables of chiral TSM states in solid-state amorphous materials.
The most well-recognized signatures of crystalline chiral TSMs are the topological surface Fermi arcs that link the surface projections of oppositely charged bulk chiral fermions~\cite{AshvinWeyl,Armitage2018,AndreiWeyl,SuyangWeyl,YulinWeylExp,AlexeyType2,CDWWeyl,IlyaIdealMagneticWeyl,ZahidLadderMultigap,chang2017large,tang2017CoSi,CoSiObserveJapan,CoSiObserveHasan,CoSiObserveChina,AlPtObserve}.
However, in amorphous chiral TSMs with average structural chirality, we find that the surface Fermi arcs are generically obscured by the surface projections of isotropic bulk spectral features that lie away from ${\bf p}={\bf 0}$ [Appendices~\ref{app:amorphousKramers},~\ref{app:amorphousCharge2}, and~\ref{app:amorphousMultifold}].
This requires us to instead search for bulk experimental signatures of amorphous chiral TSMs.

To understand the absence of surface Fermi arcs in amorphous TSMs with local chirality order, we return to the Gaussian-disordered Kramers-Weyl model previously studied in Fig.~\ref{fig:Fig2KWmainspec}.
In Fig.~\ref{fig:Fig5_KW_arcs}, we plot the $(\hat{x}+\hat{y})$-normal surface spectral function $\bar{A}_{\text{surf}}(E,{\bf p})$ [Appendix~\ref{app:surfaceGreens}] averaged over 50 replicas of the non-crystalline Kramers-Weyl model with increasing Gaussian structural disorder parameterized by $\eta$ in Eq.~(\ref{eq:GaussianDisorderMain}) and the same local frame disorder implementation, structural chirality imbalance percentages, and numerical parameters as the bulk calculations in Fig.~\ref{fig:Fig2KWmainspec} [complete calculation details for Fig.~\ref{fig:Fig5_KW_arcs} are provided in Appendix~\ref{app:amorphousKramers}].
We specifically in Fig.~\ref{fig:Fig5_KW_arcs}(a-c) plot $\bar{A}_{\text{surf}}(E,{\bf p})$ as a function of $p_{2}=(1/\sqrt{2})(p_{x}-p_{y})$ and $p_{z}$, and in Fig.~\ref{fig:Fig5_KW_arcs}(d-f) respectively plot $\bar{A}_{\text{surf}}(E,{\bf p})$ for the same system as a function of energy on counterclockwise circular paths, parameterized by $\vartheta$, surrounding $p_{2}=p_{z}=0$.

In the moderate-disorder regime [$\eta=0.2$ in Fig.~\ref{fig:Fig5_KW_arcs}(b,e)], $\bar{A}_{\text{surf}}(E,{\bf p})$ continues to exhibit two clear -- but increasingly diffuse -- Fermi-arc surface states with the same connectivity and topological chirality [positive slopes] as the right-handed enantiomer of the crystalline Kramers-Weyl model [$\eta=0$ in Fig.~\ref{fig:Fig5_KW_arcs}(a,d)], consistent with the average right-handedness of the disordered system.
More subtly, like the linearly dispersing states that comprise the disordered bulk Kramers-Weyl fermion at ${\bf p}={\bf 0}$ in Fig.~\ref{fig:Fig2KWmainspec}(b,c), the Fermi velocities of the surface Fermi arcs in 
Fig.~\ref{fig:Fig5_KW_arcs}(d-f) are also renormalized to larger values with increasing $\eta$.
This is also reminiscent of the experimentally-observed, nearly vertical, topological Dirac-cone surface states of amorphous Bi$_{2}$Se$_{3}$, whose Fermi velocities compared to those of crystalline Bi$_{2}$Se$_{3}$ are strongly upwardly renormalized by lattice disorder~\cite{corbae_evidence_2020,Ciocys2023}.

In the strong-disorder regime [$\eta=0.5$], the surface Fermi arcs of the Kramers-Weyl model become nearly indistinguishable in the constant-energy spectral function [Fig.~\ref{fig:Fig5_KW_arcs}(c)], having largely merged with the bulk Kramers-Weyl Fermi pocket at ${\bf p}={\bf 0}$, as well as the isotropic [sphere-like] bulk spectral features at $|{\bf p}|= \pi$ and larger momenta [Fig.~\ref{fig:Fig2KWmainspec}(f)].
Intriguingly, the triviality of the strong-disorder surface states can also be inferred through their spectral dispersion [Fig.~\ref{fig:Fig5_KW_arcs}(f)], in which the surface-state Fermi velocities have become renormalized to nearly infinite [singular] values.
Hence, while $\bar{A}_{\text{surf}}(E,{\bf p})$ in Fig.~\ref{fig:Fig5_KW_arcs}(f) appears to still contain surface states, the spectral features in Fig.~\ref{fig:Fig5_KW_arcs}(f) can no longer be designated \emph{topological} surface Fermi arcs, as they no longer exhibit well-defined 1D topological chirality [positive, non-singular slopes].

To reconcile the absence of topological surface Fermi arcs in Fig.~\ref{fig:Fig5_KW_arcs}(c,f) with the presence of bulk chiral [Kramers-Weyl] fermions in Figs.~\ref{fig:Fig2KWmainspec}(c,f,i) and~\ref{fig:Fig3randomKW3F}(a-d), we recall that even in the crystalline Kramers-Weyl model, surface Fermi arcs are not always present.
Specifically, the crystalline Kramers-Weyl model only exhibits surface Fermi arcs if the strength of the Dresselhaus SOC [$v_{x,y,z}$ in Eq.~(\ref{eq:HKWMaink})] exceeds the strength of orbital hopping [$t_{x,y,z}$ in Eq.~(\ref{eq:HKWMaink}), see Appendix~\ref{app:PristineKramers} and Ref.~\cite{KramersWeyl}].
This can be understood by recalling that topological surface Fermi arcs can only appear in chiral TSMs for which the bulk Fermi surface is divided into topologically distinct Fermi pockets that project to different regions of the surface BZ~\cite{AshvinWeyl,Armitage2018,AndreiWeyl,SuyangWeyl,YulinWeylExp,AlexeyType2,CDWWeyl,IlyaIdealMagneticWeyl}.
In the weak-SOC regime of the Kramers-Weyl model, each Kramers-Weyl point is surrounded by a sphere of compensating chiral charge, preventing the appearance of topological Fermi arcs for any choice of surface termination.  
We previously found in Figs.~\ref{fig:Fig2KWmainspec}(f,i) and~\ref{fig:Fig3randomKW3F}(b,c) that in the amorphous regime of the non-crystalline Kramers-Weyl model, the Kramers-Weyl fermion at ${\bf p}={\bf 0}$ similarly becomes fully surrounded by sphere-like bulk Fermi pockets at higher $|{\bf p}|$, even when SOC is strong.
Therefore, like in the weak-SOC crystalline Kramers-Weyl model, the larger-$|{\bf p}|$ isotropic spectral features in the strongly disordered Kramers-Weyl model prevent any surface termination from hosting topologically nontrivial projected bulk Fermi pockets, ensuring the absence of topological surface Fermi arcs.
More generally, we find that similar ring- and sphere-like bulk spectral features also prevent the appearance of topological surface Fermi arcs in the strongly-disordered double-Weyl and multifold fermion models in Appendices~\ref{app:amorphousCharge2} and~\ref{app:amorphousMultifold}, respectively.

\begin{figure*}[t]
\centering
\includegraphics[width=\linewidth]{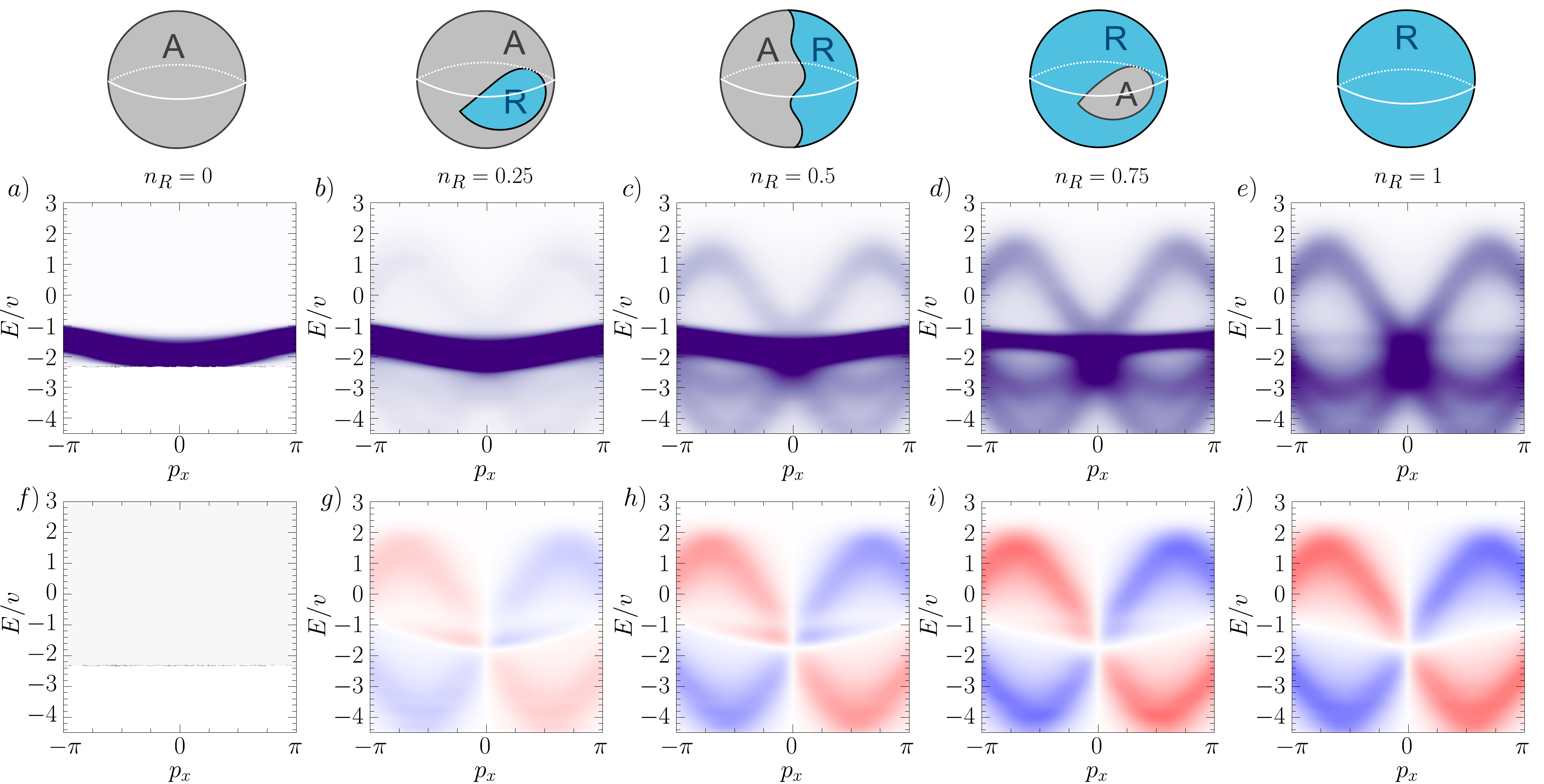}
\caption{\textbf{Bulk spectral signatures of amorphous chiral multifold fermions with coexisting chiral and achiral domains.}
(a-e) The spectral function $\bar{A}(E,\mathbf{p})$ [Eq.~(\ref{eq:SpecFuncMain})] averaged over 50 replicas of the symmorphic multifold model [Eq.~(\ref{eq:HamBloch3FMain}) and Appendix~\ref{sec:Multifold}], where each replica is subject to strong Gaussian structural and local frame disorder parameterized by $\eta=0.5$ in Eq.~(\ref{eq:GaussianDisorderMain}) and contains contiguous right-handed [$\chi_{\alpha}=1$ in Eq.~(\ref{eq:temp2siteFrameBreakdownMain}) and the surrounding text] and structurally \emph{achiral} domains [$\chi_{\alpha}=0$] with varying concentrations respectively given by $n_R=N_R/N_{\mathrm{sites}}$ and $n_{A}=1-n_{R}$ [respectively schematically depicted with blue $R$ and gray $A$ regions in the top row, see Fig.~\ref{fig:Fig1flow}(a,b) and Appendices~\ref{app:DiffTypesDisorder} and~\ref{app:PhysicalObservables}].
Even for (b) a mostly achiral system [$n_{R}=0.25$], faint dispersive spectral features corresponding to the upper and lower bands of the disordered multifold fermion at ${\bf p}={\bf 0}$ [Fig.~\ref{fig:Fig3randomKW3F}(e)] are visible on top of a large background signal of nondispersive trivial states at $E/v \approx -1.5$, and (c-e) become increasingly well-resolved for increasing $n_{R}$.
(f-j) The $\langle L^{x}(E,{\bf p})\rangle$ component of the orbital-angular-momentum- [OAM-] dependent spectral function vector [Eq.~(\ref{eq:OAMDOSMain})] for the systems in (a-e), respectively, plotted using a log-scale color map in which red is positive and blue is negative.
In all panels, the nondispersive trivial states exhibit a vanishing OAM polarization, whereas the dispersive multifold bands show a high degree of OAM polarization, as exemplified by the $n_{R}$-dependent $\langle L^{x}(E,{\bf p})\rangle$ signal in (f-j).   
$\langle L^{x}(E,{\bf p})\rangle$ in (f-j) is specifically nonvanishing for only the dispersive topological [upper and lower] multifold fermion bands, and may hence provide a signature of topological chirality in amorphous systems that is detectable through OAM-sensitive experiments like circular dichroism angle-resolved photoemission spectroscopy [CD-ARPES]~\cite{OAMmultifold2,OAMmultifold3}.
Complete calculation details for all panels are provided in Appendix~\ref{app:amorphousMultifold}.}
\label{fig:Fig6_achiralmain} 
\end{figure*}

We therefore next turn our attention to bulk signatures of amorphous chiral TSMs.
Before performing bulk calculations, we will find it useful to survey potential material realizations of amorphous chiral TSMs, from which we may then identify material-relevant simulation parameters.  
Local structural chirality has previously been proposed to occur in amorphous and glassy solid-state materials, but primarily in a manner in which some -- or even most -- local regions are structurally achiral, and only a few patches are locally chiral.
Specifically, chalcogenide glasses like As$_2$S$_3$ have been found to exhibit optical signatures of structural chirality [optical activity or gyrotropy], which has been suggested to arise from small concentrations [1-10\%] of chiral defect centers~\cite{ChiralGlass1,ChiralGlass3}.
Theoretical studies~\cite{AmorphousTeDFT,AmorphousChalcogenideNatCommH} of amorphous tellurium have also suggested the presence of ring- and chain-like local structural motifs with chiral point group symmetry that coexist with other achiral local structures.
Thin films of the amorphous phase-change material Ge$_2$Sb$_2$Te$_5$ [notably a 3D TI in its crystalline form~\cite{Wieder22}, see~\url{https://topologicalquantumchemistry.com/#/detail/42876}], have further been shown to exhibit photoinduced optical activity, which may also originate from chiral defect centers similar to those proposed in chalcogenide glasses~\cite{ChiralPhaseChange1,ChiralPhaseChange2}.

Inspired by Refs.~\cite{ChiralGlass1,ChiralGlass3,AmorphousTeDFT,AmorphousChalcogenideNatCommH,ChiralPhaseChange1,ChiralPhaseChange2}, and recognizing that the most ideal currently known crystalline chiral TSM materials host $\Gamma$-point multifold fermions~\cite{chang2017large,tang2017CoSi,CoSiObserveJapan,CoSiObserveHasan,CoSiObserveChina,AlPtObserve}, we therefore next construct a disordered multifold system that contains both chiral and \emph{achiral} domains with varying relative concentrations.
To implement a mixed chiral-achiral system, we begin by placing the symmorphic multifold model [Fig.~\ref{fig:MainBands}(e-h), Eq.~(\ref{eq:HamBloch3FMain}), and Appendix~\ref{sec:Multifold}] on a lattice with strong Gaussian structural and local frame disorder [$\eta=0.5$ in Eq.~(\ref{eq:GaussianDisorderMain}), see Fig.~\ref{fig:Fig1flow}(b) and Appendix~\ref{app:DiffTypesDisorder}].
We then generate the momentum-resolved average Green's function $\bar{\mathcal{G}}(E,\mathbf{p})$ [Eq.~(\ref{eq:averageOneMomentumGreenMain})] by averaging the system over 50 disorder replicas.
However, unlike the previous analyses in this work, each replica now contains contiguous domains of either right-handed sites [$\chi_{\alpha}=1$ in Eq.~(\ref{eq:temp2siteFrameBreakdownMain}) and the surrounding text] or achiral sites [$\chi_{\alpha}=0$] with the respective concentrations $n_R=N_R/N_{\mathrm{sites}}$ and $n_{A}=1-n_{R}$.

In Fig.~\ref{fig:Fig6_achiralmain}(a-e), we plot the spectral function $\bar{A}(E,{\bf p})$ [Eq.~(\ref{eq:SpecFuncMain})] of the mixed chiral-achiral multifold system for increasing $n_{R}$ [complete calculation details for Fig.~\ref{fig:Fig6_achiralmain} are provided in Appendix~\ref{app:amorphousMultifold}].
We observe that $\bar{A}(E,{\bf p})$ continues to exhibit dispersive spectral features that correspond to the upper and lower topological bands of the disordered multifold fermion at ${\bf p}={\bf 0}$ [Fig.~\ref{fig:Fig3randomKW3F}(e)].
However, the dispersive topological bands in Fig.~\ref{fig:Fig6_achiralmain}(a-e) also coexist with nondispersive trivial states near $E/v \approx -1.5$, which originate from both the central flat band of the multifold fermion in locally structurally chiral regions with $\chi_{\alpha}=1$, and from achiral regions in which all three bands comprising the multifold fermion are nondispersive [\emph{i.e.} regions for which $\chi_{\alpha}=\chi_{\beta}=0$ such that the chiral OAM coupling term vanishes, see Appendix~\ref{app:amorphousMultifold}].
We further observe in Fig.~\ref{fig:Fig6_achiralmain}(a-e) that the relative spectral weights of the dispersive topological and nondispersive trivial states respectively scale with $n_{R}$ and $n_{A}$.
Notably, even for a mostly achiral system [$n_{R}=0.25$ in Fig.~\ref{fig:Fig6_achiralmain}(b)], faint linearly dispersive bands are visible near ${\bf p}={\bf 0}$ on top of a large background signal of nondispersive trivial states, and become increasingly well-resolved for increasing $n_{R}$ [Fig.~\ref{fig:Fig6_achiralmain}(c-e)].

Importantly, we find that the dispersive topological multifold fermion bands are better revealed by comparing the overall energy spectrum to the OAM-dependent spectrum.
To demonstrate this, we show in Fig.~\ref{fig:Fig6_achiralmain}(f-j) the OAM textures of the mixed chiral-achiral disordered systems, focusing on the $\langle L^{x}(E,{\bf p})\rangle$ component of the OAM-dependent spectral function vector [Eq.~(\ref{eq:OAMDOSMain})]. 
While the central grouping of nondispersive states near $E/v \approx -1.5$ still carries some residual spectral weight in the $\langle L^{x}(E,{\bf p})\rangle$ OAM texture, it is much smaller than that of the dispersive topological bands, \emph{even for small $n_R$}. 
To rule out analysis artifacts due to our choice of the $\langle L^{x}(E,{\bf p})\rangle$ OAM-texture component, we have also in Appendix~\ref{app:amorphousMultifold} computed the magnitude of the OAM-dependent spectral function vector $|\langle {\bf L}(E,{\bf p})\rangle|$ and the total OAM polarization, which even for small $n_{R}$ only show appreciable weight on the two dispersive topological bands.

The pattern in Fig.~\ref{fig:Fig6_achiralmain} of faint -- but highly OAM-polarized -- dispersive topological bands that  coexist with a large background signal of nondispersive trivial bands with vanishing OAM polarization could therefore provide a signature of topological chirality in amorphous systems that is detectable through OAM-sensitive experiments, such as CD-ARPES~\cite{OAMmultifold2,OAMmultifold3}. 
In Appendix~\ref{app:amorphousKramers}, we also investigate mixed chiral-achiral Kramers-Weyl systems [Eqs.~(\ref{eq:amorphousKWTmatrixFinalChiralityMain}),~(\ref{eq:KWHeavisideMain}), and~(\ref{eq:KWmainMats})] and observe that even for large $n_{A}$, the spectrum analogously subdivides into faint, dispersive, highly \emph{spin}-polarized topological bands on top of a large background signal of nondispersive, spin-unpolarized trivial states.
In amorphous 3D materials with strong SOC, dispersive spin-polarized bands could hence provide a signature of amorphous chiral fermions in spin-ARPES, which has proven to be a powerful tool in unraveling the spin textures of crystalline~\cite{HasanKaneReview} and recently amorphous~\cite{corbae_evidence_2020} topological materials.

{
\vspace{0.2in}
\centerline{\bf B. Future directions}
\vspace{0.1in}
}
In this work, we have shown that condensed-matter realizations of chiral fermions can be stabilized by real-space geometry -- specifically local chirality order -- in amorphous solids.
Our extensive analysis bridges the crystalline and strongly disordered regimes of chiral TSMs, and suggests several promising material venues and directions for future study.
First, following the arguments of Ref.~\cite{KramersWeyl}, in real solid-state, nonmagnetic amorphous materials with strong SOC and local chirality order, the presence of large and diverse numbers of occupied and unoccupied atomic orbitals on each site guarantees that there must generically be chiral [\emph{e.g.} Kramers-Weyl] fermions at ${\bf p}={\bf 0}$ \emph{somewhere} in the energy spectrum, though potentially with weak dispersion and lying far from the Fermi level.
Aside from the amorphous monosilicides~\cite{Molinari2023,Rocchino2024}, chalcogenides~\cite{ChiralGlass1,ChiralGlass3,AmorphousTeDFT,AmorphousChalcogenideNatCommH}, and phase-change materials~\cite{ChiralPhaseChange1,ChiralPhaseChange2} discussed earlier in this work, other amorphous solids may also host previously unappreciated structural chirality, which could then serve as a new property by which to classify amorphous materials in online databases~\cite{PerssonAmorphousDatabase}.
Structural chirality has also previously been demonstrated in 3D organic quasicrystals~\cite{LocalChiralityQuasicrystalVirus}, suggesting that more familiar solid-state quasicrystals should be reexamined for structural chirality, and that it may be feasible to engineer structurally chiral quasicrystalline metamaterials.

The numerical methods developed in this work [Appendix~\ref{sec:numericalMethods}] may immediately be applied to search for chiral fermions in first-principles calculations of monatomic amorphous materials, such as amorphous selenium and tellurium~\cite{AmorphousChalcogenideNatCommH}.
In future studies, it would be advantageous to expand the methods in this work to systems with multiple atomic species, and to systematically analyze the models in this work beyond the regime of weak Anderson disorder explored in Appendices~\ref{app:amorphousKramers},~\ref{app:amorphousCharge2}, and~\ref{app:amorphousMultifold}.
Furthermore, as the model-independent topological machinery developed in this work is limited to amorphous chiral fermions near ${\bf p}={\bf 0}$  [Appendices~\ref{app:EffectiveHamiltonian} and~\ref{sec:WilsonBerry}], it would be useful in future works to develop topological invariants based solely on the exact two-momentum Green's function $\mathcal{G}(E,{\bf p},{\bf p}')$ [Eq.~(\ref{eq:MomGreenFuncMain})], whose limitations could then be directly compared to those of topological invariants computed from single-particle Green's functions in strongly correlated systems~\cite{MottFailureBradlynPhilipps,MottFailureGoldman}.
Larger-$|{\bf p}|$ topological invariants beyond the mean-field effective Hamiltonian $\mathcal{H}_\text{Eff}({\bf p})$ [Eq.~(\ref{eq:AvgHEffMain}) and Appendix~\ref{app:EffectiveHamiltonian}] may also shed light on whether the Nielsen-Ninomiya chiral fermion doubling theorem still holds on amorphous lattices, as originally questioned in Ref.~\cite{NielNino81b}.
Lastly, though we have focused on amorphous chiral fermions with the same ${\bf k}\cdot {\bf p}$ [effective] Hamiltonians as $\Gamma$-point crystalline chiral fermions, the non-crystalline group theory developed in Appendices~\ref{app:pseudoK} and~\ref{app:corepAmorphous} may also be employed to predict realizations of non-crystalline chiral fermions without crystalline counterparts, such as spin-2 chiral fermions.

Finally, beyond topology, a larger urgent question is whether structural chirality can be definitively induced and quantitatively modeled in amorphous solid-state materials.
Local chirality domains have numerically been shown to be energetically favorable in liquid crystals~\cite{LocalChiralityDomainLiquidCrystal}, and local chirality order has already been induced by explicit construction in structurally and orientationally disordered metamaterials~\cite{lindell1994electromagneticBook}.
Local chirality in first-principles calculations of amorphous materials could potentially be quantitatively identified by implementing the density-correlation-based structural chirality parameters detailed in Ref.~\cite{KamienLubenskyChiralParameter}.
Numerous industry-relevant responses have been demonstrated in crystals with intertwined structural and topological chirality~\cite{deJuanAdolfoCPGE,ReesCPGEObserve,OAMmultifold2,OAMmultifold3,ChiralCatalysisPdGa,ClaudiaChiralCatalysis,ClaudiaNatureEnergyCatalysis}, and should be revisited in candidate amorphous materials with local chirality order.
For example, amorphous metal chalcogenides have already been shown to be excellent catalysts~\cite{AmorphousMetalChalcogenide}, and could potentially become enantioselective under the presence of local chirality order.  
Structurally chiral amorphous materials may also provide a route towards realizing non-crystalline generalizations of other phenomena that are closely related to condensed-matter chiral fermions, like 3D chiral phonons~\cite{ChiralPhononQuartz}. 
Furthermore, if realized at low or zero frequency, phonon generalizations of the ${\bf p}={\bf 0}$ non-crystalline chiral fermions identified in this work would also be relevant to the jamming transition~\cite{SussmanJammedDiracWeyl}. Lastly, it would be interesting to examine whether the electrically induced amorphization recently reported in Ref.~\cite{RiteshElectricalAmorphous} could be coupled with additional external fields or chiral substrates to generate simultaneously tunable structural and local chirality order in amorphous materials.

{
\vspace{0.2in}
\centerline{\bf ACKNOWLEDGMENTS}
\vspace{0.1in}
}
We thank Selma Franca, Frances Hellman, Siddhartha Sarkar, Victor Miles, Sachin Vaidya, Niels Schr\"{o}ter, Olga Arroyo Gasc\'{o}n, and Julie Karel for helpful discussions.
We further thank Luis Elcoro and Mois Arroyo for helpful correspondence regarding non-crystallographic rod groups.  
During the long preparation of this work, non-crystalline Weyl points were also predicted in inherently 3D moir\'{e} lattices~\cite{3DMoireWeylPRL,3DMoireWeylIOP}. 
B.~J.~W. acknowledges support from the European Union’s Horizon Europe research and innovation program (ERC-StG-101117835-TopoRosetta), ANR PIA funding (ANR-20-IDEES-0002), and CNRS IRP Project NP-Strong.
B.~J.~W. further acknowledges the Laboratoire de Physique des Solides, Orsay for hosting during the preparation of this work.
A.~G.~G. and J.~S. acknowledge financial support from the European Research Council (ERC) Consolidator grant under grant agreement No. 101042707 (TOPOMORPH). 
J.~S. also acknowledges support from the program QuanTEdu-France No. ANR-22-CMAS-0001 France 2030.

{
\vspace{0.2in}
\centerline{\bf DATA AVAILABILITY}
\vspace{0.1in}
}

The data supporting the theoretical findings of this study are available within the paper, and the codes and data needed to reproduce the figures are provided in Ref.~\cite{Amorphous_zenodo2026}. 
The numerical calculations were performed using the Kwant package~\cite{Groth_2014_kwant}.

\clearpage
\onecolumngrid

\begin{appendix}

\begin{center}
{\bf Supplemental Material for ``Geometry-Enforced Topological Chiral Fermions in Amorphous Chiral Metals''}
\end{center}

\tableofcontents

\setcounter{secnumdepth}{5}
\renewcommand{\theparagraph}{\bf \thesubsubsection.\arabic{paragraph}}
\renewcommand{\theHfigure}{S\arabic{figure}}

\renewcommand{\thefigure}{S\arabic{figure}}
\setcounter{figure}{0} 

\clearpage

\section{Symmetry Group Theory and Nodal Degeneracies}
\label{app:groupTheory}

In this Appendix, we will establish terminology for the group-theoretic, geometric, and topological quantities used in this work.  
First, in Appendix~\ref{app:symDefs}, we will define system symmetry groups, which most familiarly include space groups. 
We will then establish a definition -- employed throughout the rest of this work -- for the symmetry groups of structurally chiral objects.  
Next, in Appendix~\ref{app:corepDefs}, we will define the little groups and small [co]representations [coreps] of symmetry groups with lattice translation symmetry.  
In Appendix~\ref{app:corepDefs} we will also review the well-established correspondence between little-group small coreps and robust nodal degeneracies at specific crystal momenta [${\bf k}$ points] in band structures.

We will next shift focus to systems with strong structural disorder, such as amorphous solids.  
In Appendix~\ref{app:pseudoK}, we will first introduce the concept of approximate \emph{pseudo-momenta} ${\bf p}$, which index plane-wave states that can be used to construct a [pseudo-] momentum-space spectral characterization of a solid without translation symmetry~\cite{varjas_topological_2019,marsal_topological_2020,marsal_obstructed_2022,JustinHat}.
Next, in Appendix~\ref{app:pseudoK}, we will generalize the concept of little groups at ${\bf k}$ points in translationally-invariant solids to approximate [average] little groups at ${\bf p}$ points in disordered [amorphous] solids.  
Lastly, in Appendix~\ref{app:corepAmorphous}, we will characterize the small coreps of the average little groups in amorphous systems and their relationship to [topologically chiral] nodal degeneracies in both pristine and disordered solids.

\subsection{Symmetry and Chirality Group Theory Definitions}
\label{app:symDefs}

In this section, we will first define system symmetry groups.  We will then establish a definition for the subset of symmetry groups that characterize odd-dimensional objects with structural chirality [\emph{i.e.} handedness]~\cite{FlackChiral}.  To summarize this section, \emph{in this work, we have employed a simplified nomenclature in which a chiral symmetry group is defined as a group containing the symmetries that leave invariant a structurally chiral object}.  Below, we will discuss the discrepancies between this nomenclature and other symmetry-group nomenclatures in the crystallographic literature.

To begin, we consider a system composed of objects with well-defined [\emph{i.e.} exponentially localized] positions ${\bf r}_{\alpha}$.  
For the purposes of this discussion, we consider the system to be a solid-state material or metamaterial, as opposed to a simplified tight-binding model, and divide the system into smaller units termed local regions [patches].
Each local region can then be classified as whether it lies near the system boundary, or deeper within the bulk, by comparing its distance from the physical system boundary to interaction [hopping] ranges and the correlation lengths of quasiparticle excitations [\emph{e.g.} the spatial extent of topological surface states]~\cite{ProdanKohnNearsighted}.
Each realistic system carries a \emph{dimensionality} that is defined by whether the system appears to be finite or effectively infinite from the perspective of bulk local regions.  
For now, we will also demand that the bulk system be \emph{translationally invariant} in all of the directions in which it is effectively infinite as defined above.
Later, beginning in Appendix~\ref{app:pseudoK}, we will relax this constraint to accommodate the amorphous systems studied in this work.

The Hamiltonian of a bulk system subject to the constraints above is left invariant under [\emph{i.e.} respects] a group of symmetries $G$, where throughout this work, we will follow the convention of Ref.~\cite{MTQC} and use the capital letters $G$ and $H$ to denote symmetry groups, with the meaning of each symbol defined specifically within the [sub]section in which it appears.
In the above case in which the system bulk is countably infinite and translationally invariant in some or all directions, the system is termed a \emph{crystal}, and formally respects the symmetries of a symmetry group $G$ that contains the group $G_{T}$ of integer lattice translations in $d_{T}$ linearly independent directions~\cite{BigBook,MTQC}.
As emphasized above, the bulk may also have a spatial extent [\emph{i.e.} thickness] into additional dimensions beyond its $d_{T}$-dimensional lattice translations [or more formally, be embedded in a higher-dimensional space~\cite{WiederLayers,DiracInsulator,LayerGroupDegeneracyEnumeration}], such that the system Hamiltonian respects symmetries, such as reflections and rotations, that act in a $d$-dimensional space, where $d\geq d_{T}$.
For example, in a few-layer, quasi-2D solid-state device with exact in-plane 2D lattice translation symmetries [\emph{e.g.} a device with aligned and commensurate layers, as opposed to quasicrystalline or moir\'{e} order], $d=3$, but $d_{T}=2$.
When $d=d_{T}$, $G$ is termed a [Shubnikov] space group [SG] if $d=3$, a wallpaper or plane group if $d=2$, and a line group if $d=1$~\cite{BigBook,MTQC,WiederLayers,DiracInsulator,IUCrVolASG,MagneticBook,ConwaySymmetries}. 
Conversely if $d>d_{T}$, then $G$ is termed a \emph{subperiodic group}~\cite{ConwaySymmetries,MagneticBook,subperiodicTables,WiederLayers,HingeSM,LayerGroupDegeneracyEnumeration}.
For future discussions, we will find it helpful to define the $d-d_{T}$ linearly independent finite dimensions of a crystalline system as $d_{F}$, such that:
\begin{equation}
 d = d_{T} + d_{F},
\label{eq:finiteDimNoAmorph}
\end{equation} 
for any system whose bulk is either translationally invariant or finite in every linearly-independent direction.
In the above example of a translationally-invariant few-layer device with $d=3$ and $d_{T}=2$ [and hence $d_{F}=1$], $G$ is termed a \emph{layer group}~\cite{subperiodicTables,WiederLayers,LayerGroupDegeneracyEnumeration}.
Next, if $d=3$ and $d_{T}=1$ [and hence $d_{F}=2$], $G$ is termed a \emph{rod group}~\cite{subperiodicTables,MagneticBook,HingeSM,linegroupsbook,eightfoldRodGroups}.
Lastly if $d_{T}=0$, then for any values of $d$ and $d_{F}$, $G$ is termed a \emph{point group}~\cite{ConwaySymmetries,BigBook,BilbaoPoint,PointGroupTables,SpinPointMcClarty}.

If $d$ is odd, then the position-space system may carry a discrete \emph{structural chirality} or handedness~\cite{FlackChiral,linegroupsbook}.  
For the case of $d=3$ [3D] relevant to solid-state materials, objects are considered to be structurally chiral if $G$ does not contain rotoinversion [\emph{i.e.} improper rotation] symmetries $g$ of the form:
\begin{equation}
g = \{\mathcal{I}\times C_{ni}|{\bf v}\},
\label{eq:rotoinversion}
\end{equation}
where $\mathcal{I}$ is 3D spatial inversion, $C_{ni}$ is a proper rotation by $360 / n$ degrees about the $i$-axis, and ${\bf v}$ is a translation operation that acts on the system after the [point group] operation of $\mathcal{I}\times C_{ni}$~\cite{FlackChiral}.
Well-known rotoinversion symmetries include inversion itself $\mathcal{I}=\mathcal{I}\times C_{1i}$ [more specifically $\mathcal{I}=\mathcal{I}\times C^{2}_{1i}$ in spin-orbit-coupled systems], as well  as mirror reflection $M_{i}=\mathcal{I}\times C_{2i}$~\cite{BigBook}.

A structurally chiral 3D system with a symmetry group $G_{R}$ may be arbitrarily defined as a ``right-handed enantiomer'' by fixing a convention for a handed mathematical function of the atomic coordinates ${\bf r}_{\alpha}$ [such as the sign of a cross product] that is chosen to match, for example, the handedness of helical structures within a material or the rotation of linearly polarized light in an optical activity experiment~\cite{FlackChiral,WatsonCrickDNA,KamienLubenskyChiralParameter,linegroupsbook,MurakamiChiralPhonon,ProsAndConsChiralityMeasurements,MolecularChiralityMeasurementTools}.  
Acting with a rotoinversion $g\notin G_{R}$ [Eq.~(\ref{eq:rotoinversion})] will then convert the right-handed system with the symmetry group $G_{R}$ into a left-handed [structurally-chiral] enantiomer with a symmetry group $G_{L}$.

\begin{figure}
\centering
\includegraphics[height=7cm,keepaspectratio]{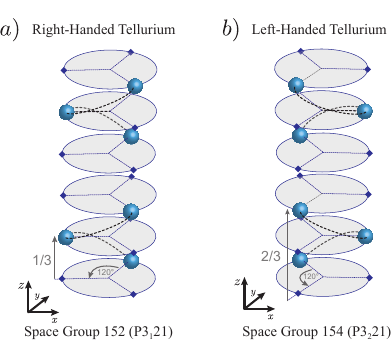}  
\caption{Example chiral space groups that are distinct for right- and left-handed material enantiomers.
(a) The crystal structure of right-handed chiral tellurium~\cite{OriginalChiralTellurium,ChiralTellurium1,ChiralTellurium2}.
The right-handed crystal structure in (a) respects the symmetries of chiral space group [SG] 152 ($P3_{1}21$), which is generated by lattice translations in the $(x,y)$-plane and a screw symmetry $\{C_{3z}|00\frac{1}{3}\}$ that consists of a $120$-degree, right-handed rotation about the $z$-axis followed by a one-third lattice translation in the $z$-direction [gray arrows in (a)].
(b) The crystal structure of left-handed chiral tellurium.
The left-handed crystal structure in (b) respects the symmetries of chiral SG 154 ($P3_{2}21$), which is generated by lattice translations in the $(x,y)$-plane and a screw symmetry $\{C_{3z}|00\frac{2}{3}\}$ that consists of a $120$-degree, right-handed rotation about the $z$-axis followed by a \emph{two-thirds} lattice translation in the $z$-direction [gray arrows in (b)].
Despite the crystal structures in (a,b) simply representing mirror- or inversion-related enantiomers of the same material, they respect the symmetries of distinct chiral SGs.}
\label{figapp:tellurium}
\end{figure}

In recent years, several competing definitions have been introduced for the symmetry groups $G_{R,L}$ of structurally chiral systems.
Within the crystallography community, the International Union of Crystallography [IUCr] refers to the 65 nonmagnetic SGs of structurally chiral crystals as the \emph{Sohncke groups}~\cite{FlackChiral,IUCrVolASG,ClaudiaChiralReview1,ClaudiaChiralReview2}.  The IUCr then terms the subset of Sohncke groups for which $G_{R}$ is not isomorphic to $G_{L}$ as the 22 chiral SGs, which come in enantiomorphic pairs such as SG 152 ($P3_{1}21$) and SG 154 ($P3_{2}21$) [see~\url{https://dictionary.iucr.org/Chiral_space_group} for further details].  For example, SG 152 ($P3_{1}21$) -- notably the SG of right-handed chiral tellurium~\cite{OriginalChiralTellurium,ChiralTellurium1,ChiralTellurium2} -- contains the screw symmetry $\{C_{3z}|00\frac{1}{3}\}$, which consists of a right-handed rotation by $120$ degrees about the $z$-axis followed by a one-third lattice translation in the $z$ direction [Fig.~\ref{figapp:tellurium}(a)].
Conversely, SG 154 ($P3_{2}21$) -- the SG of left-handed chiral tellurium~\cite{OriginalChiralTellurium,ChiralTellurium1,ChiralTellurium2} -- contains a different screw symmetry $\{C_{3z}|00\frac{2}{3}\}$, which instead consists of a right-handed rotation by $120$ degrees about the $z$-axis followed by a two-thirds lattice translation in the $z$ direction [Fig.~\ref{figapp:tellurium}(b)].
Because $\mathcal{I}$ commutes with $C_{3z}$ but exchanges the signs of translations, acting with $\mathcal{I}$ on $\{C_{3z}|00\frac{1}{3}\}$ in SG 152 ($P3_{1}21$) converts it to $\{C_{3z}|00-\frac{1}{3}\}=\{E|00-1\}\{C_{3z}|00\frac{2}{3}\}$ [where $E$ is the identity element], which is instead a symmetry element of SG 154 ($P3_{2}21$) and no longer an element of SG 152 ($P3_{1}21$).  
Hence, $G_{R}=P3_{1}21$ and $G_{L}=P3_{2}21$ are enantiomorphic pairs of Sohncke groups that are exchanged by the action of $\mathcal{I}$.

However, there also exist other Sohncke SGs for which acting with $\mathcal{I}$ sends all of the elements in the SG to elements in the \emph{same} SG, such that $G_{R}$ is isomorphic to $G_{L}$.  The most notable example of such a case is simply SG 1 ($P1$), which is only generated by integer lattice translations, but nevertheless characterizes structurally chiral crystals [with the lowest possible symmetry].
Consequently, under the IUCr definition, there exist structurally chiral crystals -- like those with only the symmetries of SG 1 ($P1$) -- whose SGs are \emph{not} termed ``chiral.''

During the recent period of rapid development in understanding the intertwined roles of topology and structural chirality in crystalline materials, some members of the topological materials community adopted a different, simplified definition in which the symmetry groups that admit structurally chiral objects [\emph{i.e.} those without rotoinversion symmetries, see the text surrounding Eq.~(\ref{eq:rotoinversion})] are instead termed \emph{chiral symmetry groups}~\cite{KramersWeyl,ZahidLadderMultigap,ZahidNatRevMatWeyl,MTQCmaterials,CDWWeyl,AndreiMaterials2,LargeScaleMultifoldSearch}.  
Hence unlike under the IUCr definition, the 65 Sohncke SGs are equivalently termed \emph{chiral SGs} under the simplified topological materials definition of chiral symmetry groups.

In this work, we will not only study structurally chiral crystals with $d=d_{T}=3$, but we will also study $d=3$ structurally chiral -- \emph{but otherwise strongly disordered [amorphous]} -- systems with $d_{T}=0,1$, for which the term ``Sohncke group'' does not apply.  
Therefore, for brevity and consistency with recent works on topological chiral crystals~\cite{KramersWeyl,ZahidLadderMultigap,ZahidNatRevMatWeyl,MTQCmaterials,CDWWeyl,AndreiMaterials2,LargeScaleMultifoldSearch}, we will \emph{not} in this work employ the IUCr nomenclature for chiral symmetry groups. 
Instead, we will adopt a purposefully simplified nomenclature in which a chiral symmetry group is succinctly defined as one without rotoinversion symmetries of the form of Eq.~(\ref{eq:rotoinversion}).
In the nomenclature employed in this work, chiral symmetry groups are therefore simply defined as the symmetry groups of structurally chiral objects with $d=3$ and varying $d_{T}$.

\subsection{Little Groups, Small Corepresentations, and Nodal Degeneracies with Translation Symmetry}
\label{app:corepDefs}

In this section, we will review the relationship, previously established in Refs.~\cite{AbrikosovNewFermion,Young3DDirac,ManesNewFermion,DDP,NewFermions,chang2017large,tang2017CoSi,KramersWeyl,Young2DDirac,Armitage2018,DiracInsulator,Bradlyn2017,MTQC,YoungMagnetic,CanoMagneticNewFermion,DoubleWeylPhonon,MurakamiChargeFour,ParticleDictionary,ExhaustiveKdotP,MagneticKramersWeyl,InigoMultifoldReview}, between the small corepresentations of little groups and nodal degeneracies in band structures. 
We begin by considering a system that respects the symmetries of a symmetry group $H$ that only contains unitary symmetry elements.
Hence, $H$ is isomorphic to a Type-I Shubnikov magnetic symmetry group~\cite{BigBook,MagneticBook,MTQC,MTQCmaterials}.
Each \emph{representation} $\sigma$ of $H$ is defined by the \emph{characters} of each unitary symmetry $h\in H$, which are themselves given by the sum of the eigenvalues of $h$ over a set of energy eigenstates that \emph{transform} in $\sigma$.
A Hamiltonian that respects the symmetries of $H$ is then also termed to transform in $\sigma$ if it is spanned by energy eigenstates with the same characters [summed unitary symmetry eigenvalues] as $\sigma$.
The \emph{dimension} of $\sigma$ is given by the character of the identity element $E\in H$.
Most importantly, the smallest [lowest-dimensional] representations of $H$ with linearly independent [unitary] symmetry characters [\emph{i.e.} eigenvalue sums] are termed the \emph{irreducible representations} [irreps] of $H$~\cite{BigBook,MTQC,MTQCmaterials,Wigner1932,wigner1959group,ZakBandrep1,ZakBandrep2,Bradlyn2017,Bandrep1,millerTables}.
If the energy eigenstates of the system characterize integer-angular-momentum [``spinless'' fermionic or bosonic] quasiparticle excitations, then $H$ is termed a \emph{single} symmetry group and $\sigma$ is termed a \emph{single-valued representation}~\cite{BigBook,Bradlyn2017,Bandrep1,MTQC}. 
Conversely, if the energy eigenstates characterize [``spinful'' fermionic] quasiparticle excitations with half-integer angular momenta, then $H$ is termed a \emph{double} symmetry group and $\sigma$ is termed a \emph{double-valued representation}.

We next consider a symmetry group $G$ that is a supergroup of $H$:
\begin{equation}
\label{eqapp:supergroup}
G = H \cup g_{A} H,
\end{equation}
where $g_{A}$ is an antiunitary symmetry, such as time-reversal [$\mathcal{T}$] symmetry [more specifically $\{\mathcal{T}|{\bf 0}\}$ in the notation of Eq.~(\ref{eq:rotoinversion})]. 
Depending on the details of $g_{A}$, $G$ may either be isomorphic to a magnetic [$g_{A}\neq \{\mathcal{T}|{\bf 0}\}$] or a nonmagnetic [$g_{A}=\{\mathcal{T}|{\bf 0}\}$] Shubnikov symmetry group.  
\emph{Whether or not their dimension is doubled compared to the irreducible representations of the subgroup $H$ of $G$}, the irreducible representations of $G$ are termed \emph{irreducible corepresentations} [coreps], because $G$ contains antiunitary symmetry elements~\cite{MTQC,millerTables,BigBook}.

Next, we consider the Fourier-transformed [crystal] momentum- [${\bf k}$-] space description of the crystalline system with the symmetry group $G$.  
At each ${\bf k}$ point, the reciprocal-space Hamiltonian $\mathcal{H}_{\bf k}$ of the system is left invariant under the symmetries of the \emph{little group} $G_{\bf k}\subseteq G$~\cite{BigBook,MTQC}.  
$G_{\bf k}$ is isomorphic to a symmetry group with the same $d$ and $d_{T}$ as $G$ [see Eq.~(\ref{eq:finiteDimNoAmorph}) and the surrounding text], and is specifically defined as the subgroup containing both the unitary and antiunitary symmetries $g\in G$ that leave ${\bf k}$ invariant up to an integer linear combination of the reciprocal lattice vectors ${\bf K}_{\mu}$:
\begin{equation}
g{\bf k} \text{ mod } {\bf K}_{\mu} = {\bf k},
\label{eq:EquivKPoints}
\end{equation}
for all $g\in G_{\bf k}$ and all of the $d_{T}$ values of $\mu$ [\emph{i.e.} the number of linearly independent lattice translation directions, and hence the number of linearly independent reciprocal lattice vectors ${\bf K}_{\mu}]$.
Along with the statement that:
\begin{equation}
\mathcal{T}{\bf k}=-{\bf k},
\label{eq:TonK}
\end{equation}
Eq.~(\ref{eq:EquivKPoints}) indicates that in a $\mathcal{T}$-invariant [nonmagnetic] crystal with $d_{T}$ lattice translations, there are $2^{d_{T}}$ $\mathcal{T}$-invariant ${\bf k}$ [TRIM] points that lie [in reduced crystal momentum units] at $k_{\mu}=0,\pi$ for each of the $d_{T}$ linearly independent components $k_{\mu}$ of ${\bf k}$~\cite{BigBook}.

Because $G_{\bf k}$ is isomorphic to a symmetry group with translations [$d_{T}>0$], then there are infinitely many irreducible coreps of $G_{\bf k}$. 
Here and below, we will take $G_{\bf k}$ to contain antiunitary symmetry elements, because we specifically in this work focus on nonmagnetic [$\mathcal{T}$-invariant] topological semimetals whose nodal degeneracies lie at $\mathcal{T}$-invariant [crystal] momenta.
Though $G_{\bf k}$ has infinitely many coreps, the infinite set of irreducible coreps of $G_{\bf k}$ can be generated from a \emph{finite} set of irreducible \emph{small coreps} to which all other irreducible coreps are unitarily related [equivalent] via the matrix representatives of lattice translations [Bloch phases]~\cite{BigBook,Bradlyn2017,Bandrep1,MTQC,millerTables}.  
At a high-symmetry ${\bf k}$ point, the degeneracy of each band multiplet is furthermore in one-to-one correspondence with the dimension of a small corep $\tilde{\sigma}_{\bf k}$ of $G_{\bf k}$~\cite{AbrikosovNewFermion,Young3DDirac,ManesNewFermion,DDP,NewFermions,chang2017large,tang2017CoSi,KramersWeyl,Young2DDirac,Armitage2018,DiracInsulator,Bradlyn2017,MTQC,YoungMagnetic,CanoMagneticNewFermion,DoubleWeylPhonon,MurakamiChargeFour,ParticleDictionary,ExhaustiveKdotP,MagneticKramersWeyl,InigoMultifoldReview}. 
To see this, consider the low-energy Hamiltonian $\mathcal{H}({\bf k})$ of a band degeneracy at a point ${\bf k}$.  For each unitary symmetry $h\in G_{\bf k}$, the \emph{unitary matrix representative} of $h$ in the small corep $\tilde{\sigma}_{\bf k}$ corresponding to $\mathcal{H}({\bf k})$ is expressed as $\Delta_{\tilde{\sigma}_{\bf k}}(h)$, and satisfies~\cite{DiracInsulator}:
\begin{equation}
\Delta_{\tilde{\sigma}_{\bf k}}(h)\mathcal{H}(h^{-1}{\bf k})\Delta^{\dag}_{\tilde{\sigma}_{\bf k}}(h)=\Delta_{\tilde{\sigma}_{\bf k}}(h)\mathcal{H}({\bf k})\Delta^{\dag}_{\tilde{\sigma}_{\bf k}}(h)=\mathcal{H}({\bf k}).
\label{eq:unitaryLittleGroup}
\end{equation}
Additionally, for each antiunitary symmetry $g_{A}\in G_{\bf k}$, there also exists an \emph{antiunitary matrix representative} of $g_{A}$ in the small corep $\tilde{\sigma}_{\bf k}$ corresponding to $\mathcal{H}({\bf k})$, which can be expressed as $\Delta_{\tilde{\sigma}_{\bf k}}(g_{A})=U_{\tilde{\sigma}_{\bf k}}(g_{A})K$, where $U_{\tilde{\sigma}_{\bf k}}(g_{A})$ is a unitary matrix and $K$ is complex conjugation.
As with the matrix representatives $\Delta_{\tilde{\sigma}_{\bf k}}(h)$ of the unitary symmetries $h\in G_{\bf k}$, the antiunitary matrix representatives $\Delta_{\tilde{\sigma}_{\bf k}}(g_{A})$ satisfy a similar relation to Eq.~(\ref{eq:unitaryLittleGroup})~\cite{DiracInsulator}:
\begin{equation}
\Delta_{\tilde{\sigma}_{\bf k}}(g_{A})\mathcal{H}(g_{A}^{-1}{\bf k})\Delta^{-1}_{\tilde{\sigma}_{\bf k}}(g_{A}) = \Delta_{\tilde{\sigma}_{\bf k}}(g_{A})\mathcal{H}({\bf k})\Delta^{-1}_{\tilde{\sigma}_{\bf k}}(g_{A}) = U_{\tilde{\sigma}_{\bf k}}(g_{A})\mathcal{H}^{*}({\bf k})U^{\dag}_{\tilde{\sigma}_{\bf k}}(g_{A}) = \mathcal{H}({\bf k}).
\label{eq:antiUnitaryLittleGroup}
\end{equation}

Importantly, if $\tilde{\sigma}_{\bf k}$ is an \emph{irreducible} small corep of $G_{\bf k}$, then there exists a strong constraint on the possible matrices $\Lambda$ that commute with $\Delta_{\tilde{\sigma}_{\bf k}}$.
Specifically, for both reducible and irreducible coreps $\tilde{\sigma}_{\bf k}$ of $G_{\bf k}$, there generically exist matrices $\Lambda$ for which:
\begin{equation}
\Delta_{\tilde{\sigma}_{\bf k}}(g)\Lambda\Delta^{-1}_{\tilde{\sigma}_{\bf k}}(g)=\Lambda,
\label{eq:irreducibleCorepDef}
\end{equation}
for all unitary and antiunitary symmetries $g\in G_{\bf k}$. 
However, in an irreducible corep -- and not in a reducible corep -- the only matrices $\Lambda$ that satisfy Eq.~(\ref{eq:irreducibleCorepDef}) are those proportional to the identity matrix $\Lambda\propto \mathds{1}$.
This implies that $\mathcal{H}({\bf k})\propto\mathds{1}\propto\Delta_{\tilde{\sigma}_{\bf k}}(E)$ where $E\in G_{\bf k}$ is the identity element.

Eq.~(\ref{eq:irreducibleCorepDef}) and the surrounding text indicate that, given a [Bloch] energy eigenstate, $|\psi^{i}_{\bf k}\rangle$ in the Hilbert space of a Hamiltonian $\mathcal{H}({\bf k})$ transforming in an irreducible small corep $\tilde{\sigma}_{\bf k}$, the matrix representative of each symmetry $\Delta_{\tilde{\sigma}_{\bf k}}(g)$ either leaves $|\psi^{i}_{\bf k}\rangle$ invariant up to a phase, or takes $|\psi^{i}_{\bf k}\rangle$ to a different [orthogonal] energy eigenstate $|\psi^{j}_{\bf k}\rangle$.
Using Eqs.~(\ref{eq:unitaryLittleGroup}) and~(\ref{eq:antiUnitaryLittleGroup}), this implies that generically, for each unitary and antiunitary symmetry $g\in G_{\bf k}$:
\begin{equation}
\mathcal{H}({\bf k})\Delta_{\tilde{\sigma}_{\bf k}}(g)|\psi^{i}_{\bf k}\rangle = \mathcal{H}({\bf k})|\psi^{j}_{\bf k}\rangle = E^{j}_{\bf k}|\psi^{j}_{\bf k}\rangle = \Delta_{\tilde{\sigma}_{\bf k}}(g)\mathcal{H}({\bf k})|\psi^{i}_{\bf k}\rangle = E^{i}_{\bf k}\Delta_{\tilde{\sigma}_{\bf k}}(g)|\psi^{i}_{\bf k}\rangle = E^{i}_{\bf k} |\psi^{j}_{\bf k}\rangle,
\label{eq:corepToEnergyEDegen}
\end{equation}
indicating that the energies are equal $E^{i}_{\bf k}=E^{j}_{\bf k}$ for all of the Bloch states in the low-energy subspace of $\mathcal{H}({\bf k})$ that together transform in the same irreducible small corep $\tilde{\sigma}_{\bf k}$.

Finally, if $G_{\bf k}$ lies at a high-symmetry ${\bf k}$ point, then the little groups $G_{{\bf k} + {\bf q}}$ at nearby crystal momenta ${\bf k} + {\bf q}$ will have lower symmetry than $G_{\bf k}$, and hence the bands at ${\bf k} + {\bf q}$ are likely [though not guaranteed] to have lower degeneracy [be split] compared to those at ${\bf k}$.  
Exploiting this reasoning, numerous previous works have established an effective equivalence between high-symmetry-${\bf k}$-point coreps and \emph{nodal band degeneracies}~\cite{AbrikosovNewFermion,Young3DDirac,ManesNewFermion,DDP,NewFermions,chang2017large,tang2017CoSi,KramersWeyl,Young2DDirac,Armitage2018,DiracInsulator,Bradlyn2017,MTQC,YoungMagnetic,CanoMagneticNewFermion,DoubleWeylPhonon,MurakamiChargeFour,ParticleDictionary,ExhaustiveKdotP,MagneticKramersWeyl,InigoMultifoldReview}.  
More generally, the set of all possible band degeneracies at all ${\bf k}$ points is almost entirely characterized by small coreps, with the remaining cases consisting of topological nodal degeneracies for which bands transforming in the same small coreps are prevented from anticrossing by a topological invariant defined on a loop or a sphere enclosing the crossing point~\cite{Bradlyn2017,MTQC,AshvinWeyl,Z2Pack,Armitage2018,YoungkukDiracLineNode,ChenWithWithoutMonopole,AhnMonopole,TomasBZDWilsonMonopole,TMDHOTI,Wieder22,BinghaiClaudiaWeylReview,ZahidNatRevMatWeyl}.
In the specific case of \emph{topologically chiral} nodal degeneracies, such as conventional Weyl fermions~\cite{AshvinWeyl,VolovikWeylMonopole,VolovikChiralAnomaly,HaldaneOriginalWeyl,MurakamiWeyl,BurkovBalents,AndreiWeyl,HasanWeylDFT,Armitage2018,SuyangWeyl,LvWeylExp,YulinWeylExp,AliWeylQPI,AlexeyType2,ZJType2,BinghaiClaudiaWeylReview,ZahidNatRevMatWeyl,CDWWeyl,IlyaIdealMagneticWeyl}, the integral of the Berry curvature flux on a sphere surrounding the degeneracy is quantized to a nonzero integer, which is termed the \emph{chiral charge} of the nodal degeneracy [see also Appendix~\ref{sec:WilsonBerry}].
It was recognized in Ref.~\cite{KramersWeyl} that because the chiral charge of a nodal point is reversed by a rotoinversion symmetry of the form of Eq.~(\ref{eq:rotoinversion}), then a topologically chiral nodal degeneracy can only exist at a point ${\bf k}$ if $G_{\bf k}$ is isomorphic to a chiral SG [as defined in Appendix~\ref{app:symDefs}].
Furthermore, because the little group $G_{\bf k}$ at any {\bf k} point is a subgroup of the full system SG $G_{\bf k}\subseteq G$, then $G_{\bf k}$ cannot contain rotoinversion symmetries if $G$ does not also contain rotoinversion symmetries.
Hence, in a structurally chiral crystal for which $G$ is a chiral SG, \emph{all} of the little groups $G_{\bf k}$ are also isomorphic to chiral SGs, and therefore topologically chiral nodal degeneracies are permitted at \emph{all} ${\bf k}$ points.
Using this result, the authors of Ref.~\cite{KramersWeyl} then further showed that \emph{every} point-like nodal degeneracy in a nonmagnetic [$\mathcal{T}$-invariant] structurally chiral crystal necessarily carries a nontrivial chiral charge, a key result that we will use in this work as a guide towards constructing and modeling amorphous topologically chiral semimetals.

\subsection{Pseudo-Momentum Space, Average Symmetry Groups, and Average Little Groups in Amorphous Solids}
\label{app:pseudoK}

Motivated by the observation of mesoscopic response and spectroscopic signatures consistent with topological insulator states in strongly disordered and amorphous samples of bismuth-based materials~\cite{TIamorphousThinFilm,corbae_evidence_2020,Ciocys2023,Banerjee:2017jd,DC2018,Ramaswamy:2019bp,Osmic2024,Andrino2025} and non-crystalline metamaterials~\cite{Mitchell2018,Liu2020,Zhou2020,Jia2023,ZhangDelplace2023scattering}, and by the appearance of topological mesoscopic response in highly non-crystalline samples of well-established topological semimetals~\cite{Li2021aWTex,Fujiwara2023kagome,KarelAmorphousBerryCMG,Asir2025,Molinari2023,Rocchino2024}, we next extend our analysis to amorphous systems.  
Having established in Appendices~\ref{app:symDefs} and~\ref{app:corepDefs} the position-space symmetry groups and momentum-space little groups of translationally-invariant crystals, we will specifically in this section establish the analogous group-theoretic quantities in amorphous solids.  
To begin, we consider a $d$-dimensional system with $d_{T}$ directions of exact translation symmetry, as we did previously in Appendix~\ref{app:symDefs}.
As discussed in the text preceding Eq.~(\ref{eq:finiteDimNoAmorph}), we again here take the system to be a realistic solid-state material or metamaterial with a finite -- but potentially large -- spatial extent in all directions.
We therefore again separate the system into local regions [patches], and specifically consider regions deep within the bulk as defined relative to interaction distances and excitation correlation lengths~\cite{ProdanKohnNearsighted}.

\begin{figure}
\centering    
\includegraphics[width=\linewidth]{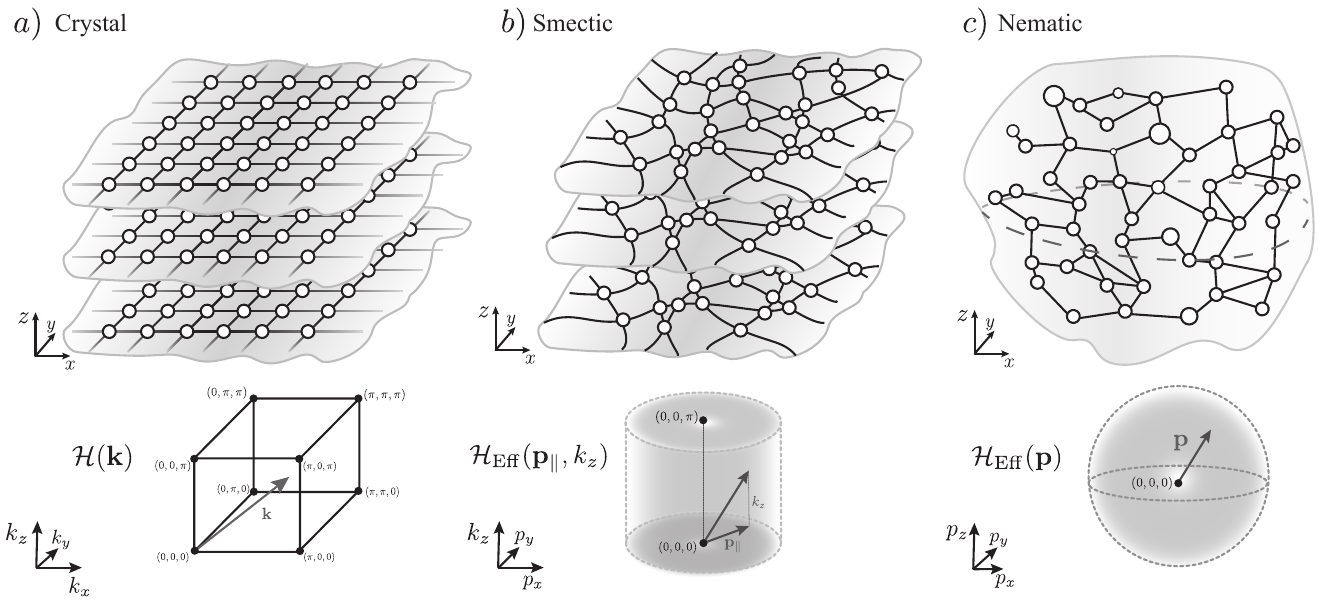}
\caption{\label{fig:dadtdf}Real and momentum space in 3D systems with and without translation symmetry.
(a-c) Examples of $d=3$, $d_{f}=0$ crystalline and non-crystalline systems with varying $d_{T}$ and $d_{A}$ in Eq.~(\ref{eq:finiteDimAmorph}).
(a) Real- and momentum-space description of a \emph{crystal} with $d_{T}=3$, $d_{A}=0$.
The system in (a) respects lattice translation symmetries in the $x$-, $y$-, and $z$-directions.
The Fourier-transformed Bloch Hamiltonian $\mathcal{H}(\bf{k})$ captures the exact energy spectrum, and is parameterized by a crystal momentum vector ${\bf k}$ lying in the $3$-torus of the first Brillouin zone [BZ, ${\bf k}\in \mathbb{T}^{3}$].
For the 3D crystal in (a), there are eight values of ${\bf k}$ [$k_{x,y,z}=0,\pi$] that are returned to themselves by reciprocal lattice vectors and time-reversal symmetry [TRIM points, see the text surrounding Eq.~(\ref{eq:TonK})].
(b) Real- and [pseudo-] momentum-space description of a 3D disordered system with $d_{T}=1$, $d_{A}=2$. 
The system in (b) respects lattice translation symmetry in only the $z$-direction, and is disordered in the remaining two directions in the $(x,y)$-plane.
As detailed in Appendix~\ref{app:SmecticNematicDisorder}, throughout this work, we refer to $d=3$, $d_{T}=1$, $d_{A}=2$, $d_{f}=0$ systems with average chirality as exhibiting \emph{smectic} disorder, owing to their combination of chirality [Ising-like orientational] order, 1D lattice translation symmetry, and [optionally] partial local reference frame order.
Unlike in (a), the effective Hamiltonian $\mathcal{H}_{\text{Eff}}(\mathbf{p}_{\parallel},{k}_z)$ in (b) only approximates the many-particle energy spectrum [see Appendix~\ref{app:EffectiveHamiltonian}], and is parameterized by one crystal momentum $k_{z}$ lying in the $1$-torus [circle] of the first 1D BZ [$k_{z}\in (\mathbb{T}^{1}=\mathbb{S}^{1})$] and two approximate pseudo-momenta ${\bf p}_{\parallel}= (p_{x},p_{y})$ with unbounded, real values [${\bf p}_{\parallel}\in\mathbb{R}^{2}$].
Further unlike in (a), the partially disordered system in (b) has only two TRIM points [${\bf p}_{\parallel}={\bf 0}$ for $k_{z}=0,\pi$], because ${\bf p}_{\parallel}$ is defined for all real values, and not modulo reciprocal lattice vectors [see the text following Eq.~(\ref{eq:TonP})].
(c) Real- and pseudo-momentum-space description of a 3D disordered system with $d_{T}=0$, $d_{A}=3$.
The system in (c) is disordered in all directions in 3D space. 
As detailed in Appendix~\ref{app:SmecticNematicDisorder}, throughout this work, we refer to $d=3$, $d_{T}=0$, $d_{A}=3$, $d_{f}=0$ systems with average chirality as exhibiting \emph{nematic} disorder, owing to their combination of chirality [``orientational''] order and complete structural disorder.
Like in (b), and unlike in (a), the effective Hamiltonian 
$\mathcal{H}_{\text{Eff}}(\mathbf{p})$ in (c) only approximates the many-particle energy spectrum.
In the system in (c) with complete structural disorder, $\mathcal{H}_{\text{Eff}}(\mathbf{p})$ is parameterized by an unbounded 3D pseudo-momentum vector ${\bf p}\in \mathbb{R}^{3}$ with only one TRIM point [${\bf p}={\bf 0}$].
}
\end{figure}

However unlike previously in Appendix~\ref{app:symDefs}, we next consider the possibility that there also exist linearly independent directions in which the bulk system is effectively infinite, but \emph{not} translationally invariant.  
Instead, we now allow the system to be infinite -- but not periodic -- in a $d_{A}$-dimensional system [sub]space of \emph{non-crystalline} directions, where we have employed the subscript $A$ in anticipation of numerically analyzing systems that are in particular \emph{amorphous} in their non-crystalline infinite directions.
In its $d_{A}$-dimensional infinite [sub]space, we now allow the system to exhibit quasicrystalline or moir\'{e} structural order, or even, as indicated above, to be strongly disordered [\emph{i.e.} amorphous].
Below, in Appendix~\ref{app:PhysicalObservables}, we will rigorously numerically define strong structural disorder as disorder strengths at which the off-diagonal-in-momentum elements of the replica-averaged Fourier-transformed matrix Green's function $\bar{\mathcal{G}}(E,{\bf p},{\bf p}')$ become much smaller than the smallest elements of the diagonal-in-momentum average Green's function $\bar{\mathcal{G}}(E,{\bf p})=\bar{\mathcal{G}}(E,{\bf p},{\bf p})$ [see the text surrounding Eq.~(\ref{eq:MomGreenFunc})].
Lastly, beyond the $d_{T}+d_{A}$ directions in which the system bulk is effectively infinite, the bulk is necessarily finite in the remaining $d-d_{T}-d_{A}$ linearly independent directions, which we designate as the $d_{f}$ finite directions of the non-crystalline system.
Hence as a refinement to Eq.~(\ref{eq:finiteDimNoAmorph}) for bulk, non-crystalline [\emph{e.g.} amorphous or quasicrystalline] systems, any $d$-dimensional system that is translationally invariant in $d_{T}$ directions and disordered and effectively infinite in $d_{A}$ directions satisfies:
\begin{equation}
 d = d_{T} + d_{A} + d_{f}.
\label{eq:finiteDimAmorph}
\end{equation}

For example, a free-standing 2D system embedded in 3D space may be translationally invariant in both in-plane directions, such that:
\begin{equation}
d=3,\ d_{T}=2,\ d_{A}=0,\ d_{f}=1.
\label{eq:2T0A1f}
\end{equation}
Alternatively, the system may be strongly disordered in one in-plane direction, such that:
\begin{equation}
d=3,\ d_{T}=1,\ d_{A}=1,\ d_{f}=1,
\label{eq:1T1A1f}
\end{equation}
or the system may be strongly disordered in both in-plane directions, such that:
\begin{equation}
d=3,\ d_{T}=0,\ d_{A}=2,\ d_{f}=1.
\label{eq:0T2A1f}
\end{equation}
In the three examples above, $d_{f}$ is restricted to be $1$, because the $d=3$ bulk system inhabits 3D space, but has only two effectively infinite directions in which it may either be translationally invariant or amorphous. 
In the remaining $d_{f}=1$ direction, the system is comparably thin [\emph{e.g.} thinner than the correlation lengths of topological boundary states on its top and bottom surfaces, or even atomically thin].
In Fig.~\ref{fig:dadtdf}, we provide additional examples of $d=3$ non-crystalline systems that are relevant to the amorphous topological semimetals studied in this work.

In a bulk system with $d_{A}\neq 0$, the exact energy bands [dispersion relation] along the $d_{A}$ system directions cannot be solved and plotted as a function of crystal momenta ${\bf k}$, because integer lattice translations are not system symmetries in the disordered $d_{A}$ system directions.
However, as shown in Refs.~\cite{varjas_topological_2019,marsal_topological_2020,marsal_obstructed_2022} and further detailed in Appendix~\ref{app:EffectiveHamiltonian}, an \emph{effective} momentum-space Hamiltonian $\mathcal{H}_{\text{Eff}}({\bf p})$ can be generated by computing the single-particle Green's function of the full system at a series of \emph{approximate [pseudo-] momenta} ${\bf p}$ drawn from a continuous space of plane waves, where the units of ${\bf p}$ are defined relative to average nearest-neighbor intersite distances~\cite{corbae_evidence_2020,Ciocys2023}.
Unlike the periodic crystal momentum ${\bf k}$, which lies in the $d_{T}$-torus of the first Brillouin zone [${\bf k}\in \mathbb{T}^{{d}_{T}}$], the plane-wave momenta ${\bf p}$ carry unbounded values in the $d_{A}$-dimensional space of real numbers [${\bf p}\in \mathbb{R}^{{d}_{A}}$, see Fig.~\ref{fig:dadtdf}].
The eigenspectrum of $\mathcal{H}_{\text{Eff}}({\bf p})$, owing to its origin from Green's functions and approximate pseudo-momenta, is only an approximation of the actual energy eigenspectrum.
Specifically, the matrix elements of the effective Hamiltonian $\mathcal{H}_{\text{Eff}}({\bf p})$ are in general quite sensitive to numerical details including the energy cut 
at which $\mathcal{H}_{\text{Eff}}({\bf p})$ is computed and the number and ${\bf p}$-dependence of the functions used to generate $\mathcal{H}_{\text{Eff}}({\bf p})$ [such as moments in the kernel polynomial method~\cite{Weisse06}, see the Supporting Information for Ref.~\cite{marsal_topological_2020} and Appendix~\ref{app:EffectiveHamiltonian} for further details].

Nevertheless, previous works have shown that the one-momentum, Fourier-transformed Green's function $\mathcal{G}(E,{\bf p})$ can still exhibit sharp and topologically meaningful spectral features that are accurately captured by $\mathcal{H}_\text{Eff}({\bf p})$, including bulk and surface [amorphous] nodal degeneracies with Dirac-like dispersion~\cite{marsal_topological_2020,corbae_evidence_2020,JustinHat,Ciocys2023}, and especially near ${\bf p}= {\bf 0}$, at which the spectral function computed from $\mathcal{G}(E,{\bf p})$ is sharpest [see Fig.~\ref{fig:HeffBreak}(a) for the corresponding result obtained in this work for the replica-averaged, one-momentum Green's function $\bar{\mathcal{G}}(E,{\bf p})$].
Below and in Appendix~\ref{app:corepAmorphous}, we will show that the approximate [average] system symmetry group still carries an approximate notion of little groups which, like their crystalline counterparts [Appendix~\ref{app:corepDefs}], are characterized by small coreps corresponding to [topological] nodal degeneracies in the spectrum of the pseudo-momentum- [${\bf p}$-] space effective Hamiltonian $\mathcal{H}_\text{Eff}({\bf p})$.
Importantly, as we will show in Appendix~\ref{app:EffectiveHamiltonian}, we specifically find that the low-energy [${\bf k}\cdot{\bf p}$] dispersion relations and topological chiral charges of ${\bf p}={\bf 0}$ nodal degeneracies in strongly disordered systems with chiral average symmetry groups are \emph{insensitive} to the reference energy cut and number of harmonic-function moments used to generate $\mathcal{H}_\text{Eff}({\bf p})$.
This provides strong evidence that the single-particle topology that we numerically extract in this work via the long-wavelength approximate [single-particle effective] Hamiltonian $\mathcal{H}_\text{Eff}({\bf p}\approx {\bf 0})$ provides an accurate descriptor of the actual [many-particle] topology of disordered chiral metals near ${\bf p} ={\bf 0}$.

\paragraph*{\bf Average Symmetry Groups} -- $\ $ We will next discuss the symmetry group structure and chirality of average symmetry groups in amorphous systems. 
We will introduce these concepts in the context of representative examples of bulk amorphous systems with $d=3$, $d_{A}=2$, and $d_{T}=0$ [such that $d_{f}=1$] in Eq.~(\ref{eq:finiteDimAmorph}), which can be interpreted as 2D disordered systems with a sense of the out-of-plane direction~\cite{WiederLayers,LayerGroupDegeneracyEnumeration}.

To begin, we consider an amorphous system with $d=3$, $d_{A}=2$, $d_{T}=0$, and $d_{f}=1$ in Eq.~(\ref{eq:finiteDimAmorph}).
We next define the 2D space of disordered atoms to be spanned by vectors in the $(x,y)$-plane.
We then note that by approximating $\mathcal{H}_{\text{Eff}}({\bf p})$ to be parameterized by a \emph{continuous, non-periodic} momentum parameter ${\bf p}$ drawn from vectors in the 2D real plane, we are implicitly approximating the effective Hamiltonian $\mathcal{H}_{\text{Eff}}({\bf p})$ to be the Fourier transform of a 2D position-space Hamiltonian with \emph{continuous} translation symmetry in $d_{A}=2$ linearly independent spatial directions.
Hence, the average symmetry group [ASG] $\tilde{G}$ necessarily contains the group $\tilde{G}_{T,d_{A}}$ of linearly independent continuous translations in $d_{A}$ real-space directions:
\begin{equation}
\tilde{G}_{T,d_{A}}\subseteq \tilde{G}.
\label{eq:ASG}
\end{equation}

In an amorphous system, the approximation of continuous translation symmetry $\tilde{G}_{T,d_{A}}$ carries a $|{\bf p}|$-dependent range of validity: the approximation that a disordered system is a continuous medium is expected to be more valid at long wavelengths and hence small momenta near $|{\bf p}|=0$ at which the system appears relatively homogeneous along the $d_{A}$ spatial directions. At short wavelengths [and hence large momenta $|{\bf p}|\rightarrow \infty$], lattice- [average-nearest-neighbor-] scale inhomogeneities will have a strong effect and likely invalidate the approximation of continuous translation symmetry.
This can be seen in the spectral function computed from $\mathcal{G}(E,{\bf p})$ in an amorphous system, which exhibits relatively sharp features near ${\bf p}={\bf 0}$, but becomes diffuse as $|{\bf p}|\rightarrow\infty$ [see Refs.~\cite{marsal_topological_2020,corbae_evidence_2020,JustinHat,Ciocys2023} and Fig.~\ref{fig:HeffBreak}(a)].
In this work, we primarily focus on ${\bf p}={\bf 0}$ chiral fermions in the amorphous setting, which we find to exhibit dispersion relations and topology that are well captured by the long-wavelength effective [single-particle] Hamiltonian $\mathcal{H}_{\text{Eff}}({\bf p})$ within a perturbative range of ${\bf p} = {\bf 0}$.

It is important to emphasize that approximate [average] $\tilde{G}_{T,d_{A}}$ symmetry does not necessarily guarantee that the system is isotropic in the $d_{A}$-dimensional space of approximate continuous translation symmetries.
For example, in a 2D nematic liquid crystal that lies in the $(x,y)$-plane and consists of rods that [on the average] point in the $y$-direction, there are still approximately two linearly independent continuous translation symmetries in the $x$- and $y$-directions, even though symmetries that relate the $x$- and $y$-directions are broken by the internal [rod-director] orientational order~\cite{chaikinLubenskyBook,CollingsGoodbyLiquidCrystalBook,KamienLiquidCrystalRMP,KamienChiralLiquidCrystal,ChiralNematicBook}.

\begin{figure}[t]
\centering    
\includegraphics[width=0.9\linewidth]{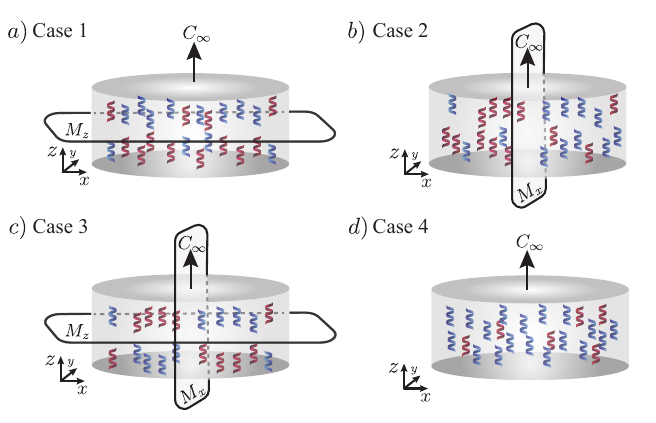}
\caption{Average symmetry and chirality in quasi-2D amorphous systems.
In this figure, we depict four representative [but not exhaustive] cases of average symmetry groups [ASGs] $\tilde{G}$ that contain only [average] unitary crystal symmetries in amorphous systems with $d=3$, $d_{A}=2$, $d_{T}=0$, $d_{f}=1$ in Eq.~(\ref{eq:finiteDimAmorph}).
(a-d) For each $d=3$ system in this figure, $d_{A}=2$ indicates that the system is structurally disordered in the $(x,y)$-plane, and $d_{f}=1$ indicates that the system bulk is relatively thin in the $z$-direction, as detailed in the text following Eq.~(\ref{eq:0T2A1f}).
Each panel schematically depicts a system slice at fixed $y$, where the red and blue helices represent local structurally chiral regions [patches] within the slice that are respectively right- and left-handed.
Cases with equal numbers of mirror-related red and blue helices are structurally achiral on the average, whereas an imbalance between red and blue helices indicates average [net] chirality [see Appendix~\ref{app:DiffTypesDisorder} and the text surrounding Eqs.~(\ref{eq:dA2case}) and~(\ref{eq:KWnumericalModelDiscreteChi})].
As discussed in the text surrounding Eq.~(\ref{eq:cInftyGenerator}), amorphous systems are isotropic on the average across their $d_{A}$ disordered directions.
(a-d) For the $d_{A}=2$ systems in this figure, the average system isotropy manifests through the statement that $\tilde{G}$ contains the group $C_{\infty}$ ($\infty$) of continuous [proper] rotation symmetry about the $z$-axis [$C_{\infty}\subset \tilde{G}$, see the text surrounding Eq.~(\ref{eq:cInftyGenerator})].
(a) Case 1: $\tilde{G}$ contains [average] mirror reflection [improper rotation symmetry, see the text surrounding Eq.~(\ref{eq:rotoinversion})] about the $z$-axis [$M_{z}$, see Eq.~(\ref{eq:case1})], as well as 3D spatial inversion symmetry [$\mathcal{I}$] due to the combination of $M_{z}$ and twofold rotation symmetry $C_{2z}\in C_{\infty}$ [Eq.~(\ref{eq:ImirrorBreakdown})].
$\tilde{G}$ in (a) is therefore achiral.
(b) Case 2: $\tilde{G}$ contains mirror reflection symmetry about the $x$-axis [$M_{x}$], as well as reflections about all normal vectors in the $(x,y)$-plane due to $C_{\infty}$ symmetry [Eq.~(\ref{eq:case2})].
$\tilde{G}$ in (b) is therefore also achiral.
(c) Case 3: $\tilde{G}$ contains the combined symmetries of (a,b) Cases 1 and 2 [Eq.~(\ref{eq:case3})], and is therefore achiral.
(d) Case 4: $\tilde{G}$ does not contain any rotoinversion symmetries, unlike the systems in (a-c) [Eq.~(\ref{eq:2DchiralAmorphous})].  
As recognized in this work and defined in the text following Eq.~(\ref{eq:dA2case}), $\tilde{G}$ in (d) Case 4 is therefore \emph{chiral}~\cite{lindell1994electromagneticBook}, and is specifically left-handed, on the average.}
\label{fig:average_sym}
\end{figure}

For amorphous systems, however, further constraints arise via the emergence of additional approximate symmetries in $\tilde{G}$.
Specifically, in addition to being homogeneous at long wavelengths due to strong structural disorder, amorphous systems without internal orientational order [such as uniform or net ferromagnetism] are also spectrally \emph{isotropic} at long wavelengths in their $d_{A}$ ${\bf p}$-space directions~\cite{zallen_physics_1998,VanMechelen:2018cy,vanMechelenNonlocal,Ciocys2023,Grushin2020,Corbae_2023}.
Importantly and central to the present work, this does not imply that the eigenstates of $\mathcal{H}_{\text{Eff}}({\bf p})$ in an amorphous system are identical along different pseudo-momentum directions.
Instead, the isotropic nature of the long-wavelength effective Hamiltonian is implemented through emergent, infinity-fold [continuous] \emph{proper} rotation symmetries~\cite{spring_amorphous_2021,springMagneticAverageTI,corbae_evidence_2020}.
For example in the case of $d_{A}=2$ [Fig.~\ref{fig:average_sym}], $\mathcal{H}_{\text{Eff}}({\bf p})$ in the vicinity of $|{\bf p}|=0$ on the average respects the symmetries of the non-crystallographic~\cite{varjas_topological_2019,QuasiCrystalHOTI1,QuasiCrystalHOTI2,QuasiCrystalHOTI3,AmorphousHOTI1,AmorphousHOTI2,Else2021} continuous rotation symmetry [point] group $C_{\infty}$ ($\infty$)~\cite{linegroupsbook,PointGroupTables,SpinPointMcClarty}, where here and throughout this work, we adopt the labeling convention of the Bilbao Crystallographic Server and list point groups by their symbol in the Schoenflies notation [\emph{e.g.} $C_{n}$] followed in parentheses by their symbol in the Hermann-Mauguin [international] notation [\emph{e.g.} ($n$)]~\cite{BilbaoPoint}.
The point group $C_{\infty}$ ($\infty$) is specifically generated by:
\begin{equation}
C_{(2\pi/\phi) z} = \{C_{(2\pi/\phi) z}|{\bf 0}\},
\label{eq:cInftyGenerator}
\end{equation}
where $\phi$ is here an infinitesimal angle about the $z$-axis.
Rather than requiring the eigenstates of $\mathcal{H}_{\text{Eff}}({\bf p})$ to be identical along different directions, the elements of $C_{\infty}$ ($\infty$) unitarily relate their spin-orbital textures and hence, as recognized in this work, also relate their Berry curvature textures in pseudo-momentum space.
For $d_{A}=2$ systems, this can be summarized by stating that the system ASG $\tilde{G}$ contains [or is isomorphic to] the supergroup generated by combining infinitesimal [continuous] proper rotations $C_{(2\pi/\phi) z}$ and the group $\tilde{G}_{T,2}$ of continuous 2D translations:
\begin{equation}
\tilde{G}_{T,2}\cup C_{(2\pi/\phi) z}\tilde{G}_{T,2}\subseteq\tilde{G}.
\label{eq:dA2case}
\end{equation}

We note that in the case of $d_{A}=3$, $\mathcal{H}_{\text{Eff}}({\bf p})$ in the vicinity of $|{\bf p}|=0$ further [on the average] respects the 3D continuous proper rotation symmetries in the supergroup SO(3) of $C_{\infty}$ ($\infty$).
We will shortly detail below in the text surrounding Eq.~(\ref{eq:nematicALGbreakdown}) an example of a $d_{A}=3$, ${\bf p}\approx{\bf 0}$ effective Hamiltonian with average SO(3) rotation symmetry.
Lastly, analogous to its exact crystal symmetry group counterpart $G$ [Appendix~\ref{app:symDefs}], we designate in this work a $d=3$ ASG $\tilde{G}$ to be a \emph{chiral ASG} if $\tilde{G}$ does not contain [average] rotoinversion symmetries of the form of Eq.~(\ref{eq:rotoinversion}) [which are notably absent in $\tilde{G}_{T,2}$ and $C_{\infty}$ ($\infty$), as well as in $\tilde{G}_{T,3}$ and SO(3)].

To provide further intuition for the above example of quasi-2D amorphous systems with $d=3$, $d_{A}=2$, $d_{T}=0$, and $d_{f}=1$, and for now taking $\tilde{G}$ to only contain unitary symmetries, we next consider four representative [but not exhaustive] cases for the system ASG $\tilde{G}$, which we have schematically depicted in Fig.~\ref{fig:average_sym}:
\begin{itemize}
\item \textbf{Case 1:} $\tilde{G}$ contains a reflection [improper rotation symmetry] with an out-of-plane [$z$-direction] normal vector [Fig.~\ref{fig:average_sym}(a)]:
\begin{equation}
\{M_{z}|{\bf 0}\}\in\tilde{G}.
\label{eq:case1}
\end{equation}  
Because the continuous rotation group $C_{\infty}$ ($\infty$) contains twofold rotation symmetry about the $z$-axis [$C_{2z}\in C_{\infty}$], then 3D spatial inversion is also an element of the ASG $\{\mathcal{I}|{\bf 0}\}\in\tilde{G}$, owing to the relationship between $M_{z}$ and $C_{2z}$~\cite{WiederLayers}: 
\begin{equation}
\mathcal{I}=M_{z}\times C^{-1}_{2z} = M_{z}C^{-1}_{2z},
\label{eq:ImirrorBreakdown}
\end{equation}
where the second equality is expressed in the more common symmetry operator product notation in which $\times$ symbols are suppressed~\cite{BigBook}.
The presence of $\mathcal{I}$ and $M_{z}$ in the ASG $\tilde{G}$ indicates that in this case, the system ASG is not chiral, because $\mathcal{I}$ and $M_{z}$ are rotoinversion symmetries of the form of Eq.~(\ref{eq:rotoinversion}).
From a topological perspective, because $M_{z}$ symmetry can protect Dirac-cone surface [edge] states on 1D edges with normal vectors lying in the $(x,y)$-plane, strongly disordered insulators in this case can host $M_{z}$-protected topological crystalline insulator [TCI] states that can be smoothly connected to well-established topologically stable 2D TCIs with mirror reflection and translation symmetry~\cite{TeoFuKaneTCI,Fu2011,LiangIsobeInteractingTCI,ZhidaHermeleTopologicalCrystals,Song:2018cj,MTQC,PengChenDiagnosisMagneticPRB,ma2023averageSPT1,AltlandBrouwerDisorderedTCI,ZhidaIntrinsicDisorderedAXI}.
\item \textbf{Case 2:}  $\tilde{G}$ contains the complete set of reflections with in-plane normal vectors, including $M_{x}$ and $M_{y}$, but $\tilde{G}$ does not contain $M_{z}$ symmetry [Fig.~\ref{fig:average_sym}(b)]:
\begin{equation}
\{M_{x}|{\bf 0}\}\in\tilde{G},\ \{M_{y}|{\bf 0}\}\in\tilde{G},\ \{M_{z}|{\bf 0}\}\notin\tilde{G},
\label{eq:case2}
\end{equation}  
noting that $M_{x,y}$ are related by the 90-degree rotation symmetry $C_{4z}\in C_{\infinity}$.
Because the system ASG $\tilde{G}$ contains [infinitely many] rotoinversion symmetries of the form of Eq.~(\ref{eq:rotoinversion}), then $\tilde{G}$ is in this case not chiral. 
Furthermore, unlike Case 1 above, $\tilde{G}$ does not contain $\mathcal{I}$ symmetry [though $\tilde{G}$ does contain $C_{2z}$ symmetry like the system in Case 1].
Previous works have noted that because the system ASG in this case contains infinitely many reflection symmetries, then any flat 1D system edge exhibits a mirror reflection symmetry [on the average], and can hence host average-symmetry-protected Dirac cones \cite{fulga_statistical_2014,spring_amorphous_2021}. 
However, caution is warranted when diagnosing the bulk of such as state as an average-symmetry-protected TCI, because the known [potentially complete] set of stable, short-range-entangled 2D TCIs~\cite{TeoFuKaneTCI,Fu2011,LiangIsobeInteractingTCI,ZhidaHermeleTopologicalCrystals,Song:2018cj,MTQC,PengChenDiagnosisMagneticPRB,ma2023averageSPT1,AltlandBrouwerDisorderedTCI,ZhidaIntrinsicDisorderedAXI} does not include any phases with edge Dirac cones that are protected by reflection symmetries with in-plane normal vectors, and because trivial [Wannierizable or ``fragile''] states can also exhibit model-dependent, in-gap Dirac-cone edge states that mimic those of stable TCIs~\cite{AshvinFragile,JenFragile,HingeSM,YoungMagnetic,CPtrivialDirac,AFMtrivialDiracPRL,ProjectiveGaugeFloatingTrivialDirac,CookMultiplicative,LukaSurfaceChiralDisorderPRX}. 
\item \textbf{Case 3:} $\tilde{G}$ contains the symmetries of both Cases 1 and 2 above [Fig.~\ref{fig:average_sym}(c)]:
\begin{equation}
\{M_{x}|{\bf 0}\}\in\tilde{G},\ \{M_{z}|{\bf 0}\}\in\tilde{G}.
\label{eq:case3}
\end{equation}  
In this case, the system is on the average centro- [$\mathcal{I}$-] symmetric like the system in Case 1, and $\tilde{G}$ in this case is similarly achiral.
Like in Cases 1 and 2, the system in this case can also exhibit reflection-protected Dirac-cone edge states through the mechanisms detailed above.
\item \textbf{Case 4:} $\tilde{G}$ \emph{only} contains the continuous 2D translations in $\tilde{G}_{T,2}$ and the continuous rotations in $C_{\infty}$ ($\infty$) [Fig.~\ref{fig:average_sym}(d)], such that:
\begin{equation}
\{M_{x}|{\bf 0}\}\notin\tilde{G},\ \{M_{y}|{\bf 0}\}\notin\tilde{G},\ \{M_{z}|{\bf 0}\}\notin\tilde{G}, 
\label{eq:2DchiralAmorphous}
\end{equation}
and hence $\{\mathcal{I}|{\bf 0}\}\notin\tilde{G}$, because $\{C_{2z}|{\bf 0}\}\in\tilde{G}$ and $\mathcal{I}=M_{z}C^{-1}_{2z}$ [see the text surrounding Eqs.~(\ref{eq:rotoinversion}) and~(\ref{eq:ImirrorBreakdown})].
Unlike all of the cases above, an amorphous $d=3$, $d_{A}=2$ system with an ASG $\tilde{G}$ satisfying Eq.~(\ref{eq:2DchiralAmorphous}) exhibits \emph{average structural chirality} [defined in the text following Eq.~(\ref{eq:dA2case})].
This case has been overlooked in previous studies of amorphous topological phases because in this case, the ASG $\tilde{G}$ does not contain reflection symmetries that can protect Dirac-cone edge states [either exactly or on the average].
Nevertheless, following the logic of Ref.~\cite{KramersWeyl}, we emphasize in this work that nontrivial topological features in disordered systems can also be enforced via the \emph{absence} of average symmetry, in this case specifically manifesting as the emergence of average structural and hence topological chirality in the bulk via the absence of rotoinversion symmetries in the system ASG $\tilde{G}$.  
\end{itemize}

\paragraph*{\bf Average Little Groups} -- $\ $ Having shown above that there exist amorphous systems both with and without chiral ASGs, we will next establish a precise notion of approximate [average] little groups in disordered systems, which we will then similarly classify as chiral or achiral.
We first recall that as detailed in Refs.~\cite{BigBook,Bradlyn2017,MTQC} and Appendix~\ref{app:corepDefs}, the little group $G_{\bf k}$ in a crystalline solid is defined as the subgroup $G_{\bf k}\subseteq G$ of the full position-space system SG $G$ that leaves the crystal momentum ${\bf k}$ invariant up to [modulo] a reciprocal lattice vector. 
As established above in the text preceding Eq.~(\ref{eq:ASG}), in a medium with [approximately] continuous translation symmetry, such as an amorphous solid, the Fourier-transformed spectrum is instead parameterized by pseudo-momenta ${\bf p}$ that take on \emph{unbounded} real values ${\bf p}\in\mathbb{R}^{d_{A}}$, as opposed to the \emph{periodic} crystal momenta ${\bf k}\in \mathbb{T}^{d_{T}}$ in a crystalline solid.
Hence the condition for determining the average little groups in an amorphous system is simpler than in a crystalline solid: the average little group [ALG] $\tilde{G}_{\bf p}$ is the subgroup of the ASG $\tilde{G}_{\bf p}\subseteq\tilde{G}$ for which each symmetry element $g\in\tilde{G}_{\bf p}$ simply leaves ${\bf p}$ invariant [without a modulo operation]:
\begin{equation}
g{\bf p}={\bf p}.
\label{eq:ALG}
\end{equation}
Importantly, the absence of a modulo operation in Eq.~(\ref{eq:ALG}) compared to the analogous translationally-invariant expression in Eq.~(\ref{eq:EquivKPoints}) implies significant differences in the pseudo-momentum-space symmetries of non-crystalline systems.  
For example, in nonmagnetic [$\mathcal{T}$-invariant], non-crystalline systems, pseudo-momentum is still odd under $\mathcal{T}$:
\begin{equation}
\mathcal{T}{\bf p}=-{\bf p}.
\label{eq:TonP}
\end{equation}
However, along with Eqs.~(\ref{eq:EquivKPoints}),~(\ref{eq:TonK}), and~(\ref{eq:ALG}), Eq.~(\ref{eq:TonP}) implies that in a nonmagnetic, non-crystalline system with $d_{T}$-dimensional exact lattice translations and $d_{A}$ infinite non-crystalline directions, there are \emph{only} $2^{d_{T}}$ $\mathcal{T}$-invariant ${\bf k}$ and ${\bf p}$ [TRIM] points.
These ${\bf k}$- and ${\bf p}$-space TRIM points together lie [in reduced momentum units] at $k_{\mu}=0,\pi$ and $p_{\nu}=0$ for each of the $d_{T}$ linearly independent crystal momentum components $k_{\mu}$ and each of the $d_{A}$ linearly independent pseudo-momentum components $p_{\nu}$ [see Fig.~\ref{fig:dadtdf}(b,c) for examples].

We emphasize that Eqs.~(\ref{eq:ALG}) and~(\ref{eq:TonP}) differ from earlier works on amorphous topological systems~\cite{marsal_topological_2020,corbae_evidence_2020,marsal_obstructed_2022,spring_amorphous_2021}, which instead claimed that average system symmetries [\emph{e.g.} $C_{\infty}$ and $\mathcal{T}$] are restored as ${\bf p}\rightarrow {\bs \infty}$, leading to a compactification of the pseudo-momentum BZ, the emergence of a modulo operation in Eqs.~(\ref{eq:ALG}) and~(\ref{eq:TonP}), and the presence of a high-symmetry ${\bf p}$ point at $|{\bf p}|=\infty$.
Specifically, the earlier works noted that at very large $|{\bf p}|$, the effective Hamiltonian $\mathcal{H}_{\text{Eff}}({\bf p})$ becomes independent of the direction of ${\bf p}$, and therefore invariant under the continuous proper rotation symmetries in $C_{\infty}$ ($\infty$) [$d_{A}=2$] or SO(3) [$d_{A}=3$].
Because the ALG $\tilde{G}_{{\bf p}={\bf 0}}$ in an amorphous system also contains continuous rotation symmetry in $d_{A}$ dimensions [see the text surrounding Eqs.~(\ref{eq:cInftyGenerator}) and~(\ref{eq:ALG})], then the authors of Refs.~\cite{marsal_topological_2020,corbae_evidence_2020,marsal_obstructed_2022,spring_amorphous_2021} concluded that the same ALG [average] crystal symmetries that are present at ${\bf p}={\bf 0}$ are also restored at ${\bf p}={\bs \infty}$, formally implying that $\tilde{G}_{{\bf p}={\bf 0}}$ and $\tilde{G}_{{\bf p}={\bs \infty}}$ are isomorphic.
The designation of ${\bf p}={\bs \infty}$ as a high-symmetry ${\bf p}$ point further allowed the authors of Refs.~\cite{marsal_topological_2020,corbae_evidence_2020,marsal_obstructed_2022,spring_amorphous_2021} to compute average-symmetry-based indicators of non-crystalline insulating topology in their model systems.

However, as discussed in the Supplemental Material for Ref.~\cite{spillage_2022}, the ${\bf p}\rightarrow{\bs \infty}$ behavior of an amorphous tight-binding model is highly sensitive to the Wannier basis states used to construct the model. 
In particular, if an amorphous tight-binding model is constructed from hard-shell or even $\delta$-function-localized Wannier orbitals, then there exists in each disordered system direction $\hat{r}$ a smallest intersite vector ${\bf a}_{\text{min},\hat{r}}$ for which plane waves with $|{\bf p}|>2\pi/|{\bf a}_{\text{min},\hat{r}}|$ cannot be distinguished from each other, because the plane waves oscillate at wavelengths shorter than the smallest intersite separation $|{\bf a}_{\text{min},\hat{r}}|$ in the direction of ${\bf p}$.
Taken over all directions in the disordered $d_{A}$-dimensional system [sub]space, this gives rise to an infinite family of pseudo-momentum cutoffs ${\bf P}_{\hat{r}}$ that satisfy:
\begin{equation}
{\bf P}_{\hat{r}}\cdot {\bf a}_{\text{min},\hat{r}} = 2\pi.
\label{eq:EmergentPinftyModulo}
\end{equation}
The pseudo-momentum cutoff vectors ${\bf P}_{\hat{r}}$ in Eq.~(\ref{eq:EmergentPinftyModulo}) act as emergent amorphous reciprocal lattice vectors that together compactify the pseudo-momentum BZ via the restoration of modulo operations in Eqs.~(\ref{eq:ALG}) and~(\ref{eq:TonP}).
Importantly, however, Eq.~(\ref{eq:EmergentPinftyModulo}) is not a general result, but is rather an artifact of the choice of highly localized tight-binding [Wannier] basis states.
Specifically, if we had instead chosen more realistic Wannier basis functions with longer, exponentially decaying tails, then there would no longer exist lower bounds on the wavelength and direction resolution of ${\bf p}$ [${\bf a}_{\text{min},\hat{r}}\rightarrow {\bf 0}$ for all $\hat{r}$], and there would correspondingly no longer exist finite-valued ${\bf P}_{\hat{r}}$ that satisfy Eq.~(\ref{eq:EmergentPinftyModulo}).
Therefore, for generality and looking ahead to real-material applications of the theory developed in this work, we will not in this work treat ${\bf p}\rightarrow {\bs \infty}$ as a high-symmetry pseudo-momentum [\emph{e.g.} TRIM] point.

Finally, having established definitions for the ASGs and ALGs of non-crystalline systems, we will lastly define the chirality and pseudo-momentum stars of ALGs.
First, because an ALG is isomorphic to an ASG, then the definition of a chiral ALG is therefore the same as that of a chiral ASG: a $d=3$ ALG is chiral if it does not contain any [average] rotoinversion symmetries of the form of Eq.~(\ref{eq:rotoinversion}).
For example, in Case 4 above, the $d=3$, $d_{A}=2$, $d_{T}=0$, $d_{f}=1$ system respects a chiral ASG given by:
\begin{equation}
\tilde{G} = \tilde{G}_{T,2}\cup\{C_{(2\pi/\phi) z}|{\bf 0}\}\tilde{G}_{T,2}.
\label{eq:chiralASG}
\end{equation}
All of the symmetries in $\tilde{G}$ in Eq.~(\ref{eq:chiralASG}) return ${\bf p}={\bf 0}$ to itself, such that the ALG at ${\bf p}={\bf 0}$ in Eq.~(\ref{eq:chiralASG}) is given by:
\begin{equation}
\tilde{G}_{{\bf p}={\bf 0}} = \tilde{G} = \tilde{G}_{T,2}\cup\{C_{(2\pi/\phi) z}|{\bf 0}\}\tilde{G}_{T,2}.
\label{eq:AmorphousBreakdownGamma}
\end{equation}
Conversely for ${\bf p}\neq{\bf 0}$, only the continuous translations in Eq.~(\ref{eq:chiralASG}) return ${\bf p}$ to itself, whereas the continuous family of rotations generated by $\{C_{(2\pi/\phi) z}|{\bf 0}\}$ do not return ${\bf p}\neq{\bf 0}$ to itself, such that:
\begin{equation}
\tilde{G}_{{\bf p}\neq{\bf 0}} = \tilde{G}_{T,2}.
\label{eq:lowSymmetryALG}
\end{equation}
Below, in the text surrounding Eq.~(\ref{eq:nematicALGbreakdown}), we will shortly also analyze the symmetry group structure of the ALG $\tilde{G}_{{\bf p}={\bf 0}}$ in a $d=3$, $d_{A}=3$, $d_{T}=0$, $d_{f}=0$ amorphous system with net [average] structural chirality, which is spectrally isotropic [on the average] in all of 3D ${\bf p}$ space [Fig.~\ref{fig:dadtdf}(c)].

Importantly, after allowing for Wannier tight-binding basis states that are not $\delta$-function localized [see the text surrounding Eq.~(\ref{eq:EmergentPinftyModulo})], the ALG at any other finite momentum ${\bf p}\neq {\bf 0}$ is still isomorphic to $\tilde{G}_{{\bf p}\neq{\bf 0}}$, even as ${\bf p}\rightarrow {\bs \infty}$.
Hence, from a group-theoretic perspective, ${\bf p}={\bs \infty}$ is not considered to be a single ``special'' momentum point at which the ALG $\tilde{G}_{{\bf p}={\bs \infty}}$ carries restored [emergent] symmetries.
Instead, ${\bf p}\rightarrow {\bs \infty}$ characterizes an infinite \emph{family} [``momentum star''~\cite{BigBook,MTQC}] of ${\bf p}$ points that are related to each other by elements of the system ASG $\tilde{G}$ [\emph{e.g.} elements of the continuous rotation group $C_{\infty}$ ($\infty$)].
Furthermore, as discussed earlier in this section, in Refs.~\cite{marsal_topological_2020,corbae_evidence_2020,JustinHat,Ciocys2023}, and in the text surrounding Fig.~\ref{fig:HeffBreak}, the ${\bf p}$-resolved spectral function computed from Green's functions in an amorphous system becomes diffuse in energy as $|{\bf p}|\rightarrow\infty$, because it is probing a strongly disordered system at the lattice scale.
This implies that the system at very large $|{\bf p}|$ is also poorly captured by a single-particle effective Hamiltonian $\mathcal{H}_{\text{Eff}}({\bf p})$, limiting our ability to characterize the large-${\bf p}$ system spectrum with ALGs and their small [co]representations.

To conclude, in this section, we established the structure of and terminology for the real- and [pseudo-] momentum-space symmetry groups of amorphous solids.  Next, in Appendix~\ref{app:corepAmorphous}, we will discuss the corepresentations [coreps] of the infinite [and continuous] ALGs near ${\bf p}\approx {\bf 0}$ in amorphous systems, focusing specifically on a subset of ALG small coreps that describe nodal spectral degeneracies with the same dispersion relations and topology as exact small coreps of crystalline little groups.

\subsection{Pseudo-Momentum Space Corepresentations and Chiral Fermions}
\label{app:corepAmorphous}

In this section, we will lastly detail how little group coreps, discussed previously for translationally-invariant systems in Appendix~\ref{app:corepDefs}, manifest in amorphous systems, as well as their relationship to the ${\bf k}\cdot{\bf p}$ form of the low-energy effective Hamiltonian.   
To begin, we first recall that, as established in Eq.~(\ref{eq:ALG}) and the preceding text, the ALG $\tilde{G}_{\bf p}$ at a given pseudo-momentum point ${\bf p}$ is isomorphic to an ASG, as opposed to an average point group [APG].  
Hence, $\tilde{G}_{\bf p}$ in general contains an infinite subgroup of [continuous] translation symmetries:
\begin{equation}
\tilde{G}_{T,d_{A}} \subseteq \tilde{G}_{\bf p}.
\end{equation}
Therefore, like its exact crystalline analog $G_{\bf k}$ [see Refs.~\cite{BigBook,MTQC,Bandrep1} and the text following Eq.~(\ref{eq:EquivKPoints})], $\tilde{G}_{\bf p}$ has infinitely many coreps.
In the crystalline case, the infinite number of coreps of $G_{\bf k}$ can be reduced to a finite subset of \emph{small} coreps $\tilde{\sigma}_{\bf k}$ to which all of the other coreps of $G_{\bf k}$ are related by the unitary matrix representatives of the discrete translations in $G_{\bf k}$~\cite{BigBook,Bradlyn2017,Bandrep1,MTQC,millerTables}.
This reduction from a [countably] infinite set of coreps to a finite set of small coreps is crucially reliant on $G_{\bf k}$ containing only discrete [crystallographic] symmetries.
For example, if $G_{\bf k}$ instead contained a non-crystallographic continuous rotation subgroup, as can occur in non-crystallographic rod groups [see Refs.~\cite{linegroupsbook,eightfoldRodGroups} and the text following Eq.~(\ref{eq:finiteDimNoAmorph})], $G_{\bf k}$ would still have an infinite number of \emph{small} coreps, which would greatly complicate our ability to meaningfully restrict consideration to a finite number of small coreps and the associated nodal degeneracies in the physical energy spectrum.

In amorphous systems, the same issue detailed above for non-crystallographic rod groups with continuous rotation symmetries necessarily occurs for the ${\bf p}={\bf 0}$ ALG $\tilde{G}_{{\bf p}={\bf 0}}$.
Specifically, in amorphous systems with $d_{A}>1$ in Eq.~(\ref{eq:finiteDimAmorph}), $\tilde{G}_{{\bf p}={\bf 0}}$ generically contains continuous rotation symmetries, such as the infinity-fold proper rotation symmetry $C_{(2\pi/\phi) z}$ defined in Eq.~(\ref{eq:cInftyGenerator}).
Because the APG $C_{\infty}$ ($\infty$) generated by $C_{(2\pi/\phi) z}$ is an infinite group, then it has infinitely many irreducible coreps~\cite{linegroupsbook,PointGroupTables,SpinPointMcClarty}.
This implies that at a ${\bf p}$ point in an amorphous system with a continuous rotation symmetry [such as ${\bf p}={\bf 0}$ for $d_{A}=2,3$, see Refs.~\cite{spring_amorphous_2021,springMagneticAverageTI,corbae_evidence_2020}, Fig.~\ref{fig:dadtdf}(b,c), and Eqs.~(\ref{eq:dA2case}) and~(\ref{eq:AmorphousBreakdownGamma})], the ALG $\tilde{G}_{\bf p}$ has both an infinite number of irreducible coreps, and an infinite number of irreducible \emph{small} coreps. 
For ALGs with infinitely many small coreps, this raises the question of which subset of the infinite small coreps are most relevant to low-energy nodal spectral degeneracies.
However, for the purposes of this work, we may use the existence of a reference crystalline system with the same low-energy spectral features to provide bounds on our analysis of ALG small coreps.
Specifically, in this work, we for simplicity limit consideration to ALG small coreps that are equivalent to exact crystalline small coreps through an isotropic deformation procedure that we will detail below.
Importantly, one need not impose this restriction, and we leave for future works the tantalizing possibility of amorphous nodal degeneracies that transform in ALG small coreps without equivalent crystalline counterparts~\cite{HigherChargeNonCrystalLasing}.

In this work we will particularly focus on a subset of ${\bf p}={\bf 0}$ ALGs that are isomorphic to chiral ASGs because we expect -- and will demonstrate for specific examples below -- that following the arguments in Ref.~\cite{KramersWeyl}, at least a subset of their small coreps correspond to the ${\bf k}\cdot {\bf p}$ effective Hamiltonians of amorphous chiral fermions.  
Specifically, as shown in Refs.~\cite{KramersWeyl,ZhijunWeakSOCKramersWeyl}, for all small coreps in nonmagnetic chiral crystals that correspond to low-energy Hamiltonians $\mathcal{H}({\bf k})$ with dispersive [split] bands, the lower bands of $\mathcal{H}({\bf k})$ carry nontrivial quantized chiral charges $C$ that are indicated by the nonvanishing, $\mathbb{Z}$-valued integral of the Berry curvature flux on a sphere $S$ surrounding ${\bf k}={\bf 0}$:
\begin{equation}
C = \frac{1}{2\pi}\oint_{S}\Tr[{\bf F}({\bf k})]\cdot d{\bf A},
\label{eq:Berrycurvintegral}
\end{equation}
where ${\bf F}({\bf k})$ is the matrix Berry curvature~\cite{Z2Pack,Armitage2018}.
To select a meaningful subset of the infinitely many small coreps of $\tilde{G}_{{\bf p}={\bf 0}}$ in an amorphous system with an ASG $\tilde{G}$, we will restrict our focus to specific examples of amorphous small coreps $\tilde{\sigma}_{{\bf p}={\bf 0}}$ whose ${\bf k}\cdot{\bf p}$ effective Hamiltonians can be obtained by deforming the ${\bf k}\cdot{\bf p}$ Hamiltonians of \emph{crystalline} $\Gamma$-point [${\bf k}={\bf 0}$] small coreps without closing a gap or changing the polynomial form of the low-energy band dispersion.
These deformations are generically permitted by symmetry, because [symmorphic] crystalline little groups with discrete translation and rotation symmetries [\emph{i.e.} those that are not generated by glide or screw symmetries~\cite{BigBook,MTQC}] are necessarily subgroups of ALGs with continuous translation and rotation symmetries.

To demonstrate the above statements in a concrete example, we consider a $\Gamma$-point [${\bf k}={\bf 0}$] crystalline Kramers-Weyl chiral fermion in nonmagnetic SG 16 ($P222$)~\cite{KramersWeyl}.
As detailed below in Appendix~\ref{app:PristineKramers}, $G=P222$ contains twofold rotation symmetries about the $x$-, $y$-, and $z$-axes $C_{2,xyz}$ and time-reversal symmetry $\mathcal{T}$:
\begin{equation}
C_{2x} = \{C_{2x}|{\bf 0}\},\ C_{2y} = \{C_{2y}|{\bf 0}\},\ C_{2z} = \{C_{2z}|{\bf 0}\},\ \mathcal{T} = \{\mathcal{T}|{\bf 0}\},
\label{eq:SG16rotations}
\end{equation}
as well the group $G_{T,3}$ of 3D orthorhombic integer [lattice] translations, where the absence of a tilde on $G_{T,3}$ denotes that it is an exact crystalline translation group, as opposed to the approximate [average] continuous translation group $\tilde{G}_{T,d_{A}}$ discussed in Eq.~(\ref{eq:ASG}) and the surrounding text.

At the $\Gamma$ point in SG 16 ($P222$), the little group $G_{{\bf k}={\bf 0}}=G_{\Gamma}$ is just given by the full SG:
\begin{equation}
G_{\Gamma}=G=P222.
\label{eq:GammaIsFullSG}
\end{equation}
The little group $G_{\Gamma}$ has only one, two-dimensional, double-valued [spinful] small corep that corresponds to a linear ${\bf k}\cdot{\bf p}$ Hamiltonian of the form:
\begin{equation}
\mathcal{H}({\bf k}) = v_{x}k_{x}\sigma^{x} + v_{y}k_{y}\sigma^{y} + v_{z}k_{z}\sigma^{z},
\label{eq:kpKWCrystalline}
\end{equation}
where the $2\times2$ $\sigma^{i}$ matrices index a low-energy degree of freedom with a half-integer angular momentum [but importantly, not necessarily the pure electron spin component $S_{i}$ itself due to spin-orbit coupling~\cite{KramersWeyl,BradlynTQCSpinTexture}].
The ${\bf k}\cdot{\bf p}$ Hamiltonian $\mathcal{H}({\bf k})$ is left invariant under the symmetries in $G_{\Gamma}$ [Eqs.~(\ref{eq:SG16rotations}) and~(\ref{eq:GammaIsFullSG})]:
\begin{equation}
C_{2i}\mathcal{H}({\bf k})C_{2i}^{-1} = \sigma^{i}\mathcal{H}(C_{2i}^{-1}{\bf k})\sigma^{i}\text{ for }i=x,y,z,\ \mathcal{T}\mathcal{H}({\bf k})\mathcal{T}^{-1} = \sigma^{y}\mathcal{H}^{*}(-{\bf k})\sigma^{y},
\label{eq:pristineKPSyms}
\end{equation}
as well as the 3D discrete translation symmetries in $G_{T,3}$, which are represented as Bloch phases multiplied by the $2\times 2$ identity matrix $\sigma^{0}$.
$\mathcal{H}({\bf k})$ in Eq.~(\ref{eq:kpKWCrystalline}) specifically is centered around the time-reversal-invariant momentum [TRIM point]~\cite{KramersWeyl,OtherKramersWeyl,AndreiTalk} ${\bf k}={\bf 0}$, and notably takes the canonical form of a $|C|=1$ conventional Weyl fermion~\cite{AshvinWeyl,VolovikWeylMonopole,VolovikChiralAnomaly,HaldaneOriginalWeyl,MurakamiWeyl,BurkovBalents,AndreiWeyl,HasanWeylDFT,Armitage2018,SuyangWeyl,LvWeylExp,YulinWeylExp,AliWeylQPI,AlexeyType2,ZJType2,BinghaiClaudiaWeylReview,ZahidNatRevMatWeyl,CDWWeyl,IlyaIdealMagneticWeyl}.  For these reasons, the nodal degeneracy at ${\bf k}={\bf 0}$ in Eq.~(\ref{eq:kpKWCrystalline}) was in Ref.~\cite{KramersWeyl} termed a \emph{Kramers-Weyl} [KW] fermion.

As first noted in Ref.~\cite{KramersWeyl}, Eq.~(\ref{eq:pristineKPSyms}) can be tuned to a more isotropic form that satisfies the symmetries of any of the chiral crystallographic $\Gamma$-point little groups by setting some or all of the $v_{i}$ coefficients to be equal.  
In this work, we further recognize that by tuning the $v_{i}$ to be equal, Eq.~(\ref{eq:pristineKPSyms}) can also be made to respect the \emph{continuous} average symmetries of an amorphous system, and hence in that limit transforms in a small corep of a \emph{non-crystallographic} ALG.

For example, if we in Eq.~(\ref{eq:pristineKPSyms}) tune $v_{x}=v_{y}=\tilde{v}_{xy}$, replace two of the crystal momenta $k_{x,y}$ with the continuous [approximate] pseudo-momenta $p_{x,y}$, and re-interpret the ${\bf k}\cdot {\bf p}$ Hamiltonian to be \emph{effective} [as opposed to exact, see Refs.~\cite{varjas_topological_2019,marsal_topological_2020,marsal_obstructed_2022} and Appendix~\ref{app:EffectiveHamiltonian}], then the low-energy Hamiltonian becomes:
\begin{equation}
\mathcal{H}_{\text{Eff}}(p_{x},p_{y},k_{z}) = \tilde{v}_{xy}\left(p_{x}\tilde{\sigma}^{x} + p_{y}\tilde{\sigma}^{y}\right) + v_{z}k_{z}\tilde{\sigma}^{z} + \tilde{\mu}\tilde{\sigma}^{0},
\label{eq:kpKWSmectic}
\end{equation}
where $\tilde{v}_{xy}$ is the disorder-renormalized Fermi velocity in the $(p_{x},p_{y})$-plane~\cite{corbae_evidence_2020,Ciocys2023}, $\tilde{\mu}$ is the disorder-renormalized chemical potential, $\tilde{\sigma}^{0}$ is the $2 \times 2$ identity matrix, and where each $\tilde{\sigma}^{i}$ is a $\mathcal{T}$-odd $2\times 2$ Pauli matrix that need not represent the same physical degree of freedom [\emph{e.g.} the $i$-th (pseudo)spin component] as $\sigma^{i}$ in Eq.~(\ref{eq:kpKWCrystalline}).
Eq.~(\ref{eq:kpKWSmectic}) instead now transforms in a two-dimensional, double-valued small corep of a $d_{A}=2$ chiral ALG that we denote as $\tilde{G}_{\Gamma,2}$, where $\tilde{G}_{\Gamma,2}$ is generated by the APG symmetries:
\begin{equation}
C_{(2\pi/\phi) z} = \{C_{(2\pi/\phi) z}|{\bf 0}\},\ C_{2x} = \{C_{2x}|{\bf 0}\},\ \mathcal{T} = \{\mathcal{T}|{\bf 0}\},
\label{eq:smecticALGsyms}
\end{equation}
and additionally contains the group $G_{T,z}$ of integer 1D translations in the $\hat{z}$-direction and the group $\tilde{G}_{T,2}$ of 2D continuous translations in the $(x,y)$-plane.  
In Eq.~(\ref{eq:smecticALGsyms}), $\phi$ denotes an infinitesimal rotation angle about the $z$-axis.
The symmetries in Eq.~(\ref{eq:smecticALGsyms}) can be represented through their action on $\mathcal{H}_{\text{Eff}}(p_{x},p_{y},k_{z})$ in Eq.~(\ref{eq:kpKWSmectic}):
\begin{eqnarray}
C_{(2\pi/\phi) z}\mathcal{H}_{\text{Eff}}(p_{x},p_{y},k_{z})C_{(2\pi/\phi) z}^{-1} &=& e^{i (\phi/2)\tilde{\sigma}^{z}}\mathcal{H}_{\text{Eff}}(C^{-1}_{(2\pi/\phi) z}{\bf p}_{x,y},k_{z})e^{-i (\phi/2)\tilde{\sigma}^{z}}, \nonumber \\
C_{2x}\mathcal{H}_{\text{Eff}}(p_{x},p_{y},k_{z})C_{2x}^{-1} &=& \tilde{\sigma}^{x}\mathcal{H}_{\text{Eff}}(p_{x},-p_{y},-k_{z})\tilde{\sigma}^{x}  \nonumber \\
\mathcal{T}\mathcal{H}_{\text{Eff}}(p_{x},p_{y},k_{z})\mathcal{T}^{-1} &=& \tilde{\sigma}^{y}\mathcal{H}^{*}_{\text{Eff}}(-p_{x},-p_{y},-k_{z})\tilde{\sigma}^{y},
\label{eq:smecticKPSyms}
\end{eqnarray}
where ${\bf p}_{x,y}=(p_{x},p_{y})$, and where we again note that the [now both continuous and discrete] translation symmetries in the ALG $\tilde{G}_{\Gamma,2}$ are represented as phases multiplied by the $2\times 2$ identity matrix $\tilde{\sigma}^{0}$, and hence do not provide further constraints on $\mathcal{H}_{\text{Eff}}(p_{x},p_{y},k_{z})$ beyond those in Eq.~(\ref{eq:smecticKPSyms}).

For completeness, we note that the chiral ALG $\tilde{G}_{\Gamma,2}$ of Eq.~(\ref{eq:kpKWSmectic}) can also be expressed in the form:
\begin{equation}
\tilde{G}_{\Gamma,2} = (p\infty 22)_{RG} \cup \{\mathcal{T}|000\}(p\infty 22)_{RG} \cup \{E|\epsilon 00\}(p\infty 22)_{RG} \cup \{\mathcal{T}|\epsilon 00\}(p\infty 22)_{RG},
\label{eq:smecticALGbreakdown}
\end{equation}
where $\{E|\epsilon 00\}$ denotes an infinitesimal translation along the $x$-axis, and where $(p\infty 22)_{RG}$ denotes the non-crystallographic magnetic [unitary] chiral rod group generated by continuous rotations about the $z$-axis $C_{(2\pi/\phi) z}$, twofold rotations about the $x$-axis $C_{2x}$, and integer lattice translations $\{E|001\}$ along the $z$-axis~\cite{MTQC,linegroupsbook,eightfoldRodGroups}.  $(p\infty 22)_{RG}$ in turn admits the decomposition:
\begin{equation}
(p\infty 22)_{RG} = D_{\infty} \cup \{E|001\}D_{\infty},
\end{equation}
where $D_{\infty}$ denotes the non-crystallographic chiral point group $D_{\infty}$ ($\infty 22$) obtained by extending $C_{\infty}$ ($\infty$) by $C_{2x}$~\cite{PointGroupTables,ChiralNematicBook}:
\begin{equation}
D_{\infty} = C_{\infty} \cup C_{2x}C_{\infty}.
\end{equation}
In this work, we employ a terminology in which $d=3$, $d_{T}=1$, $d_{A}=2$, $d_{f}=0$ partially [layered] amorphous systems [Eq.~(\ref{eq:finiteDimAmorph})] with average [net] structural chirality, such as systems with ${\bf p} = {\bf 0}$ ALGs like $\tilde{G}_{\Gamma,2}$ [Eqs.~(\ref{eq:smecticALGsyms}) and~(\ref{eq:smecticALGbreakdown})], are denoted as having \emph{smectic} disorder [see Appendix~\ref{app:SmecticNematicDisorder} for additional details].

Finally, in Eq.~(\ref{eq:kpKWCrystalline}) we may also realize a fully isotropic effective ${\bf k}\cdot {\bf p}$ Hamiltonian by tuning $v_{x}=v_{y}=v_{z}=\tilde{v}$ and replacing the full vector crystal momentum ${\bf k}$ with the 3D pseudo-momentum vector ${\bf p}$:
\begin{equation}
\mathcal{H}_{\text{Eff}}({\bf p}) = \tilde{v}\left(p_{x}\tilde{\sigma}^{x} + p_{y}\tilde{\sigma}^{y} + p_{z}\tilde{\sigma}^{z}\right) + \tilde{\mu}\tilde{\sigma}^{0},
\label{eq:kpKWnematic}
\end{equation}
where $\tilde{v}$ is the disorder-renormalized isotropic Fermi velocity, $\tilde{\mu}$ is again the disorder-renormalized chemical potential, and where each $\tilde{\sigma}^{i}$ again may not represent the same physical degrees of freedom as $\sigma^{i}$ in Eq.~(\ref{eq:kpKWCrystalline}).
Eq.~(\ref{eq:kpKWSmectic}) in this 3D isotropic limit now transforms in a two-dimensional, double-valued small corep of a $d_{A}=3$ chiral ALG that we denote as $\tilde{G}_{\Gamma,3}$:
\begin{equation}
\tilde{G}_{\Gamma,3} = \text{SO}(3) \cup \{\mathcal{T}|000\}\text{SO}(3) \cup \{E|\epsilon 00\}\text{SO}(3) \cup \{\mathcal{T}|\epsilon 00\}\text{SO}(3),
\label{eq:nematicALGbreakdown}
\end{equation}
where $\{E|\epsilon 00\}$ is again an infinitesimal translation along the $x$-axis, and employing a convention in which for consistency with the crystallographic rotation [point] groups~\cite{ConwaySymmetries,BigBook,BilbaoPoint,PointGroupTables,SpinPointMcClarty}, we refer to the spinful isotropic rotation group as the double group of SO($3$), rather than as a distinct group SU($2$).
The generating symmetries of $\tilde{G}_{\Gamma,3}$ in Eq.~(\ref{eq:nematicALGbreakdown}) can be represented through their action on $\mathcal{H}_{\text{Eff}}({\bf p})$ in Eq.~(\ref{eq:kpKWnematic}):
\begin{eqnarray}
C_{(2\pi/\phi) z}\mathcal{H}_{\text{Eff}}({\bf p})C_{(2\pi/\phi) z}^{-1} &=& e^{i (\phi/2)\tilde{\sigma}^{z}}\mathcal{H}_{\text{Eff}}(C^{-1}_{(2\pi/\phi) z}{\bf p})e^{-i (\phi/2)\tilde{\sigma}^{z}}, \nonumber \\
C_{(2\pi/\theta) x}\mathcal{H}_{\text{Eff}}({\bf p})C_{(2\pi/\theta) x}^{-1} &=& e^{i (\theta/2)\tilde{\sigma}^{x}}\mathcal{H}_{\text{Eff}}(C^{-1}_{(2\pi/\theta) x}{\bf p})e^{-i (\theta/2)\tilde{\sigma}^{x}}, \nonumber \\
\mathcal{T}\mathcal{H}_{\text{Eff}}({\bf p})\mathcal{T}^{-1} &=& \tilde{\sigma}^{y}\mathcal{H}^{*}_{\text{Eff}}(-{\bf p})\tilde{\sigma}^{y},
\label{eq:nematicKPSyms}
\end{eqnarray}
where $\theta$ denotes an infinitesimal rotation angle about the $x$-axis, and noting for completeness that the continuous translation symmetries in $\tilde{G}_{\Gamma,3}$ are again represented as phases multiplied by the $2\times 2$ identity matrix $\tilde{\sigma}^{0}$.
As detailed below in Appendix~\ref{app:SmecticNematicDisorder}, in this work, we employ a terminology in which $d=3$, $d_{T}=0$, $d_{A}=3$, $d_{f}=0$ fully amorphous systems [Eq.~(\ref{eq:finiteDimAmorph})] with average structural chirality, such as those with ${\bf p} = {\bf 0}$ ALGs like $\tilde{G}_{\Gamma,3}$ [Eq.~(\ref{eq:nematicALGbreakdown})], are denoted as having \emph{nematic} disorder.

To conclude, in this section, we have demonstrated that although ${\bf p}={\bf 0}$ ALGs in $d=3$, $d_{A}=2,3$ amorphous solids in general have an infinite number of small coreps, there exist specific $d=3$ ALGs for which a subset of the small coreps correspond to effective ${\bf k}\cdot {\bf p}$ Hamiltonians with the same dispersion relations and topology as nodal degeneracies in disorder-free crystals.  
We provided above two examples of double-valued chiral ALG small coreps that correspond to amorphous generalizations of KW fermions.
More generally, the dispersion deformation procedure employed in this section to generate amorphous chiral fermions from crystalline ${\bf k}\cdot {\bf p}$ Hamiltonians -- with examples detailed in the text preceding Eqs.~(\ref{eq:kpKWSmectic}) and~(\ref{eq:kpKWnematic}) -- can be applied to any $d=3$ crystalline nodal degeneracy with the symmetries of a symmorphic little group for which the nodal degeneracy can be made rotationally symmetric in two or all three directions without changing the polynomial form of its low-energy dispersion relation. 
Below, in Appendices~\ref{app:amorphousCharge2} and~\ref{app:amorphousMultifold}, we will respectively use this reasoning to generate amorphous models that exhibit $|C|=2$~\cite{ZahidMultiWeylSrSi2,StepanMultiWeyl,AndreiMultiWeyl,XiDaiMultiWeyl} and multifold~\cite{AbrikosovNewFermion,ManesNewFermion,DDP,NewFermions,chang2017large,tang2017CoSi,CanoMagneticNewFermion,DoubleWeylPhonon,MurakamiChargeFour,InigoMultifoldReview} chiral fermions at ${\bf p}={\bf 0}$.

\section{Numerical Methods}
\label{sec:numericalMethods}

In this Appendix, we will introduce and detail the numerical methods employed throughout this study. 
We will start in Appendix~\ref{app:lattices} by defining the non-crystalline lattices and forms of disorder implemented in our calculations.
Specifically, we will first in Appendix~\ref{app:DiffTypesDisorder} discuss the distinct methods that we employed to generate lattices without discrete translation symmetry, and our methods for implementing orientational disorder in models with additional internal [on-site] spin and orbital degrees of freedom.
Adapting terminology from the study of liquid crystals~\cite{chaikinLubenskyBook,CollingsGoodbyLiquidCrystalBook,KamienLiquidCrystalRMP,KamienChiralLiquidCrystal,ChiralNematicBook}, we will then in Appendix~\ref{app:SmecticNematicDisorder} define the smectic and nematic model disorder regimes employed throughout this study.
Next, in Appendix~\ref{app:PhysicalObservables}, we will detail the computational methods employed in this study to construct approximate continuum [pseudo-] momentum-space descriptions of disordered topological semimetals via replica-averaged, momentum-resolved matrix Green's functions.
We will specifically demonstrate in Appendix~\ref{app:PhysicalObservables} that in the strongly disordered lattice models in this work averaged over large numbers of [$\approx 20-50$] disorder realizations [\emph{i.e.} replicas~\cite{ParisiReplicaCourse,BerthierGlassAmorphousReview}], the Fourier-transformed average matrix Green's function approximately reduces to a function of just a single plane-wave pseudo-momentum, allowing us to establish a physically meaningful momentum-dependent energy spectrum in a disordered system.
Building on this framework, we will then in Appendix~\ref{app:surfaceGreens} introduce a precise method for constructing the momentum-resolved \emph{surface} spectral functions of disordered tight-binding models, which later in Appendices~\ref{app:amorphousKramers},~\ref{app:amorphousCharge2}, and~\ref{app:amorphousMultifold} will facilitate our analysis of the Fermi-arc surface states of disordered topological chiral semimetals.

Working towards a quantitative topological characterization of the disordered nodal fermions present in our momentum-resolved spectral function calculations, we will next in Appendix~\ref{app:EffectiveHamiltonian} detail our methodology for constructing pseudo-momentum-space single-particle effective Hamiltonians in non-crystalline tight-binding models~\cite{varjas_topological_2019,marsal_topological_2020,marsal_obstructed_2022}.
We will specifically in Appendix~\ref{app:EffectiveHamiltonian} show that for all of the non-crystalline topological semimetal models studied in this work, the single-particle effective Hamiltonian exhibits the same Berry phases and low-energy [${\bf k}\cdot {\bf p}$] polynomial pseudo-momentum dependence [though with highly varying dispersion relation coefficients] over a surprisingly large range of numerical implementation parameters.  
Motivated by this result, we will lastly in Appendix~\ref{sec:WilsonBerry} introduce an amorphous Wilson loop method for inferring the topological chiral charges of many-particle disordered nodal degeneracies in non-crystalline tight-binding models using the energy eigenstates of the effective Hamiltonian.

\subsection{Amorphous Lattice Models}
\label{app:lattices}

Amorphous solids, owing to their absence of long-range structural order, are frequently studied using relatively simple models with the highest possible degree of disorder~\cite{zallen_physics_1998,Grushin2020,Corbae_2023}.
This can include typical Anderson-type disorder, where the on-site or hopping energies randomly vary, as well as \emph{structural disorder}, in which the atomic connectivity [bond coordination] or positions randomly vary~\cite{weaire_electronic_1971}. 
In some works, structural disorder is alternatively termed ``topological disorder,'' especially in models and materials in which the lattice connectivity [\emph{i.e.} graph topology] is disordered [\emph{e.g.} graphene-based systems with high densities of defects consisting of 5-, 7-, and 8-membered rings~\cite{toh_synthesis_2020}].
However, because we are in this work studying the momentum-space electronic topology of amorphous metals, we will avoid referring to lattice and bond-coordination disorder as topological disorder, and will instead term them forms of structural disorder.

The above discussion of Anderson and structural disorder -- which omits details of internal or other local degrees of freedom -- is frequently considered a sufficient starting point for constructing tight-binding models of amorphous systems.
However, amorphous materials have also been shown to host phases of matter, such as magnetically ordered~\cite{AmorphousFMBookChap,ChudnovskyAmorphousFM,Gambino1974,Hellman1992} and 3D topological insulating states~\cite{corbae_evidence_2020,Ciocys2023}, that can only be captured in amorphous tight-binding models with specific local or internal degrees of freedom, such as electronic spins~\cite{PhysRevLett.95.226801,KaneMele2DZ2,Fu2007,WenZooReview}.
The study of non-crystalline topological phases has in particular been greatly advanced in recent years through the introduction of more elaborate amorphous tight-binding models with on-site [internal spin and orbital] degrees of freedom~\cite{agarwala_topological_2017,yang_topological_2019,corbae_evidence_2020,Ciocys2023,spring_amorphous_2021,springMagneticAverageTI,StructAmoTopoOrder,spillage_2022}, as well as models with repeated, randomly oriented local structural motifs~\cite{marsal_topological_2020,marsal_obstructed_2022,JustinHat}.

In these more complicated models, notions of order can derive not just from the regularity of the lattice, but also from long-range correlations between the internal or orientational degrees of freedom of short-range-interacting local units.
Though not previously emphasized in other studies of amorphous topological phases, this contrast between orientational and structural [positional] order is well-established in other condensed-matter settings.
For example, liquid crystals composed of rod-like molecules can experimentally exhibit \emph{nematic} phases in which the molecular rods lie at random positions, but on the average point in the same direction~\cite{chaikinLubenskyBook,CollingsGoodbyLiquidCrystalBook,KamienLiquidCrystalRMP,KamienChiralLiquidCrystal,ChiralNematicBook}.
In this work, we are particularly interested in modeling and predicting material realizations of topological amorphous metals in which the local units are orientationally and structurally disordered, but nevertheless exhibit local structural chirality and form domains with the same handedness [on the average, see Fig.~\ref{fig:average_sym}(d) and the text following Eq.~(\ref{eq:dA2case})].
To construct physically motivated models that exhibit contrasting structural disorder and discrete orientational [chirality] order, we must carefully separate and categorize the different possible forms of order and disorder in non-crystalline tight-binding models with internal degrees of freedom, which we will detail below in Appendices~\ref{app:DiffTypesDisorder} and~\ref{app:SmecticNematicDisorder}.
Specifically, we will first in Appendix~\ref{app:DiffTypesDisorder}, enumerate the different forms of disorder that can be realized in the topological semimetal models studied in this work [Appendix~\ref{app:models}], as well as our numerical methods for implementing model disorder. 
Then, in Appendix~\ref{app:SmecticNematicDisorder}, we will draw analogy to the study of liquid crystals to define the \emph{smectic} and \emph{nematic} disorder regimes of our structurally chiral amorphous tight-binding models [see Fig.~\ref{fig:dadtdf}(b,c)], which we will show in Appendices~\ref{app:amorphousKramers},~\ref{app:amorphousCharge2}, and~\ref{app:amorphousMultifold} generically host non-crystalline generalizations of topological chiral fermions.

\subsubsection{Numerical Implementation and Categories of Model Disorder}
\label{app:DiffTypesDisorder}

Below, we will enumerate the different forms of disorder present in our tight-binding models.
Each of the topological semimetal models studied in this work is constructed from a repeated local unit [site] with internal [on-site] spin and orbital degrees of freedom [see Appendix~\ref{app:models} for model details].
Hence, the disorder in our models can be sorted into two categories: 
\begin{itemize}
\item \textbf{Structural Disorder:} Disorder pertaining to the deviation of the atomic [site] positions from a regular crystalline lattice.
\item \textbf{Internal Degree-of-Freedom Disorder:} Disorder pertaining to the interaction [hopping] frame orientation and handedness [local chirality, see Fig.~\ref{fig:average_sym} and the surrounding text] of the internal spin and orbital degrees of freedom on each site.
\end{itemize}

\begin{figure}[t]
\centering
\includegraphics[width=\linewidth]{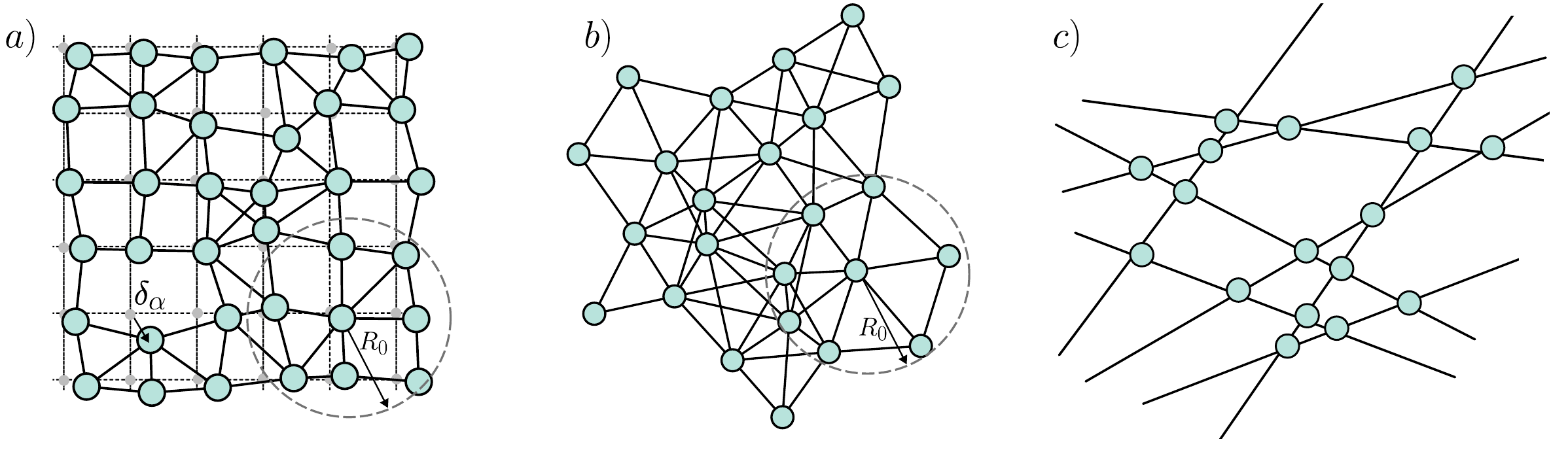}
\caption{Schematic depiction of the structurally disordered lattice models in this work.
In each panel, we show a 2D non-crystalline lattice of sites [blue circles] that only interact with other sites to which they are connected by bonds [black lines].
The 2D arrangement of sites in each panel can either be interpreted as the full space of a system with $d=2$, $d_{A}=2$, $d_{T}=0$, and $d_{f}=0$ in Eq.~(\ref{eq:finiteDimAmorph}), or as the $(x,y)$-coordinate subspace of a system with $d=3$, $d_{A}=2$, $d_{T}=1$, and $d_{f}=0$.
(a) A non-crystalline system with Gaussian structural disorder.  
To generate the disordered system in (a), we first place each site $\alpha$ at an initial position ${\bf r}^{C}_{\alpha}$ on a reference crystalline [square] lattice [gray circles and lines]. 
We then displace each site in (a) by a random vector ${\bs \delta}_{\alpha}$ [Eqs.~(\ref{eq:deltaComponents}) and~(\ref{eq:GaussianDeltaDisplacement})] for which each component $\delta_{i,\alpha}$ is drawn from a Gaussian distribution [Eq.~(\ref{eq:GaussianDisorder})].
Next, we iteratively relax all of the intersite bond lengths to narrow their distribution, as detailed in the text surrounding Eqs.~(\ref{eq:iterateDisplacement}) and~(\ref{eq:relaxationIterationProcedure}).
After the iterative bond relaxation, we designate two sites in (a) as ``connected'' [and to therefore carry nonzero hopping elements] if they are separated by less than a distance of $R_{0}$ set to the initial crystal lattice spacing.  
Gaussian-disordered lattices in general retain a ``memory'' of the crystalline lattice from which they originated, and should therefore be used with caution when modeling amorphous systems~\cite{StructAmoTopoOrder,Franca2024,peng2024structural,Wulles2022}.
(b) A random 2D lattice.
Each site in (b) is initially placed at a random position in the 2D plane, and is then iteratively displaced to narrow the distribution of intersite bond lengths, as detailed in the text surrounding Eqs.~(\ref{eq:iterateDisplacement}) and~(\ref{eq:relaxationIterationProcedure}).
After the bond relaxation process, two sites in (b) are then designated as connected if they are separated by a distance less than $R_{0}$, where $R_{0}$ is an arbitrary [user-defined] parameter.
Unlike in (a), the random lattice in (b) does not carry a unique reference crystalline limit, and is therefore inherently non-crystalline.
(c) A 2D Mikado lattice~\cite{marsal_obstructed_2022,marsal_topological_2020}.
To generate a Mikado lattice, we begin by filling space with randomly located and oriented lines [$d_{A}=2$] or planes [$d_{A}=3$].
We then place lattice sites at the intersection points between the lines or planes, and designate two sites as connected if they are nearest neighbors along the random lines [$d_{A}=2$] or along the intersection lines of the random planes [$d_{A}=3$].
Lastly, we iteratively relax the Mikado lattice intersite bond lengths, as we did in (a,b).
However unlike in (a,b), we do not change the designation of connected sites throughout the bond relaxation process in the Mikado lattice (c).
This ensures that each site on the [relaxed] Mikado lattice carries a fixed number of neighbors [\emph{i.e.} bond coordination], mimicking the behavior of real solid-state amorphous materials~\cite{weaire_electronic_1971,zallen_physics_1998,toh_synthesis_2020,AmorphousTeDFT,AmorphousChalcogenideNatCommH,Grushin2020,Corbae_2023,Fujiwara2023kagome,KarelAmorphousBerryCMG}.
For example, each site in the 2D Mikado lattice in (c) is connected to exactly four other sites.
Like the random lattice in (b), and unlike the Gaussian-disordered lattice in (a), Mikado lattices do not carry unique reference crystalline limits, and are therefore also inherently non-crystalline.}
\label{appfig:structuraldisorder}
\end{figure}

\begin{figure}[t]
    \centering
    \includegraphics[width=\linewidth]{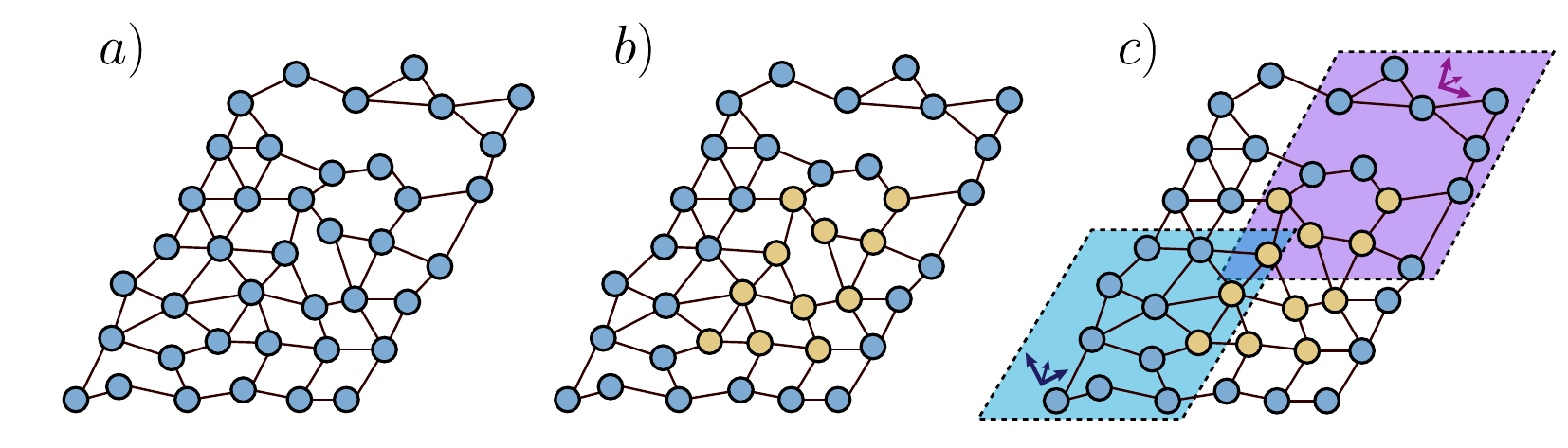}
    \caption{Internal frame disorder implementation.
    (a) A non-crystalline lattice model consisting of disordered sites with the same internal spin and orbital degrees of freedom.
    We further take each site in (a) to consist of a structurally chiral object, such as a chiral molecule [see Appendix~\ref{app:symDefs}].
    In (a), hopping interactions [bonds] between connected chiral sites carry the same dependence on the global Cartesian coordinate frame [see Eq.~(\ref{eq:frameExplicit}) and the following text].
    However in a more realistic model of a strongly disordered solid-state material, local regions should not have physical properties that depend on the global coordinate frame.  
    Hence in this work, we disorder the local hopping frame on each site by augmenting each bond with an O(3) matrix degree of freedom, which can be subdivided into a $\mathbb{Z}_{2}$ local frame chirality scalar $\chi_{\alpha}$ and an SO(3) local frame orientation matrix $R_{\alpha}$ [see the text preceding Eq.~(\ref{eq:frameModification2})].
    We specifically focus in this work on systems with long-range local chirality order in $\chi_{\alpha}$, strong structural disorder [defined below in Appendix~\ref{app:PhysicalObservables}], and without long-range order in $R_{\alpha}$.
    After disordering (a) the site positions, we therefore also implement local frame disorder via two subsequent steps.
    (b) First, we implement local chirality disorder by creating a randomly located \emph{minority chirality domain} of contiguous sites [depicted as yellow circles] that is small on the scale of the system width and in which all sites carry the same fixed value of $\chi_{\alpha}=0,\pm 1$.
    We then assign to all other sites, which comprise the \emph{majority chirality domain} [blue circles in (b)], a different constant value of $\chi_{\alpha}$.
    (c) Lastly, we select several smaller, randomly located regions throughout the system -- termed \emph{frame orientation domains} [depicted as blue and purple shaded areas] -- and assign sites within each region $R_{\alpha}$ matrices characterized by the same, randomly selected values of the frame rotation angles $\theta_{\alpha}$ and $\phi_{\alpha}$ in Eqs.~(\ref{eq:2dSO2Rmat}),~(\ref{eq:eZhatUnitVec}),~(\ref{eq:uDisorderDef}), and~(\ref{eqn:RotationMatrix}).
    If a site lies within the overlap of multiple frame orientation domains [depicted in dark blue in (c)], we assign the site $\theta_{\alpha}$ and $\phi_{\alpha}$ values respectively equal to the average of the $\theta_{\alpha}$ and $\phi_{\alpha}$ angles of sites in the overlapping frame orientation domains.}
\label{appfig:disordertypes} 
\end{figure}

We note that unlike in some other studies of disordered systems, we do not in this work systematically incorporate Anderson-type disorder, under which the chemical potentials of sites or bond signs are allowed to fluctuate randomly.
Specifically, as we will discuss below in this section and in Appendix~\ref{app:models}, the hopping matrix elements in the models in this work deterministically depend only on the tensors of Pauli matrices in the initial Hamiltonian, the [disordered] local internal reference frames of each pair of connected sites, and a hopping rangedness function that rescales the hopping magnitude -- but not its sign or the relative values of its matrix elements -- based on the distance between the two sites. 
The absence of Anderson-type disorder in the majority of our calculations can be understood from two perspectives.
First, as discussed in greater detail below in Appendix~\ref{app:amorphousKramers}, it has been previously established that ${\bf k}={\bf 0}$ chiral fermions, like those in the crystalline limits of the models studied in this work, can be pairwise gapped by periodic potentials [\emph{e.g.} Peierls charge-density waves]~\cite{KramersWeyl,LemutBeenakkerKramersWeylSupercell}.
If bond-sign [bond-order-wave] or chemical potential disorder were realized with exactly alternating signs on neighboring sites in an amorphous metal with a ${\bf p}={\bf 0}$ chiral fermion, this could similarly give rise to a disordered density-wave insulator, like the amorphous obstructed atomic limits analyzed in Ref.~\cite{marsal_obstructed_2022}. 
However in this work, we are focused on the opposite limit of gapless non-crystalline topological semimetals.
Second, in this work, we seek to model amorphous solid-state systems in which the local units largely carry chemically and electronically similar environments [up to frame rotations and chirality inversions], because such systems represent the simplest structurally chiral [on the average] generalizations of well-established and readily experimentally accessible non-crystalline solid-state materials like [achiral] amorphous silicon~\cite{weaire_electronic_1971}, Fe$_x$Sn$_{1-x}$~\cite{Fujiwara2023kagome}, and Co$_2$MnGa~\cite{KarelAmorphousBerryCMG}.
Hence because we wish to model gapless disordered systems far from the limit of average dimerization, and because we wish to construct physically motivated models of amorphous materials with electronically similar local units, then we will largely omit Anderson-type bond-sign and chemical potential disorder in our numerical analysis and in the enumeration below of model disorder categories.
Importantly, however, we have numerically confirmed that when each of the non-crystalline topological semimetal models in this work is simulated with very strong structural and internal degree-of-freedom disorder [\emph{e.g.} on random lattices with large spin and orbital frame disorder, see Figs.~\ref{appfig:structuraldisorder}(b),~\ref{appfig:disordertypes}(c),~\ref{fig:KW_Unif}(a-d),~\ref{fig:C2_Unif}, and~\ref{fig:3F_Unif}], the topological features of each model remain stable under the subsequent addition of weak Anderson [on-site chemical potential] disorder.

Lastly, before detailing our disorder implementation methods, it is also important to briefly highlight differences between the analysis in this work and previous studies of disordered Weyl semimetals.
Through a combination of analytic, numerical, and experimental methods under a large range of disorder conditions and non-crystalline settings, earlier works largely concluded that Weyl semimetal phases become diffusive metals or insulators in the presence of strong disorder~\cite{ringel2015,Chen2015,Altland2015,Altland2016,Gorbar2016,Slager2017,Buchhold2018,Buchhold2018b,Roy2018,Wilson2018,yang_topological_2019,Pixley2021,Brillaux2021,Franca2024,Franca2024b,Grossi2023b,WeylQuasicrystalSCBott,NdAlSi2024DisorderWeylExp,JedJustinCPGEWeylDisorder2024,YiBurkovDiffusiveWeyl}, with surface and bulk phase transitions potentially occurring at distinct disorder scales.
However, the previous analyses almost entirely focused on \emph{band-inversion-type} Weyl semimetals with oppositely-charged Weyl points at $\pm {\bf k}$, which are easily mixed by disorder and are expected to merge into trivial bulk Fermi rings [spheres] as the system approaches the spectrally-isotropic amorphous limit [see Appendix~\ref{app:pseudoK} and Refs.~\cite{zallen_physics_1998,VanMechelen:2018cy,vanMechelenNonlocal,Ciocys2023,Grushin2020,Corbae_2023}].
Furthermore, in contrast to the present work, the previous investigations of disordered topologically chiral semimetals were also performed using disorder realizations that were structurally achiral on the average, or overlooked the role of average or exact structural chirality in their analyses.
Conversely, in the models introduced and analyzed in this work [detailed in Appendix~\ref{app:models}], only bulk chiral fermions with the \emph{same} topological chiral charges are mixed by disorder in systems with net [average] structural chirality.
Hence the topological semimetal systems studied in this work instead remain gapless and bulk-topological up to strong disorder scales, as we will explicitly demonstrate below in Appendices~\ref{app:amorphousKramers},~\ref{app:amorphousCharge2}, and~\ref{app:amorphousMultifold}.

\paragraph*{\bf Structural Disorder Implementation} -- $\ $ At first glance, it may seem straightforward to define an amorphous system as a system with very strong structural disorder.
However in practice, it can be difficult to define the degree of structural disorder in a non-crystalline system, because this requires comparing a structurally disordered system to a set of reference crystalline lattices.
Frequently there is no unique way of identifying a single ``correct'' crystalline limit [though we will numerically establish a working definition of strong disorder below in Appendix~\ref{app:PhysicalObservables} using the two-momentum disorder-averaged Green's function $\bar{\mathcal{G}}(E,\mathbf{p},\mathbf{p'})$].
Additionally, given a system of irregularly placed lattice sites, there are also in general many different paths along which the system may be deformed to the same crystalline limit, and it can be difficult to meaningfully compare relative distances along these paths to quantify the ``degree'' of disorder.

We will seek to avoid these issues by approximating the amorphous limit of a solid-state system using several different numerical implementations of structural disorder, and then comparing system observables obtained using each disorder scheme.
As we will explicitly demonstrate below in Appendices~\ref{app:amorphousKramers},~\ref{app:amorphousCharge2}, and~\ref{app:amorphousMultifold}, each of our non-crystalline tight-binding models exhibits the same spectral and topological features across several distinct structural disorder implementation methods. 
This provides strong evidence that our structural disorder methods converge to the same amorphous limit, and that in this limit, the models are universally governed by the same average-symmetry, average-geometry [structural chirality], and topological principles identified in this work, both at long wavelengths [${\bf p}\approx{\bf 0}$], as well as at even shorter wavelengths [$|{\bf p}|\approx \pi/\bar{a}$, where $\bar{a}$ is the average nearest-neighbor spacing].

In this work, we specifically employ three distinct numerical methods for approximating the strong structural disorder present in amorphous lattices [schematically depicted in Fig.~\ref{appfig:structuraldisorder}].
Each of the methods consists of a two-step process.
We first generate a set of initial coordinates in which each site $\alpha$ carries an initial position $\tilde{\bf r}_{\alpha}$.
For each of the three methods, we then, as detailed below, subsequently apply a \emph{relaxation procedure} to narrow the distribution of intersite bond distances to better match those of real, solid-state amorphous materials~\cite{zallen_physics_1998,toh_synthesis_2020,Corbae_2023}.
Below, we enumerate the three numerical structural disorder implementations that we employ prior to bond relaxation, highlighting their respective advantages and limitations:
\begin{itemize}
\item \textbf{Gaussian structural disorder:} We begin with a 3D tight-binding model on a regular crystalline lattice with translation symmetry in three linearly independent directions.  
As depicted in Fig.~\ref{appfig:structuraldisorder}(a), we then displace each site $\alpha$ with an initial position in the crystal ${\bf r}^{C}_{\alpha}$ by a random vector ${\bs \delta}_{\alpha}$:
\begin{equation}
{\bs \delta}_{\alpha} = \begin{pmatrix}
\delta_{x,a}\\
\delta_{y,a}\\
\delta_{z,a}
\end{pmatrix},
\label{eq:deltaComponents}
\end{equation}
such that the position $\tilde{\bf r}_{\alpha}$ of each site in the disordered system [prior to bond relaxation] is given by:
\begin{equation}
\tilde{\bf r}_{\alpha} = {\bf r}^{C}_{\alpha} + {\bs \delta}_{\alpha}.
\label{eq:GaussianDeltaDisplacement}
\end{equation}
Across the disordered system, each component $\delta_{i,\alpha}$ of the displacement vector in the $d_{A}$ disordered system [sub]space [see Eq.~(\ref{eq:finiteDimAmorph})] is drawn from a Gaussian [normal] distribution $\mathcal{N}(0,\eta)$, in which $0$ denotes the mean value and $\eta$ indicates the standard deviation~\cite{StructAmoTopoOrder,Franca2024,peng2024structural,Wulles2022}. 
Conversely, each component $\delta_{i,\alpha}$ that lies outside of the $d_{A}$ disordered system [sub]space is set to zero. 
For example, in the case of a 3D system with smectic Gaussian disorder [detailed below in Appendix~\ref{app:SmecticNematicDisorder}], the $d_{A}=2$ disordered subspace lies in the $(x,y)$-plane, and the $(x,y)$ coordinates of each site are therefore displaced from the crystalline lattice by random displacements $\delta_{xy,\alpha}$ in Eqs.~(\ref{eq:deltaComponents}) and~(\ref{eq:GaussianDeltaDisplacement}).
Unlike the in-plane displacements, the out-of-plane displacements remain zero under $d_{A}=2$ smectic Gaussian disorder [$\delta_{z,\alpha}=0$ for all $\alpha$], preserving lattice translation symmetry in the $z$-direction. 
In summary, for a system in this work with Gaussian structural disorder, the components $\delta_{i,\alpha}$ of ${\bs \delta}_{\alpha}$ in the $d_{A}$-dimensional disordered system [sub]space exhibit a distribution governed by a Gaussian probability density function:
\begin{equation}
P(\delta_{i,\alpha}) = \frac{1}{\sqrt{2\pi\eta^{2}}}\exp\left(-\frac{\delta_{i,\alpha}^2}{2\eta^2}\right).
\label{eq:GaussianDisorder}
\end{equation}

From a physical perspective, Gaussian disorder can be considered akin to melting and then subsequently quenching a crystalline material close to its melting temperature~\cite{StructAmoTopoOrder,Franca2024,peng2024structural,Wulles2022}.
In this picture, the Gaussian-disordered site displacements are distributed with a standard deviation of $\eta^2 \propto k_{B}T$ in Eq.~(\ref{eq:GaussianDisorder}), where $T$ is the maximum temperature of the melt just prior to the system quench.  
From a modeling perspective, Gaussian disorder is advantageous because it allows one to continuously track system properties from a clearly defined reference crystal [$\eta=0$] into the regime of strong structural disorder.
For example, in our Gaussian disorder calculations using square and cubic lattices, we define the nearest [connected] neighbors of a particular site [both before and throughout bond relaxation iterations] as those that lie within a distance $R_{0}$ from the site, where $R_{0}$ can meaningfully be set to the $\eta=0$ crystalline lattice spacing [Fig.~\ref{appfig:structuraldisorder}(a)].
However, for the same reason -- namely that Gaussian disorder retains some ``memory'' of the $\eta=0$ crystal lattice -- numerical quantities obtained from Gaussian disorder may not always truly approximate the behavior of a maximally structurally disordered [\emph{i.e.} amorphous] system.  
Hence in this work, we have computed and compared numerical observables in all of our non-crystalline tight-binding models using both Gaussian disorder as well as the inherently non-crystalline random and Mikado lattice structural disorder schemes detailed below.

\item \textbf{Random lattices:} To generate the structurally disordered initial site positions $\tilde{\bf r}_{\alpha}$ in a 3D random lattice, we begin with a system with $d=3$ and $d_{A}=2,3$ in Eq.~(\ref{eq:finiteDimAmorph}).
We then assign each component of $\tilde{\bf r}_{\alpha}$ in the $d_{A}$ disordered system [sub]space a random value drawn from a uniform distribution across the system width [Fig.~\ref{appfig:structuraldisorder}(b)]~\cite{agarwala_topological_2017,xiao_photonic_2017,Mitchell2018,poyhonen_amorphous_2018,yang_topological_2019}.
The remaining $d-d_{A}$ components of $\tilde{\bf r}_{\alpha}$ are then assigned regularly spaced [crystalline] values.

As previously for the case of Gaussian structural disorder, the nearest neighbors of a particular site on a random lattice can be defined as the those that lie within a distance $R_{0}$ from the site [both before and throughout iterative bond relaxation, see Fig.~\ref{appfig:structuraldisorder}(b)].
However unlike for Gaussian disorder, $R_{0}$ is an \emph{arbitrary} value for the random lattice, because there does not exist a unique reference crystalline limit.
In this sense, a structurally disordered random lattice is inherently non-crystalline.
We note that in a random lattice subject to the arbitrary-range $R_{0}$ nearest-neighbor hopping cutoff detailed above, sites do not carry a consistent [fixed] bond coordination.
This notably differs from most real solid-state amorphous materials, in which most local units [atoms or molecules] exhibit the same fixed bond coordination, owing to their chemically similar local environments~\cite{weaire_electronic_1971,zallen_physics_1998,toh_synthesis_2020,AmorphousTeDFT,AmorphousChalcogenideNatCommH,Grushin2020,Corbae_2023,Fujiwara2023kagome,KarelAmorphousBerryCMG}.

\item \textbf{Mikado lattices:} For a final form of structural disorder, we can also generate a 3D non-crystalline lattice by placing lines [in a 2D position-component subspace for $d_{A}=2$ in Eq.~(\ref{eq:finiteDimAmorph})] or planes [for $d_{A}=3$ in Eq.~(\ref{eq:finiteDimAmorph})] with random locations and orientations throughout 3D space~\cite{marsal_obstructed_2022,marsal_topological_2020}.
For the purposes of the discussion below, we will find it helpful for the case of $d_{A}=2$ to take the random lines to specifically parameterize $(x,y)$-coordinate components.

To numerically implement random lines or planes, we begin by moving along the $y$-axis for $d_{A}=2$ [or $z$-axis for $d_{A}=3$] and select a set of points spaced according to a Poisson process such that within a length $R$, the 1D density of points $\rho$ satisfies the Poisson distribution $\text{Pois}\left(2R\sqrt{\pi\rho}\right)$.
For $d_{A}=2$, we then use each randomly placed point to draw a line in $(x,y)$-coordinate space with an angle from the $x$-axis randomly drawn from the uniform distribution $\mathcal{U}(0,2\pi)$.
For $d_{A}=3$, we instead use each randomly placed point to draw a 2D plane with a normal vector $\hat{n} =(\sin(\theta)\cos(\phi),\sin(\theta)\sin(\phi),\cos(\theta))$, where $\theta$ is randomly drawn from the uniform distribution $\mathcal{U}(0,\pi)$ and $\phi$ is randomly drawn from the uniform distribution $\mathcal{U}(0,2\pi)$.

Next, using the random network of lines [$d_{A}=2$] or planes [$d_{A}=3$], we then place lattice sites at the zero-dimensional intersection points between the lines [with regular spacing in the remaining $z$-component of $\tilde{\bf r}_{\alpha}$ for $d_{A}=2$] or planes [for $d_{A}=3$].
Unlike previously for the Gaussian-disordered and random lattices [Fig.~\ref{appfig:structuraldisorder}(a,b)], we next strictly define sites on this \emph{Mikado} lattice [in the disordered $d_{A}$ space] to be connected by bonds if [and only if] they are nearest neighbors along the randomly oriented lines [$d_{A}=2$, depicted in Fig.~\ref{appfig:structuraldisorder}(c)] or nearest neighbors along the intersection lines of two randomly oriented planes [for $d_{A}=3$].
Unlike in the Gaussian disorder and random lattice implementations above, the designation of two sites $\alpha$ and $\beta$ as ``connected'' on the Mikado lattice is based entirely on the initial graph of site positions $\tilde{\bf r}_{\alpha}$ and $\tilde{\bf r}_{\beta}$, and is not updated throughout the subsequent iterative bond relaxation process, as we will shortly detail below.
Like the random lattices detailed above, and unlike in the case of Gaussian structural disorder, Mikado lattices of line and plane intersection points do not carry unique reference crystalline limits, and are therefore also inherently non-crystalline.

Lastly, we note that this specific choice of nearest-neighbor bond assignments guarantees that each site on the Mikado lattice carries a fixed nearest-neighbor bond coordination in the $d_{A}$-dimensional disordered system [sub]space.
Specifically, for $d_{A}=2$, each site is connected to exactly four other [in-plane] sites [as well as two other sites by out-of plane $z$-direction crystalline bonds, see Fig.~\ref{appfig:structuraldisorder}(c) for the in-plane bond coordination]. 
Similarly, for $d_{A}=3$, each site on the Mikado lattice is connected to six other sites.
The fixed bond coordination of Mikado lattices is advantageous for modeling real solid-state materials, which, as detailed above, largely consist of local units with the same bond coordination, owing to their chemically similar local environments~\cite{weaire_electronic_1971,zallen_physics_1998,toh_synthesis_2020,AmorphousTeDFT,AmorphousChalcogenideNatCommH,Grushin2020,Corbae_2023,Fujiwara2023kagome,KarelAmorphousBerryCMG}.
For example in amorphous carbon monolayers, nearly every atom carries a fixed threefold in-plane bond coordination -- like that of crystalline graphene -- due to the $sp^{2}$ bonding in monolayer carbon~\cite{toh_synthesis_2020}.
\end{itemize}

After using one of the above three structural disorder implementations to generate a set of initial positions $\tilde{\bf r}_{\alpha}$ for a non-crystalline lattice, there will in general exist a significant variation in the intersite bond lengths in the $d_{A}$-dimensional disordered system [sub]space.
Specifically, for each site $\alpha$, we may define an initial $d_{A}$-[sub]space coordinate vector $\tilde{\bf u}_\alpha$ where:
\begin{equation}
\tilde{\bf u}_{\alpha} = \tilde{\bf r}_{\alpha} - \left[\tilde{\bf r}_{\alpha}\cdot\hat{z} 
 \right]\hat{z} = \begin{pmatrix}
\tilde{x}_a\\
\tilde{y}_a\\
0
\end{pmatrix}\quad\text{ for }d_{A}=2,
\label{eq:da2initialDvector}
\end{equation}
and:
\begin{equation}
\tilde{\bf u}_{\alpha} = \tilde{\bf r}_{\alpha} = \begin{pmatrix}
\tilde{x}_a\\
\tilde{y}_a\\
\tilde{z}_a
\end{pmatrix}\quad\text{ for }d_{A}=3.
\label{eq:da3initialDvector}
\end{equation}
Using Eqs.~(\ref{eq:da2initialDvector}) and~(\ref{eq:da3initialDvector}), we then define for each pair of sites the $d_{A}$-[sub]space intersite bond vector:
\begin{equation}
\tilde{\bf u}_{\alpha\beta} = \tilde{\bf u}_{\alpha} - \tilde{\bf u}_{\beta},
\label{eq:initialRelaxBonds}
\end{equation}
where $\alpha$ and $\beta$ denote two connected [\emph{i.e.} bonded or neighboring] sites within the specific structural disorder implementation as detailed above and in Fig.~\ref{appfig:structuraldisorder}.

However, unlike the initial values of $\tilde{\bf u}_{\alpha\beta}$ in our calculations, the distances between neighboring atomic or molecular units in real solid-state amorphous materials exhibit a comparably narrow distribution of values centered around the average nearest-neighbor spacing $\bar{a}$.
More specifically, in real solid-state amorphous materials, the local unit positions ${\bf r}_{\alpha}$ give rise to a structure factor $S({\bf p})$ that exhibits a sharp, ring-like feature  [for $d_{A}=2$] or sphere-like feature [for $d_{A}=3$] at $|{\bf p}|=2\pi/\bar{a}$~\cite{KittelIntroSSP,zallen_physics_1998,mansha_robust_2017,toh_synthesis_2020,Ciocys2023,HawatStructureFactor,mordret2024beatingaliasinglimitaperiodic}, where:
\begin{equation}
S({\bf p}) = \frac{1}{N_{\text{sites}}}\left| \sum_{\alpha=1}^{N_\text{sites}} e^{i{\bf p}\cdot{\bf r}_{\alpha}} \right|^{2}.
\label{eq:StructureFactor}
\end{equation}

To more closely align the distribution of bond lengths in our structurally disordered models with those of solid-state amorphous materials, we therefore next iteratively relax the $d_{A}$ components of the initial site positions $\tilde{\bf r}_{\alpha}$ employing the procedure detailed in Refs.~\cite{spillage_2022,marsal_obstructed_2022,Ciocys2023}.
In each iteration of our relaxation procedure, we begin with the set of $d_{A}$-[sub]space coordinate vectors $\tilde{\bf u}_{\alpha}$ in Eqs.~(\ref{eq:da2initialDvector}) and~(\ref{eq:da3initialDvector}).
We then successively displace each site $\alpha$ by renormalizing:
\begin{equation}
\tilde{\bf u}_{\alpha} \rightarrow \tilde{\bf u}_{\alpha} + {\bf D}_{\alpha},
\label{eq:iterateDisplacement}
\end{equation}
where:
\begin{equation}
\mathbf{D}_{\alpha} = -\sum_\beta  k\left(|\tilde{\bf u}_{\alpha\beta}| - l_{0}\right) \frac{\tilde{\bf u}_{\alpha\beta}}{|\tilde{\bf u}_{\alpha\beta}|},
\label{eq:relaxationIterationProcedure}
\end{equation}
in which $\tilde{\bf u}_{\alpha\beta} = \tilde{\bf u}_{\alpha}-\tilde{\bf u}_{\beta}$ [Eq.~(\ref{eq:initialRelaxBonds})], and the index $\beta$ in the sum runs only over sites that are ``connected'' to $\alpha$ as defined in Fig.~\ref{appfig:structuraldisorder} and the text following Eq.~(\ref{eq:GaussianDisorder}).
In Eq.~(\ref{eq:relaxationIterationProcedure}), $l_{0}$ is a user-defined parameter that represents a target final value for the average nearest-neighbor site spacing $\bar{a}$.
To obtain a value for $l_{0}$ in our $d=3$ system calculations, we first fix the system lengths in the Cartesian $x$-, $y$-, and $z$-directions as $L_{x}$, $L_{y}$, and $L_{z}$, respectively, and identify the total number of sites $N_{\text{sites}}$.
We then consider either a hypothetical regular tetragonal [$d_{A}=2$] or cubic [$d_{A}=3$] lattice with the same $L_{x,y,z}$ and $N_{\text{sites}}$.
For both $d_{A}=2$ [with lattice disorder in the $(x,y)$-coordinate plane] and $d_{A}=3$, we then set $l_{0}$ to the value of the intersite spacing in the $(x,y)$-plane of the hypothetical crystalline lattice.
Lastly, Eq.~(\ref{eq:relaxationIterationProcedure}) also contains a second user-defined parameter $k$ that sets the scale of the displacement distances $|{\bf D}_{\alpha}|$ in each relaxation iteration.
We chose the value $k=l_{0}/4$ for all of the calculations performed in this work.

After modifying the $d_{A}$ disordered components of the initial site positions $\tilde{\bf r}_{\alpha}$ through several iterations of the bond relaxation procedure in Eqs.~(\ref{eq:iterateDisplacement}) and~(\ref{eq:relaxationIterationProcedure}), we obtain a set of final site positions ${\bf r}_{\alpha}$ for which:
\begin{equation}
{\bf r}_{\alpha} = \tilde{\bf u}_{\alpha} + \left[\tilde{\bf r}_{\alpha}\cdot\hat{z} 
 \right]\hat{z}\quad\text{ for }d_{A}=2,
\label{eq:da2Finalvector}
\end{equation}
and:
\begin{equation}
{\bf r}_{\alpha} = \tilde{\bf u}_{\alpha}\quad\text{ for }d_{A}=3,
\label{eq:da3Finalvector}
\end{equation}
where $\tilde{\bf u}_{\alpha}$ in Eqs.~(\ref{eq:da2Finalvector}) and~(\ref{eq:da3Finalvector}) represents the final $d_{A}$-[sub]space coordinate vector for the site $\alpha$ after all bond relaxation iterations.
Importantly, two sites $\alpha$ and $\beta$ with the respective final positions ${\bf r}_{\alpha}$ and ${\bf r}_{\beta}$ are designated as connected in our calculations through criteria that are specific to the structural disorder implementation scheme employed to generate the initial site positions $\tilde{\bf r}_{\alpha}$ and $\tilde{\bf r}_{\beta}$.
Specifically, for Gaussian disorder and random lattices, we consider two sites $\alpha$ and $\beta$ to be ``connected'' or ``bonded'' [\emph{i.e.} to have nonzero hopping] if their final coordinates ${\bf r}_{\alpha}$ and ${\bf r}_{\beta}$ lie within the prespecified distance $R_{0}$ of each other [Fig.~\ref{appfig:structuraldisorder}(a,b)].
Conversely on the Mikado lattice, $\alpha$ and $\beta$ are designated as connected based entirely on whether they were connected by lines or planes in the initial Mikado lattice generation procedure [Fig.~\ref{appfig:structuraldisorder}(c)], even though their final positions in general differ from their initial positions after the bond relaxation process [${\bf r}_{\alpha,\beta}\neq\tilde{\bf r}_{\alpha,\beta}$].
For the non-crystalline tight-binding models studied in this work [detailed in Appendices~\ref{app:amorphousKramers},~\ref{app:amorphousCharge2}, and~\ref{app:amorphousMultifold}], we find that $\sim 10$ relaxation iterations are sufficient in practice to generate a system configuration with a structure factor $S({\bf p})$ [Eq.~(\ref{eq:StructureFactor})] that exhibits a qualitatively sharp peak at $|{\bf p}|=2\pi/\bar{a}$ like in real solid-state amorphous materials~\cite{KittelIntroSSP,zallen_physics_1998,toh_synthesis_2020,Ciocys2023,HawatStructureFactor,mordret2024beatingaliasinglimitaperiodic,Corbae_2023}.

\paragraph*{\bf Internal Degree-of-Freedom Disorder Implementation} -- $\ $ In structurally disordered tight-binding models consisting of point-like sites with internal spin and orbital degrees of freedom, there also exists significant freedom in defining the intersite hopping matrix elements.
For example, in a non-crystalline 3D tight-binding model with a spin-1/2 degree of freedom on each site parameterized by the Pauli matrices $\sigma^{i}$, a bond between two neighboring [connected] sites $\alpha$ and $\beta$ with the respective positions ${\bf r}_{\alpha}$ and ${\bf r}_{\beta}$ can be assigned hopping that is proportional to any linear combination of $\sigma^{i}$ [including the $2\times 2$ identity matrix $\sigma^{0}$].
More specifically, to interpret the above system as a disordered realization of a crystal that at low energies near ${\bf p}\approx {\bf 0}$ exhibits Dresselhaus spin-orbit coupling [SOC] of the form $\sum_{i=x,y,z}p_{i}\sigma^{i}$ [such as the disordered Kramers-Weyl semimetals studied in this work, see Appendix~\ref{app:amorphousKramers} and Refs.~\cite{DresselhausSOC,KramersWeyl,OtherKramersWeyl,AndreiTalk}], it is tempting to simply assign the SOC part of the hopping matrix elements $T^{\sigma}_{\alpha\beta}$ between the connected sites $\alpha$ and $\beta$ to be:
\begin{equation}
T^{\sigma}_{\alpha\beta} = iv_{x}[{\bf d}_{\alpha\beta}\cdot\hat{x}]\sigma^{x} + iv_{y}[{\bf d}_{\alpha\beta}\cdot\hat{y}]\sigma^{y} + iv_{z}[{\bf d}_{\alpha\beta}\cdot\hat{z}]\sigma^{z}, 
\label{eq:TmatrixRot}
\end{equation}
where:
\begin{equation}
{\bf d}_{\alpha\beta} = {\bf r}_{\alpha} - {\bf r}_{\beta}.
\label{eq:tempDAB}
\end{equation}

However, Eq.~(\ref{eq:TmatrixRot}) subtly contains implicit knowledge of the global Cartesian coordinate axes.
For example, if ${\bf d}_{\alpha\beta}$ in Eq.~(\ref{eq:TmatrixRot}) is oriented along the Cartesian $x$-direction, then the hopping matrix $T^{\sigma}_{\alpha\beta}$ is proportional to $\sigma^{x}$, as it is in the crystalline Kramers-Weyl model [see Appendix~\ref{app:PristineKramers} and Refs.~\cite{KramersWeyl,OtherKramersWeyl,AndreiTalk}].
This can be summarized by stating that Eq.~(\ref{eq:TmatrixRot}) exhibits an implicit ``frame'' orientational order that respectively locks hopping in the Cartesian $x,y,z$-directions to $\sigma^{x,y,z}$ bonds, despite the absence of well-defined global crystallographic axes in the disordered system.  
The internal frame of the intersite hopping $T^{\sigma}_{\alpha\beta}$ can be seen by re-expressing Eq.~(\ref{eq:TmatrixRot}) in the form:
\begin{equation}
T^{\sigma}_{\alpha\beta} = i{\bf d}_{\alpha\beta}^\mathsf{T}Q_{\sigma}{\bs \sigma},
\label{eq:frameExplicit}
\end{equation}
where:
\begin{equation}
Q_{\sigma} = \begin{pmatrix}
              v_x&0&0\\
              0&v_y&0\\
              0&0&v_z
          \end{pmatrix},
\label{eq:SOCframeQ}
\end{equation}
and:
\begin{equation}
{\bs \sigma} =  \begin{pmatrix}
\sigma^{x} \\ \sigma^{y} \\ \sigma^{z}\end{pmatrix}.
\label{eq:tempSigmaVec}
\end{equation}
In Eqs.~(\ref{eq:frameExplicit}) and~(\ref{eq:SOCframeQ}), $Q_{\sigma}$ can then be interpreted as a row vector of three frame vectors ${\bf f}_{i}$:
\begin{equation}
Q_{\sigma} = \begin{pmatrix}
              {\bf f}_{x} & {\bf f}_{y} & {\bf f}_{z}
          \end{pmatrix},
\label{eq:QframeDef}
\end{equation}
where:
\begin{equation}
{\bf f}_{x} = \begin{pmatrix}
v_{x} \\ 
0 \\ 
0\end{pmatrix},\ {\bf f}_{y} = \begin{pmatrix}
0 \\ 
v_{y} \\ 
0\end{pmatrix},\ {\bf f}_{z} = \begin{pmatrix}
0 \\ 
0 \\ 
v_{z}\end{pmatrix},
\end{equation}
such that each local hopping frame vector ${\bf f}_{i}$ is aligned with a global Cartesian coordinate axis.

In a realistic model of a strongly disordered solid-state material, however, no particular local region in position space should have physical properties that depend on the global coordinate frame, such as the $x$-direction in an arbitrary coordinate system.
Hence, to model amorphous solid-state systems, we must also for each model in this work disorder the local internal frame orientation so that the non-crystalline model does not implicitly contain orbital hopping or SOC terms that are locked in any particular region to the global coordinate frame.  
More subtly, as we will discuss below, the internal degrees of freedom on each site may also give rise to inequivalent intersite hopping if the local coordinate frame of the spins and orbitals on each site is right- or left-handed [\emph{i.e.} carries distinct local chirality~\cite{LocalChiralityMoleculeReviewNatChem,LocalChiralityLiquidCrystals,KamienLubenskyChiralParameter,LocalChiralityDomainLiquidCrystal,LocalChiralityTransfer,KamienChiralLiquidCrystal,LocalChiralityQuasicrystalVirus,lindell1994electromagneticBook,LocalChiralityVillain,LocalChiralityWenZee,LocalChiralityBaskaran,LocalChiralitySpinFrame}].

The most general, length-preserving [orthogonal] transformation that leaves invariant the \emph{magnitude} of the inner products $|{\bf f}_{i}\cdot {\bf f}_{j}|$ can be implemented by acting on each frame vector ${\bf f}_{i}$ with a matrix $M_{\alpha\beta}$ that is a representation of the group O(3).
Tellingly, O(3) can be re-expressed as the direct product of the group $\mathbb{Z}_{2\chi}$ and the proper rotation group SO(3)~\cite{BigBook,PinGroup,PointGroupTables,linegroupsbook}:
\begin{equation}
\text{O}(3) = \text{SO}(3) \cup \mathcal{I}_{\chi}\text{SO}(3) = \mathbb{Z}_{2\chi}\times\text{SO}(3),
\label{eq:tempO3}
\end{equation}
where:
\begin{equation}
\mathbb{Z}_{2\chi}=\left\{E,\mathcal{I}_{\chi}\right\},
\label{eq:tempZ2chi}
\end{equation}
in which $E$ is the identity element and $\mathcal{I}_{\chi}$ is spatial inversion in the internal coordinate space of the frame vectors ${\bf f}_{x,y,z}$.  
Importantly, in Eqs.~(\ref{eq:tempO3}) and~(\ref{eq:tempZ2chi}), the elements in SO(3) rotate the frame vectors while preserving their relative \emph{handedness} $\sgn[({\bf f}_{x}\times{\bf f}_{y})\cdot{\bf f}_{z}]$, whereas the internal rotoinversion elements in the set $\mathcal{I}_{\chi}\text{SO}(3)$ also rotate the frame vectors, but conversely reverse their handedness.
Overall, this implies that to generalize $T^{\sigma}_{\alpha\beta}$ in Eq.~(\ref{eq:frameExplicit}) to accommodate the possibility of local frame disorder, we should augment $T^{\sigma}_{\alpha\beta}$ with an additional O(3) matrix degree of freedom $M_{\alpha\beta}$ that acts on $Q_{\sigma}$ [and hence the frame vectors ${\bf f}_{x,y,z}$ through Eq.~(\ref{eq:QframeDef})].
Using the decomposition of O(3) in Eqs.~(\ref{eq:tempO3}) and~(\ref{eq:tempZ2chi}), we may then represent $M_{\alpha\beta}$ as the product of a $\mathbb{Z}_{2}$ scalar $\chi_{\alpha\beta}=\pm 1$ and an SO(3) proper rotation matrix $R_{\alpha\beta}$:
\begin{equation}
T^{\sigma}_{\alpha\beta} = i{\bf d}_{\alpha\beta}^\mathsf{T}M_{\alpha\beta}Q_{\sigma}{\bs \sigma} = i{\bf d}_{\alpha\beta}^\mathsf{T}\left[\chi_{\alpha\beta}R_{\alpha\beta}\right]Q_{\sigma}{\bs \sigma} = i\chi_{\alpha\beta}{\bf d}_{\alpha\beta}^\mathsf{T}R_{\alpha\beta}Q_{\sigma}{\bs \sigma}.
\label{eq:frameModification}
\end{equation}

Inspired by analogous local reference frame transformations in lattice gauge theory~\cite{KogutLatticeGaugeRMP}, we next choose the specific forms of $\chi_{\alpha\beta}$ and $R_{\alpha\beta}$ in Eq.~(\ref{eq:frameModification}) to be:
\begin{equation}
\chi_{\alpha\beta} = \frac{1}{2}\left(\chi_{\alpha}+\chi_{\beta}\right),\ R_{\alpha\beta} = R_{\beta}R_{\alpha}^\mathsf{T},
\label{eq:temp2siteFrameBreakdown}
\end{equation}
implying that:
\begin{equation}
T^{\sigma}_{\alpha\beta} = i\left(\frac{\chi_{\alpha}+\chi_{\beta}}{2}\right)\left[R_{\alpha}R^\mathsf{T}_{\beta}{\bf d}_{\alpha\beta}\right]^\mathsf{T}Q_{\sigma}{\bs \sigma},
\label{eq:frameModification2}
\end{equation}
such that the $\mathbb{Z}_{2}$ frame chirality [handedness] and SO(3) frame orientation of each bond originate from internal properties of the two connected lattice sites $\alpha$ and $\beta$.
Specifically, in Eq.~(\ref{eq:frameModification2}), each site $\alpha$ carries two sets of orientational parameters -- a $\mathbb{Z}_{2}$ Ising-spin-like~\cite{chaikinLubenskyBook} discrete scalar orientational parameter $\chi_{\alpha}=\pm 1$ that corresponds to ``local frame chirality''~\cite{LocalChiralityMoleculeReviewNatChem,LocalChiralityLiquidCrystals,KamienLubenskyChiralParameter,LocalChiralityDomainLiquidCrystal,LocalChiralityTransfer,KamienChiralLiquidCrystal,LocalChiralityQuasicrystalVirus,lindell1994electromagneticBook,LocalChiralityVillain,LocalChiralityWenZee,LocalChiralityBaskaran,LocalChiralitySpinFrame}, and a continuous SO(3) frame orientation parameterized by the matrix $R_{\alpha}$. 
For example, when Eq.~(\ref{eq:frameModification2}) is realized on a homogeneous crystalline lattice, $\chi_{\alpha}$ controls the handedness of hopping [here Dresselhaus SOC~\cite{DresselhausSOC,KramersWeyl}] that breaks rotoinversion symmetries and gives rise to structural chirality [see Eq.~(\ref{eq:rotoinversion})], and $R_{\alpha}$ controls the relative orientations of the spin texture and the crystallographic axes, as detailed below in the text surrounding Eq.~(\ref{eq:SpinDOS}).  
In this sense, $\chi_{\alpha}$ can also be viewed as a time-reversal-invariant generalization of the local spin chirality order parameter introduced in Refs.~\cite{LocalChiralityVillain,LocalChiralityWenZee,LocalChiralityBaskaran} to distinguish spin-liquid ground states.

The decomposition of the O(3) local frame transformations into $\chi_{\alpha}$ and $R_{\alpha}$ in Eqs.~(\ref{eq:frameModification}),~(\ref{eq:temp2siteFrameBreakdown}), and~(\ref{eq:frameModification2}) allows us to study systems in which local frame chirality and orientation are separately ordered or disordered.
The ordering of $\chi_{\alpha}$ and $R_{\alpha}$ can either be evaluated at the lattice level, or by using local averages to construct continuum order parameters~\cite{chaikinLubenskyBook,KogutLatticeGaugeRMP}.
For example, we may introduce a continuous chirality order parameter field $\chi({\bf r})$~\cite{KamienLubenskyChiralParameter,LocalChiralityVillain,LocalChiralityWenZee,LocalChiralityBaskaran,LocalChiralitySpinFrame} that is equal to the average value of $\chi_{\alpha}$ for sites $\alpha$ whose positions ${\bf r}_{\alpha}$ lie within a specified vicinity of ${\bf r}$.
We may then define \emph{chirality domains} in which all local regions parameterized by ${\bf r}$ exhibit the same continuum local chirality $\chi({\bf r})=\pm 1$
such that deep within a chirality domain, $\chi({\bf r})$ is equal to $\chi_{\alpha}$ for all sites $\alpha$ and ${\bf r}$ away from the domain boundaries.  
In this work, we are largely focused on systems in which the lattice positions and local SO(3) frame orientations are disordered, but in which the $\mathbb{Z}_{2}$ local frame chirality conversely forms domains with long-range [constant] order.  
As an aside, we note for completeness that in Appendices~\ref{app:amorphousKramers} and~\ref{app:amorphousMultifold} below, we also consider the physically-motivated case~\cite{ChiralGlass1,ChiralGlass2,ChiralGlass3,ChiralGlassGiuliaSummary,ChiralGlassMolecules,AmorphousTeDFT,AmorphousChalcogenideNatCommH,ChiralPhaseChange1,ChiralPhaseChange2,MixedChiralPhotonicAlloy} of non-crystalline systems with strong structural disorder and varying concentrations of both chiral [$\chi({\bf r})=\pm 1$] and \emph{achiral} [$\chi({\bf r})=0$] domains.

In the non-crystalline tight-binding models introduced and analyzed in this work in Appendices~\ref{app:amorphousKramers},~\ref{app:amorphousCharge2}, and~\ref{app:amorphousMultifold}, the hopping matrix elements in general have more complicated [\emph{e.g.} nonlinear] relationships to the local coordinate frame compared to $T^{\sigma}_{\alpha\beta}$ in Eq.~(\ref{eq:frameExplicit}). 
However, each hopping matrix element for each bond between two sites $\alpha$ and $\beta$ still enters the Hamiltonian via tensor contractions with the intersite hopping vector ${\bf d}_{\alpha\beta}$.
Hence rather than identify in each model below the dependence of each term on the local frame vectors ${\bf f}_{i}$, we will instead divide the O(3) frame transformations in Eqs.~(\ref{eq:frameModification}),~(\ref{eq:temp2siteFrameBreakdown}), and~(\ref{eq:frameModification2}) into two families of hopping matrix modifications, which may then be separately ordered or disordered [Fig.~\ref{appfig:disordertypes}]:
\begin{itemize}

  \item \textbf{Local chirality disorder:} Each of the tight-binding models studied in this work [enumerated in Appendix~\ref{app:models}] respects the symmetries of a chiral space group [defined in Appendix~\ref{app:symDefs}] when placed on a crystalline lattice.
  Rather than determining the explicit dependence of the hopping matrix elements in the Hamiltonian of each model on the local frame chirality [handedness], we instead more simply identify the terms that break rotoinversion symmetries of the form of Eq.~(\ref{eq:rotoinversion}) when the model is placed on a crystalline lattice, and hence give rise to structural chirality.
  Generalizing the $\mathbb{Z}_{2}$ local frame chirality introduced for $T^{\sigma}_{\alpha\beta}$ in the text surrounding Eq.~(\ref{eq:frameModification2}), we then multiply a chirality- [handedness-] controlling [sub]set of the rotoinversion-symmetry-breaking hopping terms in each model [\emph{i.e.} a set of terms that controls the sign of $C_{\mathcal{H}}$ in the text surrounding Eqs.~(\ref{eq:appStructuralChirality}),~(\ref{eq:quadraticStructuralChirality}), and~(\ref{eq:multifoldStructuralChirality})] by a scalar prefactor proportional to $\chi_{\alpha}+\chi_{\beta}$, in which $\chi_{\alpha}$ represents the local chirality [handedness] of each lattice site $\alpha$.

  In this work, we specifically analyze disordered 3D systems in which $\chi_{\alpha}$ exhibits long-range order across a \emph{majority domain} that is large on the scale of the system width.
  To implement a long-range-ordered chirality domain, we first divide the system into two regions by selecting connected sites within a randomly located contiguous region that is small on the scale of the system width [depicted as yellow circles in Fig.~\ref{appfig:disordertypes}(b)], and assign a constant value of $\chi_{\alpha}$ to each site in the region. 
  We designate this smaller region the \emph{minority chirality domain}, and choose its size such that it contains a prespecified number of sites $N_{m}$ with the same local chirality $\chi_{\alpha}=0$ or $\chi_{\alpha}=\pm 1$, where $N_{m}/N_{\mathrm{sites}}<0.5$. 
  The side lengths $l_{i}$ of the minority chirality domain in our calculations specifically satisfy:
  \begin{equation}
  l_{i}=L_{i}\left(N_m/N_{\mathrm{sites}}\right)^{1/3}\text{ for }i=x,y,z,
  \label{eq:chiralityCornerImplementation}
  \end{equation}
  where $L_{i}$ is $i$-th component of the largest intersite separation vector ${\bf d}_{\alpha\beta}$.
  We then place the lowermost corner of the minority chirality domain [that with the smallest values of $x$, $y$, and $z$] at a random position for which the $i$-th component is drawn from the uniform distribution $\mathcal{U}((L_i-l_i)/4, (L_i-l_i) - (L_i-l_i)/4)$. 
  The $(L_i-l_i)/4$ offset in the distribution of minority chirality domain corner positions ensures that for disorder realizations with open boundaries, the minority domain lies well within the system bulk, which acts to prevent chirality domain walls from obscuring spectral signatures of topological surface states [see Appendix~\ref{app:surfaceGreens} for additional open-boundary-condition calculation details]. 
  The remaining sites in the system outside of the minority domain [depicted as blue circles in Fig.~\ref{appfig:disordertypes}(b)] comprise the \emph{majority chirality} domain.
  The majority chirality domain similarly contains $N_{\mathrm{sites}} - N_{m}$ sites with the same local chirality $\chi_{\alpha}=0$ or $\chi_{\alpha}=\pm 1$, with $\chi_{\alpha}$ taking distinct values for sites within the minority and majority chirality domains.

  \item \textbf{Local frame orientation disorder:} In each non-crystalline tight-binding model in this work [Appendices~\ref{app:amorphousKramers},~\ref{app:amorphousCharge2}, and~\ref{app:amorphousMultifold}], hopping interactions between connected sites enter the Hamiltonian via contractions of vectors and tensors of Pauli matrices with the intersite displacement vector ${\bf d}_{\alpha\beta}$ [Eq.~(\ref{eq:tempDAB})].
  In each case, the Pauli matrices that enter the Hamiltonian for each bond specifically depend on the \emph{orientation} of ${\bf d}_{\alpha\beta}$.
  Hence, we may implement the SO(3) local frame disorder discussed in the text surrounding Eq.~(\ref{eq:frameModification}) by applying randomly generated local rotations to ${\bf d}_{\alpha\beta}$ via the redefinition:
  \begin{equation}
  {\bf d}_{\alpha\beta} \rightarrow R_{\alpha}R_{\beta}^{\mathsf{T}}{\bf d}_{\alpha\beta} \underset{def}{\equiv} \tilde{{\bf d}}_{\alpha\beta},
  \end{equation} 
  in which each matrix $R_{\alpha}$ is an SO(3) internal degree-of-freedom [spin and orbital] frame rotation matrix assigned to the site $\alpha$, analogous to the lattice gauge theory local frame rotations detailed in Ref.~\cite{KogutLatticeGaugeRMP}.

  Our procedure for numerically generating an ensemble of $R_{\alpha}$ matrices in a non-crystalline system depends on whether the internal degrees of freedom on each site carry a preferred axis in 3D internal-coordinate space.
  For example, in the smectic-disordered non-crystalline double-Weyl semimetals analyzed in this work [see Appendices~\ref{app:SmecticNematicDisorder} and~\ref{app:amorphousCharge2}], each local unit has a preferred internal $z$-axis direction that is locked to the global Cartesian $z$-axis, such that each $R_{\alpha}$ is reduced to an SO(2) rotation matrix about the internal $z$ axis.
  In this case, each $R_{\alpha}$ is given by:
  \begin{equation}
      R_{\alpha} = \begin{pmatrix}
          \cos(\phi_\alpha) & - \sin(\phi_\alpha) & 0\\
          \sin(\phi_\alpha) & \cos(\phi_\alpha) & 0
          \\
          0&0&1
       \end{pmatrix},
  \label{eq:2dSO2Rmat}
  \end{equation}
  where $\phi_{\alpha}\in[0,2\pi)$ is a rotation angle in the $(x,y)$-plane randomly selected from a Gaussian [normal] distribution with an average of $\phi_{\alpha}=0$ and a standard deviation of $\eta$ [denoted by $\mathcal{N}\left(0,\eta\right)$, see the text preceding Eq.~(\ref{eq:GaussianDisorder})].

  For the more general case of 3D internal degrees of freedom without preferred axes, we construct $R_{\alpha}$ by first assigning to each site a local unit vector $\hat{u}_{\alpha}$ that indicates the SO(3)-transformed $z$-axis unit vector $\hat{e}_{z}$ of the internal frame of the site $\alpha$.
  We specifically define:
  \begin{equation}
      \hat{e}_{z} = \begin{pmatrix}
  0 \\ 
  0 \\
  1\end{pmatrix},
  \label{eq:eZhatUnitVec}
  \end{equation}
  and:
  \begin{equation}
  \hat{u}_{\alpha} = \begin{pmatrix}
  \sin(\theta_{\alpha})\cos(\phi_{\alpha}) \\ 
  \sin(\theta_{\alpha})\sin(\phi_{\alpha}) \\
  \cos(\theta_{\alpha})\end{pmatrix},
  \label{eq:uDisorderDef}
  \end{equation}
  where $\theta_{\alpha} \in [0,\pi]$ and $\phi_{\alpha} \in [0,2\pi)$ are random angles assigned to each site $\alpha$, each independently drawn from the Gaussian distribution $\mathcal{N}\left(0,\eta\right)$ [see the text surrounding Eqs.~(\ref{eq:GaussianDisorder}) and~(\ref{eq:2dSO2Rmat})].
  To construct the SO(3) matrix $R_{\alpha}$ that transforms the internal frame vector $\hat{e}_{z}$ into $\hat{u}_{\alpha}$, we employ Rodrigues' formula~\cite{friedberg2022rodriguesolindedeslois}:
  \begin{equation}
  \hat{v}_{\mathrm{rot}} = \hat{v}\cos(\gamma) + \left[\hat{w}\times \hat{v}\right]\sin(\gamma) + \hat{w}(\hat{w}\cdot \hat{v})\left[1-\cos(\gamma)\right],
  \label{eq:RotatedVector}
  \end{equation}
  in which the unit vector $\hat{v}$ is transformed into the unit vector $\hat{v}_{\mathrm{rot}}$ by rotating $\hat{v}$ by angle $\gamma$ about a distinct unit vector $\hat{w}$.
  For the specific purpose of transforming $\hat{e}_{z}$ in Eq.~(\ref{eq:eZhatUnitVec}) into $\hat{u}_{\alpha}$ in Eq.~(\ref{eq:uDisorderDef}), we specialize to the case of Eq.~(\ref{eq:RotatedVector}) in which $\hat{w}$ is the unit vector that bisects the angle between $\hat{e}_{z}$ and $\hat{u}_{\alpha}$:
  \begin{equation}
  \hat{w} = \frac{\hat{u}_{\alpha}+ \hat{e}_z}{\lvert \hat{u}_{\alpha} + \hat{e}_z\rvert},
  \end{equation}
  and employ the rotation angle $\gamma=\pi$, such that:
  \begin{equation}
  \hat{u}_{\alpha} = -\hat{e}_z + 2(\hat{u}_{\alpha}+\hat{e}_z)\frac{(\hat{u}_{\alpha}+\hat{e}_z)\cdot \hat{e}_z}{|\hat{u}_{\alpha}+\hat{e}_z|^2}.
  \label{eq:rotatedEZuAlphaRelation}
  \end{equation}
  Applying the rotation matrix generation algorithm in Ref.~\cite{MollerRotationMatrix} to the implicit equality in Eq.~(\ref{eq:rotatedEZuAlphaRelation}), we lastly obtain the following matrix elements of $R_{\alpha}$:
  \begin{equation}
  \left[R_{\alpha}\right]_{ll'} =  - \delta_{ll'} + 2\frac{[\hat{u}_{\alpha}+\hat{e}_z]_l[\hat{u}_{\alpha}+\hat{e}_z]_{l'}}{|\hat{u_{\alpha}}+\hat{e}_z|^2}.
  \label{eqn:RotationMatrix}
  \end{equation}

  Having established our procedure for generating the matrix elements of the SO(2) or SO(3) local frame rotation matrix $R_{\alpha}$ on each site $\alpha$, we next detail our procedure for assigning $R_{\alpha}$ matrices to each site in a disorder realization [\emph{i.e.} replica].
  In this work, we wish to study systems with long-range local chirality order in $\chi_{\alpha}$ [see Eq.~(\ref{eq:frameModification2}) and the surrounding text], strong structural disorder [defined below in Appendix~\ref{app:PhysicalObservables}], and without long-range order in $R_{\alpha}$.
  However in our calculations, we observed that when $R_{\alpha}$ is independently randomly generated on each site, or equivalently when there is a proliferation of large-angle domain walls between regions of constant $R_{\alpha}$, then the pseudo-momentum spectrum exhibits additional features that can obscure and overwhelm spectral signatures of ${\bf p}={\bf 0}$ chiral fermions.
  Hence to balance the desired absence of long-range order in $R_{\alpha}$ with the complications arising from short-range disorder in $R_{\alpha}$, we disorder $R_{\alpha}$ by creating within each system replica a random number of \emph{frame orientation domains} of constant $R_{\alpha}$ that are small [but not infinitesimal] compared the system width [depicted as blue and purple regions in Fig.~\ref{appfig:disordertypes}(c)].  
  In each disorder replica, the number of frame orientation domains is given by an integer $N_{\mathrm{reg}}$ drawn at random from the set $\left\{2,3,4,5\right\}$.
  The side lengths $l_{i}$ of each frame orientation domain satisfy:
  \begin{equation}
  l_{i}=L_{i}r^{1/3}\text{ for }i=x,y,z,
  \end{equation}
  where $L_{i}$ is $i$-th component of the largest intersite separation vector ${\bf d}_{\alpha\beta}$ and $r$ is a number randomly drawn from the uniform distribution $\mathcal{U}(0.25,0.75)$. 
  We place the lowermost corner of each frame orientation domain [that with the smallest values of $x$, $y$, and $z$] at a random position for which the $i$-th component is drawn from the uniform distribution $\mathcal{U}((L_i-l_i)/4, (L_i-l_i) - (L_i-l_i)/4)$. 
  As with the chirality domain implementation discussed above in the text surrounding Eq.~(\ref{eq:chiralityCornerImplementation}), the $(L_i-l_i)/4$ offset in the distribution of frame orientation domain corner positions ensures that for disorder realizations with open boundaries, frame orientation domain walls lie deep within the system bulk, preventing them from obscuring spectral signatures of topological surface states [see Appendix~\ref{app:surfaceGreens} for additional open-boundary-condition calculation details].  
  Lastly, if a site lies within the overlap of multiple frame orientation domains [such as the dark blue region in  Fig.~\ref{appfig:disordertypes}(c)], we assign the site $\theta_{\alpha}$ and $\phi_{\alpha}$ values equal to the average of the $\theta_{\alpha}$ and $\phi_{\alpha}$ angles of sites in the overlapping frame orientation domains.  
\end{itemize}

\subsubsection{Smectic and Nematic Disorder}
\label{app:SmecticNematicDisorder}

In this work, we overall employ two distinct model disorder regimes, which we term ``smectic'' and ``nematic'' to draw analogy with similar order in liquid crystals~\cite{chaikinLubenskyBook,CollingsGoodbyLiquidCrystalBook,KamienLiquidCrystalRMP,KamienChiralLiquidCrystal,ChiralNematicBook}.
First, the non-crystalline tight-binding models studied in this work inhabit three spatial dimensions [$d=3$ in Eq.~(\ref{eq:finiteDimAmorph})].
As discussed above in Appendix~\ref{app:DiffTypesDisorder}, each model in this work is characterized by three independent parameters that can be separately ordered or disordered: the atomic positions, the scalar chirality [handedness] $\chi_{\alpha}$ of the local internal degree-of-freedom frame, and the SO(3) orientation $R_{\alpha}$ of the internal frame relative to the Cartesian coordinate axes [see Eq.~(\ref{eq:frameModification2}) and the surrounding text].
We are primarily focused in this work on systems with long-range  order in $\chi_{\alpha}$ and varying degrees of order in the atomic positions and $R_{\alpha}$.
In our analysis of the models below, we specifically either choose the atomic [site] positions to be fully disordered [$d_{A}=3$ in Eq.~(\ref{eq:finiteDimAmorph})], or to retain lattice translation symmetry in just the $\hat{z}$-direction [$d_{A}=2$].
Adapting terminology from the study of liquid crystals, we then term the former $d=3$, $d_{T}=0$, $d_{A}=3$, $d_{f}=0$ case  \emph{nematic} disorder, and the latter layered $d=3$, $d_{T}=1$, $d_{A}=2$, $d_{f}=0$ case \emph{smectic} disorder [schematically depicted Fig.~\ref{fig:dadtdf}(b,c), see the text surrounding Eq.~(\ref{eq:finiteDimAmorph}) for the definitions of $d_{T}$, $d_{A}$, and $d_{f}$].
In this work, both nematic and smectic disordered systems are characterized by long-range chirality order in $\chi_{\alpha}$, which may be interpreted as a $\mathbb{Z}_{2}$ Ising-like~\cite{chaikinLubenskyBook,LocalChiralityVillain,LocalChiralityWenZee,LocalChiralityBaskaran} orientational parameter [see the text surrounding Eq.~(\ref{eq:frameModification2})].
This is analogous to nematic and smectic phases of liquid crystals composed of rod-like molecules, in which the rods similarly exhibit orientational order, but are respectively fully or partially structurally disordered.
Specifically, nematic and smectic liquid crystals are distinguished by their degree of structural disorder, with $d=3$ nematic liquid crystals exhibiting complete structural disorder, and $d=3$ smectic liquid crystals instead forming layered structures with approximate 1D lattice translation symmetry~\cite{chaikinLubenskyBook,CollingsGoodbyLiquidCrystalBook,KamienLiquidCrystalRMP,KamienChiralLiquidCrystal,ChiralNematicBook}.

Lastly, in both the smectic and nematic disorder regimes of our non-crystalline tight-binding models, we have additional freedom to choose whether the SO(3) local frame orientation $R_{\alpha}$ is fully or partially ordered, or disordered.  
In this work, we primarily use smectic disorder to study a non-crystalline double-Weyl semimetal model with a preferred internal $z$-axis [see Appendix~\ref{app:amorphousCharge2}].
In the double-Weyl model, we hence choose $R_{\alpha}$ to also exhibit partial ordering along the $z$-axis, such that it is reduced to an SO(2) matrix of $(x,y)$-plane internal rotations in our smectic disorder calculations.  
From a physical perspective, preferred axes have also been shown to manifest in real solid-state amorphous materials due to interface and substrate effects~\cite{Helmann2017} and local structural motifs~\cite{Gambino1974,Hellman1992}, and can be promoted through growth methods including vapor diffusion and evaporation~\cite{Crystal_Growth}.
Finally, in our nematic disorder calculations for the remaining non-crystalline chiral fermion models in Appendices~\ref{app:amorphousKramers} and~\ref{app:amorphousMultifold}, we conversely disorder $R_{\alpha}$ across its full range of SO(3) orientational values.  
For completeness, we note that in the liquid crystal literature, the term ``chiral nematic'' [equivalently ``twisted nematic'' or cholesteric] also appears, but refers to systems composed of \emph{achiral} rod-like units that break system-wide [average] rotoinversion symmetries and become structurally chiral [on the average] by arranging in helical screw-like structures~\cite{ChiralNematicBook}.
This is distinct from the nematic disordered systems studied in this work, in which the atomic [site] positions do not form large-scale structural motifs, and in which average chirality instead emerges from an imbalance in the concentrations of oppositely handed structurally chiral local units [see Fig.~\ref{fig:average_sym} and the surrounding text].

\subsection{Momentum-Resolved Green's Functions and Disorder Averaging}
\label{app:PhysicalObservables}

In this section, we will first define the position-space tight-binding basis states of the non-crystalline tight-binding models analyzed in this work [detailed in Appendices~\ref{app:amorphousKramers},~\ref{app:amorphousCharge2}, and~\ref{app:amorphousMultifold}].
From this, we will then construct a Fourier-transformed description of each non-crystalline system using continuum pseudo-momentum plane-wave states, which we will apply to the system Green's function to establish a momentum-space description of the many-particle, two-momentum system spectrum.
Lastly, we will show below that by averaging over multiple disorder realizations [replicas], the pseudo-momentum-space Green's function of each model in this work can approximately simplify to a function of a single pseudo-momentum. 
We will then use this observation to numerically establish a definition of strong structural disorder as the regime in which each non-crystalline [\emph{i.e.} amorphous] system exhibits a well-defined one-momentum Green's function.

\paragraph*{\bf Tight-Binding Formalism} -- $\ $ To begin, each model in this work is expressed in a tight-binding formalism in which the single-particle Hilbert space is spanned by the tight-binding basis states $\lvert \mathbf{r}_{\alpha},l\rangle$, where each $\lvert \mathbf{r}_{\alpha},l\rangle$ is  localized on a site $\alpha$ at the position ${\bf r}_{\alpha}$ and carries an internal [spin and orbital] degree-of-freedom index $l$.
We represent each $\lvert \mathbf{r}_{\alpha},l\rangle$ as an $N_{\text{sites}}N_{\text{orb}}$ column vector where $N_{\text{sites}}$ is the number of system sites and $N_{\text{orb}}$ is the number of internal degrees of freedom on each site [which we take to be the same across every site within each model studied in this work].
To complete the tight-binding basis, we lastly define the orthogonality condition:
\begin{equation}
\langle \mathbf{r}_\alpha,l \rvert \mathbf{r}_\beta,l'\rangle = \int\ d^{3}x\  \psi_{\alpha,l}^{*}({\bf r})
\psi_{\beta,l'}({\bf r}) = \delta_{\alpha\beta}\delta_{ll'},
\label{eq:TBbasisStates}
\end{equation}
where $\psi_{\alpha,l}({\bf r})$ is the wavefunction of the Wannier tight-binding orbital of the site $\alpha$ with the internal degree-of-freedom index $l$ evaluated at the position ${\bf r}$.
The Kronecker delta functions in the second equality in Eq.~(\ref{eq:TBbasisStates}) indicate that we have restricted our tight-binding basis states to be orthogonal within each site and extremely compact [in practice $\delta$-function localized] on the scale of the shortest nearest-neighbor intersite distances. 
For each model in this work, we then use the tight-binding basis states to define a Hamiltonian matrix that consists of on-site potentials and hopping interactions between connected sites [as separately defined for each structural disorder implementation in Fig.~\ref{appfig:structuraldisorder} and the surrounding text].
In the basis of Eq.~(\ref{eq:TBbasisStates}), $\mathcal{H}$ for each disordered system is hence an $N_{\text{sites}}N_{\text{orb}}\times N_{\text{sites}}N_{\text{orb}}$ matrix.

\paragraph*{\bf Pseudo-Momentum-Space Green's Functions in Non-Crystalline Systems} -- $\ $ To characterize topological semimetal states, it is necessary to establish a momentum-space system description in which spectrally isolated nodal degeneracies can be assigned topological invariants~\cite{Armitage2018,Wieder22,BinghaiClaudiaWeylReview,ZahidNatRevMatWeyl}.
We will specifically accomplish this for the amorphous topological semimetals studied in this work through [pseudo-] momentum space Green's functions.
To begin, in a crystalline or non-crystalline system, we first define the real-space (retarded) Green's function at a given energy $E$ to be:
\begin{equation}
\mathcal{G}\left(E\right) = \left(\left(E + i \varepsilon\right)\mathbb{1}_{N_{\text{sites}}N_{\text{orb}}}-\mathcal{H}\right)^{-1},
\label{eq:RealSpaceGreen}
\end{equation}
where the matrix elements of $\mathcal{G}(E)$ in the tight-binding basis are denoted as:
\begin{equation}
\left[\mathcal{G}\left(E\right) \right]^{\alpha\beta,ll'} = \langle \mathbf{r}_\alpha,l \rvert \mathcal{G}\left(E\right) \lvert \mathbf{r}_\beta,l'\rangle.
\label{appeq:Greens}
\end{equation}
In Eq.~(\ref{eq:RealSpaceGreen}), $\varepsilon$ is a spectral broadening parameter that emerges as a consequence of our computational methods.
In our calculations, $\varepsilon$ can be approximated by dividing the system bandwidth [in energy] by the number of moments used in the kernel polynomial method [KPM] expansion of the Fourier-transformed Green's function, which we will shortly define below [see Refs.~\cite{Weisse06,varjas_topological_2019} for further details]. 
For each of the disordered model calculations in this work, we used 128 moments in the KPM expansion.
In combination with the tight-binding parameters used for each model, this approximately yields $\varepsilon\approx 0.04$ for the amorphous Kramers-Weyl model in Appendix~\ref{app:amorphousKramers}, $\varepsilon\approx 0.04$ for the amorphous double-Weyl model in Appendix~\ref{app:amorphousCharge2}, and $\varepsilon\approx 0.06$ for the amorphous chiral multifold fermion model in Appendix~\ref{app:amorphousMultifold}.

Though amorphous materials lack lattice translation symmetry, previous works have shown that when the position-space Green's function $\mathcal{G}(E)$ is Fourier-transformed using continuum plane-wave states parameterized by the pseudo-momentum ${\bf p}$, the Fourier-transformed Green's function can still exhibit well-resolved topological spectral features that are physically observable in angle-resolved photoemission experiments~\cite{marsal_topological_2020,corbae_evidence_2020,JustinHat,Ciocys2023}. 
As emphasized throughout this work and quantitatively explored in Appendix~\ref{app:EffectiveHamiltonian}, topological features in the Fourier-transformed Green's function are particularly well defined near ${\bf p}= {\bf 0}$, where the ${\bf p}$-resolved spectral function is sharpest.
Conversely, caution is warranted when extracting spectral and topological features from the Fourier-transformed Green's function far from ${\bf p}={\bf 0}$, because the use of hard-shell or $\delta$-function-localized Wannier orbitals, such as in Eq.~(\ref{eq:TBbasisStates}), can give rise to model-dependent features at $|{\bf p}|$ greater than the largest inverse bond length in a non-crystalline system [see the text surrounding Eq.~(\ref{eq:EmergentPinftyModulo})].

To construct the Fourier-transformed Green's function of a non-crystalline system, we begin by introducing a normalized set of plane-wave states~\cite{spillage_2022}:
\begin{equation} 
\lvert \mathbf{p},l\rangle = \frac{1}{\sqrt{N_{\text{sites}}}}\sum_{\alpha=1}^{N_{\text{sites}}} \exp(i \mathbf{p}\cdot \mathbf{r}_\alpha)\lvert \mathbf{r}_\alpha,l\rangle,
\label{appeq:planewaves}
\end{equation}
where $\lvert \mathbf{r}_\alpha,l\rangle$ is defined in the text preceding Eq.~(\ref{eq:TBbasisStates}), the plane-wave pseudo-momentum ${\bf p}$ takes unbounded values from ${\bf 0}$ to ${\bs \infty}$, and where the units and hence meaningful values of ${\bf p}$ respectively derive from the units and relative values of ${\bf r}_{\alpha}$. 
Along with the orthogonality of the position-space basis states $\lvert \mathbf{r}_\alpha,l\rangle$ in Eq.~(\ref{eq:TBbasisStates}), Eq.~(\ref{appeq:planewaves}) implies that the projection of each plane-wave state onto each position-space basis state is given by:
\begin{equation}
\langle \mathbf{r}_\alpha,l'\lvert \mathbf{p},l \rangle = \dfrac{1}{\sqrt{N_{\text{sites}}}} \exp(i\mathbf{p}\cdot \mathbf{r}_\alpha) \delta_{ll'}, 
\label{appeq:plane}
\end{equation} 
and that the inner products between plane-wave states are given by:
\begin{equation}
\langle \mathbf{p},l \rvert \mathbf{p}',l'\rangle = \frac{1}{N_{\text{sites}}}\sum_{\alpha,\beta=1}^{N_{\text{sites}}}\exp(-i\mathbf{p}\cdot\mathbf{r}_\alpha)\exp(i\mathbf{p}'\cdot\mathbf{r}_\beta)\langle \mathbf{r}_\alpha,l\lvert \mathbf{r}_\beta,l'\rangle =
\frac{1}{N_{\text{sites}}}\delta_{ll'}\sum^{N_{\text{sites}}}_{\alpha=1} \exp(-i\mathbf{r}_\alpha \cdot (\mathbf{p}-\mathbf{p}')).
\label{eq:OrthoPlaneWaves}
\end{equation}
We note that in the limit of a crystal in which lattice translations are system symmetries and ${\bf p}$ is replaced with the periodic crystal momentum ${\bf k}$, the summation and exponential in the right-most equality in Eq.~(\ref{eq:OrthoPlaneWaves}) together simplify to the delta function $\delta({\bf p}-{\bf p}')$, such that $\rvert \mathbf{p},l\rangle$ and $\rvert \mathbf{p}',l'\rangle$ are orthogonal~\cite{BigBook}.
However in a system with an irregular lattice, Eq.~(\ref{eq:OrthoPlaneWaves}) does not further simplify.
Throughout this work, we will use the symbol ${\bf k}$ to only index exact crystal momentum in translationally-invariant systems, and will otherwise use the symbol ${\bf p}$ to denote the more general case of plane-wave pseudo-momenta in non-crystalline [\emph{i.e.} amorphous] systems.

Having established a set of pseudo-momentum-space states $\lvert \mathbf{p},l\rangle$ [Eq.~(\ref{appeq:planewaves})] that remain applicable in a non-crystalline system, we next use $\lvert \mathbf{p},l\rangle$ to Fourier-transform the position-space system Green's function $\mathcal{G}(E)$ [Eq.~(\ref{eq:RealSpaceGreen})].
We specifically define the $N_{\text{orb}}\times N_{\text{orb}}$ \emph{[pseudo-] momentum-resolved matrix Green's function} $\mathcal{G}(E,{\bf p},{\bf p}')$ via its matrix elements using Eqs.~(\ref{appeq:Greens}) and~(\ref{appeq:plane}):
\begin{equation}
\begin{split}
    \left[\mathcal{G}(E,\mathbf{p},\mathbf{p}')\right]^{ll'} &= \left<\mathbf{p},l\right|\mathcal{G}(E)|\mathbf{p}',l'\rangle \\
    &= \sum_{\alpha,\beta=1}^{N_{\text{sites}}}\sum_{l'',l'''=1}^{N_{\text{orb}}}\left<\mathbf{p},l\lvert {\bf r}_{\alpha},l''\right>\left< {\bf r}_{\alpha},l''\right|   \mathcal{G}(E)\left|{\bf r}_{\beta}, l'''\right>\left<{\bf r}_{\beta}, l'''\lvert\mathbf{p}',l'\right> \\
    &= \frac{1}{N_{\text{sites}}}\sum_{\alpha,\beta=1}^{N_{\text{sites}}}\exp(-i\mathbf{p}\cdot\mathbf{r}_\alpha)\exp(i\mathbf{p}'\cdot\mathbf{r}_\beta)\langle \mathbf{r}_\alpha,l |\mathcal{G}(E)|\mathbf{r}_\beta,l' \rangle \\
    &= \frac{1}{N_{\mathrm{sites}}}\sum_{\alpha,\beta=1}^{N_{\text{sites}}}\exp(i(\mathbf{p}'\cdot\mathbf{r}_\beta  - \mathbf{p}\cdot\mathbf{r}_\alpha))\left[\mathcal{G}(E)\right]^{\alpha\beta,ll'}.
\end{split}
\label{eq:MomGreenFunc}
\end{equation}
As previously with Eq.~(\ref{eq:OrthoPlaneWaves}), the summation and exponential in the final equality of Eq.~(\ref{eq:MomGreenFunc}) do not further simplify in the absence of lattice regularity.
This implies that unlike in a crystal, the Fourier-transformed Green's function $\mathcal{G}(E,{\bf p},{\bf p}')$ generically carries both diagonal-in-momentum [${\bf p}={\bf p}'$] and off-diagonal-in-momentum [${\bf p}\neq{\bf p}'$] matrix elements in a structurally disordered system.

In practice, however, we find that for the non-crystalline models in this work, the off-diagonal-in-momentum elements of $\mathcal{G}(E,{\bf p},{\bf p}')$ become much smaller than the diagonal-in-momentum elements when each model is averaged over many [$\approx 20-50$] individually generated disorder realizations.
Specifically, our computational resources only permit simulating relatively small individual disorder realizations with $\sim 20^{3}$ [$\sim 8000$] atoms.
However, we seek to model solid-state systems that have thermodynamically large numbers of atoms [or short-range-interacting patches~\cite{ProdanKohnNearsighted}] and are self-averaging. 
Hence, to generate a momentum-space Green's function that is more representative of a real solid-state amorphous material, we employ a variant of the replica method~\cite{ParisiReplicaCourse,BerthierGlassAmorphousReview} to construct a two-momentum \emph{average Green's function} $\bar{\mathcal{G}}(E,\mathbf{p},\mathbf{p'})$.
We specifically implement an \emph{averaging procedure} in which we approximate the spectral and topological properties of large self-averaging systems by averaging the Fourier-transformed matrix Green's function $\mathcal{G}_{i}(E,{\bf p},{\bf p}')$ [Eq.~(\ref{eq:MomGreenFunc})] over an ensemble of $\approx 20-50$ \emph{replicas} that each contain independently generated structural and internal degree-of-freedom disorder and are indexed below by $i$:
\begin{equation}
\left[\bar{\mathcal{G}}(E,\mathbf{p},\mathbf{p'})\right]^{ll'} = \frac{1}{N_{\text{rep}}}\sum_{i=1}^{N_{\text{rep}}} \left[\mathcal{G}_{i}(E,\mathbf{p},\mathbf{p'})\right]^{ll'},
\label{eq:averageTWoMomentumGreen}
\end{equation}
where $N_{\text{rep}}$ is the number of replicas, and where the same structural disorder implementation scheme from Appendix~\ref{app:DiffTypesDisorder} is used for each replica in the ensemble
.
From Eq.~(\ref{eq:averageTWoMomentumGreen}), we then define the $N_{\text{orb}}\times N_{\text{orb}}$ one-momentum, Fourier-transformed average matrix Green's function $\bar{\mathcal{G}}(E,{\bf p})$ from the diagonal-in-momentum elements of $\bar{\mathcal{G}}(E,\mathbf{p},\mathbf{p'})$:
\begin{equation}
\left[\bar{\mathcal{G}}(E,\mathbf{p})\right]^{ll'} = \frac{1}{N_{\text{rep}}}\sum_{i=1}^{N_{\text{rep}}} \left[\mathcal{G}_{i}(E,\mathbf{p},\mathbf{p})\right]^{ll'}.
\label{eq:averageOneMomentumGreen}
\end{equation}

To demonstrate that the two-momentum average Green's function $\bar{\mathcal{G}}(E,\mathbf{p},\mathbf{p'})$ in practice reduces to the diagonal-in-momentum average Green's function $\bar{\mathcal{G}}(E,\mathbf{p})$ for the models and disorder ensembles employed in this work, we first introduce the symbol $\lVert M \rVert$ to indicate the magnitude of the largest element of the matrix $M$:
\begin{equation}
\lVert M \rVert = \max_{ij} \left\{\bigg|\left[M\right]^{ij}\bigg|\right\},
\label{eq:NormMatrix}
\end{equation}
where $\left[M\right]^{ij}$ are the matrix elements of $M$.
We then analyze the relative magnitudes of the matrix elements of $\bar{\mathcal{G}}(E,\mathbf{p},\mathbf{p'})$, $\bar{\mathcal{G}}(E,\mathbf{p})$, and $\bar{\mathcal{G}}(E,\mathbf{p'})$ for each non-crystalline model in Appendix~\ref{app:models} using the following procedure:

\begin{enumerate}
    \item For each pair of pseudo-momenta ${\bf p}$ and ${\bf p}'$, we use Eq.~(\ref{eq:NormMatrix}) to construct a dimensionless function $\Lambda(E,{\bf p},{\bf p}')$ that quantifies the magnitude of the off-diagonal-in-momentum matrix elements of $\bar{\mathcal{G}}(E,\mathbf{p},\mathbf{p'})$ relative to the smaller of the two quantities $\lVert \bar{\mathcal{G}}(E,\mathbf{p})\rVert$ and $\lVert \bar{\mathcal{G}}(E,\mathbf{p}')\rVert$:
    \begin{equation}
    \Lambda(E,{\bf p},{\bf p}') = \frac{\lVert \bar{\mathcal{G}}(E,\mathbf{p},\mathbf{p}') \rVert}{\min\left(\left\{ \lVert \bar{\mathcal{G}}(E,\mathbf{p})\rVert,\lVert \bar{\mathcal{G}}(E,\mathbf{p}')\rVert\right\}\right)}.
    \label{eq:EqGreen1D}
    \end{equation}

    \item We next group pseudo-momenta by their magnitudes to reduce the high-dimensional data from the previous computation of $\Lambda(E,{\bf p},{\bf p}')$ in Eq.~(\ref{eq:EqGreen1D}) into a form that can be succinctly plotted.     
    We begin by selecting a pair of pseudo-momenta magnitudes $p$ and $p'$.
    For each $p$ and $p'$, we then identify all values of ${\bf p}$ and ${\bf p}'$ for which $\lvert \mathbf{p}\rvert=p$ and $\lvert \mathbf{p}'\rvert=p'$.
    Lastly, we compare all of the values of $\Lambda(E,{\bf p},{\bf p}')$ with the same pseudo-momenta magnitudes $p$ and $p'$, and assign the maximum value of $\Lambda(E,{\bf p},{\bf p}')$ across these inputs to the dimensionless function $\Lambda(E,p,p')$. 
    In the notation of Eqs.~(\ref{eq:NormMatrix}) and~(\ref{eq:EqGreen1D}), this procedure can be summarized through the expression:
    \begin{equation}
    \Lambda(E,p,p') = \max_{{\substack{\mathbf{p}',\lvert \mathbf{p}'\rvert=p' \\ \mathbf{p},\lvert \mathbf{p}\rvert=p}}}\left[\Lambda(E,{\bf p},{\bf p}')\right] = \max_{{\substack{\mathbf{p}',\lvert \mathbf{p}'\rvert=p' \\ \mathbf{p},\lvert \mathbf{p}\rvert=p}}}\left[ \frac{\lVert \bar{\mathcal{G}}(E,\mathbf{p},\mathbf{p}') \rVert}{\min\left(\left\{ \lVert \bar{\mathcal{G}}(E,\mathbf{p})\rVert,\lVert \bar{\mathcal{G}}(E,\mathbf{p}')\rVert\right\}\right)}\right].
    \label{eq:GreenTwoMom3D}
    \end{equation}
\end{enumerate}

\begin{figure}[t]
\centering   
\includegraphics[width=\linewidth]{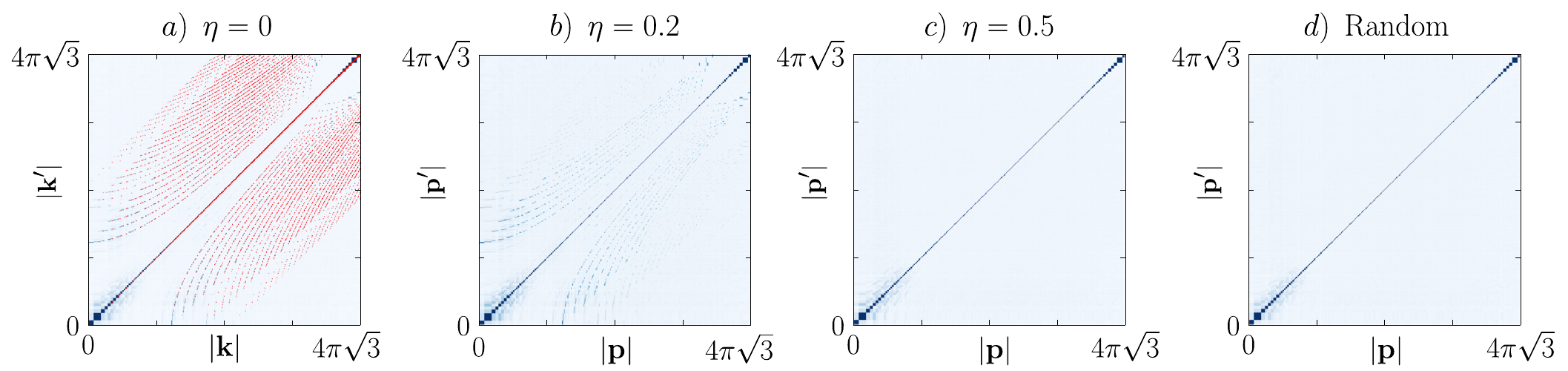}
\caption{Disappearance of the off-diagonal-in-momentum Green's function elements in the strongly disordered Kramers-Weyl model.
(a-d) Maximal matrix elements of the two-momentum average Green's function $\bar{\mathcal{G}}(E,\mathbf{p},\mathbf{p'})$ for the Kramers-Weyl model [Appendix~\ref{app:Kramers}], visualized through $\Lambda(E,p,p')$ in Eq.~(\ref{eq:GreenTwoMom3D}).
In (a-d), we specifically plot $\Lambda(E,p,p')$ with $p=|{\bf p}|$ and $p'=|{\bf p}'|$ and $E$ set to the energy of the Kramers-Weyl node at ${\bf k}={\bf 0}$ or ${\bf p}={\bf 0}$ using the procedure detailed in Appendix~\ref{app:EffectiveHamiltonian}.
(a) $\Lambda(E,p,p')$ for the Kramers-Weyl model in the crystalline limit [Eqs.~(\ref{eq:AmoKW}),~(\ref{eq:KWTmatrix}), and~(\ref{eq:appHKW})] with the approximate pseudo-momentum ${\bf p}$ replaced by the exact crystal momentum ${\bf k}$.
In (a), ${\bf k}$ is expressed in units of the exact nearest-neighbor lattice spacing $a=1$, whereas in (b-d), ${\bf p}$ is expressed in units of the average nearest-neighbor spacing $\bar{a}=1$.
The Fourier-transformed Green's function in (a) exhibits a series of parallel sharp lines [blue curves] at ${\bf k}={\bf k}' + {\bf K}_{\mu}$, where ${\bf K}_{\mu}$ is a reciprocal lattice vector [see Eq.~(\ref{eq:DiracDeltaTwoMom}) and the following text].
To better highlight the parallel curves in (a), we have superposed the data with exact red lines governed by Eq.~(\ref{eq:3DguideLines}).  
(b,c) The non-crystalline Kramers-Weyl model [Eq.~(\ref{eq:amorphousKWTmatrixFinalChirality})] with the tight-binding parameters in Eq.~(\ref{eq:disorderedKWparams}), and with nematic Gaussian structural and frame disorder parameterized by the standard deviations $\eta=0.2$ in (b) and $\eta=0.5$ in (c) [see Appendix~\ref{app:lattices}].
In (b,c) the data were additionally obtained by averaging over 20 disorder realizations [\emph{i.e.} replicas] with $20^{3}=8000$ sites each and contiguous domains of right- and left-handed sites within each replica with the respective concentrations $n_R=0.7$ and $n_L=1-N_R/N_{\mathrm{sites}}=0.3$ [see Fig.~\ref{appfig:disordertypes}(b)].
(b) As $\eta$ is increased, the parallel blue lines vanish away from ${\bf p}={\bf p}'$ in $\Lambda(E,p,p')$, beginning at larger $|{\bf p}|$ and $|{\bf p}'|$ and working inwards towards smaller pseudo-momenta until (c) in the vicinity of $\eta\approx 0.5$, only the ${\bf p}={\bf p}'$ line remains visible.
We designate the disorder regime in (c), in which the off-diagonal-in-momentum matrix elements of $\bar{\mathcal{G}}(E,\mathbf{p},\mathbf{p'})$ vanish, as the \emph{strong structural disorder} limit of the non-crystalline Kramers-Weyl model.  
(d) $\Lambda(E,p,p')$ computed over 20 $d_{A}=3$ [nematic disordered] random lattices [Appendix~\ref{app:DiffTypesDisorder}] with $20^{3}$ sites each, random frame disorder parameterized by $\eta=0.5$, and contiguous chirality domains with $n_R=0.7$ and $n_L=0.3$.
Like in the large-$\eta$ Gaussian disorder calculation in (c), the off-diagonal-in-momentum elements of $\bar{\mathcal{G}}(E,\mathbf{p},\mathbf{p'})$ vanish in the random-lattice KW model in (d), consistent with the absence of ${\bf p}\neq{\bf p}'$ lines in $\Lambda(E,p,p')$.
This indicates that the non-crystalline models in (c) and (d) lie in the same regime of strong structural disorder from the perspective of $\bar{\mathcal{G}}(E,\mathbf{p},\mathbf{p'})$.}
\label{fig:3DGreenKW}
\end{figure}

\begin{figure}[t]
\centering   
\includegraphics[width=\linewidth]{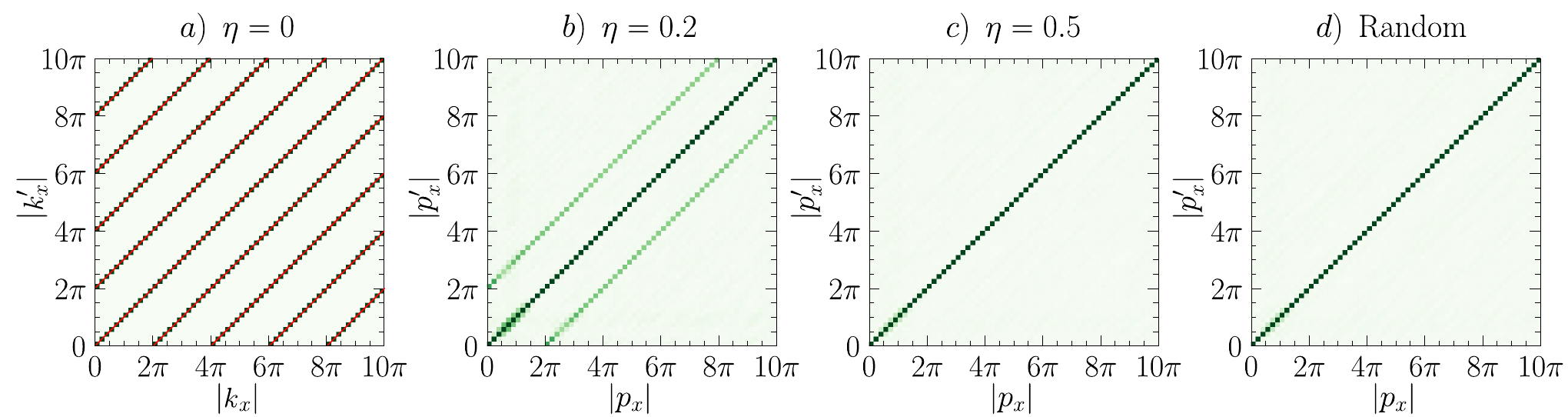}
\caption{Disappearance of the off-diagonal-in-momentum Green's function elements in the strongly disordered double-Weyl model.
(a-d) Maximal matrix elements of the two-momentum average Green's function 
$\bar{\mathcal{G}}(E,\mathbf{p},\mathbf{p'})$ for the quadratic double-Weyl model [Appendix~\ref{sec:Charge2}], visualized through
$\Lambda(E,{\bf p},{\bf p}')$ 
[Eq.~(\ref{eq:EqGreen1D})].
In (a-d), we specifically plot 
$\Lambda(E,{\bf p},{\bf p}')$ 
for ${\bf p}=(p_{x},0,0)$ and ${\bf p}'=(p'_{x},0,0)$ with $E$ set to the energy of the double-Weyl node at ${\bf k}={\bf 0}$ or ${\bf p}={\bf 0}$ using the procedure detailed in Appendix~\ref{app:EffectiveHamiltonian}.
(a) 
$\Lambda(E,{\bf p},{\bf p}')$ 
for the double-Weyl model in the crystalline limit [Eqs.~(\ref{eq:AmoC2}),~(\ref{eq:TmatrixQuadraticDef}), and~(\ref{eq:pristineDoubleWeyl})] with the approximate pseudo-momentum $p_{x}$ replaced by the exact crystal momentum $k_{x}$.
In (a), $k_{x}$ is expressed in units of the exact in-plane nearest-neighbor lattice spacing $a_{\parallel}$ [Eq.~(\ref{eq:quadraticWeylFixedLatticeParams})], whereas in (b-d), $p_{x}$ is expressed in units of the average in-plane nearest-neighbor spacing $\bar{a}_{\parallel}=1$.
The Fourier-transformed Green's function in (a) exhibits a series of parallel sharp features [green lines] at $k_{x} = k_{x}' + 2\pi n_{1}$, where $n_{1}\in\mathbb{Z}$ and ${\bf K}_{1} = 2\pi\hat{x}$ is an in-plane reciprocal lattice vector [see Eq.~(\ref{eq:KEqSumReciprocal}) and the following text].
To better highlight the parallel sharp features in (a), we have superposed the data with exact red lines governed by Eq.~(\ref{eq:2DguideLines}).  
(b,c) The non-crystalline double-Weyl model [Eq.~(\ref{eq:HopChiralC2})] with the tight-binding parameters in Eq.~(\ref{eq:paramsC2amo}), and with smectic Gaussian structural and frame disorder parameterized by the standard deviations $\eta=0.2$ in (b) and $\eta=0.5$ in (c) [see Appendix~\ref{app:lattices}].
In (b,c) the data were additionally obtained by averaging over 20 disorder realizations [replicas] with $20^{3}=8000$ sites each [\emph{i.e.} 20 layers with 400 sites each] and contiguous domains of right- and left-handed sites within each replica with the respective concentrations $n_R=0.7$ and $n_L=1-N_R/N_{\mathrm{sites}}=0.3$ [see Fig.~\ref{appfig:disordertypes}(b)].
(b) As $\eta$ is increased, the parallel green lines vanish away from $p_{x,y}=p'_{x,y}$ in the $d_{A}=2$ $(p_{x},p_{y})$ subspace of 
$\Lambda(E,{\bf p},{\bf p}')$ 
[see the text following Eq.~(\ref{eq:GaussianDisorder})], beginning at larger $p_{x,y}$ and $p_{x,y}'$ and working inwards towards smaller pseudo-momenta until (c) in the vicinity of $\eta\approx 0.5$, only the $p_{x,y}=p'_{x,y}$ line remains visible.
We designate the disorder regime in (c), in which the off-diagonal-in-momentum matrix elements of $\bar{\mathcal{G}}(E,\mathbf{p},\mathbf{p'})$ vanish in the $d_{A}=2$ disordered system subspace, as the \emph{strong structural disorder} limit of the non-crystalline double-Weyl model.  
(d) 
$\Lambda(E,{\bf p},{\bf p}')$ 
computed over 20 $d_{A}=2$ [smectic disordered] random lattices [Appendix~\ref{app:DiffTypesDisorder}] with $20^{3}$ sites each, random SO(2) smectic frame disorder [Eq.~(\ref{eq:2dSO2Rmat})] parameterized by $\eta=0.5$, and contiguous chirality domains with $n_R=0.7$ and $n_L=0.3$.
Like in the large-$\eta$ Gaussian disorder calculation in (c), the off-diagonal-in-momentum elements of $\bar{\mathcal{G}}(E,\mathbf{p},\mathbf{p'})$ vanish in the $d_{A}=2$ random-lattice double-Weyl model in (d), consistent with the absence of $p_{x,y}\neq p_{x,y}'$ lines in 
$\Lambda(E,{\bf p},{\bf p}')$.
This indicates that the non-crystalline models in (c) and (d) lie in the same regime of strong structural disorder from the perspective of $\bar{\mathcal{G}}(E,\mathbf{p},\mathbf{p'})$.}
\label{fig:3DGreenC2}
\end{figure}

\begin{figure}[t]
\centering   
\includegraphics[width=\linewidth]{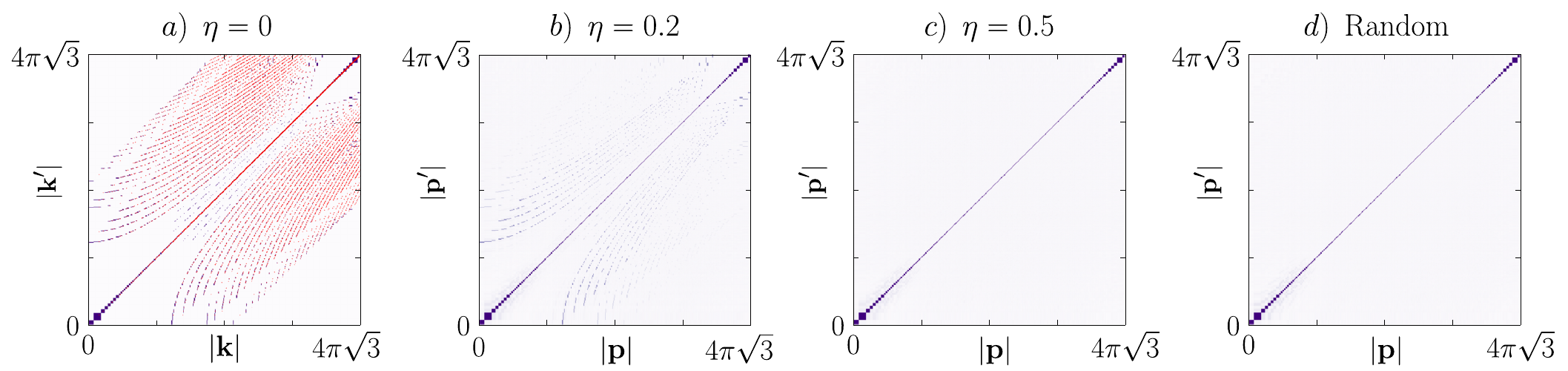}
\caption{Disappearance of the off-diagonal-in-momentum Green's function elements in the strongly disordered chiral multifold fermion model.
(a-d) Maximal matrix elements of the two-momentum average Green's function $\bar{\mathcal{G}}(E,\mathbf{p},\mathbf{p'})$ for the spin-1 chiral multifold fermion model [Appendix~\ref{sec:Multifold}], visualized through $\Lambda(E,p,p')$ in Eq.~(\ref{eq:GreenTwoMom3D}).
In (a-d), we specifically plot $\Lambda(E,p,p')$ with $p=|{\bf p}|$ and $p'=|{\bf p}'|$ and $E$ set to the energy of the threefold nodal degeneracy at ${\bf k}={\bf 0}$ or ${\bf p}={\bf 0}$ using the procedure detailed in Appendix~\ref{app:EffectiveHamiltonian}.
(a) $\Lambda(E,p,p')$ for the multifold fermion model in the crystalline limit [Eqs.~(\ref{eq:3FTmatrix}),~(\ref{eq:3FMmatrix}), and~(\ref{eq:multifoldBandV})] with the approximate pseudo-momentum ${\bf p}$ replaced by the exact crystal momentum ${\bf k}$.
In (a), ${\bf k}$ is expressed in units of the exact nearest-neighbor lattice spacing $a=1$, whereas in (b-d), ${\bf p}$ is expressed in units of the average nearest-neighbor spacing $\bar{a}=1$.
The Fourier-transformed Green's function in (a) exhibits a series of parallel sharp lines [purple curves] at ${\bf k}={\bf k}' + {\bf K}_{\mu}$, where ${\bf K}_{\mu}$ is a reciprocal lattice vector [see Eq.~(\ref{eq:DiracDeltaTwoMom}) and the following text].
To better highlight the parallel curves in (a), we have superposed the data with exact red lines governed by Eq.~(\ref{eq:3DguideLines}).  
(b,c) The non-crystalline multifold model [Eq.~(\ref{eq:amorphous3FTmatrixFinalChirality})] with the tight-binding parameters in Eq.~(\ref{eq:disordered3Fparams}), and with nematic Gaussian structural and frame disorder parameterized by the standard deviations $\eta=0.2$ in (b) and $\eta=0.5$ in (c) [see Appendix~\ref{app:lattices}].
In (b,c) the data were additionally obtained by averaging over 20 disorder realizations [\emph{i.e.} replicas] with $20^{3}=8000$ sites each and contiguous domains of right- and left-handed sites within each replica with the respective concentrations $n_R=0.7$ and $n_L=1-N_R/N_{\mathrm{sites}}=0.3$ [see Fig.~\ref{appfig:disordertypes}(b)].
(b) As $\eta$ is increased, the parallel purple lines vanish away from ${\bf p}={\bf p}'$ in $\Lambda(E,p,p')$, beginning at larger $|{\bf p}|$ and $|{\bf p}'|$ and working inwards towards smaller pseudo-momenta until (c) in the vicinity of $\eta\approx 0.5$, only the ${\bf p}={\bf p}'$ line remains visible.
We designate the disorder regime in (c), in which the off-diagonal-in-momentum matrix elements of $\bar{\mathcal{G}}(E,\mathbf{p},\mathbf{p'})$ vanish, as the \emph{strong structural disorder} limit of the non-crystalline multifold fermion model.  
(d) $\Lambda(E,p,p')$ computed over 20 $d_{A}=3$ [nematic disordered] random lattices [Appendix~\ref{app:DiffTypesDisorder}] with $20^{3}$ sites each, random frame disorder parameterized by $\eta=0.5$, and contiguous chirality domains with $n_R=0.7$ and $n_L=0.3$.
Like in the large-$\eta$ Gaussian disorder calculation in (c), the off-diagonal-in-momentum elements of $\bar{\mathcal{G}}(E,\mathbf{p},\mathbf{p'})$ vanish in the random-lattice multifold model in (d), consistent with the absence of ${\bf p}\neq{\bf p}'$ lines in $\Lambda(E,p,p')$.
This indicates that the non-crystalline models in (c) and (d) lie in the same regime of strong structural disorder from the perspective of $\bar{\mathcal{G}}(E,\mathbf{p},\mathbf{p'})$.}
\label{fig:3DGreen3F}
\end{figure}

Using Eqs.~(\ref{eq:EqGreen1D}) and~(\ref{eq:GreenTwoMom3D}), we have computed $\Lambda(E,{\bf p},{\bf p}')$ and $\Lambda(E,p,p')$ for each of the non-crystalline models in Appendix~\ref{app:models} and structural disorder implementation schemes in Appendix~\ref{app:DiffTypesDisorder}.
For all of the non-crystalline models in this work subject to either large-$\eta$ Gaussian structural disorder [$\eta\approx 0.5$] or placed on random or Mikado lattices, we observe that $\Lambda(E,p,p')$ rapidly decays as a function of $\left|\mathbf{p}-\mathbf{p}'\right|$ in the $d_{A}$-dimensional system [sub]space of non-crystalline directions [see Eq.~(\ref{eq:finiteDimAmorph}) and the surrounding text], with representative examples shown in Figs.~\ref{fig:3DGreenKW},~\ref{fig:3DGreenC2}, and~\ref{fig:3DGreen3F}.
More generally, for all of the models studied in this work subject to large-$\eta$ Gaussian structural disorder or placed on random or Mikado lattices, we observe that the off-diagonal-in-momentum elements of $\bar{\mathcal{G}}(E,\mathbf{p},\mathbf{p}')$ are nearly vanishing on the scale of the initial tight-binding [non-crystalline] lattice parameters.

First, to generate the data in Figs.~\ref{fig:3DGreenKW}(a-c),~\ref{fig:3DGreenC2}(a-c), and~\ref{fig:3DGreen3F}(a-c), we place each of the three non-crystalline topological semimetal models in this work on a lattice with $N_{\mathrm{sites}}=20^{3}=8000$ sites, increasing Gaussian structural and local frame orientation disorder parameterized by the same standard deviation $\eta$, and chirality domains of unequal volume [70\% right-handed, 30\% left-handed, see Fig.~\ref{appfig:disordertypes}(b)].
We then compute either $\Lambda(E,{\bf p},{\bf p}')$ [Eq.~(\ref{eq:EqGreen1D})] or $\Lambda(E,p,p')$ [Eq.~(\ref{eq:GreenTwoMom3D})] by averaging the system over 20 disorder realizations [replicas].
The results of this calculation are plotted in Fig.~\ref{fig:3DGreenKW}(a-c) for the non-crystalline Kramers-Weyl model in Appendix~\ref{app:amorphousKramers} with nematic disorder [$d_{A}=3$], Fig.~\ref{fig:3DGreenC2}(a-c) for the non-crystalline double-Weyl model in Appendix~\ref{app:amorphousCharge2} with smectic disorder [$d_{A}=2$], and Fig.~\ref{fig:3DGreen3F}(a-c) for the chiral multifold fermion model in Appendix~\ref{app:amorphousMultifold} with nematic disorder [$d_{A}=3$, see Appendix~\ref{app:SmecticNematicDisorder} for the definitions of nematic and smectic model disorder employed in this work].

For each of the models in this work, $\Lambda(E,p,p')$ in the crystalline limit [$\eta=0$] exhibits a series of high-intensity lines along:
\begin{equation}
{\bf k}' = {\bf k} + n_{1}{\bf K}_{1} + n_{2}{\bf K}_{2} + n_{3}{\bf K}_{3},\quad n_{1,2,3}\in\mathbb{Z},
\label{eq:KEqSumReciprocal}
\end{equation}
where ${\bf K}_{\mu}$ is the $\mu$-th reciprocal lattice vector:
\begin{equation}
{\bf K}_{1} = 2\pi\hat{x},\ {\bf K}_{2} = 2\pi\hat{y},\ {\bf K}_{3} = 2\pi\hat{z}.
\label{eq:reciprocalLatVecs}
\end{equation}
This occurs because in the crystalline limit in which the approximate pseudo-momentum ${\bf p}$ reduces to the exact crystal momentum ${\bf k}$, $\mathcal{G}(E,\mathbf{p},\mathbf{p}')$ in Eq.~(\ref{eq:MomGreenFunc}) [and hence $\bar{\mathcal{G}}(E,\mathbf{p},\mathbf{p'})$ in Eq.~(\ref{eq:averageTWoMomentumGreen})] simplifies to a function of a single crystal momentum ${\bf k}$:
\begin{eqnarray}
\left[\mathcal{G}(E,\mathbf{k},\mathbf{k}')\right]^{ll'} &=& \frac{1}{N_{\mathrm{sites}}}\sum_{\alpha,\beta=1}^{N_{\text{sites}}}\exp(i(\mathbf{k}'\cdot\mathbf{r}_\beta  - \mathbf{k}\cdot\mathbf{r}_\alpha))\left[\mathcal{G}(E)\right]^{\alpha\beta,ll'} \nonumber \\
&=& \frac{1}{N_{\mathrm{sites}}}\sum_{\alpha,\beta=1}^{N_{\text{sites}}}\exp(i(\mathbf{k}'\cdot\mathbf{r}_\beta  - \mathbf{k}\cdot\mathbf{r}_\alpha))\left[\left(\left(E + i \varepsilon\right)\mathbb{1}_{N_{\text{sites}}N_{\text{orb}}}-\mathcal{H}\right)^{-1}\right]^{\alpha\beta,ll'} \nonumber \\
&=& \frac{1}{(N_{\text{Sites}})^2}\sum_{\alpha,\beta=1}^{N_{\text{sites}}}\exp(i(\mathbf{k}'\cdot\mathbf{r}_\beta  - \mathbf{k}\cdot\mathbf{r}_\alpha))\sum_{{\bf k}''\in\text{BZ}}\exp(i{\bf k}''({\bf r_{\alpha}-{\bf r}_{\beta}}))\left[\left(\left(E + i \varepsilon\right)\mathbb{1}_{N_{\text{orb}}}-\mathcal{H}({\bf k}'')\right)^{-1}\right]^{ll'} \nonumber \\
&=& \frac{1}{(N_{\text{sites}})^2}\sum_{\alpha,\beta=1}^{N_{\text{sites}}}\sum_{{\bf k}''\in\text{BZ}}\exp(i({\bf k}'' - {\bf k})\cdot{\bf r}_{\alpha})\exp(i({\bf k}' - {\bf k}'')\cdot{\bf r}_{\beta})\left[\left(\left(E + i \varepsilon\right)\mathbb{1}_{N_{\text{orb}}}-\mathcal{H}({\bf k}'')\right)^{-1}\right]^{ll'} \nonumber \\
&=& \sum_{{\bf k}''\in\text{BZ}}\delta({\bf k}'' - {\bf k})\delta({\bf k}' - {\bf k}'')\left[\left(\left(E + i \varepsilon\right)\mathbb{1}_{N_{\text{orb}}}-\mathcal{H}({\bf k}'')\right)^{-1}\right]^{ll'} \nonumber \\
&=& \delta({\bf k}'-{\bf k})\left[\left(\left(E + i \varepsilon\right)\mathbb{1}_{N_{\text{orb}}}-\mathcal{H}({\bf k})\right)^{-1}\right]^{ll'} \nonumber \\
&=& \delta({\bf k}'-{\bf k})\left[\mathcal{G}(E, {\bf k})\right]^{ll'},
\label{eq:DiracDeltaTwoMom}
\end{eqnarray}
where $\mathcal{H}({\bf k})$ is the Bloch Hamiltonian, $\mathcal{G}(E,{\bf k})$ is the crystal-momentum-resolved single-particle Green's function~\cite{ChenBernevigQPI,Saito_TB_2016,Kohsaka_2017}, and where for simplicity we have specialized to the case of a crystalline system without sublattice degrees of freedom.
The strong [$\delta$-function] localization to the ${\bf k}={\bf k}'$ line of $\Lambda(E,{\bf p},{\bf p}')$ in Fig.~\ref{fig:3DGreenC2}(a) and $\Lambda(E,p,p')$ in Figs.~\ref{fig:3DGreenKW}(a) and~\ref{fig:3DGreen3F}(a) can then be attributed to the simplification of $\mathcal{G}(E,\mathbf{k},\mathbf{k}')$ in Eq.~(\ref{eq:DiracDeltaTwoMom}).

We next recognize that in the crystalline limit, ${\bf k}$ and ${\bf k}'$ are only well-defined modulo the reciprocal lattice vectors ${\bf K}_{1,2,3}$ in Eqs.~(\ref{eq:KEqSumReciprocal}) and~(\ref{eq:reciprocalLatVecs}).
This implies that exact copies of the ${\bf k}={\bf k}'$ line of high-intensity $\Lambda(E,{\bf p},{\bf p}')$ values should also appear more generally in $\Lambda(E,{\bf p},{\bf p}')$ for ${\bf k}={\bf k}'\text { mod }{\bf K}_{1,2,3}$, as seen in Figs.~\ref{fig:3DGreenKW}(a),~\ref{fig:3DGreenC2}(a), and~\ref{fig:3DGreen3F}(a).
To highlight the reciprocal lattice periodicity of $\Lambda(E,{\bf p},{\bf p}')$ and $\Lambda(E,p,p')$, we have drawn red guide lines over the data in Figs.~\ref{fig:3DGreenKW}(a),~\ref{fig:3DGreenC2}(a), and~\ref{fig:3DGreen3F}(a).  
For $\Lambda(E,{\bf p},{\bf p}')$ plotted with ${\bf k}=(k_{x},0,0)$ and ${\bf k}'=(k'_{x},0,0)$ computed for the smectic-disordered double-Weyl model in Fig.~\ref{fig:3DGreenC2}(a), the red guide lines are given by:
\begin{equation}
k_{x}' = k_{x}+2\pi n_{1},\quad n_{1}\in\mathbb{Z},
\label{eq:2DguideLines}
\end{equation}
where the $2\pi n_{1}$ term originates from the definition of the  reciprocal lattice vector ${\bf K}_{1}$ in Eqs.~(\ref{eq:KEqSumReciprocal}) and~(\ref{eq:reciprocalLatVecs}).
For $\Lambda(E,p,p')$ in the nematic-disordered Kramers-Weyl model in Fig.~\ref{fig:3DGreenKW}(a) and the spin-1 chiral multifold model in Fig.~\ref{fig:3DGreen3F}(a), the guide lines and data instead satisfy:
\begin{equation}
k'_{x} = k_{x} + 2\pi n_{1},\ k'_{y} = k_{y} + 2\pi n_{2},\ k'_{z} = k_{z} + 2\pi n_{3},\quad (n_1,n_2,n_3)\in\mathbb{Z}^3, 
\label{eq:3DguideLines}
\end{equation}
where the $2\pi n_{1,2,3}$ terms originate from the definitions of the reciprocal lattice vectors ${\bf K}_{1,2,3}$ in Eqs.~(\ref{eq:KEqSumReciprocal}) and~(\ref{eq:reciprocalLatVecs}).  
For $k=|{\bf k}|$ and $k'=|{\bf k}'|$, Eq.~(\ref{eq:3DguideLines}) then indicates the presence of guide lines and data along paths that satisfy:
\begin{equation}
(k')^{2} = (k_{x}+2\pi n_{1})^{2} + (k_{y}+2\pi n_{2})^{2} + (k_{z}+2\pi n_{3})^{2},\quad (n_1,n_2,n_3)\in\mathbb{Z}^3.
\label{eq:guideLineConsitency3D}
\end{equation}

Unlike for the double-Weyl model in Fig.~\ref{fig:3DGreenC2}(a), the guide lines and data for the Kramers-Weyl model in Fig.~\ref{fig:3DGreenKW}(a) and the multifold fermion model in Fig.~\ref{fig:3DGreen3F}(a) appear curved at small $k$ and $k'$ for $k\neq k'$, and only appear straight for $k\neq k'$ at larger momenta.
To understand this, we recognize that the trend lines in Figs.~\ref{fig:3DGreenKW}(a) and~\ref{fig:3DGreen3F}(a) fall into two cases depending on the integers $n_{1,2,3}$.
First, if $n_{1}=n_{2}=n_{3}=0$ in Eq.~(\ref{eq:guideLineConsitency3D}), then there exists a straight trend line for all values of $k$ and $k'$ that satisfies:
\begin{equation}
k'=k,\quad\text{ for }(n_1,n_2,n_3)=(0,0,0).
\end{equation}
However if any of the integers $n_{1,2,3}$ are nonzero [denoted below as $(n_1,n_2,n_3)\in\mathbb{Z}^3 \setminus \{(0,0,0)\}$], Eq.~(\ref{eq:guideLineConsitency3D}) instead characterizes curved trend lines at small $k$ and $k'$.
For small $k\propto k_{x}$ [which produces the small-$k$ data points in $\Lambda(E,{\bf p},{\bf p}')$ through Eq.~(\ref{eq:GreenTwoMom3D})], and recalling that $k$ and $k'$ are positive semidefinite, expanding Eq.~(\ref{eq:guideLineConsitency3D}) produces an upper branch of trend lines governed by:
\begin{equation}
k' \approx 2\pi\sqrt{n_{1}^{2}+n_{2}^{2}+n_{3}^2} + \frac{k^{2}}{4\pi\sqrt{n_{1}^{2}+n_{2}^{2}+n_{3}^2}},\quad (n_1,n_2,n_3)\in\mathbb{Z}^3 \setminus \{(0,0,0)\}.
\label{eq:quadratic3DGuides}
\end{equation}
Next, because Eq.~(\ref{eq:guideLineConsitency3D}) is invariant under the exchange of $k_{x,y,z}$ and $k_{x,y,z}'$ [up to the signs of the integers $n_{1,2,3}$], then there also exists a second, lower branch of trend lines in Figs.~\ref{fig:3DGreenKW}(a) and~\ref{fig:3DGreen3F}(a) that are governed at small $k$ by the mirror image of Eq.~(\ref{eq:quadratic3DGuides}) about the $k=k'$ line:
\begin{equation}
k' \approx \sqrt{4\pi\sqrt{n_{1}^{2}+n_{2}^{2}+n_{3}^2}\left[k - 2\pi\sqrt{n_{1}^{2}+n_{2}^{2}+n_{3}^2}\right]},\quad (n_1,n_2,n_3)\in\mathbb{Z}^3 \setminus \{(0,0,0)\}.
\end{equation}
Finally for large $k_{x,y,z}$, and hence large $k$, Eq.~(\ref{eq:guideLineConsitency3D}) instead saturates in a linear relationship between $k$ and $k'$:
\begin{equation}
k' \approx k \pm 2\pi\sqrt{n_{1}^{2}+n_{2}^{2}+n_{3}^2},\quad (n_1,n_2,n_3)\in\mathbb{Z}^3, 
\end{equation}
consistent with the linear trend lines at large momenta in Figs.~\ref{fig:3DGreenKW}(a) and~\ref{fig:3DGreen3F}(a).

Having analyzed $\Lambda(E,{\bf p},{\bf p}')$ in the $\eta=0$ [crystalline] limit of the topological semimetal models studied in this work [Appendix~\ref{app:models}], we next track the evolution of $\Lambda(E,{\bf p},{\bf p}')$ for Gaussian structural disorder with increasing $\eta>0$.
As $\eta$ is tuned away from the crystalline limit, we observe that the parallel lines in $\Lambda(E,{\bf p},{\bf p}')$ and $\Lambda(E,p,p')$ in Figs.~\ref{fig:3DGreenKW}(b,c),~\ref{fig:3DGreenC2}(b,c), and~\ref{fig:3DGreen3F}(b,c) begin to vanish.
Specifically, the parallel lines in $\Lambda(E,{\bf p},{\bf p}')$ in Eq.~(\ref{eq:EqGreen1D}) [and hence in $\Lambda(E,p,p')$ in Eq.~(\ref{eq:GreenTwoMom3D})] first vanish for small values of $\eta$ at the largest values of $p=|{\bf p}|$ and $p'=|{\bf p}'|$ satisfying Eq.~(\ref{eq:3DguideLines}) [Figs.~\ref{fig:3DGreenKW}(b,c) and~\ref{fig:3DGreen3F}(b,c)] or satisfying Eq.~(\ref{eq:2DguideLines}) [Fig.~\ref{fig:3DGreenC2}(b,c)], and then successively vanish at smaller $p$ and $p'$ as $\eta$ is increased.
This can be understood by recognizing that each parallel line in $\Lambda(E,{\bf p},{\bf p}')$ lies at a value:
\begin{equation}
{\bf p}' = {\bf p} + \tilde{\bf P}_{\mu},
\end{equation} 
for which
\begin{equation}
\tilde{\bf P}_{\mu}\cdot\tilde{\bf a}_{\mu} = 2\pi,
\label{eq:approximateLatticeTranslationGpp}
\end{equation}
where $|\tilde{\bf a}_{\mu}|$ represents a length scale at which real-space lattice translation symmetry remains approximately preserved, because $\tilde{\bf P}_{\mu}$ remains an approximate reciprocal lattice vector of $\bar{\mathcal{G}}(E,\mathbf{p},\mathbf{p'})$.
For weak structural disorder [$\eta\sim 0$], the effects of lattice translation symmetry breaking are in this picture only felt at the shortest [lattice-scale] wavelengths, whereas for larger $\eta$, lattice translation symmetry is effectively broken in the two-momentum average Green's function $\bar{\mathcal{G}}(E,\mathbf{p},\mathbf{p'})$ at larger $|\tilde{\bf a}_{\mu}|$, and hence smaller $|\tilde{\bf P}_{\mu}|$ via Eq.~(\ref{eq:approximateLatticeTranslationGpp}).

We further observe that at a critical value of $\eta$, which occurs in the vicinity of $\eta\approx 0.5$ for all of the models in this work, the high-intensity lines away from $p=p'$ in $\Lambda(E,p,p')$ all vanish [see Figs.~\ref{fig:3DGreenKW}(c),~\ref{fig:3DGreenC2}(c), and~\ref{fig:3DGreen3F}(c)].
We use this observation to in this work quantitatively define \emph{strong structural disorder} as the model regime in which the off-diagonal-in-momentum matrix elements of $\bar{\mathcal{G}}(E,\mathbf{p},\mathbf{p'})$ vanish, as indicated by the disappearance of the ${\bf p}\neq{\bf p}'$ lines in $\Lambda(E,{\bf p},{\bf p}')$ in Eq.~(\ref{eq:EqGreen1D}) [and hence in $\Lambda(E,p,p')$ in Eq.~(\ref{eq:GreenTwoMom3D})].  
The limit of strong structural disorder can therefore heuristically be understood  as the regime in which discrete [crystalline] lattice translation symmetries are effectively broken on all but the longest system length scales, leaving only continuous translations as approximate [average] long-wavelength system symmetries [in addition to average point group symmetries, see the text surrounding Eq.~(\ref{eq:ASG}) for a more formal group-theoretic discussion of approximate translation symmetry in strongly disordered systems].
Throughout this work, we will use the regime of strong structural disorder to approximate the spectral and topological properties of real solid-state amorphous materials.

Lastly, to show that the disappearance of the off-diagonal-in-momentum elements of $\bar{\mathcal{G}}(E,\mathbf{p},\mathbf{p'})$ is a more generic feature of strongly disordered lattices, and not limited to large-$\eta$ Gaussian structural disorder, we have also computed $\Lambda(E,{\bf p},{\bf p}')$ for all of the models in this work on inherently non-crystalline random and Mikado lattices [see Appendix~\ref{app:DiffTypesDisorder}].
In Figs.~\ref{fig:3DGreenKW}(d),~\ref{fig:3DGreenC2}(d), and~\ref{fig:3DGreen3F}(d), we show the result of this calculation for each model in this work placed on a lattice with randomly generated positions.
To generate the data in Figs.~\ref{fig:3DGreenKW}(d),~\ref{fig:3DGreenC2}(d), and~\ref{fig:3DGreen3F}(d), we specifically place each of the three non-crystalline topological semimetal models in this work on a random lattice with $N_{\mathrm{sites}}=20^{3}=8000$ sites, local frame orientation disorder parameterized by $\eta=0.5$, and contiguous chirality domains with $n_R=0.7$ and $n_L=1-N_R/N_{\mathrm{sites}}=0.3$ [see Fig.~\ref{appfig:disordertypes}(b)].
We then again compute either $\Lambda(E,{\bf p},{\bf p}')$ [Eq.~(\ref{eq:EqGreen1D})] or $\Lambda(E,p,p')$ [Eq.~(\ref{eq:GreenTwoMom3D})] by averaging the system over 20 disorder replicas.
For the non-crystalline Kramers-Weyl model in Appendix~\ref{app:amorphousKramers} [Fig.~\ref{fig:3DGreenKW}(d)] and the chiral multifold fermion model in Appendix~\ref{app:amorphousMultifold} [Fig.~\ref{fig:3DGreen3F}(d)], our calculations were performed using $d_{A}=3$ [nematic disordered] random lattices and SO(3) Gaussian internal frame disorder [Eq.~(\ref{eqn:RotationMatrix})], and for the non-crystalline double-Weyl model in Appendix~\ref{app:amorphousCharge2} [Fig.~\ref{fig:3DGreenC2}(d)], our calculations were performed with $d_{A}=2$ [smectic disordered] random lattices with preferred-$z$-axis SO(2) Gaussian internal frame disorder [Eq.~(\ref{eq:2dSO2Rmat})].
Like in the large-$\eta$ Gaussian disorder calculations shown in Figs.~\ref{fig:3DGreenKW}(c),~\ref{fig:3DGreenC2}(c), and~\ref{fig:3DGreen3F}(c), the random lattice calculations in Figs.~\ref{fig:3DGreenKW}(d),~\ref{fig:3DGreenC2}(d), and~\ref{fig:3DGreen3F}(d) only exhibit sharp lines in $\Lambda(E,{\bf p},{\bf p}')$ or $\Lambda(E,p,p')$ along ${\bf p}={\bf p}'$, indicating that both sets of calculations lie in the same regime of strong structural disorder from the perspective of $\bar{\mathcal{G}}(E,\mathbf{p},\mathbf{p'})$.
However we note that in general, calculations for the same model with large-$\eta$ Gaussian structural disorder or on random lattices may still exhibit more subtle differences, as discussed in further detail in the text following Eq.~(\ref{eq:GaussianDisorder}).

\begin{figure}[t]
\centering   \includegraphics[width=\linewidth]{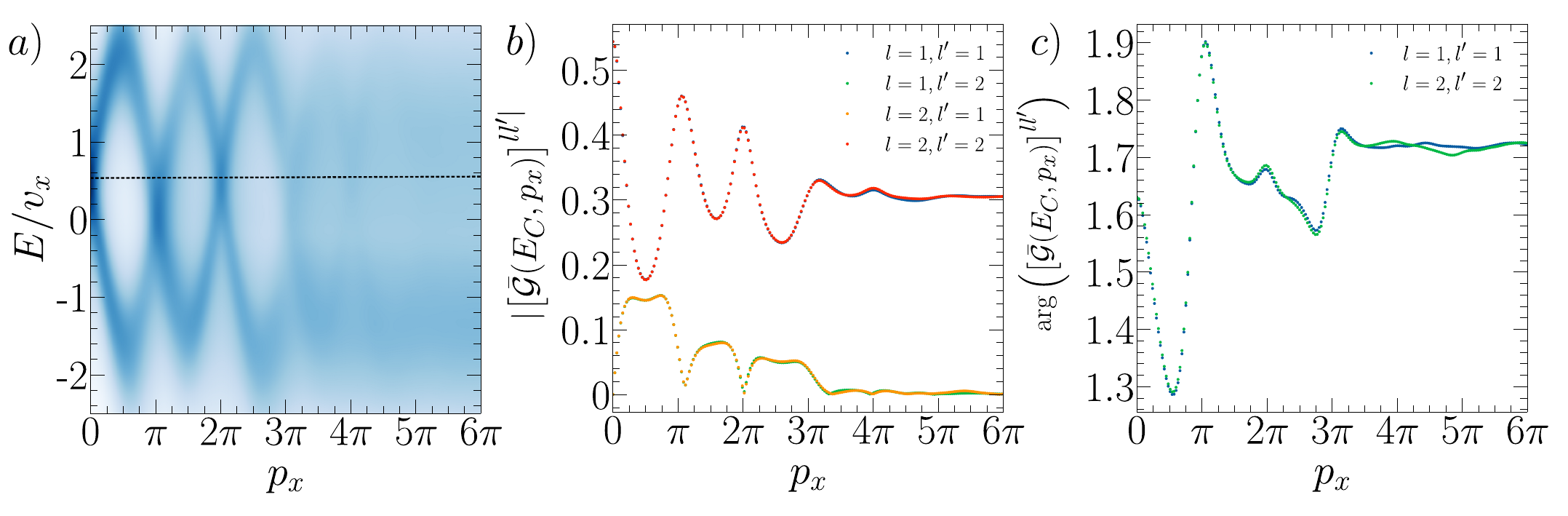} 
\caption{Momentum dependence of the average spectral function sharpness and Green's function matrix elements.
To generate the data shown in this figure, we place the non-crystalline Kramers-Weyl model [Eq.~(\ref{eq:amorphousKWTmatrixFinalChirality})] with the tight-binding parameters in Eq.~(\ref{eq:disorderedKWparams}) on a lattice with moderate nematic Gaussian structural and frame disorder [$\eta=0.15$, see Fig.~\ref{fig:3DGreenKW}].
We then average the system over 50 disorder realizations [replicas, see Eq.~(\ref{eq:averageTWoMomentumGreen})] with $20^{3}=8000$ sites each and contiguous domains of right- and left-handed sites within each replica with the respective concentrations $n_R=0.7$ and $n_L=1-N_R/N_{\mathrm{sites}}=0.3$ [see Fig.~\ref{appfig:disordertypes}(b)].
(a) The disorder-averaged spectral function $\bar{A}(E,\mathbf{p})$ [Eq.~(\ref{eq:SpecFunc})] for the moderately-disordered Kramers-Weyl model with ${\bf p}=(p_{x},0,0)$.
For all panels in this figure, ${\bf p}$ is expressed in units of the average nearest-neighbor spacing $\bar{a}=1$.
At small $|{\bf p}|$, $\bar{A}(E,\mathbf{p})$ in (a) exhibits sharp, band-structure-like features with Brillouin-zone-like repetitions. 
Conversely for larger $|{\bf p}|$, the spectral features in $\bar{A}(E,\mathbf{p})$ become broadened by disorder.
At a critical range of ${\bf p}$, which occurs at $|{\bf p}|\approx 3\pi$ in (a), the spectral features in $\bar{A}(E,\mathbf{p})$ blur into a continuous smear of spectral weight that is nearly independent of ${\bf p}$ and $E$.
(b,c) The magnitude and complex phase [argument function in radian units] of the matrix elements of the one-momentum average Green's function $\bar{\mathcal{G}}(E,{\bf p})$ [Eq.~(\ref{eq:averageOneMomentumGreen})] of the moderately disordered Kramers-Weyl model in (a), taken at $E_{C} = 0.57$ [the dashed line in (a)].
For all values of the internal spin degree-of-freedom indices $l,l'=1,2$ [defined in the text following Eq.~(\ref{eq:AmoKW})], $\bar{\mathcal{G}}(E,{\bf p})$ in (b,c) saturates at a constant complex value.
We have also computed $\bar{\mathcal{G}}(E,{\bf p})$ for the system in (a) at other energies beyond $E_{C} = 0.57$ in (b,c), and observe similar large-$|{\bf p}|$ saturation at those energies as well.
This indicates that overall, the disorder-averaged, momentum-resolved matrix Green's function $\bar{\mathcal{G}}(E,{\bf p})$ itself, and not just its trace [\emph{e.g.} $\bar{A}(E,\mathbf{p})$, see Eq.~(\ref{eq:SpecFunc})], exhibits numerical saturation at large $|{\bf p}|$ in the non-crystalline Kramers-Weyl model.
However, as discussed in this section and in the text surrounding Eq.~(\ref{eq:EmergentPinftyModulo}), caution is warranted when interpreting the large-$|{\bf p}|$ spectral features of disordered tight-binding models -- such as the saturated spectral weight and average Green's function matrix elements in (a-c) -- because the large-$|{\bf p}|$  properties of disordered systems are highly sensitive to model-dependent details of the underlying Wannier basis states.}
\label{fig:HeffBreak}
\end{figure}

\paragraph*{\bf Energy Spectra in Non-Crystalline Systems} -- $\ $ Finally, as shown in previous works~\cite{marsal_topological_2020,corbae_evidence_2020,JustinHat,Ciocys2023}, we may use the [disorder-averaged] one-momentum Green's function $\bar{\mathcal{G}}(E,\mathbf{p})$ in Eq.~(\ref{eq:averageOneMomentumGreen}) to define the disorder-averaged, momentum-resolved spectral function $\bar{A}(E,\mathbf{p})$ of a non-crystalline system:
\begin{equation}
\bar{A}(E,\mathbf{p}) = -\frac{1}{\pi}\text{Im}\left\{\Tr\left[\bar{\mathcal{G}}\left(E,\mathbf{p}\right)\right]\right\} = -\frac{1}{\pi}\text{Im}\left\{\sum_{l=1}^{N_{\text{orb}}}\left[\bar{\mathcal{G}}\left(E,\mathbf{p}\right)\right]^{ll}\right\} .
\label{eq:SpecFunc}
\end{equation}
In the crystalline limit [Eq.~(\ref{eq:DiracDeltaTwoMom})], $\bar{A}(E,\mathbf{p})$ reduces to a Lorentzian function with a width of $\varepsilon$ centered on the Fermi surface [\emph{i.e.} the momentum-resolved density of states]:
\begin{eqnarray}
A(E,{\bf k}) &=& -\frac{1}{\pi}\text{Im}\left\{\Tr\left[\mathcal{G}\left(E,\mathbf{k}\right)\right]\right\} \nonumber \\
&=& -\frac{1}{\pi}\text{Im}\left\{\Tr\left[\left(\left(E + i \varepsilon\right)\mathbb{1}_{N_{\text{orb}}}-\mathcal{H}({\bf k})\right)^{-1}\right]\right\} \nonumber \\
&=& \frac{1}{\pi}\mathlarger{\sum}_{j=1}^{N_{\text{orb}}}\frac{\varepsilon}{\left( E-E^{j}_{\bf k}\right)^{2} + \varepsilon^{2}},
\label{eq:crystallineAfunc}
\end{eqnarray}
where $E^{j}_{\bf k}$ is the $j$-th energy eigenvalue of the Bloch Hamiltonian $\mathcal{H}({\bf k})$~\cite{ChenBernevigQPI,Saito_TB_2016,Kohsaka_2017}.
Hence, the spectral function $A(E,{\bf k})$ is typically plotted on a logarithmic scale.
Throughout this work, we therefore similarly use a logarithmic scale for all spectral function plots, both crystalline and non-crystalline.

Unlike in a crystalline system, however, $\bar{A}(E,\mathbf{p})$ in a non-crystalline system exhibits a ${\bf p}$-dependent sharpness, because the pseudo-momentum ${\bf p}$ is only an approximate system quantum number due to the absence of lattice regularity.
In a non-crystalline system with strong structural disorder [as defined in the text following Eq.~(\ref{eq:approximateLatticeTranslationGpp})], $\bar{A}(E,\mathbf{p})$ is specifically sharpest at the smallest values of ${\bf p}$.
This can be understood by recognizing that the ${\bf p}\approx {\bf 0}$ spectrum characterizes the system at its smallest momenta and hence longest wavelengths, and that strongly disordered [\emph{e.g.} amorphous] systems can exhibit approximate continuous translation symmetry when viewed on length scales much longer than the average-nearest-neighbor intersite spacing [see the text surrounding Eq.~(\ref{eq:ASG})].
Together, this implies the existence of a well-defined dispersion relation in an amorphous system near ${\bf p}={\bf 0}$.
Conversely at larger $|{\bf p}|$, and hence shorter wavelengths, the absence of continuous and exact lattice translation symmetry has a stronger effect on the energy spectrum, leading to significant disorder broadening in $\bar{A}(E,\mathbf{p})$.

To demonstrate these effects, we show in Fig.~\ref{fig:HeffBreak}(a) a typical $\bar{A}(E,\mathbf{p})$ plot for a moderately disordered non-crystalline topological semimetal model.
The data in Fig.~\ref{fig:HeffBreak} were specifically obtained by placing the non-crystalline Kramers-Weyl model [Eq.~(\ref{eq:amorphousKWTmatrixFinalChirality})] with the tight-binding parameters in Eq.~(\ref{eq:disorderedKWparams}) on a lattice with moderate nematic Gaussian structural and frame disorder [$\eta=0.15$, see Fig.~\ref{fig:3DGreenKW}].
We then average the system over 50 disorder realizations [replicas, see Eq.~(\ref{eq:averageTWoMomentumGreen})] with $20^{3}=8000$ sites each and contiguous domains of right- and left-handed sites within each replica with the respective concentrations $n_R=0.7$ and $n_L=1-N_R/N_{\mathrm{sites}}=0.3$ [see Fig.~\ref{appfig:disordertypes}(b)].
In the vicinity of ${\bf p}\approx {\bf 0}$, the spectral function in Fig.~\ref{fig:HeffBreak}(a) exhibits sharp features with a band-structure-like dispersion relation and Brillouin-zone-like repetitions, like those previously observed for the topological Dirac-cone surface states of amorphous Bi$_2$Se$_3$ in angle-resolved photoemission experiments~\cite{Ciocys2023}.
At larger $|{\bf p}|$, however, $\bar{A}(E,\mathbf{p})$ in Fig.~\ref{fig:HeffBreak}(a) becomes increasingly diffuse until, in the vicinity of $p_{x}=3\pi$ [in units of the average nearest-neighbor spacing $\bar{a}=1$], the previously sharp spectral features in $\bar{A}(E,\mathbf{p})$ become indistinguishable.
Instead, the large-$|{\bf p}|$ spectral function in Fig.~\ref{fig:HeffBreak}(a) exhibits a continuous smear of non-vanishing spectral weight that is nearly independent of ${\bf p}$ and $E$.
We further find that the pseudo-momentum independence of the moderately disordered Kramers-Weyl model at large $|{\bf p}|$ occurs not just in the spectral function [Fig.~\ref{fig:HeffBreak}(a)], but also in each complex matrix element of the one-momentum average Green's function $\bar{\mathcal{G}}(E,{\bf p})$ itself, whose magnitudes and phases [argument functions] are respectively shown in Fig.~\ref{fig:HeffBreak}(b,c).

It is important to emphasize, however, that large-$|{\bf p}|$ spectral features in disordered tight-binding models are in general model-dependent, and should hence be analyzed with caution.
Specifically, as previously discussed in the text surrounding Eq.~(\ref{eq:EmergentPinftyModulo}), in amorphous tight-binding models with hard-shell or $\delta$-function-localized Wannier basis states like those employed in this work [Eq.~(\ref{eq:TBbasisStates})], there exists in each disordered system direction $\hat{r}$ a smallest intersite separation vector ${\bf a}_{\text{min},\hat{r}}$.
This implies that the system will appear identical from the perspective of all plane waves with $|{\bf p}|>2\pi/|{\bf a}_{\text{min},\hat{r}}|$, because the plane waves oscillate at wavelengths shorter than the smallest intersite separation $|{\bf a}_{\text{min},\hat{r}}|$ in the direction of ${\bf p}$.
Hence for each disordered system modeled in this work, both the disorder-averaged momentum-resolved matrix Green's function $\bar{\mathcal{G}}(E,{\bf p})$ and spectral function $\bar{A}(E,\mathbf{p})$ become ${\bf p}$-independent for each $E$ at a model-dependent set of ${\bf p}$, which specifically occurs for the moderately disordered Kramers-Weyl model in Fig.~\ref{fig:HeffBreak} in the vicinity of $|{\bf p}|\approx 3\pi$.

Conversely, the smallest-momentum features of amorphous tight-binding models are considerably more model-independent than the large-$|{\bf p}|$ features, because they represent longer-wavelength system properties that are far less sensitive to lattice-scale details, including the exponential [or $\delta$-function] tails of Wannier tight-binding basis states and the choice of model disorder implementation scheme [see Appendix~\ref{app:DiffTypesDisorder}].
We will shortly use the sharpness and generality of the small-momentum features in $\bar{\mathcal{G}}(E,{\bf p})$ and $\bar{A}(E,\mathbf{p})$ to construct approximate [effective] single-particle Hamiltonians for the ${\bf p}={\bf 0}$ nodal degeneracies in non-crystalline nodal semimetals [Appendix~\ref{app:EffectiveHamiltonian}], which we will then show exhibit quantized and tunable topological invariants [Appendix~\ref{sec:WilsonBerry}].

\subsection{Non-Crystalline Surface Spectra from Surface Green's Functions}
\label{app:surfaceGreens}

The 2D surface states of a 3D crystalline topological material can be simulated with a tight-binding model by cutting the model into a slab geometry and computing the Fourier-transformed slab Green's function projected to the top surface slab layer~\cite{WannierToolsSoluyanov}.
However in the strongly disordered [amorphous] tight-binding models analyzed in this work, there do not exist clear notions of ``top'' and ``bottom'' layers when the systems are terminated in directions lying in their $d_{A}$-dimensional disordered system [sub]space [see Eq.~(\ref{eq:finiteDimAmorph}) and the surrounding text].
Hence to extract and analyze the disordered 2D surface states of a $d=3$ bulk disordered system, we must introduce a method for approximately determining the set of tight-binding basis states that occupy the ``top'' region of the system.
The projector onto the sites designed as lying within the top region may then be used to construct surface Green's and spectral functions.

To compute the 2D surface-state spectrum of a non-crystalline 3D topological semimetal, we begin by considering an initially bulk [large] $d=3$ system.
Without loss of generality, we take the system to be non-crystalline in all directions [\emph{i.e.} to have nematic disorder, see Appendix~\ref{app:SmecticNematicDisorder}], such that it has $d_{A}=3$, $d_{T}=0$, and $d_{f}=0$ in the language of Eq.~(\ref{eq:finiteDimAmorph}).
We next place our system on hybrid [slab] boundary conditions in which the system is periodic [though not necessarily regular] in two directions in the 2D subspace parameterized by $x_{\parallel 1,2}$, and open in the $x_{\perp}$-direction in a coordinate basis in which the position ${\bf r}_{\alpha}$ of each site $\alpha$ is given by:
\begin{equation}
{\bf r}_{\alpha} = \begin{pmatrix} 
\left(x_{\perp}\right)_{\alpha}\\
\left(x_{\parallel 1}\right)_{\alpha}\\
\left(x_{\parallel 2}\right)_{\alpha}
\end{pmatrix},
\label{eq:slabCoords}
\end{equation}
where $(x_{\perp})_\alpha$ is the position component of ${\bf r}_{\alpha}$ along the direction of the slab normal vector $\hat{x}_{\perp}$, and where $(x_{\parallel 1})_{\alpha}$ and $(x_{\parallel 2})_{\alpha}$ are components of ${\bf r}_{\alpha}$ in the remaining in-plane slab coordinate subspace spanned by the unit vectors $\hat{x}_{\parallel 1}$ and $\hat{x}_{\parallel 2}$.

Initially, the slab model has a number of sites given by the integer $N_{\text{sites}}$, where each site has $N_{\text{orb}}$ internal spin and orbital degrees of freedom [see Appendix~\ref{app:DiffTypesDisorder}]. 
Even after performing the iterative bond [intersite distance] relaxation procedure detailed in the text surrounding Eq.~(\ref{eq:iterateDisplacement}), a disordered $d=3$ system cut into a slab geometry will in general still host ``dangling'' sites that are effectively undercoordinated compared to other sites deep within the slab bulk.
In our calculations, we specifically consider sites to be dangling when they are connected to fewer than two other sites, using the disorder-implementation-specific definition of connected [\emph{i.e.} bonded or neighboring] sites established in Fig.~\ref{appfig:structuraldisorder} and the surrounding text.
However, dangling surface atoms have been extensively shown to produce trivial, model-dependent flat-band-like surface states~\cite{ShockleyStates}, which in topological materials can obscure signatures of intrinsic topological surface states.
Furthermore in real amorphous materials, inter-atom forces in general provide a surface confining potential that can bind undercoordinated surface atoms, and atoms that only feel this potential weakly will in practice dissipate from a sample during its preparation via processes like evaporation.
Therefore, to form a more realistic model without trivial dangling-bond surface states, we next delete ``dangling'' sites from our slab model, resulting in a remaining slab-geometry system with $\tilde{N}_{\text{Sites}}$, where $\tilde{N}_{\text{Sites}}\leq N_{\text{Sites}}$.

Next, we designate the position-space tight-binding Hamiltonian of the disordered slab system with deleted [``evaporated''] dangling surface atoms to be $\mathcal{H}_{\text{slab}}$.
The single-particle tight-binding Hilbert space of $\mathcal{H}_{\text{slab}}$, like in the bulk systems previously discussed in Appendix~\ref{app:PhysicalObservables}, are spanned by the Wannier basis states $\lvert \mathbf{r}_{\alpha},l\rangle$, where each $\lvert \mathbf{r}_{\alpha},l\rangle$ is again localized on a site $\alpha$ at the position ${\bf r}_{\alpha}$ and carries an internal [spin and orbital] degree-of-freedom index $l$.
In our calculations, each $\lvert \mathbf{r}_{\alpha},l\rangle$ in the slab is represented as an $\tilde{N}_{\text{sites}}N_{\text{orb}}$ column vector where $\tilde{N}_{\text{sites}}$ is the number of slab system sites after the removal of dangling surface atoms, and $N_{\text{orb}}$ is the number of internal degrees of freedom on each site [which we again take to be the same across every site in the slab].
As detailed in the text surrounding Eq.~(\ref{eq:TBbasisStates}), the real-space tight-binding basis states $\lvert \mathbf{r}_{\alpha},l\rangle$ are importantly orthonormal, and are in practice in this work $\delta$-function localized.

Like the bulk systems previously discussed in Appendix~\ref{app:PhysicalObservables}, the slab system at each energy $E$ can also be characterized by a position-space Green's function matrix:
\begin{equation}
\mathcal{G}_{\text{slab}}\left(E\right) = \left(\left(E + i \varepsilon\right)\mathds{1}_{\tilde{N}_{\text{sites}}N_{\text{orb}}}-\mathcal{H}_{\text{slab}}\right)^{-1},
\label{eq:RealSpaceGreenSlab}
\end{equation}
where $\varepsilon$ is again a broadening parameter that determines the energy resolution of the position-space Green's function [see the text following Eq.~(\ref{appeq:Greens})].
The matrix elements of the real-space slab Green's function are then given by:
\begin{equation}
\left[\mathcal{G}_{\text{slab}}\left(E\right) \right]^{\alpha\beta,ll'} = \langle \mathbf{r}_\alpha,l \rvert \mathcal{G}_{\text{slab}}\left(E\right) \lvert \mathbf{r}_\beta,l'\rangle.
\label{appeq:GreensSlab}
\end{equation}

In a slab model of a pristine crystal, the ``surface'' atoms can straightforwardly be chosen by identifying sites within a prespecified number of slab layers [system unit cells in the direction of the slab Miller index].
However in a disordered slab system, the absence of lattice translation symmetry [and hence slab layer indices] necessitates instead using an alterative method to designate atoms [sites] as lying within the ``top'' system region.
To numerically define the top-surface region of our disordered slab model, we therefore next construct the $\tilde{N}_{\text{sites}}N_{\text{orb}}\times\tilde{N}_{\text{sites}}N_{\text{orb}}$ top-surface matrix projector:
\begin{equation}
P_{\text{surf}}=\sum_{\alpha \in \mathrm{Window}}\sum_{l=1}^{N_{\text{orb}}}|{\bf r}_{\alpha},l\rangle\langle {\bf r}_{\alpha},l|,
\label{eq:topSurfaceProj}
\end{equation}
where the position-space cutoff window in the first summation in Eq.~(\ref{eq:topSurfaceProj}) designates a site $\alpha$ as lying within the top-surface slab region if its coordinate component $(x_{\perp})_{\alpha}$ in the basis of Eq.~(\ref{eq:slabCoords}) satisfies:
\begin{equation}
\label{eq:WindowCondition}
x_{\perp}^{\mathrm{max}} - \xi \leq (x_{\perp})_{\alpha} \leq x_{\perp}^{\mathrm{max}},
\end{equation}
where $x_{\perp}^{\mathrm{max}}$ is the $x_{\perp}$ coordinate of the furthest site along the $x_{\perp}$-direction, and $\xi$ is a user-defined top surface depth.
For all of the slab model calculations performed in this work, we specifically choose $\xi$ such that the designated ``top-surface'' region contains $\tilde{N}_{\text{sites}}^{2/3}$ sites [\emph{i.e.} $\Tr[P_{\text{surf}}] \approx \tilde{N}_{\text{sites}}^{2/3}N_{\text{orb}}$ in Eq.~(\ref{eq:topSurfaceProj})].
We then use Eqs.~(\ref{eq:RealSpaceGreenSlab}) and~(\ref{eq:topSurfaceProj}) to define the top-surface position-space Green's function:
\begin{equation}
\mathcal{G}_{\mathrm{surf}}\left(E\right) = P_{\text{surf}}^{\phantom{}}\mathcal{G}_\text{slab}\left(E\right)P_{\text{surf}}^{\dagger} =P_{\text{surf}}\mathcal{G}_\text{slab}\left(E\right)P_{\text{surf}},
\label{eq:GreenTop}
\end{equation}
where $\mathcal{G}_{\mathrm{surf}}(E)$, like $\mathcal{G}_\text{slab}(E)$ and $P_{\text{surf}}$, here remains an $\tilde{N}_{\text{sites}}N_{\text{orb}}\times\tilde{N}_{\text{sites}}N_{\text{orb}}$ matrix.

In pristine models of 3D crystals cut into a slab geometries, the position-space [surface and bulk] slab Green's function can be partially Fourier-transformed into a hybrid coordinate system with one residual discrete position-space [slab layer] index and two continuous in-plane crystal momenta [in addition to internal and sublattice degree-of-freedom indices]~\cite{VDBoriginalAxion,VDBNicoAxionHybrid,MTQC,PozoCecileAdolfoAxion,PartialAxionHOTINumerics}.
This is functionally accomplished by treating the slab as a [greatly] thickened $d=3$, $d_{A}=0$, $d_{T}=2$, $d_{f}=1$ system as defined in Eq.~(\ref{eq:finiteDimAmorph}) and the surrounding text, projecting the slab Green's function onto the top surface layer[s], and then Fourier-transforming the crystalline slab Green's function in the translationally-invariant $d_{T}=2$ in-plane directions.
Specifically, in a numerical crystalline slab model with a large number of layers, there exists some ambiguity in whether the model is interpreted as a $d=3$, $d_{A}=0$, $d_{T}=3$, $d_{f}=0$ model with open [slab] boundary conditions, or as a $d=3$, $d_{A}=0$, $d_{T}=2$, $d_{f}=1$ model that is just very thick [but not approximately infinitely so] in its $d_{f}=1$ finite direction.
From a physical perspective, both interpretations must yield the same observables, but the latter $d_{T}=2$ interpretation in particular facilitates the computation of topological surface states via surface projectors like Eq.~(\ref{eq:topSurfaceProj}) that isolate atoms in the outermost ``surface'' layers.

By analogy, we therefore next similarly treat our $d_{A}=3$ disordered slab model as if it were instead a thickened $d=3$, $d_{A}=2$, $d_{f}=1$ [$d_{T}=0$] system, and then apply the continuous plane-wave Fourier transform previously detailed in Eq.~(\ref{appeq:planewaves}), but now \emph{only} for the two in-plane non-crystalline slab directions.
Specifically, in the thick 3D slab system with just two $d_{A}$ in-plane non-crystalline directions, we construct \emph{hybrid} Fourier-transformed plane-wave states with only two pseudo-momentum components $p_{\parallel 1,2}$, while allowing the spatial coordinates ${\bf r}_{\alpha}$ of each site $\alpha$ to remain three-dimensional:
\begin{eqnarray}
\lvert p_{\parallel 1}, p_{\parallel 2},l\rangle &=& \frac{1}{\sqrt{\tilde{N}_{\text{sites}}}}\sum_{\alpha=1}^{\tilde{N}_{\text{sites}}} \exp\left[i p_{\parallel 1}\left(\hat{x}_{\parallel 1}\cdot \mathbf{r}_\alpha\right) + i p_{\parallel 2}\left(\hat{x}_{\parallel 2}\cdot \mathbf{r}_\alpha\right)\right]\lvert \mathbf{r}_\alpha,l\rangle \nonumber\\ &=& \frac{1}{\sqrt{\tilde{N}_{\text{sites}}}}\sum_{\alpha=1}^{\tilde{N}_{\text{sites}}}\exp\left[ip_{\parallel1}(x_{\parallel 1})_{\alpha} + ip_{\parallel 2}(x_{\parallel 2})_{\alpha}\right]\lvert \mathbf{r}_\alpha,l\rangle,
\label{appeq:planewavesSlab}
\end{eqnarray}
where we have simplified using Eq.~(\ref{eq:slabCoords}).
Eq.~(\ref{appeq:planewavesSlab}), along with the orthogonality of the position-space basis states $\lvert \mathbf{r}_\alpha,l\rangle$ in Eq.~(\ref{eq:TBbasisStates}) and the definition of the slab coordinate basis in Eq.~(\ref{eq:slabCoords}), together imply that the projection of each hybrid slab plane-wave state onto each position-space basis state is given by:
\begin{eqnarray}
    \langle \mathbf{r}_\alpha,l'\lvert p_{\parallel 1}, p_{\parallel 2},l\rangle &=& \frac{1}{\sqrt{\tilde{N}_{\text{sites}}}} \exp\left[i p_{\parallel 1}\left(\hat{x}_{\parallel 1}\cdot \mathbf{r}_\alpha\right) + i p_{\parallel 2}\left(\hat{x}_{\parallel 2}\cdot \mathbf{r}_\alpha\right)\right] \delta_{ll'} \nonumber \\
    &=& \frac{1}{\sqrt{\tilde{N}_{\text{sites}}}}\exp\left[ip_{\parallel1}(x_{\parallel 1})_{\alpha} + ip_{\parallel 2}(x_{\parallel 2})_{\alpha}\right]\delta_{ll'},
\label{appeq:planeSlab}
\end{eqnarray}
and that the inner products between hybrid plane-wave states are given by:
\begin{eqnarray}
\langle p_{\parallel 1}, p_{\parallel 2},l &\rvert& p'_{\parallel 1}, p'_{\parallel 2},l'\rangle \nonumber \\
&=& \frac{1}{\tilde{N}_{\text{sites}}}\sum_{\alpha,\beta=1}^{\tilde{N}_{\text{sites}}}\exp\left[-i p_{\parallel 1}\left(\hat{x}_{\parallel 1}\cdot \mathbf{r}_\alpha\right) - i p_{\parallel 2}\left(\hat{x}_{\parallel 2}\cdot \mathbf{r}_\alpha\right)\right]\exp\left[i p'_{\parallel 1}\left(\hat{x}_{\parallel 1}\cdot \mathbf{r}_\beta\right) + i p'_{\parallel 2}\left(\hat{x}_{\parallel 2}\cdot \mathbf{r}_\beta\right)\right]\langle \mathbf{r}_\alpha,l\lvert \mathbf{r}_\beta,l'\rangle \nonumber \\
&=& \frac{1}{\tilde{N}_{\text{sites}}}\delta_{ll'}\sum^{\tilde{N}_{\text{sites}}}_{\alpha=1}\exp\left[-i\left(\hat{x}_{\parallel 1}\cdot \mathbf{r}_\alpha\right)\left(p_{\parallel 1} - p'_{\parallel 1}\right)\right]\exp\left[-i\left(\hat{x}_{\parallel 2}\cdot \mathbf{r}_\alpha\right)\left(p_{\parallel 2} - p'_{\parallel 2}\right)\right] \nonumber \\
&=& \frac{1}{\tilde{N}_{\text{sites}}}\delta_{ll'}\sum^{\tilde{N}_{\text{sites}}}_{\alpha=1}\exp\left[-i\left(x_{\parallel 1}\right)_{\alpha}\left(p_{\parallel 1} - p'_{\parallel 1}\right)\right]\exp\left[-i\left(x_{\parallel 2}\right)_{\alpha}\left(p_{\parallel 2} - p'_{\parallel 2}\right)\right] \nonumber \\
&=& \frac{1}{\tilde{N}_{\text{sites}}}\delta_{ll'}\sum^{\tilde{N}_{\text{sites}}}_{\alpha=1}\exp\left[-i\left(x_{\parallel 1}\right)_{\alpha}\left(p_{\parallel 1} - p'_{\parallel 1}\right)-i\left(x_{\parallel 2}\right)_{\alpha}\left(p_{\parallel 2} - p'_{\parallel 2}\right)\right].
\label{eq:OrthoPlaneWavesSlab}
\end{eqnarray}
Importantly, as with the continuous plane-wave states that we previously used to characterize bulk 3D non-crystalline systems [Eqs.~(\ref{appeq:plane}) and~(\ref{eq:OrthoPlaneWaves})], Eqs.~(\ref{appeq:planeSlab}) and~(\ref{eq:OrthoPlaneWavesSlab}) do not further simplify in slab systems with irregular lattices.

Having established a set of pseudo-momentum-space plane-wave states for the disordered slab, we next construct a Fourier-transformed description of the top surface of the slab system.
Following the bulk construction in Eq.~(\ref{eq:MomGreenFunc}), we specifically next use Eqs.~(\ref{eq:slabCoords}),~(\ref{appeq:GreensSlab}),~(\ref{eq:GreenTop}), and~(\ref{appeq:planeSlab}) to define the \emph{[pseudo-] momentum-resolved surface Green's function} $\mathcal{G}_{\text{surf}}(E,p_{\parallel 1},p_{\parallel,2}, p'_{\parallel 1},p'_{\parallel,2})$ via its matrix elements:
\begin{equation}
\begin{split}
&\left[\mathcal{G}_{\text{surf}}\left(E,p_{\parallel 1},p_{\parallel,2}, p'_{\parallel 1},p'_{\parallel,2}\right)\right]^{ll'} \\ \\
&\ = \langle p_{\parallel 1},p_{\parallel,2},l \lvert \mathcal{G}_{\text{surf}}\left(E\right)\rvert p'_{\parallel 1},p'_{\parallel,2},l' \rangle, \\ 
&\ = \frac{1}{\tilde{N}_{\text{sites}}}\sum_{\alpha,\beta=1}^{\tilde{N}_{\text{sites}}}\exp\left[-i p_{\parallel 1}\left(\hat{x}_{\parallel 1}\cdot \mathbf{r}_\alpha\right) - i p_{\parallel 2}\left(\hat{x}_{\parallel 2}\cdot \mathbf{r}_\alpha\right)\right]\exp\left[i p'_{\parallel 1}\left(\hat{x}_{\parallel 1}\cdot \mathbf{r}_\beta\right) + i p'_{\parallel 2}\left(\hat{x}_{\parallel 2}\cdot \mathbf{r}_\beta\right)\right]\langle \mathbf{r}_\alpha,l | \mathcal{G}_{\text{surf}}\left(E\right)|\mathbf{r}_\beta,l' \rangle \\
&\ = \frac{1}{\tilde{N}_{\mathrm{sites}}}\sum_{\alpha,\beta=1}^{\tilde{N}_{\text{sites}}}\exp\left(i p'_{\parallel 1}\left(\hat{x}_{\parallel 1}\cdot \mathbf{r}_\beta\right) + i p'_{\parallel 2}\left(\hat{x}_{\parallel 2}\cdot \mathbf{r}_\beta\right) - i p_{\parallel 1}\left(\hat{x}_{\parallel 1}\cdot \mathbf{r}_\alpha\right) - i p_{\parallel 2}\left(\hat{x}_{\parallel 2}\cdot \mathbf{r}_\alpha\right)\right)\left[\mathcal{G}_{\text{surf}}\left(E\right)\right]^{\alpha\beta,ll'} \\
&\ = \frac{1}{\tilde{N}_{\mathrm{sites}}}\sum_{\alpha,\beta=1}^{\tilde{N}_{\text{sites}}}\exp\left(i p'_{\parallel 1}\left(x_{\parallel 1}\right)_{\beta} + i p'_{\parallel 2}\left(x_{\parallel 2}\right)_{\beta} - i p_{\parallel 1}\left(x_{\parallel 1}\right)_{\alpha} - i p_{\parallel 2}\left(x_{\parallel 2}\right)_{\alpha}\right)\left[\mathcal{G}_{\text{surf}}\left(E\right)\right]^{\alpha\beta,ll'}.
\end{split}
\label{eq:GreenSurfaceMomentum}
\end{equation}
As previously with the Fourier-transformed bulk Green's function $\mathcal{G}(E,\mathbf{p},\mathbf{p}')$ in Eq.~(\ref{eq:MomGreenFunc}), the surface Green's function $\mathcal{G}_{\text{surf}}(E,p_{\parallel 1},p_{\parallel,2}, p'_{\parallel 1},p'_{\parallel,2})$ is an $N_{\text{orb}}\times N_{\text{orb}}$ matrix that is parameterized by two pseudo-momentum vectors $(p_{\parallel 1}, p_{\parallel 2})$ and $(p'_{\parallel 1}, p'_{\parallel 2})$, and does not further simplify in the absence of lattice regularity.

In this work, our open- [slab-] boundary-condition calculations are particularly limited by our inability to simulate individual slab systems with greater than $\sim 20^{3}$ [$\sim$ 8000] sites.  
However, as previously with the disordered bulk systems analyzed in this work [see the text surrounding Eq.~(\ref{eq:averageTWoMomentumGreen})], we may gain better numerical convergence of \emph{surface} spectral and topological properties -- and more accurately model the surface states of real solid-state amorphous systems with thermodynamically large numbers of atoms -- by using the replica method~\cite{ParisiReplicaCourse,BerthierGlassAmorphousReview} to construct a two-momentum \emph{average surface Green's function} $\bar{\mathcal{G}}_{\text{surf}}(E,p_{\parallel 1},p_{\parallel,2}, p'_{\parallel 1},p'_{\parallel,2})$.
To implement $\bar{\mathcal{G}}_{\text{surf}}(E,p_{\parallel 1},p_{\parallel,2}, p'_{\parallel 1},p'_{\parallel,2})$, we specifically average the Fourier-transformed surface Green's function matrix $\mathcal{G}_{i,\text{surf}}(E,p_{\parallel 1},p_{\parallel,2}, p'_{\parallel 1},p'_{\parallel,2})$ [Eq.~(\ref{eq:GreenSurfaceMomentum})] over an ensemble of $\approx 50$ slab system replicas that each contain independently generated structural and internal degree-of-freedom disorder and are indexed below by $i$:
\begin{equation}
\left[\bar{\mathcal{G}}_{\text{surf}}(E,p_{\parallel 1},p_{\parallel,2}, p'_{\parallel 1},p'_{\parallel,2})\right]^{ll'} = \frac{1}{N_{\text{rep}}}\sum_{i=1}^{N_{\text{rep}}} \left[\mathcal{G}_{i,\text{surf}}(E,p_{\parallel 1},p_{\parallel,2}, p'_{\parallel 1},p'_{\parallel,2})\right]^{ll'},
\label{eq:averageTWoMomentumGreenSlab}
\end{equation}
where $N_{\text{rep}}$ is the number of slab replicas, and where the same structural disorder implementation scheme from Appendix~\ref{app:lattices} is used for each slab replica in the ensemble
.

In Appendix~\ref{app:PhysicalObservables}, we previously observed that for all of the non-crystalline topological semimetal models [Appendix~\ref{app:models}] and structural disorder implementation schemes [Appendix~\ref{app:DiffTypesDisorder}] employed in this work, the off-diagonal-in-momentum matrix elements of the two-momentum bulk average Green's function $\bar{\mathcal{G}}(E,\mathbf{p},\mathbf{p}')$ are nearly vanishing on the scale of the initial tight-binding [non-crystalline] lattice parameters [see Figs.~\ref{fig:3DGreenKW},~\ref{fig:3DGreenC2}, and~\ref{fig:3DGreen3F}].
This allowed us to approximate $\bar{\mathcal{G}}(E,\mathbf{p},\mathbf{p}')$ using a simpler, \emph{one-momentum} average Green's function $\bar{\mathcal{G}}(E,\mathbf{p})$ [Eq.~(\ref{eq:averageOneMomentumGreen})], from which the bulk spectral functions and topology can then be computed.
Motivated by this observation, we next similarly define the one-momentum, Fourier-transformed average \emph{surface} Green's function matrix $\bar{\mathcal{G}}_{\text{surf}}(E,p_{\parallel 1} ,p_{\parallel 2})$ from the diagonal-in-momentum elements of $\bar{\mathcal{G}}_{\text{surf}}(E,p_{\parallel 1},p_{\parallel,2}, p'_{\parallel 1},p'_{\parallel,2})$:
\begin{equation}
\left[\bar{\mathcal{G}}_{\text{surf}}(E,p_{\parallel 1} ,p_{\parallel 2})\right]^{ll'} = \frac{1}{N_{\text{rep}}}\sum_{i=1}^{N_{\text{rep}}} \left[\mathcal{G}_{i,\text{surf}}(E,p_{\parallel 1},p_{\parallel,2}, p_{\parallel 1},p_{\parallel,2})\right]^{ll'}.
\label{eq:averageOneMomentumGreenSurface}
\end{equation}

Lastly, following our previous definition of the disorder-averaged, momentum-resolved bulk spectral function $\bar{A}(E,{\bf p}) \propto \text{Im}\{\Tr[\bar{\mathcal{G}}(E,{\bf p})]\}$ [Eq.~(\ref{eq:SpecFunc})], we similarly use Eq.~(\ref{eq:averageOneMomentumGreenSurface}) to define the disorder-averaged, momentum-resolved \emph{surface} spectral function $\bar{A}_{\text{surf}}(E,p_{\parallel 1},p_{\parallel 2})$ of our disordered slab system:
\begin{equation}
\bar{A}_{\text{surf}}\left(E,p_{\parallel 1},p_{\parallel 2}\right) = -\frac{1}{\pi}\text{Im}\left\{\Tr\left[\bar{\mathcal{G}}_{\text{surf}}\left(E,p_{\parallel 1},p_{\parallel,2}\right) \right] \right\} = -\frac{1}{\pi}\text{Im}\left\{\sum_{l=1}^{N_{\text{orb}}}\left[\bar{\mathcal{G}}_{\text{surf}}\left(E,p_{\parallel 1},p_{\parallel,2}\right) \right]^{ll} \right\}.
\label{eq:averageSpectrumSurface}
\end{equation}
Below, in Appendices~\ref{app:amorphousKramers},~\ref{app:amorphousCharge2}, and~\ref{app:amorphousMultifold}, we will use $\bar{A}_{\text{surf}}(E,p_{\parallel 1},p_{\parallel 2})$ in Eq.~(\ref{eq:averageSpectrumSurface}) to track the evolution of the Fermi-arc surface states of structurally chiral topological semimetals under increasingly strong lattice and frame disorder ranging from the nearly crystalline to the amorphous regimes.

\subsection{Effective Hamiltonians}
\label{app:EffectiveHamiltonian}

In disordered systems, there does not exist an exact single-particle dispersion relation [band structure], due to the absence of exact lattice or continuum translation symmetries.
However as previously discussed in Appendix~\ref{app:PhysicalObservables}, an approximate, Fourier-transformed, one-momentum [disorder-averaged] bulk Green's function $\bar{\mathcal{G}}(E,\mathbf{p})$ can separately be computed at each energy $E$ using continuum plane-wave states indexed by the pseudo-momentum ${\bf p}$ [Eqs.~(\ref{appeq:planewaves}) and~(\ref{eq:averageOneMomentumGreen})].
A disorder-averaged, momentum-resolved spectral function $\bar{A}(E,{\bf p})$ can then be computed from $\bar{\mathcal{G}}(E,\mathbf{p})$ [$\bar{A}(E,{\bf p}) \propto \text{Im}\{\Tr[\bar{\mathcal{G}}(E,{\bf p})]\}$, see Eq.~(\ref{eq:SpecFunc})].
However unlike in a crystalline system, $\bar{A}(E,{\bf p})$ is not uniformly sharp, but rather exhibits ${\bf p}$-dependent disorder broadening.
Nevertheless, $\bar{A}(E,{\bf p})$ can still exhibit sharp, model-independent features with well-defined dispersion relations~\cite{marsal_topological_2020,corbae_evidence_2020,JustinHat,Ciocys2023}, especially in the vicinity of ${\bf p}\approx {\bf 0}$ [\emph{i.e.} at the longest system wavelengths, see Fig.~\ref{fig:HeffBreak} and the surrounding text].

Previous works~\cite{varjas_topological_2019,marsal_topological_2020,marsal_obstructed_2022} have shown that the sharpest features in $\bar{A}(E,{\bf p})$ can in turn be reproduced by constructing an approximate single-particle \emph{effective Hamiltonian} $\mathcal{H}_\text{Eff}({\bf p})$ using the [average] one-momentum Green's function matrix $\bar{\mathcal{G}}(E,\mathbf{p})$.
Specifically, in non-crystalline tight-binding models that consist only of sites with the same $N_{\text{orb}}$ internal spin and orbital degrees of freedom -- like those analyzed in this work [Appendices~\ref{app:lattices} and~\ref{app:models}] -- an $N_{\text{orb}}\times N_{\text{orb}}$ Hermitian effective Hamiltonian matrix $\mathcal{H}_\text{Eff}({\bf p})$ can be defined by restricting $\bar{\mathcal{G}}(E,\mathbf{p})$ [Eq.~(\ref{eq:averageOneMomentumGreen})] to a reference energy cut $E_{C}$ and computing:
\begin{equation}
\mathcal{H}_{\text{Eff}}(\mathbf{p}) \underset{\text{def}}{\equiv} \mathcal{H}_{\text{Eff}}(E_{C},{\bf p}) = -\frac{1}{2}\left(\bar{\mathcal{G}}^{-1}(E_C,\mathbf{p})  + \left[\bar{\mathcal{G}}^{-1}(E_C,\mathbf{p}) \right]^{\dagger}\right) - E_{\text{C}}.
\label{eq:AvgHEff}
\end{equation}
Single-particle topological invariants computed from $\mathcal{H}_\text{Eff}({\bf p})$ have then been shown to accurately capture the many-particle topology of gapped [insulating] amorphous systems~\cite{corbae_evidence_2020,marsal_topological_2020,marsal_obstructed_2022,springMagneticAverageTI}.
Hermitian linear combinations of Green's functions similar to Eq.~(\ref{eq:AvgHEff}) have also recently been employed to construct and compute single-particle topological invariants in strongly correlated systems, such as Mott insulators~\cite{CanoMottGreens1,CanoMottGreens2,MottFailureBradlynPhilipps,MottFailureGoldman}.

\begin{figure}
\centering   
\includegraphics[width=\linewidth]{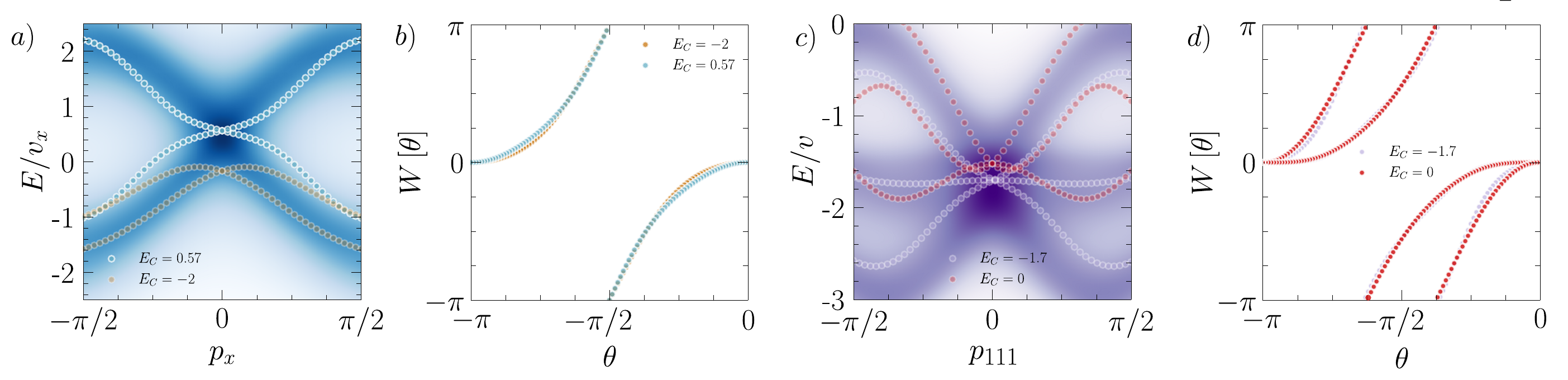} 
\caption{Benchmarking the spectrum and wavefunctions of the effective Hamiltonian.
To generate this figure, we first place (a,b) the non-crystalline Kramers-Weyl model [Eq.~(\ref{eq:amorphousKWTmatrixFinalChirality}) with the parameters in Eq.~(\ref{eq:disorderedKWparams})] and (c,d) multifold fermion model [Eq.~(\ref{eq:amorphous3FTmatrixFinalChirality}) with the parameters in Eq.~(\ref{eq:disordered3Fparams})] on lattices with $N_{\mathrm{sites}}=20^{3}=8000$ sites and moderate nematic Gaussian structural and frame disorder parameterized by the standard deviation $\eta = 0.15$ [see Appendix~\ref{app:lattices}].
We then average each model over 50 disorder realizations [replicas, see Eq.~(\ref{eq:averageTWoMomentumGreen})] with contiguous domains of right- and left-handed sites within each replica with the respective concentrations $n_R=0.7$ and $n_L=1-N_R/N_{\mathrm{sites}}=0.3$ [see Fig.~\ref{appfig:disordertypes}(b)].
(a) The average spectral function $\bar{A}(E,\mathbf{p})$ [Eq.~(\ref{eq:SpecFunc})] for the moderately-disordered Kramers-Weyl model with ${\bf p}=(p_{x},0,0)$ and the energy eigenvalues of the single-particle effective Hamiltonian $\mathcal{H}_{\text{Eff}}(\mathbf{p}) \equiv \mathcal{H}_{\text{Eff}}(E_{C},{\bf p})$ obtained from the average Green's function $\bar{\mathcal{G}}(E,{\bf p})$ [Eq.~(\ref{eq:averageOneMomentumGreen})] for $E_{C}=-2$ [orange circles] and $E_{C}=0.57$ [light blue circles].
For both choices of $E_{C}$ in (a), $\mathcal{H}_{\text{Eff}}({\bf p})$ exhibits a linearly dispersing twofold nodal degeneracy at ${\bf p}={\bf 0}$, the canonical form of a Kramers-Weyl fermion~\cite{KramersWeyl,OtherKramersWeyl,AndreiTalk}.
We find that in the non-crystalline Kramers-Weyl model, the energy spectrum of $\mathcal{H}_{\text{Eff}}({\bf p})$ most closely qualitatively matches the many-particle spectrum of $\bar{A}(E,{\bf p}\approx {\bf 0})$ when $E_{C}$ is set to the value of the maximum spectral weight at ${\bf p}={\bf 0}$.
(c) The average spectral function $\bar{A}(E,\mathbf{p})$ for the moderately disordered multifold fermion model with ${\bf p}$ taken along $p_{111} = (1/\sqrt{3})(p_{x}+p_{y}+p_{z})$ and the energy eigenvalues of $\mathcal{H}_{\text{Eff}}(\mathbf{p})$ constructed with $E_{C}= -1.7$ [light purple circles] and $E_{C}=0$ [red circles].
For both choices of $E_{C}$ in (c), $\mathcal{H}_{\text{Eff}}({\bf p})$ exhibits a spin-1 [chiral] fermion with a threefold nodal degeneracy at ${\bf p}={\bf 0}$ [Appendix~\ref{sec:Multifold}].
Like in (a), we find that for the non-crystalline multifold fermion model in (c), the energy spectrum of $\mathcal{H}_{\text{Eff}}({\bf p})$ most closely qualitatively matches $\bar{A}(E,{\bf p}\approx {\bf 0})$ when $E_{C}$ is set to lie at the largest spectral weight of the many-particle multifold degeneracy at ${\bf p}={\bf 0}$ [\emph{i.e.} the highest spectral weight away from the nondispersing feature at larger $E$, see Fig.~\ref{fig:3F_figBulk}(b)].  
To determine whether the eigenstates of $\mathcal{H}_{\text{Eff}}(E_{C},\mathbf{p})$ are similarly affected by the choice of $E_{C}$, we next compute the Wilson loop spectrum on a sphere surrounding the nodal degeneracy at ${\bf p}={\bf 0}$ in each model [Appendix~\ref{sec:WilsonBerry}], and plot the results in (b,d) as functions of the sphere polar angle $\theta$ for the same $E_{C}$ used to construct $\mathcal{H}_{\text{Eff}}(E_{C},\mathbf{p})$ in (a,c).  
For both (b) the Abelian [scalar] Wilson loop over the lower band of $\mathcal{H}_{\text{Eff}}(\mathbf{p})$ in the Kramers-Weyl model in (a), and (d) the non-Abelian [matrix] Wilson loop over the lowest two bands of $\mathcal{H}_{\text{Eff}}(\mathbf{p})$ in the multifold fermion model in (c), we surprisingly find \emph{almost no variation} in the Wilson loop eigenvalues over a large range of $E_{C}$ surrounding each nodal degeneracy.
This implies that for the models studied in this work, the eigenstates of $\mathcal{H}_{\text{Eff}}(\mathbf{p})$ are considerably less sensitive to the choice of $E_{C}$ than their corresponding energy eigenvalues. 
Hence, this indicates that the single-particle wavefunctions of $\mathcal{H}_{\text{Eff}}(\mathbf{p})$ provide a numerically stable basis for computing the many-particle topology of ${\bf p}={\bf 0}$ nodal degeneracies in amorphous systems.}
\label{fig:H_eff_benchmark}
\end{figure}

Though it is often notationally suppressed, $\mathcal{H}_\text{Eff}({\bf p})$ is in general highly dependent on the choice of $E_{C}$ in Eq.~(\ref{eq:AvgHEff}).
However for the ${\bf p}\approx{\bf 0}$ non-crystalline [amorphous] nodal degeneracies in the models studied in this work [Appendices~\ref{app:amorphousKramers},~\ref{app:amorphousCharge2}, and~\ref{app:amorphousMultifold}], we find that $\mathcal{H}_\text{Eff}({\bf p})$ exhibits the same low-energy [${\bf k}\cdot {\bf p}$] polynomial dispersion relation, non-Abelian Berry phases [Wilson loop eigenvalues, see Appendix~\ref{sec:WilsonBerry} and Refs.~\cite{Fidkowski2011,AndreiXiZ2,ArisInversion,Cohomological,HourglassInsulator,DiracInsulator,Z2Pack,BarryFragile,AdrienFragile,HOTIBernevig,HingeSM,WiederAxion,KoreanFragile,ZhidaBLG,TMDHOTI,KooiPartialNestedBerry,PartialAxionHOTINumerics,GunnarSpinFragileWilson,BinghaiOscillationWilsonLoop,Wieder22}], and therefore overall topological classification over a large range of $E_{C}$.
In Fig.~\ref{fig:H_eff_benchmark} we show calculations supporting this observation performed on the non-crystalline Kramers-Weyl model [Appendix~\ref{app:amorphousKramers}] and multifold fermion model [Appendix~\ref{app:amorphousMultifold}].
To generate the data shown in Fig.~\ref{fig:H_eff_benchmark}, we specifically place the non-crystalline Kramers-Weyl model [Eq.~(\ref{eq:amorphousKWTmatrixFinalChirality}) with the parameters in Eq.~(\ref{eq:disorderedKWparams})] and multifold fermion model [Eq.~(\ref{eq:amorphous3FTmatrixFinalChirality}) with the parameters in Eq.~(\ref{eq:disordered3Fparams})] on lattices with $N_{\mathrm{sites}}=20^{3}=8000$ sites and moderate nematic Gaussian structural and frame disorder parameterized by the standard deviation $\eta = 0.15$ [see Appendix~\ref{app:lattices}].
We then average each model over 50 disorder realizations [replicas, see Eq.~(\ref{eq:averageTWoMomentumGreen})] with contiguous domains of right- and left-handed sites within each replica with the respective concentrations $n_R=0.7$ and $n_L=1-N_R/N_{\mathrm{sites}}=0.3$ [see Fig.~\ref{appfig:disordertypes}(b)].
For each model, we then benchmark the accuracy and numerical stability of the energy spectrum and wavefunctions [specifically Berry phases] of the single-particle effective Hamiltonian by constructing $\mathcal{H}_{\text{Eff}}(\mathbf{p}) \equiv \mathcal{H}_{\text{Eff}}(E_{C},{\bf p})$ over a range of $E_{C}$.

\paragraph*{\bf Benchmarking the Energy Spectrum of the Effective Hamiltonian} -- $\ $ To evaluate the accuracy of the energy eigenvalues of $\mathcal{H}_{\text{Eff}}(\mathbf{p})$ for varying $E_{C}$, we respectively plot in Fig.~\ref{fig:H_eff_benchmark}(a,c) the disorder-averaged spectral functions $\bar{A}(E,{\bf p})$ [Eq.~(\ref{eq:SpecFunc})] of the non-crystalline Kramers-Weyl and multifold fermion models detailed above, and overlay the band structure of $\mathcal{H}_{\text{Eff}}(\mathbf{p})$ constructed from $\bar{\mathcal{G}}(E,\mathbf{p})$ [Eq.~(\ref{eq:AvgHEff})] for each model using two different choices of $E_{C}$.
We specifically in Fig.~\ref{fig:H_eff_benchmark}(a) first plot the
average spectral function $\bar{A}(E,\mathbf{p})$ of the Kramers-Weyl model with moderate [$\eta=0.15$] nematic Gaussian disorder overlaid with the energy eigenvalues of $\mathcal{H}_{\text{Eff}}(\mathbf{p})$ constructed at $E_{C}=-2$ [orange circles] and $E_{C}=0.57$ [light blue circles].
For both choices of $E_{C}$ in Fig.~\ref{fig:H_eff_benchmark}(a), $\mathcal{H}_{\text{Eff}}({\bf p})$ exhibits a linearly dispersing twofold nodal degeneracy at ${\bf p}={\bf 0}$, the canonical form of a Kramers-Weyl fermion [see Refs.~\cite{KramersWeyl,OtherKramersWeyl,AndreiTalk} and Appendix~\ref{app:Kramers}].
We further observe that for values of $E_{C}$ beyond those in Fig.~\ref{fig:H_eff_benchmark}(a), the non-crystalline Kramers-Weyl model [Eq.~(\ref{eq:amorphousKWTmatrixFinalChirality})] for all of the structural disorder implementation schemes in this work continues to exhibit a Kramers-Weyl fermion at ${\bf p}={\bf 0}$, albeit with strong variations in the velocity of the Kramers-Weyl cone.
Overall, we observe that for the non-crystalline Kramers-Weyl model, the energy spectrum of $\mathcal{H}_{\text{Eff}}({\bf p})$ most closely qualitatively matches the many-particle spectrum of $\bar{A}(E,{\bf p}\approx {\bf 0})$ when $E_{C}$ is set to the value of the maximum spectral weight at ${\bf p}={\bf 0}$.
Hence for the disordered Kramers-Weyl model calculations performed in this work [detailed in Appendix~\ref{app:amorphousKramers}], we consistently construct the effective Hamiltonian with $E_{C}$ centered at the maximum value of $\bar{A}(E,{\bf p})$ at ${\bf p}={\bf 0}$ to maximize the spectral accuracy of $\mathcal{H}_{\text{Eff}}({\bf p})$.

Similarly, we next in Fig.~\ref{fig:H_eff_benchmark}(c) plot the average spectral function $\bar{A}(E,\mathbf{p})$ of the non-crystalline multifold fermion model with moderate [$\eta=0.15$] nematic Gaussian disorder overlaid with the energy eigenvalues of $\mathcal{H}_{\text{Eff}}(\mathbf{p})$ constructed at $E_{C}= -1.7$ [light purple circles] and $E_{C}=0$ [red circles].
For both choices of $E_{C}$ in Fig.~\ref{fig:H_eff_benchmark}(c), $\mathcal{H}_{\text{Eff}}({\bf p})$ exhibits a spin-1 [chiral] fermion with a threefold nodal degeneracy at ${\bf p}={\bf 0}$ [see Refs.~\cite{ManesNewFermion,chang2017large,tang2017CoSi,KramersWeyl,DoubleWeylPhonon,CoSiObserveJapan,CoSiObserveHasan,CoSiObserveChina,AlPtObserve,PdGaObserve,PtGaObserve,AltlandSpin1Light,DingARPESReversal,ZahidLadderMultigap,Sanchez2023} and 
Appendix~\ref{sec:Multifold}].
Furthermore, over a large range of $E_{C}$ values beyond those selected in Fig.~\ref{fig:H_eff_benchmark}(c), the non-crystalline multifold fermion model [Eq.~(\ref{eq:amorphous3FTmatrixFinalChirality})] for all of the structural disorder implementation schemes in this work continues to exhibit a spin-1 fermion with an approximate rotation-symmetry-enforced threefold degeneracy at ${\bf p}={\bf 0}$ for comparably large system sizes and numbers of disorder replicas [Eq.~(\ref{eq:averageTWoMomentumGreen})] as those employed to generate Fig.~\ref{fig:H_eff_benchmark}(c).   
Like in Fig.~\ref{fig:H_eff_benchmark}(a), we overall find that for the non-crystalline multifold fermion model in Fig.~\ref{fig:H_eff_benchmark}(c), the energy spectrum of $\mathcal{H}_{\text{Eff}}({\bf p})$ most closely qualitatively matches $\bar{A}(E,{\bf p}\approx {\bf 0})$ when $E_{C}$ is set to lie at the largest spectral weight of the many-particle multifold degeneracy at ${\bf p}={\bf 0}$ [\emph{i.e.} the highest spectral weight away from the nondispersing spectral feature at larger $E$, see Fig.~\ref{fig:3F_figBulk}(b)].
Therefore for all of the disordered multifold fermion model calculations performed in this work [detailed in Appendix~\ref{app:amorphousMultifold}], we construct the effective Hamiltonian with $E_{C}$ centered at the maximum value of $\bar{A}(E,{\bf p})$ at ${\bf p}={\bf 0}$ away from higher-energy spectral features to maximize the spectral accuracy of $\mathcal{H}_{\text{Eff}}({\bf p})$.

Finally, though not shown in Fig.~\ref{fig:H_eff_benchmark} above, we have also performed analogous benchmarking of the effective Hamiltonian $\mathcal{H}_{\text{Eff}}({\bf p})$ for varying $E_{C}$ in the non-crystalline double-Weyl model analyzed in Appendix~\ref{app:amorphousCharge2}.
Like the Kramers-Weyl and multifold fermion models in Fig.~\ref{fig:H_eff_benchmark}(a,c), we observe that over a large range of $E_{C}$, $\mathcal{H}_{\text{Eff}}({\bf p})$ for the double-Weyl model exhibits an average rotation-symmetry-enforced, quadratically dispersing double-Weyl fermion~\cite{ZahidMultiWeylSrSi2,StepanMultiWeyl,AndreiMultiWeyl,XiDaiMultiWeyl} at ${\bf p}={\bf 0}$.
We similarly further observe that in the non-crystalline double-Weyl model, the band structure of the single-particle effective Hamiltonian $\mathcal{H}_{\text{Eff}}({\bf p})$ most closely matches the sharp features in $\bar{A}(E,{\bf p}\approx {\bf 0})$ when $E_{C}$ is set to lie at the largest spectral weight of the many-particle double-Weyl point at ${\bf p}={\bf 0}$.
Hence for the disordered quadratic double-Weyl model calculations performed in this work [detailed in Appendix~\ref{app:amorphousCharge2}], we consistently construct the effective Hamiltonian with $E_{C}$ centered at the maximum value of $\bar{A}(E,{\bf p})$ at ${\bf p}={\bf 0}$ to maximize the spectral accuracy of $\mathcal{H}_{\text{Eff}}({\bf p})$.

\paragraph*{\bf Benchmarking the Wavefunctions of the Effective Hamiltonian} -- $\ $ Having shown that the energy eigenvalues [band structure] of the single-particle effective Hamiltonian $\mathcal{H}_{\text{Eff}}(\mathbf{p}) \equiv \mathcal{H}_{\text{Eff}}(E_{C},{\bf p})$ [Eq.~(\ref{eq:AvgHEff})] can reliably be fit to the longest-wavelength sharp features of $\bar{A}(E,{\bf p})$ near ${\bf p}={\bf 0}$ [see Figs.~\ref{fig:HeffBreak}(a) and~\ref{fig:H_eff_benchmark}(a,c)], we next quantitatively evaluate the numerical stability of the wavefunctions of $\mathcal{H}_{\text{Eff}}({\bf p})$.
To specifically address whether the eigenstates of $\mathcal{H}_{\text{Eff}}({\bf p})$ provide a reliable basis for calculating single-particle topological invariants, we compute for each non-crystalline model the amorphous sphere Wilson loop [non-Abelian Berry phase, see Appendix~\ref{sec:WilsonBerry} and Refs.~\cite{Fidkowski2011,AndreiXiZ2,ArisInversion,Cohomological,HourglassInsulator,DiracInsulator,Z2Pack,BarryFragile,AdrienFragile,HOTIBernevig,HingeSM,WiederAxion,KoreanFragile,ZhidaBLG,TMDHOTI,KooiPartialNestedBerry,PartialAxionHOTINumerics,GunnarSpinFragileWilson,BinghaiOscillationWilsonLoop,Wieder22}] of the occupied bands of $\mathcal{H}_{\text{Eff}}({\bf p})$ for varying numerical implementation parameters.
In Fig.~\ref{fig:H_eff_benchmark}(b,d) we respectively plot the Wilson loop eigenvalues of the occupied band[s] of the effective Hamiltonians in Fig.~\ref{fig:H_eff_benchmark}(a,c), with the colors of the circles in each panel indicating the reference $E_{C}$ used to construct $\mathcal{H}_{\text{Eff}}({\bf p})$.
First, for both choices of $E_{C}$ in the non-crystalline Kramers-Weyl model in Fig.~\ref{fig:H_eff_benchmark}(b), we observe nearly the \emph{same} eigenvalues in the Wilson loop spectrum computed over the lower band of $\mathcal{H}_{\text{Eff}}({\bf p})$ on a sphere surrounding the nodal degeneracy at ${\bf p}={\bf 0}$ [which reduces to the Abelian or scalar Berry phase because there is only one occupied band, see the text surrounding Eqs.~(\ref{eq:nonAbelianBerry1}) and~(\ref{eq:nonAbelianBerry2})].
Next, we further find that for the multifold fermion model in Fig.~\ref{fig:H_eff_benchmark}(c), \emph{both} of the non-Abelian [matrix] Wilson loop eigenvalues computed over the lower two bands of $\mathcal{H}_{\text{Eff}}({\bf p})$ remain nearly constant across the two choices of $E_{C}$ in Fig.~\ref{fig:H_eff_benchmark}(c,d).

More generally, we surprisingly find that for the non-crystalline topological semimetals studied in this work [Appendices~\ref{app:amorphousKramers},~\ref{app:amorphousCharge2}, and~\ref{app:amorphousMultifold}], the [non-] Abelian Wilson loop eigenvalues of the occupied bands of $\mathcal{H}_{\text{Eff}}({\bf p})$ exhibit \emph{almost no variation} over a large range of $E_{C}$ surrounding the topological nodal degeneracy at ${\bf p}={\bf 0}$ in each model.
Though not shown in Fig.~\ref{fig:H_eff_benchmark}(b,d), we also further observe that the amorphous sphere Wilson loop eigenvalues of each model are largely insensitive to the number of harmonic-function moments employed in the kernel polynomial method expansion used to generate the average Green's function matrix $\bar{\mathcal{G}}(E,\mathbf{p})$, and hence $\mathcal{H}_{\text{Eff}}({\bf p})$ [see Ref.~\cite{Weisse06,varjas_topological_2019} and the text following Eqs.~(\ref{appeq:Greens}) and~(\ref{eq:AvgHEff})].
In Appendices~\ref{app:amorphousKramers} and~\ref{app:amorphousMultifold}, we further use $\bar{\mathcal{G}}(E,\mathbf{p})$ to respectively compute the spin and orbital-angular-momentum textures of the non-crystalline Kramers-Weyl and multifold fermion models [see the text surrounding Eqs.~(\ref{eq:SpinDOS}) and~(\ref{eq:OAMDOS})].
In both models, we observe that the ${\bf p}={\bf 0}$ disordered chiral fermions exhibit the same structural-chirality-locked, monopole-like angular momentum textures as their crystalline counterparts, which serves as an approximate indicator of their many-particle topological chiral charges~\cite{KramersWeyl,BradlynTQCSpinTexture,OAMWeyl2,OAMmultifold1,OAMmultifold2,OAMmultifold3,OAMmultifold4,OAMmultifold5}.
Along with the insensitivity of the Wilson loop eigenvalues to the choice of $E_{C}$, this provides strong evidence that the wavefunctions of the effective Hamiltonian $\mathcal{H}_{\text{Eff}}(\mathbf{p})$ [Eq.~(\ref{eq:AvgHEff})] serve as both an accurate and numerically stable basis for computing single-particle topological invariants that capture the exact [many-particle] topology of ${\bf p}={\bf 0}$ nodal degeneracies in amorphous topological semimetals.
Lastly, for the models studied in this work, we observed that the eigenstates of $\mathcal{H}_{\text{Eff}}(\mathbf{p})$ are considerably less sensitive to the choice of $E_{C}$ than their corresponding energy eigenvalues.
This suggests the intriguing possibility, which we leave for future work, that the accuracy of $\mathcal{H}_{\text{Eff}}(E_{C},\mathbf{p})$ in capturing the full many-body properties of amorphous systems in the vicinity of an energy $E$ and the pseudo-momentum ${\bf p}={\bf 0}$ is controlled by a convergent quantum perturbation theory expansion in $E-E_{C}$ and ${\bf p}$.

\subsection{Amorphous Wilson Loops}
\label{sec:WilsonBerry}

Like band insulators, gapless [semimetallic] crystalline systems may also be classified as topological~\cite{Armitage2018,Wieder22,BinghaiClaudiaWeylReview,ZahidNatRevMatWeyl}.
In gapless systems with lattice translation symmetry, this is first accomplished by determining the set of nodal degeneracies that connect the valence [occupied] and conduction [unoccupied] bands.
For each nodal degeneracy, one may then compute topological invariants on surfaces surrounding the nodal degeneracy [such as spheres in 3D crystals], and further classify the system as a \emph{topological semimetal} if any of the computed topological invariants are nontrivial.  
The most general tool for computing topological invariants in nodal semimetals is the non-Abelian Wilson loop [holonomy] matrix~\cite{Fidkowski2011,AndreiXiZ2,ArisInversion,Cohomological,HourglassInsulator,DiracInsulator,Z2Pack,BarryFragile,AdrienFragile,HOTIBernevig,HingeSM,WiederAxion,KoreanFragile,ZhidaBLG,TMDHOTI,KooiPartialNestedBerry,PartialAxionHOTINumerics,GunnarSpinFragileWilson,BinghaiOscillationWilsonLoop,Wieder22}.
For example, the spectral flow of the Wilson loop eigenvalues on a sphere can be used to deduce the $\mathbb{Z}_{2}$ monopole charge of a Dirac nodal line in a 3D crystal~\cite{AhnMonopole,TomasBZDWilsonMonopole,TMDHOTI}. 
More relevant to the present work, a point-like nodal degeneracy in a crystalline system can be identified as a condensed-matter realization of a chiral fermion -- such as a conventional Weyl fermion~\cite{AshvinWeyl,VolovikWeylMonopole,VolovikChiralAnomaly,HaldaneOriginalWeyl,MurakamiWeyl,BurkovBalents,AndreiWeyl,HasanWeylDFT,Armitage2018,SuyangWeyl,LvWeylExp,YulinWeylExp,AliWeylQPI,AlexeyType2,ZJType2,BinghaiClaudiaWeylReview,ZahidNatRevMatWeyl,CDWWeyl,IlyaIdealMagneticWeyl} -- if the sphere Wilson loop eigenvalues exhibit a nontrivial winding number [\emph{i.e.} chiral spectral flow], which corresponds to the chiral charge $C$ of the nodal degeneracy~\cite{Z2Pack}.
Specifically, given a nodal degeneracy at ${\bf k}={\bf k}_{W}$ in a crystal with a low-energy ${\bf k}\cdot{\bf p}$ Hamiltonian of the form $\mathcal{H}({\bf q})$ where ${\bf q}={\bf k}-{\bf k}_{W}$, the $\mathbb{Z}$-valued topological chiral charge $C$ of the nodal point is given by the quantized integral of the total Berry curvature flux $\Tr[{\bf F}({\bf q})]\cdot d{\bf A}$ of the occupied [lower] bands of $\mathcal{H}({\bf q})$ on a sphere $S$ surrounding ${\bf k}_{W}$:
\begin{equation}
C = \frac{1}{2\pi}\oint_{S}\Tr[{\bf F}({\bf q})]\cdot d{\bf A},
\label{eq:crystallineWilsonSphere}
\end{equation}
where $C$ is in turn indicated by the sphere Wilson loop winding number~\cite{Z2Pack,Armitage2018}.

In a $d$-dimensional non-crystalline system, an exact Bloch Hamiltonian parameterized by the $d$-component crystal momentum vector ${\bf k}$ cannot be defined, owing to the absence of lattice translation symmetry in all $d$ system directions. 
Hence as discussed throughout this work, it has historically been challenging to quantitatively classify gapless non-crystalline systems as nodal semimetals, or to compute the topological invariants of suspected nodal degeneracies in models without lattice translation symmetry.
Recent works~\cite{Grossi2023b,WeylQuasicrystalSCBott} have partially overcome this issue in 3D layered quasicrystalline semimetal models with exact lattice translation symmetry in the stacking direction [$d=3$, $d_{A}=2$, $d_{T}=1$, and $d_{f}=0$ in the notation of Eq.~(\ref{eq:finiteDimAmorph})].
In Refs.~\cite{Grossi2023b,WeylQuasicrystalSCBott}, the authors specifically constructed time-reversal-broken, band-inversion-driven topological quasicrystalline Weyl semimetals and nodal superconductors where the nodal degeneracies lay along the $k_{z}$ axis [taking $z$ to be the real-space quasicrystal stacking direction].
The authors then identified the topology of the nodal points by computing the 2D [charge or superconducting] real-space Chern marker or Bott index of the occupied states of a series of 2D hybrid-coordinate-space Hamiltonians in the $(x,y)$-coordinate-plane parameterized by the crystal momentum $k_{z}$.  
However for the fully structurally disordered [$d_{A}=3$] nodal semimetals studied in this work, the approach employed in Refs.~\cite{Grossi2023b,WeylQuasicrystalSCBott} fails, because there does not exist an exact crystal momentum $k_{z}$ with which to parameterize 2D real- [hybrid-] space topological invariants.

Previous works have partially overcome this issue by focusing on the spectral localizer, which conversely can be computed in models with all values of $d$ and $d_{A}$ in Eq.~(\ref{eq:finiteDimAmorph}).
Introduced in Ref.~\cite{LORING2015}, the spectral localizer is an operator that quantifies whether the position and Hamiltonian operators can simultaneously be deformed to commute without closing an energy gap at a particular reference energy.
Because insulators for which the Hamiltonian and projected position operator commute are necessarily trivial atomic limits, this implies that the spectral localizer can indicate whether a crystalline or non-crystalline model lies in a stable topological phase~\cite{LORING2015}.
More recently, the spectral localizer was extended to nodal semimetals, in which the localizer zeroes can approximately indicate the number and dispersion of well-isolated nodal degeneracies that lie close to the Fermi level~\cite{Schulz-Baldes21,Schulz-Baldes2022,Franca2024b,Franca2024}.
In Ref.~\cite{Franca2024}, the authors specifically computed the spectral localizer in a structurally disordered multifold fermion model with a sublattice structure and an \emph{achiral} average symmetry group [see Appendices~\ref{app:pseudoK} and~\ref{app:amorphousMultifold}], and observed that the in-gap localizer modes merged with the continuum of trivial localizer states at the same Gaussian disorder strength at which the topological surface Fermi arcs vanished in the surface spectral function. 
However, the spectral localizer only provides a qualitative indicator of topology in nodal semimetals, and has not yet been associated to an exact topological semimetal index theorem outside of idealized models.

To overcome these issues in $d_{A}=3$ non-crystalline systems, we first recognize that as shown in Appendix~\ref{app:corepAmorphous}, strongly disordered [amorphous] systems can host average-symmetry-enforced nodal degeneracies at ${\bf p}={\bf 0}$.
In Appendix~\ref{app:EffectiveHamiltonian}, we then showed that the polynomial dispersion relation and wavefunctions [specifically Berry phases] of the single-particle effective Hamiltonian $\mathcal{H}_{\text{Eff}}(\mathbf{p}) \equiv \mathcal{H}_{\text{Eff}}(E_{C},{\bf p})$ constructed from the disorder-averaged one-momentum Green's function $\bar{\mathcal{G}}(E,{\bf p})$ [Eqs.~(\ref{eq:averageOneMomentumGreen}) and~(\ref{eq:AvgHEff})] are highly numerically stable over a wide range of numerical simulation parameters and reference energy cuts $E_{C}$ for pseudo-momenta ${\bf p}$ near ${\bf p}={\bf 0}$ [\emph{i.e.} at the longest system wavelengths].
This motivates us to in this work introduce an \emph{amorphous Wilson loop method} in which the eigenstates of $\mathcal{H}_{\text{Eff}}(\mathbf{p})$ are used to infer the chiral charges of many-particle nodal degeneracies at ${\bf p}={\bf 0}$ in strongly disordered 3D topological semimetals.

\begin{figure}[t]
\centering   
\includegraphics[width=\linewidth]{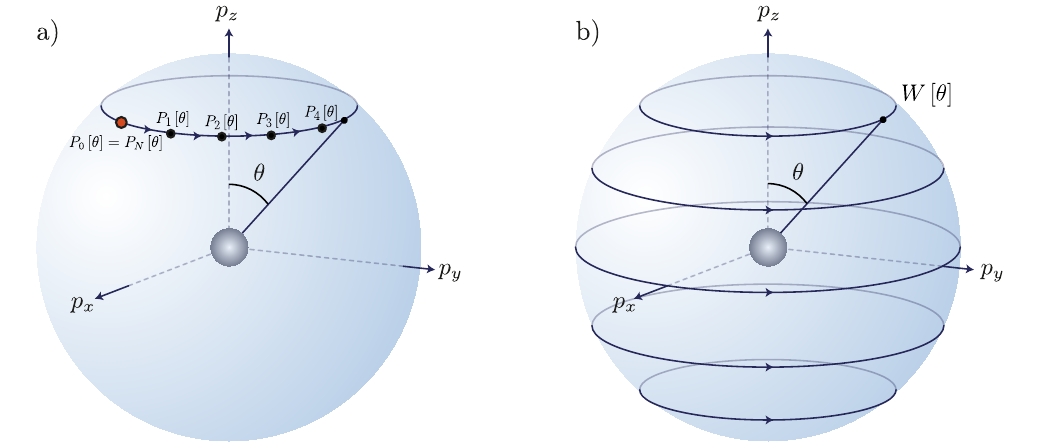}
\caption{Amorphous Wilson loop method schematic.
(a) Schematic representation of the discretized Wilson loop matrix computation on a closed path [black line] on a sphere surrounding a nodal degeneracy at ${\bf p} = {\bf 0}$ [gray circle].
The eigenvalues of the non-Abelian Wilson loop matrix $W[\theta]$ on a closed contour are specifically obtained via the path-ordered product of projectors $P_{i}[\theta]$ beginning and ending at the base point ${\bf p}_{0}[\theta]={\bf p}_{N}[\theta]$ [the red circle in (a), see Eqs.~(\ref{eq:tempPathOrderedW}),~(\ref{eq:refinedWilsonMatrix}), and~(\ref{eq:WilsonElements}) and the surrounding text].
In the amorphous Wilson loop method introduced in this work for ${\bf p}={\bf 0}$ nodal degeneracies in 3D non-crystalline topological semimetals, the chiral charge of a many-particle disordered ${\bf p}={\bf 0}$ nodal point is determined by first constructing a single-particle effective Hamiltonian $\mathcal{H}_{\text{Eff}}(\mathbf{p})$ for the nodal degeneracy as detailed in Appendix~\ref{app:EffectiveHamiltonian} [Eq.~(\ref{eq:AvgHEff})].
We then diagonalize $\mathcal{H}_{\text{Eff}}(\mathbf{p})$ for ${\bf p}$ lying on a sphere enclosing ${\bf p}={\bf 0}$ [blue transparent sphere].
(b) We next identify a series of counterclockwise parallel curves on the enclosing sphere parameterized by the polar angle $\theta$.
Lastly, we compute the eigenvalues of $W[\theta]$ for the occupied bands of $\mathcal{H}_{\text{Eff}}(\mathbf{p})$ at each polar angle $\theta$.
As detailed in the text surrounding Eq.~(\ref{eq:amorphousWilsonSphere}), the topological chiral charge $C\in\mathbb{Z}$ of the ${\bf p}={\bf 0}$ nodal degeneracy is specifically given by the winding number of the phase angles $\gamma_{n}[\theta]$ of the eigenvalues of $W[\theta]$ [\emph{i.e.} the non-Abelian Berry phases, see Eqs.~(\ref{eq:nonAbelianBerry1}) and~(\ref{eq:nonAbelianBerry2})].
We hence in this work identify ${\bf p}={\bf 0}$ nodal spectral features with $C\neq 0$ in disordered tight-binding models as condensed-matter realizations of disordered 3D chiral fermions.}
\label{fig:Wilson_schema}
\end{figure}

We next detail our method for computing the Wilson loop spectrum of a ${\bf p}={\bf 0}$ nodal degeneracy in a $d=3$ non-crystalline nodal semimetal model composed only of sites with the same $N_{\text{orb}}$ internal spin and orbital degrees of freedom, like those introduced and analyzed in this work [Appendices~\ref{app:lattices} and~\ref{app:models}]. 
To begin, we first construct the $N_{\text{orb}}\times N_{\text{orb}}$ effective Hamiltonian matrix $\mathcal{H}_{\text{Eff}}(\mathbf{p})$ using Eq.~(\ref{eq:AvgHEff}) with $E_{C}$ set to lie at the largest spectral weight of the many-particle nodal point at ${\bf p}={\bf 0}$ [see Appendix~\ref{app:EffectiveHamiltonian} for further $\mathcal{H}_{\text{Eff}}(\mathbf{p})$ calculation details].  
We then diagonalize $\mathcal{H}_{\text{Eff}}(\mathbf{p})$ at each pseudo-momentum point ${\bf p}[\theta]$ at the polar angle $\theta$ on a sphere surrounding ${\bf p}={\bf 0}$ [blue transparent sphere in Fig.~\ref{fig:Wilson_schema}(a,b)].
Next, we construct the $N_{\text{orb}}\times N_{\text{orb}}$ matrix projector onto the $N_{\text{occ}}$ occupied eigenstates $|u^{j}_{{\bf p}[\theta]}\rangle$ of $\mathcal{H}_{\text{Eff}}(\mathbf{p})$:
\begin{equation}
P_{\bf p}[\theta] = \sum_{j=1}^{N_{\text{occ}}}|u^{j}_{{\bf p}[\theta]}\rangle\langle u^{j}_{{\bf p}[\theta]}|.
\label{eq:OccupiedProjectorWilson}
\end{equation}

The topological chiral charge of a ${\bf p}={\bf 0}$ nodal degeneracy can be obtained from the eigenvalues of the Wilson loop matrix $W[\theta]$ [non-Abelian Berry phases] computed on a series of counterclockwise parallel curves indexed by the polar angle $\theta$ of a pseudo-momentum sphere enclosing ${\bf p}={\bf 0}$ [see Fig.~\ref{fig:Wilson_schema}(b) and Refs.~\cite{Z2Pack,Armitage2018}].
To compute the Wilson loop matrix $W[\theta]$ for each parallel closed contour on the sphere, we first introduce the $N_{\text{orb}}\times N_{\text{orb}}$ matrix $\tilde{W}[\theta]$, whose form originates from approximating the continuum non-Abelian Berry phase on a discretized ${\bf p}$ mesh with $N$ pseudo-momentum points ${\bf p}_{i}[\theta]$ at each polar angle $\theta$ [see Fig.~\ref{fig:Wilson_schema}(a,b) and Refs.~\cite{Fidkowski2011,AndreiXiZ2,ArisInversion,Cohomological,HourglassInsulator,DiracInsulator,Z2Pack,BarryFragile,AdrienFragile,HOTIBernevig,HingeSM,WiederAxion,KoreanFragile,ZhidaBLG,TMDHOTI,KooiPartialNestedBerry,PartialAxionHOTINumerics,GunnarSpinFragileWilson,BinghaiOscillationWilsonLoop,Wieder22}]:
\begin{equation}
\tilde{W}[\theta] = P_{0}[\theta]P_{1}[\theta]P_{2}[\theta]\ldots P_{N}[\theta] = P_{0}[\theta]P_{1}[\theta]P_{2}[\theta]\ldots P_{0}[\theta] = \prod_{i=0}^{N}P_{i}[\theta] = \prod_{i=0}^{N}\left(\sum_{j=1}^{N_{\text{occ}}}|u^{j}_{{\bf p}_{i}[\theta]}\rangle\langle u^{j}_{{\bf p}_{i}[\theta]}|\right).
\label{eq:tempPathOrderedW}
\end{equation}
Unlike straight-line Wilson loops in crystalline systems~\cite{Fidkowski2011,AndreiXiZ2,ArisInversion,Cohomological,HourglassInsulator,DiracInsulator,Z2Pack,BarryFragile,AdrienFragile,HOTIBernevig,HingeSM,WiederAxion,KoreanFragile,ZhidaBLG,TMDHOTI,KooiPartialNestedBerry,PartialAxionHOTINumerics,GunnarSpinFragileWilson,BinghaiOscillationWilsonLoop,Wieder22}, the path-ordered product of projectors in Eq.~(\ref{eq:tempPathOrderedW}) does not contain a sewing [embedding] matrix $V({\bf K})$, because $\tilde{W}[\theta]$ is instead here evaluated on a closed contour in the pseudo-momentum Brillouin zone that begins and ends at the same [base] point ${\bf p}_{0}[\theta]={\bf p}_{N}[\theta]$ [the red circle in Fig.~\ref{fig:Wilson_schema}(a)].

We next introduce the $N_{\text{orb}}\times N_{\text{orb}}$ overlap matrix:
\begin{equation}
G_{i}[\theta] = P_{i}[\theta]P_{i+1}[\theta],
\label{eq:overlapProductDef}
\end{equation}
which along with Eqs.~(\ref{eq:OccupiedProjectorWilson}) and~(\ref{eq:tempPathOrderedW}) and the idempotence of projectors $P_{i}^{2}[\theta]=P_{i}[\theta]$ implies that:
\begin{equation}
\tilde{W}[\theta] = \prod_{i=0}^{N-1}G_{i}[\theta].
\label{eq:overlapPrimeDefinition}
\end{equation}
To interpret the eigenvalues of $\tilde{W}[\theta]$ as [the complex exponentials of] non-Abelian Berry phases, it is necessary that the eigenvalues of each $G_{i}[\theta]$ in Eqs.~(\ref{eq:overlapProductDef}) and~(\ref{eq:overlapPrimeDefinition}) are either zero or complex phases of magnitude $1$.  
However, this condition is only strictly satisfied for $N\rightarrow \infinity$, and particularly fails for ${\bf p}$ meshes with relatively small values of $N$~\cite{multipole,WladTheory}. 
To overcome the deviation of the magnitude of the eigenvalues of each $G_{i}[\theta]$ from $0$ and $1$, we follow Refs.~\cite{multipole,WladTheory} and next perform a singular value decomposition [SVD] of each overlap matrix:
\begin{equation}
G_{i}[\theta]=U_{i}[\theta]D_{i}[\theta]V_{i}^{\dag}[\theta].
\label{eq:overlapSVD}
\end{equation}
From Eq.~(\ref{eq:overlapSVD}), we then define the $N_{\text{orb}}\times N_{\text{orb}}$ matrix:
\begin{equation}
F_{i}[\theta] = U_{i}[\theta]V^{\dag}_{i}[\theta],
\label{eq:FfromSVD}
\end{equation}
where the eigenvalues of each $F_{i}[\theta]$, unlike the eigenvalues of each overlap matrix $G_{i}[\theta]$, are now either zero or complex phases of magnitude $1$ for all finite ${\bf p}$-mesh sizes $N$.  
Lastly, we use the $F_{i}[\theta]$ matrices established in Eqs.~(\ref{eq:overlapSVD}) and~(\ref{eq:FfromSVD}) to construct a refined variant of $\tilde{W}[\theta]$ in Eqs.~(\ref{eq:tempPathOrderedW}) and~(\ref{eq:overlapPrimeDefinition}):
\begin{equation}
\tilde{W}'[\theta] = \prod_{i=0}^{N-1}F_{i}[\theta].
\label{eq:refinedWilsonMatrix}
\end{equation}

The eigenvalues of the refined $N_{\text{orb}}\times N_{\text{orb}}$ Wilson loop matrix $\tilde{W}'[\theta]$ in Eq.~(\ref{eq:refinedWilsonMatrix}) consist of $N_{\text{orb}} - N_{\text{occ}}$ zeroes and $N_{\text{occ}}$ gauge-invariant complex phases with magnitude $1$.
To focus on the nonzero eigenvalues of $\tilde{W}'[\theta]$, we next restrict consideration to the subspace of $\tilde{W}'[\theta]$ within the image of the occupied-state projectors $P_{i}[\theta]$ in Eqs.~(\ref{eq:OccupiedProjectorWilson}) and~(\ref{eq:tempPathOrderedW}).
This is functionally accomplished by constructing an $N_{\text{occ}}\times N_{\text{occ}}$ unitary Wilson loop matrix $W[\theta]$ whose elements are given by the inner products of the occupied eigenstates $|u^{j}_{{\bf p}_{0}[\theta]}\rangle$ at each Wilson loop base point ${\bf p}_{0}[\theta]$ with respect to $\tilde{W}'[\theta]$ in Eq.~(\ref{eq:refinedWilsonMatrix}): 
\begin{equation}
\bigg[W[\theta]\bigg]^{jk} = \langle u^{j}_{{\bf p}_{0}[\theta]}|\tilde{W}'[\theta]|u^{k}_{{\bf p}_{0}[\theta]}\rangle.
\label{eq:WilsonElements}
\end{equation}

Finally, the eigenvalues of the unitary Wilson loop matrix $W[\theta]$ take the form of $N_{\text{occ}}$ gauge-invariant complex phases $\exp(i\gamma_{n}[\theta])$ that are smooth and continuous functions of $\theta$ for energetically isolated nodal degeneracies at ${\bf p}={\bf 0}$. 
The phase angles $\gamma_{n}[\theta]$ are well-defined modulo $2\pi$ and are termed the non-Abelian Berry phases.
The non-Abelian Berry phases $\gamma_{n}[\theta]$ are then related to the total [Abelian] Berry phase $\gamma[\theta]$ by:
\begin{equation}
\det\left(W[\theta]\right) = \prod_{n=1}^{N_{\text{occ}}} e^{i\gamma_{n}[\theta]} = e^{i\gamma[\theta]},
\label{eq:nonAbelianBerry1}
\end{equation}
such that:
\begin{equation}
\gamma[\theta] = \left(\sum_{n=1}^{N_{\text{occ}}}\gamma_{n}[\theta]\right)\text{ mod }2\pi.
\label{eq:nonAbelianBerry2}
\end{equation}
The winding number of the total Berry phase $\gamma[\theta]$ [and hence the set of non-Abelian Berry phases $\{\gamma_{n}[\theta]\}$ via Eqs.~(\ref{eq:nonAbelianBerry1}) and~(\ref{eq:nonAbelianBerry2})] on a sphere $S$ surrounding a nodal point at ${\bf p}={\bf 0}$ indicates the quantized topological chiral charge $C\in\mathbb{Z}$ of the nodal degeneracy, as it corresponds to the surface integral of the Berry curvature flux:
\begin{equation}
C = \frac{1}{2\pi}\oint_{S}\Tr[{\bf F}({\bf p})]\cdot d{\bf A},
\label{eq:amorphousWilsonSphere}
\end{equation}
where Eq.~(\ref{eq:amorphousWilsonSphere}) represents the non-crystalline, pseudo-momentum-space generalization of Eq.~(\ref{eq:crystallineWilsonSphere}).
We hence in this work identify ${\bf p}={\bf 0}$ nodal spectral features with $C\neq 0$ in disordered tight-binding models as condensed-matter realizations of disordered 3D chiral fermions.

\clearpage

\section{Structurally Chiral Lattice Models with $\Gamma$-Point Chiral Fermions with and without Disorder}
\label{app:models}

In this section, we will introduce and analyze lattice models of structurally chiral crystals [see Appendix~\ref{app:symDefs}] that exhibit symmetry-enforced, topologically chiral nodal degeneracies at the $\Gamma$ point [${\bf k}={\bf 0}$] in the absence of disorder, and that \emph{continue} to exhibit topological chiral fermions at ${\bf p}={\bf 0}$ in the amorphous regime.
We will first study each topological semimetal model in the crystalline limit, specifically computing the bulk band structure and Fermi-arc surface states and employing symmetry group theory to analytically characterize the chirality-controlled model topology and structurally achiral critical phases.
We will then analyze each model in the presence of strong structural disorder, using a variety of disorder implementation schemes [detailed in Appendix~\ref{app:lattices}] to approximate the amorphous regime.
In our analysis below, we will employ a range of numerical and analytic methods to demonstrate that each model in the amorphous regime continues to exhibit topological chiral fermions for disorder realizations that carry a net [average] structural chirality [see Fig.~\ref{fig:average_sym} and the surrounding text]. 
For each disordered model in this section, we will specifically below compute the bulk and surface spectrum, demonstrate the role of average chirality in controlling the model topology, and analyze the low-energy theory near ${\bf p}={\bf 0}$ from the perspective of non-crystalline [average] symmetry group theory [see Appendices~\ref{app:pseudoK} and~\ref{app:corepAmorphous}].
Where admitted by the internal model degrees of freedom [see the text following Eq.~(\ref{eq:paramsC2amo})], we will also compute and analyze the angular momentum textures of each model under  physically motivated conditions.

In Appendix~\ref{app:Kramers}, we will begin by studying the crystalline limit [Appendix~\ref{app:PristineKramers}] and amorphous regime [Appendix~\ref{app:amorphousKramers}] of the Kramers-Weyl tight-binding model previously introduced in Refs.~\cite{KramersWeyl,OtherKramersWeyl}. 
We will then in Appendix~\ref{sec:Charge2} introduce a tight-binding model with a $\Gamma$-point double-Weyl fermion~\cite{ZahidMultiWeylSrSi2,StepanMultiWeyl,AndreiMultiWeyl,XiDaiMultiWeyl}, which we will then analyze in the crystalline limit [Appendix~\ref{app:PristineQuadratic}] and the amorphous regime [Appendix~\ref{app:amorphousCharge2}].
Lastly, in Appendix~\ref{sec:Multifold}, we will introduce a tight-binding model that exhibits a symmetry-enforced, threefold-degenerate spin-1 chiral [multifold] fermion~\cite{ManesNewFermion,chang2017large,tang2017CoSi,KramersWeyl,DoubleWeylPhonon,CoSiObserveJapan,CoSiObserveHasan,CoSiObserveChina,AlPtObserve,PdGaObserve,PtGaObserve,AltlandSpin1Light,DingARPESReversal,ZahidLadderMultigap,Sanchez2023} at the $\Gamma$ point. 
Importantly, unlike in earlier multifold fermion models~\cite{chang2017large}, the multifold model introduced in Appendix~\ref{sec:Multifold} is constrained by symmorphic crystal symmetries and only contains on-site [internal orbital] degrees of freedom, allowing us to more straightforwardly disorder the model while enforcing a sense of local [average] chirality [see Ref.~\cite{Franca2024} and Appendix~\ref{app:DiffTypesDisorder}].
Finally, we will conclude this section by analyzing the symmorphic multifold model introduced in this work both in the crystalline limit [Appendix~\ref{app:PristineMulifold}] and in the amorphous regime [Appendix~\ref{app:amorphousMultifold}].

\subsection{Kramers-Weyl Models}
\label{app:Kramers}

\subsubsection{Symmetry and Chirality in the Crystalline Kramers-Weyl Model}
\label{app:PristineKramers}

In this section, we will analyze the bulk electronic structure, symmetry group theory, topology, surface states, and critical phases of the structurally chiral [see Appendix~\ref{app:symDefs}] crystalline Kramers-Weyl [KW] tight-binding model~\cite{KramersWeyl,OtherKramersWeyl,AndreiTalk,Volovik1985scKW}.
To begin, the KW tight-binding model introduced in Refs.~\cite{KramersWeyl,OtherKramersWeyl} can be written in position space as:
\begin{equation}
    \mathcal{H}_{\mathrm{KW}} = \sum_{\langle \alpha\beta \rangle}\sum_{l,l'\in \left\{1,2\right\}}c_{\alpha,l}^{\dagger}\langle {\bf r}_\alpha,l | \mathcal{H} | {\bf r}_\beta,l'\rangle c_{\beta,l'}^{\phantom{}},
\label{eq:AmoKW}
\end{equation}
where the operator $c^{\dagger}_{\alpha,l}$ creates a particle at the site $\alpha$ with an internal spin-1/2 degree of freedom indexed by $l$, and the tight-binding Hilbert space consists of a Kramers pair of spinful $s$ orbitals [indexed by $l,l'=1,2$] at each lattice site [indexed by $\alpha$ and $\beta$].
Following the notation of Ref.~\cite{PhysRevLett.95.226801}, the $\langle$ and $\rangle$ symbols in the sum over sites in Eq.~(\ref{eq:AmoKW}) indicate that pairs of sites $\alpha,\beta$ are only included within the summation if they lie within a specified distance, denoted $R_0$, of each other. 
In the crystalline limit discussed in this section, $\left\langle\alpha\beta\right\rangle$ in Eq.~(\ref{eq:AmoKW}) will reduce to nearest-neighbor lattice sites.
In Eq.~(\ref{eq:AmoKW}), the tight-binding basis states $|{\bf r}_\alpha, l\rangle$ satisfy the orthogonality relation $\langle {\bf r}_\alpha, l|{\bf r}_\beta, l'\rangle = \delta_{\alpha\beta}\delta_{ll'}$, where $l$ and $l'$ belong to the set $\left\{1, 2\right\}$, representing the internal spin degrees of freedom, and $\alpha$ and $\beta$ denote the lattice sites [see Eq.~(\ref{eq:TBbasisStates}) and the surrounding text].

We define the elements of the intersite hopping matrix $T_{\alpha\beta}$ for $\mathcal{H}_{\mathrm{KW}}$ in Eq.~(\ref{eq:AmoKW}) through inner products in the following manner:
\begin{equation}
         T_{\alpha\beta} \underset{\text{def}}{\equiv}\begin{pmatrix}
            \langle {\bf r}_\alpha,1 | \mathcal{H} | {\bf r}_\beta,1\rangle & \langle {\bf r}_\alpha,1 | \mathcal{H} | {\bf r}_\beta,2\rangle\\
            \langle {\bf r}_\alpha,2 | \mathcal{H} | {\bf r}_\beta,1\rangle & 
            \langle {\bf r}_\alpha,2 | \mathcal{H} | {\bf r}_\beta,2\rangle
          \end{pmatrix}=\frac{1}{2}f(|\mathbf{d}_{\alpha\beta}|)\left(i\mathbf{d}_{\alpha\beta}^\mathsf{T}\begin{pmatrix}
              v_x&0&0\\
              0&v_y&0\\
              0&0&v_z
          \end{pmatrix}\bm{\sigma} + \left[\mathbf{d}_{\alpha\beta}^\mathsf{T}\begin{pmatrix}
              t_x&0&0\\
              0&t_y&0\\
              0&0&t_z
          \end{pmatrix}\mathbf{d}_{\alpha\beta}\right]\sigma^0\right),
\label{eq:KWTmatrix}
\end{equation}
where the parameters $t_{i}$ indicate the strength of nearest-neighbor spinless [$s$-orbital-like] hopping, and the $v_{i}$ terms correspond to nearest-neighbor Dresselhaus spin-orbit coupling [SOC]~\cite{DresselhausSOC}.  
In Eq.~(\ref{eq:KWTmatrix}), the intersite separation vector ${\bf d}_{\alpha\beta}$ is given by:
\begin{equation}
    \mathbf{d}_{\alpha\beta}
= \begin{pmatrix}
x_{\alpha}-x_{\beta}\\y_{\alpha}-y_{\beta}\\z_{\alpha}-z_{\beta}
\end{pmatrix},
\label{eq:dVectorDef}
\end{equation} 
and the overall radial hopping function strength $f(|\mathbf{d}_{\alpha\beta}|)$ is given by:
 \begin{equation}
      f(|\mathbf{d}_{\alpha\beta}|) = \Theta(R_0-|\mathbf{d}_{\alpha\beta}|)\exp(a-|\mathbf{d}_{\alpha\beta}|),
\label{eq:KWHeaviside}
 \end{equation}
where $\Theta$ denotes the Heaviside function, $|\mathbf{d}_{\alpha\beta}|$ is the distance between the sites $\alpha$ and $\beta$, $a$ is a length that becomes the lattice spacing in each of the $x,y,z$ directions when the sites are regularly spaced [crystalline], and $R_0$ is the maximum distance allowed for hoppings.
In Eq.~(\ref{eq:KWTmatrix}), ${\bm \sigma}$ is a vector of $2\times 2$ Pauli matrices:
\begin{equation}
{\bm \sigma} = \begin{pmatrix}
\sigma^{x} \\ \sigma^{y}\\\sigma^{z}\end{pmatrix},
\end{equation}
where each Pauli matrix $\sigma^{x,y,z}$ acts on the spin-1/2 internal degree of freedom at each lattice site, such that $\sigma^{0}$ is correspondingly the $2\times 2$ identity matrix.
For the remainder of this section,  we then set the parameters of the function $f(|\mathbf{d}_{\alpha\beta}|)$ to be:
\begin{equation}
a=1, R_0 = 1.3a = 1.3,
\end{equation}
such that $R_0= 1.3$ to ensure that only nearest-neighbor hoppings appear in Eq.~(\ref{eq:KWHeaviside}) in the pristine limit.

\begin{figure}[t]
\centering
\includegraphics[width=\linewidth]{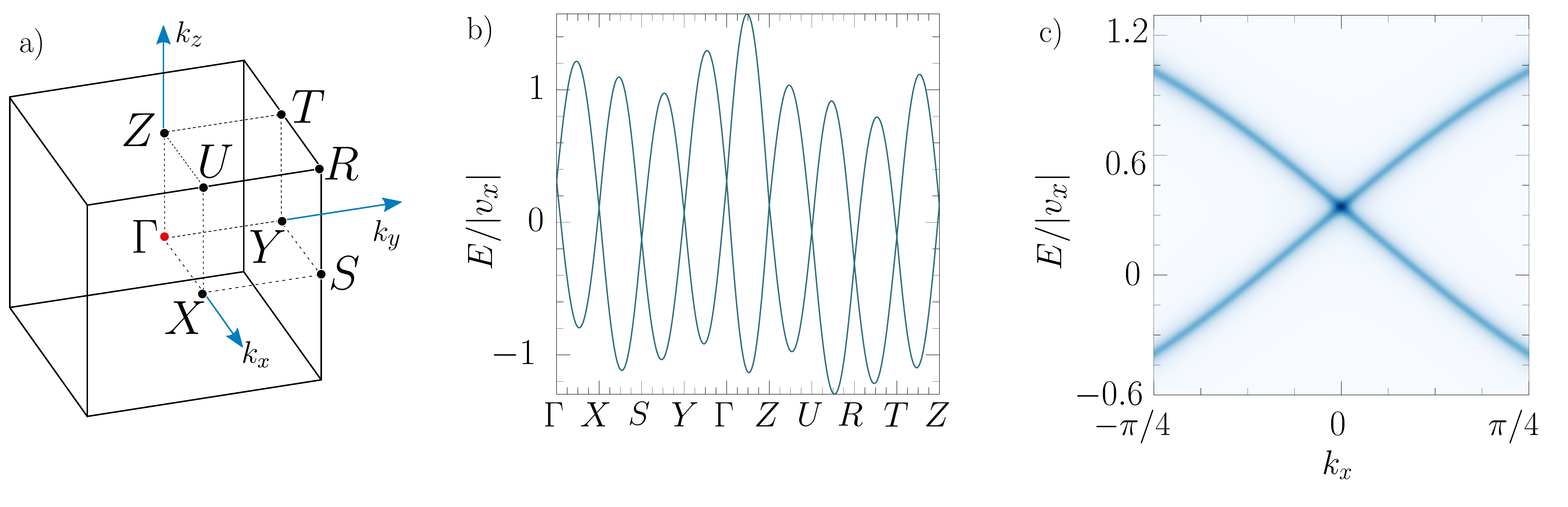}
\caption{Bulk spectrum of the crystalline Kramers-Weyl model. 
(a) The Brillouin zone [BZ] of chiral SG 16 ($P222$), the SG of the Kramers-Weyl [KW] model [see Eqs.~(\ref{eq:appHKW}) and~(\ref{eq:pristineSyms}) and Refs.~\cite{KramersWeyl,OtherKramersWeyl}].
(b) Bulk band structure of the crystalline KW model using the parameters in Eq.~(\ref{eq:appTBparams}).
(c) The spectral function $A(E,\mathbf{p})$ [see Eq.~(\ref{eq:SpecFunc}) and the surrounding text] of the KW chiral fermion [Eq.~(\ref{eq:appHKP})] at the $\Gamma$ point in (b).}
\label{fig:KW_figBulk}
\end{figure}

\paragraph*{\bf Symmetries and Group Representations} -- $\ $ When defined on a regular orthorhombic lattice, the tight-binding model in Eqs.~(\ref{eq:AmoKW}) and~(\ref{eq:KWTmatrix}) takes the form of a $2\times 2$ Bloch Hamiltonian in ${\bf k}$-space:
\begin{equation}
    \mathcal{H}({\bf k})= \sum_{i=x,y,z}t_{i}\cos(k_i)\sigma^0 + \sum_{i=x,y,z}v_{i}\sin(k_i)\sigma^i,
\label{eq:appHKW}
\end{equation}
As emphasized in Ref.~\cite{KramersWeyl}, when the $t_{i}$ or $v_{i}$ parameters are chosen to be different in the $i=x,y,z$ directions, and when $v_{i}\neq 0$ for all $i$ [\emph{i.e.} SOC is nonvanishing in all directions], $\mathcal{H}({\bf k})$ respects the symmetries of nonmagnetic [Shubnikov~\cite{BigBook,MagneticBook,MTQC,MTQCmaterials}] \emph{chiral} [see Appendix~\ref{app:symDefs}] SG 16 ($P222$), which is generated by twofold rotations around the $i=x,y,z$ axes $C_{2i}$, $\mathcal{T}
$ symmetry, and 3D orthorhombic lattice translations.  Aside from translations, which act on ${\bf k}$-space Hamiltonians as phases, the symmetries of SG 16 ($P222$) can be represented through their action on Eq.~(\ref{eq:appHKW}):
\begin{eqnarray}
C_{2i}\mathcal{H}({\bf k})C_{2i}^{-1} &=& \sigma^{i}\mathcal{H}(C^{-1}_{2i}{\bf k})\sigma^{i}, \nonumber \\
\mathcal{T}\mathcal{H}({\bf k})\mathcal{T}^{-1} &=& \sigma^{y}\mathcal{H}^{*}(-{\bf k})\sigma^{y}.
\label{eq:pristineSyms}
\end{eqnarray}
At the lattice scale, the structural chirality [handedness] $C_{\mathcal{H}}$ of the crystalline KW model [Eqs.~(\ref{eq:AmoKW}),~(\ref{eq:KWTmatrix}), and~(\ref{eq:appHKW})] is specifically encoded in the signs of the $v_{i}$ parameters~\cite{KramersWeyl}:
\begin{equation}
C_{\mathcal{H}} = \text{sgn}\left(\prod_{i=x,y,z} v_{i}\right),
\label{eq:appStructuralChirality}
\end{equation}
where $C_{\mathcal{H}}=1$ [$C_{\mathcal{H}}=-1$] corresponds to the right-handed or $R$ [left-handed or $L$] enantiomer of the KW tight-binding model.
In a crystalline lattice, $C_{\mathcal{H}}$ is a global geometric property, because each bond in the $i=x,y,z$ direction carries the same sign of $v_{i}$.
However, by allowing the signs of the $v_{i}$ hopping parameters in Eq.~(\ref{eq:KWTmatrix}) to vary for bonds originating from each site $\alpha$, we may also promote $C_{\mathcal{H}}$ in Eq.~(\ref{eq:appStructuralChirality}) to a spatially varying \emph{local chirality} $\chi_{\alpha}$ [see Refs.~\cite{LocalChiralityMoleculeReviewNatChem,LocalChiralityLiquidCrystals,KamienLubenskyChiralParameter,LocalChiralityDomainLiquidCrystal,LocalChiralityTransfer,KamienChiralLiquidCrystal,LocalChiralityQuasicrystalVirus,lindell1994electromagneticBook,LocalChiralityVillain,LocalChiralityWenZee,LocalChiralityBaskaran,LocalChiralitySpinFrame} and the text surrounding Eq.~(\ref{eq:temp2siteFrameBreakdown})].
Below, in the text following Eqs.~(\ref{eq:KWnumericalModelDiscreteChi}) and~(\ref{eq:amorphousKWTmatrixFinalChirality}), we will shortly implement a non-crystalline generalization of the KW model in which the local chirality $\chi_{\alpha}$ represents a parameter that can be ordered or disordered independent from the lattice structural order, and can specifically be tuned to control non-crystalline spin textures and topology.

In Fig.~\ref{fig:KW_figBulk}(a,b), we plot the band structure of $\mathcal{H}({\bf k})$ [Eq.~(\ref{eq:appHKW})] with the choice of parameters:
\begin{equation}
t_{x} = 0.1,\ t_{y} = 0.12,\ t_{z} = 0.09,\ v_{x} = \pm 1,\ v_{y} = 1.1,\ v_{z} = 1.35,
\label{eq:appTBparams}
\end{equation}
where the $+$ [$-$] sign for $v_{x}$ is used to model the $R$ [$L$] enantiomer [Eq.~(\ref{eq:appStructuralChirality})].
When $v_{i}\neq 0$ for all $i=x,y,z$, $\mathcal{H}({\bf k})$ at half filling realizes a topological semimetal state with eight topologically chirally charged nodal [Weyl~\cite{AshvinWeyl,VolovikWeylMonopole,VolovikChiralAnomaly,HaldaneOriginalWeyl,MurakamiWeyl,BurkovBalents,AndreiWeyl,HasanWeylDFT,Armitage2018,SuyangWeyl,LvWeylExp,YulinWeylExp,AliWeylQPI,AlexeyType2,ZJType2,BinghaiClaudiaWeylReview,ZahidNatRevMatWeyl,CDWWeyl,IlyaIdealMagneticWeyl}] degeneracies [Fig.~\ref{fig:KW_figBulk}(c)] in each Brillouin Zone [BZ].  The topological [semi]metallic state of $\mathcal{H}({\bf k})$ at half filling represents an example of filling-enforced gaplessness, in which the combination of $\mathcal{T}$ and lattice translation symmetry requires that the system be gapless [or exhibit symmetry-preserving topological order] at odd electronic fillings per unit cell through an SG-symmetric generalization of the Lieb-Schultz-Mattis theorem~\cite{LSM,WPVZ,WiederLayers,WPVZfollowUp,DDP,chang2017large,YoungMagnetic,WatanabeMagneticWPVZ,ma2023averageSPT2,BradlynDDPqsl}.  
Correspondingly, the valence and conduction bands of $\mathcal{H}({\bf k})$ together transform in the two-dimensional, double-valued elementary band corep $(\bar{E})_{1a}\uparrow P222$, which represents the smallest, minimally connected elementary band corep in SG 16 $(P222)$
~\cite{Bradlyn2017,MTQC,Bandrep1}.
As shown in Ref.~\cite{KramersWeyl}, the eight Weyl points are specifically pinned to the $\mathcal{T}$-invariant ${\bf k}$ [TRIM] points and -- unlike conventional Weyl fermions at generic crystal momenta -- can be labeled by two-dimensional small coreps of the TRIM-point little groups, and are hence termed KW fermions [see Appendix~\ref{app:corepDefs}].  
For future calculations in amorphous systems [see Appendices~\ref{app:corepAmorphous} and~\ref{app:amorphousKramers}], we will find it useful to note that the little group $G_{\Gamma}$ at the $\Gamma$ point is isomorphic to SG 16 ($P222$) itself [see the text preceding Eq.~(\ref{eq:EquivKPoints})].
The KW fermion at $\Gamma$ then specifically transforms in the two-dimensional, double-valued small corep $\bar{\Gamma}_{5}$ of $G_{\Gamma}$ in the labeling convention of the~\href{https://www.cryst.ehu.es/cgi-bin/cryst/programs/corepresentations.pl}{Corepresentations} tool on the Bilbao Crystallographic Server [in which SG 16 is alternatively denoted as Shubnikov SG 16.2 ($P2221'$)]~\cite{MTQC,MTQCmaterials}.

\paragraph*{\bf Topology} -- $\ $ We next detail the distribution of topological chiral charges $C$ across the eight KW points, where $C$ is equal to the integral of the Berry curvature flux on a sphere surrounding the Weyl point~\cite{AshvinWeyl}.  At each of the TRIM points, the low-energy ${\bf k}\cdot {\bf p}$ Hamiltonian $\mathcal{H}_{{\bf k}_{\mathcal{T}}}({\bf q})$ can be obtained by expanding Eq.~(\ref{eq:appHKW}) to linear order about the TRIM point ${\bf k}_{\mathcal{T}}$:
\begin{equation}
\mathcal{H}_{{\bf k}_{\mathcal{T}}}({\bf q}) = \sum_{i=x,y,z} e^{i\left({\bf k}_{\mathcal{T}}\cdot \hat{\bf r}_{i}\right)}v_{i}q_{i}\sigma^{i}= {\bf d}_{{\bf k}_{\mathcal{T}}}({\bf q})\cdot{\bm \sigma},
\label{eq:appHKP}
\end{equation}
where ${\bf q} \approx {\bf k} - {\bf k}_{\mathcal{T}}$, $\hat{\bf r}_{i}$ is a unit vector in the $i=x,y,z$ direction, and:
\begin{equation}    
{\bf d}_{\mathbf{k}_{\mathcal{T}}}({\bf q})= 
\begin{pmatrix}
v_x q_x e^{ik_{x,\mathcal{T}}}\\
v_y q_y e^{ik_{y,\mathcal{T}}}\\
v_z q_z e^{ik_{z,\mathcal{T}}}
\end{pmatrix}.
\label{eq:tempKWdvector}
\end{equation}
We note that in Eqs.~(\ref{eq:appHKP}) and~(\ref{eq:tempKWdvector}), we have suppressed factors of the $2\times 2$ identity matrix $\sigma^{0}$, because they correspond to effective chemical potentials on the TRIM-point KW fermions and do not affect the calculation of the topological chiral charge, which only depends on the band structure and is independent of the Fermi level.

\begin{figure}[t]
\centering
\includegraphics[width=\linewidth]{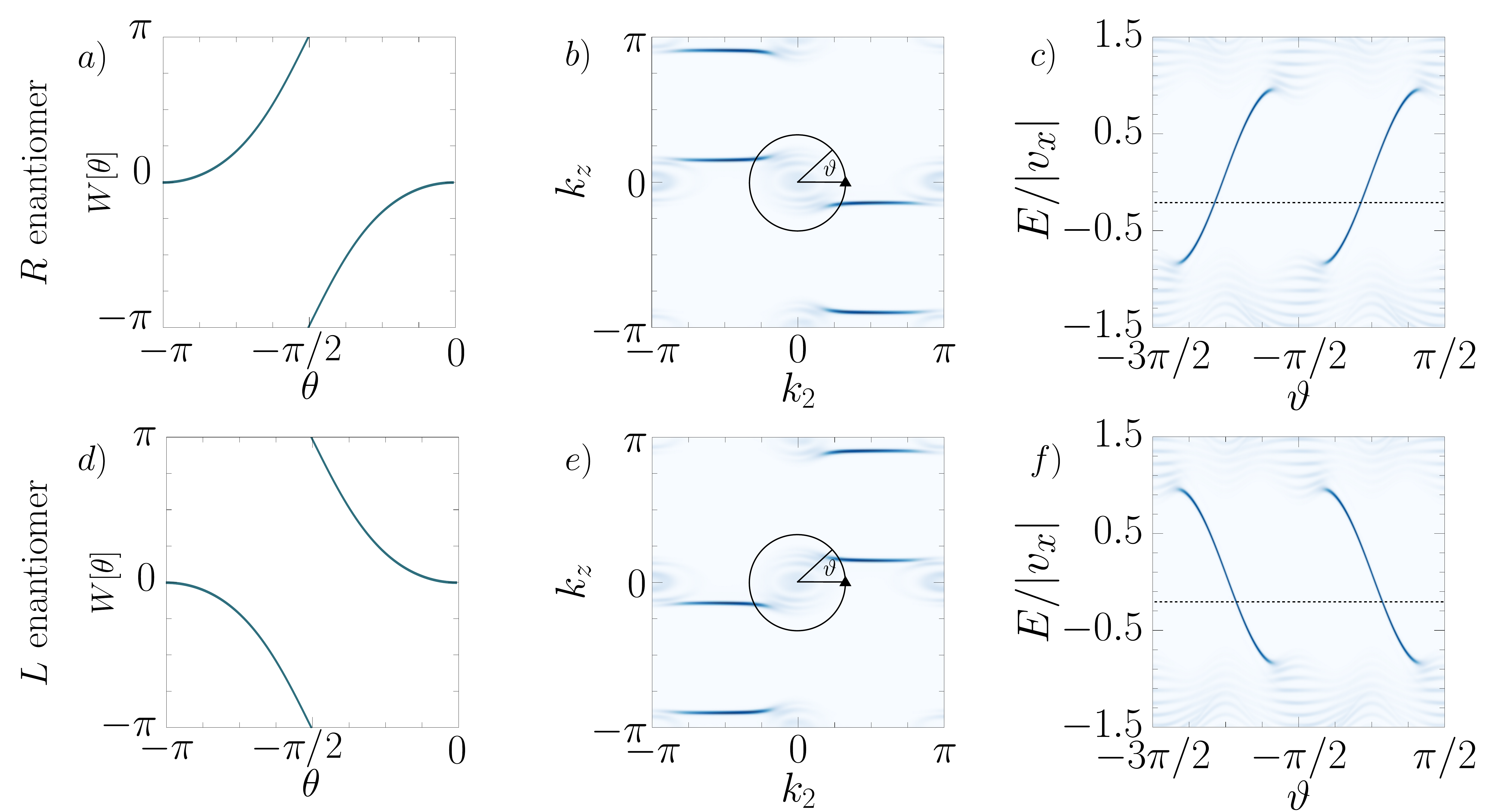}
\caption{Chirality and topology of the crystalline Kramers-Weyl model.
Panels (a-c)  and (d-f) respectively show topological bulk and $(110)$-surface spectral properties of the right-handed [$R$] and left-handed [$L$] enantiomers of the Kramers-Weyl [KW] tight-binding model [Eqs.~(\ref{eq:AmoKW}),~(\ref{eq:KWTmatrix}), and~(\ref{eq:appHKW})] obtained using the parameters in Eq.~(\ref{eq:appTBparams}).
In (a) and (d), we show the Wilson loop spectrum [Appendix~\ref{sec:WilsonBerry}] of the lowest band, computed respectively for the $R$ and $L$ model enantiomer on a sphere surrounding the KW nodal degeneracy at the $\Gamma$ point [see Fig.~\ref{fig:KW_figBulk}].
The Wilson loop spectra in (a) and (d) respectively wind once in the positive and negative directions, indicating that the KW fermion at the $\Gamma$ point carries a chiral charge of $C=1$ for the $R$ enantiomer and $C=-1$ for the $L$ enantiomer, consistent with $C$ at each TRIM point being directly determined by the lattice-scale structural chirality $C_{\mathcal{H}}$ through Eq.~(\ref{eq:appChiralCharge})~\cite{KramersWeyl}.
(b) and (e) respectively show the $(110)$-surface Fermi arcs for the $R$ and $L$ KW model enantiomers computed from surface Green's functions at $E/|v_x|=-0.4$ and plotted as functions of $k_{2} = (1/\sqrt{2})(k_{x}-k_{y})$ and $k_{z}$. 
In (c) and (f), we then respectively plot the $(110)$-surface spectral functions of the $R$ and $L$ model enantiomers computed as functions of energy on counterclockwise circular paths surrounding $k_{2}=k_{z}=0$ in (b,e). 
The Fermi pockets at $k_{2}=k_{z}=0$ in (b) and (e) contain the superposed projections of the KW fermions at $\Gamma$ and $S$ in Fig.~\ref{fig:KW_figBulk}(b), which through Eq.~(\ref{eq:appHKP}) carry the same chiral charges $C=1$ for the $R$ enantiomer and $C=-1$ for the $L$ enantiomer.
This gives rise to a net surface chiral charge projection of $C=2$ for (b,c) the $R$ enantiomer and $C=-2$ for (e,f) the $L$ enantiomer.
The surface Fermi arcs in (c) and (f) respectively cross the dashed horizontal line in each panel two times with positive and negative slopes, confirming that the bulk Fermi pocket projections in (b) and (e) at $k_{2}=k_{z}=0$ respectively carry net chiral charges of $C=2$ for the $R$ enantiomer and $C=-2$ for the $L$ enantiomer. 
The data in panels (b,d,e,f) were obtained from the tight-binding model in Eqs.~(\ref{eq:AmoKW}),~(\ref{eq:KWTmatrix}) and~(\ref{eq:appHKW}) placed on a regular lattice that was infinite in the $\hat{z}$-direction and finite with 150 sites and periodic boundary conditions in the $\hat{x}-\hat{y}$-direction [reciprocal to $k_{2}$] and finite with 20 sites and open boundary conditions in the $\hat{x}+\hat{y}$- [$(110)$-] direction.}
\label{fig:KW_figWilson}
\end{figure}

Eq.~(\ref{eq:appHKP}) takes the canonical form of a $|C|=1$ Weyl point~\cite{AshvinWeyl}, and hence the specific value of the chiral charge $C_{{\bf k}_{\mathcal{T}}}$ of each KW point can straightforwardly be analytically computed as follows.  
First, we term $\mathbf{d}_r$ a \emph{regular point} if the Jacobian $\left|\partial d_{{\bf k}_{\mathcal{T}}}^{i} /\partial q_j\right|$ is nonzero, where ${\bf d}_{{\bf k}_{\mathcal{T}}}$ is defined in Eq.~(\ref{eq:appHKP}). We then define the degree of a continuous mapping between two compact oriented manifolds of the same dimension as a number that represents the number of times that the domain manifold wraps around the range manifold under the mapping. The degree of such a map is termed a \emph{homotopy invariant}~\cite{cho2006topological} and can be used to characterize the topology of the system.  
In the case of Eq.~(\ref{eq:appHKP}), the map is from ${\bf q} = (q_{x},q_{y},q_{z})$, which is a vector in the real plane $\mathbb{R}^{3} \backslash \{(0,0,0)\}$, to the vector ${\bf d}({\bf q}) = {\bf d}_{\mathbf{k}_{\mathcal{T}}}({\bf q})$, which lies on $\mathds{R}^{3}\backslash \{(0,0,0)\}$.
We say that $\mathbf{q} = \left(q_x,q_y,q_z\right)$ is a preimage of $\mathbf{d}_r$ if $\mathbf{d}(q_x,q_y,q_z) = \mathbf{d}_r$.
A regular point can have $N$ preimages, denoted as $\mathbf{q}^n$, such that $\mathbf{d}_{\mathbf{k}_{\mathcal{T}}}(\mathbf{q}^n) = \mathbf{d}_r$ with $n = 1,\dots,N$. The chiral charge at ${\bf q} = {\bf 0}$ is then given by:
\begin{equation}
     C_{{\bf k}_{\mathcal{T}}} = \sum_{n=1}^{N}\mathrm{sgn}\left(\left|\frac{\partial d_{\mathbf{k}_{\mathcal{T}}}^{i}}{\partial q_j}\right|_{\mathbf{q} = \mathbf{q }^n}\right).
\label{eq:KWchiralJacobianRelation}
\end{equation}
For the KW model linearly expanded at each TRIM point [Eq.~(\ref{eq:appHKP})], the Jacobian reads as:
\begin{equation}
    \left|\frac{\partial d_{\mathbf{k}_{\mathcal{T}}}^{i}}{\partial q_j}\right| = v_x v_y v_z e^{ik^{x}_{\mathcal{T}}}e^{ik^{y}_{\mathcal{T}}}e^{ik^{z}_{\mathcal{T}}} = \prod_{i=x,y,z} e^{i\left({\bf k}_{\mathcal{T}}\cdot \hat{\bf r}_{i}\right)}v_{i}.
\end{equation}
Brouwer's lemma~\cite{MilnorTopology1965} guarantees that the degree of the map is the same for all regular points. For the specific choice $\mathbf{d}_r = (v_x e^{ik_{\mathcal{T}}^x},0,0)$, this yields a single preimage $q^1 = (1,0,0)$. Consequently, the chiral charge of the KW fermion at each TRIM point ${\bf k}_{\mathcal{T}}$ in the KW model can be expressed as:
\begin{equation}
    C_{{\bf k}_{\mathcal{T}}} = \text{sgn}\left(\prod_{i=x,y,z} e^{i\left({\bf k}_{\mathcal{T}}\cdot \hat{\bf r}_{i}\right)}v_{i}\right).
\label{eq:imageProofKWCharge}
\end{equation}

Most importantly, Eq.~(\ref{eq:imageProofKWCharge}) can be rewritten as follows:
\begin{equation}
C_{{\bf k}_{\mathcal{T}}} = \text{sgn}\left(\prod_{i=x,y,z} e^{i\left({\bf k}_{\mathcal{T}}\cdot \hat{\bf r}_{i}\right)}v_{i}\right) = \left(\prod_{i=x,y,z} e^{i\left({\bf k}_{\mathcal{T}}\cdot \hat{\bf r}_{i}\right)}\right)\text{sgn}\left(\prod_{i=x,y,z} v_{i}\right) = \left(\prod_{i=x,y,z} e^{i\left({\bf k}_{\mathcal{T}}\cdot \hat{\bf r}_{i}\right)}\right)C_{\mathcal{H}}.
\label{eq:appChiralCharge}
\end{equation}
Eq.~(\ref{eq:appChiralCharge}) represents the simplest example of the statement, theoretically established in Ref.~\cite{KramersWeyl} and subsequently experimentally demonstrated in Refs.~\cite{AlPtObserve,PdGaObserve,DingARPESReversal,SessiPdGaQPIReversal}, that in structurally chiral metals, the low-energy topological chirality of nodal degeneracies [here the chiral charges $C_{{\bf k}_{\mathcal{T}}}$ of the eight KW fermions] is directly inherited from and controlled by the lattice-scale geometry [here the structural chirality $C_{\mathcal{H}}$ from Eq.~(\ref{eq:appStructuralChirality}) through the signs of the $v_{i}$ parameters in Eqs.~(\ref{eq:appHKW}) and~(\ref{eq:appHKP})].  
As shown below in Appendices~\ref{app:amorphousKramers},~\ref{app:amorphousCharge2}, and~\ref{app:amorphousMultifold}, we find in this work that the relationship between structural and topological chirality established in Ref.~\cite{KramersWeyl} can remarkably survive in highly disordered -- and even amorphous -- metals.

To numerically confirm Eq.~(\ref{eq:appChiralCharge}), we have computed in Fig.~\ref{fig:KW_figWilson} the $\Gamma$-point sphere Wilson loop spectrum [Appendix~\ref{sec:WilsonBerry} and Refs.~\cite{Fidkowski2011,AndreiXiZ2,ArisInversion,Cohomological,HourglassInsulator,DiracInsulator,Z2Pack,BarryFragile,AdrienFragile,HOTIBernevig,HingeSM,WiederAxion,KoreanFragile,ZhidaBLG,TMDHOTI,KooiPartialNestedBerry,PartialAxionHOTINumerics,GunnarSpinFragileWilson,BinghaiOscillationWilsonLoop,Wieder22}] and $(110)$- [($\hat{x} + \hat{y}$)-normal] surface Fermi arcs of the $R$ [$C_{\mathcal{H}}=1$] and $L$ [$C_{\mathcal{H}}=-1$] enantiomers of the KW tight-binding model [Eqs.~(\ref{eq:AmoKW}),~(\ref{eq:KWTmatrix}), and~(\ref{eq:appHKW})].  
First, in Fig.~\ref{fig:KW_figWilson}(a) [Fig.~\ref{fig:KW_figWilson}(d)] we show the Wilson loop spectrum of the lower band of the KW model computed on a sphere surrounding the nodal degeneracy at the $\Gamma$ point [${\bf k}={\bf 0}$] in Fig.~\ref{fig:KW_figBulk}(b) for the $R$ [$L$] model enantiomer.
The Wilson loop spectra in Fig.~\ref{fig:KW_figWilson}(a,d) respectively wind once in the positive direction for the $R$ enantiomer and once in the negative direction for the $L$ enantiomer, indicating that the nodal degeneracy at the ${\bf k}={\bf 0}$ TRIM point carries a chiral charge of $C=1$ for the $R$ enantiomer and $C=-1$ for the $L$ enantiomer, consistent with its designation as a KW chiral fermion whose chiral topology is tunable via the lattice-scale structural chirality~\cite{KramersWeyl}.

\paragraph*{\bf Surface Fermi Arcs} -- $\ $ To provide corroborating evidence for the bulk KW chiral charge distribution in Eq.~(\ref{eq:appChiralCharge}), we next compute the surface states of the KW tight-binding model.
In Fig.~\ref{fig:KW_figWilson}(b,c) and Fig.~\ref{fig:KW_figWilson}(e,f) we respectively plot the $(110)$-surface states of the $R$ and $L$ enantiomers of the KW model as functions of two surface momenta and energy, where $k_{1,2}$ in  Fig.~\ref{fig:KW_figWilson}(b,e) are defined as:
\begin{equation}
k_{1,2} = \frac{1}{\sqrt{2}}\left(k_{x}\pm k_{y}\right).
\label{appeq:surfacemomenta}
\end{equation}
In Fig.~\ref{fig:KW_figWilson}(b,c,e,f), the Fermi pocket at $k_{2}=k_{z}=0$ contains the projected bulk Fermi pockets of the nodal degeneracies at $\Gamma$ and $S$ in Fig.~\ref{fig:KW_figBulk}(b), which through Eq.~(\ref{eq:appHKP}) carry the same chiral charges $C=1$ for the $R$ enantiomer and $C=-1$ for the $L$ enantiomer.
We correspondingly observe a time-reversed pair of $(110)$-surface Fermi arcs emanating from the $k_{2}=k_{z}=0$ point in Fig.~\ref{fig:KW_figWilson}(b,e), with the arcs roughly exhibiting opposite constant-energy helicities for opposite enantiomers [\emph{i.e.} the arcs on the top and bottom of the projected bulk Fermi pockets at $k_{2}=k_{z}=0$ respectively bend to the left and right for the $R$ enantiomer and respectively bend to the right and left for the $L$ enantiomer].
Though the qualitative helicity of surface Fermi arcs in constant-energy spectral functions like Fig.~\ref{fig:KW_figWilson}(b,e) is \emph{not} a topological property, we note that changes in surface Fermi-arc helicity associated to structural-chirality-mediated changes in the topological chiral charges of bulk nodal degeneracies have similarly been observed in numerous previous experimental and first-principles investigations of structurally chiral topological semimetals~\cite{AlPtObserve,PdGaObserve,DingARPESReversal,SessiPdGaQPIReversal,InternalChiralTheory,InternalChiralExp}.  
Unlike the constant-energy surface Fermi-arc helicity, the winding of the surface Fermi arcs on a closed contour as a function of energy \emph{is} a bulk topological property~\cite{AshvinWeyl,VolovikWeylMonopole,VolovikChiralAnomaly,HaldaneOriginalWeyl,MurakamiWeyl,BurkovBalents,AndreiWeyl,HasanWeylDFT,Armitage2018,SuyangWeyl,LvWeylExp,YulinWeylExp,AliWeylQPI,AlexeyType2,ZJType2,BinghaiClaudiaWeylReview,ZahidNatRevMatWeyl,CDWWeyl,IlyaIdealMagneticWeyl}.
In Fig.~\ref{fig:KW_figWilson}(c,f), we therefore plot the $(110)$-surface Fermi arcs as functions of energy on the counterclockwise paths encircling $k_{2}=k_{z}=0$ in Fig.~\ref{fig:KW_figWilson}(b,e).
The surface Fermi arcs respectively cross the dashed horizontal line in Fig.~\ref{fig:KW_figWilson}(c,f) two times with positive and negative slopes [velocities], confirming that the bulk Fermi pocket projections at $k_{2}=k_{z}=0$ in Fig.~\ref{fig:KW_figWilson}(b,e) respectively carry net chiral charges of $C=2$ for the $R$ enantiomer and $C=-2$ for the $L$ enantiomer.

\paragraph*{\bf Chirality Transitions and Achiral Critical Phases} -- $\ $ Lastly, we note that both the lattice structural chirality $C_{\mathcal{H}}$ and the topological chiral charges $C_{{\bf k}_{\mathcal{T}}}$ of the KW fermions at each TRIM point become ill-defined if one or more of the $v_{i}$ parameters in the KW tight-binding model [Eqs.~(\ref{eq:AmoKW}),~(\ref{eq:KWTmatrix}), and~(\ref{eq:appHKW})] goes to zero.  In the specific case in which just one of the $v_{i}$ parameters is taken to vanish, pairs of KW points become connected by nodal lines that are protected by a combination of lattice and spin symmetries contained within a structurally \emph{achiral} ``spin SG''~\cite{BrinkmanSpinSpace,corticelli2022spin,liu2022spin,SpinSpaceChina1,SpinSpaceChina2,SpinSpaceChina3,SpinSpaceChina4,spinSpaceChenNoSOC}.  For example, if $v_{x}\rightarrow 0$ while $v_{y,z}$ remain nonzero, $\mathcal{H}({\bf k})$ gains an additional spinless (spin SG) mirror [twofold rotoinversion] symmetry $\tilde{M}_{x}$, which can be represented through its action on Eq.~(\ref{eq:appHKW}):
\begin{equation}
\tilde{M}_{x}\mathcal{H}(k_{x},k_{y},k_{z})\tilde{M}_{x}^{-1} = \mathcal{H}(-k_{x},k_{y},k_{z}).
\label{eq:appSpinSG}
\end{equation}
Because $\tilde{M}_{x}$ is a rotoinversion symmetry, $\mathcal{H}({\bf k})$ with $v_{x}=0$ no longer exhibits a well-defined structural chirality [see Appendix~\ref{app:symDefs}], consistent with its ill-defined TRIM-point topological chirality [Eq.~(\ref{eq:appChiralCharge})].  This can be seen by recognizing that $\tilde{M}_{x}$ in Eq.~(\ref{eq:appSpinSG}) converts the right-handed form of $\mathcal{H}({\bf k})$ [taken with $v_{i}\neq 0$ for all $i=x,y,z$] to its left-handed form and vice versa [Eq.~(\ref{eq:appStructuralChirality})], and hence also reverses the signs of the chiral charges of the eight bulk KW fermions [Eqs.~(\ref{eq:appHKP}) and~(\ref{eq:appChiralCharge})].

Eqs.~(\ref{eq:pristineSyms}) and~(\ref{eq:appSpinSG}) further indicate that in the limit of $v_{x}=0$, $\mathcal{H}({\bf k})$ also respects the spin SG antiunitary symmetry $\tilde{M}_{x}\times\mathcal{T}$, which is represented through the action:
\begin{equation}
\left(\tilde{M}_{x}\times\mathcal{T}\right)\mathcal{H}(k_{x},k_{y},k_{z})\left(\tilde{M}_{x}\times\mathcal{T}\right)^{-1} = \sigma^{y}\mathcal{H}^{*}(k_{x},-k_{y},-k_{z})\sigma^{y}.
\label{eq:appSpinSGTR}
\end{equation}
Because $(\tilde{M}_{x}\times\mathcal{T})^{2}=-1$, then bands along the $\tilde{M}_{x}\times\mathcal{T}$-invariant BZ lines $k_{y,z}=0,\pi$ [Fig.~\ref{fig:KW_figBulk}(b)]  must be at least twofold degenerate~\cite{WiederLayers}.  In the case of $v_{x}=0$, $v_{y,z}\neq 0$ in the KW Hamiltonian [Eq.~(\ref{eq:appHKW})], this indicates that the bands along $k_{y,z}=0,\pi$ will form twofold-degenerate nodal lines that linearly disperse in the $k_{y,z}$ directions.  Previous works have termed the $v_{x}=0$, $v_{y,z}\neq 0$ structurally and topologically achiral critical phase a ``Kramers nodal-line semimetal'' state~\cite{KramersNodalLineTh,KramersNodalLineSC}, as it is the parent phase of two KW metals with opposite structural and topological chirality, as discussed in the text following Eq.~(\ref{eq:appSpinSG}).

For completeness, we also note that the action of spatial inversion ($\mathcal{I}$) can be represented on Eq.~(\ref{eq:appHKW}) as:
\begin{equation}
\mathcal{I}\mathcal{H}({\bf k})\mathcal{I}^{-1} = \mathcal{H}(-{\bf k}).
\label{eq:IKW}
\end{equation}
Rather than acting as a symmetry of $\mathcal{H}({\bf k})$, $\mathcal{I}$ -- like the critical-phase symmetry $\tilde{M}_{x}$ in Eq.~(\ref{eq:appSpinSG}) and the surrounding text -- interconverts the right- and left-handed enantiomers of the KW model [Eq.~(\ref{eq:appStructuralChirality})]. Acting with $\mathcal{I}$ therefore also reverses the signs of the chiral charges of the eight bulk KW fermions [Eqs.~(\ref{eq:appHKP}) and~(\ref{eq:appChiralCharge})].  

\clearpage

\subsubsection{Non-Crystalline Kramers-Weyl Fermions}
\label{app:amorphousKramers}

In this section, we will next demonstrate that the Kramers-Weyl [KW] tight-binding model [Eqs.~(\ref{eq:AmoKW}) and~(\ref{eq:KWTmatrix})] continues to exhibit topological chiral [Weyl] fermions when it is realized in amorphous systems [approximated by strongly Gaussian disordered, random, and Mikado lattices] that carry a net [average] structural chirality.
To begin, the crystalline KW model in Eqs.~(\ref{eq:AmoKW}) and~(\ref{eq:KWTmatrix}) can be characterized, as detailed in Appendix~\ref{app:DiffTypesDisorder}, by three distinct sources of disorder: lattice disorder, local frame disorder, and local handedness or \emph{chirality} disorder.
As we will shortly discuss below, the latter two sources of disorder can be assigned to internal or local degrees of freedom on each lattice site. 
Hence, from a physical perspective, they can be understood either via their lattice implementation [regularization], or instead via slowly varying continuum order parameters like those in Landau theories of superconducting, magnetic, and liquid-crystal phase transitions~\cite{chaikinLubenskyBook,CollingsGoodbyLiquidCrystalBook,KamienLiquidCrystalRMP,KamienChiralLiquidCrystal,ChiralNematicBook}.

First, the most obvious way in which the KW tight-binding model can be disordered is via the regularity of its atomic positions [sites].
This \emph{structural disorder} can either be implemented by introducing random site-displacement disorder or by constructing fully random lattices.
In this section, we will consider both of these lattice-disorder possibilities.

More subtly, the hopping terms in Eqs.~(\ref{eq:AmoKW}) and~(\ref{eq:KWTmatrix}) also implicitly contain an SO(3) ``frame'' orientational order parameter that for example locks hopping in the Cartesian $x$-direction to $\sigma^{x}$ terms in the Dresselhaus SOC.
In a realistic model of a disordered solid-state material, no particular region should have physical properties that depend on the global coordinate frame, such as the $x$-direction in an arbitrary coordinate system.
Hence, we must also disorder the local internal frame-locking so that the non-crystalline model does not implicitly contain orbital hopping or SOC terms that are locked in any particular region to the global coordinate frame.  
From a continuum perspective, we can introduce a local SO(3) unit vector field $\hat{R}({\bf r})$ that is given by the average orbital and SOC frame orientation of the sites within a specified vicinity of the position ${\bf r}$.
From this perspective, a system can be defined as ``fully frame-disordered'' via the absence of long-range correlations in $\hat{R}({\bf r})$.
To implement frame disorder at the lattice scale, we assign a rotation matrix $R_{\alpha}$ [detailed in Eq.~\eqref{eqn:RotationMatrix}] to the hopping frame at each lattice site $\alpha$, such that the intersite separation [bond] vector ${\bf d}_{\alpha\beta}$ that links the lattice sites $\alpha$ and $\beta$ [Eq.~(\ref{eq:dVectorDef})] is transformed to a rotated bond vector $\tilde{\bf d}_{\alpha\beta}$ as follows:
\begin{equation}
{\bf d}_{\alpha\beta} \rightarrow R_{\alpha}R_{\beta}^{\mathsf{T}}{\bf d}_{\alpha\beta} \underset{def}{\equiv} \tilde{{\bf d}}_{\alpha\beta}.
\label{appeq:rotaKW}
\end{equation}
Eq.~(\ref{appeq:rotaKW}) is inspired by, and analogous to, local reference frame rotation transformations in lattice gauge theory~\cite{KogutLatticeGaugeRMP}.
After applying Eq.~(\ref{appeq:rotaKW}), the KW model hopping matrix elements in Eq.~(\ref{eq:KWTmatrix}) become transformed to:
\begin{equation}
    T_{\alpha\beta} = \frac{1}{2}f(|\tilde{\mathbf{d}}_{\alpha\beta}|)\left(i\left(\tilde{\mathbf{d}}_{\alpha\beta}\right)^\mathsf{T}\begin{pmatrix}
              v_x&0&0\\
              0&v_y&0\\
              0&0&v_z
          \end{pmatrix}\bm{\sigma} + \left[\left(\tilde{\mathbf{d}}_{\alpha\beta}\right)^\mathsf{T}\begin{pmatrix}
              t_x&0&0\\
              0&t_y&0\\
              0&0&t_z
\end{pmatrix}\tilde{\mathbf{d}}_{\alpha\beta}\right]\sigma^0\right).
\label{appeq:hopchargeKW}
\end{equation}
We further note that, when placed on a disordered or random lattice, the hopping amplitude [rangedness] function $f(|\tilde{\mathbf{d}}_{\alpha\beta}|)$ in Eq.~(\ref{appeq:hopchargeKW}) correspondingly varies from bond to bond, with its values for each bond determined through Eqs.~\eqref{eq:dVectorDef} and~\eqref{eq:KWHeaviside}.

In addition to structural and  local-frame order, the KW tight-binding model can also be characterized by spatially varying \emph{chirality order} that arises from the handedness of short-range bonds [hopping interactions] within the vicinity of a position ${\bf r}$~\cite{LocalChiralityMoleculeReviewNatChem,LocalChiralityLiquidCrystals,KamienLubenskyChiralParameter,LocalChiralityDomainLiquidCrystal,LocalChiralityTransfer,KamienChiralLiquidCrystal,LocalChiralityQuasicrystalVirus,lindell1994electromagneticBook,LocalChiralityVillain,LocalChiralityWenZee,LocalChiralityBaskaran,LocalChiralitySpinFrame}.
As detailed in Appendix~\ref{app:DiffTypesDisorder}, to implement lattice-scale chirality disorder, we assign each site $\alpha$ at the position ${\bf r}_{\alpha}$ a local discrete chirality:
\begin{equation}
\chi_{\alpha} = 0,\pm 1,
\label{eq:KWnumericalModelDiscreteChi}
\end{equation}
where $\chi_{\alpha}=\pm 1$ respectively for right- and left-handed sites and $\chi_{\alpha}=0$ for achiral sites.
In this work, we specifically consider disorder realizations in which there exist \emph{domains} in which all sites carry the same values of $\chi_{\alpha}$.
From a physical perspective, we may gain intuition for the discrete site chirality $\chi_{\alpha}$ by introducing a continuous chirality order parameter field $\chi({\bf r})$~\cite{KamienLubenskyChiralParameter,LocalChiralityVillain,LocalChiralityWenZee,LocalChiralityBaskaran,LocalChiralitySpinFrame}, where $\chi({\bf r})$ is equal to the average value of $\chi_{\alpha}$ for sites $\alpha$ whose positions ${\bf r}_{\alpha}$ lie within a specified vicinity of ${\bf r}$.
Hence deep within a chirality domain, $\chi({\bf r})$ is pinned to $0$ or $\pm 1$ and equal to $\chi_{\alpha}$ for all sites $\alpha$ and ${\bf r}$ away from the domain boundaries.  
The presence of spatially varying local chirality leads to a final modification of Eq.~(\ref{appeq:hopchargeKW}) as follows:
\begin{equation}
    T_{\alpha\beta} = \frac{1}{2}f(|\tilde{\mathbf{d}}_{\alpha\beta}|)\left(i\left(\frac{\chi_{\alpha}+\chi_{\beta}}{2}\right)\left(\tilde{\mathbf{d}}_{\alpha\beta}\right)^\mathsf{T}\begin{pmatrix}
              v_x&0&0\\
              0&v_y&0\\
              0&0&v_z
          \end{pmatrix}\bm{\sigma} + \left[\left(\tilde{\mathbf{d}}_{\alpha\beta}\right)^\mathsf{T}\begin{pmatrix}
              t_x&0&0\\
              0&t_y&0\\
              0&0&t_z          \end{pmatrix}\tilde{\mathbf{d}}_{\alpha\beta}\right]\sigma^0\right).
\label{eq:amorphousKWTmatrixFinalChirality}
\end{equation}
When $\chi_{\alpha}=-\chi_{\beta}$ or $\chi_{\alpha}=\chi_{\beta}=0$, the $v_{i}$ SOC terms in Eq.~(\ref{eq:amorphousKWTmatrixFinalChirality}) vanish, and only the achiral $t_{i}$ orbital hopping terms remain nonvanishing.
For the remainder of this section, we will refer to Eq.~(\ref{eq:amorphousKWTmatrixFinalChirality}) as the ``disordered'' or ``non-crystalline'' KW tight-binding model.

\begin{figure}[t]
\centering
\includegraphics[width=\linewidth]{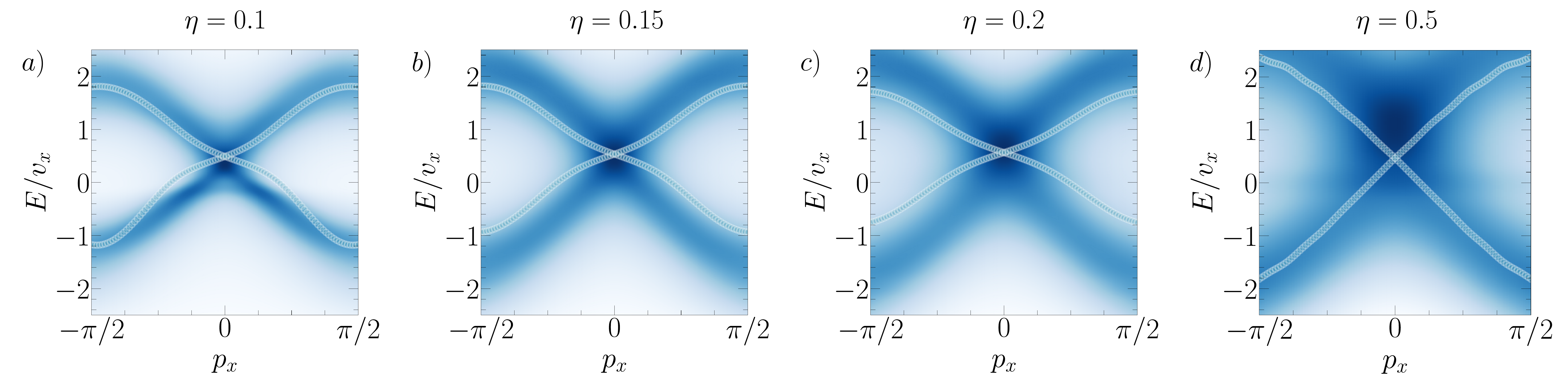}
\caption{Spectral function and effective Hamiltonian spectrum of the Kramers-Weyl model with nematic Gaussian structural disorder.
(a-d) The average spectral function $\bar{A}(E,{\bf p})$ [Eq.~(\ref{eq:SpecFunc})] of the Kramers-Weyl [KW] tight-binding model [Eq.~(\ref{eq:amorphousKWTmatrixFinalChirality})] on a lattice with increasing random nematic structural disorder [$d_{A}=3$ in Eq.~(\ref{eq:finiteDimAmorph})] parameterized by the standard deviation $\eta$, as well as random frame disorder with the same standard deviation $\eta$ and chirality domains of unequal volume, as detailed in Appendix~\ref{app:lattices}.
The data were generated using Eq.~(\ref{eq:amorphousKWTmatrixFinalChirality}) with the parameters in Eq.~(\ref{eq:disorderedKWparams}) implemented with a single domain in each replica of right-handed sites with $n_R=N_R/N_{\mathrm{sites}}=0.7$, and with the remaining volume in each replica containing a contiguous domain of left-handed sites with a corresponding concentration of $n_L=1-N_R/N_{\mathrm{sites}}=0.3$.
Each panel shows data generated by averaging over 50 disorder realizations [\emph{i.e.} replicas] with $20^{3}=8000$ sites each, as detailed in Appendix~\ref{app:PhysicalObservables}.
For all values of $\eta$, $\bar{A}(E,{\bf p})$ exhibits a linearly dispersing feature centered around ${\bf p}={\bf 0}$ that becomes increasingly diffuse as $\eta$ is increased, as well as upwardly shifted in energy by a disorder-renormalized chemical potential $\tilde{\mu}$.
Like the experimentally-observed Dirac-cone surface state of amorphous Bi$_2$Se$_3$~\cite{corbae_evidence_2020,Ciocys2023}, the Fermi velocity of the linear dispersion is also renormalized to larger values with increasing $\eta$.
In the remainder of this section, we will precisely show that the linear spectral features at ${\bf p}={\bf 0}$ in (a-d) represent increasingly disordered generalizations of the crystalline $\Gamma$-point KW fermion in Fig.~\ref{fig:KW_figBulk}(c).
As shown in Fig.~\ref{fig:3DGreenKW}, the off-diagonal-in-momentum matrix elements of the average Green's function $\bar{\mathcal{G}}(E,{\bf p},{\bf p}')$ begin to vanish at moderate disorder scales [$\eta=0.2$ in panel (c)], and are nearly vanishing for strong disorder [$\eta=0.5$ in panel (d)].
We may therefore in each panel limit our focus to the \emph{momentum-diagonal} average Green's function $\bar{\mathcal{G}}(E,{\bf p})$ [Eq.~(\ref{eq:averageOneMomentumGreen})], from which we construct an effective Hamiltonian $\mathcal{H}_{\mathrm{Eff}}(E_{C},{\bf p})$ [Eq.~(\ref{eq:AvgHEff})] whose bands are plotted with light blue circles.
$\mathcal{H}_{\mathrm{Eff}}(E_{C},{\bf p})$ in each panel was specifically constructed using a reference energy cut $E_{C}$ centered at the maximum value of the spectral function at ${\bf p}={\bf 0}$ [approximately $\tilde{\mu}$] to maximize its accuracy [see Appendix~\ref{app:EffectiveHamiltonian}], where $E_{C}$ for each panel is respectively given by (a) $E_C / v_{x} =0.56$, (b) $E_C / v_{x} =0.57$, (c) $E_C / v_{x} =0.66$, and (d) $E_C/ v_{x} = 0.71$.
In all panels, and for all choices of $E_{C}$ [see Fig.~\ref{fig:H_eff_benchmark} and the surrounding text], the effective Hamiltonian exhibits a linear nodal degeneracy at ${\bf p}={\bf 0}$; we will shortly in Fig.~\ref{fig:KW_WL} use Wilson loops to show that this degeneracy is a $|C|=1$ KW fermion.
}
\label{fig:KW_Bands_amo}
\end{figure}

As discussed in Appendix~\ref{app:PhysicalObservables}, we wish to model solid-state systems that have thermodynamically large numbers of atoms [or short-range-interacting patches~\cite{ProdanKohnNearsighted}] and are self-averaging.  
However using our computational hardware, we can only simulate individual disorder realizations with $\sim 20^{3}$ [$\sim 8000$] atoms.
We therefore additionally implement an \emph{averaging procedure} in which we approximate the spectral and topological properties of large self-averaging systems by constructing a replica-averaged~\cite{ParisiReplicaCourse,BerthierGlassAmorphousReview} momentum-resolved Green's function $\bar{\mathcal{G}}(E,\mathbf{p})$ [Eq.~(\ref{eq:averageOneMomentumGreen})] by averaging the momentum-resolved matrix Green's function over multiple [$\approx 20-50$] \emph{replicas} that each contain distinct randomly generated lattice distortions or random lattices.
As detailed in Appendix~\ref{app:PhysicalObservables}, we obtain $\bar{\mathcal{G}}(E,\mathbf{p})$ by first computing the diagonal-in-momentum piece of the momentum-resolved matrix Green's function $\mathcal{G}(E,{\bf p},{\bf p})$, while numerically establishing that the off-diagonal-in-momentum elements of $\bar{\mathcal{G}}(E,{\bf p},{\bf p}')$ vanish [on the average] for strongly disordered lattices.
Specifically, for the non-crystalline KW model in Eq.~(\ref{eq:amorphousKWTmatrixFinalChirality}) with Gaussian structural disorder, we found in Fig.~\ref{fig:3DGreenKW} that the off-diagonal-in-momentum matrix elements of $\bar{\mathcal{G}}(E,{\bf p},{\bf p}')$ begin to vanish at $\eta=0.2$, which we in this section term ``moderate'' disorder, and are nearly vanishing for $\eta=0.5$, which we hence term ``strong'' disorder.
We may therefore limit consideration to the diagonal-in-momentum matrix $\bar{\mathcal{G}}(E,\mathbf{p})$, which we compute by elementwise averaging $\mathcal{G}(E,{\bf p},{\bf p})$ at the same ${\bf p}$ [Eq.~(\ref{eq:averageOneMomentumGreen})].

\begin{figure}[t]
\centering   
\includegraphics[width=\linewidth]{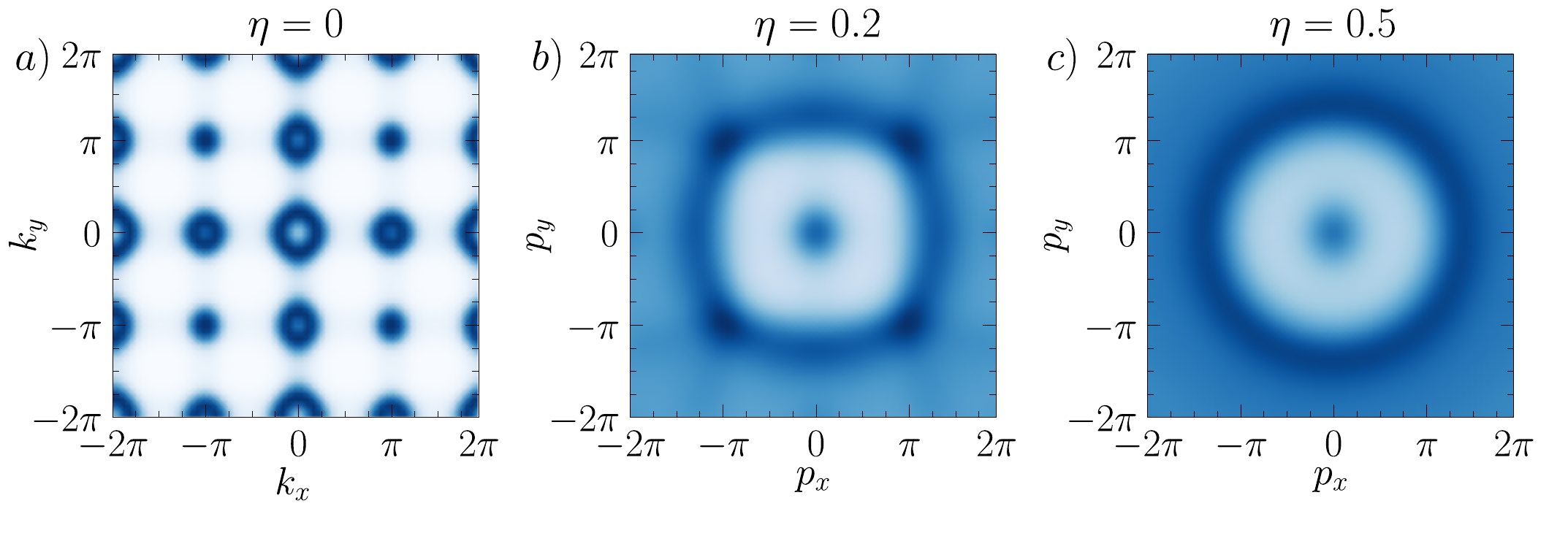}
\caption{Constant-energy spectrum of the disordered Kramers-Weyl model.
(a-c) Constant-energy cuts of the average spectral function $\bar{A}(E,{\bf p})$ [Eq.~(\ref{eq:SpecFunc})] of the non-crystalline Kramers-Weyl [KW] model [Eq.~(\ref{eq:amorphousKWTmatrixFinalChirality})] with the parameters in Eq.~(\ref{eq:disorderedKWparams}), contiguous domains of right- and left-handed sites with the respective concentrations $n_R=N_R/N_{\mathrm{sites}}=0.7$ and $n_L=1-N_R/N_{\mathrm{sites}}=0.3$, and increasing random nematic structural and frame disorder parameterized by the standard deviation $\eta$, as detailed in Appendix~\ref{app:lattices}.
To generate each panel, we first compute the momentum-resolved Green's function $\bar{\mathcal{G}}(E,{\bf p})$ at $p_{z}=0.1$ [$k_{z}=0.1$ in (a)] averaged over 20 disorder realizations [replicas] with $20^{3}$ sites each, as detailed in Appendix~\ref{app:PhysicalObservables}.
Because increasing $\eta$ generates a disorder-renormalized shift in the chemical potential at ${\bf p}={\bf 0}$ [see Fig.~\ref{fig:KW_Bands_amo}], then for each panel in this figure, we first obtain a reference energy $E_{C}$ at which the spectral weight $\bar{A}(E_{C},{\bf p})$ at ${\bf p}={\bf 0}$ is maximized.
We then compute $\bar{A}(E,{\bf p}) \propto \text{Im}\{\Tr[\bar{\mathcal{G}}(E,{\bf p})]\}$ at $E/v_x=E_{C}/v_x - 0.71$ in order to approximately capture the same cross section of the linear nodal degeneracy at ${\bf p}={\bf 0}$ in Fig.~\ref{fig:KW_Bands_amo} for varying $\eta$.
In addition to the linear spectral feature at ${\bf p}={\bf 0}$, the systems with moderate [$\eta=0.2$ in (b)] and strong [$\eta=0.5$ in (c)] disorder exhibit broadened, ring- [sphere-] like spectral features in the vicinity of $|{\bf p}|=\pi/\bar{a}$ and $|{\bf p}|=\pi\sqrt{2}/\bar{a}$ where $\bar{a}=1$ is the average nearest-neighbor spacing.
As we will further motivate by calculating the system spin texture in Fig.~\ref{fig:SpinTextureKW}, we find that the ring-like features in (b,c) originate from rotationally averaging and disorder-broadening the KW fermions at $|{\bf k}|=\pi$ and $|{\bf k}|=\pi\sqrt{2}$ in the crystalline case [$\eta=0$ in (a)], analogously to the ring-like, higher-Brillouin-zone Dirac surface states recently observed in amorphous Bi$_{2}$Se$_{3}$~\cite{corbae_evidence_2020,Ciocys2023}.
}
\label{fig:DOSKW}
\end{figure}

\begin{figure}[t]
\centering   
\includegraphics[width=\linewidth]{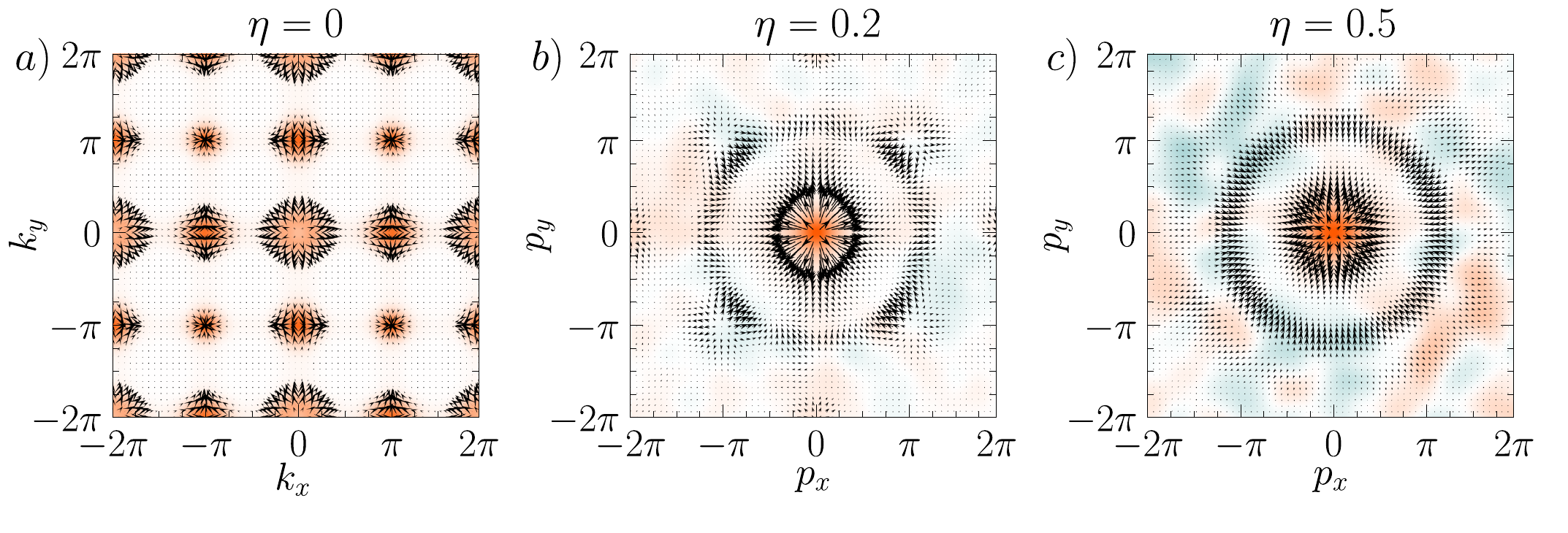}
\caption{Spin texture of the disordered Kramers-Weyl model.
(a-c) The spin texture [Eq.~\eqref{eq:SpinDOS}] of the non-crystalline Kramers-Weyl [KW] systems in Fig.~\ref{fig:DOSKW}(a-c), respectively, plotted at $k_{z}=0.1$ in (a) and $p_{z}=0.1$ in (b,c) at the constant-energy cuts detailed in Fig.~\ref{fig:DOSKW}.
In all panels, the in-plane components of the spin texture $\langle S^{x,y}(E,\mathbf{p})\rangle$ are represented as arrows with log-scale lengths, while the out-of-plane component $\langle S^{z}(E,\mathbf{p})\rangle$ is represented through a log-scale color map in which orange is positive and teal is negative.
(a) In the crystalline limit, the KW fermions at each time-reversal-invariant ${\bf k}$ point exhibit perfect monopole-like spin textures that are locked to their chiral charges~\cite{KramersWeyl}.
(b,c) As the disorder scale $\eta$ is increased, the KW fermions away from ${\bf p}={\bf 0}$ become merged into ring- [sphere-] like spectral features with largely isotropic spin textures that are inherited from their $\eta=0$ spin textures, and hence chiral charges.
Specifically, the crystalline KW fermions at $|{\bf k}|=\pi$ in (a) merge into a ring-like feature at $|{\bf p}|=\pi$ in (b,c) with an inward-pointing spin texture and high spectral weight, and the crystalline KW fermions at $|{\bf k}|=\pi\sqrt{2}$ in (a) merge into a ring-like feature at $|{\bf p}|=\pi\sqrt{2}$ in (b,c) with an outward-pointing spin texture and relatively weaker spectral weight.
This suggests a picture in which the ring-like feature in (b,c) at $|{\bf p}|=\pi$ [$|{\bf p}|=\pi\sqrt{2}$] is a many-particle disordered KW fermion with the opposite [same] chiral charge as the linear nodal degeneracy at ${\bf p}={\bf 0}$.
}
\label{fig:SpinTextureKW}
\end{figure}

\paragraph*{\bf Energy Spectrum} -- $\ $ In Fig.~\ref{fig:KW_Bands_amo} we plot the disorder-averaged, momentum-resolved spectral function $\bar{A}(E,{\bf p}) \propto \text{Im}\{\Tr[\bar{\mathcal{G}}(E,{\bf p})]\}$ [Eq.~(\ref{eq:SpecFunc})] for the disordered KW tight-binding model [Eq.~(\ref{eq:amorphousKWTmatrixFinalChirality})] with the parameters:
\begin{equation}
v_x=1,\ v_y=1.1,\ v_z=1.35,\ t_x=0.1,\ t_y=0.12,\ t_z=0.09,
\label{eq:disorderedKWparams}
\end{equation}
placed on a lattice with increasing random nematic [$d_{A}=3$ in Eq.~(\ref{eq:finiteDimAmorph})] Gaussian structural disorder parameterized by a standard deviation $\eta$, as detailed in Appendix~\ref{app:lattices}.
The spectra in Fig.~\ref{fig:KW_Bands_amo} were also computed using random local frame disorder [Eqs.~\eqref{appeq:rotaKW} and~\eqref{appeq:hopchargeKW}] implemented with the same standard deviation $\eta$ as the lattice disorder, as well as chirality domains of unequal volume [70\% right-handed, 30\% left-handed] within each disorder replica [see Appendix~\ref{app:DiffTypesDisorder}].
In all panels of Fig.~\ref{fig:KW_Bands_amo}, we observe a linearly dispersing feature centered around ${\bf p}={\bf 0}$ that exhibits increased spectral broadening for increasing disorder, but nevertheless strongly resembles the crystalline $\Gamma$-point KW fermion in Fig.~\ref{fig:KW_figBulk}(c) for all values of $\eta$ in Fig.~\ref{fig:KW_Bands_amo}. 
Notably, the nodal spectral features in Fig.~\ref{fig:KW_Bands_amo} at ${\bf p}={\bf 0}$ shift upwards in energy as $\eta$ is increased, reflecting the presence of an increasingly strong disorder-renormalized chemical potential $\tilde{\mu}$.
Like the Dirac-cone surface state of amorphous Bi$_2$Se$_3$ observed in recent angle-resolved photoemission experiments~\cite{corbae_evidence_2020,Ciocys2023}, the Fermi velocity of the linear spectral feature in Fig.~\ref{fig:KW_Bands_amo} is also renormalized to larger values as $\eta$ is increased.

Using the disorder-averaged, momentum resolved Green's function $\bar{\mathcal{G}}(E,{\bf p})$, we can also compute an approximate [effective] single-particle Hamiltonian $\mathcal{H}_{\mathrm{Eff}}(E_{C},{\bf p})$~\cite{varjas_topological_2019,marsal_topological_2020,marsal_obstructed_2022} for the spectral features at ${\bf p}={\bf 0}$ in the disordered KW model [Eq.~(\ref{eq:AvgHEff})].
Though $\mathcal{H}_{\mathrm{Eff}}(E_{C},{\bf p})$ only represents a mean-field approximation of the many-particle, momentum-dependent Hamiltonian, and is therefore generally dependent on the energy cut $E_{C}$ at which it is obtained, we have shown in Appendix~\ref{app:EffectiveHamiltonian} that the nodal degeneracies and topology of $\mathcal{H}_{\mathrm{Eff}}(E_{C},{\bf p})$ near ${\bf p}={\bf 0}$ are numerically stable and surprisingly insensitive to $E_{C}$  in the models studied in this work.
Specifically, as shown in Appendix~\ref{app:EffectiveHamiltonian}, though the explicit dependence of the effective Hamiltonian on the reference energy $E_{C}$ at which it is constructed is often notationally suppressed $\mathcal{H}_{\text{Eff}}(\mathbf{p}) = \mathcal{H}_{\text{Eff}}(E_{C},{\bf p})$, the energy eigenvalues of $\mathcal{H}_{\text{Eff}}(\mathbf{p})$ in fact \emph{strongly depend} on $E_{C}$. 
However at least in the models studied in this work, the energy \emph{eigenstates} of $\mathcal{H}_{\text{Eff}}(\mathbf{p})$ are considerably less sensitive to the choice of $E_{C}$ [see Fig.~\ref{fig:H_eff_benchmark} and the surrounding text].
We specifically found in Appendix~\ref{app:EffectiveHamiltonian} that the energy eigenvalues of $\mathcal{H}_{\text{Eff}}(\mathbf{p})$ most closely qualitatively match the low-energy, small-momentum spectral features of $\bar{A}(E,{\bf p})$ when $\mathcal{H}_{\text{Eff}}(\mathbf{p})$ is constructed with $E_{C}$ set to the value of the maximum spectral weight of $\bar{A}(E,{\bf p})$ at ${\bf p}={\bf 0}$ [with $E_{C}$ lying away from other ${\bf p}={\bf 0}$ spectral features in multiband models, see Appendix~\ref{app:amorphousMultifold}].
The insensitivity of the eigenstates [specifically Berry phases] of $\mathcal{H}_{\mathrm{Eff}}({\bf p})$ to the choice of $E_{C}$ near ${\bf p}={\bf 0}$ strongly suggests that $\mathcal{H}_{\mathrm{Eff}}({\bf p})$ accurately captures the many-particle, small-momentum spectral features and topology of the amorphous chiral metals studied in this work.
In each panel of Fig.~\ref{fig:KW_Bands_amo}, we show the bands of $\mathcal{H}_{\mathrm{Eff}}({\bf p})$ in light blue circles, in each case computed using a reference energy cut $E_{C}$ centered at the maximum value of the spectral function at ${\bf p}={\bf 0}$ to maximize the spectral accuracy of the effective Hamiltonian.
For all disorder strengths $\eta$, the effective Hamiltonian bands in Fig.~\ref{fig:KW_Bands_amo} exhibit linear nodal degeneracies at ${\bf p}={\bf 0}$.
Below, we will provide evidence that the linear spectral feature in Fig.~\ref{fig:KW_Bands_amo} at ${\bf p}={\bf 0}$ in $\bar{A}(E,{\bf p})$ [approximated by the linear nodal degeneracy at ${\bf p}={\bf 0}$ in $\mathcal{H}_{\mathrm{Eff}}({\bf p})]$ is in fact precisely a disordered KW fermion whose topology is controlled by \emph{average} structural chirality, analogous to its crystalline counterpart.

We next construct constant-energy spectral cuts of the disordered KW tight-binding model to explore spectral features at higher momenta.
In Fig.~\ref{fig:DOSKW} we plot the average spectral function $\bar{A}(E,{\bf p})$ of a disordered KW system [Eq.~(\ref{eq:amorphousKWTmatrixFinalChirality})] at fixed $E$ and $p_{z}$ with the parameters in Eq.~(\ref{eq:disorderedKWparams}) and increasing nematic structural and frame disorder averaged over 20 replicas that each contain contiguous domains of right- and left-handed sites with the respective concentrations $n_R=N_R/N_{\mathrm{sites}}=0.7$ and $n_L=1-N_R/N_{\mathrm{sites}}=0.3$.
Interestingly, as $\eta$ is increased, $\bar{A}(E,{\bf p})$ develops 3D sphere-like spectral features that appear as 2D rings in the vicinity of $|{\bf p}|=\pi/\bar{a}$ and $|{\bf p}|=\pi\sqrt{2}/\bar{a}$ in the constant-energy and fixed-$p_{z}$ spectral function in Fig.~\ref{fig:DOSKW}(b,c), where $\bar{a}=1$ is the average nearest-neighbor spacing.
The ring-like features in Fig.~\ref{fig:DOSKW}(b,c) can be understood as originating from averaging the crystalline KW fermions at $|{\bf k}|=\pi$ and $|{\bf k}|=\pi\sqrt{2}$ in Fig.~\ref{fig:DOSKW}(a) over random orientations and lattice spacings, giving rise to the characteristic isotropic spectral features of an amorphous system~\cite{spring_amorphous_2021,springMagneticAverageTI,corbae_evidence_2020,Ciocys2023}.
The appearance of 3D sphere-like KW features in Fig.~\ref{fig:DOSKW}(b,c) is also analogous to the 2D surface states of the amorphous 3D strong topological insulator Bi$_{2}$Se$_{3}$.
Specifically, in crystalline Bi$_{2}$Se$_{3}$, topological Dirac fermions lie at $\bar{\Gamma}$ [${\bf k}={\bf 0}$] in the surface Brillouin zone [BZ], and at all surface ${\bf k}$ points related to $\bar{\Gamma}$ by reciprocal lattice vectors with $|{\bf K}|=2\pi/a$ where $a$ is the surface lattice spacing.  
In the experimentally-obtained surface spectrum of amorphous Bi$_{2}$Se$_{3}$, one Dirac-cone surface state continues to lie at ${\bf p}={\bf 0}$, but the other surface Dirac cones in higher surface BZs merge into ring-like spectral features, with the first Dirac ring specifically lying at $|{\bf p}|=2\pi/\bar{a}$ where $\bar{a}$ is the average in-plane [surface] nearest-neighbor atomic spacing~\cite{corbae_evidence_2020,Ciocys2023}.

We briefly pause to note that importantly, an infinitesimally weak Peierls charge-density wave [CDW] -- or a stronger near-Peierls CDW like that in the Weyl-CDW material (TaSe$_4$)$_2$I~\cite{Monceau1984TSICDW,CDWWeyl,WeylCDWSpinTexture,KuansenBarryCDW,BenKuansenCDW,JiabinBenCDW} -- can fully gap a crystalline KW semimetal by coupling and backfolding the bulk KW fermions~\cite{KramersWeyl,LemutBeenakkerKramersWeylSupercell}.
If a single domain of the resulting Weyl-CDW insulator were then strongly disordered, the system would remain an insulator.
This raises the question of how a strongly gapped and disordered KW-CDW insulator differs from the disordered KW metal in Fig.~\ref{fig:KW_Bands_amo} -- which is instead gapless at half filling -- and whether we have in our analysis overlooked an important source of system disorder.
The resolution of this question is that for most of the disordered and non-crystalline systems studied in this work, we have not included local bond-sign [bond-order-wave] or chemical potential [on-site] disorder, which if realized with exactly alternating signs on neighboring sites in an amorphous system could give rise to a disordered density-wave insulator, like the amorphous obstructed atomic limits analyzed in Ref.~\cite{marsal_obstructed_2022}.
We have largely ignored these potential sources of disorder because we wish in this work to model chiral amorphous solid-state systems in which the local units carry chemically and electronically similar environments [up to frame rotations and chirality inversions], as such systems represent the simplest structurally chiral [on the average] generalizations of well-established and readily experimentally accessible non-crystalline solid-state materials like \emph{achiral} amorphous silicon~\cite{weaire_electronic_1971}, Fe$_x$Sn$_{1-x}$~\cite{Fujiwara2023kagome}, and Co$_2$MnGa~\cite{KarelAmorphousBerryCMG}.
However for both the non-crystalline KW model studied in this section and the other non-crystalline topological semimetal models analyzed in this work, we have numerically confirmed that when each model is simulated with very strong structural and internal degree-of-freedom disorder [\emph{e.g.} on random lattices with large spin and orbital frame disorder, see Figs.~\ref{appfig:structuraldisorder}(b),~\ref{appfig:disordertypes}(c),~\ref{fig:KW_Unif}(a-d),~\ref{fig:C2_Unif}, and~\ref{fig:3F_Unif}], the topological features of each model remain stable under the subsequent addition of weak Anderson [on-site chemical potential] disorder.

\paragraph*{\bf Spin Texture} -- $\ $ Previous works have highlighted that energetically isolated KW fermions can exhibit monopole-like spin textures~\cite{ChiralTellurium1,ChiralTellurium2,KramersWeyl,BradlynTQCSpinTexture,2DMoireKWSpinTexture}.  
To draw a connection with crystalline KW fermions, we will therefore next compute the spin texture of the spectral features in Fig.~\ref{fig:DOSKW}.
We first note that in real materials and more complicated tight-binding models with large numbers of bands, neither the total angular momentum ${\bf J}$, nor the orbital angular momentum [OAM] ${\bf L}$, nor the spin ${\bf S}$ of a momentum-space energy eigenstate is generically quantized, due to the combined effects of OAM and SOC terms and interband matrix elements~\cite{KramersWeyl,BradlynTQCSpinTexture,chang2017large}.  
Nevertheless, for a region in energy and momentum space with high spectral weight [\emph{i.e.} a large spectral function $\bar{A}(E,{\bf p})$ corresponding to Bloch eigenstates in the crystalline limit], we may still determine the \emph{degree} of ${\bf J}$, ${\bf L}$, or ${\bf S}$ polarization -- as well as the angular momentum orientations [textures] of highly polarized states -- by computing the \emph{angular-momentum-dependent spectral function}.  
For the case of spin ${\bf S}$ most relevant to the disordered KW model [Eq.~(\ref{eq:amorphousKWTmatrixFinalChirality})], we can compute the \emph{spin-dependent spectral function vector} $\langle\mathbf{S}(E,\mathbf{p})\rangle$~\cite{chang2017large,Saito_TB_2016,Kohsaka_2017}, whose components are given by:
\begin{equation}
\left\langle S^{i} \left(E,\mathbf{p}\right)\right\rangle = -\frac{1}{\pi} \text{Im}\left\{\text{Tr}\left[\hat{S}^{i} \bar{\mathcal{G}}(E,\mathbf{p})\right]\right\} = -\frac{1}{\pi} \text{Im}\left\{\text{Tr}\left[\sigma^{i} \bar{\mathcal{G}}(E,\mathbf{p})\right]\right\},
\label{eq:SpinDOS}
\end{equation}
where $\hat{S}^{i}$ is the $i$-th component of the spin operator vector $\hat{\mathbf{S}}=(\hat{S}^x,\hat{S}^y,\hat{S}^z)$, $\sigma^{i}$ is the matrix representative of the spin component operator $\hat{S}^{i}$ acting on the internal degrees of freedom of the disordered KW model, $\bar{\mathcal{G}}(E,{\bf p})$ is the disorder- [replica-] averaged momentum-resolved matrix Green's function [Eq.~\eqref{eq:averageOneMomentumGreen}], and where the trace operation is taken over the internal spin-1/2 degrees of freedom parameterized by $\sigma^{i}$ in Eq.~(\ref{eq:amorphousKWTmatrixFinalChirality}).
Using Eqs.~(\ref{eq:SpecFunc}) and~(\ref{eq:SpinDOS}), the degree of spin polarization $P_{\bf S}(E,{\bf p})$ for a spectral feature at a given $E$ and ${\bf p}$ is then given by:
\begin{equation}
P_{\bf S}(E,{\bf p}) = \frac{\left|\left\langle\mathbf{S} \left(E,\mathbf{p}\right)\right\rangle\right|}{\bar{A}(E,{\bf p})} =  \frac{\sqrt{\sum\limits_{i=x,y,z}\langle S^{i}(E,{\bf p})\rangle^{2}}}{\bar{A}(E,{\bf p})} = -\frac{\sqrt{\sum\limits_{i=x,y,z}\left(\text{Im}\left\{\Tr\left[\sigma^{i}\bar{\mathcal{G}}(E,\mathbf{p})\right]\right\}\right)^{2}}}{\text{Im}\left\{\Tr\left[\bar{\mathcal{G}}(E,\mathbf{p})\right]\right\}},
\label{eq:KWspinPolarization}
\end{equation}
in which $P_{\bf S}(E,{\bf p})$ takes values between $0$ and $1$, such that spectral features with $P_{\bf S}(E,{\bf p})\approx 1$ are termed ``highly spin-polarized''~\cite{chang2017large}.

In Fig.~\ref{fig:SpinTextureKW}, we plot the spin texture [Eq.~(\ref{eq:SpinDOS})] of the disordered KW tight-binding model with the same parameters previously employed to generate the constant-energy spectra in Fig.~\ref{fig:DOSKW}.
As discussed in Refs.~\cite{KramersWeyl,BradlynTQCSpinTexture}, in the simplified two-band KW model taken in the crystalline limit [$\eta=0$], each KW fermion exhibits a perfect monopole-like spin texture with a monopole charge that matches its topological chiral charge.
The perfect monopole-like spin texture of Eq.~(\ref{eq:SpinDOS}) at $\eta=0$ can be seen in Fig.~\ref{fig:SpinTextureKW}(a), in which the $C=+1$ KW fermions at $k_{x}=k_{y}=0,\pi$ exhibit outward-pointing spin textures along odd numbers of principal axes and inward-pointing spin textures among the remaining [even numbers of] principal axes, while the $C=-1$ KW fermions at $(k_{x},k_{y})=(\pi,0)$ and $(k_{x},k_{y})=(0,\pi)$ exhibit spin textures that respectively point inward in the $k_{x}$- and $k_{y}$-directions and outward along the other two principal axes. 
For all values of $\eta$, the central spectral feature -- which corresponds to the nodal degeneracy at ${\bf p}={\bf 0}$ in Fig.~\ref{fig:KW_Bands_amo} -- retains its nearly perfect outward-pointing, monopole-like spin texture.
The nearly perfect monopole-like spin texture near ${\bf p}={\bf 0}$ in Fig.~\ref{fig:SpinTextureKW}(b,c) likely originates [at least partially] from the relative simplicity of [\emph{i.e.} lack of additional bands and hence interband matrix elements in] the original crystalline KW tight-binding model~\cite{KramersWeyl,BradlynTQCSpinTexture} from which Eq.~(\ref{eq:amorphousKWTmatrixFinalChirality}) was derived.
Nevertheless, the topological similarity between the spin textures near ${\bf p}={\bf 0}$ at increasing $\eta$ in Fig.~\ref{fig:SpinTextureKW} suggests a picture [which we will shortly use Wilson loops to make precise] in which the linear nodal degeneracy at ${\bf p}={\bf 0}$ in Figs.~\ref{fig:KW_Bands_amo}(d),~\ref{fig:DOSKW}(c), and~\ref{fig:SpinTextureKW}(c) is a strongly disordered KW fermion with a low-energy sense of handedness [spin-texture monopole charge] that is inherited from the lattice-scale chirality imbalance [\emph{i.e.} average chirality] $n_{R}>n_{L}$ of the disordered system.

Interestingly, as $\eta$ is increased while keeping $n_{R,L}$ constant, we find that the spin texture [anti]monopoles at $|{\bf p}|=\pi$ and $|{\bf p}|=\pi\sqrt{2}$ merge into 3D sphere-like spectral features with largely isotropic spin textures [which appear as 2D rings in the constant-energy, fixed-$p_{z}$ spin-dependent spectra in Fig.~\ref{fig:SpinTextureKW}(b,c)].
Specifically, the crystalline KW fermions at $|{\bf k}|=\pi$ in Fig.~\ref{fig:SpinTextureKW}(a) merge into a ring-like feature at $|{\bf p}|=\pi$ with an inward-pointing spin texture and high spectral weight, and the crystalline KW fermions at $|{\bf k}|=\pi\sqrt{2}$ in Fig.~\ref{fig:SpinTextureKW}(a) merge into a ring-like feature at $|{\bf p}|=\pi\sqrt{2}$ with an outward-pointing spin texture and relatively weaker spectral weight.
Because the energy spectrum is more diffuse at larger momenta [Fig.~\ref{fig:HeffBreak}(a)], and because we have only shown that the effective Hamiltonian method is numerically stable near ${\bf p}={\bf 0}$ [Appendix~\ref{app:EffectiveHamiltonian}], we are unable to directly apply the average-symmetry-group representation theory [Appendices~\ref{app:pseudoK} and~\ref{app:corepAmorphous}] and amorphous Wilson loop method [Appendix~\ref{sec:WilsonBerry}] introduced in this work to diagnose the topology of the disorder-broadened, ring-like spectral features in Figs.~\ref{fig:DOSKW} and~\ref{fig:SpinTextureKW}.
Nevertheless, the alternating inward- and outward-pointing spin textures of the spectral rings in Fig.~\ref{fig:SpinTextureKW}(b,c) suggest that the ring-like feature at $|{\bf p}|=\pi$ [$|{\bf p}|=\pi\sqrt{2}$] is a many-particle disordered KW fermion with the opposite [same] chiral charge as the linear nodal degeneracy at ${\bf p}={\bf 0}$.
We will shortly show in this section [Fig.~\ref{fig:KW_Unif}] that the appearance of isotropic spectral features is a generic feature of the KW model with strong structural disorder, and also appears for disordered lattice choices [\emph{e.g.} random and Mikado, see Refs.~\cite{marsal_obstructed_2022,marsal_topological_2020} and Appendix~\ref{app:DiffTypesDisorder}] that cannot be deformed to uniquely defined crystalline limits.

\begin{figure}[t]
\centering
\includegraphics[width=\linewidth]{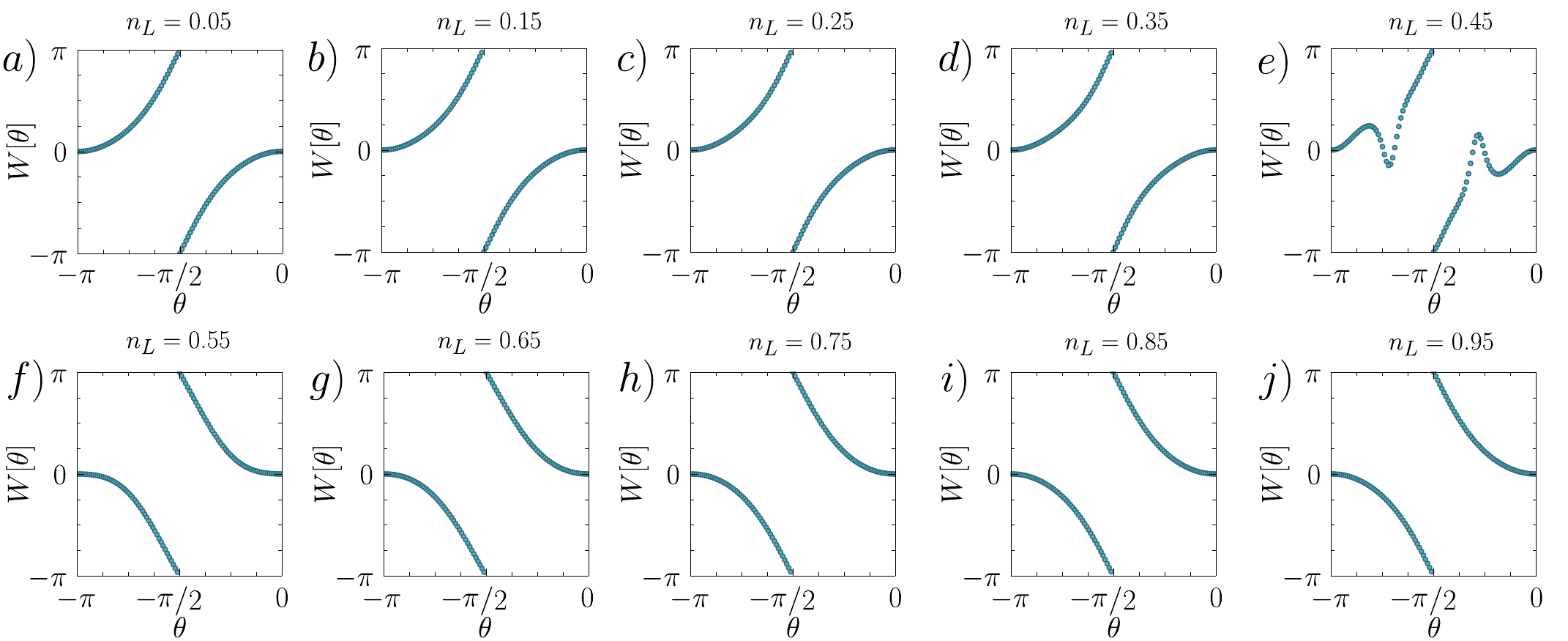}
\caption{Sphere Wilson loop spectrum of the disordered Kramers-Weyl model for varying chirality concentrations.
To generate each panel in this figure, we place the non-crystalline Kramers-Weyl [KW] model [Eq.~(\ref{eq:amorphousKWTmatrixFinalChirality})] with the parameters in Eq.~(\ref{eq:disorderedKWparams}) on a lattice with $N_{\mathrm{sites}}=15^3$ and random nematic Gaussian structural and local frame disorder parameterized by the fixed standard deviation $\eta=0.2$, as detailed in Appendix~\ref{app:lattices}. 
We then generate the replica-averaged momentum-resolved Green's function $\bar{\mathcal{G}}(E,\mathbf{p})$ [Eq.~(\ref{eq:averageOneMomentumGreen})] by averaging the system over 50 disorder realizations [replicas] that each contain contiguous domains of right- and left-handed sites with the respective concentrations $n_R=N_R/N_{\mathrm{sites}}$ and $n_{L}=1-n_{R}$.
(a-j) For 10 disorder ensembles with increasing values of $n_{L}$, we construct for each ensemble an effective Hamiltonian $\mathcal{H}_{\mathrm{Eff}}({\bf p})=\mathcal{H}_{\mathrm{Eff}}(E_{C},{\bf p})$ [Eq.~(\ref{eq:AvgHEff})] using a reference energy $E_{C}$ corresponding to the largest spectral weight $\bar{A}(E,{\bf p})$ at ${\bf p}={\bf 0}$ [Eq.~(\ref{eq:SpecFunc})], which as shown in Fig.~\ref{fig:H_eff_benchmark} maximizes the spectral accuracy of $\mathcal{H}_{\mathrm{Eff}}({\bf p})$.
We then use the eigenstates of $\mathcal{H}_{\mathrm{Eff}}({\bf p})$ to compute the amorphous [disordered] Wilson loop spectrum introduced in this work [Appendix~\ref{sec:WilsonBerry}] on a sphere surrounding the nodal degeneracy at ${\bf p}={\bf 0}$ in each disorder ensemble.
Beginning with a moderately disordered system [$\eta=0.2$, see Fig.~\ref{fig:3DGreenKW}] with (a) almost entirely right-handed sites [$n_{L}=0.05$] and continuing in increasing $n_{L}$ to a system (j) with almost entirely left-handed sites [$n_{L}=0.95$], we observe a quantized Wilson loop winding as a function of the sphere polar angle $\theta$ of (a-d) $C=1$ for $n_{L}<0.5$, (f-j) $C=-1$ for $n_{L}>0.5$, and (e) a region in the vicinity of $n_{L}\approx 0.5$ with a non-smooth Wilson spectrum.
This provides a precise indicator that the nodal degeneracy at ${\bf p}={\bf 0}$ in Fig.~\ref{fig:KW_Bands_amo} is a disordered [non-crystalline] chiral [KW] fermion that, analogous to its crystalline counterpart~\cite{KramersWeyl,AlPtObserve,PdGaObserve,DingARPESReversal,SessiPdGaQPIReversal}, carries a topological chirality that is tunable via the average system structural chirality.}
\label{fig:KW_WL}
\end{figure}

\paragraph*{\bf Wilson Loops} -- $\ $ Having shown that the non-crystalline KW model [Eq.~(\ref{eq:amorphousKWTmatrixFinalChirality})] continues to exhibit a linear nodal degeneracy at ${\bf p}={\bf 0}$ [Fig.~\ref{fig:KW_Bands_amo}(c,d)] with a monopole-like spin texture [Fig.~\ref{fig:SpinTextureKW}(c)] for strong, chirality-imbalanced disorder, we will now use an amorphous generalization of the Wilson loop method [see Appendix~\ref{sec:WilsonBerry} and Refs.~\cite{Fidkowski2011,AndreiXiZ2,ArisInversion,Cohomological,HourglassInsulator,DiracInsulator,Z2Pack,BarryFragile,AdrienFragile,HOTIBernevig,HingeSM,WiederAxion,KoreanFragile,ZhidaBLG,TMDHOTI,KooiPartialNestedBerry,PartialAxionHOTINumerics,GunnarSpinFragileWilson,BinghaiOscillationWilsonLoop,Wieder22}] to precisely show that the ${\bf p}={\bf 0}$ nodal degeneracy is a disordered [non-crystalline] KW fermion with a nontrivial topological chiral charge.

To begin, in a crystalline system, the chiral charge of a nodal degeneracy can be obtained by computing the winding number of the Wilson loop spectrum [non-Abelian Berry phases] evaluated over the occupied bands on a sphere surrounding the nodal degeneracy~\cite{Z2Pack}.
At each value of the polar angle $\theta$ of the sphere [see Fig.~\ref{fig:Wilson_schema}(b)], the Wilson loop is computed using the Bloch wavefunctions of the occupied states.
To perform an analogous calculation for a given disorder ensemble of the non-crystalline KW model [Eq.~(\ref{eq:amorphousKWTmatrixFinalChirality})], we therefore first construct an effective Hamiltonian $\mathcal{H}_{\text{Eff}}(\mathbf{p})$ [see Refs.~\cite{varjas_topological_2019,marsal_topological_2020,marsal_obstructed_2022} and Appendix~\ref{app:EffectiveHamiltonian}] for the nodal degeneracy at ${\bf p}={\bf 0}$ using the replica-averaged momentum-resolved Green's function $\bar{\mathcal{G}}(E,\mathbf{p})$ [Eq.~(\ref{eq:averageOneMomentumGreen})]. 
For each disorder ensemble of the non-crystalline KW model, we specifically first identify an energy $E_{max}$ where $\bar{A}(E,{\bf p})$ is largest at ${\bf p}={\bf 0}$, and then we construct $\mathcal{H}_{\text{Eff}}(\mathbf{p})$ using the reference energy cut $E_{C}=E_{max}$, which maximizes the spectral accuracy of $\mathcal{H}_{\text{Eff}}(\mathbf{p})$ [see Fig.~\ref{fig:H_eff_benchmark} and the surrounding text].
Finally, we use the eigenstates of $\mathcal{H}_{\mathrm{Eff}}({\bf p})$ to compute the amorphous [disordered] Wilson loop spectrum [Appendix~\ref{sec:WilsonBerry}] on a sphere surrounding the nodal degeneracy at ${\bf p}={\bf 0}$.

In Fig.~\ref{fig:KW_WL}, we show the sphere Wilson loop spectrum for the disordered KW model for 10 disorder ensembles with 50 replicas each, where each disorder replica has $N_{\mathrm{sites}}=15^{3}=3375$ sites, nematic Gaussian lattice and local frame disorder with the standard deviation $\eta=0.2$ [see Appendix~\ref{app:lattices}], and contiguous domains of right- and left-handed sites with varying chirality concentrations respectively given by $n_R=N_R/N_{\mathrm{sites}}$ and $n_{L}=1-n_{R}$.
Beginning in Fig.~\ref{fig:KW_WL}(a) with a moderately disordered system [$\eta=0.2$, see Fig.~\ref{fig:3DGreenKW}] containing almost entirely right-handed sites [$n_{L}=0.05$] and continuing in increasing $n_{L}$ to the system in Fig.~\ref{fig:KW_WL}(j) with almost entirely left-handed sites [$n_{L}=0.95$], we observe a quantized Wilson loop winding of $C=1$ for $n_{L}<0.5$ [Fig.~\ref{fig:KW_WL}(a-d)] and $C=-1$ for $n_{L}>0.5$ [Fig.~\ref{fig:KW_WL}(f-j)].
In the vicinity of $n_{L}\approx 0.5$,
the Wilson loop eigenvalues become non-smooth, indicating that the sphere Wilson loop is within the close vicinity of a topological quantum critical point [energy gap closure]; the onset of this behavior can be seen in Fig.~\ref{fig:KW_WL}(e).
The Wilson loop spectra in Fig.~\ref{fig:KW_WL} overall represent a central result of the present work, as they for the first time provide precise \emph{quantized} indicators of nodal [gapless] topology in fully structurally disordered 3D metals.
In the case of the disordered KW model introduced in this section [Eq.~(\ref{eq:amorphousKWTmatrixFinalChirality})], the Wilson loop spectra in Fig.~\ref{fig:KW_WL} specifically indicate that the nodal degeneracy at ${\bf p}={\bf 0}$ in Fig.~\ref{fig:KW_Bands_amo} is a disordered chiral [KW] fermion that, analogous to its crystalline counterpart~\cite{KramersWeyl,AlPtObserve,PdGaObserve,DingARPESReversal,SessiPdGaQPIReversal}, carries a topological chirality that is tunable via the average system structural chirality.  
Though we have only demonstrated quantized and tunable Wilson loop winding in Fig.~\ref{fig:KW_WL} for systems with moderate structural disorder [$\eta=0.2$], we will soon below show that the link between average structural chirality and low-energy topological chirality also holds for the non-crystalline KW model [Eq.~(\ref{eq:amorphousKWTmatrixFinalChirality})] on fully-disordered [\emph{e.g.} random] lattices that lack well-defined crystalline limits [see Fig.~\ref{fig:KW_Unif}].

\paragraph*{\bf Non-Crystalline Group Theory} -- $\ $ We will next make further direct connection between the disordered KW fermion in Fig.~\ref{fig:KW_Bands_amo} and its crystalline counterpart by employing a ${\bf k}\cdot {\bf p}$ deformation procedure based in symmetry group theory.  
First, as discussed in Eq.~(\ref{eq:appHKP}) and the surrounding text, when Eq.~(\ref{eq:amorphousKWTmatrixFinalChirality}) is placed on a regular orthorhombic lattice and expanded in crystal momentum ${\bf k}$ about the $\Gamma$ point [${\bf k}={\bf 0}$], the resulting ${\bf k}\cdot{\bf p}$ Hamiltonian characterizes a crystalline KW fermion:
\begin{equation}
\mathcal{H}({\bf k}) = v_{x}k_{x}\sigma^{x} + v_{y}k_{y}\sigma^{y} + v_{z}k_{z}\sigma^{z}.
\label{eq:KWnumericsKP}
\end{equation}
The KW ${\bf k}\cdot {\bf p}$ Hamiltonian in Eq.~(\ref{eq:KWnumericsKP}) specifically transforms in a two-dimensional, double-valued small corep of the little group at $\Gamma$, which is isomorphic to orthorhombic SG 16 ($P222$).  
As discussed in Appendices~\ref{app:pseudoK} and~\ref{app:corepAmorphous} and established in Refs.~\cite{zallen_physics_1998,VanMechelen:2018cy,vanMechelenNonlocal,Ciocys2023,Grushin2020,Corbae_2023}, systems with strong structural disorder are spectrally isotropic at long wavelengths in their $d_{A}$ disordered ${\bf p}$-space directions [Eq.~(\ref{eq:finiteDimAmorph})].
Under the long-wavelength deformation and averaging procedure for $\Gamma$-point Hamiltonians introduced in this work [see Eq.~(\ref{eq:SG16rotations}) and the following text], the exact Hamiltonian $\mathcal{H}({\bf k})$ in Eq.~(\ref{eq:KWnumericsKP}) can be recast as an approximate, isotropic, \emph{effective} ${\bf k}\cdot {\bf p}$ Hamiltonian expanded about ${\bf p}={\bf 0}$ in the non-crystalline KW model [Eq.~(\ref{eq:amorphousKWTmatrixFinalChirality})] on a $d_{A}=3$ strongly disordered or random lattice:
\begin{equation}
\mathcal{H}_{\text{Eff}}({\bf p}) = \tilde{v}\left(p_{x}\tilde{\sigma}^{x} + p_{y}\tilde{\sigma}^{y} + p_{z}\tilde{\sigma}^{z}\right) + \tilde{\mu}\tilde{\sigma}^{0},
\label{eq:KWdisorderKP}
\end{equation}
where $\tilde{v}$ indicates the strength of disorder-renormalized Dresselhaus SOC, $\tilde{\mu}$ is the disorder-renormalized chemical potential [see Fig.~\ref{fig:KW_Bands_amo}], $\tilde{\sigma}^{0}$ is the $2\times 2$ identity matrix, and where each $2\times 2$ Pauli matrix $\tilde{\sigma}^{i}$ is generically equal to a linear combination of $\sigma^{x,y,z}$ in Eq.~(\ref{eq:KWnumericsKP}) due to SOC reference frame disorder [see the text surrounding Eq.~(\ref{appeq:rotaKW})].

\begin{figure}
\centering
\includegraphics[width=\linewidth]{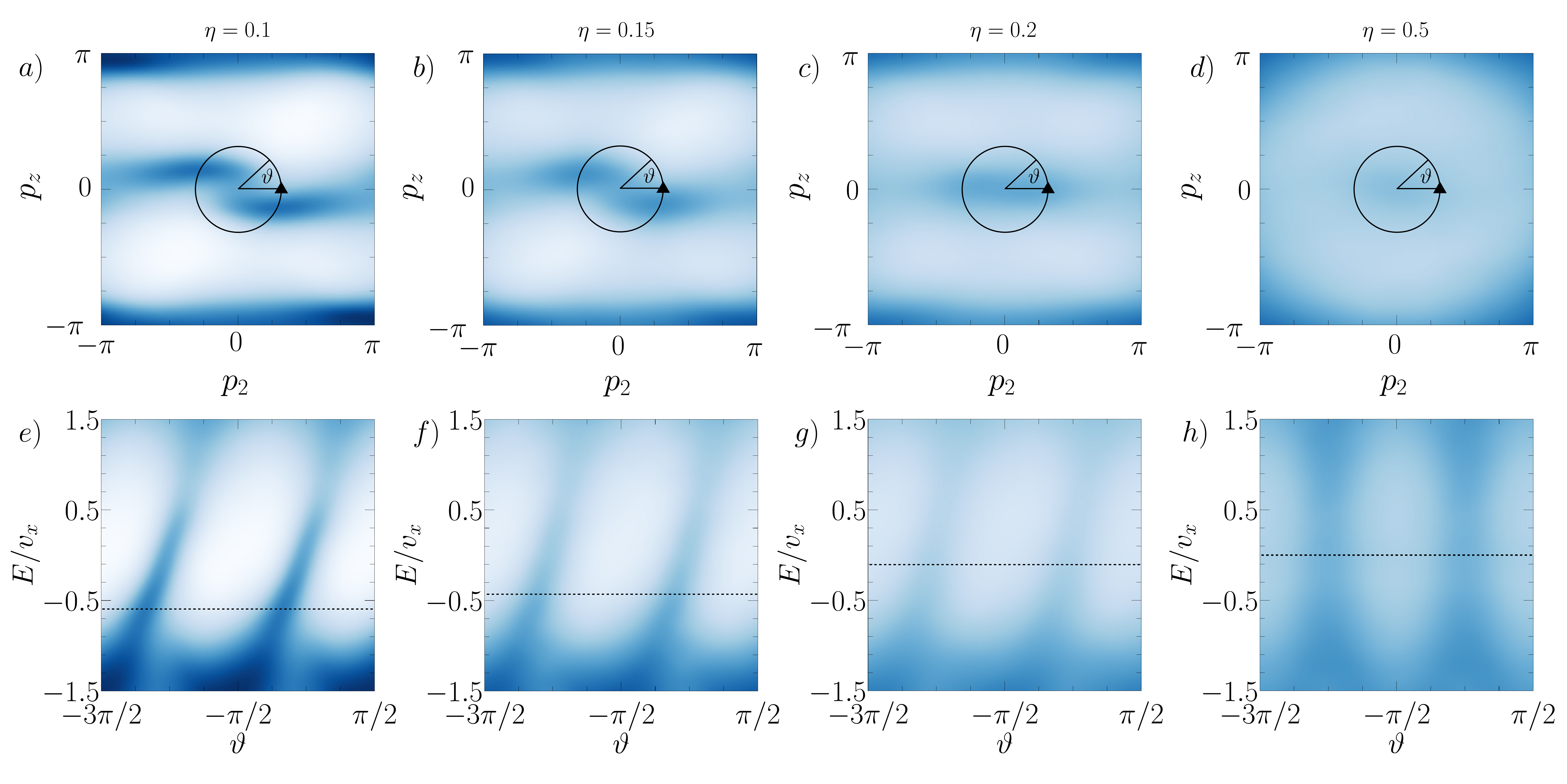}
\caption{Fermi-arc surface states of the disordered Kramers-Weyl model.
In this figure, we show the $(\hat{x}+\hat{y})$-normal surface-projected, disorder-averaged spectral function $\bar{A}_{\text{surf}}(E,{\bf p})$ [Eq.~(\ref{eq:averageSpectrumSurface})] of the non-crystalline Kramers-Weyl [KW] tight-binding model [Eq.~(\ref{eq:amorphousKWTmatrixFinalChirality})] on a lattice with increasing random nematic structural disorder [$d_{A}=3$ in Eq.~(\ref{eq:finiteDimAmorph})] parameterized by the standard deviation $\eta$, using the same chirality imbalance percentages and system parameters as in Fig.~\ref{fig:DOSKW}.
Each panel was generated by averaging over 50 disorder replicas, each with $N_{\text{sites}}=20^{3}$ prior to the removal [``evaporation''] of accidental dangling [decoupled] surface atoms generated by the disorder realization [see the text preceding Eq.~(\ref{eq:RealSpaceGreenSlab})]. 
Because surface chirality domain walls can bind flat-band-like states that overwhelm and obscure spectral signatures of topological surface Fermi arcs~\cite{InternalChiralTheory,InternalChiralExp,SussmanJammedDiracWeyl}, we have only implemented random chirality domain walls within the bulk of the finite [slab] systems used to generate this figure.
In panels (a-d), we plot $\bar{A}_{\text{surf}}(E,{\bf p})$ as a function of $p_{2}=(1/\sqrt{2})(p_{x}-p_{y})$ and $p_{z}$ for increasing $\eta$ at a fixed relative energy $E/v_{x}$ [respectively the dashed line in (e-h)] set to 0.71 below the maximum bulk spectral weight at ${\bf p}={\bf 0}$, in order to account for the disorder-renormalized chemical potential of the KW point [see Figs.~\ref{fig:KW_Bands_amo} and~\ref{fig:DOSKW}].  
In (e-h), we plot $\bar{A}_{\text{surf}}(E,{\bf p})$ as a function of energy on counterclockwise circular paths, parameterized by $\vartheta$, surrounding $p_{2}=p_{z}=0$ for the increasingly disordered KW systems in (a-d), respectively.
In (a-c) and (e-g), $\bar{A}_{\text{surf}}(E,{\bf p})$ continues to exhibit two clear -- but increasingly diffuse -- Fermi-arc surface states with the same connectivity and topological chirality [positive slopes] as the right-handed enantiomer of the crystalline KW model [Fig.~\ref{fig:KW_figWilson}(b,c)].
More subtly, like the bulk KW fermion at ${\bf p}={\bf 0}$ [Fig.~\ref{fig:KW_Bands_amo}] and the Dirac-cone surface states of amorphous Bi$_{2}$Se$_{3}$~\cite{corbae_evidence_2020,Ciocys2023}, the surface Fermi-arc Fermi velocities are also renormalized to larger values with increasing $\eta$.
In the strong-disorder regime [$\eta=0.5$], the Fermi arcs are nearly indistinguishable in (d) the constant-energy spectral function, having largely merged with the bulk KW Fermi pocket at ${\bf p}={\bf 0}$, as well as the isotropic [sphere-like] bulk spectral features at $|{\bf p}|=\pi$ and larger momenta [Figs.~\ref{fig:DOSKW}(c) and~\ref{fig:SpinTextureKW}(c)].
Intriguingly, the triviality of the strong-disorder surface states can also be seen in (h) through their spectral dispersion, in which the surface-state Fermi velocities have been renormalized to nearly infinite [singular] values, such that the surface spectral features no longer exhibit well-defined 1D topological chirality [positive, non-singular slopes].}
\label{fig:KWamorphousArcs}
\end{figure}

Like the ${\bf k}\cdot{\bf p}$ Hamiltonian of a crystalline KW fermion [see Ref.~\cite{KramersWeyl} and the text following Eq.~(\ref{eq:appTBparams})], $\mathcal{H}_{\text{Eff}}({\bf p})$ in Eq.~(\ref{eq:KWdisorderKP}) transforms in a two-dimensional, double-valued small corep of a $d=3$ chiral little group $\tilde{G}_{\Gamma,3}$.
However unlike for crystalline KW fermions, $\tilde{G}_{\Gamma,3}$ is not an exact, discrete little group, but is now instead a \emph{continuous} and \emph{approximate} [average] little group [ALG] $\tilde{G}_{\Gamma,3}$ that can be decomposed as:
\begin{equation}
\tilde{G}_{\Gamma,3} = \text{SO}(3) \cup \{\mathcal{T}|000\}\text{SO}(3) \cup \{E|\epsilon 00\}\text{SO}(3) \cup \{\mathcal{T}|\epsilon 00\}\text{SO}(3),
\label{eq:KWnumericsALG}
\end{equation}
where $\mathcal{T}$ is time-reversal and $\{E|\epsilon 00\}$ is an infinitesimal translation along the $x$-axis.
Importantly, for consistency with the crystallographic rotation [point] groups~\cite{ConwaySymmetries,BigBook,BilbaoPoint,PointGroupTables,SpinPointMcClarty}, we employ a convention in Eq.~(\ref{eq:KWnumericsALG}) in which the spinful isotropic rotation group is denoted as the double group of SO($3$), rather than as a distinct group SU($2$).
The generating symmetries of $\tilde{G}_{\Gamma,3}$ in Eq.~(\ref{eq:KWnumericsALG}) can be represented through their action on $\mathcal{H}_{\text{Eff}}({\bf p})$ in Eq.~(\ref{eq:KWdisorderKP}):
\begin{eqnarray}
C_{(2\pi/\phi) z}\mathcal{H}_{\text{Eff}}({\bf p})C_{(2\pi/\phi) z}^{-1} &=& e^{i (\phi/2)\tilde{\sigma}^{z}}\mathcal{H}_{\text{Eff}}(C^{-1}_{(2\pi/\phi) z}{\bf p})e^{-i (\phi/2)\tilde{\sigma}^{z}}, \nonumber \\
C_{(2\pi/\theta) x}\mathcal{H}_{\text{Eff}}({\bf p})C_{(2\pi/\theta) x}^{-1} &=& e^{i (\theta/2)\tilde{\sigma}^{x}}\mathcal{H}_{\text{Eff}}(C^{-1}_{(2\pi/\theta) x}{\bf p})e^{-i (\theta/2)\tilde{\sigma}^{x}}, \nonumber \\
\mathcal{T}\mathcal{H}_{\text{Eff}}({\bf p})\mathcal{T}^{-1} &=& \tilde{\sigma}^{y}\mathcal{H}^{*}_{\text{Eff}}(-{\bf p})\tilde{\sigma}^{y},
\label{eq:NumericsSecKWsymmetryAction}
\end{eqnarray}
where $\phi$ denotes an infinitesimal rotation angle about the $z$-axis [such that $\phi=\pi$ is consistent with the matrix representative of $C_{2z}$ in Eq.~(\ref{eq:pristineSyms})], $\theta$ denotes an infinitesimal rotation angle about the $x$-axis [such that $\theta=\pi$ is consistent with the matrix representative of $C_{2x}$ in Eq.~(\ref{eq:pristineSyms})], and where we note that the continuous translation symmetries in $\tilde{G}_{\Gamma,3}$ are represented as phases multiplied by the $2\times 2$ identity matrix $\tilde{\sigma}^{0}$, and have hence been suppressed for notational simplicity.

\paragraph*{\bf Disordered Fermi-Arc Surface States} -- $\ $ Having precisely established that the bulk linear spectral feature at ${\bf p}={\bf 0}$ in Fig.~\ref{fig:KW_Bands_amo} is a non-crystalline KW fermion, we will next explore the fate of its topological surface Fermi arcs under increasing disorder.
We first note that many previous works have employed a combination of analytic, numerical, and experimental methods to investigate the survival of bulk topological chiral semimetal phases and their associated surface Fermi arcs under a large range of disorder conditions and non-crystalline settings~\cite{ringel2015,Chen2015,Altland2015,Altland2016,Gorbar2016,Slager2017,Buchhold2018,Buchhold2018b,Roy2018,Wilson2018,yang_topological_2019,Pixley2021,Brillaux2021,Franca2024,Franca2024b,Grossi2023b,WeylQuasicrystalSCBott,NdAlSi2024DisorderWeylExp,JedJustinCPGEWeylDisorder2024,YiBurkovDiffusiveWeyl}.
The previous works largely concluded that both bulk topological semimetal phases and their associated surface Fermi arcs appear to be generically converted by disorder to diffusive metals or insulators, with the surface and bulk phase transitions potentially occurring at distinct disorder scales.
However, nearly all of the previous disorder analyses were specifically performed on band-inversion-type Weyl semimetals with oppositely-charged Weyl points at $\pm {\bf k}$, which are easily mixed by disorder and are expected to merge into trivial bulk Fermi rings [spheres] as the system approaches the spectrally-isotropic amorphous limit [see Appendix~\ref{app:pseudoK} and Refs.~\cite{zallen_physics_1998,VanMechelen:2018cy,vanMechelenNonlocal,Ciocys2023,Grushin2020,Corbae_2023}].
More generally, the previous investigations of disordered topologically chiral semimetals were also performed using disorder realizations that were structurally achiral, on the average, or overlooked the role of average or exact structural chirality in their analyses.
This stands in stark contrast to the models and theoretical analysis in the present work, in which only bulk chiral fermions with the \emph{same} topological chiral charges are mixed by disorder, and in which the disordered systems carry average [net or imbalanced] structural chirality, and therefore remain gapless and bulk-topological up to strong disorder scales.

However, the presence of disorder-robust bulk chiral fermions does not itself guarantee that topological surface Fermi arcs will remain well-defined and resolvable spectral features in the strong-disorder regime.
For example even in crystalline KW semimetals, if orbital hopping [$t_{x,y,z}$ in Eqs.~(\ref{eq:KWTmatrix}) and~(\ref{eq:appHKW})] is much stronger than SOC [$v_{x,y,z}$ in Eqs.~(\ref{eq:KWTmatrix}) and~(\ref{eq:appHKW})], then each bulk KW point will exhibit a 3D Rashba-like dispersion without associated 2D surface Fermi arcs, due to the absence of well-isolated, surface-projecting bulk Fermi pockets with net-nontrivial topological chiral charges~\cite{KramersWeyl,WeylNodalSurfaceFollowupPRB}.
To explore disorder-driven shifts in spectral signatures of the bulk-boundary correspondence in KW semimetals, we begin by placing the non-crystalline KW tight-binding model [Eq.~(\ref{eq:amorphousKWTmatrixFinalChirality})] on a lattice with increasing random nematic structural disorder [$d_{A}=3$ in Eq.~(\ref{eq:finiteDimAmorph})] parameterized by the standard deviation $\eta$, using the same structural chirality imbalance percentages and system parameters as the bulk calculations in Fig.~\ref{fig:DOSKW}.
Unlike in our previous calculations, we now take the system to have periodic boundary conditions in the Cartesian $(\hat{x}-\hat{y})$- and $\hat{z}$-directions and open boundary conditions in the $(\hat{x}+\hat{y})$-direction. 
We next compute the $(\hat{x}+\hat{y})$-normal surface-projected spectral function $\bar{A}_{\text{surf}}(E,{\bf p})$ [Eq.~(\ref{eq:averageSpectrumSurface})] averaged over 50 disorder replicas [noting that the Cartesian $(\hat{x}+\hat{y})$-normal surface can no longer be designated the $(110)$-surface in a lattice-disordered system].
Each disorder replica in our system is initially constructed with $20^{3}=8000$ sites, which we then reduce by removing [``evaporating''] dangling [decoupled] surface atoms that represent numerical artifacts of the disorder implementation process [see the text preceding Eq.~(\ref{eq:RealSpaceGreenSlab})]. 
Additionally, because surface chirality domain walls can bind flat-band-like states that overwhelm and obscure spectral signatures of topological surface Fermi arcs~\cite{InternalChiralTheory,InternalChiralExp,SussmanJammedDiracWeyl}, then we only place chirality domain walls deep within the bulk of each disorder replica.

In Fig.~\ref{fig:KWamorphousArcs}, we plot the disorder-averaged surface spectral function $\bar{A}_{\text{surf}}(E,{\bf p})$ of the non-crystalline KW model for increasing disorder parameterized by $\eta$.
Specifically, in Fig.~\ref{fig:KWamorphousArcs}(a-d), we plot $\bar{A}_{\text{surf}}(E,{\bf p})$ as a function of $p_{2}=(1/\sqrt{2})(p_{x}-p_{y})$ and $p_{z}$ for increasing $\eta$ at a fixed energy, and in Fig.~\ref{fig:KWamorphousArcs}(e-h), we respectively plot $\bar{A}_{\text{surf}}(E,{\bf p})$ at the same $\eta$ as a function of energy on counterclockwise circular paths, parameterized by $\vartheta$, surrounding $p_{2}=p_{z}=0$.
In the weak-to-moderate disorder regime [$\eta=0.1-0.2$ in panels (a-c) and (e-g) of Fig.~\ref{fig:KWamorphousArcs}], $\bar{A}_{\text{surf}}(E,{\bf p})$ continues to exhibit two clear -- but increasingly diffuse -- Fermi-arc surface states with the same connectivity and topological chirality [positive slopes] as the right-handed enantiomer of the crystalline KW model [Fig.~\ref{fig:KW_figWilson}(b,c)], consistent with the average right-handedness of the disordered system.
More subtly, like the linearly dispersing states that comprise the disordered bulk KW fermion at ${\bf p}={\bf 0}$ in Fig.~\ref{fig:KW_Bands_amo}, the Fermi velocities of the surface Fermi arcs in 
Fig.~\ref{fig:KWamorphousArcs}(e-h) are also renormalized to larger values with increasing $\eta$.
This is also reminiscent of the experimentally-observed, nearly vertical, topological Dirac-cone surface states of amorphous Bi$_{2}$Se$_{3}$, whose Fermi velocities compared to those of crystalline Bi$_{2}$Se$_{3}$ are strongly upwardly renormalized by lattice disorder~\cite{corbae_evidence_2020,Ciocys2023}.
In the strong-disorder regime [$\eta=0.5$], the surface Fermi arcs of the non-crystalline KW model become nearly indistinguishable in the constant-energy spectral function [Fig.~\ref{fig:KWamorphousArcs}(d)], having largely merged with the bulk KW Fermi pocket at ${\bf p}={\bf 0}$, as well as the isotropic [sphere-like] bulk spectral features at $|{\bf p}|=\pi$ and larger momenta [Figs.~\ref{fig:DOSKW}(c) and~\ref{fig:SpinTextureKW}(c)].
Intriguingly, the triviality of the strong-disorder surface states can also be inferred through their spectral dispersion [Fig.~\ref{fig:KWamorphousArcs}(h)], in which the surface-state Fermi velocities have been renormalized to nearly infinite [singular] values, such that the surface spectral features no longer exhibit well-defined 1D topological chirality [positive, non-singular slopes].

Overall, we conclude that though the bulk KW semimetal state remains topological and gapless for strong disorder, it no longer exhibits topological Fermi arcs in the amorphous regime.
In fact, this is not an unexpected result.
Specifically, topological surface Fermi arcs can only appear in topological semimetals for which the bulk Fermi surface is divided into topologically distinct Fermi pockets that project to different regions of the surface BZ~\cite{AshvinWeyl,VolovikWeylMonopole,VolovikChiralAnomaly,HaldaneOriginalWeyl,MurakamiWeyl,BurkovBalents,AndreiWeyl,HasanWeylDFT,Armitage2018,SuyangWeyl,LvWeylExp,YulinWeylExp,AliWeylQPI,AlexeyType2,ZJType2,BinghaiClaudiaWeylReview,ZahidNatRevMatWeyl,CDWWeyl,IlyaIdealMagneticWeyl}.
However, we have already shown that in the strong-disorder [amorphous] regime of the non-crystalline KW model, the higher-momenta KW fermions become merged by disorder into sphere-like bulk Fermi pockets that fully surround the KW point at ${\bf p}={\bf 0}$ [see Figs.~\ref{fig:DOSKW}(c) and~\ref{fig:SpinTextureKW}(c)].
Like in the aforementioned limit of KW fermions with 3D Rashba dispersion, and in similar cases of topologically chiral nodal-surface semimetals~\cite{KramersWeyl,WeylNodalSurfaceFollowupPRB}, this ensures that there do not exist topological surface Fermi arcs, due to the absence of topologically nontrivial projected bulk Fermi pockets.

\begin{figure}
\centering
\includegraphics[width=\linewidth]{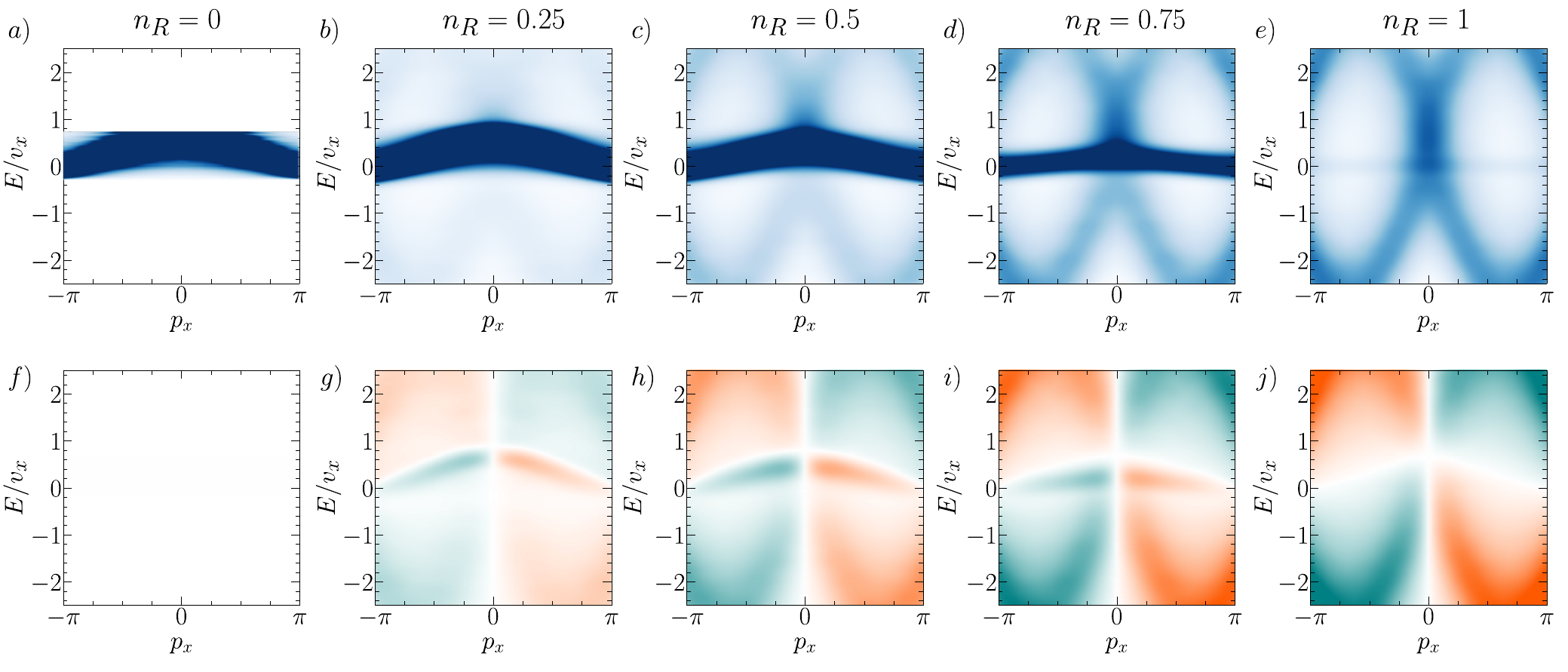}
\caption{Spectral function and spin texture of the disordered Kramers-Weyl model with coexisting chiral and achiral domains.
To generate each panel in this figure, we place the non-crystalline Kramers-Weyl [KW] model [Eq.~(\ref{eq:amorphousKWTmatrixFinalChirality})] with the parameters in Eq.~(\ref{eq:disorderedKWparams}) on a lattice with $N_{\mathrm{sites}}=15^3=3375$ and strong random nematic Gaussian structural and local frame disorder parameterized by the fixed standard deviation $\eta=0.5$, as detailed in Appendix~\ref{app:lattices}. 
We then generate the replica-averaged momentum-resolved Green's function $\bar{\mathcal{G}}(E,\mathbf{p})$ [Eq.~(\ref{eq:averageOneMomentumGreen})] by averaging the system over 50 disorder realizations [replicas] that -- unlike the previous analyses in this section -- each contain contiguous domains of either right-handed sites [$\chi_{\alpha}=1$] or \emph{achiral} sites [$\chi_{\alpha}=0$] with the respective concentrations $n_R=N_R/N_{\mathrm{sites}}$ and $n_{A}=1-n_{R}$.
(a-e) The disorder-averaged system spectral function $\bar{A}(E,{\bf p}) \propto \text{Im}\{\Tr[\bar{\mathcal{G}}(E,{\bf p})]\}$ [Eq.~(\ref{eq:SpecFunc})] for increasing $n_{R}$.  Even for (b) a mostly achiral system [$n_{R}=0.25$], faint linear Kramers-Weyl [KW] bands are visible near ${\bf p}={\bf 0}$ on top of a large background signal of nondispersive trivial states at $E/v_{x} \approx 0$, and (c-e) become increasingly well-resolved for increasing $n_{R}$.
(f-j) The $\langle S^{x}(E,{\bf p})\rangle$ component of the spin-dependent spectral function vector [Eq.~\eqref{eq:SpinDOS}] for the systems in (a-e), respectively, plotted using a log-scale color map in which orange is positive and teal is negative.
As shown below in Fig.~\ref{fig:KW_Spin_Pol}, the nondispersive trivial states are completely spin-unpolarized [$P_{\bf S}(E,{\bf p})\approx 0$ in Eq.~(\ref{eq:KWspinPolarization}) for all values of $n_{R}$], whereas the dispersive KW states exhibit a high degree of spin polarization.  
$\langle S^{x}(E,{\bf p})\rangle$ in particular is only nonvanishing for the topological disordered KW states, and may hence provide a signature of topological chirality in a strongly disordered system that can in principle be detected in spin- and angle-resolved photoemission spectroscopy [spin-ARPES] experiments~\cite{HasanKaneReview,ZahidHsieh3DTISpinTexture,SuYangWeylSpinTexture,NielsChiralSpinTexture,ChiralTellurium1,ChiralTellurium2,BiBrSpinARPES,KagomeSpinARPES,2DWeylTayRongSpinARPES,corbae_evidence_2020}.}
\label{fig:KW_Spin_Bands}
\end{figure}

\paragraph*{\bf Experimental Signatures of Coexisting Chiral and Achiral Domains} -- $\ $ 
We have thus far in this section only considered systems in which all sites within each domain are either right- or left-handed [$\chi_{\alpha}=\pm 1$ in Eq.~(\ref{eq:amorphousKWTmatrixFinalChirality})], such that each system above is in general a mix of right- and left-handed chirality domains of varying concentrations.
However, one may also consider systems in which some -- or even most -- sites are \emph{achiral} [$\chi_{\alpha}=0$ in Eq.~(\ref{eq:amorphousKWTmatrixFinalChirality})].
Theoretical models with coexisting chiral and achiral regions have previously been introduced to explain the nonvanishing optical activity [gyrotropy] of so-called ``chiral'' glasses like As$_2$S$_3$, which can exhibit optical signatures of structural chirality despite strong structural disorder~\cite{ChiralGlass1,ChiralGlass2,ChiralGlass3,ChiralGlassGiuliaSummary,ChiralGlassMolecules}.
Specifically, the authors of Refs.~\cite{ChiralGlass2,ChiralGlass3} suggested modeling inorganic chiral glasses as networks of short-range-interacting molecular units in which most structural centers are achiral, but in which a small concentration [1-10\%] of centers [local units] are conversely chiral and carry the same handedness. 
Theoretical studies~\cite{AmorphousTeDFT,AmorphousChalcogenideNatCommH} of amorphous tellurium~\cite{AmorphousTeSynthesis1,AmorphousTeSynthesis2} have also suggested the presence of ring- and chain-like local structural motifs with chiral point group symmetry that coexist with other achiral local structures.
Optical activity [chirality] has further been induced in thin films of the amorphous phase-change material Ge$_2$Sb$_2$Te$_5$, and may there also arise from chiral defect centers similar to those proposed in chiral chalcogenide glasses~\cite{ChiralPhaseChange1,ChiralPhaseChange2}.
Additionally, topological chirality in the $2+1$-D quantum Hall sense has recently been experimentally induced in a photonic metamaterial in which most sites are nonmagnetic [achiral by analogy], but where a small number [$\sim 10\%$] of random sites break time-reversal in the same ``direction'' [are chiral with the same handedness, by analogy]~\cite{MixedChiralPhotonicAlloy}.
Inspired by Refs.~\cite{ChiralGlass1,ChiralGlass2,ChiralGlass3,ChiralGlassGiuliaSummary,ChiralGlassMolecules,AmorphousTeDFT,AmorphousChalcogenideNatCommH,ChiralPhaseChange1,ChiralPhaseChange2,MixedChiralPhotonicAlloy}, we therefore next construct a disordered KW system that contains coexisting chiral and achiral domains with varying relative concentrations.
To implement a mixed chiral-achiral system, we begin by placing the non-crystalline KW model on a lattice with $N_{\mathrm{sites}}=15^3=3375$ and strong [$\eta=0.5$] random nematic Gaussian structural and local frame disorder [see Appendix~\ref{app:lattices}].
We then generate the replica-averaged momentum-resolved Green's function $\bar{\mathcal{G}}(E,\mathbf{p})$ [Eq.~(\ref{eq:averageOneMomentumGreen})] by averaging the system over 50 disorder realizations [replicas].
However, unlike the previous analyses in this section, each replica now contains contiguous domains of either right-handed sites [$\chi_{\alpha}=1$] or \emph{achiral} sites [$\chi_{\alpha}=0$], whose concentrations are respectively given by $n_R=N_R/N_{\mathrm{sites}}$ and $n_{A}=1-n_{R}$.

In Fig.~\ref{fig:KW_Spin_Bands}(a-e), we plot the disorder-averaged spectral function $\bar{A}(E,{\bf p}) \propto \text{Im}\{\Tr[\bar{\mathcal{G}}(E,{\bf p})]\}$ [Eq.~(\ref{eq:SpecFunc})] of the mixed chiral-achiral system for increasing $n_{R}$.
We find that $\bar{A}(E,{\bf p})$ exhibits topological disordered KW bands that coexist with nondispersive trivial states near $E/v_{x} \approx 0$, where the relative spectral weights of the topological and trivial states respectively scale with $n_{R}$ and $n_{A}$.
Notably, even for a mostly achiral system [$n_{R}=0.25$ in Fig.~\ref{fig:KW_Spin_Bands}(b)], faint linear KW bands are visible near ${\bf p}={\bf 0}$ on top of a large background signal of nondispersive trivial states, and become increasingly well-resolved for increasing $n_{R}$ [Fig.~\ref{fig:KW_Spin_Bands}(c-e)].

\begin{figure}
\centering
\includegraphics[width=\linewidth]{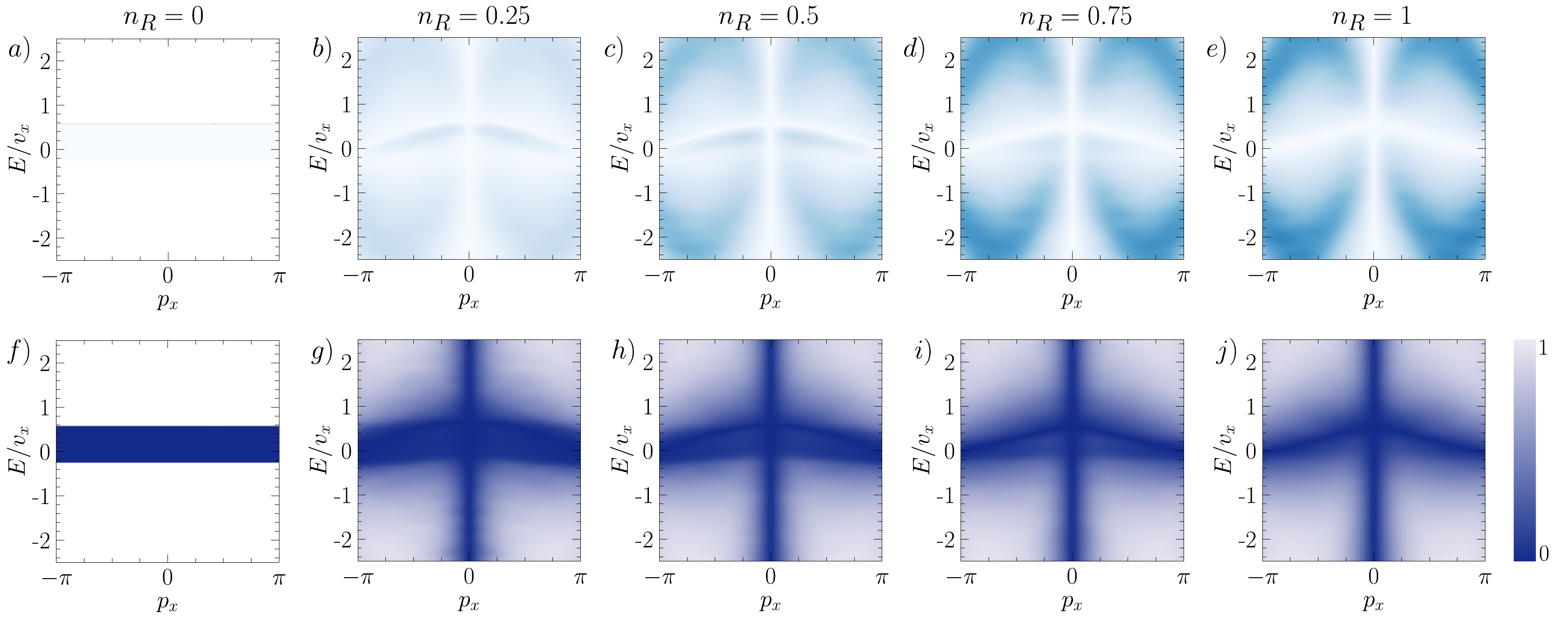}
\caption{Spin polarization of the disordered Kramers-Weyl model with coexisting chiral and achiral domains.  
(a-e) Magnitude of the spin-dependent spectral function vector $|\langle {\bf S}(E,{\bf p})\rangle|$ [Eq.~(\ref{eq:SpinDOS})] for the strongly disordered Kramers-Weyl [KW] model in Fig.~\ref{fig:KW_Spin_Bands} in which each replica contains contiguous domains of either right-handed or achiral sites with the respective concentrations $n_R=N_R/N_{\mathrm{sites}}$ and $n_{A}=1-n_{R}$, where $n_{R}$ is successively increased from (a) $n_R=0$ to (e) $n_R=1$.  
(f-j) Spin polarization $P_{\bf S}(E,{\bf p})$ [Eq.~(\ref{eq:KWspinPolarization})] of the disordered KW systems in (a-e), respectively.  
In the system spectral function [Fig.~\ref{fig:KW_Spin_Bands}(a-e)], topological disordered KW bands coexist with nondispersive trivial states near $E/v_{x} \approx 0$, where the relative spectral weights of the topological and trivial states respectively scale with $n_{R}$ and $n_{A}$.
In the disordered KW model [Eq.~(\ref{eq:amorphousKWTmatrixFinalChirality})], only the topological KW states are spin-polarized, whereas the nondispersive trivial states exhibit vanishing $|\langle {\bf S}(E,{\bf p})\rangle|$ and $P_{\bf S}(E,{\bf p})$ for all $n_{R}$. 
}
\label{fig:KW_Spin_Pol}
\end{figure}

\begin{figure}
\centering
\includegraphics[width=\linewidth]{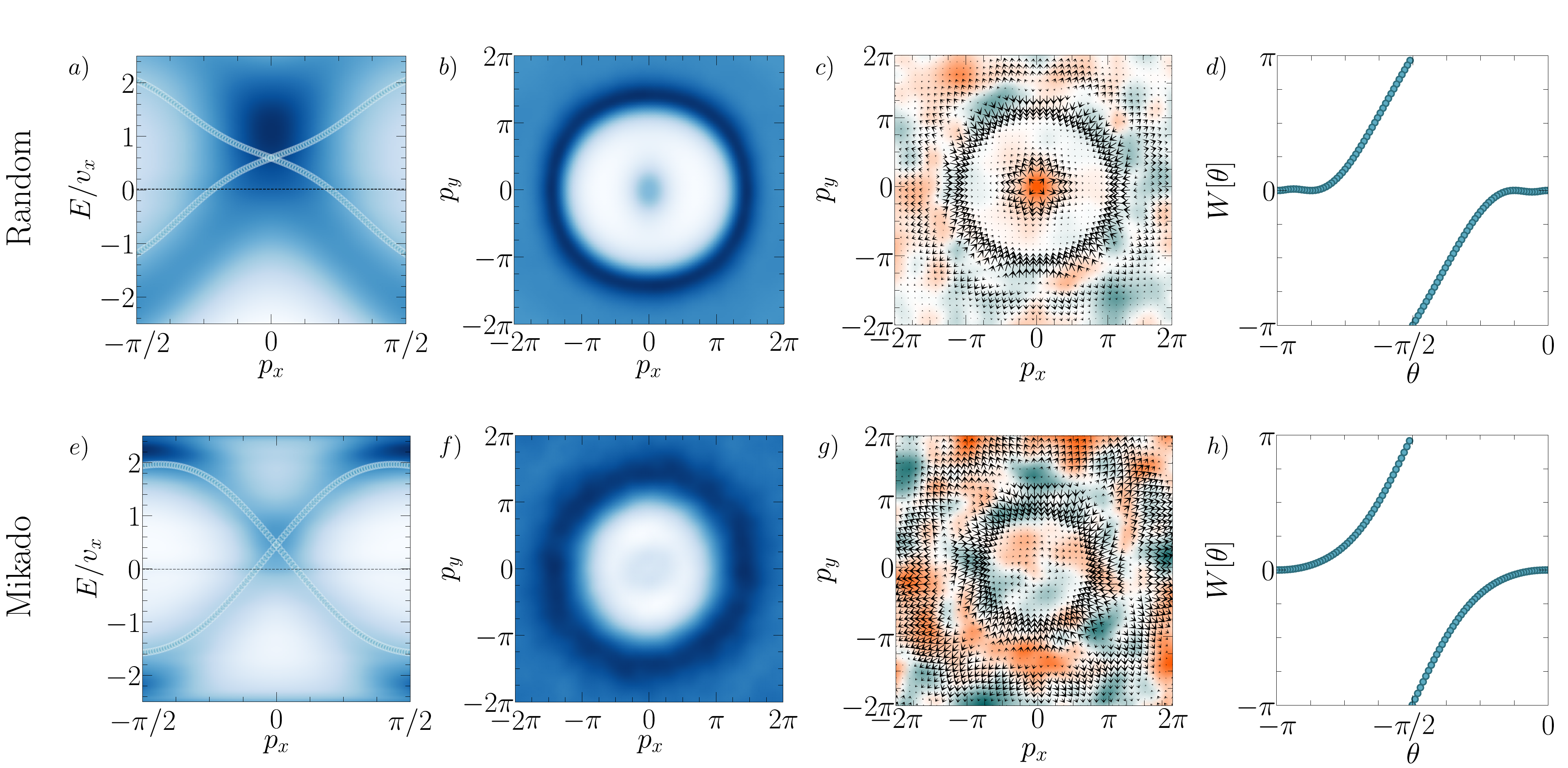}
\caption{Kramers-Weyl fermions on random and Mikado lattices.
(a-d) Bulk spectrum, spin texture, and topology of the non-crystalline Kramers-Weyl [KW] model [Eq.~(\ref{eq:amorphousKWTmatrixFinalChirality})] on a lattice with randomly located sites in $d=d_a=3$ dimensions, random frame disorder parameterized by the standard deviation $\eta=0.5$, and chirality domains of unequal volume [Appendix~\ref{app:lattices}].
(e-h) Bulk spectrum, spin texture, and topology of the non-crystalline KW model on a 3D Mikado lattice [Appendix~\ref{app:DiffTypesDisorder}] with random frame disorder parameterized by the standard deviation $\eta=0.5$ and chirality domains of unequal volume.
For all panels in this figure, the data were generated using Eq.~(\ref{eq:amorphousKWTmatrixFinalChirality}) with the tight-binding parameters in Eq.~(\ref{eq:disorderedKWparams}) implemented with a single domain in each replica of right-handed sites with $n_R=N_R/N_{\mathrm{sites}}=2/3$, and with the remaining volume in each replica containing a contiguous domain of left-handed sites with a corresponding concentration of $n_L=1-N_R/N_{\mathrm{sites}}=1/3$.
Each panel shows data generated by averaging over 50 randomly-generated replicas with $\sim 20^3$ sites each, as detailed in Appendix~\ref{app:PhysicalObservables}.
In both the (a) random and (e) Mikado lattice, the bulk average spectral function $\bar{A}(E,{\bf p})$ [Eq.~(\ref{eq:SpecFunc})] exhibits a linearly dispersing amorphous KW fermion at ${\bf p}={\bf 0}$.
The Mikado-lattice KW states at ${\bf p}={\bf 0}$ in (e) carry a lower spectral weight than those in the random lattice in (a), and coexist with faint, nondispersing trivial states near $E/v_{x} \approx 0$.
This occurs because the Mikado lattice [Fig.~\ref{appfig:structuraldisorder}] contains a significant number of long bonds which, when combined with the exponential decay contribution $\exp(a-|\mathbf{d}_{\alpha\beta}|)$ to the hopping rangedness function $f(|\mathbf{d}_{\alpha\beta}|)$ in Eq.~(\ref{eq:KWHeaviside}), gives rise to a significant number of effectively undercoordinated sites~\cite{marsal_obstructed_2022}. 
(b,f) $\bar{A}(E,{\bf p})$ at $p_{z}=0.1$ respectively at (b) $E/v_{x} = 0$ for the random lattice and (f) $E/v_{x}=0$ for the Mikado lattice [the dashed lines in (a,e), respectively].
In addition to the KW fermion at ${\bf p}={\bf 0}$, broad, ring- [sphere-] like spectral features appear in both (b,e) at $|{\bf p}|=\pi/\bar{a}$ and $|{\bf p}|=\pi\sqrt{2}/\bar{a}$, where $\bar{a}=1$ is the average nearest-neighbor spacing.
As shown in Figs.~\ref{fig:DOSKW} and~\ref{fig:SpinTextureKW}, the ring-like features in (b,e) originate from the crystalline KW fermions at $|{\bf k}|=\pi$ and $|{\bf k}|=\pi\sqrt{2}$, analogously to the ring-like, higher-Brillouin-zone Dirac surface states recently observed in amorphous Bi$_{2}$Se$_{3}$~\cite{corbae_evidence_2020,Ciocys2023}.
(c,g) The spin texture [Eq.~(\ref{eq:SpinDOS})] of the spectra in (b,e), respectively.  
In both the (c) random and (g) Mikado lattices, the KW fermion at ${\bf p}={\bf 0}$ exhibits a nearly perfect outward-pointing, monopole-like spin texture like the Gaussian-disordered KW fermion at ${\bf p}={\bf 0}$ in Fig.~\ref{fig:SpinTextureKW}, albeit with significantly weaker spectral weight for the Mikado lattice in (g).
Also like in the Gaussian-disordered KW model [Fig.~\ref{fig:SpinTextureKW}(b,c)], the ring-like features at $|{\bf p}|=\pi/\bar{a}$ [$|{\bf p}|=\pi\sqrt{2}/\bar{a}$] in (c,g) exhibit inward- [outward-] pointing spin textures.
(d,h) The amorphous Wilson loop spectra of the KW fermion at ${\bf p}={\bf 0}$ in (a,e), respectively, generated in each system from an effective Hamiltonian [light blue circles in (a,e)] constructed with its reference energy cut $E_{C}$ centered at the maximum value of the spectral function at ${\bf p}={\bf 0}$ [$E_c/v_x = 0.7$ in (a,e), see Appendix~\ref{app:EffectiveHamiltonian} for further details].
In both (d,h), the Wilson loop spectrum winds once in the positive direction, indicating that the non-crystalline KW fermion at ${\bf p}={\bf 0}$ carries a chiral charge of $C=1$, consistent with the average right-handedness of both systems [$n_{R}>n_{L}$, see Fig.~\ref{fig:KW_WL}].}
\label{fig:KW_Unif}
\end{figure}

However, we importantly find that the KW bands are better revealed by comparing the overall energy spectrum to the spin-dependent spectrum.
To demonstrate this, we show in Fig.~\ref{fig:KW_Spin_Bands}(f-j) the spin textures of the mixed chiral-achiral disordered systems, focusing on the $\langle S^{x}(E,{\bf p})\rangle$ component of the spin-dependent spectral function vector [Eq.~\eqref{eq:SpinDOS}]. While the central band near $E/v_{x} \approx 0$ still carries some residual spectral weight in the $\langle S^{x}(E,{\bf p})\rangle$ spin texture, it is much smaller than that of the KW bands, \emph{even for small $n_R$}. 
To rule out analysis artifacts due to our choice of the $\langle S^{x}(E,{\bf p})\rangle$ spin-texture component, we also compute in Fig.~\ref{fig:KW_Spin_Pol}(a-e) the magnitude of the spin-dependent spectral function vector $|\langle {\bf S}(E,{\bf p})\rangle|$ [Eq.~(\ref{eq:SpinDOS})]. 
As with $\langle S^{x}(E,{\bf p})\rangle$ in Fig.~\ref{fig:KW_Spin_Bands}(f-j), $|\langle {\bf S}(E,{\bf p})\rangle|$ in Fig.~\ref{fig:KW_Spin_Pol}(a-e) shows appreciable spectral weight on the KW bands, and only shows a much smaller weight on the central trivial band compared to the overall spectral function in Fig.~\ref{fig:KW_Spin_Bands}(a-e).

Finally, to properly account for the difference in spectral weight between the central trivial band and the topological KW bands in Fig.~\ref{fig:KW_Spin_Bands}(a-e), we calculate the total spin polarization $P_{\bf S}(E,{\bf p})$ [Eq.~(\ref{eq:KWspinPolarization})] of the mixed-achiral system, which is shown in Fig.~\ref{fig:KW_Spin_Pol}(f-j).
Crucially, we see in Fig.~\ref{fig:KW_Spin_Pol}(f-j) that \emph{only} the topological disordered KW bands are highly spin-polarized, whereas the nondispersive trivial bands near $E/v_{x} \approx 0$ are completely spin-unpolarized [$P_{\bf S}(E,{\bf p})\approx 0$ in Eq.~(\ref{eq:KWspinPolarization})], even for small values of $n_{R}$.
This pattern of faint -- but highly spin-polarized -- dispersive topological bands on top of a large background signal of largely spin-unpolarized, nondispersive, trivial bands can in principle be detected via spin- and angle-resolved photoemission spectroscopy [spin-ARPES], which has proven to be a powerful tool in unraveling the spin textures of crystalline~\cite{HasanKaneReview,ZahidHsieh3DTISpinTexture,SuYangWeylSpinTexture,NielsChiralSpinTexture,ChiralTellurium1,ChiralTellurium2,BiBrSpinARPES,KagomeSpinARPES,2DWeylTayRongSpinARPES} and recently amorphous~\cite{corbae_evidence_2020} topological materials.

\paragraph*{\bf Random Lattices} -- $\ $ Lastly, one might be concerned that the topological and spectral properties of non-crystalline KW fermions obtained in this section are specific to our use of Gaussian structural disorder, which admits a well-defined crystalline limit.
To show that this is not the case, we will conclude this section by computing the bulk energy spectrum, spin texture, and ${\bf p}={\bf 0}$ amorphous Wilson loops of the non-crystalline KW model [Eq.~(\ref{eq:amorphousKWTmatrixFinalChirality})] on randomly generated lattices without well-defined [unique] crystalline limits.
To begin, we place the non-crystalline KW model on a $d=d_{A}=3$ lattice [Eq.~(\ref{eq:finiteDimAmorph})] with randomly located sites in three dimensions, strong random frame disorder parameterized by the standard deviation $\eta=0.5$, and contiguous chirality domains of unequal volume with the respective concentrations of right-handed sites $n_R=N_R/N_{\mathrm{sites}}=2/3$ and left-handed sites $n_L=1-N_R/N_{\mathrm{sites}}=1/3$ [see Appendix~\ref{app:lattices}].
To simulate a large, self-averaging system, we further randomly generate 50 lattices [replicas] with $20^3 = 8000$ sites each, and then compute the replica-averaged momentum-resolved Green's function $\bar{\mathcal{G}}(E,{\bf p},{\bf p}')$ [Eq.~(\ref{eq:averageTWoMomentumGreen})].
As previously for the KW model with strong Gaussian disorder [Fig.~\ref{fig:3DGreenKW}(c)], we find that the off-diagonal-in-momentum elements $\bar{\mathcal{G}}(E,{\bf p},{\bf p}')$ vanish [on the average] as well for the non-crystalline KW model on random lattices [Fig.~\ref{fig:3DGreenKW}(d)].
We may therefore again limit consideration to the diagonal-in-momentum replica-averaged momentum-resolved Green's function $\bar{\mathcal{G}}(E,\mathbf{p})$ [Eq.~(\ref{eq:averageOneMomentumGreen})], from which we below compute the spectral and topological properties of random-lattice KW fermions.

In Fig.~\ref{fig:KW_Unif}(a,b), we show the disorder-averaged, momentum-resolved spectral function $\bar{A}(E,{\bf p}) \propto \text{Im}\{\Tr[\bar{\mathcal{G}}(E,{\bf p})]\}$ [Eq.~(\ref{eq:SpecFunc})] of the random-lattice KW model respectively plotted as a function of energy and momentum [$p_{x}$] and at a fixed energy and $p_{z}$ as a function of the remaining two momenta $p_{x,y}$.
Like in the Gaussian-disordered lattice [Fig.~\ref{fig:KW_Bands_amo}], $\bar{A}(E,{\bf p})$ for the random lattice exhibits a linearly dispersing KW fermion at ${\bf p}={\bf 0}$ [Fig.~\ref{fig:KW_Unif}(a)].
Interestingly, $\bar{A}(E,{\bf p})$ also exhibits broadened, ring- [sphere-] like spectral features at  $|{\bf p}|=\pi/\bar{a}$ and $|{\bf p}|=\pi\sqrt{2}/\bar{a}$, where $\bar{a}=1$ is the average nearest-neighbor spacing [Fig.~\ref{fig:KW_Unif}(b)].
The same ring-like spectral features at $|{\bf p}|=\pi$ and $|{\bf p}|=\pi\sqrt{2}$ also appear in the non-crystalline KW model with Gaussian lattice disorder [Fig.~\ref{fig:DOSKW}(b,c)], and were there shown to originate from disorder-driven rotational averaging of the crystalline KW fermions at $|{\bf k}|=\pi$ and $|{\bf k}|=\pi\sqrt{2}$ [Fig.~\ref{fig:DOSKW}(a)], which gives rise to the characteristic isotropic spectral features of amorphous systems~\cite{spring_amorphous_2021,springMagneticAverageTI,corbae_evidence_2020,Ciocys2023}.

To further analyze the random-lattice KW spectral features in Fig.~\ref{fig:KW_Unif}(a,b), we next use $\bar{\mathcal{G}}(E,\mathbf{p})$ to compute the spin texture [Eq.~(\ref{eq:SpinDOS})], which is plotted in Fig.~\ref{fig:KW_Unif}(c) at the same energy and $p_{z}$ values as the bulk spectral function $\bar{A}(E,{\bf p})$ in Fig.~\ref{fig:KW_Unif}(b).  
Like in the non-crystalline KW model with Gaussian lattice disorder [Fig.~\ref{fig:SpinTextureKW}], the random-lattice KW fermion at ${\bf p}={\bf 0}$ in Fig.~\ref{fig:KW_Unif}(c) exhibits a nearly perfect outward-pointing, monopole-like spin texture, consistent with the average system right-handedness [see Ref.~\cite{KramersWeyl} and the text following Eq.~(\ref{eq:KWspinPolarization})].  
Also like in the Gaussian-disordered KW model [Fig.~\ref{fig:SpinTextureKW}(b,c)], the ring-like spectral feature at $|{\bf p}|=\pi$ [$|{\bf p}|=\pi\sqrt{2}$] in Fig.~\ref{fig:KW_Unif}(c) exhibits an inward- [outward-] pointing spin texture.
As discussed in the text surrounding Fig.~\ref{fig:SpinTextureKW}, this is consistent with the intuition above that the ring-like features originate from rotationally averaging the KW fermions at $|{\bf k}|=\pi$ and $|{\bf k}|=\pi\sqrt{2}$ in the crystalline KW model while preserving the link between their topology and spin texture.
This picture suggests that the ring-like spectral feature at $|{\bf p}|=\pi$ [$|{\bf p}|=\pi\sqrt{2}$] in Fig.~\ref{fig:KW_Unif}(c) represents a many-particle disordered KW fermion with a negative [positive] chiral charge.

In this regard, it is interesting to note that in the original works by Nielsen and Ninomiya~\cite{NielNino81a,NielNino81b}, it was suggested that chiral fermion doubling could potentially be circumvented on amorphous lattices.
Our spin-texture calculations for the non-crystalline KW model [Eq.~(\ref{eq:amorphousKWTmatrixFinalChirality})] on Gaussian-disordered [Fig.~\ref{fig:SpinTextureKW}], random [Fig.~\ref{fig:KW_Unif}(c)] and Mikado [Fig.~\ref{fig:KW_Unif}(g)] lattices hint that on an amorphous lattice with average structural chirality, fermion doubling may still hold via a sequence of sphere-like Fermi pockets with alternatingly signed, many-particle topological chiral charges that continue to appear down to the smallest system wavelengths [largest $|{\bf p}|$].
However, addressing this possibility would require a more careful treatment of tight-binding basis-state tails at small wavelengths [see the text surrounding Fig.~\ref{fig:dadtdf} and Eq.~(\ref{eq:TBbasisStates})], as well as formulating a Green's-function-based method for computing the many-particle chiral charges of disordered, sphere-like Fermi pockets at larger $|{\bf p}|$ than the ${\bf p}\approx {\bf 0}$ amorphous Wilson-loop method introduced in this work [Appendix~\ref{sec:WilsonBerry}].
We therefore leave this intriguing direction for future study.

Lastly, to precisely confirm that the linear nodal degeneracy at ${\bf p}={\bf 0}$ in Fig.~\ref{fig:KW_Unif}(a) is indeed a topological chiral [KW] fermion, we compute the sphere Wilson loop spectrum of the random-lattice KW model at ${\bf p}={\bf 0}$.  
To obtain the Wilson loop spectrum, we first construct an effective Hamiltonian $\mathcal{H}_{\mathrm{Eff}}({\bf p})=\mathcal{H}_{\mathrm{Eff}}(E_{C},{\bf p})$ [Eq.~(\ref{eq:AvgHEff}), light blue circles in Fig.~\ref{fig:KW_Unif}(a)] using a reference energy cut $E_{C}$ set to the energy of the largest spectral weight $\bar{A}(E,{\bf p})$ at ${\bf p}={\bf 0}$, which as shown in Fig.~\ref{fig:H_eff_benchmark} maximizes the spectral accuracy of $\mathcal{H}_{\mathrm{Eff}}({\bf p})$.
We then use the occupied [lower] bands of $\mathcal{H}_{\mathrm{Eff}}({\bf p})$ to compute the sphere Wilson spectrum.
For the random-lattice KW system with $n_{R} =2/3$ and $n_{L}=1/3$, the amorphous Wilson loop at ${\bf p}={\bf 0}$ winds once in the positive direction [Fig.~\ref{fig:KW_Unif}(d)], indicating that the nodal spectral feature at ${\bf p}={\bf 0}$ is a $C=1$ amorphous KW fermion.
This is consistent with our earlier Wilson loop analysis in Fig.~\ref{fig:KW_WL}, in which we demonstrated that the Gaussian-disordered KW model hosts a non-crystalline KW fermion at ${\bf p}={\bf 0}$ whose low-energy topological chirality is controlled by the average system chirality [\emph{i.e.} the ratio $n_{R}/n_{L}$].
Though not shown in Fig.~\ref{fig:KW_Unif}, we have numerically confirmed that the topological chiral charge of the random-lattice KW fermion at ${\bf p}={\bf 0}$ is also controlled by the ratio $n_{R}/n_{L}$, and specifically also undergoes a topological phase transition between $C = \pm 1$ when $n_{R}/n_{L} \approx 1$.

As a final test that the results in this section are system-independent and hence truly representative of $d_{A}=3$ [Eq.~(\ref{eq:finiteDimAmorph})] amorphous KW fermions, we again place the non-crystalline KW model on a $d=d_{A}=3$ random lattice, but instead now use a \emph{Mikado} lattice, which is generated by placing random planes in 3D space, rather than 3D points [see Fig.~\ref{appfig:structuraldisorder}(c) and Refs.~\cite{marsal_obstructed_2022,marsal_topological_2020}].
Importantly, in the 3D Mikado lattice models in this work, we enforce that each site carries a predefined [fixed] sixfold bonding coordination, \emph{even} if there exist closer sites [see Appendix~\ref{app:DiffTypesDisorder}].
In practice, this is implemented in our Mikado lattice models by modifying the hopping rangedness function $f(|\mathbf{d}_{\alpha\beta}|)$ [Eq.~(\ref{eq:KWHeaviside})] to only contain the exponential decay contribution $\exp(a-|\mathbf{d}_{\alpha\beta}|)$, and to only be nonzero for pairs of sites that lie along prespecified Mikado lattice bonds.
We next randomly generate $50$ Mikado lattice replicas with $\sim 20^{3}$ sites each and compute the replica-averaged momentum-resolved Green's function $\bar{\mathcal{G}}(E,{\bf p})$ [Eq.~(\ref{eq:averageOneMomentumGreen})].
Within each replica, we also incorporate strong random frame disorder parameterized by the standard deviation $\eta=0.5$, and implement contiguous chirality domains of unequal volume with the respective concentrations of right-handed sites $n_R=N_R/N_{\mathrm{sites}}=2/3$ and left-handed sites $n_L=1-N_R/N_{\mathrm{sites}}=1/3$ [see Appendix~\ref{app:lattices}].

In Fig.~\ref{fig:KW_Unif}(e-h), we show the bulk spectral function $\bar{A}(E,{\bf p})$, spin texture, and sphere Wilson spectrum of the Mikado-lattice KW model.  
The Mikado-lattice KW model in Fig.~\ref{fig:KW_Unif}(e-h) exhibits the same spectral features and topology as the corresponding properties of the random-lattice KW model in Fig.~\ref{fig:KW_Unif}(a-d), with the exception that on the Mikado lattice, the KW states at ${\bf p}={\bf 0}$ carry a lower spectral weight than those in the random lattice and coexist with faint, nondispersing trivial states near $E/v_{x} \approx 0$ [which are partially obscured by the dashed horizontal line in Fig.~\ref{fig:KW_Unif}(e)].
We can understand the appearance of flat-band-like trivial states in the Mikado lattice in Fig.~\ref{fig:KW_Unif}(e) by recognizing that unlike the random lattice, the $d_{A}=3$ Mikado lattice contains a significant number of long bonds, which when combined with the exponential decay contribution $\exp(a-|\mathbf{d}_{\alpha\beta}|)$ to the hopping rangedness function $f(|\mathbf{d}_{\alpha\beta}|)$ in Eq.~(\ref{eq:KWHeaviside}) gives rise to a correspondingly significant number of effectively undercoordinated sites~\cite{marsal_obstructed_2022}.
We note that if we had instead taken $f(|\mathbf{d}_{\alpha\beta}|)$ to be constant [independent of the intersite distance $|{\bf d}_{\alpha\beta}|$], as it is for example in the Weaire-Thorpe model of amorphous silicon~\cite{weaire_electronic_1971}, each site would instead carry an effective sixfold coordination of same-strength bonds [see Fig.~\ref{appfig:structuraldisorder}(c) and Refs.~\cite{marsal_obstructed_2022,marsal_topological_2020}], and flat-band-like trivial states would no longer appear in the bulk energy spectrum.   
Finally, as previously with the random-lattice KW model, we have numerically confirmed that the topological chiral charge of the Mikado-lattice KW fermion at ${\bf p}={\bf 0}$ is also controlled by $n_{R}/n_{L}$, and specifically also undergoes a topological phase transition between $C = \pm 1$ when $n_{R}/n_{L} \approx 1$.\\

To conclude, we have employed several different lattice realizations and numerical methods to demonstrate that on strongly disordered and inherently non-crystalline [random] lattices, there still exist amorphous generalizations of ${\bf p}={\bf 0}$ KW fermions with nontrivial topological chiral charges that are controlled by the average [net] system chirality.
Though we have focused in this section on systems with $d_{A}=3$ nematic disorder [see Appendix~\ref{app:SmecticNematicDisorder}], we have also confirmed that our conclusions remain valid for ${\bf p}={\bf 0}$ KW fermions with $d_{A}=2$, $d_{T}=1$ smectic disorder.
Below, in Appendix~\ref{app:amorphousCharge2}, we will explicitly numerically demonstrate the existence of amorphous generalizations of quadratic double-Weyl fermions~\cite{ZahidMultiWeylSrSi2,StepanMultiWeyl,AndreiMultiWeyl,XiDaiMultiWeyl} in systems with smectic disorder. 
Lastly, we have also confirmed that the random-lattice KW model in Fig.~\ref{fig:KW_Unif}(a-d) continues to exhibit a linearly dispersing non-crystalline KW fermion at ${\bf p}={\bf 0}$ with a quantized $|C|=1$ sphere Wilson loop winding number under the subsequent addition of weak Anderson [on-site chemical potential] disorder.

\newpage

\subsection{Double-Weyl Fermion Models}
\label{sec:Charge2}

\subsubsection{Symmetry and Chirality in a Crystalline Model with a $\Gamma$-Point Double-Weyl Fermion}
\label{app:PristineQuadratic}

We will next in this section analyze the bulk electronic structure, symmetry group theory, topology, surface states, and critical phases of a structurally chiral [see Appendix~\ref{app:symDefs}] crystalline tight-binding model with a symmetry-enforced $|C|=2$ chiral [double-Weyl] fermion~\cite{ZahidMultiWeylSrSi2,StepanMultiWeyl,AndreiMultiWeyl,XiDaiMultiWeyl} at the $\Gamma$ point [${\bf k}={\bf 0}$]. 
To construct a double-Weyl tight-binding model, we begin by placing a time-reversed pair of spinless, angular-momentum-one $p_{x} \pm i p_{y}$ magnetic atomic orbitals~\cite{MTQC} at each site of a lattice.
In real space, the atomic orbitals are coupled via a Hamiltonian of the form:
\begin{equation}
    \mathcal{H}_{\mathrm{C2}} = \sum_{\langle \alpha\beta \rangle}\sum_{l,l'\in \left\{1,2\right\}}c_{\alpha,l}^{\dagger}\langle {\bf r}_\alpha,l | \mathcal{H} | {\bf r}_\beta,l'\rangle c^{\phantom{}}_{\beta,l'},
    \label{eq:AmoC2}
\end{equation}
where the operator $c^{\dagger}_{\alpha,l}$ creates a particle at the site $\alpha$ with an internal orbital degree of freedom given by $l$ [indexed by $l,l'=1,2$] at each lattice site [indexed by $\alpha$ and $\beta$].
Following the notation of Ref.~\cite{PhysRevLett.95.226801} previously employed for the Kramers-Weyl [KW] model in Appendix~\ref{app:PristineKramers}, the $\langle$ and $\rangle$ symbols in the sum over sites in Eq.~(\ref{eq:AmoC2}) indicate that pairs of sites $\alpha,\beta$ are only included within the summation if they lie within a specified distance of each other that varies for different hopping interactions and directions, as we will shortly detail below.
Specifically, in the crystalline limit discussed in this section, $\left\langle\alpha\beta\right\rangle$ in Eq.~(\ref{eq:AmoC2}) will reduce to nearest-neighbor lattice sites along the $\hat{z}$-axis and second-nearest-neighbor sites in the $(x,y)$-plane. This choice will give rise to leading-order quadratic terms for the low-energy in-plane dispersion of the nodal degeneracy at the $\Gamma$ point, and low-energy linear dispersion in the $z$- [out-of-plane] direction.
As previously for the KW model in Appendix~\ref{app:PristineKramers}, the tight-binding basis states $|{\bf r}_\alpha, l\rangle$ in Eq.~(\ref{eq:AmoC2}) satisfy the orthogonality relation $\langle {\bf r}_\alpha, l|{\bf r}_\beta, l'\rangle = \delta_{\alpha\beta}\delta_{ll'}$, where $l$ and $l'$ belong to the set $\left\{1, 2\right\}$, representing the internal orbital degrees of freedom, and $\alpha$ and $\beta$ denote the lattice sites [see Eq.~(\ref{eq:TBbasisStates}) and the surrounding text].

We next define the elements of the intersite hopping matrix $T_{\alpha\beta}$ for $\mathcal{H}_{\mathrm{C2}}$ in Eq.~(\ref{eq:AmoC2}) through inner products in the following manner:
\begin{equation}
\begin{split}
T_{\alpha\beta} &\underset{def}{\equiv}\begin{pmatrix}
            \langle {\bf r}_\alpha,1 | \mathcal{H} | {\bf r}_\beta,1\rangle & \langle {\bf r}_\alpha,1 | \mathcal{H} | {\bf r}_\beta,2\rangle\\
            \langle {\bf r}_\alpha,2 | \mathcal{H} | {\bf r}_\beta,1\rangle & 
            \langle {\bf r}_\alpha,2 | \mathcal{H} | {\bf r}_\beta,2\rangle
          \end{pmatrix} \\
          &= \frac{1}{2}\left[ \left({\bf d}^\mathsf{T}_{\alpha\beta} t_1\left(\left|\mathbf{d}_{\alpha\beta}\right|\right)Q_{x^2-y^2}{\bf d}_{\alpha\beta}\right)\tau^y + \left({\bf d}_{\alpha\beta}^{\mathsf{T}}t_1\left(\left|\mathbf{d}_{\alpha\beta}\right|\right)Q_{xy}{\bf d}_{\alpha\beta}\right)\tau^x  + \left({\bf d}^{\mathsf{T}}_{\alpha\beta} t_1\left(\left|\mathbf{d}_{\alpha\beta}\right|\right)\bm{V}_z\right)\tau^z \right] \\ &\quad+ \frac{1}{2}\left({\bf d}_{\alpha\beta}^{\mathsf{T}}t_2\left(\left|\mathbf{d}_{\alpha\beta}\right|\right)Q_{0}{\bf d}_{\alpha\beta}\right)\tau^0,
          \end{split}
\label{eq:TmatrixQuadraticDef}
\end{equation}
where the intersite separation vector ${\bf d}_{\alpha\beta}$ is given by:
\begin{equation}
    \mathbf{d}_{\alpha\beta}
= \begin{pmatrix}
x_{\alpha}-x_{\beta}\\y_{\alpha}-y_{\beta}\\z_{\alpha}-z_{\beta}
\end{pmatrix}.
\label{eq:dVectorDefC2}
\end{equation} 
In Eq.~(\ref{eq:TmatrixQuadraticDef}), each $\tau^{i}$ is a $2\times 2$ Pauli matrix that acts on the internal orbital degree of freedom at each lattice site [such that $\tau^{0}$ is the identity matrix in the internal orbital space], the contracted vectors and matrices are respectively defined as: 
\begin{equation}
\bm{V}_{z} = v_z\begin{pmatrix}0 \\0\\ i\end{pmatrix}, \quad
    Q_{x^2 -y^2} = v_1\begin{pmatrix}
    1&0&0\\
    0&-1&0\\
    0&0&0
    \end{pmatrix},\quad
    Q_{xy} = \frac{v_2}{2}\begin{pmatrix}
        0&-1&0\\
        -1&0&0\\
        0&0&0
    \end{pmatrix},\quad\text{and}\quad Q_0 = \begin{pmatrix}
        t_{xy}&0&0\\
        0&t_{xy}&0\\
        0&0&t_z
    \end{pmatrix},
\label{eq:charge2mostGeneralLandQ}
\end{equation}
and the strengths of the radial hopping functions $t_{1,2}\left(\left|\mathbf{d}_{\alpha\beta}\right|\right)$ are given by:
\begin{align}
&t_1\left(\left|\mathbf{d}_{\alpha\beta}\right|\right) = \begin{pmatrix}
\Theta(R_1-|\mathbf{d}_{\alpha\beta}|)\exp(\sqrt{2}a_{\parallel}-|\mathbf{d}_{\alpha\beta}|) &0 & 0 \\ 0 & \Theta(R_1-|\mathbf{d}_{\alpha\beta}|)\exp(\sqrt{2}a_{\parallel}-|\mathbf{d}_{\alpha\beta}|) & 0 \\0 & 0 & \Theta(R_1-|\mathbf{d}_{\alpha\beta}|)\exp(a_{\perp}-|\mathbf{d}_{\alpha\beta}|)
    \end{pmatrix}, \nonumber \\
&t_2\left(\left|\mathbf{d}_{\alpha\beta}\right|\right) = \begin{pmatrix}
 \Theta(R_2-|\mathbf{d}_{\alpha\beta}|)\exp(a_{\parallel}-|\mathbf{d}_{\alpha\beta}|) & 0 & 0 \\ 0 &\Theta(R_2-|\mathbf{d}_{\alpha\beta}|)\exp(a_{\parallel}-|\mathbf{d}_{\alpha\beta}|) & 0 \\ 0 & 0 &\Theta(R_1-|\mathbf{d}_{\alpha\beta}|)\exp(a_{\perp}-|\mathbf{d}_{\alpha\beta}|)
\end{pmatrix}.
\label{eq:doubleWeylTparams}
\end{align}
In Eq.~(\ref{eq:doubleWeylTparams}), $a_{\parallel
}=a_x=a_y$ denotes the lattice spacing in the $(x,y)$-plane, and $a_z=a_\perp$ denotes the lattice spacing in the $z$-direction.
For the remainder of this section,  we  will set the lattice spacing parameters of the functions $t_{1,2}\left(\left|\mathbf{d}_{\alpha\beta}\right|\right)$ in Eq.~(\ref{eq:doubleWeylTparams}) to be:
\begin{equation}
a_z = a_{\perp} = 1.4a_{\parallel},\ R_1 = 1.5a_{\parallel},\ R_2 = 1.3a_{\parallel},
\label{eq:quadraticWeylFixedLatticeParams}
\end{equation}
where the choice of $R_1= 1.5a_{\parallel}$ ensures that only first-nearest-neighbor hoppings are included in the out-of-plane direction and that only second-nearest-neighbor hoppings are included in the $(x,y)$-plane for terms proportional to $t_{1}\left(\left|\mathbf{d}_{\alpha\beta}\right|\right)$ in Eq.~(\ref{eq:doubleWeylTparams}).
In Eq.~(\ref{eq:quadraticWeylFixedLatticeParams}), the choice of $R_2 = 1.3a_{\parallel}$ ensures that only first-nearest-neighbor hoppings are allowed in the $(x,y)$-plane for terms proportional to $t_{2}\left(\left|\mathbf{d}_{\alpha\beta}\right|\right)$ in Eq.~(\ref{eq:doubleWeylTparams}).

\begin{figure}
\centering
\includegraphics[width=\linewidth]{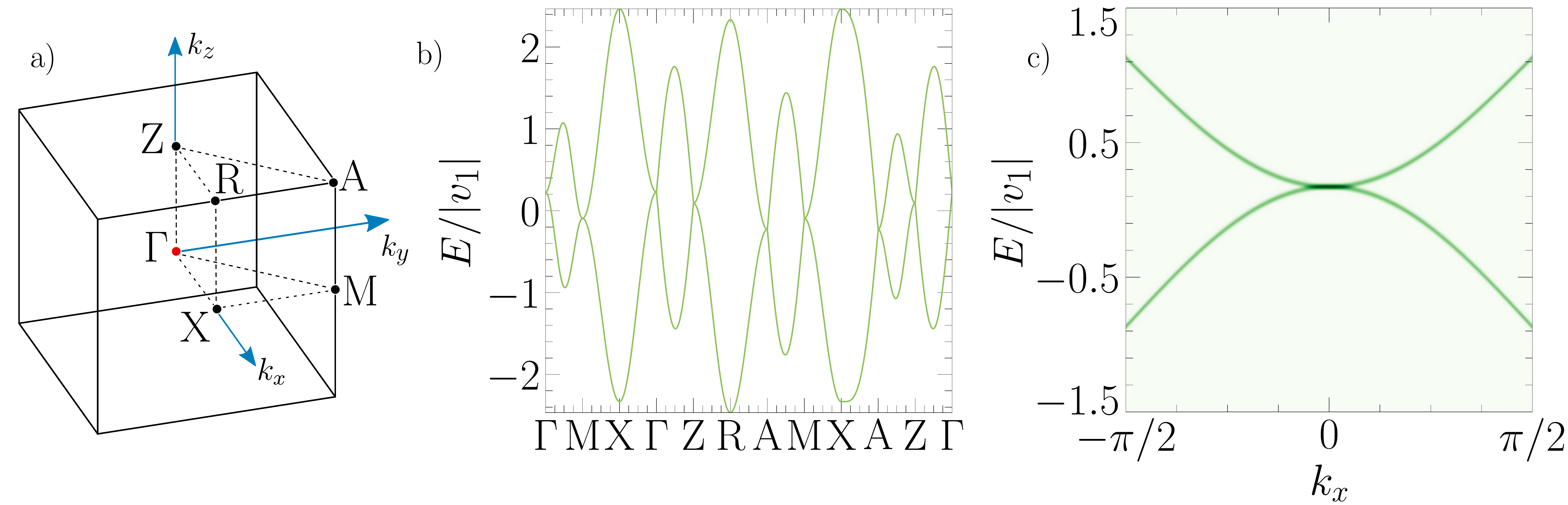}
\caption{Bulk spectrum of the crystalline quadratic double-Weyl model. 
(a) The Brillouin zone [BZ] of chiral SG 89 ($P422$), the SG of the double-Weyl tight-binding model [Eqs.~\eqref{eq:AmoC2},~\eqref{eq:TmatrixQuadraticDef}, and~\eqref{eq:pristineDoubleWeyl}].
(b) Bulk band structure of the crystalline double-Weyl model using the parameters in Eq.~(\ref{eq:appTBparamsC2}).
(c) The spectral function $A(E,\mathbf{p})$ [see Eq.~(\ref{eq:SpecFunc}) and the surrounding text] of the double-Weyl fermion at the $\Gamma$ point in (b) [Eq.~(\ref{eq:appHKPC2})].}
\label{fig:charge2_figBulk}
\end{figure}

\paragraph*{\bf Symmetries and Group Representations} -- $\ $ When defined on a regular tetragonal lattice with the hopping ranges discussed in the text surrounding Eq.~(\ref{eq:quadraticWeylFixedLatticeParams}), the tight-binding model in Eqs.~(\ref{eq:AmoC2}) and~(\ref{eq:TmatrixQuadraticDef}) can be Fourier transformed into a $2\times2$ Bloch Hamiltonian of the form:
\begin{equation}
 \mathcal{H}({\bf k}) =  t_{xy}\left[\cos(k_{x}) + \cos(k_{y})\right]\tau^{0} + t_{z}\cos(k_{z})\tau^{0} + v_{1}\sin(k_x)\sin(k_y)\tau^x + v_{2}\left[\cos(k_x)-\cos(k_y)\right]\tau^y + v_{z}\sin(k_z)\tau^z.
\label{eq:pristineDoubleWeyl}
\end{equation}
In Eqs.~(\ref{eq:TmatrixQuadraticDef}),~(\ref{eq:charge2mostGeneralLandQ}), and~(\ref{eq:pristineDoubleWeyl}), the $t_{i}$ parameters indicate the strength of nearest-neighbor hopping between the same spinless $p_{x}\pm ip_{y}$ orbitals, and the $v_{i}$ terms correspond to effective orbital-angular-momentum [OAM]~\cite{OAMCDArpes1,OAMCDArpes2,OAMWeyl1,OAMWeyl2,OAMmultifold1,OAMmultifold2,OAMmultifold3,OAMmultifold4,OAMmultifold5,OAMnodalLine,OAMzxShenARPES3DTI} coupling between first-nearest neighbors in the $z$-direction and second-nearest neighbors in the $(x,y)$-plane. 
The Bloch Hamiltonian $\mathcal{H}({\bf k})$ in Eq.~(\ref{eq:pristineDoubleWeyl}) respects the spinless [single-group] symmetries of symmorphic chiral SG 89 ($P422$), for which the generating elements can be represented through their action on $\mathcal{H}({\bf k})$:
\begin{eqnarray}
\tilde{\mathcal{T}}\mathcal{H}(k_{x},k_{y},k_{z})\tilde{\mathcal{T}}^{-1} &=& \tau^{x}\mathcal{H}^{*}(-k_{x},-k_{y},-k_{z})\tau^{x}, \nonumber \\ 
\tilde{C}_{4z}\mathcal{H}(k_{x},k_{y},k_{z})\tilde{C}^{-1}_{4z} &=& \tau^{z}\mathcal{H}(k_{y},-k_{x},k_{z})\tau^{z}, \nonumber \\
\tilde{C}_{2x}\mathcal{H}(k_{x},k_{y},k_{z})\tilde{C}_{2x}^{-1} &=& \tau^{x}\mathcal{H}(k_{x},-k_{y},-k_{z})\tau^{x},
\label{eq:PristineCharge2TBsyms}
\end{eqnarray}
where we have used tildes to denote that the symmetries are elements of single SGs, as they act on internal integer-angular-momentum [$p_{x}\pm ip_{y}$] degrees of freedom~\cite{BigBook,MTQC}. 
We have not included lattice translations in Eq.~(\ref{eq:PristineCharge2TBsyms}), because they act as pure Bloch phases in momentum space~\cite{BigBook}, and hence do not provide further constraints on $\mathcal{H}({\bf k})$.
Formally, the occupied and unoccupied bands of $\mathcal{H}({\bf k})$ in Eq.~(\ref{eq:pristineDoubleWeyl}) together transform in the two-dimensional, single-valued elementary band corep $(E)_{1a}\uparrow P422$ of SG 89 ($P422$)~\cite{Bradlyn2017,MTQC,Bandrep1}.

As we will motivate below by the rotoinversion symmetries that emerge when the $v_{1,2,z}$ parameters are taken to vanish, the lattice structural chirality of the quadratic double-Weyl model [Eqs.~(\ref{eq:AmoC2}),~(\ref{eq:TmatrixQuadraticDef}), and~(\ref{eq:pristineDoubleWeyl})] is determined by the product of the $v_{1}$, $v_{2}$, and $v_{z}$ model parameters:
\begin{equation}
C_{\mathcal{H}} = \text{sgn}\left(v_{1}v_{2}v_{z}\right),
\label{eq:quadraticStructuralChirality}
\end{equation}
where $C_{\mathcal{H}}=1$ [$C_{\mathcal{H}}=-1$] corresponds to the $R$ [$L$] model enantiomer.
As in the crystalline Kramers-Weyl model previously analyzed in Appendix~\ref{app:PristineKramers}, $C_{\mathcal{H}}$ in Eq.~(\ref{eq:quadraticStructuralChirality}) is a global geometric property of the quadratic double-Weyl model on a crystalline lattice, because the OAM coupling between each unit cell carries the same signs of $v_{1,2,z}$.
However, by allowing the signs of the $v_{1}$, $v_{2}$, or $v_{z}$ hopping parameters in Eqs.~(\ref{eq:TmatrixQuadraticDef}) and~(\ref{eq:charge2mostGeneralLandQ}) to vary for bonds originating from each site $\alpha$, we may also promote $C_{\mathcal{H}}$ in Eq.~(\ref{eq:quadraticStructuralChirality}) to a spatially varying local chirality $\chi_{\alpha}$ [see Refs.~\cite{LocalChiralityMoleculeReviewNatChem,LocalChiralityLiquidCrystals,KamienLubenskyChiralParameter,LocalChiralityDomainLiquidCrystal,LocalChiralityTransfer,KamienChiralLiquidCrystal,LocalChiralityQuasicrystalVirus,lindell1994electromagneticBook,LocalChiralityVillain,LocalChiralityWenZee,LocalChiralityBaskaran,LocalChiralitySpinFrame} and the text surrounding Eq.~(\ref{eq:temp2siteFrameBreakdown})].
Below, in the text following Eqs.~(\ref{eq:doubleWeylLocalAlpha}) and~(\ref{eq:HopChiralC2}), we will shortly implement a non-crystalline generalization of the double-Weyl model in which the local chirality $\chi_{\alpha}$ represents a parameter that can be ordered or disordered independent from the lattice structural order, and can specifically be tuned to control the topology of non-crystalline double-Weyl fermions.

For all values of the hopping parameters in Eq.~(\ref{eq:pristineDoubleWeyl}), the energy spectrum exhibits four nodal degeneracies at $k_{x}=k_{y}=0,\pi$ and $k_{z}=0,\pi$ with quadratic dispersion in the $k_{x,y}$ directions and linear dispersion in the $k_{z}$ direction.  
We emphasize that unlike previously for the crystalline KW model in Appendix~\ref{app:PristineKramers}, the minimal insulating filling [single-particle band connectivity] for spinless fermions in SG 89 ($P422$) is just $1$.
This indicates that the four twofold nodal degeneracies of $\mathcal{H}({\bf k})$ [Eq.~(\ref{eq:pristineDoubleWeyl})] at half filling are non-minimal [not filling-enforced]~\cite{WPVZ,WiederLayers,WPVZfollowUp}, and instead arise due to the higher-angular-momentum $p_{x}\pm ip_{y}$ atomic orbital basis states used to construct $\mathcal{H}_{\mathrm{C2}}$ in Eq.~(\ref{eq:AmoC2}).
If we had instead constructed a minimal tight-binding model in SG 89 ($P422$) with spinless $s$ orbitals at the $1a$ Wyckoff position [\emph{i.e.} the origin], the model would just have a single band, and would hence not exhibit symmetry-pinned quadratic degeneracies at $k_{x}=k_{y}=0,\pi$ for $k_{z}=0,\pi$.
For future calculations in amorphous systems [see Appendices~\ref{app:corepAmorphous} and~\ref{app:amorphousCharge2}], we will find it useful to note that the little group $G_{\Gamma}$ at the $\Gamma$ point is isomorphic to SG 89 ($P422$) itself [see the text preceding Eq.~(\ref{eq:EquivKPoints})].
The nodal degeneracy at $\Gamma$ then specifically transforms in the two-dimensional, single-valued small corep $\Gamma_{5}$ of $G_{\Gamma}$ in the labeling convention of the~\href{https://www.cryst.ehu.es/cgi-bin/cryst/programs/corepresentations.pl}{Corepresentations} tool on the Bilbao Crystallographic Server [in which SG 89 is alternatively denoted as Shubnikov SG 89.88 ($P4221'$)]~\cite{MTQC,MTQCmaterials}.

\paragraph*{\bf Topology} -- $\ $ In Fig.~\ref{fig:charge2_figBulk}(a,b), we plot the band structure of $\mathcal{H}({\bf k})$ [Eq.~(\ref{eq:pristineDoubleWeyl})] with the choice of parameters:
\begin{equation}
t_{xy} = 0.1,\
t_z = 0.12,\
v_1 = \pm 2.5,\
v_z= 2,\
v_2 = 3,
\label{eq:appTBparamsC2}
\end{equation}
where the $+$ [$-$] sign for $v_{1}$ is used to model the $R$ [$L$] enantiomer via Eq.~(\ref{eq:quadraticStructuralChirality}).
Similar to the earlier demonstration in Ref.~\cite{KramersWeyl} and the text surrounding Eq.~(\ref{eq:appChiralCharge}) of the relationship between lattice structural chirality and the low-energy topological chirality of KW fermions, we will next prove that the four symmetry-enforced TRIM-point nodal degeneracies at $k_{x}=k_{y}=0,\pi$ and $k_{z}=0,\pi$ in Fig.~\ref{fig:charge2_figBulk}(b) each carry a chiral charge $C = \pm 2$ whose overall sign is inherited from the lattice-scale structural chirality $C_{\mathcal{H}}$ [Eq.~(\ref{eq:quadraticStructuralChirality})].
To demonstrate this relation, we begin by expanding Eq.~\eqref{eq:pristineDoubleWeyl} to quadratic order about the TRIM point $\mathbf{k}_\mathcal{T}$:
\begin{equation}
\mathcal{H}_{{\bf k}_{\mathcal{T}}}({\bf q}) = {\bf d}_{{\bf k}_{\mathcal{T}}}({\bf q})\cdot{\bm \tau},
\label{eq:appHKPC2}
\end{equation}
where ${\bf q} \approx {\bf k} - {\bf k}_{\mathcal{T}}$, and in which:
\begin{equation}
    \mathbf{d}_{\mathbf{k}_{\mathcal{T}}}({\bf q})= 
\begin{pmatrix}
v_1q_xq_ye^{ik^{x}_{\mathcal{T}}}_{\phantom{}}e^{ik^{y}_{\mathcal{T}}}_{\phantom{}}\\
v_2\left(q_y^2e^{ik^{y}_{\mathcal{T}}}_{\phantom{}}-q_x^2e^{ik^{x}_{\mathcal{T}}}_{\phantom{}}\right)/2\\
v_zq_ze^{ik^{z}_{\mathcal{T}}}   
\end{pmatrix},
\label{eq:dvectorPristineDoubleWeyl}
\end{equation}
and:
\begin{equation}
{\bm \tau} = \begin{pmatrix}
\tau_x\\
\tau_y\\
\tau_z
\end{pmatrix}.
\label{eq:tempQuadraticTau}
\end{equation}
In Eqs.~(\ref{eq:appHKPC2}) and~(\ref{eq:dvectorPristineDoubleWeyl}), we have suppressed factors of the $2\times 2$ identity matrix $\tau^{0}$, because they correspond to effective chemical potentials on the TRIM-point quadratic degeneracies and do not affect the calculation of the topological chiral charge, which only depends on the band structure and is independent of the Fermi level.

Eq.~\eqref{eq:dvectorPristineDoubleWeyl} takes the previously established form of a $|C|=2$ double-Weyl point~\cite{ZahidMultiWeylSrSi2,StepanMultiWeyl,AndreiMultiWeyl,XiDaiMultiWeyl}.
The topological chiral charge $C_{\mathbf{k}_{\mathcal{T}}}$ of each of the four nodal degeneracies can analytically be computed as follows. 
First, as previously detailed in the text surrounding Eq.~(\ref{eq:KWchiralJacobianRelation}), the chiral charge of each of the four degeneracies is given by:
\begin{equation}
     C_{{\bf k}_{\mathcal{T}}} = \sum_{n=1}^{N}\mathrm{sgn}\left(\left|\frac{\partial d_{\mathbf{k}_{\mathcal{T}}}^{i}}{\partial q_j}\right|_{\mathbf{q} = \mathbf{q }^n}\right).
\label{eq:doubleWeylChargeDerivatives}
\end{equation}
For Eq.~(\ref{eq:pristineDoubleWeyl}) expanded to quadratic order for each of the four nodal degeneracies at half filling [Eq.~\eqref{eq:appHKPC2}], the Jacobian in Eq.~(\ref{eq:doubleWeylChargeDerivatives}) then reads as:
\begin{equation}
    \left|\frac{\partial d_i}{\partial q_j}\right| = (v_1v_2v_z)e^{ik^{z}_{\mathcal{T}}}_{\phantom{}}\left(q_x^2e^{ik^{y}_{\mathcal{T}}}_{\phantom{}} + q_y^2e^{ik^{x}_{\mathcal{T}}}_{\phantom{}}\right).
\label{eq:tempQuadPreimage}
\end{equation}
Notice that when \( e^{i k^{x}_{\mathcal{T}}} \neq e^{i k^{y}_{\mathcal{T}}} \), the sign of the Jacobian determinant becomes dependent on the direction in momentum space. As a consequence, the associated topological charge would not be well defined, because it would vary with the orientation of the enclosing surface. 
This would be inconsistent with the presence of an isolated band crossing, which requires a well-defined integer topological charge. 
However, all four of the double-Weyl points in Eq.~(\ref{eq:pristineDoubleWeyl}) occur at TRIM points for which $k^{x}_{\mathcal{T}}=k^{y}_{\mathcal{T}}$ [Fig.~\ref{fig:charge2_figBulk}(a,b)], such that their topological charges remain well defined.
For the specific choice $\mathbf{d}_r = (0,v_2e^{ik^{y}_{\mathcal{T}}}/2,0)$, Eq.~(\ref{eq:tempQuadPreimage}) yields two preimages: $q^1 = (0,1,0)$ and $q^2 = (0,-1,0)$. 
Following Brouwer's lemma~\cite{MilnorTopology1965} as discussed in the text preceding Eq.~(\ref{eq:imageProofKWCharge}), this, along with the restriction to the four TRIM points with $k^{x}_{\mathcal{T}}=k^{y}_{\mathcal{T}}$, implies that the chiral charge of the quadratic degeneracy at the TRIM point ${\bf k}_{\mathcal{T}}$ is given by:
\begin{equation}
C_{{\bf k}_{\mathcal{T}}} = 2\left(e^{ik^{y}_{\mathcal{T}}}_{\phantom{}}e^{ik^{z}_{\mathcal{T}}}\right)\text{sgn}\left(v_{1}v_{2}v_{z}\right).
\label{eq:imageProofC2Charge}
\end{equation}

\begin{figure}[t]
\centering
\includegraphics[width=\linewidth]{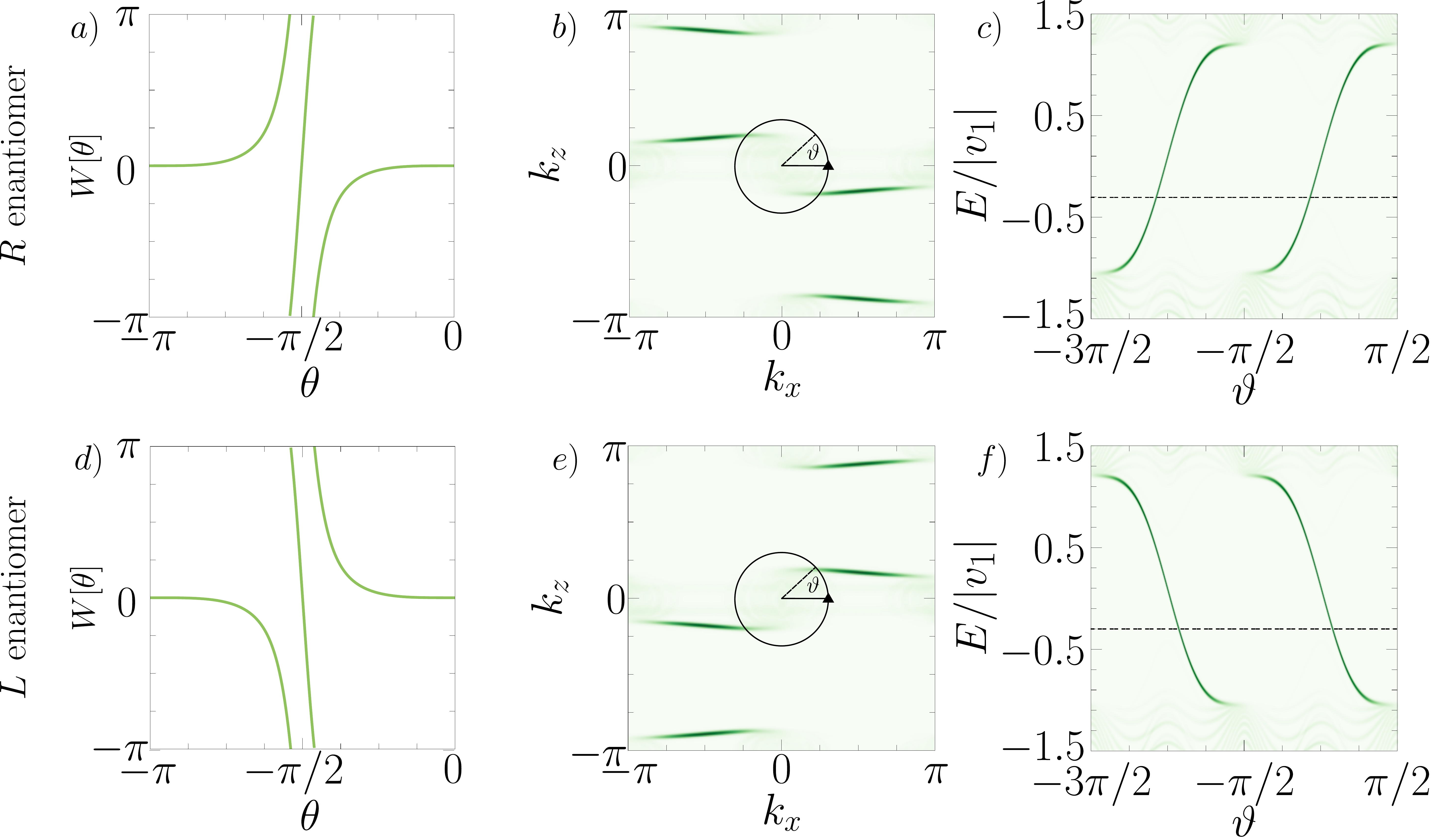}
\caption{Chirality and topology of the crystalline double-Weyl model.
Panels (a-c)  and (d-f) respectively show bulk topological and surface spectral details of the right-handed [$R$] and left-handed [$L$] enantiomers of the double-Weyl tight-binding model [Eqs.~(\ref{eq:AmoC2}),~(\ref{eq:TmatrixQuadraticDef}), and~(\ref{eq:appHKPC2})] obtained using the parameters in Eq.~(\ref{eq:appTBparamsC2}).
In (a) and (d), we show the Wilson loop spectrum of the lowest band, defined in Appendix~\ref{sec:WilsonBerry}, computed respectively for the $R$ and $L$ model enantiomer on a sphere surrounding the quadratically dispersing nodal degeneracy at the $\Gamma$ point of Eq.~(\ref{eq:pristineDoubleWeyl}) [see Fig.~\ref{fig:charge2_figBulk}].
The Wilson loop spectra in (a) and (d) respectively wind twice in the positive and negative directions, indicating that the double-Weyl fermion at the $\Gamma$ point carries a chiral charge of $C=2$ for the $R$ enantiomer and $C=-2$ for the $L$ enantiomer, such that $C$ is directly determined by the lattice-scale structural chirality $C_{\mathcal{H}}$ through Eq.~(\ref{eq:appChiralChargeC2}).
(b) and (e) respectively show the $(010)$-surface Fermi arcs for the $R$ and $L$ double-Weyl model enantiomers computed from surface Green's functions at $E/|v_1|=-0.4$. 
In (c) and (f), we then respectively plot the $(010)$-surface spectral functions of the $R$ and $L$ model enantiomers computed as functions of energy on counterclockwise circular paths, parameterized by $\vartheta$, surrounding $k_{x}=k_{z}=0$ in (b,e). 
The Fermi pocket at $k_{x}=k_{z}=0$ in (b,e) only contains the projection of the double-Weyl fermion at $\Gamma$ in Fig.~\ref{fig:charge2_figBulk}(b), which through Eq.~(\ref{eq:appHKPC2}) carries a chiral charge of $C=2$ for (b) the $R$ enantiomer and $C=-2$ for (e) the $L$ enantiomer.
The surface Fermi arcs in (c) and (f) respectively cross the dashed horizontal line two times with positive and negative slopes, confirming that the bulk Fermi pocket projections in (b) and (e) at $k_{x}=k_{z}=0$ respectively carry a net chiral charge of $C=2$ for the $R$ enantiomer and $C=-2$ for the $L$ enantiomer. 
The data in panels (b,d,e,f) were obtained from the tight-binding model in Eqs.~(\ref{eq:AmoC2}) and~(\ref{eq:TmatrixQuadraticDef}) placed on a regular lattice that was infinite in the $\hat{z}$-direction and finite with 150 sites and periodic boundary conditions in the $\hat{x}$-direction and finite with 20 sites and open boundary conditions in the $\hat{y}$- [$(010)$-] direction.}
\label{fig:charge2_figWilson}
\end{figure}

Importantly, we may then use Eq.~(\ref{eq:quadraticStructuralChirality}) to re-express Eq.~(\ref{eq:imageProofC2Charge}) in the form:
\begin{equation}
C_{{\bf k}_{\mathcal{T}}} = 2\left(e^{ik^{y}_{\mathcal{T}}}_{\phantom{}}e^{ik^{z}_{\mathcal{T}}}\right)C_{\mathcal{H}}.
\label{eq:appChiralChargeC2}
\end{equation}
Hence, like the KW model analyzed in Ref.~\cite{KramersWeyl} and in the text surrounding Eq.~(\ref{eq:appChiralCharge}), the low-energy topological chirality of the quadratic nodal
degeneracies in the tight-binding model introduced in this section [the chiral charge $C_{\mathbf{k}_{\mathcal{T}}}$ of each double-Weyl fermion in Fig.~\ref{fig:charge2_figBulk}(b,c)] is also directly inherited from and controlled by the lattice-scale
geometry [here the structural chirality $C_{\mathcal{H}}$ in Eq.~(\ref{eq:quadraticStructuralChirality})] through the signs of the $v_i$ parameters in Eqs.~\eqref{eq:charge2mostGeneralLandQ} and~\eqref{eq:appTBparamsC2}.

To numerically confirm Eq.~\eqref{eq:appChiralChargeC2}, we have computed in Fig.~\ref{fig:charge2_figWilson} the $\Gamma$-point sphere Wilson loop spectrum [Appendix~\ref{sec:WilsonBerry}] and $(010)$-surface Fermi arcs of the $R$ [$C_{\mathcal{H}} = 1$] and $L$ [$C_{\mathcal{H}} = -1$] enantiomers
of the double-Weyl tight-binding model [Eqs.~\eqref{eq:AmoC2},~\eqref{eq:TmatrixQuadraticDef}, and~\eqref{eq:pristineDoubleWeyl}]. First, in Fig.~\ref{fig:charge2_figWilson}(a) [Fig.~\ref{fig:charge2_figWilson}(d)] we show the Wilson loop spectrum of the lower band of the double-Weyl model computed on a sphere surrounding the nodal degeneracy at the $\Gamma$ point [$\mathbf{k} = \mathbf{0}$] in Fig.~\ref{fig:charge2_figBulk}(b)
for the $R$ [$L$] model enantiomer. 
The Wilson loop spectrum in Fig.~\ref{fig:charge2_figWilson}(a) winds twice in the positive direction for the $R$ enantiomer, and the Wilson loop spectrum in Fig.~\ref{fig:charge2_figWilson}(d) winds twice in the negative direction for the $L$ enantiomer, indicating that the nodal degeneracy at the $\Gamma$ point carries a chiral charge of $C = 2$ for the $R$ enantiomer and $C = -2$ for the $L$ enantiomer.
This is consistent with Eq.~(\ref{eq:appChiralChargeC2}) and the surrounding text, in which we identified the quadratically dispersing $\Gamma$-point nodal degeneracy in Eq.~(\ref{eq:pristineDoubleWeyl}) as a double-Weyl [charge-$|2|$ chiral] fermion whose topology is tunable via the lattice-scale structural chirality.

\paragraph*{\bf Surface Fermi Arcs} -- $\ $ To provide corroborating evidence for the bulk chiral charge distribution in Eq.~(\ref{eq:appChiralChargeC2}), we next compute the surface states of the double-Weyl tight-binding model.
In Fig.~\ref{fig:charge2_figWilson}(b,c) and Fig.~\ref{fig:charge2_figWilson}(e,f) we respectively plot the
$(010)$-surface states of the $R$ and $L$ enantiomer of the double-Weyl model as functions of two surface momenta and energy.
In Fig.~\ref{fig:charge2_figWilson}(b,c,e,f), the Fermi pocket at $k_{x}=k_{z}=0$ only contains the projected bulk Fermi pocket of the nodal degeneracy at $\Gamma$ in Fig.~\ref{fig:charge2_figWilson}(b), which through Eq.~(\ref{eq:appHKPC2}) carries a chiral charge of $C=2$ for the $R$ enantiomer and $C=-2$ for the $L$ enantiomer. 
We correspondingly observe a time-reversed pair of $(010)$-surface Fermi arcs emanating from the $k_{2}=k_{z}=0$ point in Fig.~\ref{fig:charge2_figWilson}(b,e), with the arcs roughly exhibiting opposite constant-energy helicities for opposite enantiomers, as they did previously for the KW model in Fig.~\ref{fig:KW_figWilson}(b,e) [\emph{i.e.} the arcs on the top and bottom of the projected bulk Fermi pockets at $k_{2}=k_{z}=0$ respectively bend to the left and right for the $R$ enantiomer and respectively bend to the right and left for the $L$ enantiomer].
However, as previously discussed in the text following Eq.~(\ref{appeq:surfacemomenta}), the qualitative helicity of surface Fermi arcs in constant-energy spectral functions like Fig.~\ref{fig:charge2_figWilson}(b,e) 
is not a topological property.
We therefore additionally in Fig.~\ref{fig:charge2_figWilson}(c,f) compute the $(010)$-surface Fermi arcs as functions of energy on counterclockwise paths encircling $k_{x}=k_{z}=0$ in Fig.~\ref{fig:charge2_figWilson}(b,e), which conversely provides a \emph{quantitative} indicator of the bulk topology~\cite{AshvinWeyl,VolovikWeylMonopole,VolovikChiralAnomaly,HaldaneOriginalWeyl,MurakamiWeyl,BurkovBalents,AndreiWeyl,HasanWeylDFT,Armitage2018,SuyangWeyl,LvWeylExp,YulinWeylExp,AliWeylQPI,AlexeyType2,ZJType2,BinghaiClaudiaWeylReview,ZahidNatRevMatWeyl,CDWWeyl,IlyaIdealMagneticWeyl}.
The surface Fermi arcs respectively cross the dashed horizontal line  in Fig.~\ref{fig:charge2_figWilson}(c,f) two times with
positive and negative slopes [velocities], confirming that the bulk Fermi pocket projections at $k_{x}=k_{z}=0$ in  Fig.~\ref{fig:charge2_figWilson}(b,e) respectively carry a chiral charge of $C=2$ for the $R$ enantiomer and $C=-2$ for the $L$ enantiomer.

\paragraph*{\bf Achiral Critical Phases} -- $\ $ Lastly, as previously for the eight Weyl points in the KW model in Appendix~\ref{app:PristineKramers}, the topological chiral charges $C_{{\bf k}_{\mathcal{T}}}$ of the four double-Weyl fermions at $k_{x}=k_{y}=0,\pi$ and $k_{z}=0,\pi$ become ill-defined if one or more of $v_{1,2}$ or $v_{z}$ in Eq.~(\ref{eq:pristineDoubleWeyl}) goes to zero.
As we will show below, this precisely occurs because when $v_{1}$, $v_{2}$, or $v_{z}$ goes to zero, the system SG [and therefore the little group at each TRIM point, see Appendix~\ref{app:corepDefs}] gains additional rotoinversion symmetries, and hence becomes achiral [see Eq.~(\ref{eq:rotoinversion}) and the surrounding text]. 
For example if $v_{z}\rightarrow 0$ in Eq.~(\ref{eq:pristineDoubleWeyl}), then $\mathcal{H}({\bf k})$ gains an additional spatial inversion [$\mathcal{I}$] symmetry, which can be represented through its symmetry action:
\begin{equation}
\mathcal{I}\mathcal{H}({\bf k})\mathcal{I}^{-1} = \mathcal{H}(-{\bf k}).
\label{eq:charge2inversionLimit}
\end{equation}
Next, if $v_{1}\rightarrow 0$ while the other parameters in $\mathcal{H}({\bf k})$ remain nonzero, then $\mathcal{H}({\bf k})$ develops additional mirror-like [twofold rotoinversion] symmetries about the $x$-axis and $y$-axis $\bar{M}_{x,y}$:
\begin{eqnarray}
\bar{M}_{x}\mathcal{H}(k_{x},k_{y},k_{z})\bar{M}_{x}^{-1} = \mathcal{H}(-k_{x},k_{y},k_{z}), \nonumber \\
\bar{M}_{y}\mathcal{H}(k_{x},k_{y},k_{z})\bar{M}_{y}^{-1} = \mathcal{H}(k_{x},-k_{y},k_{z}).
\label{eq:charge2mXlimit}
\end{eqnarray}
Formally, when acting on the $p_{x}\pm ip_{y}$ basis states of Eq.~(\ref{eq:AmoC2}), $\bar{M}_{x}$ [$\bar{M}_{y}$] is equivalent to the combination of mirror reflection $\tilde{M}_{x}$ [$\tilde{M}_{y}$] about the $x$- [$y$-] axis followed by a $\tau^{x}$ rotation [exchange] of the internal orbital degrees of freedom.
Finally, if $v_{2}\rightarrow 0$ while the other parameters in $\mathcal{H}({\bf k})$ remain nonzero, then $\mathcal{H}({\bf k})$ develops additional mirror-like symmetries about the $(x\pm y)$-axes $\bar{M}_{x \pm y}$:
\begin{eqnarray}
\bar{M}_{x+y}\mathcal{H}(k_{x},k_{y},k_{z})\bar{M}_{x+y}^{-1} = \mathcal{H}(k_{y},k_{x},k_{z}), \nonumber \\
\bar{M}_{x-y}\mathcal{H}(k_{x},k_{y},k_{z})\bar{M}_{x-y}^{-1} = \mathcal{H}(-k_{y},-k_{x},k_{z}).
\label{eq:charge2mXpmYlimit}
\end{eqnarray}
Formally, when acting on the $p_{x}\pm ip_{y}$ basis states of Eq.~(\ref{eq:AmoC2}), $\bar{M}_{x\pm y}$ is equivalent to the combination of mirror reflection $\tilde{M}_{x \pm y}$ about the $(x\pm y)$-axis followed by a $\tau^{y}$ rotation [signed exchange] of the internal orbital degrees of freedom.
For completeness, we note that though there do not exist standard-setting achiral SGs with fourfold rotation symmetry, integer lattice translations in the $x$- and $y$-directions, and mirror reflections about the $(x\pm y)$-axes like in Eqs.~(\ref{eq:PristineCharge2TBsyms}) and~(\ref{eq:charge2mXpmYlimit}), this combination of symmetries is contained within SG 99 in the \emph{nonstandard} setting $C4mm$~\cite{AroyoBCS1}.

To conclude, Eqs.~(\ref{eq:charge2inversionLimit}),~(\ref{eq:charge2mXlimit}), and~(\ref{eq:charge2mXpmYlimit}) together show that the quadratic double-Weyl fermion model introduced in this section [Eqs.~(\ref{eq:AmoC2}),~(\ref{eq:TmatrixQuadraticDef}), and~(\ref{eq:pristineDoubleWeyl})] becomes structurally achiral if \emph{any} of the parameters $v_{1}$, $v_{2}$, or $v_{z}$ go to zero.
This supports our earlier statement in this section [Eq.~(\ref{eq:quadraticStructuralChirality})] that the structural chirality $C_{\mathcal{H}}$ of the quadratic double-Weyl fermion model is determined by the sign of the product of $v_{1}$, $v_{2}$, and $v_{z}$, as opposed to more simply the individual signs of any of the tight-binding parameters in Eq.~(\ref{eq:pristineDoubleWeyl}). 

\clearpage

\subsubsection{Non-Crystalline Double-Weyl Fermions}
\label{app:amorphousCharge2}

In this section, we will next demonstrate that the double-Weyl fermion model previously introduced in Appendix~\ref{app:PristineQuadratic} [Eqs.~\eqref{eq:AmoC2} and~\eqref{eq:TmatrixQuadraticDef}] also exhibits topological double-Weyl fermions when it is realized in amorphous systems [approximated by strongly Gaussian disordered and random lattices] that carry a net [average] structural chirality.

To begin, unlike the Kramers-Weyl [KW] model previously analyzed in Appendix~\ref{app:Kramers}, the low-energy theory of the crystalline double-Weyl model exhibits nodal degeneracies with quadratic dispersion in the $k_{x,y}$ directions and linear dispersion in the $k_{z}$ direction. 
Microscopically, this can be understood by recognizing that the double-Weyl model in Eqs.~\eqref{eq:AmoC2} and~\eqref{eq:TmatrixQuadraticDef} -- unlike the KW model [Eqs.~(\ref{eq:AmoKW}) and~(\ref{eq:KWTmatrix})] or the multifold fermion model that we will shortly analyze below [Eqs.~\eqref{eq:Amo3F} and~\eqref{eq:3FTmatrix}] -- inhabits a reference frame in which the real-space $z$-axis is a preferred direction along which hopping interactions carry a functional form that is distinct from those in the $(x,y)$-plane. 
Unlike previously for the disordered KW model analyzed in Appendix~\ref{app:amorphousKramers}, we will therefore in this section analyze the non-crystalline double-Weyl model in the presence of \emph{smectic} disorder in which the internal orbital reference frame and atomic positions are only disordered in the $(x,y)$-plane and remain ordered in the $z$-direction [$d=3$, $d_{T}=1$, $d_{A}=2$, $d_{f}=0$ in the notation of Eq.~(\ref{eq:finiteDimAmorph}), see Fig.~\ref{fig:dadtdf}(b) and Appendix~\ref{app:SmecticNematicDisorder}].

In addition to the atomic positions, the hopping terms in Eqs.~\eqref{eq:AmoC2} and~\eqref{eq:TmatrixQuadraticDef} also depend on the local coordinate reference frame.
For example, the hoppings in Eq.~(\ref{eq:TmatrixQuadraticDef}) along the Cartesian $x$- and $y$- directions always include OAM terms proportional to $\tau^{y}$ in the internal space of $p_{x}\pm ip_{y}$ orbitals.
However in a realistic model of a disordered solid-state system with a preferred $z$-axis direction, no particular region should have physical properties that depend on the $(x,y)$-plane subspace of the global coordinate frame.
Hence, we must also disorder the SO(2) in-plane part of the local coordinate frame in the double-Weyl model.
From a continuum perspective, this can be accomplished by introducing a local SO(2) unit vector field $\hat{R}({\bf r})$ that is given by the average orbital frame orientation in the $(x,y)$-plane of the sites within a specified vicinity of the position ${\bf r}$.
From this perspective, a system can then be defined as ``fully frame-disordered'' via the absence of long-range correlations in $\hat{R}({\bf r})$ [see Fig.~\ref{appfig:disordertypes}(b) and the surrounding text]. 
To implement frame disorder at the lattice scale, we specifically assign an SO(2) rotation matrix $R_{\alpha}$ [Eq.~(\ref{eq:2dSO2Rmat})] to the hopping frame at each lattice site $\alpha$, such that the intersite separation [bond] vector ${\bf d}_{\alpha\beta}$ that links the lattice sites $\alpha$ and $\beta$ [Eq.~(\ref{eq:dVectorDefC2})] is transformed to a rotated bond vector $\tilde{\bf d}_{\alpha\beta}$ as follows:
\begin{equation}
{\bf d}_{\alpha\beta} \rightarrow R_{\alpha}R_{\beta}^{\mathsf{T}}{\bf d}_{\alpha\beta} \underset{def}{\equiv} \tilde{{\bf d}}_{\alpha\beta}.
\label{appeq:rotaC2}
\end{equation}
After applying Eq.~(\ref{appeq:rotaC2}), the double-Weyl model hopping matrix elements in Eq.~(\ref{eq:TmatrixQuadraticDef}) become transformed to:
\begin{equation}
    \begin{split}
                  T_{\alpha\beta} 
          = &\frac{1}{2}\left[\left\{\left[\left(\tilde{{\bf d}}_{\alpha\beta}\right)^\mathsf{T}t_1\left(\left|\mathbf{d}_{\alpha\beta}\right|\right) Q_{x^2-y^2}\tilde{{\bf d}}_{\alpha\beta}\right]\tau^y + \left[\left(\tilde{{\bf d}}_{\alpha\beta}\right)^{\mathsf{T}}t_1\left(\left|\mathbf{d}_{\alpha\beta}\right|\right)Q_{xy}\tilde{{\bf d}}_{\alpha\beta}\right]\tau^x  + \left[\left(\tilde{{\bf d}}_{\alpha\beta}\right)^{\mathsf{T}} t_1\left(\left|\mathbf{d}_{\alpha\beta}\right|\right)\bm{V}_{z}\right]\tau^z\right\}\right] \\
          &+\frac{1}{2}\left[\left(\tilde{{\bf d}}_{\alpha\beta}\right)^{\mathsf{T}}t_2\left(\left|\tilde{{\bf d}}_{\alpha\beta}\right|\right)Q_{0}\tilde{{\bf d}}_{\alpha\beta}\right]\tau^0.
\end{split}
\label{eq:HopC2}
\end{equation}

In addition to structural and  local-frame order, the double-Weyl tight-binding model can also be characterized by spatially varying \emph{chirality order} that arises from the handedness of hopping interactions within the vicinity of a position ${\bf r}$~\cite{LocalChiralityMoleculeReviewNatChem,LocalChiralityLiquidCrystals,KamienLubenskyChiralParameter,LocalChiralityDomainLiquidCrystal,LocalChiralityTransfer,KamienChiralLiquidCrystal,LocalChiralityQuasicrystalVirus,lindell1994electromagneticBook,LocalChiralityVillain,LocalChiralityWenZee,LocalChiralityBaskaran,LocalChiralitySpinFrame}.
As detailed in Appendix~\ref{app:DiffTypesDisorder}, to implement lattice-scale chirality disorder, we assign each site $\alpha$ at the position ${\bf r}_{\alpha}$ a local discrete chirality:
\begin{equation}
\chi_{\alpha} = 0,\pm 1,
\label{eq:doubleWeylLocalAlpha}
\end{equation}
where $\chi_{\alpha}=\pm 1$ respectively for right- and left-handed sites and $\chi_{\alpha}=0$ for achiral sites.
In this work, we specifically consider disorder realizations in which there exist \emph{domains} in which all sites carry the same values of $\chi_{\alpha}$ [see Fig.~\ref{appfig:disordertypes}(c) and the surrounding text].
Implementing lattice-scale variations in the local chirality leads to a final modification of Eq.~(\ref{eq:HopC2}) as follows:
\begin{equation}
\begin{split}
T_{\alpha\beta} 
    = &\frac{1}{2}\bigg[ \left(\frac{\chi_{\alpha}+\chi_{\beta}}{2}\right)\bigg\{\left[\left(\tilde{{\bf d}}_{\alpha\beta}\right)^\mathsf{T}t_1\left(\left|\mathbf{d}_{\alpha\beta}\right|\right) Q_{x^2-y^2}\tilde{{\bf d}}_{\alpha\beta}\right]\tau^y + \left[\left(\tilde{{\bf d}}_{\alpha\beta}\right)^{\mathsf{T}}t_1\left(\left|\mathbf{d}_{\alpha\beta}\right|\right)Q_{xy}\tilde{{\bf d}}_{\alpha\beta}\right]\tau^x \\ 
    \ &\ + \left[\left(\tilde{{\bf d}}_{\alpha\beta}\right)^{\mathsf{T}} t_1\left(\left|\mathbf{d}_{\alpha\beta}\right|\right)\bm{V}_{z}\right]\tau^z\bigg\}\bigg] + \frac{1}{2}\left[\left(\tilde{{\bf d}}_{\alpha\beta}\right)^{\mathsf{T}}t_2\left(\left|\tilde{{\bf d}}_{\alpha\beta}\right|\right)Q_{0}\tilde{{\bf d}}_{\alpha\beta}\right]\tau^0.
\end{split}
\label{eq:HopChiralC2}
\end{equation}
We emphasize that following our analysis of structural chirality in the crystalline double-Weyl model [Eq.~(\ref{eq:quadraticStructuralChirality}) and the text following Eq.~(\ref{eq:charge2inversionLimit})], we have implemented Eq.~(\ref{eq:HopChiralC2}) such that the local chirality of each site $\chi_{\alpha}$ determines the signs of all of the terms arising from contractions of the matrices $Q_{x^2-y^2}$ and $Q_{xy}$ and the vector $\bm{V}_{z}$, rather than just a subset of those terms.
For the remainder of this section, we will refer to Eq.~(\ref{eq:HopChiralC2}) as the ``disordered double-Weyl tight-binding model.''

\begin{figure}[t]
\centering
\includegraphics[width=\linewidth]{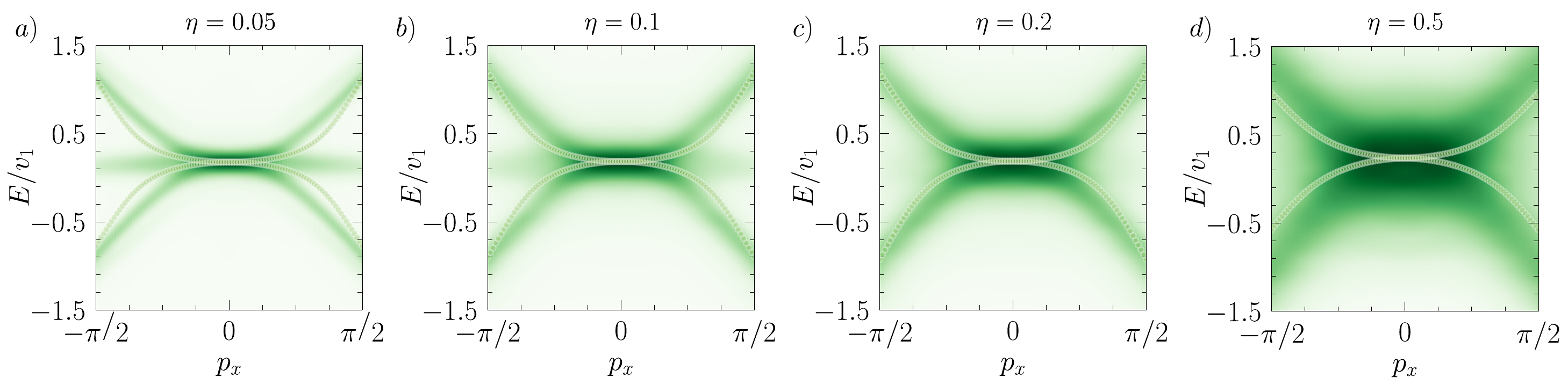}
\caption{ Spectral function and effective Hamiltonian spectrum of the double-Weyl model with smectic Gaussian structural disorder.
(a-d) The average spectral function $\bar{A}(E,{\bf p})$ [Eq.~(\ref{eq:SpecFunc})] of the double-Weyl tight-binding model [Eq.~(\ref{eq:AmoC2})] on a lattice with increasing random smectic structural disorder [$d_{A}=2$ in Eq.~(\ref{eq:finiteDimAmorph})] parameterized by the standard deviation $\eta$, as well as random frame disorder with the same standard deviation $\eta$ and chirality domains of unequal volume [area per layer], as detailed in Appendix~\ref{app:lattices}.
The data were generated using Eq.~(\ref{eq:HopChiralC2}) with the parameters in Eq.~(\ref{eq:paramsC2amo}) implemented with a single domain in each replica of right-handed sites with $n_R=N_R/N_{\mathrm{sites}}=0.7$, and with the remaining volume in each replica containing a contiguous domain of left-handed sites with a corresponding concentration of $n_L=1-N_R/N_{\mathrm{sites}}=0.3$.
Each panel shows data generated by averaging over 50 disorder realizations [replicas] with $20^{3}=8000$ sites [\emph{i.e.} 20 identical layers with 400 sites each], as detailed in Appendix~\ref{app:PhysicalObservables}.
For all values of $\eta$, $\bar{A}(E,{\bf p})$ exhibits a quadratically dispersing feature centered around ${\bf p}={\bf 0}$ that becomes increasingly diffuse as $\eta$ is increased, as well as upwardly shifted in energy by a [weakly] disorder-renormalized chemical potential $\tilde{\mu}$.
Even at small $\eta$, we also observe a flat-band-like feature centered at the nodal degeneracy, which we attribute to chirality domain walls in the double-Weyl model subject to $d_{A}=2$ smectic disorder.
In the remainder of this section, we will precisely show that the quadratic spectral features at ${\bf p}={\bf 0}$ in (a-d) represent increasingly disordered generalizations of the crystalline $\Gamma$-point double-Weyl fermion in Fig.~\ref{fig:charge2_figBulk}(c).
As shown in Fig.~\ref{fig:3DGreenC2}, the off-diagonal-in-momentum matrix elements of the average Green's function $\bar{\mathcal{G}}(E,{\bf p},{\bf p}')$ begin to vanish in the $(p_{x},p_{y})$-plane at moderate disorder scales [$\eta=0.2$ in panel (c)], and are nearly vanishing for strong disorder [$\eta=0.5$ in panel (d)].
We may therefore in each panel limit consideration to the \emph{momentum-diagonal} average Green's function $\bar{\mathcal{G}}(E,{\bf p})$ [Eq.~(\ref{eq:averageOneMomentumGreen})], from which we construct an effective Hamiltonian $\mathcal{H}_{\mathrm{Eff}}(E_{C},{\bf p})$ [Eq.~(\ref{eq:AvgHEff})] whose bands are plotted with light green circles.
$\mathcal{H}_{\mathrm{Eff}}(E_{C},{\bf p})$ in each panel was specifically constructed using a reference energy cut $E_{C}$ centered at the maximum value of the spectral function at ${\bf p}={\bf 0}$ [approximately $\tilde{\mu}$] to maximize its accuracy [see Appendix~\ref{app:EffectiveHamiltonian}], where $E_{C}$ for each panel is respectively given by (a) $E_C /v_1 = 0.37$, (b)
$E_C /v_1 = 0.38$, (c) $E_C /v_1 = 0.38$, and (d) $E_C /v_1 = 0.39$.
In all panels, and for all choices of $E_{C}$ [see Fig.~\ref{fig:H_eff_benchmark} and the surrounding text], the effective Hamiltonian exhibits a quadratic nodal degeneracy at ${\bf p}={\bf 0}$; we will shortly in Fig.~\ref{fig:C2_WL} use Wilson loops to show that this degeneracy is a $|C|=2$ double-Weyl fermion.}
\label{fig:C2_bands_amo}
\end{figure}

As discussed in Appendix~\ref{app:PhysicalObservables}, we wish to model solid-state systems that have thermodynamically large numbers of atoms [or short-range-interacting patches~\cite{ProdanKohnNearsighted}] and are self-averaging.  
However using our computational hardware, we can only simulate individual disorder realizations with $\sim 20^{3}$ [$\sim 8000$] atoms.
We therefore additionally analyze the non-crystalline double-Weyl model using an \emph{averaging procedure} in which we construct a replica-averaged~\cite{ParisiReplicaCourse,BerthierGlassAmorphousReview} momentum-resolved Green's function $\bar{\mathcal{G}}(E,\mathbf{p})$ [Eq.~(\ref{eq:averageOneMomentumGreen})] by averaging the momentum-resolved matrix Green's function over multiple [$\approx 20-50$] \emph{replicas} that each contain distinct randomly generated lattice distortions or random lattices.
As detailed in Appendix~\ref{app:PhysicalObservables}, we obtain $\bar{\mathcal{G}}(E,\mathbf{p})$ by first computing the diagonal-in-momentum piece of the momentum-resolved matrix Green's function $\mathcal{G}(E,{\bf p},{\bf p})$, while numerically establishing that the off-diagonal-in-momentum elements of $\bar{\mathcal{G}}(E,{\bf p},{\bf p}')$ vanish in the $(p_{x},p_{y})$-coordinate plane [on the average] for lattices with strong smectic disorder.
Specifically, for the non-crystalline double-Weyl model in Eq.~(\ref{eq:HopChiralC2}) with Gaussian structural disorder, we found in Fig.~\ref{fig:3DGreenC2} that the off-diagonal-in-momentum matrix elements of $\bar{\mathcal{G}}(E,{\bf p},{\bf p}')$ begin to vanish in the $(p_{x},p_{y})$-plane at $\eta=0.2$, which we in this section term ``moderate'' disorder, and are nearly vanishing for $\eta=0.5$, which we hence term ``strong'' disorder.
We therefore limit consideration to the diagonal-in-momentum matrix $\bar{\mathcal{G}}(E,\mathbf{p})$, which we compute by elementwise averaging $\mathcal{G}(E,{\bf p},{\bf p})$ at the same ${\bf p}$ [Eq.~(\ref{eq:averageOneMomentumGreen})].

\paragraph*{\bf Energy Spectrum} -- $\ $ In Fig.~\ref{fig:C2_bands_amo} we plot the disorder-averaged, momentum-resolved spectral function for the disordered double-Weyl tight-binding model with the parameters:
\begin{equation}
t_{xy} = 0.1,\
t_z = 0.12,\
v_1 =  2.5,\
v_z= 2,\
v_2 = 3,
\label{eq:paramsC2amo}
\end{equation}
placed on a lattice with increasing random smectic [$d_{A}=2$ in Eq.~(\ref{eq:finiteDimAmorph})] Gaussian structural disorder parameterized by a standard deviation $\eta$, as detailed in Appendix~\ref{app:lattices}.
The spectra in Fig.~\ref{fig:C2_bands_amo} were also computed using random local frame disorder [Eqs.~\eqref{appeq:rotaC2} and~\eqref{eq:HopChiralC2}] implemented with the same standard deviation $\eta$ as the lattice disorder, as well as chirality domains of unequal volume [70\% right-handed, 30\% left-handed] within each disorder replica [see Appendix~\ref{app:DiffTypesDisorder}].
In all panels of Fig.~\ref{fig:C2_bands_amo}, we observe a quadratically dispersing feature centered around ${\bf p}={\bf 0}$ that exhibits increased spectral broadening for increasing disorder, but nevertheless strongly resembles the crystalline $\Gamma$-point double-Weyl fermion in Fig.~\ref{fig:charge2_figBulk}(c) for all values of $\eta$ in Fig.~\ref{fig:C2_bands_amo}.
We also, even for very weak disorder [$\eta\sim 0.05$], observe a flat-band-like feature centered at the quadratic nodal degeneracy in Fig.~\ref{fig:C2_bands_amo}, which we attribute to chirality domain walls in the double-Weyl model subject to $d_{A}=2$ smectic disorder.
This can be contrasted with the non-crystalline KW model previously analyzed in Appendix~\ref{app:amorphousKramers}, for which flat-band-like features pinned to the nodal degeneracy are notably absent for nematic Gaussian disorder [Fig.~\ref{fig:KW_Bands_amo}], and are only weakly visible on Mikado lattices [Fig.~\ref{fig:KW_Unif}(e)].
The nodal spectral features in Fig.~\ref{fig:C2_bands_amo} at ${\bf p}={\bf 0}$ also shift upwards in energy as $\eta$ is increased, reflecting the presence of a disorder-renormalized chemical potential $\tilde{\mu}$.
However unlike in the disordered KW model [Fig.~\ref{fig:KW_Bands_amo}], both $\tilde{\mu}$ and the strength of the [quadratic] dispersion of the nodal degeneracy at ${\bf p}={\bf 0}$ in the disordered double-Weyl model in Fig.~\ref{fig:C2_bands_amo} vary much more weakly with $\eta$.

\begin{figure}[t]
\centering   
\includegraphics[width=\linewidth]{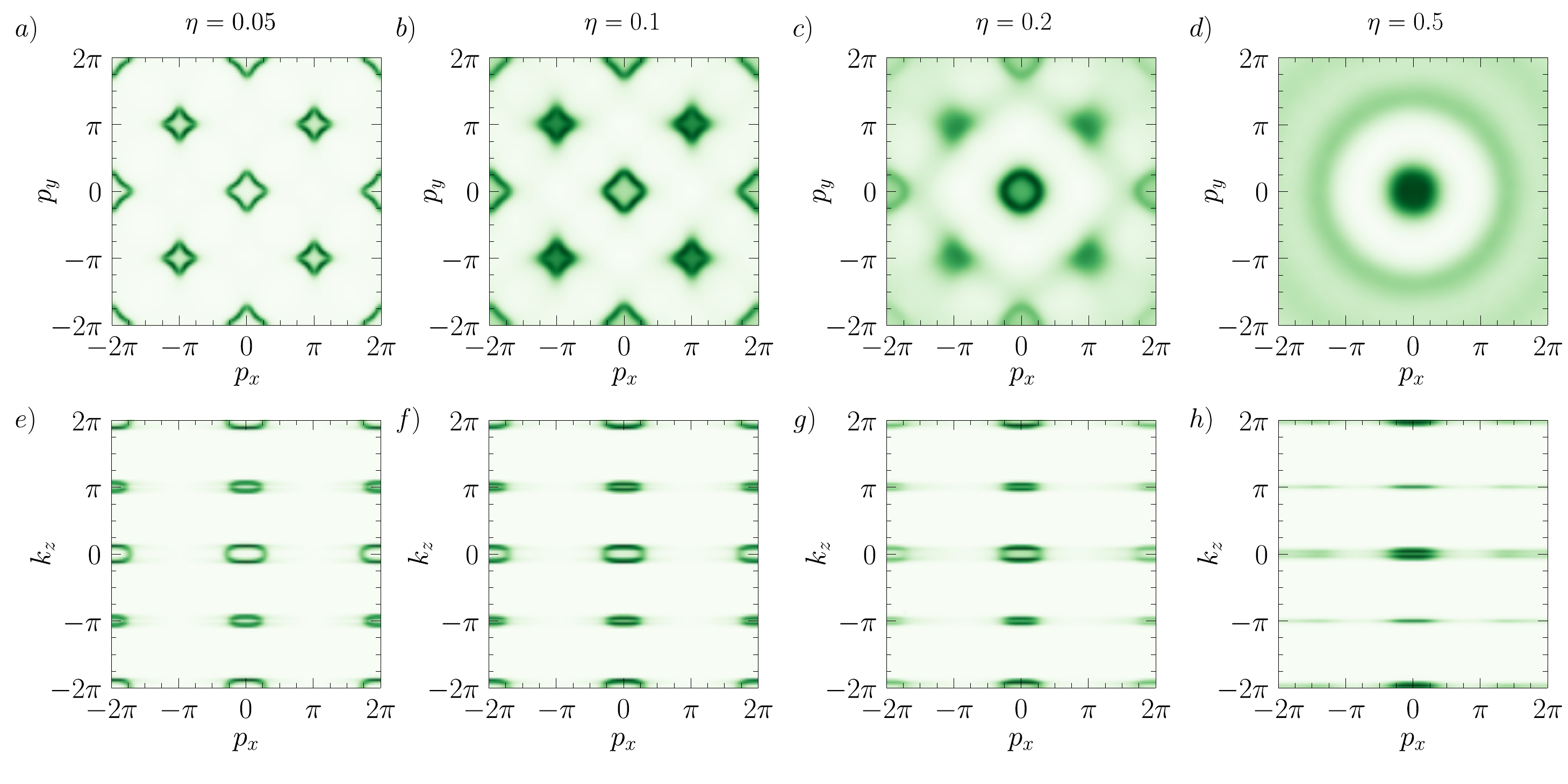}
\caption{Constant-energy spectrum of the disordered double-Weyl model.
(a-f) Constant-energy cuts of the average spectral function $\bar{A}(E,{\bf p})$ [Eq.~(\ref{eq:SpecFunc})] of the non-crystalline double-Weyl model [Eq.~(\ref{eq:HopChiralC2})] with the parameters in Eq.~(\ref{eq:appTBparamsC2}), contiguous domains of right- and left-handed sites with the respective concentrations $n_R=N_R/N_{\mathrm{sites}}=0.7$ and $n_L=1-N_R/N_{\mathrm{sites}}=0.3$, and increasing random smectic structural and frame disorder parameterized by the standard deviation $\eta$, as detailed in Appendix~\ref{app:lattices}.
Panels (a-d) show $\bar{A}(E,{\bf p})$ at $k_{z}=0.1$ and panels (e-h) show $\bar{A}(E,{\bf p})$ at $p_{y}=0.1$.
To generate each panel, we first compute the momentum-resolved Green's function $\bar{\mathcal{G}}(E,{\bf p})$ averaged over 50 disorder realizations [replicas] with $20^{3}$ sites each, as detailed in Appendix~\ref{app:PhysicalObservables}.
Because increasing $\eta$ generates a [weak] disorder-renormalized shift in the chemical potential at ${\bf p}={\bf 0}$ [see Fig.~\ref{fig:C2_bands_amo}], then for each panel in this figure, we first obtain a reference energy $E_{C}$ at which the spectral weight $\bar{A}(E_{C},{\bf p})$ at ${\bf p}={\bf 0}$ is maximized.
We then compute $\bar{A}(E,{\bf p}) \propto \text{Im}\{\Tr[\bar{\mathcal{G}}(E,{\bf p})]\}$ at $E/v_1=E_{C}/v_1 - 0.41$ in order to approximately capture the same cross section of the nodal feature at ${\bf p}={\bf 0}$ in Fig.~\ref{fig:C2_bands_amo} for varying $\eta$.
In addition to the quadratically dispersing spectral feature at ${\bf p}={\bf 0}$, the systems with strong disorder [$\eta=0.5$ in (d,h)] exhibit broadened, ring-like spectral features at $k_{z}=0,\pi$ in the vicinity of $\sqrt{p_{x}^{2} + p_{y}^{2}}=\pi\sqrt{2}/\bar{a}$ and $\sqrt{p_{x}^{2} + p_{y}^{2}}=2\pi/\bar{a}$, where $\bar{a}=1$ is the average nearest-neighbor spacing in the real-space $(x,y)$-plane. 
We find that the ring-like features in (d,h) originate from rotationally averaging and disorder-broadening the double-Weyl fermions at $k_{z}=0,\pi$ and $\sqrt{k_{x}^{2} + k_{y}^2}=\pi\sqrt{2}, 2\pi$ in the crystalline case, analogously to the ring-like, higher-Brillouin-zone Dirac surface states recently observed in amorphous Bi$_{2}$Se$_{3}$~\cite{corbae_evidence_2020,Ciocys2023}.
Unlike the $\bar{A}(E,{\bf p})$ cuts at constant energy and $k_{z}$ in (a-d), the $\bar{A}(E,{\bf p})$ cuts at constant energy and $p_{y}$ in (e-h) only vary as functions of $\eta$ in the $k_{z}=0,\pi$ planes, because each non-crystalline double-Weyl system in this figure is only subject to $d_{A}=2$ smectic disorder in the $(x,y)$-plane and remains translationally invariant in the $z$-direction [see Fig.~\ref{fig:dadtdf}(b) and Appendix~\ref{app:SmecticNematicDisorder}]. 
We specifically observe that the spectral features at $|p_{x}| = 0, 2\pi$ and $k_{z} = 0, \pi$ in (e-g) become broadened with increasing $\eta$ and eventually (h) loosely connected to each other in the $k_{z}=0,\pi$ planes by spectral weight from the ring-like feature at $\sqrt{p_{x}^{2} + p_{y}^{2}}=\pi\sqrt{2}$ in (d).
Interestingly, $\bar{A}(E,{\bf p})$ in (g,h) also closely resembles the experimentally observed Fermi surface of the quasi-1D structurally chiral Weyl semimetal (TaSe$_4$)$_2$I in its high-temperature state~\cite{TypeIIIWeylARPES,WujunZhijunTSIARPES,OpticalChangesTSIARPES}.}
\label{fig:DOSC2}
\end{figure}

Using the disorder-averaged, momentum resolved Green's function $\bar{\mathcal{G}}(E,{\bf p})$, we can also compute an approximate [effective] single-particle Hamiltonian $\mathcal{H}_{\text{Eff}}(E_{C},{\bf p}) \equiv \mathcal{H}_{\text{Eff}}(\mathbf{p})$~\cite{varjas_topological_2019,marsal_topological_2020,marsal_obstructed_2022} for the spectral features at ${\bf p}={\bf 0}$ in the disordered double-Weyl model [Eq.~(\ref{eq:AvgHEff})].
Though $\mathcal{H}_{\mathrm{Eff}}(E_{C},{\bf p})$ only represents a mean-field approximation of the many-particle, momentum-dependent Hamiltonian, and is therefore generally dependent on the energy cut $E_{C}$ at which it is obtained, we have shown in Appendix~\ref{app:EffectiveHamiltonian} that the nodal degeneracies and topology of $\mathcal{H}_{\mathrm{Eff}}(E_{C},{\bf p})$ near ${\bf p}={\bf 0}$ are numerically stable and surprisingly insensitive to $E_{C}$  in the models studied in this work.
In each panel of Fig.~\ref{fig:C2_bands_amo}, we show the bands of $\mathcal{H}_{\mathrm{Eff}}({\bf p})$ in light green circles, in each case computed using a reference energy cut $E_{C}$ centered at the maximum value of the spectral function at ${\bf p}={\bf 0}$ to maximize the spectral accuracy of the effective Hamiltonian.
For all disorder strengths $\eta$, the effective Hamiltonian bands in Fig.~\ref{fig:C2_bands_amo} exhibit quadratic nodal degeneracies at ${\bf p}={\bf 0}$.
Below, we will provide evidence that the quadratically dispersing spectral feature in Fig.~\ref{fig:C2_bands_amo} at ${\bf p}={\bf 0}$ in $\bar{A}(E,{\bf p})$ [approximated by the quadratic nodal degeneracy at ${\bf p}={\bf 0}$ in $\mathcal{H}_{\mathrm{Eff}}({\bf p})]$ is in fact precisely a disordered double-Weyl fermion whose topology is controlled by \emph{average} structural chirality, analogous to its crystalline counterpart

We next construct constant-energy spectral cuts of the disordered double-Weyl tight-binding model to explore spectral features at higher momenta.
In Fig.~\ref{fig:DOSC2} we plot the average spectral function $\bar{A}(E,{\bf p})$ of a disordered double-Weyl system [Eq.~(\ref{eq:HopChiralC2})] at fixed $E$ with the parameters in Eq.~(\ref{eq:paramsC2amo}) and increasing smectic structural and frame disorder averaged over 50 replicas that each contain contiguous domains of right- and left-handed sites with the respective concentrations $n_R=N_R/N_{\mathrm{sites}}=0.7$ and $n_L=1-N_R/N_{\mathrm{sites}}=0.3$.
Fig.~\ref{fig:DOSC2}(a-d) show $\bar{A}(E,{\bf p})$ cuts at constant $k_{z}$ and $E$ [$k_{z}=0.1$], and Fig.~\ref{fig:DOSC2}(e-h) show $\bar{A}(E,{\bf p})$ cuts at constant $p_{y}$ and $E$ [$p_{y}=0.1$].
Interestingly, as $\eta$ is increased, $\bar{A}(E,{\bf p})$ develops 2D ring-like spectral features at $k_{z}=0,\pi$ in the vicinity of $\sqrt{p_{x}^{2} + p_{y}^{2}}=\pi\sqrt{2}/\bar{a}$ and $\sqrt{p_{x}^{2} + p_{y}^{2}}=2\pi/\bar{a}$, where $\bar{a}=1$ is the average nearest-neighbor spacing in the real-space $(x,y)$-plane. 
The ring-like features in Fig.~\ref{fig:DOSC2}(d,h) can be understood as respectively originating from averaging the crystalline double-Weyl fermions at $\sqrt{k_{x}^{2} + k_{y}^2}=\pi\sqrt{2}, 2\pi$ and $k_{z}=0,\pi$ in Fig.~\ref{fig:DOSC2}(a) over random in-plane SO(2) frame orientations and lattice spacings, giving rise to the characteristic isotropic spectral features of an amorphous system [with $d_{A}=2$ in the notation of Eq.~(\ref{eq:finiteDimAmorph})]~\cite{spring_amorphous_2021,springMagneticAverageTI,corbae_evidence_2020,Ciocys2023}.
Like the larger-$|{\bf p}|$ ring- [sphere-] like spectral features in the non-crystalline KW model in Fig.~\ref{fig:DOSKW}(b,c), the 2D ring-like double-Weyl features in Fig.~\ref{fig:DOSC2}(d,h) are also analogous to the 2D surface states of the amorphous 3D strong topological insulator Bi$_{2}$Se$_{3}$~\cite{corbae_evidence_2020,Ciocys2023}.
However, unlike in the constant-energy spectra of the disordered KW model previously analyzed in Fig.~\ref{fig:DOSKW} and the surrounding text, the $\bar{A}(E,{\bf p})$ cuts at constant energy and $p_{y}$ in Fig.~\ref{fig:DOSC2}(e-h) only vary as functions of $\eta$ in the $k_{z}=0,\pi$ planes, because each non-crystalline double-Weyl system analyzed in this section is only subject to $d_{A}=2$ smectic disorder in the $(x,y)$-plane, and remains translationally invariant in the $z$-direction [see Fig.~\ref{fig:dadtdf}(b) and Appendix~\ref{app:SmecticNematicDisorder}].
We specifically observe that the spectral features at $|p_{x}| = 0, 2\pi$ and $k_{z} = 0, \pi$ in Fig.~\ref{fig:DOSC2}(e-g) broaden with increasing $\eta$, and eventually at $\eta=0.5$ in Fig.~\ref{fig:DOSC2}(h) become loosely connected to each other in the $k_{z}=0,\pi$ planes by spectral weight from the ring-like feature at $\sqrt{p_{x}^{2} + p_{y}^{2}}=\pi\sqrt{2}$ in Fig.~\ref{fig:DOSC2}(d).
Intriguingly, the weakly connected, topologically distinct, flattened Fermi pockets in the constant-$p_{y}$ non-crystalline double-Weyl spectra in Fig.~\ref{fig:DOSC2}(g,h) also closely resemble experimental ARPES measurements of the Fermi surface of the quasi-1D structurally chiral Weyl semimetal (TaSe$_4$)$_2$I~\cite{TypeIIIWeylARPES,WujunZhijunTSIARPES,OpticalChangesTSIARPES} above the onset temperature of its charge-density wave state~\cite{Monceau1984TSICDW,CDWWeyl,WeylCDWSpinTexture}.

\paragraph*{\bf Absence of Angular Momentum Texture} -- $\ $ Unlike the non-crystalline KW model previously analyzed in Appendix~\ref{app:amorphousKramers}, the non-crystalline double-Weyl model [Eqs.~(\ref{eq:HopChiralC2})] does not admit the computation of angular momentum textures.
To understand this, we first recognize that a spin or orbital angular momentum [OAM] texture in a [semi]metal arises from the [crystal or continuum pseudo-] momentum variation of two or more components of an angular-momentum-dependent spectral function vector [\emph{e.g.} $\langle\mathbf{S}(E,\mathbf{p})\rangle$ in Eq.~(\ref{eq:SpinDOS})]~\cite{chang2017large,Saito_TB_2016,Kohsaka_2017,OAMCDArpes1,OAMmultifold2,OAMnodalLine,OAMzxShenARPES3DTI}.
In Eqs.~\eqref{eq:AmoC2} and~\eqref{eq:TmatrixQuadraticDef}, we constructed the non-crystalline double-Weyl model by placing a time-reversed pair of spinless, angular-momentum-one $p_{x}\pm ip_{y}$ orbital basis states on each lattice site.
In the $2\times 2$ internal model subspace of the $p_{x}\pm ip_{y}$ basis states, we may represent the $\hat{L}^{z}$ operator as a diagonal matrix $L^z$ with eigenvalues $l^{z}=\pm 1$. 
We may then use $L^{z}$ to define the $z$-component of the [``local''~\cite{OAMCDArpes1,OAMmultifold2,OAMnodalLine,OAMzxShenARPES3DTI}] OAM texture $\langle L^{z}(E,\mathbf{p})\rangle \propto \text{Im}\{\Tr[L^{z}\bar{\mathcal{G}}(E,{\bf p})]\}$ by analogy to the spin texture defined in Eq.~(\ref{eq:SpinDOS}) and the surrounding text.

However, because each site does not also contain an internal $l^{z}=0$ [\emph{e.g.} spinless $p_{z}$] degree of freedom, then we cannot define matrix representatives of $\hat{L}^{x}$ and $\hat{L}^{y}$ that both transform consistently under the [exact or average] system symmetries and satisfy the canonical angular momentum commutation relations $\left[L^a,L^b\right] = i \epsilon^{abc}L^c$, where $\epsilon^{abc}$ is the Levi-Civita symbol~\cite{GriffithsBook}.
This can also be seen by contradiction: if there were instead well-defined matrix representatives for $\hat{L}^{x,y}$ in the double-Weyl model, this would require the raising and lowering [ladder] operators $\hat{L}^{\pm}=\hat{L}^x\pm i \hat{L}^y$ to also have well-defined matrix representatives.
Acting with $\hat{L}^{\pm}$ on an $l^{z}=\pm 1$ $p_{x}\pm ip_{y}$ basis state would, however, result in an even-$l^{z}$ state outside of the $2\times 2$ internal model subspace of $p_{x}\pm ip_{y}$ orbitals.  
This can be contrasted with the non-crystalline multifold model that we will shortly analyze in Appendix~\ref{app:amorphousMultifold}, which instead contains a full shell of spinless $p$ orbitals at each site [in addition to a spinless $s$ orbital, see the text surrounding Eqs.~\eqref{eq:Amo3F} and~\eqref{eq:3FTmatrix}], and hence admits the computation of an OAM texture [Eqs.~(\ref{eq:OAMDOS}) and~(\ref{eq:OAMlMatricies})].

\paragraph*{\bf Wilson Loops} -- $\ $ Having shown that the non-crystalline double-Weyl model [Eq.~(\ref{eq:HopChiralC2})] continues to exhibit a quadratic nodal degeneracy at ${\bf p}={\bf 0}$ for strong, chirality-imbalanced disorder [Fig.~\ref{fig:C2_bands_amo}(c,d)], we will now use an amorphous generalization of the Wilson loop method [see Appendix~\ref{sec:WilsonBerry} and Refs.~\cite{Fidkowski2011,AndreiXiZ2,ArisInversion,Cohomological,HourglassInsulator,DiracInsulator,Z2Pack,BarryFragile,AdrienFragile,HOTIBernevig,HingeSM,WiederAxion,KoreanFragile,ZhidaBLG,TMDHOTI,KooiPartialNestedBerry,PartialAxionHOTINumerics,GunnarSpinFragileWilson,BinghaiOscillationWilsonLoop,Wieder22}] to precisely show that the ${\bf p}={\bf 0}$ nodal degeneracy is a disordered [non-crystalline] double-Weyl fermion with a nontrivial topological chiral charge.

\begin{figure}[t]
\centering
\includegraphics[width=\linewidth]{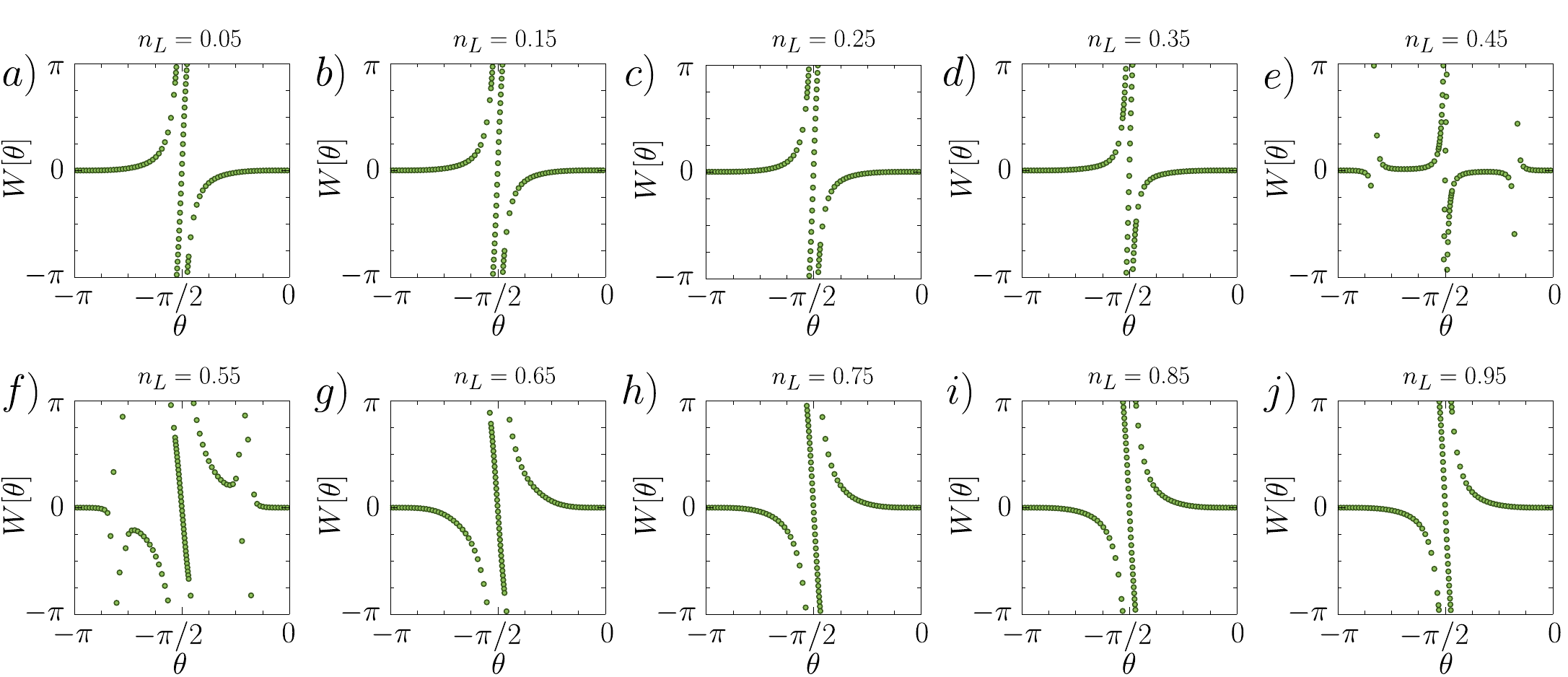}
\caption{Sphere Wilson loop spectrum of the disordered double-Weyl model for varying chirality concentrations.
To generate each panel in this figure, we place the non-crystalline double-Weyl  model [Eq.~(\ref{eq:HopChiralC2})] with the parameters in Eq.~(\ref{eq:appTBparamsC2}) on a lattice with $N_{\mathrm{sites}}=15^3$ [\emph{i.e.} 15 identical layers with 225 sites each] and random smectic Gaussian structural and local frame disorder parameterized by the fixed standard deviation $\eta=0.2$, as detailed in Appendix~\ref{app:lattices}. 
We then generate the replica-averaged momentum-resolved Green's function $\bar{\mathcal{G}}(E,\mathbf{p})$ [Eq.~(\ref{eq:averageOneMomentumGreen})] by averaging the system over 50 disorder realizations [replicas] that each contain contiguous domains of right- and left-handed sites with the respective concentrations $n_R=N_R/N_{\mathrm{sites}}$ and $n_{L}=1-n_{R}$.
(a-j) For 10 disorder ensembles with increasing values of $n_{L}$, we construct for each ensemble an effective Hamiltonian $\mathcal{H}_{\mathrm{Eff}}({\bf p})=\mathcal{H}_{\mathrm{Eff}}(E_{C},{\bf p})$ [Eq.~(\ref{eq:AvgHEff})] using a reference energy $E_{C}$ corresponding to the largest spectral weight $\bar{A}(E,{\bf p})$ at ${\bf p}={\bf 0}$ [Eq.~(\ref{eq:SpecFunc})], which as discussed in the text surrounding Fig.~\ref{fig:H_eff_benchmark} maximizes the spectral accuracy of $\mathcal{H}_{\mathrm{Eff}}({\bf p})$.
We then use the eigenstates of $\mathcal{H}_{\mathrm{Eff}}({\bf p})$ to compute the amorphous [disordered] Wilson loop spectrum introduced in this work [Appendix~\ref{sec:WilsonBerry}] on a sphere surrounding the nodal degeneracy at ${\bf p}={\bf 0}$ in each disorder ensemble.
Beginning with a moderately disordered system [$\eta=0.2$, see Fig.~\ref{fig:3DGreenC2}] with (a) almost entirely right-handed sites [$n_{L}=0.05$] and continuing in increasing $n_{L}$ to a system (j) with almost entirely left-handed sites [$n_{L}=0.95$], we observe a quantized Wilson loop winding as a function of the sphere polar angle $\theta$ of (a-d) $C=2$ for $n_{L}<0.5$, (g-j) $C=-2$ for $n_{L}>0.5$, and (e,f) a region in the vicinity of $n_{L}\approx 0.5$ with a non-smooth Wilson spectrum.
This provides a precise indicator that the nodal degeneracy at ${\bf p}={\bf 0}$ in Fig.~\ref{fig:C2_bands_amo} is a disordered [non-crystalline] double-Weyl fermion that, analogous to its crystalline counterpart [Fig.~\ref{fig:charge2_figWilson}], carries a topological chirality that is tunable via the average system structural chirality.}
\label{fig:C2_WL}
\end{figure}

To begin, in a crystalline system, the chiral charge of a nodal degeneracy can be obtained by computing the winding number of the Wilson loop spectrum [non-Abelian Berry phases] evaluated over the occupied bands on a sphere surrounding the nodal degeneracy~\cite{Z2Pack}.
At each value of the polar angle $\theta$ of the sphere [see Fig.~\ref{fig:Wilson_schema}(b)], the Wilson loop is computed using the Bloch wavefunctions of the occupied states.
To perform an analogous calculation for a given disorder ensemble of the non-crystalline double-Weyl model [Eq.~(\ref{eq:HopChiralC2})], we therefore first construct an effective Hamiltonian $\mathcal{H}_{\text{Eff}}(\mathbf{p})$ [see Refs.~\cite{varjas_topological_2019,marsal_topological_2020,marsal_obstructed_2022} and Appendix~\ref{app:EffectiveHamiltonian}] for the nodal degeneracy at ${\bf p}={\bf 0}$ using the replica-averaged momentum-resolved Green's function $\bar{\mathcal{G}}(E,\mathbf{p})$ [Eq.~(\ref{eq:averageOneMomentumGreen})]. 
For each disorder ensemble of the non-crystalline double-Weyl model, we specifically first identify an energy $E_{max}$ where $\bar{A}(E,{\bf p})$ is largest at ${\bf p}={\bf 0}$, and then we construct $\mathcal{H}_{\text{Eff}}(\mathbf{p})$ using the reference energy cut $E_{C}=E_{max}$, which maximizes the spectral accuracy of $\mathcal{H}_{\text{Eff}}(\mathbf{p})$ [see Fig.~\ref{fig:H_eff_benchmark} and the surrounding text].
Finally, we use the eigenstates of $\mathcal{H}_{\mathrm{Eff}}({\bf p})$ to compute the amorphous [disordered] Wilson loop spectrum [Appendix~\ref{sec:WilsonBerry}] on a sphere surrounding the nodal degeneracy at ${\bf p}={\bf 0}$.

In Fig.~\ref{fig:C2_WL}, we show the sphere Wilson loop spectrum for the disordered double-Weyl model for 10 disorder ensembles with 50 replicas each, where each disorder replica has $N_{\mathrm{sites}}=15^{3}$ sites [subdivided into 15 identical layers with 225 sites each], smectic Gaussian lattice and local frame disorder with the standard deviation $\eta=0.2$ [see Appendix~\ref{app:lattices}], and contiguous domains of right- and left-handed sites with varying chirality concentrations respectively given by $n_R=N_R/N_{\mathrm{sites}}$ and $n_{L}=1-n_{R}$.
Beginning in Fig.~\ref{fig:C2_WL}(a) with a moderately disordered system [$\eta=0.2$, see Fig.~\ref{fig:3DGreenC2}] containing almost entirely right-handed sites [$n_{L}=0.05$] and continuing in increasing $n_{L}$ to the system in Fig.~\ref{fig:C2_WL}(j) with almost entirely left-handed sites [$n_{L}=0.95$], we observe a quantized Wilson loop winding of $C=2$ for $n_{L}<0.5$ [Fig.~\ref{fig:C2_WL}(a-d)] and $C=-2$ for $n_{L}>0.5$ [Fig.~\ref{fig:C2_WL}(g-j)].
In the vicinity of $n_{L}\approx 0.5$, the Wilson loop eigenvalues become non-smooth, indicating that the sphere Wilson loop is within the close vicinity of a topological quantum critical point [energy gap closure]; this behavior can be seen across Fig.~\ref{fig:C2_WL}(e,f).
In the case of the disordered double-Weyl model introduced in this section [Eq.~(\ref{eq:HopChiralC2})], the Wilson loop spectra in Fig.~\ref{fig:C2_WL} specifically indicate that the nodal degeneracy at ${\bf p}={\bf 0}$ in Fig.~\ref{fig:C2_bands_amo} is a disordered chiral [double-Weyl] fermion that, analogous to its crystalline counterpart [Fig.~\ref{fig:charge2_figWilson}], carries a topological chirality that is tunable via the average system structural chirality.  
Though we have only demonstrated quantized and tunable Wilson loop winding in Fig.~\ref{fig:C2_WL} for systems with moderate structural disorder [$\eta=0.2$], we will soon below show that the link between average structural chirality and low-energy topological chirality also holds for the non-crystalline double-Weyl model [Eq.~(\ref{eq:HopChiralC2})] on fully-disordered [random] lattices that lack well-defined crystalline limits [see Fig.~\ref{fig:C2_Unif}].

\paragraph*{\bf Non-Crystalline Group Theory} -- $\ $ We will next make further direct connection between the disordered double-Weyl fermion in Fig.~\ref{fig:C2_bands_amo} and its crystalline counterpart by employing a ${\bf k}\cdot {\bf p}$ deformation procedure based in symmetry group theory.  
First, as discussed in Eq.~(\ref{eq:appHKPC2}) and the surrounding text, when Eq.~(\ref{eq:HopChiralC2}) is placed on a regular tetragonal lattice and expanded in crystal momentum ${\bf k}$ about the $\Gamma$ point [${\bf k}={\bf 0}$], the resulting ${\bf k}\cdot{\bf p}$ Hamiltonian characterizes a crystalline double-Weyl fermion:
\begin{equation}
\mathcal{H}({\bf k}) = v_1 k_x k_y \tau^{x} + \frac{1}{2}v_2\left(k_y^2-k_x^2\right)\tau^{y} + v_z k_z \tau^{z}.
\label{eq:C2numericsKP}
\end{equation}
The double-Weyl ${\bf k}\cdot {\bf p}$ Hamiltonian in Eq.~(\ref{eq:C2numericsKP}) specifically transforms in a two-dimensional, single-valued small corep of the little group at $\Gamma$, which is isomorphic to tetragonal SG 89 ($P422$).  
As discussed in Appendices~\ref{app:pseudoK} and~\ref{app:corepAmorphous} and established in Refs.~\cite{zallen_physics_1998,VanMechelen:2018cy,vanMechelenNonlocal,Ciocys2023,Grushin2020,Corbae_2023}, systems with strong structural disorder are spectrally isotropic at long wavelengths in their $d_{A}$ disordered ${\bf p}$-space directions [Eq.~(\ref{eq:finiteDimAmorph})].
Under the long-wavelength deformation and averaging procedure for $\Gamma$-point Hamiltonians introduced in this work, the exact Hamiltonian $\mathcal{H}({\bf k})$ in Eq.~(\ref{eq:C2numericsKP}) can be recast as an approximate, rotationally-invariant, \emph{effective} ${\bf k}\cdot {\bf p}$ Hamiltonian expanded about ${\bf p}={\bf 0}$ in the non-crystalline double-Weyl model [Eq.~(\ref{eq:HopChiralC2})] on a $d_{A}=2$ strongly disordered or random lattice:
\begin{equation}
\mathcal{H}_{\text{Eff}}(p_x,p_y,k_z) = \tilde{v}_{xy}\left[ p_xp_y\tilde{\tau}^x + \frac{1}{2}\left(p_y^2 - p_x^2\right)\tilde{\tau}^y\right] + v_z k_z \tau^z + \tilde{\mu}\tilde{\tau}^{0},
\label{eq:C2disorderKP}
\end{equation}
where $\tilde{v}_{xy}$ indicates the strength of disorder-renormalized OAM coupling in the real-space $(x,y)$-plane, $\tilde{\mu}$ is the disorder-renormalized chemical potential [see Fig.~\ref{fig:C2_bands_amo}], $\tilde{\tau}^{0}$ is the $2\times 2$ identity matrix, and where the $2\times 2$ Pauli matrices $\tilde{\tau}^{x,y}$ are generically equal to linear combinations of $\tau^{x,y}$ in Eq.~(\ref{eq:C2numericsKP}) due to in-plane OAM reference frame disorder [see the text surrounding Eqs.~(\ref{eq:pristineDoubleWeyl}) and~(\ref{appeq:rotaC2})].

Like the ${\bf k}\cdot{\bf p}$ Hamiltonian of a crystalline double-Weyl fermion, $\mathcal{H}_{\text{Eff}}(p_x,p_y,k_z)$ in Eq.~(\ref{eq:C2disorderKP}) transforms in a two-dimensional, single-valued small corep of a $d=3$ chiral little group $\tilde{G}_{\Gamma,2}$ [see the text preceding Eq.~(\ref{eq:appTBparamsC2})].
However unlike for crystalline double-Weyl fermions, $\tilde{G}_{\Gamma,2}$ is not an exact, discrete little group, but is now instead a \emph{continuous} and \emph{approximate} [average] little group [ALG] $\tilde{G}_{\Gamma,2}$ that can be decomposed as:
\begin{equation}
\tilde{G}_{\Gamma,2} = (p\infty 22)_{RG} \cup \{\mathcal{T}|000\}(p\infty 22)_{RG} \cup \{E|\epsilon 00\}(p\infty 22)_{RG} \cup \{\mathcal{T}|\epsilon 00\}(p\infty 22)_{RG},
\label{eq:C2numericsALG}
\end{equation}
where $\{E|\epsilon 00\}$ denotes an infinitesimal translation along the $x$-axis, and where $(p\infty 22)_{RG}$ denotes the non-crystallographic chiral rod group generated by continuous rotations about the $z$-axis $\tilde{C}_{(2\pi/\phi) z}$, twofold rotations about the $x$-axis $\tilde{C}_{2x}$, and integer lattice translations $\{E|001\}$ along the $z$-axis [see Refs.~\cite{MTQC,linegroupsbook,eightfoldRodGroups} and the text following Eq.~(\ref{eq:smecticALGbreakdown}) for further details of $\tilde{G}_{\Gamma,2}$].
The generating symmetries of $\tilde{G}_{\Gamma,2}$ in Eq.~(\ref{eq:C2numericsALG}) can be represented through their action on $\mathcal{H}_{\text{Eff}}(p_x,p_y,k_z)$ in Eq.~(\ref{eq:C2disorderKP}):
\begin{eqnarray}
\tilde{C}_{(2\pi/\phi) z}\mathcal{H}_{\text{Eff}}(p_{x},p_{y},k_{z})\tilde{C}_{(2\pi/\phi) z}^{-1} &=& e^{i\phi\tau^{z}}\mathcal{H}_{\text{Eff}}(\tilde{C}^{-1}_{(2\pi/\phi) z}{\bf p}_{x,y},k_{z})e^{-i\phi\tau^{z}}, \nonumber \\
\tilde{C}_{2x}\mathcal{H}_{\text{Eff}}(p_{x},p_{y},k_{z})\tilde{C}_{2x}^{-1} &=& \tilde{\tau}^{x}\mathcal{H}_{\text{Eff}}(p_{x},-p_{y},-k_{z})\tilde{\tau}^{x},  \nonumber \\
\tilde{\mathcal{T}}\mathcal{H}_{\text{Eff}}(p_{x},p_{y},k_{z})\tilde{\mathcal{T}}^{-1} &=& \tilde{\tau}^{x}\mathcal{H}^{*}_{\text{Eff}}(-p_{x},-p_{y},-k_{z})\tilde{\tau}^{x},
\label{eq:disorderDoubleWeylKPSyms}
\end{eqnarray}
where ${\bf p}_{x,y}=(p_{x},p_{y})$, $\phi$ denotes an infinitesimal rotation angle about the $z$-axis [such that $\phi=\pi/2$ is consistent with the matrix representative of $\tilde{C}_{4z}$ in Eq.~(\ref{eq:PristineCharge2TBsyms})], and where we note that the continuous translation symmetries in $\tilde{G}_{\Gamma,2}$ are represented as phases multiplied by the $2\times 2$ identity matrix $\tilde{\tau}^{0}$, and have hence been suppressed for notational simplicity.
In Eq.~(\ref{eq:disorderDoubleWeylKPSyms}), the tildes on $\tilde{C}_{(2\pi/\phi) z}$, $\tilde{C}_{2x}$, and $\tilde{\mathcal{T}}$ denote that the symmetries are elements of the single ALG $\tilde{G}_{\Gamma,2}$, because the symmetries act on integer-angular-momentum $p_{x}\pm ip_{y}$ internal degrees of freedom [see Refs.~\cite{BigBook,MTQC} and the text surrounding Eq.~(\ref{eq:AmoC2})].
The symmetry action of the smectic-disordered non-crystalline double-Weyl ${\bf k}\cdot {\bf p}$ Hamiltonian in Eq.~(\ref{eq:disorderDoubleWeylKPSyms}) can hence be contrasted with that of the $d_{A}=2$ smectic-disordered Kramers-Weyl ${\bf k}\cdot {\bf p}$ Hamiltonian in Eq.~(\ref{eq:smecticKPSyms}).
In both cases, the ${\bf k}\cdot {\bf p}$ Hamiltonians respect the symmetries of the chiral ALG $\tilde{G}_{\Gamma,2}$ [Eq.~(\ref{eq:C2numericsALG})] -- however, the non-crystalline Kramers-Weyl fermion in Eq.~(\ref{eq:kpKWSmectic}) transforms in a two-dimensional, double-valued [spinful] small corep of the double ALG $\tilde{G}_{\Gamma,2}$, whereas the non-crystalline double-Weyl fermion in Eq.~(\ref{eq:C2disorderKP}) transforms in a two-dimensional, single-valued [spinless] small corep of the single ALG $\tilde{G}_{\Gamma,2}$.
This importantly demonstrates that even in amorphous systems with the same average symmetry group, different local degrees of freedom and microscopic interactions can give rise to spectral features that transform in different \emph{irreducible coreps} of the average symmetry group, leading to the appearance of topologically distinct states~\cite{marsal_topological_2020}.

\paragraph*{\bf Disordered Fermi-Arc Surface States} -- $\ $ Having established that the bulk quadratic spectral feature at ${\bf p}={\bf 0}$ in Fig.~\ref{fig:C2_bands_amo} is a non-crystalline double-Weyl fermion, we will next explore disorder-driven shifts in spectral signatures of its bulk-boundary correspondence.
To begin, we place the non-crystalline double-Weyl tight-binding model [Eq.~(\ref{eq:HopChiralC2})] on a lattice with increasing random smectic structural disorder [$d_{A}=2$ in Eq.~(\ref{eq:finiteDimAmorph})] parameterized by the standard deviation $\eta$, using the same structural chirality imbalance percentages and system parameters as the bulk calculations in Fig.~\ref{fig:DOSC2}.
Unlike in our previous calculations, we now take the system to have periodic boundary conditions in the Cartesian $x$- and $z$-directions and open boundary conditions in the $y$-direction. 
We next compute the $\hat{y}$-normal surface-projected spectral function $\bar{A}_{\text{surf}}(E,{\bf p})$ [Eq.~(\ref{eq:averageSpectrumSurface})] averaged over 50 disorder replicas [noting that the Cartesian $\hat{y}$-normal surface can no longer be designated the $(010)$-surface in a lattice-disordered system].
Each disorder replica in our system is initially constructed with $20^{3}=8000$ sites subdivided into $20$ parallel, identical layers, where each layer contain $400$ sites and has a normal vector oriented in the Cartesian $z$-direction [see Appendix~\ref{app:SmecticNematicDisorder}].
We then further reduce the number of system sites by removing [``evaporating''] dangling [decoupled] surface atoms on the $\hat{y}$-normal surface that represent numerical artifacts of the disorder implementation process [see the text preceding Eq.~(\ref{eq:RealSpaceGreenSlab})]. 
Additionally, because surface chirality domain walls can bind flat-band-like states that overwhelm and obscure spectral signatures of topological surface Fermi arcs~\cite{InternalChiralTheory,InternalChiralExp,SussmanJammedDiracWeyl}, then we only place chirality domain walls deep within the bulk of each disorder replica.

\begin{figure}[t]
\centering
\includegraphics[width=\linewidth]{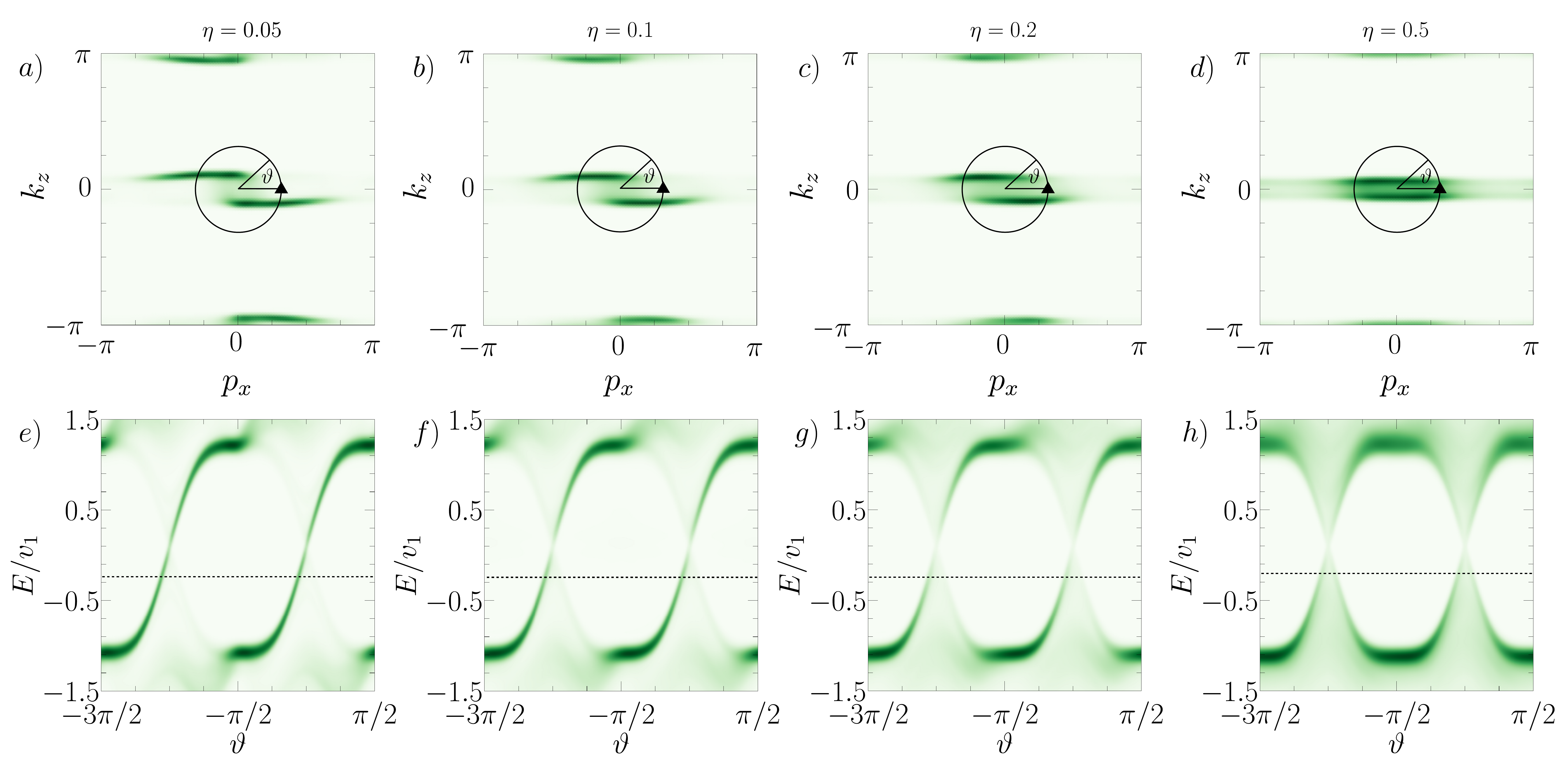}
\caption{Fermi-arc surface states of the disordered double-Weyl model.
In this figure, we show the $\hat{y}$-normal surface-projected, disorder-averaged spectral function $\bar{A}_{\text{surf}}(E,{\bf p})$ [Eq.~(\ref{eq:averageSpectrumSurface})] of the non-crystalline double-Weyl tight-binding model [Eq.~(\ref{eq:HopChiralC2})] on a lattice with increasing random smectic structural disorder [$d_{A}=2$ in Eq.~(\ref{eq:finiteDimAmorph})] parameterized by the standard deviation $\eta$, using the same chirality imbalance percentages and system parameters as in Fig.~\ref{fig:DOSC2}.
Each panel was generated by averaging over 50 disorder replicas, each with $N_{\text{sites}}=20^{3}$ prior to the removal [``evaporation''] of accidental dangling [decoupled] surface atoms generated by the disorder realization [see the text preceding Eq.~(\ref{eq:RealSpaceGreenSlab})]. 
Because surface chirality domain walls can bind flat-band-like states that overwhelm and obscure spectral signatures of topological surface Fermi arcs~\cite{InternalChiralTheory,InternalChiralExp,SussmanJammedDiracWeyl}, we have only implemented random chirality domain walls within the bulk of the finite [slab] systems used to generate this figure.
In panels (a-d), we plot $\bar{A}_{\text{surf}}(E,{\bf p})$ as a function of $p_x$ and $k_{z}$ for increasing $\eta$ at a fixed relative energy $E/v_{1}$ [respectively the dashed line in (e-h)] set to 0.41 below the maximum bulk spectral weight at ${\bf p}={\bf 0}$, in order to account for the disorder-renormalized chemical potential of the double-Weyl point [see Figs.~\ref{fig:C2_bands_amo} and~\ref{fig:DOSC2}].  
In (e-h), we plot $\bar{A}_{\text{surf}}(E,{\bf p})$ as a function of energy on counterclockwise circular paths, parameterized by $\vartheta$, surrounding $p_{x}=k_{z}=0$ for the increasingly disordered double-Weyl systems in (a-d), respectively.
In (a-c) and (e-g), $\bar{A}_{\text{surf}}(E,{\bf p})$ continues to exhibits two clear Fermi-arc surface states with the same connectivity and topological chirality [positive slopes] as the right-handed enantiomer of the crystalline double-Weyl model [Fig.~\ref{fig:charge2_figWilson}(b,c)].
Unlike the increasingly positively sloped and diffuse surface Fermi arcs in the disordered Kramers-Weyl model [Fig.~\ref{fig:KWamorphousArcs}], the surface Fermi arcs in the double-Weyl model remain sharp and surface-localized into the moderate-disorder regime [$\eta=0.2$ in (c,g)].
The Fermi-arc surface states in the double-Weyl model instead become trivial with increasing $\eta$ by retracting to $p_{x}=k_{z}=0$ across the constant energy cuts in (c,d), and by losing their well-defined chiral dispersion via merging with disorder-broadened bulk states in the circular contours in (g,h).
In the strong-disorder regime [$\eta=0.5$ in (d,h)], the Fermi arcs can specifically no longer be distinguished as surface-localized, right-moving states [sharp, dark green spectral features with positive slopes], such that the circle-contour energy spectrum in (h) no longer exhibits features with well-defined topological chirality.}
\label{fig:C2_arcs_amo}
\end{figure}

In Fig.~\ref{fig:C2_arcs_amo}, we plot the disorder-averaged surface spectral function $\bar{A}_{\text{surf}}(E,{\bf p})$ of the non-crystalline double-Weyl model for increasing disorder parameterized by $\eta$.
Specifically, in Fig.~\ref{fig:C2_arcs_amo}(a-d), we plot $\bar{A}_{\text{surf}}(E,{\bf p})$ as a function of $p_{x}$ and $k_{z}$ for increasing $\eta$ at a fixed energy, and in Fig.~\ref{fig:C2_arcs_amo}(e-h), we respectively plot $\bar{A}_{\text{surf}}(E,{\bf p})$ at the same $\eta$ as a function of energy on counterclockwise circular paths, parameterized by $\vartheta$, surrounding $p_{x}=k_{z}=0$.
In the weak-to-moderate disorder regime [$\eta=0.05-0.2$ in panels (a-c) and (e-g) of Fig.~\ref{fig:C2_arcs_amo}], $\bar{A}_{\text{surf}}(E,{\bf p})$ continues to exhibits two clear Fermi-arc surface states with the same connectivity and topological chirality [positive slopes] as the right-handed enantiomer of the crystalline double-Weyl model [Fig.~\ref{fig:charge2_figWilson}(b,c)], consistent with the average right-handedness of the disordered system.
Unlike the increasingly positively sloped and diffuse surface Fermi arcs in the disordered Kramers-Weyl model [Fig.~\ref{fig:KWamorphousArcs}], the surface Fermi arcs in the double-Weyl model remain sharp and surface-localized into the moderate-disorder regime [$\eta=0.2$ in Fig.~\ref{fig:C2_arcs_amo}(c,g)].
The Fermi-arc surface states in the double-Weyl model instead become trivial with increasing $\eta$ by retracting to $p_{x}=k_{z}=0$ across the constant energy cuts in Fig.~\ref{fig:C2_arcs_amo}(c,d), and by losing their well-defined chiral dispersion via merging with disorder-broadened bulk states in the circular contours in Fig.~\ref{fig:C2_arcs_amo}(g,h).
In the strong-disorder regime [$\eta=0.5$ in Fig.~\ref{fig:C2_arcs_amo}(d,h)], the Fermi arcs can specifically no longer be distinguished as surface-localized, right-moving states [sharp, dark green spectral features with positive slopes], such that the circle-contour energy spectrum in Fig.~\ref{fig:C2_arcs_amo}(h) no longer exhibits features with well-defined topological chirality.

Overall, we conclude that like the non-crystalline KW model previously analyzed in Fig.~\ref{fig:KWamorphousArcs} and the surrounding text, the non-crystalline double-Weyl model [Eq.~(\ref{eq:HopChiralC2})] remains topological and gapless for strong disorder, but no longer exhibits topological Fermi arcs in the amorphous regime.  
This can be understood by recognizing that the double-Weyl model with strong smectic lattice disorder is spectrally isotropic in the $(p_{x},p_{y})$-coordinate plane [Fig.~\ref{fig:DOSC2}(d)], and specifically exhibits larger-$|{\bf p}|$ ring-like spectral features in the $k_{z}=0$ plane that surround the double-Weyl fermion at ${\bf p}={\bf 0}$.
Similar to the strongly disordered KW model in Fig.~\ref{fig:KWamorphousArcs}(d,h) and topologically chiral nodal-surface semimetals~\cite{KramersWeyl,WeylNodalSurfaceFollowupPRB}, the ring-like spectral features in the strongly disordered double-Weyl model ensure that there do not exist topological Fermi arcs on surfaces whose normal vectors lie in the real-space $(x,y)$-plane, due to the absence of topologically nontrivial projected bulk Fermi pockets.

\begin{figure}[t]
\centering
\includegraphics[width=\linewidth]{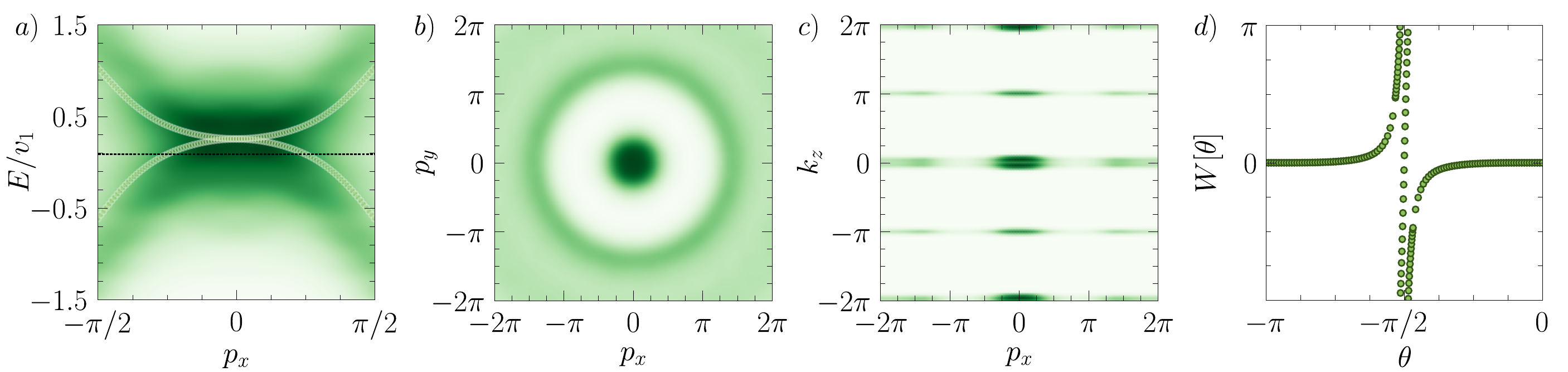}
\caption{Double-Weyl model on layered random lattices. 
(a-d) Bulk spectrum and topology of the non-crystalline double-Weyl model [Eq.~(\ref{eq:HopChiralC2})] on a smectic lattice [Appendix~\ref{app:SmecticNematicDisorder}] consisting of identical layers, each with randomly located sites in the $(x,y)$-coordinate subspace of a system with $d=3$, $d_{A}=2$, $d_{T}=1$, and $d_{f}=0$ in the notation of Eq.~(\ref{eq:finiteDimAmorph}).
Each layer also contains random frame disorder parameterized by the standard deviation $\eta=0.5$ and chirality domains of unequal area [Appendix~\ref{app:lattices}]. 
For all panels in this figure, the data were generated using Eq.~(\ref{eq:AmoC2}) with the tight-binding parameters in Eq.~(\ref{eq:paramsC2amo}) implemented with a single domain in each replica of right-handed sites with $n_R=N_R/N_{\mathrm{sites}}=2/3$, and with the remaining volume in each replica containing a contiguous domain of left-handed sites with a corresponding concentration of $n_L=1-N_R/N_{\mathrm{sites}}=1/3$.
Each panel shows data generated by averaging over 50 randomly-generated replicas with $\sim 20^3$ sites [subdivided into $20$ identical layers with $400$ sites each], as detailed in Appendix~\ref{app:PhysicalObservables}. 
In (a) the bulk average spectral function $\bar{A}(E,{\bf p})$ [Eq.~(\ref{eq:SpecFunc})] exhibits a quadratically dispersing amorphous double-Weyl fermion at ${\bf p}={\bf 0}$.
(b,c) $\bar{A}(E,{\bf p})$ at $E/v_1 = 0.1$ [the dashed line in (a)], plotted in panel (b) at $k_{z}=0.1$ as a function of $p_{x}$ and $p_{y}$ and in panel (c) at $p_{y}=0.1$ as a function of $p_{x}$ and $k_{z}$.
In addition to the double-Weyl fermion at ${\bf p}={\bf 0}$, broad, ring-like spectral features appear in (b) the $k_{z}=0,\pi$ planes at $\sqrt{p_{x}^{2} + p_{y}^{2}}=\pi\sqrt{2}/\bar{a}$ and $\sqrt{p_{x}^{2} + p_{y}^{2}}=2\pi/\bar{a}$, where $\bar{a}=1$ is the average nearest-neighbor spacing in the real-space $(x,y)$-plane. 
As shown in Fig.~\ref{fig:DOSC2}, the ring-like features in (b) originate from the double-Weyl fermions at $\sqrt{k_{x}^{2} + k_{y}^2}=\pi\sqrt{2}, 2\pi$ and $k_{z}=0,\pi$ in the crystalline case, analogously to the ring-like, higher-Brillouin-zone Dirac surface states recently observed in amorphous Bi$_{2}$Se$_{3}$~\cite{corbae_evidence_2020,Ciocys2023}.
As in the strongly Gaussian disordered double-Weyl model in Fig.~\ref{fig:DOSC2}(h), the $\bar{A}(E,{\bf p})$ cuts at constant energy and $p_{y}$ in (c) exhibit spectral weight that is confined to the $k_{z}=0,\pi$ planes, as opposed to appearing isotropic in all three ${\bf p}$ directions like in the random-lattice Kramers-Weyl model [Fig.~\ref{fig:KW_Unif}(b)].
This occurs because unlike the $d_{A}=3$ random-lattice Kramers-Weyl model in Fig.~\ref{fig:KW_Unif}(a-d), the random-lattice double-Weyl model in this figure is only subject to $d_{A}=2$ smectic disorder in the $(x,y)$-coordinate plane, and remains translationally invariant in the $z$-direction [see Fig.~\ref{fig:dadtdf}(b) and Appendix~\ref{app:SmecticNematicDisorder}]. 
(d) The amorphous Wilson loop spectrum of the double-Weyl fermion at ${\bf p}={\bf 0}$ in (a), generated from an effective Hamiltonian [light green circles in (a)] constructed with its reference energy $E_{C}$ centered at the maximum value of the spectral function at ${\bf p}={\bf 0}$ [$E_{C}/v_1 = 0.41$ in (a), see Appendix~\ref{app:EffectiveHamiltonian} for further details].
In (d), the Wilson loop spectrum winds twice in the positive direction, indicating that the non-crystalline double-Weyl fermion at ${\bf p}={\bf 0}$ carries a chiral charge of $C=2$, consistent with the average right-handedness of the disordered system [$n_{R}>n_{L}$, see Fig.~\ref{fig:C2_WL}].}
\label{fig:C2_Unif}
\end{figure}

\paragraph*{\bf Random Lattices} -- $\ $ Lastly, one might be concerned that the topological and spectral properties of non-crystalline double-Weyl fermions obtained in this section are specific to our use of Gaussian structural disorder, which admits a well-defined crystalline limit.
To show that this is not the case, we will conclude this section by computing the bulk energy spectrum and ${\bf p}={\bf 0}$ amorphous Wilson loops of the non-crystalline double-Weyl model [Eq.~(\ref{eq:HopChiralC2})] on randomly generated lattices without well-defined [unique] crystalline limits.
To begin, we place the non-crystalline double-Weyl model on a smectic lattice [Appendix~\ref{app:SmecticNematicDisorder}] with $d=3$, $d_{A}=2$, $d_{T}=1$, and $d_{f}=0$ [Eq.~(\ref{eq:finiteDimAmorph})] consisting of identical layers, each with randomly located sites in the $(x,y)$-coordinate plane.
Each layer also contains strong random frame disorder parameterized by the standard deviation $\eta=0.5$ and contiguous chirality domains of unequal area with the respective concentrations of right-handed sites $n_R=N_R/N_{\mathrm{sites}}=2/3$ and left-handed sites $n_L=1-N_R/N_{\mathrm{sites}}=1/3$ [see Appendix~\ref{app:lattices}].
To simulate a large, self-averaging system, we further randomly generate 50 lattices [replicas] with $20^3 = 8000$ sites [subdivided into $20$ identical layers with $400$ sites each], and then compute the replica-averaged momentum-resolved Green's function $\bar{\mathcal{G}}(E,{\bf p},{\bf p}')$ [Eq.~(\ref{eq:averageTWoMomentumGreen})].
As previously for the double-Weyl model with strong Gaussian disorder [Fig.~\ref{fig:3DGreenC2}(c)], we find that the off-diagonal-in-momentum elements $\bar{\mathcal{G}}(E,{\bf p},{\bf p}')$ vanish [on the average] as well for the non-crystalline double-Weyl model on random lattices [Fig.~\ref{fig:3DGreenC2}(d)].
We may therefore again limit consideration to the diagonal-in-momentum replica-averaged momentum-resolved Green's function $\bar{\mathcal{G}}(E,\mathbf{p})$ [Eq.~(\ref{eq:averageOneMomentumGreen})], from which we below compute the spectral and topological properties of random-lattice double-Weyl fermions.

In Fig.~\ref{fig:C2_Unif}(a,b,c), we respectively show the disorder-averaged, momentum-resolved spectral function $\bar{A}(E,{\bf p}) \propto \text{Im}\{\Tr[\bar{\mathcal{G}}(E,{\bf p})]\}$ [Eq.~(\ref{eq:SpecFunc})] of the random-lattice double-Weyl model plotted as a function of energy and momentum [$p_{x}$], at a fixed energy and $k_{z}$ as a function of the remaining two momenta $p_{x,y}$, and at a fixed energy and $p_y$ as a function of the remaining two momenta $p_x$ and $k_z$.
Like in the Gaussian-disordered lattice [Fig.~\ref{fig:C2_bands_amo}], $\bar{A}(E,{\bf p})$ for the random lattice exhibits a quadratically dispersing double-Weyl fermion at ${\bf p}={\bf 0}$ [Fig.~\ref{fig:C2_Unif}(a)].
Interestingly, $\bar{A}(E,{\bf p})$ also exhibits broadened, ring-like spectral features in the $k_{z}=0,\pi$ planes at $\sqrt{p_{x}^{2} + p_{y}^{2}}=\pi\sqrt{2}/\bar{a}$ and $\sqrt{p_{x}^{2} + p_{y}^{2}}=2\pi/\bar{a}$, where $\bar{a}=1$ is the average nearest-neighbor spacing in the real-space $(x,y)$-plane [Fig.~\ref{fig:C2_Unif}(b)].
The same ring-like spectral features at $\sqrt{p_{x}^{2} + p_{y}^{2}}=\pi\sqrt{2}$ and $\sqrt{p_{x}^{2} + p_{y}^{2}}=2\pi$ also appear in the non-crystalline double-Weyl model with Gaussian lattice disorder [Fig.~\ref{fig:DOSC2}(c,d)], and were there shown to originate from disorder-driven rotational averaging~\cite{spring_amorphous_2021,springMagneticAverageTI,corbae_evidence_2020,Ciocys2023} of the crystalline double-Weyl fermions at $k_{z}=0,\pi$ and respectively $\sqrt{k_{x}^{2} + k_{y}^2}=\pi\sqrt{2}$ and $\sqrt{k_{x}^{2} + k_{y}^2} = 2\pi$.
For completeness, we lastly note that as in the strongly Gaussian disordered double-Weyl model in Fig.~\ref{fig:DOSC2}(h), the $\bar{A}(E,{\bf p})$ cuts at constant energy and $p_{y}$ in Fig.~\ref{fig:C2_Unif}(c) exhibit spectral weight that is confined to the $k_{z}=0,\pi$ planes, as opposed to appearing isotropic in all three ${\bf p}$ directions like in the random-lattice Kramers-Weyl model [Fig.~\ref{fig:KW_Unif}(b)].
This occurs because unlike the $d_{A}=3$ random-lattice Kramers-Weyl model previously analyzed in Fig.~\ref{fig:KW_Unif}(a-d), the random-lattice double-Weyl model in Fig.~\ref{fig:C2_Unif} is only subject to $d_{A}=2$ smectic disorder in the $(x,y)$-coordinate plane, and remains translationally invariant in the $z$-direction [see Fig.~\ref{fig:dadtdf}(b) and Appendix~\ref{app:SmecticNematicDisorder}].

Finally, to precisely confirm that the quadratic nodal degeneracy at ${\bf p}={\bf 0}$ in Fig.~\ref{fig:C2_Unif}(d) is indeed a topological double-Weyl fermion, we compute the sphere Wilson loop spectrum of the random-lattice double-Weyl model at ${\bf p}={\bf 0}$.  
To obtain the Wilson loop spectrum, we first construct an effective Hamiltonian $\mathcal{H}_{\mathrm{Eff}}({\bf p})=\mathcal{H}_{\mathrm{Eff}}(E_{C},{\bf p})$ [Eq.~(\ref{eq:AvgHEff}), light green circles in Fig.~\ref{fig:C2_Unif}(a)] using a reference energy cut $E_{C}$ set to the energy of the largest spectral weight $\bar{A}(E,{\bf p})$ at ${\bf p}={\bf 0}$, which we found to maximize the spectral accuracy of $\mathcal{H}_{\mathrm{Eff}}({\bf p})$ [see the text surrounding Fig.~\ref{fig:H_eff_benchmark}].
We then use the occupied [lower] bands of $\mathcal{H}_{\mathrm{Eff}}({\bf p})$ to compute the sphere Wilson loop spectrum.
For the random-lattice double-Weyl system with $n_{R} =2/3$ and $n_{L}=1/3$, the amorphous Wilson loop at ${\bf p}={\bf 0}$ winds twice in the positive direction [Fig.~\ref{fig:C2_Unif}(d)], indicating that the nodal spectral feature at ${\bf p}={\bf 0}$ is a $C=2$ amorphous double-Weyl fermion.
This is consistent with our earlier Wilson loop analysis in Fig.~\ref{fig:C2_WL}, in which we demonstrated that the Gaussian-disordered double-Weyl model hosts a non-crystalline double-Weyl fermion at ${\bf p}={\bf 0}$ whose low-energy topological chirality is controlled by the average system chirality [\emph{i.e.} the ratio $n_{R}/n_{L}$].
Though not shown in Fig.~\ref{fig:C2_Unif}, we have numerically confirmed that the topological chiral charge of the random-lattice double-Weyl fermion at ${\bf p}={\bf 0}$ is also controlled by the ratio $n_{R}/n_{L}$, and specifically also undergoes a topological phase transition between $C = \pm 2$ when $n_{R}/n_{L} \approx 1$.
Additionally, though not shown in Fig.~\ref{fig:C2_Unif}, we have further confirmed that the non-crystalline double-Weyl model [Eq.~(\ref{eq:HopChiralC2})] on smectic \emph{Mikado} lattices [see Fig.~\ref{appfig:structuraldisorder}(c) and Refs.~\cite{marsal_obstructed_2022,marsal_topological_2020}] also exhibits ${\bf p}={\bf 0}$ double-Weyl fermions with nontrivial topological chiral charges that are controlled by the average system chirality.
Lastly, we have confirmed that the random-lattice double-Weyl model in Fig.~\ref{fig:C2_Unif} continues to exhibit a quadratically dispersing non-crystalline double-Weyl fermion at ${\bf p}={\bf 0}$ with a quantized $|C|=2$ sphere Wilson loop winding number under the subsequent addition of weak Anderson [on-site chemical potential] disorder.

\newpage
\subsection{Chiral Multifold Fermion Models}
\label{sec:Multifold}

\subsubsection{Symmetry and Chirality in a Crystalline Model with a $\Gamma$-Point Spin-1 Chiral Multifold Fermion}
\label{app:PristineMulifold}

In this section, we will introduce and analyze the bulk electronic structure, symmetry group theory, topology, surface states, and achiral critical phase of a structurally chiral [see Appendix~\ref{app:symDefs}] crystalline tight-binding model with a threefold-degenerate spin-1 chiral fermion~\cite{ManesNewFermion,chang2017large,tang2017CoSi,KramersWeyl,DoubleWeylPhonon,CoSiObserveJapan,CoSiObserveHasan,CoSiObserveChina,AlPtObserve,PdGaObserve,PtGaObserve,AltlandSpin1Light,DingARPESReversal,ZahidLadderMultigap,Sanchez2023} at its $\Gamma$ point [${\bf k}={\bf 0}$].
To begin, we note that $\Gamma$-point chiral multifold fermions are already featured in well-established models of B20-structure chiral cubic materials, such as RhSi~\cite{chang2017large}.
However, these earlier models are nonsymmorphic, and specifically contain screw symmetries that give rise to sublattice degrees of freedom within each unit cell.
Hence, the earlier $\Gamma$-point chiral multifold fermion models are more difficult to disorder while maintaining a sense of local [average] chirality than models with only internal spin and orbital degrees of freedom [see Ref.~\cite{Franca2024} and Appendix~\ref{app:DiffTypesDisorder}]. 
This in turn makes it more difficult to numerically identify disorder regimes in which the earlier models exhibit non-crystalline topological chiral fermions at ${\bf p}={\bf 0}$ [see Ref.~\cite{KramersWeyl} and Appendices~\ref{app:pseudoK} and~\ref{app:corepAmorphous}].
We will therefore in this section introduce and analyze a new, symmorphic model with a $\Gamma$-point chiral multifold fermion that only arises from on-site [internal] orbital degrees of freedom, which we will then straightforwardly disorder in Appendix~\ref{app:amorphousMultifold} following the same numerical procedure employed for the other non-crystalline models studied in this work.

To construct our symmorphic multifold fermion model, we begin by placing four spinless orbitals at each site of a lattice.  
In order, the four basis states specifically consist of spinless $s$, $p_{x}$, $p_{y}$, and $p_{z}$ orbitals.
The four basis states can also alternatively be understood as the \emph{molecular} orbitals of tightly bound tetrahedral molecules with four atoms [or four states on $sp^{3}$ bonds]~\cite{Bradlyn2017,chang2017large,mcQuarriePchem}, such as those appearing in the decoupled tetrahedron limit of the Weaire-Thorpe model of amorphous silicon~\cite{weaire_electronic_1971}.
In real space, the four atomic orbitals are then coupled via a Hamiltonian of the form:
\begin{equation}
    \mathcal{H}_{\mathrm{3F}} = \sum_{\langle \alpha\beta \rangle}\sum_{l,l'\in \left\{1,2,3,4\right\}}c_{\alpha,l}^{\dagger}\langle {\bf r}_\alpha,l | \mathcal{H} | {\bf r}_\beta,l'\rangle c^{\phantom{}}_{\beta,l'} + \sum_{\alpha}\sum_{l,l'\in\left\{1,2,3,4\right\}}c_{\alpha,l}^{\dagger}\langle {\bf r}_\alpha,l|\mathcal{H}|{\bf r}_\alpha,l'\rangle c^{\phantom{}}_{\alpha,l'},
    \label{eq:Amo3F}
\end{equation}
where the operator $c^{\dagger}_{\alpha,l}$ creates a particle at the site $\alpha$ with an internal orbital degree of freedom given by $l$ [indexed by $l=1,2,3,4$].
Following the notation of Ref.~\cite{PhysRevLett.95.226801} previously employed for the other models studied in this work, the $\langle$ and $\rangle$ symbols in the sum over sites in Eq.~(\ref{eq:Amo3F}) indicate that pairs of sites $\alpha,\beta$ are only included within the summation if they lie within a specified distance, denoted $R_0$, of each other. 
Unlike the crystalline Kramers-Weyl [KW] and double-Weyl models respectively analyzed in the text following Eqs.~(\ref{eq:AmoKW}) and~(\ref{eq:AmoC2}), the crystalline multifold fermion model in Eq.~(\ref{eq:Amo3F}) additionally contains a second, on-site interaction [mass] term, which we will below show corresponds to symmetry-allowed crystal field splitting.
In the crystalline limit discussed in this section, $\left\langle\alpha\beta\right\rangle$ in Eq.~(\ref{eq:Amo3F}) will reduce to nearest-neighbor lattice sites.
In Eq.~(\ref{eq:Amo3F}), the tight-binding basis states $|{\bf r}_\alpha, l\rangle$ satisfy the orthogonality relation $\langle {\bf r}_\alpha, l|{\bf r}_\beta, l'\rangle = \delta_{\alpha\beta}\delta_{ll'}$, where $l$ and $l'$ belong to the set $\left\{1,2,3,4\right\}$, representing the internal orbital degrees of freedom, and $\alpha$ and $\beta$ denote the lattice sites [see Eq.~(\ref{eq:TBbasisStates}) and the surrounding text].

We next define the intersite hopping matrix $T_{\alpha\beta}$ and on-site mass matrix $M_{\alpha}$ for $\mathcal{H}_{\mathrm{3F}}$ in Eq.~(\ref{eq:Amo3F}) through inner products in the following manner:
\begin{align}
T_{\alpha\beta} &\underset{def}{\equiv}
\begin{pmatrix}
\langle {\bf r}_\alpha,1 | \mathcal{H} | {\bf r}_\beta,1\rangle &
\langle {\bf r}_\alpha,1 | \mathcal{H} | {\bf r}_\beta,2\rangle &
\langle {\bf r}_\alpha,1 | \mathcal{H} | {\bf r}_\beta,3\rangle &
\langle {\bf r}_\alpha,1 | \mathcal{H} | {\bf r}_\beta,4\rangle\\
\langle {\bf r}_\alpha,2 | \mathcal{H} | {\bf r}_\beta,1\rangle &
\langle {\bf r}_\alpha,2 | \mathcal{H} | {\bf r}_\beta,2\rangle &
\langle {\bf r}_\alpha,2 | \mathcal{H} | {\bf r}_\beta,3\rangle &
\langle {\bf r}_\alpha,2 | \mathcal{H} | {\bf r}_\beta,4\rangle\\
\langle {\bf r}_\alpha,3| \mathcal{H} | {\bf r}_\beta,1\rangle &
\langle {\bf r}_\alpha,3 | \mathcal{H} | {\bf r}_\beta,2\rangle &
\langle {\bf r}_\alpha,3 | \mathcal{H} | {\bf r}_\beta,3\rangle &
\langle {\bf r}_\alpha,3 | \mathcal{H} | {\bf r}_\beta,4\rangle\\
\langle {\bf r}_\alpha,4 | \mathcal{H} | {\bf r}_\beta,1\rangle &
\langle {\bf r}_\alpha,4 | \mathcal{H} | {\bf r}_\beta,2\rangle &
\langle {\bf r}_\alpha,4 | \mathcal{H} | {\bf r}_\beta,3\rangle &
\langle {\bf r}_\alpha,4| \mathcal{H} | {\bf r}_\beta,4\rangle
\end{pmatrix} \nonumber \\
&= \frac{1}{2}f\left(\left|{\bf d}_{\alpha\beta}\right|\right) \left[\sum_{\gamma = 1,2,3} t_{\gamma}\left( {\bf d}_{\alpha\beta}^{\mathsf{T}}B_{\gamma}{\bf d}_{\alpha\beta}  \right) + i v \left({\bf d}^{\mathsf{T}}_{\alpha\beta}\bm{V}\right)\right],
\label{eq:3FTmatrix}
\end{align}
and:
\begin{equation}
M_{\alpha} \underset{def}{\equiv}
\begin{pmatrix}
\langle {\bf r}_\alpha,1 | \mathcal{H} | {\bf r}_\alpha,1\rangle &
\langle {\bf r}_\alpha,1 | \mathcal{H} | {\bf r}_\alpha,2\rangle &
\langle {\bf r}_\alpha,1 | \mathcal{H} | {\bf r}_\alpha,3\rangle &
\langle {\bf r}_\alpha,1 | \mathcal{H} | {\bf r}_\alpha,4\rangle\\
\langle {\bf r}_\alpha,2 | \mathcal{H} | {\bf r}_\alpha,1\rangle &
\langle {\bf r}_\alpha,2 | \mathcal{H} | {\bf r}_\alpha,2\rangle &
\langle {\bf r}_\alpha,2 | \mathcal{H} | {\bf r}_\alpha,3\rangle &
\langle {\bf r}_\alpha,2 | \mathcal{H} | {\bf r}_\alpha,4\rangle\\
\langle {\bf r}_\alpha,3| \mathcal{H} | {\bf r}_\alpha,1\rangle &
\langle {\bf r}_\alpha,3 | \mathcal{H} | {\bf r}_\alpha,2\rangle &
\langle {\bf r}_\alpha,3 | \mathcal{H} | {\bf r}_\alpha,3\rangle &
\langle {\bf r}_\alpha,3 | \mathcal{H} | {\bf r}_\alpha,4\rangle\\
\langle {\bf r}_\alpha,4 | \mathcal{H} | {\bf r}_\alpha,1\rangle &
\langle {\bf r}_\alpha,4 | \mathcal{H} | {\bf r}_\alpha,2\rangle &
\langle {\bf r}_\alpha,4 | \mathcal{H} | {\bf r}_\alpha,3\rangle &
\langle {\bf r}_\alpha,4 | \mathcal{H} | {\bf r}_\alpha,4\rangle
\end{pmatrix} = m(\mu^z+\tau^z\mu^z + \tau^z),
\label{eq:3FMmatrix}
\end{equation}
where the intersite separation vector ${\bf d}_{\alpha\beta}$ in Eq.~(\ref{eq:3FTmatrix}) is given by:
\begin{equation}
    \mathbf{d}_{\alpha\beta}
= \begin{pmatrix}
x_{\alpha}-x_{\beta}\\y_{\alpha}-y_{\beta}\\z_{\alpha}-z_{\beta}
\end{pmatrix},
\label{eq:dVectorDef3F}
\end{equation}
and the overall radial hopping strength $f\left(\lvert \mathbf{d}_{\alpha\beta}\rvert\right)$ is given by Eq~\eqref{eq:KWHeaviside}.
In Eq.~(\ref{eq:3FTmatrix}), each $B_{\gamma}$ is a $(3\times3)\times (4\times4)$ tensor and $\bm{V}$ is a $(3\times 1)\times(4\times4)$ tensor, where $B_{\gamma}$ and $\bm{V}$ take the following forms:
\begin{equation}
    B_{1} = \begin{pmatrix}\mu^z&0&0\\0& \tau^z&0\\ 0&0&\tau^z\mu^z \end{pmatrix},\quad B_{2} = \begin{pmatrix}\tau^z&0&0\\ 0&\tau^z\mu^z&0 \\ 0&0&\mu^z\end{pmatrix},\quad B_{3} = \begin{pmatrix}\tau^z\mu^z&0&0\\ 0&\mu^z&0\\0&0&\tau^z\end{pmatrix},\quad\bm{V} = \begin{pmatrix}\tau^y\\ \tau^z\mu^y\\ \tau^x\mu^y\end{pmatrix},
\label{eq:multifoldBandV}
\end{equation}
in which:
\begin{equation}
\tau^{i}\mu^{j} = s_j \otimes s_i,
\label{eq:PauliMatrixProdDef}
\end{equation}
where $s_i$, $i\in\left\{1,2,3\right\}$ are the $2\times 2$ Pauli matrices, $s_0$ is the $2\times 2$ identity matrix, and where factors of $\tau^{0}$ and $\mu^{0}$ have been suppressed for notational simplicity. 
For the remainder of this section, we set the parameters of the function $f(|\mathbf{d}_{\alpha\beta}|)$ in Eq.~(\ref{eq:KWHeaviside}) to be:
\begin{equation}
    a=1,\ R_0 = 1.3a = 1.3,
\end{equation}
to ensure that only nearest-neighbor hoppings appear in Eq.~\eqref{eq:KWHeaviside} in the pristine [crystalline] limit.

\begin{figure}[t]
\centering
\includegraphics[width=\linewidth]{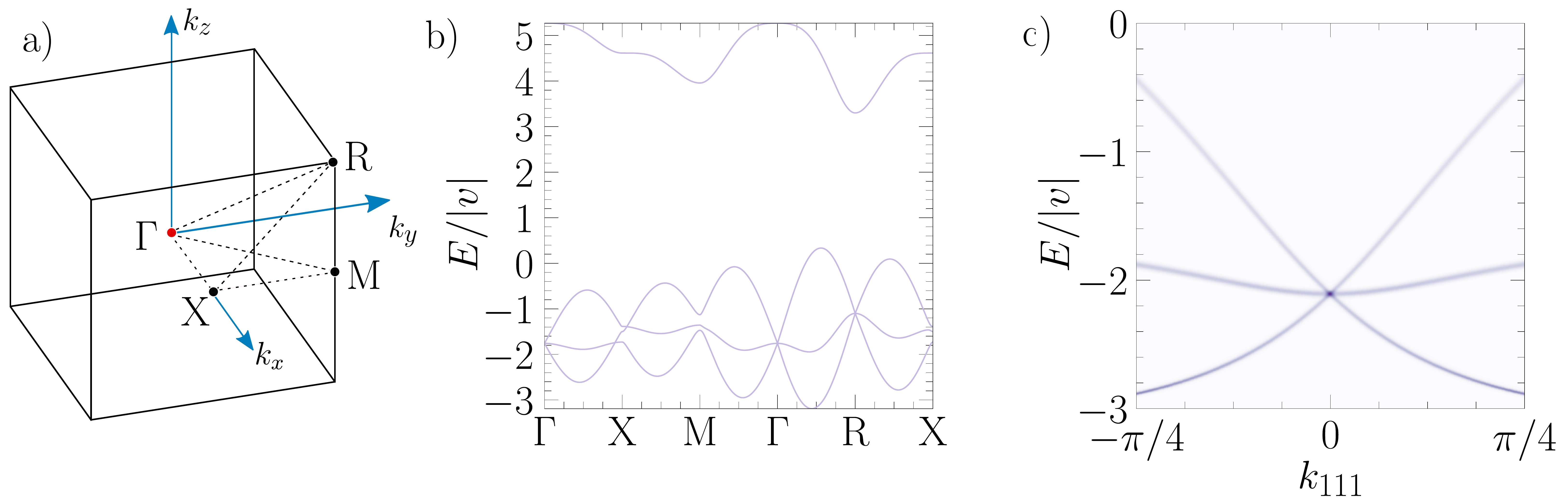}
\caption{Bulk spectrum of the crystalline multifold fermion model. 
(a) The Brillouin zone [BZ] of chiral SG 195 ($P23$), the SG of the symmorphic spin-1 chiral multifold fermion model introduced in this work [Eqs.~(\ref{eq:3FTmatrix}),~(\ref{eq:3FMmatrix}),~(\ref{eq:multifoldBandV}), and~(\ref{eq:HamBloch3F})].
(b) Bulk band structure of the crystalline multifold fermion model using the parameters in Eq.~(\ref{eq:appTBparams3F}).
(c) The spectral function $A(E,\mathbf{p})$ [Eqs.~(\ref{eq:SpecFunc}) and~(\ref{eq:crystallineAfunc})] of the threefold-degenerate spin-1 chiral fermion at the $\Gamma$ point in (b), plotted as a function of energy and $k_{111} = (1/\sqrt{3})(k_{x}+k_{y}+k_{z})$ [see Eq.~(\ref{eq:appHKPmultifold}) and the following text].}
\label{fig:3F_figBulk}
\end{figure}

\paragraph*{\bf Symmetries and Group Representations} -- $\ $ When defined on a regular cubic lattice with nearest-neighbor hoppings, the tight-binding model in Eqs.~(\ref{eq:3FTmatrix}),~(\ref{eq:3FMmatrix}), and~(\ref{eq:multifoldBandV}) can be Fourier transformed into a $4\times 4$ Bloch Hamiltonian of the form:
\begin{eqnarray}
\mathcal{H}({\bf k}) &=& t_1\left[\cos(k_x)\mu^z + \cos(k_y)\tau^z + \cos(k_z)\tau^z\mu^z\right]\nonumber \\
&+& t_2\left[\cos(k_x)\tau^z  + \cos(k_y)\tau^z\mu^z + \cos(k_z)\mu^z\right]\nonumber \\
&+& t_3\left[\cos(k_x)\tau^z\mu^z + \cos(k_y)\mu^z + \cos(k_z)\tau^z\right]\nonumber \\
&+& v\left[\sin(k_x)\tau^y + \sin(k_y)\tau^z\mu^y + \sin(k_z)\tau^x\mu^y\right] \nonumber \\
&+& m \left[\mu^z+\tau^z\mu^z + \tau^z\right].
\label{eq:HamBloch3F}
\end{eqnarray}
In Eq.~(\ref{eq:HamBloch3F}), the intersite $t_{i}$ terms correspond to positively and negatively signed, nearest-neighbor, same-orbital [$s-s$ and $p_{x,y,z}-p_{x,y,z}$] achiral spinless hoppings that are equivalent to linear combinations of inter-tetrahedron hopping terms in the Weaire-Thorpe model of amorphous silicon taken close to the limit of decoupled tetrahedra~\cite{weaire_electronic_1971}.
Conversely, the intersite $v$ term in Eq.~(\ref{eq:HamBloch3F}) corresponds to a new ``inter-tetrahedron'' orbital angular momentum [OAM] coupling that breaks rotoinversion symmetries in the crystalline limit, and hence renders the system structurally chiral [see Eq.~(\ref{eq:rotoinversion}) and the surrounding text].
Notably, unlike in more simplified multifold fermion models, the $v$ term in Eq.~(\ref{eq:HamBloch3F}) does not induce a pure monopole-like OAM texture in the three-band subspace of the threefold degeneracy at $\Gamma$ in Fig.~\ref{fig:3F_figBulk}(a,b). 
The intersite $v$ term instead implements the combined effects of OAM-crystal-momentum-locking and interband matrix elements that remove OAM quantization, which are generically permitted in many-band models and present in real materials~\cite{KramersWeyl,BradlynTQCSpinTexture,OAMWeyl2,OAMmultifold1,OAMmultifold2,OAMmultifold3,OAMmultifold4,OAMmultifold5} [see the text surrounding Fig.~\ref{fig:OAMTexture3F} for explicit numerical OAM texture calculations].

Lastly, Eq.~(\ref{eq:HamBloch3F}) also contains an $m$ mass term that corresponds to a symmetry-allowed crystal field that splits the four Bloch eigenstates at ${\bf k}={\bf 0}$ [the $\Gamma$ point in Fig.~\ref{fig:3F_figBulk}(a,b)] into a single and a threefold degeneracy.
This can most directly be seen by recognizing that the $4\times 4$ matrices  proportional to $m$ in Eq.~(\ref{eq:HamBloch3F}) satisfy the following identity:
\begin{equation}
    \mu^z+\tau^z\mu^z + \tau^z = \begin{pmatrix}
        3&0&0&0\\
        0&-1&0&0\\
        0&0&-1&0\\
        0&0&0&-1
    \end{pmatrix},
\label{eq:multifoldMidentity}
\end{equation}
indicating that in the limit of decoupled sites [$v=t_{1,2,3}=0$, $m\neq 0$], the spectrum of $\mathcal{H}({\bf k})$ consists of a singly degenerate flat band at $E=3m$ and three degenerate flat bands at $E=-m$.
Hence when $v,t_{1,2,3}\rightarrow 0$, $\mathcal{H}({\bf k})$ in Eq.~(\ref{eq:HamBloch3F}) becomes isospectral to the decoupled-tetrahedron limit of the Weaire-Thorpe model of amorphous silicon discussed earlier in this section~\cite{weaire_electronic_1971}.

The Bloch Hamiltonian $\mathcal{H}({\bf k})$ in Eq.~(\ref{eq:HamBloch3F}) respects the spinless [single-group] symmetries of symmorphic chiral cubic SG 195 ($P23$), for which the generating elements can be represented through their action on $\mathcal{H}({\bf k})$:
\begin{eqnarray}
\label{eq:PristineMultifoldsymsT}
\tilde{\mathcal{T}}\mathcal{H}(k_{x},k_{y},k_{z})\tilde{\mathcal{T}}^{-1} &=& \mathcal{H}^{*}(-k_{x},-k_{y},-k_{z}), \nonumber \\ 
\tilde{C}_{2x}\mathcal{H}(k_{x},k_{y},k_{z})\tilde{C}_{2x} &=& \mu^{z}\mathcal{H}(k_{x},-k_{y},-k_{z})\mu^{z}, \nonumber \\
\tilde{C}_{3,111}\mathcal{H}(k_{x},k_{y},k_{z})\tilde{C}_{3,111}^{-1} &=& U_{3}
\mathcal{H}(k_{y},k_{z},k_{x})U_{3}^{\dag},
\label{eq:PristineMultifoldsyms}
\end{eqnarray}
where we have used tildes to denote that the symmetries are elements of single SGs, as they act on internal integer-angular-momentum [$s$ and $p_{x,y,z}$] degrees of freedom, and where we have again not included lattice translations in Eq.~(\ref{eq:PristineMultifoldsyms}), because they act as pure Bloch phases in momentum space~\cite{BigBook,MTQC}.
In Eq.~(\ref{eq:PristineMultifoldsyms}), the unitary matrix $U_{3}$ implements the action of the threefold cubic rotation symmetry $\tilde{C}_{3,111}$ on the internal orbital degrees of freedom, and is given by:
\begin{equation}
    U_{3} = \begin{pmatrix}
        1&0&0&0\\
        0&0&0&1\\
        0&1&0&0\\
        0&0&1&0
    \end{pmatrix}.
\label{eq:cubicMultifoldC3}
\end{equation}
For completeness, we note that $\mathcal{H}({\bf k})$ also respects the spinless twofold rotation symmetries $\tilde{C}_{2y}$ and $\tilde{C}_{2z}$ of cubic SG 195 ($P23$), which can similarly be defined through their action on $\mathcal{H}({\bf k})$:
\begin{eqnarray}
\tilde{C}_{2y}\mathcal{H}(k_{x},k_{y},k_{z})\tilde{C}_{2y} &=& \tau^{z}\mathcal{H}(-k_{x},k_{y},-k_{z})\tau^{z}, \nonumber \\
\tilde{C}_{2z}\mathcal{H}(k_{x},k_{y},k_{z})\tilde{C}_{2z} &=& \tau^{z}\mu^{z}\mathcal{H}(-k_{x},-k_{y},k_{z})\tau^{z}\mu^{z}, 
\label{eq:PristineMultifoldExtraSyms}
\end{eqnarray}
such that the matrix representatives of the twofold rotation symmetries $\tilde{C}_{2i}$, $i=x,y,z$ are consistently permuted under the action of the cubic rotation symmetry $\tilde{C}_{3,111}$ [\emph{i.e.} $U_{3}$] in Eqs.~(\ref{eq:PristineMultifoldsyms}) and~(\ref{eq:cubicMultifoldC3}):
\begin{equation}
U_{3}\mu^{z}U_{3}^{\dag} = \tau^{z},\ U_{3}\tau^{z}U_{3}^{\dag} = \tau^{z}\mu^{z},\ U_{3}\tau^{z}\mu^{z}U_{3}^{\dag} = \mu^{z}.
\end{equation}

Continuing to characterize $\mathcal{H}({\bf k})$ in Eq.~(\ref{eq:HamBloch3F}), we next note that the four $s$ and $p_{x,y,z}$ orbital basis states together transform in a sum of irreducible coreps of the $1a$ tetrahedral site-symmetry group in SG 195 ($P23$), which is isomorphic to chiral point group $T$ ($23$)~\cite{Bradlyn2017,MTQC,Bandrep1,BilbaoPoint,PointGroupTables}.
Specifically, the one spinless $s$ orbital transforms in the one-dimensional $A$ corep and the three spinless $p$ orbitals together transform in the three-dimensional $T$ corep of the $1a$ site-symmetry group.
The four total bands of $\mathcal{H}({\bf k})$ hence transform in a sum of the one-dimensional elementary band corep $(A)_{1a}\uparrow P23$ and the three-dimensional elementary band corep $(T)_{1a}\uparrow P23$~\cite{Bradlyn2017,MTQC,Bandrep1}. This is consistent with the band splitting induced by the $m$ mass term previously discussed in the text surrounding Eq.~(\ref{eq:multifoldMidentity}).

\paragraph*{\bf Topology} -- $\ $ In Fig.~\ref{fig:3F_figBulk}(b), we plot the band structure of $\mathcal{H}({\bf k})$ [Eq.~(\ref{eq:HamBloch3F})] with the choice of parameters:
\begin{equation}
v=\pm 3.5,\ m=5,\ t_1=0.55,\ t_2=0.35,\ t_3=0.25.
\label{eq:appTBparams3F}
\end{equation}
For all nonzero values of the hopping parameters in Eq.~(\ref{eq:HamBloch3F}), the energy spectrum exhibits two threefold nodal degeneracies at $k_{x}=k_{y}=k_{z}=0,\pi$ [$\Gamma$ and $R$] that to leading order each consist of two linearly dispersing bands and one band with quadratic or higher dispersion.  
The energy spectrum in Fig.~\ref{fig:3F_figBulk}(b) also exhibits a singly degenerate band at much higher energies [$E/|v| \sim 4$] that is disconnected from the other three bands in the spectrum at all ${\bf k}$ points.

\begin{figure}[t]
\centering
\includegraphics[width=\linewidth]{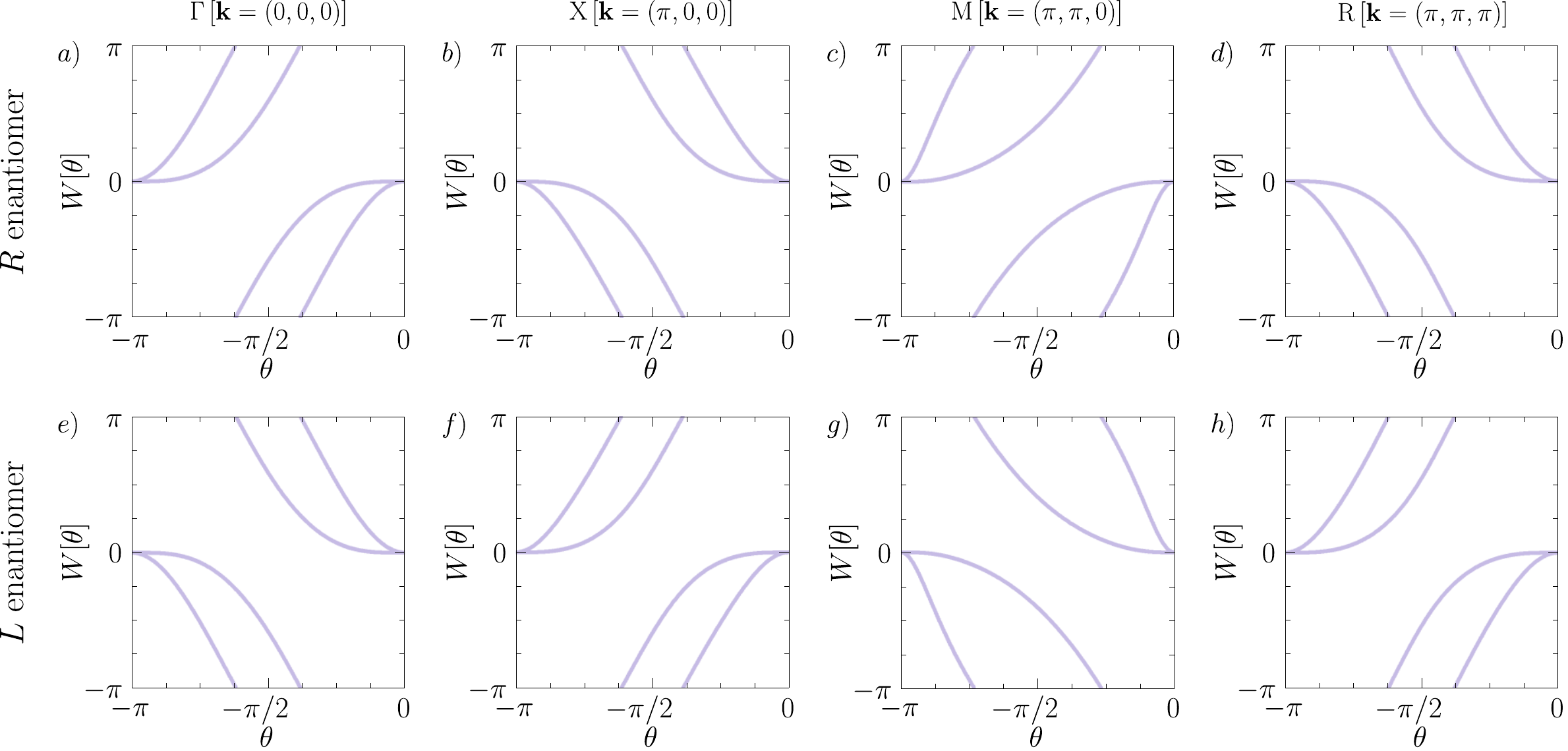}
\caption{Chirality and topology of the crystalline multifold fermion model.
Panels (a-d) and (e-h) respectively show the sphere Wilson loop spectrum of the right-handed [$R$] and left-handed [$L$] enantiomers of the multifold fermion tight-binding model [Eqs.~(\ref{eq:3FTmatrix}),~(\ref{eq:3FMmatrix}),~(\ref{eq:multifoldBandV}), and~(\ref{eq:HamBloch3F})] obtained using the parameters in Eq.~(\ref{eq:appTBparams3F}).
In each panel of this figure, we specifically show the Wilson loop eigenvalues [non-Abelian Berry phases, see Appendix~\ref{sec:WilsonBerry}] of the lowest two bands of the chiral multifold fermion model computed on a sphere surrounding one of the time-reversal-invariant momenta [TRIM points] in Fig.~\ref{fig:3F_figBulk}(a,b).
As discussed in the text surrounding Eq.~(\ref{eq:multifoldStructuralChirality}), the second and third bands of the chiral multifold fermion model with the tight-binding parameters in Eq.~(\ref{eq:appTBparams3F}) are linked by symmetry-enforced threefold degeneracies [spin-1 chiral multifold fermions] at $\Gamma$ and $R$ and clusters of conventional Weyl points along high-symmetry lines and at generic ${\bf k}$ points lying within a close vicinity of $X$ and $M$.
First, the Wilson loop spectra in (a) and (e) respectively wind twice in the positive and negative directions, indicating that the lower two bands of the chiral multifold fermion at the $\Gamma$ point carry a chiral charge of $C=2$ for the $R$ enantiomer and $C=-2$ for the $L$ enantiomer.
We similarly observe that in (d) and (h), the Wilson loop eigenvalues respectively wind twice in the negative and positive directions, indicating that the lower two bands of the chiral multifold fermion at $R$ carry a chiral charge of $C=-2$ for the $R$ enantiomer and $C=2$ for the $L$ enantiomer.
Next, in (b) and (f), the Wilson loop eigenvalues respectively wind twice in the negative and positive directions, indicating that the cluster of Weyl points linking the second and third bands near $X$ carries a net chiral charge of $C=-2$ for the $R$ enantiomer and $C=2$ for the $L$ enantiomer.
Lastly, in (c) and (g), the Wilson loop eigenvalues respectively wind twice in the positive and negative directions, indicating that the cluster of Weyl points linking the second and third bands near $M$ carries a net chiral charge of $C=2$ for the $R$ enantiomer and $C=-2$ for the $L$ enantiomer.
For both the chiral multifold fermions at $\Gamma$ and $R$ and the clusters of Weyl points near $X$ and $M$, the Wilson loop spectra in this figure overall indicate that the chiral charges $C$ of the lower two bands are directly determined by the lattice-scale structural chirality $C_{\mathcal{H}}$ in Eq.~(\ref{eq:multifoldStructuralChirality}).
Finally, though not shown in this figure, we have also computed the Wilson loop spectrum of lowest band of the chiral multifold model on spheres enclosing the nodal degeneracies at or near each TRIM point for the $R$ and $L$ model enantiomers, and respectively for each TRIM point and model handedness we observe the same Wilson loop winding numbers as those of the two-band Wilson spectra plotted in this figure.}
\label{fig:3F_extraCrossings}
\end{figure}

As shown in previous works~\cite{ManesNewFermion,NewFermions,chang2017large,tang2017CoSi,KramersWeyl,DoubleWeylPhonon}, the threefold degeneracies at $\Gamma$ and $R$ are cubic-symmetry-enforced condensed-matter realizations of spin-1 [multifold] chiral fermions.
To understand the band splitting and topology of the chiral multifold model, and for calculations that we will shortly perform in Appendix~\ref{app:amorphousMultifold} of the chiral multifold fermion model [Eqs.~(\ref{eq:3FTmatrix}),~(\ref{eq:3FMmatrix}), and~(\ref{eq:multifoldBandV})] with strong lattice disorder, we will find it useful to expand $\mathcal{H}({\bf k})$ in Eq.~(\ref{eq:HamBloch3F}) to linear order about each TRIM point $\mathbf{k}_\mathcal{T}$:
\begin{eqnarray}
\mathcal{H}_{{\bf k}_{\mathcal{T}}}({\bf q}) &=& t_1\left[\left(e^{ik_{x,\mathcal{T}}}\right)\mu^z + \left(e^{ik_{y,\mathcal{T}}}\right)\tau^z + \left(e^{ik_{z,\mathcal{T}}}\right)\tau^z\mu^z\right]\nonumber \\
&+& t_2\left[\left(e^{ik_{x,\mathcal{T}}}\right)\tau^z  + \left(e^{ik_{y,\mathcal{T}}}\right)\tau^z\mu^z + \left(e^{ik_{z,\mathcal{T}}}\right)\mu^z\right]\nonumber \\
&+& t_3\left[\left(e^{ik_{x,\mathcal{T}}}\right)\tau^z\mu^z + \left(e^{ik_{y,\mathcal{T}}}\right)\mu^z + \left(e^{ik_{z,\mathcal{T}}}\right)\tau^z\right]\nonumber \\
&+& v\left[q_x \left(e^{ik_{x,\mathcal{T}}}\right)\tau^y + q_y \left(e^{ik_{y,\mathcal{T}}}\right)\tau^z\mu^y + q_z \left(e^{ik_{z,\mathcal{T}}}\right)\tau^x\mu^y\right] \nonumber \\
&+& m \left[\mu^z+\tau^z\mu^z + \tau^z\right],
\label{eq:appHKPmultifold}
\end{eqnarray}
where ${\bf q} \approx {\bf k} - {\bf k}_{\mathcal{T}}$.
At the TRIM points $\Gamma$ [$k_{x,y,z}=0$] and $R$ [$k_{x,y,z}=\pi$], $e^{ik_{x,\mathcal{T}}}=e^{ik_{y,\mathcal{T}}}=e^{ik_{z,\mathcal{T}}}$, such that $\mathcal{H}_{{\bf k}_{\mathcal{T}}}({\bf q})$ respects the action of the threefold cubic rotation symmetry $\tilde{C}_{3,111}$.
Conversely at the $X$ [${\bf k}=(\pi,0,0)$] and $M$ [${\bf k}=(\pi,\pi,0)$] TRIM points, the exponential phase factors in Eq.~(\ref{eq:appHKPmultifold}) take different signs from each other, and $\tilde{C}_{3,111}$ is not a symmetry of the little group [Eq.~(\ref{eq:EquivKPoints})].
As we will explore below, the relative presence and absence of $\tilde{C}_{3,111}$ symmetry in the TRIM point little group has a significant effect on the symmetry-enforced degeneracy and topology at each TRIM point in Eq.~(\ref{eq:appHKPmultifold}).

First, at $\Gamma$ and $R$, the little groups $G_{\Gamma}$ and $G_{R}$ are isomorphic to SG 195 ($P23$) itself [see the text preceding Eq.~(\ref{eq:EquivKPoints})]. 
The singly-degenerate, high-energy states at $\Gamma$ and $R$ respectively transform in the one-dimensional, single-valued small coreps $\Gamma_{1}$ of $G_{\Gamma}$ and $R_{1}$ of $G_{R}$, and the threefold [multifold] degeneracies respectively transform in the three-dimensional, single-valued small coreps $\Gamma_{4}$ of $G_{\Gamma}$ and $R_{4}$ of $G_{R}$, using the labeling convention of the~\href{https://www.cryst.ehu.es/cgi-bin/cryst/programs/corepresentations.pl}{Corepresentations} tool on the Bilbao Crystallographic Server [in which SG 195 is alternatively denoted as Shubnikov SG 195.2 ($P231'$)]~\cite{MTQC,MTQCmaterials}.
Previous works~\cite{NewFermions,chang2017large,tang2017CoSi,KramersWeyl,DoubleWeylPhonon} have shown that when realized with parameters that give rise to upper and lower linearly dispersing bands and a central flat band [to leading order], the threefold degeneracies at $\Gamma$ and $R$ in Eqs.~(\ref{eq:HamBloch3F}) and~(\ref{eq:appHKPmultifold}) each carry chiral charges of $|C|=2$ when computed with both one or two filled bands, implying that the central nondispersing band carries a chiral charge of $C=0$ in this parameter regime.
For this reason, the threefold degeneracies have been termed condensed-matter realizations of spin-1 chiral fermions~\cite{ManesNewFermion,NewFermions,chang2017large,tang2017CoSi,KramersWeyl,DoubleWeylPhonon}.
To confirm this result, we respectively plot in Fig.~\ref{fig:3F_extraCrossings}(a,d) the Wilson loop eigenvalues [non-Abelian Berry phases, see Appendix~\ref{sec:WilsonBerry} and Refs.~\cite{Fidkowski2011,AndreiXiZ2,ArisInversion,Cohomological,HourglassInsulator,DiracInsulator,Z2Pack,BarryFragile,AdrienFragile,HOTIBernevig,HingeSM,WiederAxion,KoreanFragile,ZhidaBLG,TMDHOTI,KooiPartialNestedBerry,PartialAxionHOTINumerics,GunnarSpinFragileWilson,BinghaiOscillationWilsonLoop,Wieder22}] of the lower two bands of Eq.~(\ref{eq:HamBloch3F}) [Fig.~\ref{fig:3F_figBulk}(b)] on spheres surrounding the $\Gamma$ and $R$ points using the tight-binding parameters in Eq.~(\ref{eq:appTBparams3F}) with the choice $\text{sgn}(v)=+1$.
The two Wilson loop eigenvalues as a set wind twice in the positive direction in Fig.~\ref{fig:3F_extraCrossings}(a) and twice in the negative direction in Fig.~\ref{fig:3F_extraCrossings}(d), indicating that the multifold degeneracies at $\Gamma$ and $R$ respectively carry chiral charges of $C=\pm 2$ for their lower two bands when $\text{sgn}(v)=+1$.  
Importantly, if we instead choose the model parameters in Eq.~(\ref{eq:appTBparams3F}) with $\text{sgn}(v)=-1$, then the sphere Wilson loop eigenvalue winding numbers at $\Gamma$ and $R$ \emph{reverse in sign, but not in magnitude} [Fig.~\ref{fig:3F_extraCrossings}(e,h)].
Lastly, though not shown in Fig.~\ref{fig:3F_extraCrossings}, we have also computed the Wilson loop spectrum of only the lowest band of Eq.~(\ref{eq:HamBloch3F}) [Fig.~\ref{fig:3F_figBulk}(b)] on spheres surrounding $\Gamma$ and $R$ for $\text{sgn}(v)=\pm 1$ in Eqs.~(\ref{eq:appTBparams3F}) and~(\ref{eq:multifoldStructuralChirality}), and respectively observe the same Wilson loop winding numbers [chiral charges] as those of the two-band Wilson spectra in Fig.~\ref{fig:3F_extraCrossings}(a,d,e,h).

Overall, the dependence of the signs of the Wilson loop winding numbers in Fig.~\ref{fig:3F_extraCrossings}(a,d,e,h) on $\text{sgn}(v)$ implies that like the KW model analyzed in Ref.~\cite{KramersWeyl} and in the text surrounding Eq.~(\ref{eq:appChiralCharge}), and like the chiral multifold fermions experimentally studied in Refs.~\cite{AlPtObserve,PdGaObserve,DingARPESReversal,SessiPdGaQPIReversal}, the chiral multifold fermions at $\Gamma$ and $R$ in the symmorphic multifold model introduced in this work [Eqs.~(\ref{eq:3FTmatrix}),~(\ref{eq:3FMmatrix}),~(\ref{eq:multifoldBandV}), and~(\ref{eq:HamBloch3F})] carry a low-energy topological chirality that is directly inherited from and controlled by the lattice-scale structural chirality:
\begin{equation}
C_{\mathcal{H}} = \text{sgn}(v),
\label{eq:multifoldStructuralChirality}
\end{equation}
where $C_{\mathcal{H}}=1$ corresponds to the right-handed [$R$] model enantiomer, and $C_{\mathcal{H}}=-1$ corresponds to the left-handed [$L$] model enantiomer.
Below, in the text surrounding Eq.~(\ref{eq:multifoldInversionLimit}), we will further justify Eq.~(\ref{eq:multifoldStructuralChirality}) by examining the rotoinversion symmetries that emerge in Eq.~(\ref{eq:HamBloch3F}) when $v$ is taken to vanish.
Additionally, as in the crystalline KW and double-Weyl models respectively analyzed in Appendix~\ref{app:PristineKramers} and~\ref{app:PristineQuadratic}, $C_{\mathcal{H}}$ in Eq.~(\ref{eq:multifoldStructuralChirality}) is a global geometric property of the multifold fermion model on a crystalline lattice, because the OAM coupling between each unit cell carries the same sign of $v$.
However, by allowing the signs of the $v$ hopping parameter in Eqs.~(\ref{eq:3FTmatrix}) and~(\ref{eq:multifoldBandV}) to vary for bonds originating from each site $\alpha$, we may also promote $C_{\mathcal{H}}$ in Eq.~(\ref{eq:multifoldStructuralChirality}) to a spatially varying local chirality $\chi_{\alpha}$ [see Refs.~\cite{LocalChiralityMoleculeReviewNatChem,LocalChiralityLiquidCrystals,KamienLubenskyChiralParameter,LocalChiralityDomainLiquidCrystal,LocalChiralityTransfer,KamienChiralLiquidCrystal,LocalChiralityQuasicrystalVirus,lindell1994electromagneticBook,LocalChiralityVillain,LocalChiralityWenZee,LocalChiralityBaskaran,LocalChiralitySpinFrame} and the text surrounding Eq.~(\ref{eq:temp2siteFrameBreakdown})].
We will shortly below in the text following Eqs.~(\ref{eq:3FlLocalAlpha}) and~(\ref{eq:amorphous3FTmatrixFinalChirality}) implement a non-crystalline generalization of the multifold fermion model in which the local chirality $\chi_{\alpha}$ represents a parameter that can be ordered or disordered independent from the lattice structural order, and can specifically be tuned to control the OAM textures and topology of non-crystalline spin-1 multifold fermions.

Unlike at the $\Gamma$ and $R$ points, the threefold cubic symmetry $\tilde{C}_{3,111}$ does not leave the $X$ and $M$ TRIM points invariant, and is hence not an element of the little groups $G_{X}$ and $G_{M}$ [see Eq.~(\ref{eq:EquivKPoints}) and the surrounding text].
The little groups $G_{X}$ and $G_{M}$ instead only contain the twofold rotation symmetries around the $i=x,y,z$ axes $\tilde{C}_{2i}$ [as well as $\tilde{\mathcal{T}}$ and lattice translation symmetries], and are hence isomorphic to the orthorhombic subgroup SG 16 ($P222$) of the full system symmetry group SG 195 ($P23$).
Using the~\href{https://www.cryst.ehu.es/cgi-bin/cryst/programs/corepresentations.pl}{Corepresentations} tool on the Bilbao Crystallographic Server~\cite{MTQC,MTQCmaterials}, we observe that unlike $G_{\Gamma}$ and $G_{R}$, $G_{X}$ and $G_{M}$ only have one-dimensional, single-valued small coreps.
This implies that for tight-binding models in SG 195 ($P23$) with spinless orbital basis states and sufficiently large and diverse sets of symmetry-allowed hopping terms -- like the chiral multifold fermion model introduced in this section [Eqs.~(\ref{eq:3FTmatrix}),~(\ref{eq:3FMmatrix}),~(\ref{eq:multifoldBandV}), and~(\ref{eq:HamBloch3F})] -- the Bloch eigenstates at $X$ and $M$ will only be singly degenerate [see Fig.~\ref{fig:3F_figBulk}(b) the text surrounding Eq.~(\ref{eq:corepToEnergyEDegen})].

To fully characterize the nodal degeneracies and topology of the chiral multifold fermion model, it is necessary to note that though there are no enforced chiral fermions exactly at $X$ and $M$ for the tight-binding parameters in Eq.~(\ref{eq:appTBparams3F}), there are still \emph{clusters} of conventional Weyl fermions with nonvanishing net chiral charges that lie \emph{within a close vicinity} of $X$ and $M$.
To understand this, we first recognize that in the limit that $t_{1}=t_{2}=t_{3}=0$, the lower three bands of Eqs.~(\ref{eq:HamBloch3F}) and~(\ref{eq:appHKPmultifold}) exhibit topologically chiral threefold degeneracies at \emph{every} TRIM point.
Through Eq.~(\ref{eq:appHKPmultifold}), we find that the threefold degeneracies in the $t_{1,2,3}=0$ limit exhibit the same ${\bf k}\cdot {\bf p}$ Hamiltonian at each TRIM point, up to the sign of the $v$ spin-1 dispersion term.  
Following our previous analysis of the chiral charges of the multifold degeneracies at $\Gamma$ and $R$ [see Fig.~\ref{fig:3F_extraCrossings}(a,d,e,g) and the text surrounding Eq.~(\ref{eq:multifoldStructuralChirality})], this implies that the artificial threefold degeneracies at $X$ and $M$ in the $t_{1,2,3}=0$ limit also carry chiral charges of $|C|=2$ for their lower two bands, subdivided into a lowest band with $|C|=2$ and a central nondispersing band with $C=0$.

Importantly, as discussed above, only the spin-1 chiral threefold [multifold] fermions at $\Gamma$ and $R$ are protected by symmetry [\emph{i.e.} transform in little group small coreps, see Appendix~\ref{app:corepDefs}], whereas the threefold fermions at $X$ and $M$ only appear in the artificial [fine-tuned] limit in which $t_{1,2,3}$ vanish in Eqs.~(\ref{eq:HamBloch3F}) and~(\ref{eq:appHKPmultifold}).
The symmetry-allowed band splitting at $X$ and $M$ can in particular be seen from the ${\bf k}\cdot {\bf p}$ Hamiltonian $\mathcal{H}_{{\bf k}_{\mathcal{T}}}({\bf q})$ [Eq.~(\ref{eq:appHKPmultifold})], in which the exponential phase factors $e^{ik_{x,y,z;\mathcal{T}}}$ in the symmetry-allowed $t_{1,2,3}$ terms are unequal at $X$ [$k_{x}=\pi$, $k_{y}=k_{z}=0$] and $M$ [$k_{x}=k_{y}=\pi$, $k_{z}=0$]. 
The unequal exponential phase factors and independent $t_{1,2,3}$ hopping parameters then give rise to arbitrary linear combinations of the matrices $\mu^{z}$, $\tau^{z}$, and $\tau^{z}\mu^{z}$ in $\mathcal{H}_{{\bf k}_{\mathcal{T}}}({\bf q})$ that differ from the $\tilde{C}_{3,111}$-invariant combination $\mu^{z}+\tau^{z}\mu^{z}+\tau^{z}$ that is the coefficient of the $m$ mass term, and hence shift all four Bloch eigenstates at $X$ and $M$ to different energies from each other [Fig.~\ref{fig:3F_figBulk}(b)].

However, without breaking translation symmetry, the chiral charge of a nodal degeneracy can only be eliminated by interactions that merge the degeneracy with other nodal degeneracies that carry compensating chiral charges~\cite{AshvinWeyl,VolovikWeylMonopole,VolovikChiralAnomaly,HaldaneOriginalWeyl,MurakamiWeyl,BurkovBalents,AndreiWeyl,HasanWeylDFT,Armitage2018,SuyangWeyl,LvWeylExp,YulinWeylExp,AliWeylQPI,AlexeyType2,ZJType2,BinghaiClaudiaWeylReview,ZahidNatRevMatWeyl,CDWWeyl,IlyaIdealMagneticWeyl}.
Because the chiral multifold model [Eqs.~(\ref{eq:HamBloch3F}) and~(\ref{eq:appHKPmultifold})] in the limit that $t_{1,2,3}=0$ exhibits artificial threefold degeneracies at $X$ and $M$ with chiral charges of $|C|=2$ for their lower two bands, then weakly tuning $t_{1,2,3}$ away from zero [relative to $v$ and $m$] cannot remove the net chiral charges of nodal degeneracies that lie in the close vicinity of $X$ and $M$.
Consistent with this recognition, for the tight-binding parameters in Eq.~(\ref{eq:appTBparams3F}) in which $t_{1,2,3}$ are nonzero, unequal, and much weaker than $v$ and $m$, we observe that the second and third bands remain linked by \emph{clusters} of conventional Weyl fermions along high-symmetry lines and at generic ${\bf k}$ points in the close vicinity of $X$ and $M$.
We further find that each Weyl cluster carries a net chiral charge of $|C|=2$ inherited from the artificial chiral multifold fermion that appeared at the same TRIM point in the $t_{1,2,3}=0$ limit [Fig.~\ref{fig:3F_figBulk}(b)].
Specifically, beginning with the right-handed model enantiomer defined by $C_{\mathcal{H}} = \text{sgn}(v) = +1$ [Eq.~(\ref{eq:multifoldStructuralChirality})], we compute the Wilson loop spectrum [Appendix~\ref{sec:WilsonBerry}] of the lower two bands of Eq.~(\ref{eq:HamBloch3F}) on spheres that enclose all of the nodal degeneracies near $X$ and $M$ and plot the results in Fig.~\ref{fig:3F_extraCrossings}(b,c) for the $R$ enantiomer [$C_{\mathcal{H}}=+1$] and in Fig.~\ref{fig:3F_extraCrossings}(f,g) for the $L$ enantiomer [$C_{\mathcal{H}}=-1$].
The Wilson loop spectra wind twice in Fig.~\ref{fig:3F_extraCrossings}(b,c,f,g), indicating that the conventional Weyl points near $X$ and $M$ as a set each carry net chiral charges of $|C|=2$.
We specifically observe net chiral charges for the nodal degeneracies near $X$ and $M$ of $C_{X}=-C_{M}=-2$ for the $R$ enantiomer, and $C_{X}=-C_{M}=+2$ for the $L$ enantiomer.
This indicates that like the cubic-symmetry-enforced chiral multifold degeneracies at $\Gamma$ and $R$ [Fig.~\ref{fig:3F_extraCrossings}(a,d,e,h)], the clusters of Weyl points near $X$ and $M$ also exhibit a net topological chirality that is controlled by the real-space lattice structural chirality $C_{\mathcal{H}}$ [Eq.~(\ref{eq:multifoldStructuralChirality})].
Lastly, for completeness, we note that we also similarly observe clusters of conventional Weyl fermions that link the first and second bands of Eq.~(\ref{eq:HamBloch3F}) near $X$ and $M$ [Fig.~\ref{fig:3F_figBulk}(b)].
Though not shown in Fig.~\ref{fig:3F_extraCrossings}, we have also computed the Wilson loop spectrum of only the lowest band of Eq.~(\ref{eq:HamBloch3F}) on spheres enclosing all of the nodal degeneracies near $X$ and $M$ for $\text{sgn}(v)=\pm 1$ in Eqs.~(\ref{eq:appTBparams3F}) and~(\ref{eq:multifoldStructuralChirality}), and respectively observe the same Wilson loop winding numbers [net chiral charges] as those of the two-band Wilson spectra in Fig.~\ref{fig:3F_extraCrossings}(b,c,f,g).

\begin{figure}[t]
    \centering
\includegraphics[width=0.95\linewidth]{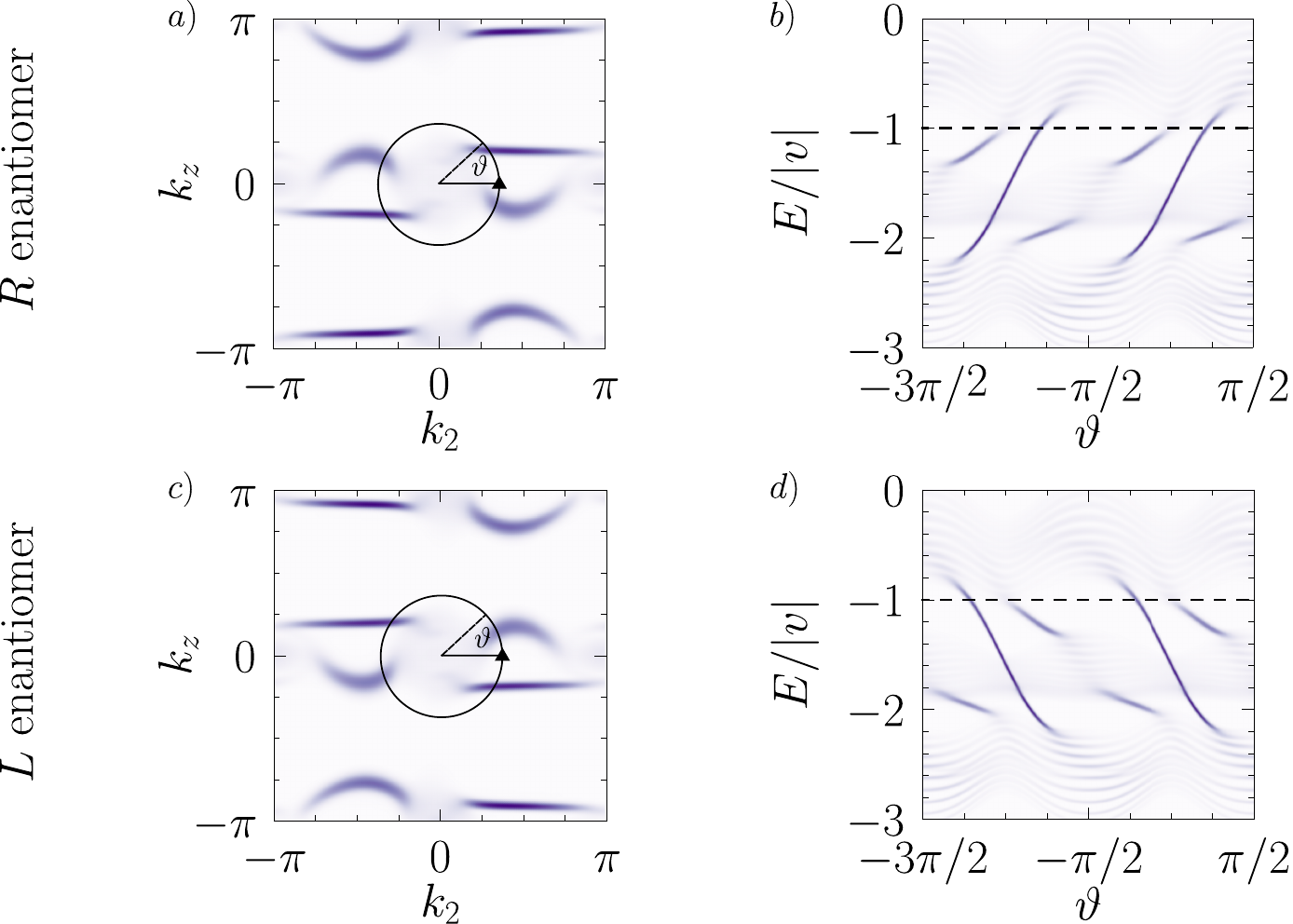}
\caption{Fermi-arc surface states of the crystalline spin-1 multifold model.
Panels (a,b) and (c,d) respectively show the $(110)$-surface spectrum of the right-handed [$R$] and left-handed [$L$] enantiomers of the crystalline multifold model [Eqs.~(\ref{eq:3FTmatrix}),~(\ref{eq:3FMmatrix}),~(\ref{eq:multifoldBandV}), and~(\ref{eq:HamBloch3F})] obtained using the parameters in Eq.~(\ref{eq:appTBparams3F}).
(a) and (c) respectively show the $(110)$-surface Fermi arcs for the $R$ and $L$ multifold model enantiomers computed from 
surface Green's functions at $E/|v|=-1$ and plotted as functions of $k_{2} = (1/\sqrt{2})(k_{x}-k_{y})$ and $k_{z}$. 
In (b) and (d), we then respectively plot the $(110)$-surface spectral functions of the $R$ and $L$ model enantiomers computed as functions of energy on counterclockwise circular paths surrounding $k_{2}=k_{z}=0$ in (a,c). 
The Fermi pockets at $k_{2}=k_{z}=0$ in (a,c) each contain the superposed projections of the spin-1 multifold fermion at $\Gamma$ in Fig.~\ref{fig:3F_figBulk}(b) and a cluster of Weyl points that link the second and third bands in the close vicinity of $M$ [see Fig.~\ref{fig:3F_extraCrossings}(b,g) and the surrounding text].
By direct computation in Fig.~\ref{fig:3F_extraCrossings}, we specifically find that the bulk Fermi pockets at $\Gamma$ and $M$ each carry a net chiral charge of $C=2$ for the $R$ enantiomer and $C=-2$ for the $L$ enantiomer.
This gives rise to a net surface chiral charge projection of $C=4$ for (a,b) the $R$ enantiomer and $C=-4$ for (c,d) the $L$ enantiomer.
The surface Fermi arcs in (b) and (d) respectively cross the dashed horizontal line in each panel four times with positive and negative slopes, confirming that the bulk Fermi pocket projections in (a) and (b) at $k_{2}=k_{z}=0$ carry net chiral charges of $C=4$ for the $R$ enantiomer and $C=-4$ for the $L$ enantiomer.
For completeness, we note that four Fermi arcs with positive velocities for the $R$ enantiomer and negative velocities for the $L$ enantiomer also cross the gap at $E/|v|=-2$ in (b) and (d), respectively.  
The lower four Fermi arcs in (b,d) originate from the projected Fermi pockets of chiral fermions linking the first and second model bands at $\Gamma$ and $M$, which, as discussed in Fig.~\ref{fig:3F_extraCrossings} and the surrounding text, each carry the same net chiral charges as the crossings between the second and third bands at $\Gamma$ and $M$  [$C=2$ for the $R$ enantiomer and $C=-2$ for the $L$ enantiomer, for a total projected charge of $C=\pm 4$ at $E/|v|=-2$].
The topological surface Fermi arcs crossing consecutive projected bulk gaps in (b,d) closely resemble -- and arise from the same bulk multifold fermions as -- the ladder-like Fermi-arc surface states recently experimentally observed in the topological semimetal alloy Rh$_{1-x}$Ni$_{x}$Si~\cite{ZahidLadderMultigap}.
The data plotted in this figure were obtained from the tight-binding model in Eqs.~(\ref{eq:3FTmatrix}),~(\ref{eq:3FMmatrix}),~(\ref{eq:multifoldBandV}), and~(\ref{eq:HamBloch3F}) placed on a regular lattice that was infinite in the $\hat{z}$-direction and finite with 150 sites and 
periodic boundary conditions in the $\hat{x}-\hat{y}$-direction [reciprocal to $k_{2}$] and with 20 sites and open boundary 
conditions in the $\hat{x}+\hat{y}$- [$(110)$-] direction.}
\label{fig:3F_figWL}
\end{figure}

\paragraph*{\bf Surface Fermi Arcs} -- $\ $ We next compute the surface spectrum of the chiral multifold fermion model to explore the bulk-boundary correspondence of the chiral fermion distribution in Figs.~\ref{fig:3F_figBulk}(b) and~\ref{fig:3F_extraCrossings}.
In Fig.~\ref{fig:3F_figWL}(a,b) and Fig.~\ref{fig:3F_figWL}(c,d), we respectively plot the $(110)$- [($\hat{x} + \hat{y}$)-normal] surface Fermi arcs of the $R$ and $L$ enantiomers of the multifold model as functions of $k_{2} = (1/\sqrt{2})(k_{x}-k_{y})$ [Eq.~\eqref{appeq:surfacemomenta}], $k_{z}$, and energy.  
To understand the $(110)$-surface Fermi-arc connectivity, we first establish the net chiral charges of the projected Fermi pockets in Fig.~\ref{fig:3F_figWL}(a,c), which each surround one of the four surface TRIM points $k_{2},k_{z}=0,\pi$.
To begin, the surface Fermi pocket at $k_{2}=k_{z}=0$ and $E/|v|=-1$ in Fig.~\ref{fig:3F_figWL} contains the projected bulk Fermi pockets of the third band of the spin-1 chiral multifold fermion at $\Gamma$ in Fig.~\ref{fig:3F_figBulk}(b) [\emph{i.e.} occupying the lower two multifold fermion bands] and the cluster of Weyl points between the second and third bands in the close vicinity of $M$ [see Fig.~\ref{fig:3F_extraCrossings}(c,g) and the surrounding text].
We previously in Fig.~\ref{fig:3F_extraCrossings}(a,c,e,g) used the two-band sphere Wilson loop spectrum to show by direct computation that the bulk Fermi pockets at $\Gamma$ and $M$ at $E/|v|=-1$ [\emph{i.e.} with two occupied bands] each carry a net chiral charge of $C=2$ for the $R$ enantiomer and $C=-2$ for the $L$ enantiomer, leading to an overall $(110)$-surface-projected chiral charge at $k_{2}=k_{z}=0$ of $C=4$ for the $R$ enantiomer and $C=-4$ for the $L$ enantiomer.

Next, it is important to establish that the $X$ and $M$ TRIM points in SG 195 ($P23$) lie in multiplicity-3 momentum stars, such that $X$ and $M$ are each related to two other TRIM points by the threefold cubic rotation symmetry $\tilde{C}_{3,111}$ [see Fig.~\ref{fig:3F_figBulk}(a) and Eq.~(\ref{eq:PristineMultifoldsyms})].
Uniquely in structurally chiral crystals -- like the chiral multifold fermion model introduced in this work -- [Eqs.~(\ref{eq:3FTmatrix}),~(\ref{eq:3FMmatrix}),~(\ref{eq:multifoldBandV}), and~(\ref{eq:HamBloch3F})] -- all topologically chiral nodal degeneracies in the same momentum star carry the same chiral charges~\cite{KramersWeyl,CDWWeyl,deJuanAdolfoCPGE}.
This occurs because topologically chiral nodal degeneracies in the same momentum star of a structurally chiral crystal are by definition only related to each other by time-reversal symmetry and proper rotation symmetries [see Eq.~(\ref{eq:rotoinversion}) and the surrounding text], which do not invert the sign of the chiral charge $C$ [Eq.~(\ref{eq:crystallineWilsonSphere})].
Hence, in addition to the Weyl points that link the second and third bands in the close vicinity of $X$ [${\bf k}=(\pi,0,0)$] and carry a net chiral charge of $C_{X}=-2$ for the $R$ enantiomer and $C_{X}=2$ for the $L$ enantiomer [Fig.~\ref{fig:3F_extraCrossings}(b,f)], there are also symmetry-related clusters of Weyl points with the same net chiral charges $C=C_{X}=-2$ for the $R$ enantiomer and $C=C_{X}=2$ for the $L$ enantiomer that lie within a close vicinity of the TRIM points ${\bf k}=(0,\pi,0)$ and ${\bf k}=(0,0,\pi)$ in the momentum star of $X$.
Similarly, in addition to the Weyl points that link the second and third bands in the close vicinity of $M$ [${\bf k}=(\pi,\pi,0)$] and carry a net chiral charge of $C_{M}=2$ for the $R$ enantiomer and $C_{M}=-2$ for the $L$ enantiomer [Fig.~\ref{fig:3F_extraCrossings}(c,g)], there are also symmetry-related clusters of Weyl points with the same net chiral charges $C=C_{M}=2$ for the $R$ enantiomer and $C=C_{M}=-2$ for the $L$ enantiomer that lie within a close vicinity of the TRIM points ${\bf k}=(\pi,0,\pi)$ and ${\bf k}=(0,\pi,\pi)$ in the momentum star of $M$.

Having established the Fermi pocket chiral charges in the vicinity of all eight bulk TRIM points [Fig.~\ref{fig:3F_figBulk}(a)], we next return to the $(110)$-surface spectrum in Fig.~\ref{fig:3F_figWL}(a,c) and use this information to establish the net chiral charges of the surface projections of the bulk Fermi pockets.
First, we previously established above that the projected bulk Fermi pocket at $k_{2}=k_{z}=0$ carries a net chiral charge of $C=4$ for the $R$ enantiomer and $C=-4$ for the $L$ enantiomer at $E/|v|=-1$.
We then recognize that the $(110)$-surface Fermi pocket at $k_{2}=\pi$, $k_{z}=0$ in Fig.~\ref{fig:3F_figWL}(a,c) contains the superposed projections of clusters of Weyl fermions that link the second and third bands near $X$ and ${\bf k}=(0,\pi,0)$, which lies in the same momentum star as $X$.
As established above and in Ref.~\cite{KramersWeyl}, because the multifold fermion model is structurally chiral, this implies that each bulk Fermi pocket that contributes to the $(110)$-surface-projected Fermi pocket at $k_{2}=\pi$, $k_{z}=0$ in Fig.~\ref{fig:3F_figWL}(a,c) exhibits a chiral charge of $C_{X}=-2$ for the $R$ enantiomer and $C_{X}=2$ for the $L$ enantiomer [Fig.~\ref{fig:3F_extraCrossings}(b,f)], leading to a total projected chiral charge of $C=-4$ for the $R$ enantiomer and $C=4$ for the $L$ enantiomer.
Next, the $(110)$-surface Fermi pocket at $k_{2}=k_{z}=\pi$ in Fig.~\ref{fig:3F_figWL}(a,c) contains the superposed projections of clusters of Weyl fermions that link the second and third bands near ${\bf k}=(\pi,0,\pi)$ and ${\bf k}=(0,\pi,\pi)$, which lie in the same momentum star as $M$.
This similarly implies that each bulk Fermi pocket that contributes to the $(110)$-surface-projected Fermi pocket at $k_{2}=k_{z}=\pi$ in Fig.~\ref{fig:3F_figWL}(a,c) exhibits a chiral charge of $C_{M}=2$ for the $R$ enantiomer and $C_{M}=-2$ for the $L$ enantiomer [Fig.~\ref{fig:3F_extraCrossings}(c,g)], leading to a total projected chiral charge of $C=4$ for the $R$ enantiomer and $C=-4$ for the $L$ enantiomer.
Lastly, like the surface TRIM point $k_{2}=k_{z}=0$, the surface TRIM point $k_{2}=0$, $k_{z}=\pi$ contains at $E/|v|=-1$ the projected bulk Fermi pockets of the third band of the spin-1 multifold fermion at $R$ in Fig.~\ref{fig:3F_figBulk}(b) and the cluster of Weyl points linking the second and third bands in the close vicinity of ${\bf k}=(0,0,\pi)$, which lies in the same momentum star as $X$ [Fig.~\ref{fig:3F_figBulk}(a)].
From the two-band sphere Wilson loop calculations in Fig.~\ref{fig:3F_extraCrossings}(b,d,f,h) and the statement established in Ref.~\cite{KramersWeyl} that all chiral fermions in the same momentum star of a structurally chiral crystal carry the same chiral charges, we find that the bulk Fermi pockets at $R$ and ${\bf k}=(0,0,\pi)$ and $E/|v|=-1$ each carry a net chiral charge of $C=-2$ for the $R$ enantiomer and $C=2$ for the $L$ enantiomer.
We thus conclude that the $(110)$-surface Fermi pocket at $k_{2}=0$, $k_{z}=\pi$ and $E/|v|=-1$ carries a total projected chiral charge of $C=-4$ for the $R$ enantiomer and $C=4$ for the $L$ enantiomer.

Now that we have computed the total projected chiral charges of the four $(110)$-surface Fermi pockets at $E/|v|=-1$, we are finally in a position to analyze the surface Fermi-arc connectivity.
In Fig.~\ref{fig:3F_figWL}(a,c), we observe four surface Fermi arcs [two time-reversed pairs] connecting the projected Fermi pocket at $k_{2}=k_{z}=0$ to the projected Fermi pocket at $k_{2}=\pi$, $k_{z}=0$, consistent with our determination that the two Fermi pockets carry the respective chiral charges $C=\pm 4$ for the $R$ enantiomer and $C=\mp 4$ for the $L$ enantiomer.
Additionally, in Fig.~\ref{fig:3F_figWL}(a,c), we also observe four surface Fermi arcs connecting the projected Fermi pocket at $k_{2}=0$, $k_{z}=\pi$ to the projected Fermi pocket at $k_{2}=k_{z}=\pi$, consistent with our determination that the two Fermi pockets carry the respective chiral charges $C=\mp 4$ for the $R$ enantiomer and $C=\pm 4$ for the $L$ enantiomer.
Interestingly, the surface spectra in Fig.~\ref{fig:3F_figWL}(a,c) are mirror images of each other, with half of the arcs roughly exhibiting opposite constant-energy helicities for opposite enantiomers, as they did previously for the KW model in Fig.~\ref{fig:KW_figWilson}(b,e) and the double-Weyl model in Fig.~\ref{fig:charge2_figWilson}(b,e).
However, as previously discussed in the text following Eq.~(\ref{appeq:surfacemomenta}), the qualitative helicity of surface Fermi arcs in constant-energy spectral functions like Fig.~\ref{fig:3F_figWL}(a,c) is not a topological property.
We therefore in Fig.~\ref{fig:3F_figWL}(b,d) limit our focus to $k_{2}=k_{z}=0$ [the surface Fermi pocket of greatest relevance to the disordered multifold calculations that we will perform below in Appendix~\ref{app:amorphousMultifold}], and additionally compute the $(110)$-surface Fermi arcs as functions of energy on closed counterclockwise paths encircling $k_{2}=k_{z}=0$, which conversely provides a quantitative indicator of the bulk topology~\cite{AshvinWeyl,VolovikWeylMonopole,VolovikChiralAnomaly,HaldaneOriginalWeyl,MurakamiWeyl,BurkovBalents,AndreiWeyl,HasanWeylDFT,Armitage2018,SuyangWeyl,LvWeylExp,YulinWeylExp,AliWeylQPI,AlexeyType2,ZJType2,BinghaiClaudiaWeylReview,ZahidNatRevMatWeyl,CDWWeyl,IlyaIdealMagneticWeyl}.
The surface Fermi arcs respectively cross the dashed horizontal line  in Fig.~\ref{fig:3F_figWL}(b,d) four times with
positive and negative slopes [velocities], confirming that the bulk Fermi pocket projections at $k_{2}=k_{z}=0$ and $E/|v|=-1$ in Fig.~\ref{fig:3F_figWL}(a,c) respectively carry a chiral charge of $C=4$ for the $R$ enantiomer and $C=-4$ for the $L$ enantiomer.

For completeness, we note that four Fermi arcs with positive velocities for the $R$ enantiomer and negative velocities for the $L$ enantiomer also cross the lower gap at $E/|v|=-2$ in Fig.~\ref{fig:3F_figWL}(b,d), respectively.  
The lower four Fermi arcs in Fig.~\ref{fig:3F_figWL}(b,d) originate from the projected Fermi pockets of chiral fermions linking the first and second model bands at $\Gamma$ and $M$, which we previously found to each carry the same net chiral charges as the crossings between the second and third bands at $\Gamma$ and $M$  [$C=2$ for the $R$ enantiomer and $C=-2$ for the $L$ enantiomer, see Fig.~\ref{fig:3F_extraCrossings} and the surrounding text]. 
This gives rise to a total projected chiral charge at $k_{2}=k_{z}=0$ and $E/|v|=-2$ of $C=4$ for the $R$ enantiomer and $C=-4$ for the $L$ enantiomer, consistent with the four positively and negatively sloped Fermi arcs that respectively cross the lower spectral gaps in Fig.~\ref{fig:3F_figWL}(b,d).
The appearance of topological surface Fermi arcs crossing consecutive projected bulk gaps in Fig.~\ref{fig:3F_figWL}(b,d) closely resembles -- and arises from the same bulk multifold fermions as -- the ladder-like Fermi-arc surface states recently observed in ARPES experiments on the topological semimetal alloy Rh$_{1-x}$Ni$_{x}$Si~\cite{ZahidLadderMultigap}.

\paragraph*{\bf Structural Chirality and Achiral Critical Phase} -- $\ $ We conclude by characterizing the structural chirality of the multifold fermion model via the rotoinversion symmetries that emerge when $v\rightarrow 0$ in Eq.~(\ref{eq:HamBloch3F}).
First, as with the other crystalline chiral fermion models discussed in this work [Appendices~\ref{app:PristineKramers} and~\ref{app:PristineQuadratic}], when $v\rightarrow 0$ in Eq.~(\ref{eq:HamBloch3F}), the topological chiral charges of the spin-1 chiral fermions at $\Gamma$ and $R$ become ill-defined and the Weyl points near $X$ and $M$ in Figs.~\ref{fig:3F_figBulk}(b) and~\ref{fig:3F_extraCrossings}(b,c,f,g) either pairwise annihilate or merge into topologically achiral degeneracies.
The absence of well-defined chiral charges for nodal degeneracies linking the first and second and second and third bands in the multifold fermion model in the $v\rightarrow 0$ limit can be understood by recognizing that as previously for the KW model in Appendix~\ref{app:PristineKramers} and the double-Weyl model in Appendix~\ref{app:PristineQuadratic}, the system SG [and therefore the little group at each TRIM point, see Appendix~\ref{app:corepDefs}] gains additional rotoinversion symmetries in the $v\rightarrow 0$ limit, and hence becomes achiral [see Eq.~(\ref{eq:rotoinversion}) and the surrounding text]. 
Of the rotoinversion symmetries that emerge when $v\rightarrow 0$ in Eq.~(\ref{eq:HamBloch3F}), the simplest is spatial inversion symmetry $\mathcal{I}$, which can be represented through its symmetry action on $\mathcal{H}({\bf k})$ in the basis of the $s$ and $p_{x,y,z}$ orbitals on each lattice site [see Eq.~(\ref{eq:Amo3F}) and the surrounding text]:
\begin{equation}
\mathcal{I}\mathcal{H}({\bf k})\mathcal{I}^{-1} = \begin{pmatrix}
        1&0&0&0\\
        0&-1&0&0\\
        0&0&-1&0\\
        0&0&0&-1
    \end{pmatrix}\mathcal{H}(-{\bf k})\begin{pmatrix}
        1&0&0&0\\
        0&-1&0&0\\
        0&0&-1&0\\
        0&0&0&-1
    \end{pmatrix}.
\label{eq:multifoldInversionLimit}
\end{equation}
Beyond $\mathcal{I}$ in Eq.~(\ref{eq:multifoldInversionLimit}), $\mathcal{H}({\bf k})$ with $v=0$ in Eq.~(\ref{eq:HamBloch3F}) is also left invariant under the action of any inversion symmetry $\overline{\mathcal{I}}$ of the form: 
\begin{equation}
\overline{\mathcal{I}}\mathcal{H}({\bf k})\overline{\mathcal{I}}^{-1} = U\mathcal{H}(-{\bf k})U^{\dagger},
\label{eq:extraInversionMultifold}
\end{equation}
or any mirror reflection symmetry $M_{i}$ about the $i=x,y,z$ axis of the form:
\begin{equation}
M_{i}\mathcal{H}({\bf k})M_{i}^{-1} = U\mathcal{H}(M_{i}^{-1}{\bf k})U^{\dag},
\label{eq:extraMirrorMultifold}
\end{equation}
where $U$ in Eqs.~(\ref{eq:extraInversionMultifold}) and~(\ref{eq:extraMirrorMultifold}) is a diagonal matrix. 
This occurs because $\mathcal{H}({\bf k})$ with $v\rightarrow 0$ in Eq.~(\ref{eq:HamBloch3F}) only contains terms consisting of even functions of ${\bf k}$ proportional to the diagonal matrices $\mu^{z}$, $\tau^{z}$, and $\tau^{z}\mu^{z}$, and because all diagonal matrices commute.
Overall, the appearance of multitudinous rotoinversion symmetries in the $v\rightarrow 0$ limit of the symmorphic chiral multifold model introduced in this section [Eqs.~(\ref{eq:3FTmatrix}),~(\ref{eq:3FMmatrix}),~(\ref{eq:multifoldBandV}), and~(\ref{eq:HamBloch3F})], along with the absence of rotoinversion symmetries when $v\neq 0$, further confirms our assignment of the real-space structural chirality [handedness] $C_{\mathcal{H}}$ to $\text{sgn}(v)$ in Eq.~(\ref{eq:multifoldStructuralChirality}) and the surrounding text.

Finally, the above discussion indicates that in the chiral multifold fermion model introduced in this work, there is freedom to consider model enantiomers of opposite handedness [defined by $C_{\mathcal{H}}=\text{sgn}(v)$ in Eq.~(\ref{eq:multifoldStructuralChirality})] to be related [\emph{i.e.} ``twinned''] under the action of either $\mathcal{I}$ [Eq.~(\ref{eq:multifoldInversionLimit})], $\overline{\mathcal{I}}$ [Eq.~(\ref{eq:extraInversionMultifold})], $M_{i}$ [Eq.~(\ref{eq:extraMirrorMultifold})], or another rotoinversion symmetry of the $v\rightarrow 0$ limit of Eq.~(\ref{eq:HamBloch3F}) not highlighted above.
Given the relative simplicity of the symmorphic multifold model in Eqs.~(\ref{eq:3FTmatrix}),~(\ref{eq:3FMmatrix}),~(\ref{eq:multifoldBandV}), and~(\ref{eq:HamBloch3F}), one may in future works wish to build upon the model by incorporating additional hopping interactions or degrees of freedom.
From a chemical perspective and looking ahead to future works, it is perhaps most realistic to enforce that the $R$ and $L$ multifold model enantiomers are ``twinned'' under the action of $\mathcal{I}$ in Eq.~(\ref{eq:multifoldInversionLimit}), because single-crystal samples of B20 CoSi -- which exhibit the same weak-SOC spin-1 multifold fermions at $\Gamma$ as Eq.~(\ref{eq:HamBloch3F}) -- have been found to exhibit chirality domains [Fig.~\ref{appfig:disordertypes}(b)] that are inversion twinned, rather than mirror twinned~\cite{InternalChiralExp,InternalChiralTheory}.

\clearpage

\subsubsection{Non-Crystalline Chiral Multifold Fermions}
\label{app:amorphousMultifold}

In this section, we will lastly demonstrate that the spin-1 chiral multifold fermion tight-binding model previously introduced in Appendix~\ref{app:PristineMulifold} [Eqs.~(\ref{eq:3FTmatrix}),~(\ref{eq:3FMmatrix}), and~(\ref{eq:multifoldBandV})] continues to exhibit chiral multifold fermions when it is realized in amorphous systems [approximated by strongly Gaussian disordered and random lattices] that carry a net [average] structural chirality.

To begin, the crystalline multifold fermion model in Eqs.~(\ref{eq:3FTmatrix}),~(\ref{eq:3FMmatrix}), and~(\ref{eq:multifoldBandV}) can be characterized, as detailed in Appendix~\ref{app:DiffTypesDisorder}, by three distinct sources of disorder: lattice disorder, local
frame disorder, and local handedness or \emph{chirality} disorder. 
First, the most straightforward way in which the multifold fermion tight-binding model can be disordered is via the regularity of its atomic positions [sites]. 
This \emph{structural disorder} can either be implemented by introducing random site-displacement disorder or by constructing fully random lattices [Fig.~\ref{appfig:structuraldisorder}]; we will consider both of these lattice-disorder possibilities in our analysis below.

Next, the hopping terms in Eqs.~(\ref{eq:3FTmatrix}),~(\ref{eq:3FMmatrix}), and~(\ref{eq:multifoldBandV}) also depend on the local coordinate reference frame, with OAM coupling in the Cartesian $x$-direction for example being consistently implemented by $\tau^{y}$.
In a realistic model of a disordered solid-state material, however, no particular region should have physical properties that depend on the global coordinate frame, such as the $x$-direction in an arbitrary coordinate system. 
Hence, we must also disorder the local internal frame-locking so that the non-crystalline model does not implicitly contain orbital hopping terms that are locked in any particular region to the global coordinate frame. From a continuum perspective, this can be accomplished in the multifold fermion model by introducing a local SO(3) unit vector field $\hat{R}({\bf r})$ that is given by the average orbital frame orientation of the sites within a specified vicinity of the position ${\bf r}$.
From this perspective, the model can then be defined as ``fully frame-disordered'' via the absence of long-range correlations in $\hat{R}({\bf r})$. 
To implement frame disorder at the lattice scale, we assign an SO(3) rotation matrix $R_{\alpha}$ [detailed in Eq.~(\ref{eqn:RotationMatrix}) and the preceding text] to the hopping frame at each lattice site $\alpha$, such that the intersite separation [bond] vector ${\bf d}_{\alpha\beta}$ that links the lattice sites $\alpha$ and $\beta$ [Eq.~(\ref{eq:dVectorDef3F})] is transformed to a rotated bond vector $\tilde{\bf d}_{\alpha\beta}$ as follows:
\begin{equation}
{\bf d}_{\alpha\beta} \rightarrow R_{\alpha}R_{\beta}^{\mathsf{T}}{\bf d}_{\alpha\beta} \underset{def}{\equiv} \tilde{{\bf d}}_{\alpha\beta}.
\label{appeq:rota3F}
\end{equation}
After applying Eq.~(\ref{appeq:rota3F}), the multifold model hopping matrix elements in Eq.~(\ref{eq:3FTmatrix}) become transformed to:
\begin{equation}
\label{eq:Hop3F}
   T_{\alpha\beta} = \frac{1}{2}f\left(\left|\tilde{{\bf d}}_{\alpha\beta}\right|\right) \left[\sum_{\gamma = 1,2,3} t_{\gamma}\left( \tilde{{\bf d}}_{\alpha\beta}^{\mathsf{T}}B_{\gamma}\tilde{{\bf d}}_{\alpha\beta}  \right) + i v \left(\tilde{{\bf d}}_{\alpha\beta}^{\mathsf{T}}\bm{V}\right)\right],
\end{equation}
whereas the mass term $M_{\alpha}$ in Eq.~(\ref{eq:3FMmatrix}) remains unchanged.

\begin{figure}[t]
\centering
\includegraphics[width=\linewidth]{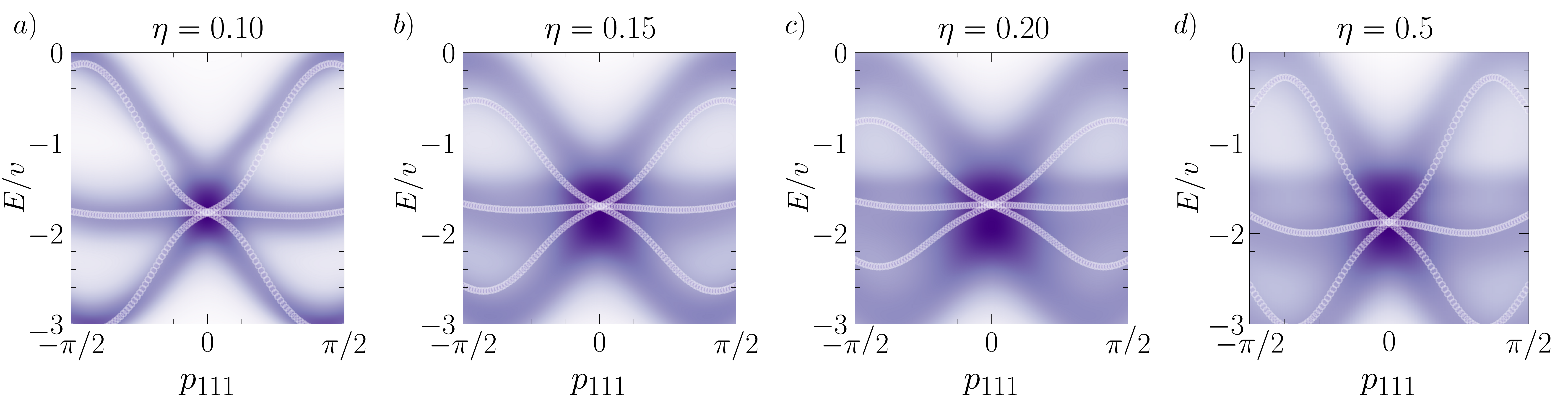}
\caption{Spectral function and effective Hamiltonian spectrum of the chiral multifold fermion model with nematic Gaussian structural disorder.
(a-d) The average spectral function $\bar{A}(E,{\bf p})$ [Eq.~(\ref{eq:SpecFunc})] of the chiral multifold tight-binding model [Eq.~(\ref{eq:amorphous3FTmatrixFinalChirality})] on a lattice with increasing random nematic structural disorder [$d_{A}=3$ in Eq.~(\ref{eq:finiteDimAmorph})] parameterized by the standard deviation $\eta$, as well as random frame disorder with the same standard deviation $\eta$ and chirality domains of unequal volume, as detailed in Appendix~\ref{app:lattices}.
The data are plotted along $p_{111} = (1/\sqrt{3})(p_{x}+p_{y}+p_{z})$, and were generated using Eq.~(\ref{eq:amorphous3FTmatrixFinalChirality}) with the parameters in Eq.~(\ref{eq:appTBparams3F}) implemented with a single domain in each replica of right-handed sites with $n_R=N_R/N_{\mathrm{sites}}=0.7$, and with the remaining volume in each replica containing a contiguous domain of left-handed sites with a corresponding concentration of $n_L=1-N_R/N_{\mathrm{sites}}=0.3$.
Each panel shows data generated by averaging over 50 disorder realizations [\emph{i.e.} replicas] with $20^{3}=8000$ sites each, as detailed in Appendix~\ref{app:PhysicalObservables}.
For all values of $\eta$, $\bar{A}(E,{\bf p})$ exhibits a threefold-degenerate feature consisting of one central set of flat states and two sets of dispersive states centered around ${\bf p}={\bf 0}$, where all three sets of states become increasingly diffuse as $\eta$ is increased, but only weakly vary in energy at ${\bf p}={\bf 0}$
with increasing $\eta$.
Though not shown in this figure, we also observe nondispersing states at much higher energies for all $\eta$, which arise from disordering the singly-degenerate trivial band at $E/|v| \approx 4.5$ in the crystalline multifold model [Fig.~\ref{fig:3F_figBulk}(b)].
In the remainder of this section, we will precisely show that the threefold-degenerate spectral features at ${\bf p}={\bf 0}$ in (a-d) represent increasingly disordered generalizations of the crystalline $\Gamma$-point spin-1 chiral multifold fermion in Fig.~\ref{fig:3F_figBulk}(c).
As shown in Fig.~\ref{fig:3DGreen3F}, the off-diagonal-in-momentum matrix elements of the average Green's function $\bar{\mathcal{G}}(E,{\bf p},{\bf p}')$ begin to vanish at moderate disorder scales [$\eta=0.2$ in panel (c)], and are nearly vanishing for strong disorder [$\eta=0.5$ in panel (d)].
We may therefore in each panel limit consideration to the \emph{momentum-diagonal} average Green's function $\bar{\mathcal{G}}(E,{\bf p})$ [Eq.~(\ref{eq:averageOneMomentumGreen})], from which we construct an effective Hamiltonian $\mathcal{H}_{\mathrm{Eff}}(E_{C},{\bf p})$ [Eq.~(\ref{eq:AvgHEff})] whose bands are plotted with light purple circles.
$\mathcal{H}_{\mathrm{Eff}}(E_{C},{\bf p})$ in each panel was specifically constructed using a reference energy cut $E_{C}$ centered at the maximum value of the spectral function at ${\bf p}={\bf 0}$ away from the nondispersing feature at larger $E$ to maximize its accuracy [see Appendix~\ref{app:EffectiveHamiltonian}], where $E_{C}$ for each panel is respectively given by (a) $E_C /v = -1.8$, (b) $E_C /v = -1.7$, (c) $E_C /v = -1.7$, and (d) $E_C /v = -1.9$.
In all panels, and for all choices of $E_{C}$ [see Fig.~\ref{fig:H_eff_benchmark} and the surrounding text], the effective Hamiltonian exhibits a threefold nodal degeneracy at ${\bf p}={\bf 0}$; we will shortly in Fig.~\ref{fig:3F_WL} use Wilson loops to show that this degeneracy is a spin-1 chiral multifold fermion.}
\label{fig:3F_bands_amo}
\end{figure}

In addition to structural and local-frame order, the multifold tight-binding model can also be characterized by spatially varying \emph{chirality order} that arises from the handedness of hopping interactions within the vicinity of a position ${\bf r}$~\cite{LocalChiralityMoleculeReviewNatChem,LocalChiralityLiquidCrystals,KamienLubenskyChiralParameter,LocalChiralityDomainLiquidCrystal,LocalChiralityTransfer,KamienChiralLiquidCrystal,LocalChiralityQuasicrystalVirus,lindell1994electromagneticBook,LocalChiralityVillain,LocalChiralityWenZee,LocalChiralityBaskaran,LocalChiralitySpinFrame}.
As detailed in Appendix~\ref{app:DiffTypesDisorder}, to implement lattice-scale chirality disorder, we assign each site $\alpha$ at the position ${\bf r}_{\alpha}$ a local discrete chirality:
\begin{equation}
\chi_{\alpha} = 0,\pm 1,
\label{eq:3FlLocalAlpha}
\end{equation}
where $\chi_{\alpha}=\pm 1$ respectively for right- and left-handed sites and $\chi_{\alpha}=0$ for achiral sites.
In this work, we specifically consider disorder realizations in which there exist \emph{domains} in which all sites carry the same values of $\chi_{\alpha}$ [see Fig.~\ref{appfig:disordertypes}(c) and the surrounding text].
Implementing lattice-scale variations in the local chirality of the multifold model leads to a final modification of Eq.~(\ref{eq:Hop3F}) as follows:
\begin{eqnarray}
T_{\alpha\beta} &=& \frac{1}{2}f\left(\left|\tilde{{\bf d}}_{\alpha\beta}\right|\right) \left[\sum_{\gamma = 1,2,3} t_{\gamma}\left[ \left(\tilde{{\bf d}}_{\alpha\beta}\right)^{\mathsf{T}}B_{\gamma}\tilde{{\bf d}}_{\alpha\beta}  \right] + i v \left(\frac{\chi_{\alpha}+\chi_{\beta}}{2}\right) \left[\left(\tilde{{\bf d}}_{\alpha\beta}\right)^{\mathsf{T}}\bm{V}\right]\right], \nonumber \\
M_{\alpha} &=& m(\mu^z+\tau^z\mu^z + \tau^z),
\label{eq:amorphous3FTmatrixFinalChirality}
\end{eqnarray}
where we have included the unmodified, achiral $M_{\alpha}$ mass term [Eqs.~(\ref{eq:Amo3F}) and~(\ref{eq:3FMmatrix})] to emphasize that $M_{\alpha}$ is also generically present in the non-crystalline system Hamiltonian.
When $\chi_{\alpha}=-\chi_{\beta}$ or $\chi_{\alpha}=\chi_{\beta} =0$, the $v$ term in Eq.~\eqref{eq:amorphous3FTmatrixFinalChirality} vanishes, and only the achiral orbital $t_{\gamma}$ hopping and $m$ mass terms remain nonvanishing. 
For the remainder of this section, we will refer to Eq.~(\ref{eq:amorphous3FTmatrixFinalChirality}) as the ``disordered'' or ``non-crystalline'' chiral multifold fermion tight-binding model.

As discussed in Appendix~\ref{app:PhysicalObservables}, we wish to model solid-state systems that have thermodynamically large numbers of atoms [or short-range-interacting patches~\cite{ProdanKohnNearsighted}] and are self-averaging.  
However using our computational hardware, we can only simulate individual disorder realizations with $\sim 20^{3}$ [$\sim 8000$] atoms.
We therefore additionally analyze the non-crystalline multifold model using an \emph{averaging procedure} in which we construct a replica-averaged~\cite{ParisiReplicaCourse,BerthierGlassAmorphousReview} momentum-resolved Green's function $\bar{\mathcal{G}}(E,\mathbf{p})$ [Eq.~(\ref{eq:averageOneMomentumGreen})] by averaging the momentum-resolved matrix Green's function over multiple [$\approx 20-50$] \emph{replicas} that each contain distinct randomly generated lattice distortions or random lattices.
As detailed in Appendix~\ref{app:PhysicalObservables}, we obtain $\bar{\mathcal{G}}(E,\mathbf{p})$ by first computing the diagonal-in-momentum piece of the momentum-resolved matrix Green's function $\mathcal{G}(E,{\bf p},{\bf p})$, while numerically establishing that the off-diagonal-in-momentum elements of $\bar{\mathcal{G}}(E,{\bf p},{\bf p}')$ vanish [on the average] for strongly disordered lattices.
Specifically, for the non-crystalline chiral multifold model in Eq.~(\ref{eq:amorphous3FTmatrixFinalChirality}) with Gaussian structural disorder, we found in Fig.~\ref{fig:3DGreen3F} that the off-diagonal-in-momentum matrix elements of $\bar{\mathcal{G}}(E,{\bf p},{\bf p}')$ begin to vanish at $\eta=0.2$, which we in this section term ``moderate'' disorder, and are nearly vanishing for $\eta=0.5$, which we hence term ``strong'' disorder.
We therefore limit consideration to the diagonal-in-momentum matrix $\bar{\mathcal{G}}(E,\mathbf{p})$, which we compute by elementwise averaging $\mathcal{G}(E,{\bf p},{\bf p})$ at the same ${\bf p}$ [Eq.~(\ref{eq:averageOneMomentumGreen})].

\paragraph*{\bf Energy Spectrum} -- $\ $ In Fig.~\ref{fig:3F_bands_amo} we plot the disorder-averaged, momentum-resolved spectral function $\bar{A}(E,{\bf p}) \propto \text{Im}\{\Tr[\bar{\mathcal{G}}(E,{\bf p})]\}$ [Eq.~(\ref{eq:SpecFunc})] for the disordered multifold tight-binding model [Eq.~(\ref{eq:amorphous3FTmatrixFinalChirality})] with the parameters:
\begin{equation}
v=3.5,\ m=5,\ t_1=0.55,\ t_2=0.35,\ t_3=0.25,
\label{eq:disordered3Fparams}
\end{equation}
placed on a lattice with increasing random nematic [$d_{A}=3$ in Eq.~(\ref{eq:finiteDimAmorph})] Gaussian structural disorder parameterized by a standard deviation $\eta$, as detailed in Appendix~\ref{app:lattices}.
The spectra in Fig.~\ref{fig:3F_bands_amo} were also computed using random local frame disorder [Eq.~\eqref{appeq:rota3F}] implemented with the same standard deviation $\eta$ as the lattice disorder, as well as chirality domains of unequal volume [70\% right-handed, 30\% left-handed] within each disorder replica [see Appendix~\ref{app:DiffTypesDisorder}].
In all panels of Fig.~\ref{fig:3F_bands_amo}, we observe a threefold-degenerate spectral feature at ${\bf p}={\bf 0}$ consisting of two sets of linearly dispersing states and one central set of flat-band-like states.
The ${\bf p}={\bf 0}$ spectral feature exhibits spectral broadening with increasing disorder, but nevertheless strongly resembles the crystalline $\Gamma$-point spin-1 chiral multifold fermion in Fig.~\ref{fig:3F_figBulk}(c) for all values of $\eta$ shown in Fig.~\ref{fig:3F_bands_amo}. 
Unlike in the non-crystalline Kramers-Weyl model [Fig.~\ref{fig:KW_Bands_amo}] and like in the non-crystalline double-Weyl model [Fig.~\ref{fig:C2_bands_amo}], the threefold spectral feature at ${\bf p}={\bf 0}$ in the non-crystalline multifold model only weakly varies in energy with increasing $\eta$.
Though not shown in Fig.~\ref{fig:3F_bands_amo}, the spectral function $\bar{A}(E,{\bf p})$ of the non-crystalline multifold model [Eq.~(\ref{eq:amorphous3FTmatrixFinalChirality})] also exhibits nondispersing states at much higher energies for all $\eta$, which arise from disordering the singly-degenerate trivial band at $E/|v| \approx 4.5$ in the crystalline multifold model [Fig.~\ref{fig:3F_figBulk}(b)].

\begin{figure}[t]
\centering   
\includegraphics[width=\linewidth]{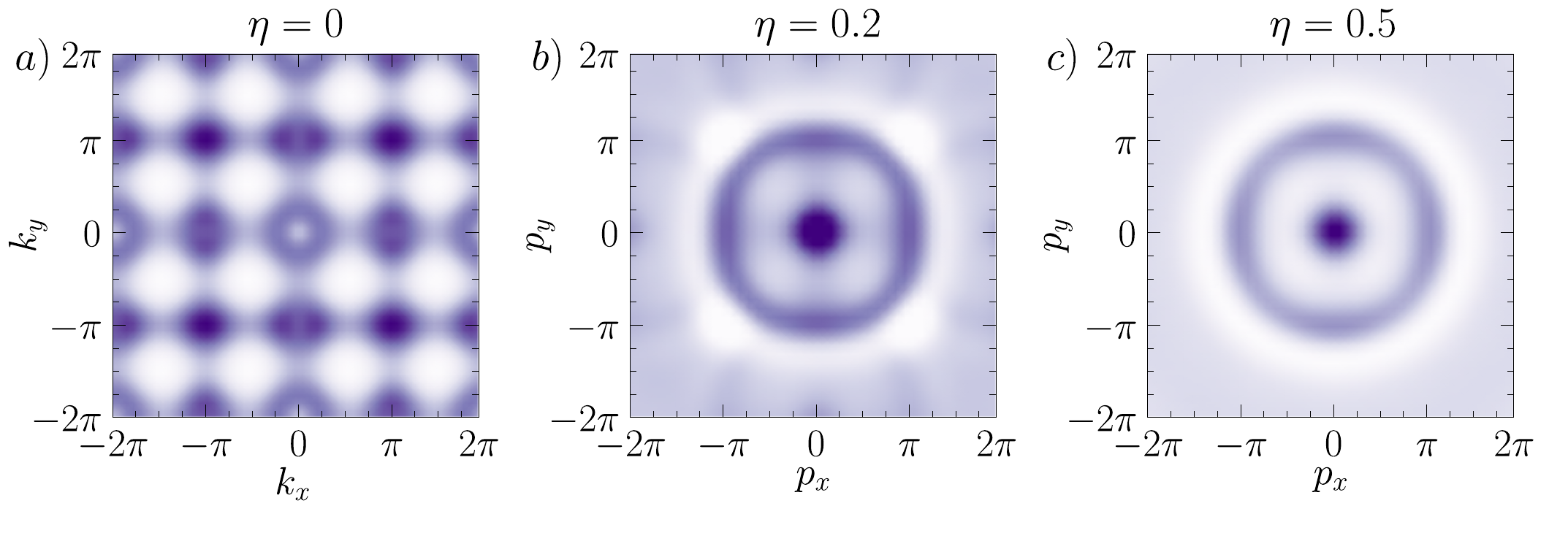}
\caption{Constant-energy spectrum of the disordered multifold fermion model.
(a-c) Constant-energy cuts of the average spectral function $\bar{A}(E,{\bf p})$ [Eq.~(\ref{eq:SpecFunc})] of the non-crystalline multifold model [Eq.~(\ref{eq:amorphous3FTmatrixFinalChirality})] with the parameters in Eq.~(\ref{eq:disordered3Fparams}), contiguous domains of right- and left-handed sites with the respective concentrations $n_R=N_R/N_{\mathrm{sites}}=0.7$ and $n_L=1-N_R/N_{\mathrm{sites}}=0.3$, and increasing random nematic structural and frame disorder parameterized by the standard deviation $\eta$, as detailed in Appendix~\ref{app:lattices}.
To generate each panel, we first compute the momentum-resolved Green's function $\bar{\mathcal{G}}(E,{\bf p})$ at $p_{z}=0.1$ [$k_{z}=0.1$ in (a)] averaged over 20 disorder realizations [replicas] with $20^{3}$ sites each, as detailed in Appendix~\ref{app:PhysicalObservables}.
To account for $\eta$-dependent variations in the energy of the threefold spectral feature at ${\bf p}={\bf 0}$ [see Fig.~\ref{fig:3F_bands_amo}], we first obtain for each $\eta$ in this figure a reference energy $E_{C}$ at which the spectral weight $\bar{A}(E_{C},{\bf p})$ of the threefold feature at ${\bf p}={\bf 0}$ is maximized [\emph{i.e.} the maximum value of $\bar{A}(E_{C},{\bf p})$ away from higher-energy nondispersing states, see Fig.~\ref{fig:3F_figBulk}(b)].
We then compute $\bar{A}(E,{\bf p}) \propto \text{Im}\{\Tr[\bar{\mathcal{G}}(E,{\bf p})]\}$ at $E/v=E_{C}/v - 0.81$ in order to approximately capture the same cross section of the threefold [multifold] nodal degeneracy at ${\bf p}={\bf 0}$ in Fig.~\ref{fig:3F_bands_amo} for varying $\eta$.
In addition to the multifold spectral feature at ${\bf p}={\bf 0}$, the systems with moderate [$\eta=0.2$ in (b)] and strong [$\eta=0.5$ in (c)] disorder exhibit broadened, ring- [sphere-] like spectral features in the vicinity of $|{\bf p}|=\pi/\bar{a}$ and $|{\bf p}|=\pi\sqrt{3}/\bar{a}$ where $\bar{a}=1$ is the average nearest-neighbor spacing.
Unlike in the previous non-crystalline Kramers-Weyl [Fig.~\ref{fig:DOSKW}] and double-Weyl [Fig.~\ref{fig:DOSC2}] models, we find that the ring-like features in (b,c) originate from a more complicated disorder-driven process than rotationally averaging and broadening the bulk chiral fermions at ${\bf k}\neq {\bf 0}$ in the crystalline case [$\eta=0$ in (a), see Figs.~\ref{fig:3F_figBulk}(b) and~\ref{fig:3F_extraCrossings} and the text following Eq.~(\ref{eq:multifoldStructuralChirality})].
We will shortly further clarify the origin of the ring-like features in (b,c) by computing the orbital angular momentum texture in Fig.~\ref{fig:OAMTexture3F}.}
\label{fig:DOS_3F}
\end{figure}

Using the disorder-averaged, momentum resolved Green's function $\bar{\mathcal{G}}(E,{\bf p})$, we can also compute an approximate [effective] single-particle Hamiltonian $\mathcal{H}_{\mathrm{Eff}}(E_{C},{\bf p}) \equiv \mathcal{H}_{\mathrm{Eff}}({\bf p})$~\cite{varjas_topological_2019,marsal_topological_2020,marsal_obstructed_2022} for the spectral features at ${\bf p}={\bf 0}$ in the disordered multifold fermion model [Eq.~(\ref{eq:AvgHEff})].
Though $\mathcal{H}_{\mathrm{Eff}}(E_{C},{\bf p})$ only represents a mean-field approximation of the many-particle, momentum-dependent Hamiltonian, and is therefore generally dependent on the energy cut $E_{C}$ at which it is obtained, we have shown in Appendix~\ref{app:EffectiveHamiltonian} that the nodal degeneracies and topology of $\mathcal{H}_{\mathrm{Eff}}(E_{C},{\bf p})$ near ${\bf p}={\bf 0}$ are numerically stable and surprisingly insensitive to $E_{C}$  in the models studied in this work.
For the non-crystalline multifold model [Eq.~(\ref{eq:amorphous3FTmatrixFinalChirality})], we specifically found in Appendix~\ref{app:EffectiveHamiltonian} that the energy eigenvalues of $\mathcal{H}_{\text{Eff}}(\mathbf{p})$ most closely qualitatively match the low-energy spectrum of $\bar{A}(E,{\bf p})$ in the vicinity of the threefold spectral feature at ${\bf p}={\bf 0}$ [Fig.~\ref{fig:3F_bands_amo}] when $\mathcal{H}_{\text{Eff}}(\mathbf{p})$ is constructed with $E_{C}$ set to lie at the maximum spectral weight of $\bar{A}(E,{\bf p})$ at ${\bf p}={\bf 0}$ away from higher-energy nondispersing spectral features [\emph{i.e.} excluding the higher-energy states that arise from disordering the singly-degenerate trivial band at $E/|v| \approx 4.5$ in Fig.~\ref{fig:3F_figBulk}(b)].
Therefore, in each panel of Fig.~\ref{fig:3F_bands_amo}, we show the bands of $\mathcal{H}_{\mathrm{Eff}}({\bf p})$ in light purple circles with $E_{C}$ centered at the threefold spectral feature in $\bar{A}(E,{\bf p})$ at ${\bf p}={\bf 0}$ to maximize its spectral accuracy.
For all disorder strengths $\eta$, we find that the effective Hamiltonian spectrum in Fig.~\ref{fig:3F_bands_amo} exhibits threefold [multifold] nodal degeneracies at ${\bf p}={\bf 0}$ with two linearly dispersing bands and a central band that is flat to leading order.
Below, we will provide evidence that the threefold spectral feature in Fig.~\ref{fig:3F_bands_amo} at ${\bf p}={\bf 0}$ in $\bar{A}(E,{\bf p})$ [approximated by the threefold nodal degeneracy at ${\bf p}={\bf 0}$ in $\mathcal{H}_{\mathrm{Eff}}({\bf p})]$ is in fact precisely a disordered spin-1 chiral multifold fermion whose topology is controlled by \emph{average} structural chirality, analogous to its crystalline counterpart.

We next construct constant-energy spectral cuts of the disordered multifold fermion tight-binding model to explore spectral features at higher momenta.
In Fig.~\ref{fig:DOS_3F} we plot the average spectral function $\bar{A}(E,{\bf p})$ of the disordered multifold model [Eq.~(\ref{eq:amorphous3FTmatrixFinalChirality})] at fixed $E$ and $p_{z}$ with the parameters in Eq.~(\ref{eq:disordered3Fparams}) and increasing nematic structural and frame disorder averaged over 20 replicas that each contain contiguous domains of right- and left-handed sites with the respective concentrations $n_R=N_R/N_{\mathrm{sites}}=0.7$ and $n_L=1-N_R/N_{\mathrm{sites}}=0.3$.
Like in the disordered Kramers-Weyl model [Figs.~\ref{fig:DOSKW}], $\bar{A}(E,{\bf p})$ in the disordered multifold model develops 3D sphere-like spectral features that appear as 2D rings in the vicinity of $|{\bf p}|=\pi/\bar{a}$ and $|{\bf p}|=\pi\sqrt{3}/\bar{a}$ in the constant-energy and fixed-$p_{z}$ spectral function in Fig.~\ref{fig:DOS_3F}(b,c), where $\bar{a}=1$ is the average nearest-neighbor spacing.
However, unlike in the other non-crystalline models previously analyzed in this work [Figs.~\ref{fig:DOSKW} and~\ref{fig:DOSC2}], we find that the ring-like features in Fig.~\ref{fig:DOS_3F}(b,c) do not simply originate from averaging the bulk ${\bf k}\neq {\bf 0}$ chiral fermions in the crystalline case over random orientations and lattice spacings, because the radii $|{\bf p}|$ of the ring-like features are not in one-to-one correspondence with the crystal momenta magnitudes $|{\bf k}|$ of the crystalline-limit bulk chiral fermions [see Figs.~\ref{fig:3F_figBulk}(b),~\ref{fig:3F_extraCrossings}, and~\ref{fig:DOS_3F} and the text following Eq.~(\ref{eq:multifoldStructuralChirality})].

\paragraph*{\bf Orbital Angular Momentum Texture} -- $\ $
To gain further clarity on the spectral features in Figs.~\ref{fig:3F_bands_amo} and~\ref{fig:DOS_3F}, we next compute the orbital angular momentum [OAM] texture of the non-crystalline multifold model.
By analogy to the spin-dependent spectral function vector in Eq.~(\ref{eq:SpinDOS}), we define the OAM texture in this work by first constructing an \emph{OAM-dependent spectral function vector} $\langle\mathbf{L}(E,\mathbf{p})\rangle$, whose components are given by:
\begin{equation}
     \left\langle L^i \left(E,\mathbf{p}\right)\right\rangle = -\frac{1}{\pi}\text{Im}\left\{\text{Tr}\left[\hat{L}^i \bar{\mathcal{G}}(E,\mathbf{p})\right]\right\} = -\frac{1}{\pi}\text{Im}\left\{\text{Tr}\left[L^i \bar{\mathcal{G}}(E,\mathbf{p})\right]\right\},
\label{eq:OAMDOS}
\end{equation}
where $\hat{L}^i$ is the $i$-th component of the OAM operator vector $\hat{\bf L}=(\hat{L}^x,\hat{L}^y,\hat{L}^z)$, $L^{i}$ is the matrix representative of the OAM component operator $\hat{L}^{i}$ acting on the internal $s$ and $p_{x,y,z}$ orbital degrees of freedom on each site, $\bar{\mathcal{G}}(E,{\bf p})$ is the one-momentum average Green's function [Eq.~(\ref{eq:averageOneMomentumGreen})], and where the trace operation is taken over the internal orbital degrees of freedom parameterized by the $4\times 4$ matrices $\tau^{i}\mu^{j}$ in Eqs.~(\ref{eq:multifoldBandV}),~(\ref{eq:PauliMatrixProdDef}), and~(\ref{eq:amorphous3FTmatrixFinalChirality}).
We may then use Eqs.~(\ref{eq:SpecFunc}) and~(\ref{eq:OAMDOS}) to define the \emph{degree} of OAM polarization $P_{\bf L}(E,{\bf p})$ for a spectral feature at a given $E$ and ${\bf p}$:
\begin{equation}
P_{\bf L}(E,{\bf p}) = \frac{\left|\left\langle\mathbf{L} \left(E,\mathbf{p}\right)\right\rangle\right|}{\bar{A}(E,{\bf p})} =  \frac{\sqrt{\sum\limits_{i=x,y,z}\langle L^{i}(E,{\bf p})\rangle^{2}}}{\bar{A}(E,{\bf p})} = -\frac{\sqrt{\sum\limits_{i=x,y,z}\left(\text{Im}\left\{\Tr\left[\ L^{i}\bar{\mathcal{G}}(E,\mathbf{p})\right]\right\}\right)^{2}}}{\text{Im}\left\{\Tr\left[\bar{\mathcal{G}}(E,\mathbf{p})\right]\right\}}.
\label{eq:3FOAMPolarization}
\end{equation}
In Eq.~(\ref{eq:3FOAMPolarization}), $P_{\bf L}(E,{\bf p})$ takes values between $0$ and $1$, allowing us to designate spectral features with $P_{\bf L}(E,{\bf p})\approx 1$ as ``highly OAM-polarized.''
We note that in the literature, there exist several distinct definitions of OAM -- of these, Eqs.~(\ref{eq:OAMDOS}) and~(\ref{eq:3FOAMPolarization}) are most closely related to the ``local'' OAM probed in circular dichroism angle-resolved photoemission spectroscopy [CD-ARPES], as opposed to the ``global'' OAM, which is instead related to the Berry curvature~\cite{OAMCDArpes1,OAMmultifold2,OAMnodalLine,OAMzxShenARPES3DTI}.

In the basis of the $s$ and $p_{x,y,z}$ orbitals on each site of the multifold fermion model [see Eq.~(\ref{eq:Amo3F}) and the surrounding text], the matrix representatives $L^{x,y,z}$ of the OAM component operators $\hat{L}^{x,y,z}$ in Eqs.~(\ref{eq:OAMDOS}) and~(\ref{eq:3FOAMPolarization}) are given by:
\begin{equation}
    L^x = \begin{pmatrix}
    0&0&0&0\\
    0&0&0&0\\
    0&0&0&-i\\
    0&0&i&0
    \end{pmatrix}, \quad
    L^y = \begin{pmatrix}
    0&0&0&0 \\
    0&0&0&i\\
    0&0&0&0\\
    0&-i&0&0
    \end{pmatrix}, \quad
    L^z = \begin{pmatrix}
    0&0&0&0\\
    0&0&-i&0\\
    0&i&0&0\\
    0&0&0&0
    \end{pmatrix},
\label{eq:OAMlMatricies}
\end{equation}
where the form of each $L^{i}$ originates from the direct sum of an angular-momentum-zero [spinless $s$-orbital] and an angular-momentum-one [spinless $p$-orbital] representation of $\hat{\bf L}$~\cite{GriffithsBook}.
The OAM component matrix representatives $L^{x,y,z}$ in Eq.~(\ref{eq:OAMlMatricies}) correspondingly satisfy the algebra:
\begin{equation}
\left[L^a,L^b\right] = i \epsilon^{abc}L^c,
\end{equation}
where $\epsilon^{abc}$ is the Levi-Civita tensor and $a$, $b$, and $c$ are drawn from $x$, $y$, and $z$.
$L^{x,y,z}$ also satisfy the total angular momentum summation relation:
\begin{equation}
L^{2} = \sum_{a=x,y,z}\left(L^{a}\right)^{2} = \begin{pmatrix}
    0&0&0&0\\
    0&2&0&0\\
    0&0&2&0\\
    0&0&0&2
    \end{pmatrix} = \begin{pmatrix}
    l_{s}(l_{s}+1)&0&0&0\\
    0&l_{p}(l_{p}+1)&0&0\\
    0&0&l_{p}(l_{p}+1)&0\\
    0&0&0&l_{p}(l_{p}+1)
    \end{pmatrix},
\end{equation}
where $l_{s}=0$ and $l_{p}=1$ are respectively the azimuthal [\emph{i.e.} OAM] quantum numbers of the $s$ and $p_{x,y,z}$ orbital basis states [see Refs.~\cite{GriffithsBook,mcQuarriePchem} and the text surrounding Eq.~(\ref{eq:Amo3F})].

\begin{figure}[t]
\centering   
\includegraphics[width=\linewidth]{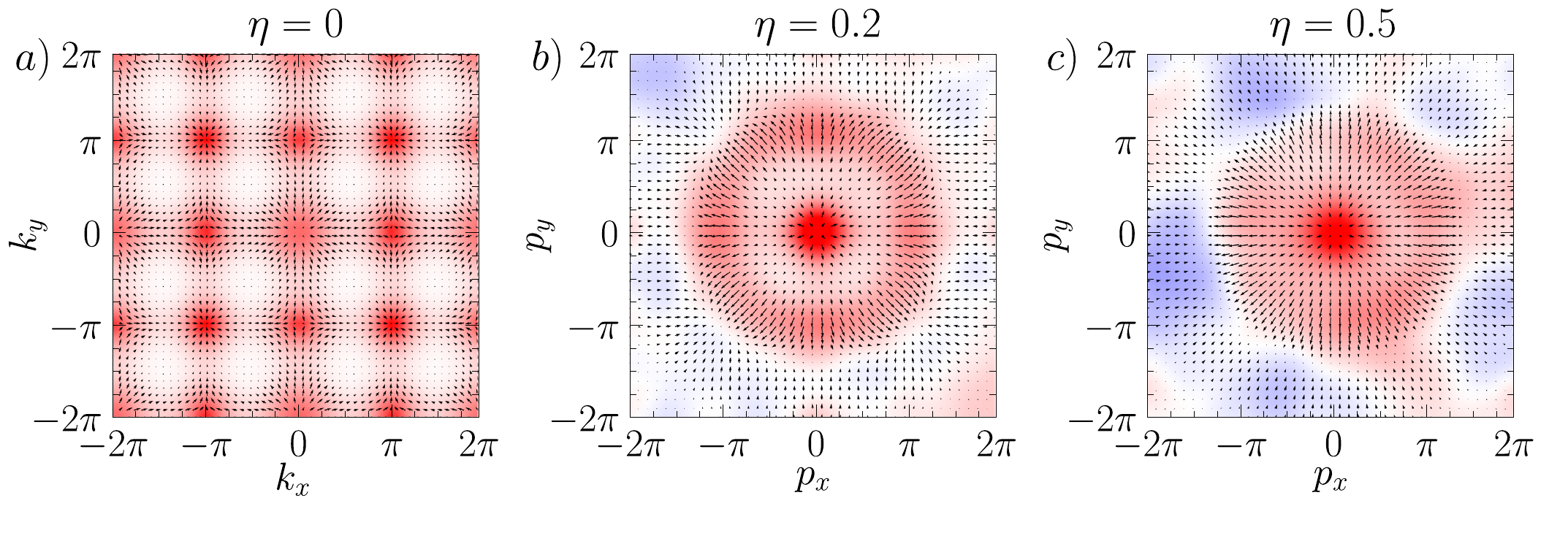}
\caption{Orbital angular momentum texture of the disordered multifold fermion model.
(a-c) The orbital angular momentum [OAM] texture [Eqs.~\eqref{eq:OAMDOS} and~(\ref{eq:OAMlMatricies})] of the non-crystalline chiral multifold systems in Fig.~\ref{fig:DOS_3F}(a-c), respectively.
In all panels, the in-plane components of the OAM texture $\langle L^{x,y}(E,\mathbf{p})\rangle$ are represented as arrows with log-scale lengths, while the out-of-plane component $\langle L^{z}(E,\mathbf{p})\rangle$ is represented through a log-scale color map in which red is positive and blue is negative.
(a) In the structurally right-handed crystalline limit [Figs.~\ref{fig:3F_figBulk}(b,c) and~\ref{fig:3F_extraCrossings}], the Fermi pockets near $k_{z}=0$ originate from a spin-1 chiral multifold fermion at $\Gamma$ [$k_{x}=k_{y}=0$] with a chiral charge of $C_{\Gamma}=2$ and clusters of conventional Weyl fermions in the vicinity of $X$ [$k_{x}=\pi$, $k_{y}=0$] and $M$ [$k_{x}=k_{y}=\pi$] with the respective chiral charges $C_{X}=-C_{M}=-2$.
We observe that the crystalline chiral multifold fermion at $\Gamma$ and the conventional Weyl fermion clusters at $X$ and $M$ in (a) exhibit nearly perfect monopole-like OAM textures in which the sign of the monopole charge is locked to sign of the chiral charge $C$.
(b,c) As the disorder scale $\eta$ is increased, the chiral fermions away from ${\bf p}={\bf 0}$ become merged into ring- [sphere-] like spectral features with largely isotropic OAM textures.
However, unlike previously in the non-crystalline Kramers-Weyl model [Fig.~\ref{fig:SpinTextureKW}(b,c)], the radii $|{\bf p}|$ and angular momentum textures of the ring-like features in (b,c) are not in one-to-one correspondence with the momenta and angular momentum textures of the chiral fermions in the $\eta=0$ crystalline limit in (a).
This likely occurs because in addition to the symmetry-enforced chiral multifold fermions pinned in the crystalline limit to $\Gamma$ and $R$ [$k_{x}=k_{y}=k_{z}=\pi$] with $C_{\Gamma}=-C_{R}=2$ [see Figs.~\ref{fig:3F_figBulk}(b) and~\ref{fig:3F_extraCrossings} and text surrounding Eq.~(\ref{eq:appHKPmultifold})], the crystalline-limit system also contains numerous conventional Weyl fermions at \emph{unpinned} ${\bf k}$, which may more easily be moved and coupled by disorder.
We specifically observe that the ${\bf k}\neq{\bf 0}$ chiral fermions in (a) merge in (b,c) into a ring-like feature at $|{\bf p}|=\pi$ with an outward-pointing OAM texture and high spectral weight and a ring-like feature at $|{\bf p}|=\pi\sqrt{3}$ with an inward-pointing OAM texture and relatively weaker spectral weight.
While the specific origin of the outward-pointing ring-like feature at $|{\bf p}|=\pi$ is more uncertain, the inward-pointing ring-like feature at $|{\bf p}|=\pi\sqrt{3}$ likely originates from the negatively charged multifold fermion at $R$ in the crystalline limit, and may hence represent a many-particle disordered chiral multifold fermion with the opposite chiral charge as the multifold nodal degeneracy at ${\bf p}={\bf 0}$.
We leave a detailed investigation of this intriguing possibility for future works.}
\label{fig:OAMTexture3F}
\end{figure}

In Fig.~\ref{fig:OAMTexture3F}, we plot the OAM texture [Eqs.~(\ref{eq:OAMDOS}) and~(\ref{eq:OAMlMatricies})] of the disordered multifold tight-binding model with the same parameters previously employed to generate the constant-energy spectra in Fig.~\ref{fig:DOS_3F}.
Previous theoretical studies and circular dichroism photoemission experiments have shown that crystalline chiral fermions can exhibit monopole-like OAM textures~\cite{OAMWeyl2,OAMmultifold1,OAMmultifold2,OAMmultifold3,OAMmultifold4,OAMmultifold5}.
We correspondingly observe that in the crystalline [$\eta=0$] limit in Fig.~\ref{fig:OAMTexture3F}(a), the multifold fermion at $\Gamma$ [$k_{x}=k_{y}=k_{z}=0$] exhibits a nearly perfect monopole-like OAM texture with outward-pointing $\langle\mathbf{L}(E,\mathbf{p})\rangle$ at all ${\bf p}$ near ${\bf p}={\bf 0}$.
In addition to symmetry-enforced multifold fermions at $\Gamma$ and $R$ [$k_{x}=k_{y}=k_{z}=\pi$] with the respective chiral charges $C_{\Gamma}=-C_{R}=2$, the right-handed enantiomer of the crystalline multifold model exhibits clusters of conventional Weyl fermions in the close vicinity of $X$ [$k_{x}=\pi$, $k_{y}=k_{z}=0$] and $M$ [$k_{x}=k_{y}=\pi$, $k_{z}=0$] with the respective net chiral charges $C_{X}=-C_{M}=-2$ [see Figs.~\ref{fig:3F_figBulk}(b) and~\ref{fig:3F_extraCrossings} and text surrounding Eq.~(\ref{eq:multifoldStructuralChirality})].
We observe in Fig.~\ref{fig:OAMTexture3F}(a) that the Weyl clusters near $X$ and $M$ also exhibit nearly perfect monopole-like OAM textures for which the sign of the OAM-texture monopole charge matches the net chiral charge of each Weyl-point cluster.
Specifically in Fig.~\ref{fig:OAMTexture3F}(a), the $C_{X}=-2$ Weyl clusters near $(k_{x},k_{y})=(\pi,0)$ and $(k_{x},k_{y})=(0,\pi)$ exhibit OAM textures that respectively point inward in the $k_{x}$- and $k_{y}$-directions and outward along the other two principal axes, and the $C_{M}=+2$ Weyl cluster near $k_{x}=k_{y}=\pi$ exhibits an inward-pointing OAM texture in the $k_{x}$- and $k_{y}$-directions and an outward-pointing OAM texture in the $k_{z}$-direction.
This can be understood by recalling that the clusters of conventional Weyl fermions near $X$ and $M$ in the crystalline multifold model with the tight-binding parameters in Eq.~(\ref{eq:appTBparams3F}) are perturbatively related to spin-1 chiral multifold fermions that respectively carry the opposite and same chiral charges as the symmetry-enforced multifold fermion at $\Gamma$ [see the text following Eq.~(\ref{eq:multifoldStructuralChirality})].
For all values of $\eta$, the central spectral feature -- which corresponds to the nodal degeneracy at ${\bf p}={\bf 0}$ in Fig.~\ref{fig:3F_bands_amo} -- retains its nearly perfect outward-pointing, monopole-like OAM texture. The topological similarity between the OAM textures near ${\bf p}={\bf 0}$ at increasing $\eta$ in Fig.~\ref{fig:OAMTexture3F} suggests a picture [which we will shortly use Wilson loops to make precise] in which the threefold [multifold] nodal degeneracy at ${\bf p}={\bf 0}$ in Figs.~\ref{fig:3F_bands_amo}(d),~\ref{fig:DOS_3F}(c), and~\ref{fig:OAMTexture3F}(c) is a strongly disordered chiral multifold fermion with a low-energy sense of handedness [OAM-texture monopole charge] that is inherited from the lattice-scale chirality imbalance [\emph{i.e.} average chirality] $n_{R}>n_{L}$ of the disordered system.

\begin{figure}[t]
\centering
\includegraphics[width=\linewidth]{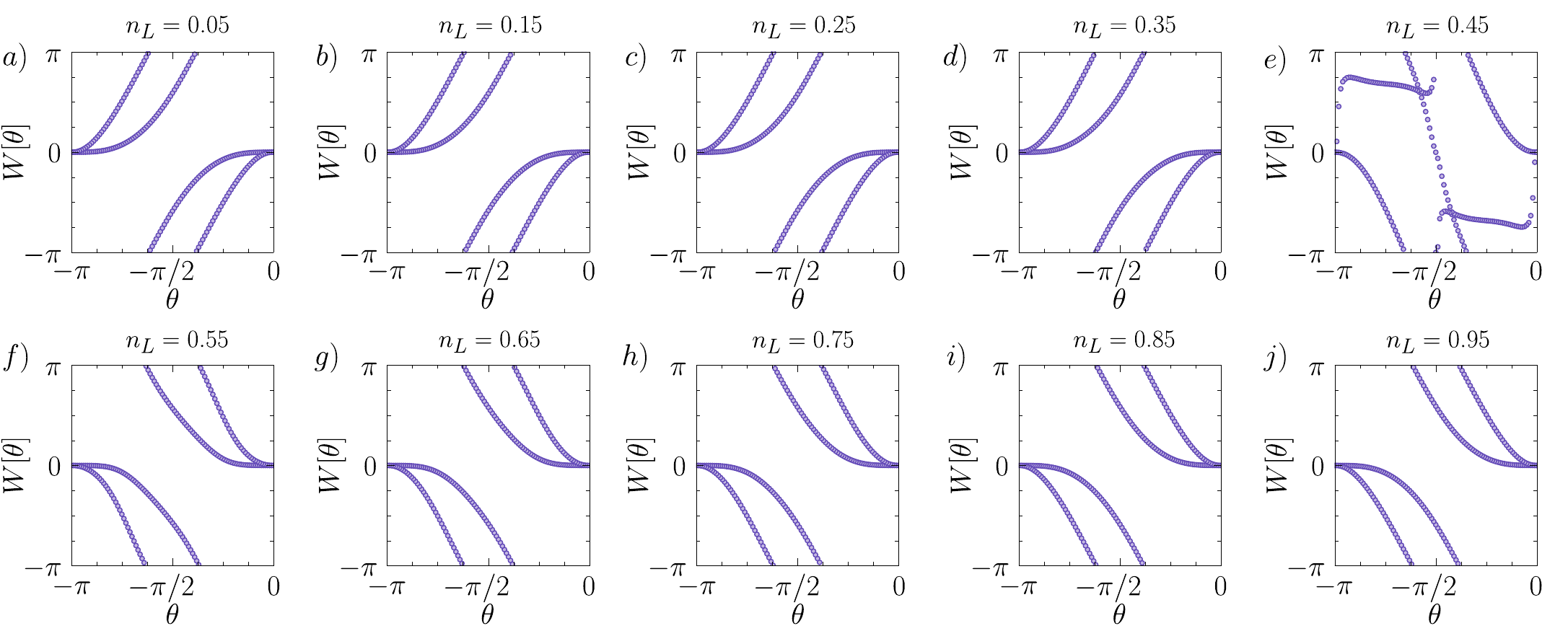}
\caption{Sphere Wilson loop spectrum of the disordered multifold model for varying chirality concentrations.
To generate each panel in this figure, we place the non-crystalline chiral multifold model [Eq.~(\ref{eq:amorphous3FTmatrixFinalChirality})] with the parameters in Eq.~(\ref{eq:disordered3Fparams}) on a lattice with $N_{\mathrm{sites}}=15^3$ and random nematic Gaussian structural and local frame disorder parameterized by the fixed standard deviation $\eta=0.2$, as detailed in Appendix~\ref{app:lattices}. 
We then generate the replica-averaged momentum-resolved Green's function $\bar{\mathcal{G}}(E,\mathbf{p})$ [Eq.~(\ref{eq:averageOneMomentumGreen})] by averaging the system over 50 disorder realizations [replicas] that each contain contiguous domains of right- and left-handed sites with the respective concentrations $n_R=N_R/N_{\mathrm{sites}}$ and $n_{L}=1-n_{R}$.
(a-j) For 10 disorder ensembles with increasing values of $n_{L}$, we construct for each ensemble an effective Hamiltonian $\mathcal{H}_{\mathrm{Eff}}({\bf p})=\mathcal{H}_{\mathrm{Eff}}(E_{C},{\bf p})$ [Eq.~(\ref{eq:AvgHEff})] using a reference energy $E_{C}$ corresponding to the largest spectral weight $\bar{A}(E,{\bf p})$ [Eq.~(\ref{eq:SpecFunc})] at ${\bf p}={\bf 0}$ [excluding the higher-energy states that arise from disordering the singly-degenerate trivial band at $E/|v| \approx 4.5$ in Fig.~\ref{fig:3F_figBulk}(b)]. 
As shown in Fig.~\ref{fig:H_eff_benchmark}(c) and the surrounding text, this procedure maximizes the spectral accuracy of $\mathcal{H}_{\mathrm{Eff}}({\bf p})$ near ${\bf p}={\bf 0}$ in the non-crystalline multifold model.
We next use the lower two eigenstates of $\mathcal{H}_{\mathrm{Eff}}({\bf p})$ in energy to compute the two-band [non-Abelian] amorphous Wilson loop spectrum introduced in this work [Appendix~\ref{sec:WilsonBerry}] on a sphere surrounding the nodal degeneracy at ${\bf p}={\bf 0}$ in each disorder ensemble.
Beginning with a moderately disordered system [$\eta=0.2$, see Fig.~\ref{fig:3DGreen3F}] with (a) almost entirely right-handed sites [$n_{L}=0.05$] and continuing in increasing $n_{L}$ to a system (j) with almost entirely left-handed sites [$n_{L}=0.95$], we observe that the two Wilson loop eigenvalues as a set exhibit quantized winding numbers as functions of the sphere polar angle $\theta$ of (a-d) $C=2$ for $n_{L}<0.5$ and (f-j) $C=-2$ for $n_{L}>0.5$, outside of (e) a region in the vicinity of $n_{L}\approx 0.5$ with a non-smooth Wilson spectrum.
This provides a precise indicator that the nodal degeneracy at ${\bf p}={\bf 0}$ in Fig.~\ref{fig:3F_bands_amo} is a disordered [non-crystalline] chiral multifold fermion that, analogous to its crystalline counterpart~\cite{AlPtObserve,PdGaObserve,DingARPESReversal,SessiPdGaQPIReversal}, carries a topological chirality that is tunable via the average system structural chirality.}
\label{fig:3F_WL}
\end{figure}

Interestingly, as $\eta$ is increased while keeping $n_{R,L}$ constant, we find that the OAM texture [anti]monopoles become broadened and merged into 3D sphere-like spectral features with largely isotropic OAM textures [which appear as 2D rings in the constant-energy, fixed-$p_{z}$ OAM-dependent spectra in Fig.~\ref{fig:OAMTexture3F}(b,c)].
Unlike previously in the non-crystalline Kramers-Weyl model [Fig.~\ref{fig:SpinTextureKW}(b,c)], however, lattice disorder in the multifold model significantly rearranges the locations of features in the spectral function $\bar{A}(E,{\bf p})$ [Eq.~(\ref{eq:SpecFunc})] and angular momentum texture [Eq.~(\ref{eq:OAMDOS})] throughout the broadening and merging process.
Specifically, in the crystalline limit in Figs.~\ref{fig:DOS_3F}(a) and~\ref{fig:OAMTexture3F}(a), the bulk Fermi pockets originate from symmetry-enforced chiral multifold fermions at $\Gamma$ [$|{\bf k}|=0$] and $R$ [$|{\bf k}|=\pi\sqrt{3}$] and conventional Weyl fermions that lie at unpinned momenta ${\bf k}$ within the close vicinity of $X$ [$|{\bf k}|=\pi$] and $M$ [$|{\bf k}|=\pi\sqrt{2}$].
However, in the moderately and strongly disordered constant-energy cuts of $\bar{A}(E,{\bf p})$ and $\langle\mathbf{L}(E,\mathbf{p})\rangle$ respectively shown in Figs.~\ref{fig:DOS_3F}(b,c) and~\ref{fig:OAMTexture3F}(b,c), ring-like spectral features only appear at $|{\bf p}|=\pi$ and $|{\bf p}|=\pi\sqrt{3}$ [with the latter exhibiting considerably weaker spectral weight], and are absent at $|{\bf p}|=\pi\sqrt{2}$.
Furthermore, the ring-like feature at $|{\bf p}|=\pi$ in Fig.~\ref{fig:OAMTexture3F}(b,c) exhibits an outward-pointing OAM texture suggestive of a positive chiral charge, whereas the crystalline chiral fermions at $|{\bf k}|=\pi$ in Fig.~\ref{fig:OAMTexture3F}(a) carry net-negative chiral and OAM-texture monopole charges.
Conversely, the ring-like spectral feature at $|{\bf p}|=\pi\sqrt{3}$ in Fig.~\ref{fig:OAMTexture3F}(b,c) exhibits an inward pointing OAM texture that is consistent with the negative chiral charge of the chiral multifold fermion at $R$ in the crystalline limit [Figs.~\ref{fig:3F_figBulk}(b) and~\ref{fig:3F_extraCrossings}].
Overall, this suggests that the Weyl fermions at unpinned [free] momenta ${\bf k}$ near $X$ and $M$ in the crystalline limit are rearranged and coupled by lattice disorder, whereas the symmetry-pinned, negatively charged multifold fermion at $R$ becomes rotationally averaged by disorder into a sphere-like feature that remains centered at $|{\bf p}|=\pi\sqrt{3}$.
Because the energy spectrum is more diffuse at larger momenta [Fig.~\ref{fig:HeffBreak}(a)], and because we have only shown that the effective Hamiltonian method is numerically stable near ${\bf p}={\bf 0}$ [Appendix~\ref{app:EffectiveHamiltonian}], we are unable to directly apply the average-symmetry-group representation theory [Appendices~\ref{app:pseudoK} and~\ref{app:corepAmorphous}] and amorphous Wilson loop method [Appendix~\ref{sec:WilsonBerry}] introduced in this work to diagnose the topology of the disorder-broadened, ring-like spectral features in Figs.~\ref{fig:DOS_3F}(b,c) and~\ref{fig:OAMTexture3F}(b,c).
Nevertheless, the OAM textures in Fig.~\ref{fig:OAMTexture3F} suggest the intriguing possibility, which we leave for exploration in future works, that the ring-like spectral feature at $|{\bf p}|=\pi\sqrt{3}$ in Figs.~\ref{fig:DOS_3F}(c) and~\ref{fig:OAMTexture3F}(c) represents a many-particle disordered chiral multifold fermion with the opposite chiral charge as the multifold nodal degeneracy at ${\bf p}={\bf 0}$.
We will shortly show in this section [Fig.~\ref{fig:3F_Unif}] that the appearance of isotropic spectral features at $|{\bf p}|=\pi$ and $|{\bf p}|=\pi\sqrt{3}$ is a more generic feature of the multifold model with the tight-binding parameters in Eq.~(\ref{eq:disordered3Fparams}) and strong structural disorder, and also appears for disorder implementations [\emph{e.g.} random lattices, see Appendix~\ref{app:DiffTypesDisorder}] that cannot be deformed to uniquely defined crystalline limits.

\paragraph*{\bf Wilson Loops} -- $\ $
Having shown that the non-crystalline multifold model [Eq.~(\ref{eq:amorphous3FTmatrixFinalChirality})] continues to exhibit a threefold nodal degeneracy at ${\bf p}={\bf 0}$ [Fig.~\ref{fig:3F_bands_amo}(c,d)] with a monopole-like OAM texture [Fig.~\ref{fig:OAMTexture3F}(c)] for strong, chirality-imbalanced disorder, we will now use an amorphous generalization of the Wilson loop method [see Appendix~\ref{sec:WilsonBerry} and Refs.~\cite{Fidkowski2011,AndreiXiZ2,ArisInversion,Cohomological,HourglassInsulator,DiracInsulator,Z2Pack,BarryFragile,AdrienFragile,HOTIBernevig,HingeSM,WiederAxion,KoreanFragile,ZhidaBLG,TMDHOTI,KooiPartialNestedBerry,PartialAxionHOTINumerics,GunnarSpinFragileWilson,BinghaiOscillationWilsonLoop,Wieder22}] to precisely show that the ${\bf p}={\bf 0}$ nodal degeneracy is a disordered [non-crystalline] spin-1 chiral multifold fermion whose lower two bands together carry a nontrivial topological chiral charge.

To begin, in a crystalline system, the chiral charge of a nodal degeneracy can be obtained by computing the winding number of the Wilson loop spectrum [non-Abelian Berry phases] evaluated over the occupied bands on a sphere surrounding the nodal degeneracy~\cite{Z2Pack}.
At each value of the polar angle $\theta$ of the sphere [see Fig.~\ref{fig:Wilson_schema}(b)], the Wilson loop is computed using the Bloch wavefunctions of the occupied states.
To perform an analogous calculation for a given disorder ensemble of the non-crystalline chiral multifold model [Eq.~(\ref{eq:amorphous3FTmatrixFinalChirality})], we therefore first construct an effective Hamiltonian $\mathcal{H}_{\text{Eff}}(\mathbf{p})$ [see Refs.~\cite{varjas_topological_2019,marsal_topological_2020,marsal_obstructed_2022} and Appendix~\ref{app:EffectiveHamiltonian}] for the nodal degeneracy at ${\bf p}={\bf 0}$ using the replica-averaged momentum-resolved Green's function $\bar{\mathcal{G}}(E,\mathbf{p})$ [Eq.~(\ref{eq:averageOneMomentumGreen})]. 
For each disorder ensemble of the non-crystalline multifold model, we specifically first identify an energy $E_{max}$ where $\bar{A}(E,{\bf p})$ is largest at ${\bf p}={\bf 0}$ [excluding the higher-energy states that arise from disordering the singly-degenerate trivial band at $E/|v| \approx 4.5$ in Fig.~\ref{fig:3F_figBulk}(b)].
We then construct $\mathcal{H}_{\text{Eff}}(\mathbf{p})$ using the reference energy cut $E_{C}=E_{max}$, which maximizes the spectral accuracy of $\mathcal{H}_{\text{Eff}}(\mathbf{p})$ in the multifold model [see Fig.~\ref{fig:H_eff_benchmark}(c) and the surrounding text].
Finally, we use the lower two eigenstates of $\mathcal{H}_{\mathrm{Eff}}({\bf p})$ in energy to compute the non-Abelian amorphous [disordered] Wilson loop spectrum [Appendix~\ref{sec:WilsonBerry}] on a sphere surrounding the nodal degeneracy at ${\bf p}={\bf 0}$.

In Fig.~\ref{fig:3F_WL}, we show the sphere Wilson loop spectrum for the disordered chiral multifold model for 10 disorder ensembles with 50 replicas each, where each disorder replica has $N_{\mathrm{sites}}=15^{3}=3375$ sites, nematic Gaussian lattice and local frame disorder with the standard deviation $\eta=0.2$ [see Appendix~\ref{app:lattices}], and contiguous domains of right- and left-handed sites with varying chirality concentrations respectively given by $n_R=N_R/N_{\mathrm{sites}}$ and $n_{L}=1-n_{R}$.
Beginning in Fig.~\ref{fig:3F_WL}(a) with a moderately disordered system [$\eta=0.2$, see Fig.~\ref{fig:3DGreen3F}] containing almost entirely right-handed sites [$n_{L}=0.05$] and continuing in increasing $n_{L}$ to the system in Fig.~\ref{fig:3F_WL}(j) with almost entirely left-handed sites [$n_{L}=0.95$], we observe that the two Wilson loop eigenvalues as a set exhibit quantized winding numbers of $C=2$ for $n_{L}<0.5$ [Fig.~\ref{fig:3F_WL}(a-d)] and $C=-2$ for $n_{L}>0.5$ [Fig.~\ref{fig:3F_WL}(f-j)].
In the vicinity of $n_{L}\approx 0.5$, the Wilson loop eigenvalues become non-smooth, indicating that the sphere Wilson loop is within the close vicinity of a topological quantum critical point [energy gap closure between the second and third effective Hamiltonian bands]; the onset of this behavior can be seen in Fig.~\ref{fig:3F_WL}(e).
The Wilson loop spectra in Fig.~\ref{fig:3F_WL} overall represent a central result of the present work, as they for the first time provide precise \emph{quantized} indicators of nodal [gapless] topology in fully structurally disordered 3D metals, and represent the first calculation of non-Abelian Berry phases [multiband Wilson loop eigenvalues] in a 3D amorphous system.  
In the case of the disordered multifold model introduced in this section [Eq.~(\ref{eq:amorphous3FTmatrixFinalChirality})], the Wilson loop spectra in Fig.~\ref{fig:3F_WL} specifically indicate that the nodal degeneracy at ${\bf p}={\bf 0}$ in Fig.~\ref{fig:3F_bands_amo} is a disordered spin-1 chiral multifold fermion that, analogous to its crystalline counterpart~\cite{AlPtObserve,PdGaObserve,DingARPESReversal,SessiPdGaQPIReversal}, carries a topological chirality that is tunable via the average system structural chirality.
Though we have only demonstrated quantized and tunable two-band Wilson loop winding in Fig.~\ref{fig:3F_WL} for multifold systems with moderate structural disorder [$\eta=0.2$], we will soon below show that the link between average structural chirality and low-energy topological chirality also holds for the non-crystalline multifold model [Eq.~(\ref{eq:amorphous3FTmatrixFinalChirality})] on fully-disordered [random] lattices that lack well-defined crystalline limits [see Fig.~\ref{fig:3F_Unif}].

\paragraph*{\bf Non-Crystalline Group Theory} -- $\ $ We will next make further direct connection between the disordered chiral multifold fermion in Fig.~\ref{fig:3F_bands_amo} and its crystalline counterpart by employing symmetry group theory.  
First, as discussed in Eq.~(\ref{eq:appHKPmultifold}) and the surrounding text, when Eq.~(\ref{eq:amorphous3FTmatrixFinalChirality}) is placed on a regular cubic lattice and expanded in crystal momentum ${\bf k}$ about the $\Gamma$ point [${\bf k}={\bf 0}$], the resulting ${\bf k}\cdot{\bf p}$ Hamiltonian takes the form:
\begin{equation}
\mathcal{H}({\bf k}) = v\left(k_x\tau^y + k_y\tau^z\mu^y + k_z\tau^x\mu^y\right) + m'\left(\mu^z+\tau^z\mu^z + \tau^z\right),
\label{eq:3FnumericsKP}
\end{equation}
where $m'=m + t_{1}+t_{2}+t_{3}$ in Eq.~(\ref{eq:appHKPmultifold}).
As discussed in the text following Eqs.~(\ref{eq:multifoldMidentity}) and~(\ref{eq:appHKPmultifold}), the energy eigenvalues of the ${\bf k}\cdot{\bf p}$ Hamiltonian in Eq.~(\ref{eq:3FnumericsKP}) at ${\bf k}={\bf 0}$ consist of a singly degenerate trivial state at $E=3m'$ and a threefold degeneracy at $E=-m'$ that corresponds to the node of a spin-1 chiral multifold fermion~\cite{ManesNewFermion,chang2017large,tang2017CoSi,KramersWeyl,DoubleWeylPhonon,CoSiObserveJapan,CoSiObserveHasan,CoSiObserveChina,AlPtObserve,PdGaObserve,PtGaObserve,AltlandSpin1Light,DingARPESReversal,ZahidLadderMultigap,Sanchez2023}.
The little group $G_{\Gamma}$ at the $\Gamma$ point [see Eq.~(\ref{eq:EquivKPoints})] is isomorphic to the system SG itself [cubic SG 195 ($P23)$], such that the single degeneracy transforms in a one-dimensional, single-valued small corep of $G_{\Gamma}$ and the threefold degeneracy transforms in a three-dimensional, single-valued small corep of $G_{\Gamma}$.
As established in Refs.~\cite{zallen_physics_1998,VanMechelen:2018cy,vanMechelenNonlocal,Ciocys2023,Grushin2020,Corbae_2023} and discussed in Appendices~\ref{app:pseudoK} and~\ref{app:corepAmorphous}, systems with strong structural disorder are spectrally isotropic at long wavelengths in their $d_{A}$ disordered ${\bf p}$-space directions [Eq.~(\ref{eq:finiteDimAmorph})].
Unlike the $\Gamma$-point ${\bf k}\cdot{\bf p}$ Hamiltonians for the Kramers-Weyl [Eq.~(\ref{eq:KWnumericsKP})] and double-Weyl [Eq.~(\ref{eq:C2numericsKP})] models, the multifold model ${\bf k}\cdot{\bf p}$ Hamiltonian is \emph{already} spectrally isotropic in 3D ${\bf k}$ space when expanded to linear order [Eq.~(\ref{eq:3FnumericsKP})].
Hence, unlike the previous topological semimetal ${\bf k}\cdot{\bf p}$ models in Eqs.~(\ref{eq:KWnumericsKP}) and~(\ref{eq:C2numericsKP}), we do not need to tune model parameters in Eq.~(\ref{eq:3FnumericsKP}) to apply the long-wavelength deformation and averaging procedure for disordered $\Gamma$-point Hamiltonians detailed in Eq.~(\ref{eq:SG16rotations}) and the following text.
Instead, we may simply reinterpret [recast] Eq.~(\ref{eq:3FnumericsKP}) as the approximate, isotropic, \emph{effective} ${\bf p}\approx{\bf 0}$ ${\bf k}\cdot {\bf p}$ Hamiltonian of the non-crystalline multifold model [Eq.~(\ref{eq:amorphous3FTmatrixFinalChirality})] on a $d_{A}=3$ strongly disordered or random lattice:
\begin{equation}
\mathcal{H}_{\text{Eff}}({\bf p}) = \tilde{v}\left(k_x\tilde{\tau}^y + k_y\tilde{\tau}^z\tilde{\mu}^y + k_z\tilde{\tau}^x\tilde{\mu}^y\right) + \tilde{m}'\left(\tilde{\mu}^z+\tilde{\tau}^z\tilde{\mu}^z + \tilde{\tau}^z\right) + \tilde{\mu}'\tilde{\tau}^{0}\tilde{\mu}^{0},
\label{eq:3FdisorderKP}
\end{equation}
where $\tilde{v}$ indicates the strength of disorder-renormalized OAM coupling [see the text following Eq.~(\ref{eq:HamBloch3F})], $\tilde{m}'$ is the disorder-renormalized crystal field splitting at ${\bf p}={\bf 0}$ [Eq.~(\ref{eq:multifoldMidentity})], $\tilde{\mu}'$ is a disorder-renormalized chemical potential that may be present depending on the details of the disorder implementation, and $\tilde{\tau}^{0}\tilde{\mu}^{0}$ is the $4\times 4$ identity matrix.
In Eq.~(\ref{eq:3FdisorderKP}), $\tilde{\tau}^{y}$, $\tilde{\tau}^{z}\tilde{\mu}^{y}$, and $\tilde{\tau}^{x}\tilde{\mu}^{y}$ are generically each equal to a linear combination of $\tau^{y}$, $\tau^{z}\mu^{y}$, and $\tau^{x}\mu^{y}$ in Eq.~(\ref{eq:3FnumericsKP}) due to OAM reference frame disorder [see the text surrounding Eq.~(\ref{appeq:rota3F})], and $\tilde{\mu}^{z}$, $\tilde{\tau}^{z}\tilde{\mu}^{z}$, and $\tilde{\tau}^{z}$ are generically each equal to a linear combination of $\mu^{z}$, $\tau^{z}\mu^{z}$, and $\tau^{z}$ in Eq.~(\ref{eq:3FnumericsKP}) due to achiral orbital hopping mixing, which is also implemented by Eq.~(\ref{appeq:rota3F}) via the rotated bond vector $\tilde{\bf d}_{\alpha\beta}$ in the contraction of the $B_{\gamma}$ tensors in Eqs.~(\ref{eq:multifoldBandV}) and~(\ref{eq:amorphous3FTmatrixFinalChirality}).

The ${\bf k}\cdot{\bf p}$ effective Hamiltonian $\mathcal{H}_{\text{Eff}}({\bf p})$ in Eq.~(\ref{eq:3FdisorderKP}) respects the continuous symmetries of the approximate [average] little group [ALG] $\tilde{G}_{\Gamma,3}$, which admits the decomposition:
\begin{equation}
\tilde{G}_{\Gamma,3} = \text{SO}(3) \cup \{\mathcal{T}|000\}\text{SO}(3) \cup \{E|\epsilon 00\}\text{SO}(3) \cup \{\mathcal{T}|\epsilon 00\}\text{SO}(3),
\label{eq:3FnumericsALG}
\end{equation}
where $\mathcal{T}$ is time-reversal and $\{E|\epsilon 00\}$ is an infinitesimal translation along the $x$-axis.
The generating symmetries of $\tilde{G}_{\Gamma,3}$ in Eq.~(\ref{eq:3FnumericsALG}) can be represented through their action on $\mathcal{H}_{\text{Eff}}({\bf p})$ in Eq.~(\ref{eq:3FdisorderKP}):
\begin{eqnarray}
\tilde{C}_{(2\pi/\phi) z}\mathcal{H}_{\text{Eff}}({\bf p})\tilde{C}_{(2\pi/\phi) z}^{-1} &=& e^{i (\phi/2)\tilde{\tau}^{z}\tilde{\mu}^{z}}\mathcal{H}_{\text{Eff}}(\tilde{C}^{-1}_{(2\pi/\phi) z}{\bf p})e^{-i (\phi/2)\tilde{\tau}^{z}\tilde{\mu}^{z}}, \nonumber \\
\tilde{C}_{(2\pi/\theta) x}\mathcal{H}_{\text{Eff}}({\bf p})\tilde{C}_{(2\pi/\theta) x}^{-1} &=& e^{i (\theta/2)\tilde{\mu}^{z}}\mathcal{H}_{\text{Eff}}(\tilde{C}^{-1}_{(2\pi/\theta) x}{\bf p})e^{-i (\theta/2)\tilde{\mu}^{z}}, \nonumber \\
\tilde{\mathcal{T}}\mathcal{H}_{\text{Eff}}({\bf p})\tilde{\mathcal{T}}^{-1} &=& \mathcal{H}^{*}_{\text{Eff}}(-{\bf p}),
\label{eq:disorder3FKPSyms}
\end{eqnarray}
where $\phi$ denotes an infinitesimal rotation angle about the $z$-axis [such that $\phi=\pi$ is consistent with the matrix representative of $\tilde{C}_{2z}$ in Eq.~(\ref{eq:PristineMultifoldExtraSyms})], $\theta$ denotes an infinitesimal rotation angle about the $x$-axis [such that $\theta=\pi$ is consistent with the matrix representative of $\tilde{C}_{2x}$ in Eq.~(\ref{eq:PristineMultifoldsyms})], and where we note that the continuous translation symmetries in $\tilde{G}_{\Gamma,3}$ are represented as phases multiplied by the $4\times 4$ identity matrix $\tilde{\tau}^{0}\tilde{\mu}^{0}$, and have hence been suppressed for notational simplicity.
In Eq.~(\ref{eq:disorder3FKPSyms}), the tildes on $\tilde{C}_{(2\pi/\phi) z}$, $\tilde{C}_{(2\pi/\theta) x}$, and $\tilde{\mathcal{T}}$ denote that the symmetries are elements of the single ALG $\tilde{G}_{\Gamma,3}$, because the symmetries act on integer-angular-momentum $s$ and $p_{x,y,z}$ internal degrees of freedom [see Refs.~\cite{BigBook,MTQC} and the text surrounding Eq.~(\ref{eq:Amo3F})].

\begin{figure}[t]
\centering
\includegraphics[width=\linewidth]{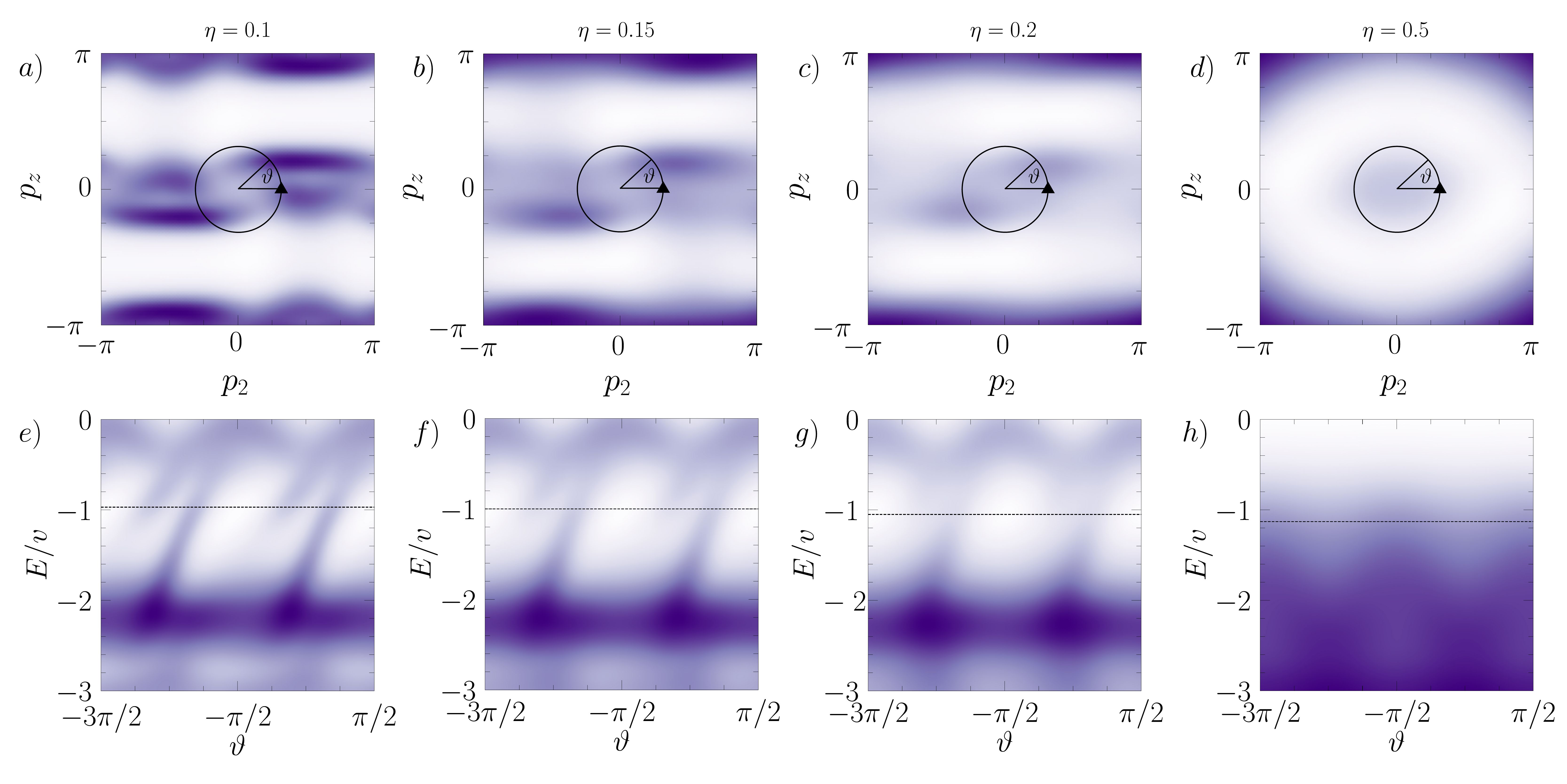}
\caption{Fermi-arc surface states of the disordered multifold fermion model.
In this figure, we show the $(\hat{x}+\hat{y})$-normal surface-projected, disorder-averaged spectral function $\bar{A}_{\text{surf}}(E,{\bf p})$ [Eq.~(\ref{eq:averageSpectrumSurface})] of the non-crystalline chiral multifold tight-binding model [Eq.~(\ref{eq:amorphous3FTmatrixFinalChirality})] on a lattice with increasing random nematic structural disorder [$d_{A}=3$ in Eq.~(\ref{eq:finiteDimAmorph})] parameterized by the standard deviation $\eta$, using the same chirality imbalance percentages and system parameters as in Fig.~\ref{fig:DOS_3F}.
Each panel was generated by averaging over 50 disorder replicas, each with $N_{\text{sites}}=20^{3}$ prior to the removal [``evaporation''] of accidental dangling [decoupled] surface atoms generated by the disorder realization [see the text preceding Eq.~(\ref{eq:RealSpaceGreenSlab})]. 
Because surface chirality domain walls can bind flat-band-like states that overwhelm and obscure spectral signatures of topological surface Fermi arcs~\cite{InternalChiralTheory,InternalChiralExp,SussmanJammedDiracWeyl}, we have only implemented random chirality domain walls within the bulk of the finite [slab] systems used to generate this figure.
In panels (a-d), we plot $\bar{A}_{\text{surf}}(E,{\bf p})$ as a function of $p_{2}=(1/\sqrt{2})(p_{x}-p_{y})$ and $p_{z}$ for increasing $\eta$ at a fixed relative energy $E/v$ [respectively the dashed line in (e-h)] set to 0.81 below the threefold [multifold] nodal degeneracy at ${\bf p}={\bf 0}$ in Figs.~\ref{fig:3F_bands_amo} and~\ref{fig:DOS_3F} in order to approximately capture the same cross section of the bulk Fermi surface for varying $\eta$, and hence any associated topological surface states.
In (e-h), we plot $\bar{A}_{\text{surf}}(E,{\bf p})$ as a function of energy on counterclockwise circular paths, parameterized by $\vartheta$, surrounding $p_{2}=p_{z}=0$ for the increasingly disordered multifold systems in (a-d), respectively.
In (a-c) and (e-g), $\bar{A}_{\text{surf}}(E,{\bf p})$ continues to exhibit two clear -- but increasingly diffuse -- Fermi-arc surface states with the same connectivity and topological chirality [positive slopes] as two of the four surface Fermi arcs in the right-handed enantiomer of the crystalline chiral multifold model [Fig.~\ref{fig:3F_figWL}(a,b)].
Conversely, between $\eta=0.1$ in (a,e) and $\eta=0.2$ in (c,g), one time-reversed pair of Fermi arcs has disappeared from $\bar{A}_{\text{surf}}(E,{\bf p})$ [the upper-left and bottom-right arcs in (a)].
This is consistent with the constant-energy bulk spectral function [Fig.~\ref{fig:DOS_3F}] and OAM texture [Fig.~\ref{fig:OAMTexture3F}], in which the positively charged cluster of Weyl fermions at $M$ [$|{\bf k}|=\pi\sqrt{2}$] in the crystalline limit [see Figs.~\ref{fig:3F_figBulk}(b) and~\ref{fig:3F_extraCrossings} and text surrounding Eq.~(\ref{eq:multifoldStructuralChirality})] disappears from the spectrum, rather than growing into an isotropic [sphere- or ring-like] spectral feature.
Specifically, in the weak-disorder regime [$\eta=0.1$ in (a,e)], the surface Fermi pocket at $p_{2}=p_{z}=0$ contains the superposed projections of the bulk $C=+2$ chiral multifold fermion at ${\bf p}={\bf 0}$ and the $C=+2$ cluster of conventional Weyl fermions at $M$, consistent with the four surface Fermi arcs visible in (a,e).
The coincident disappearance in $\eta$ of two $(\hat{x}+\hat{y})$-normal surface Fermi arcs and the bulk spectral features at $|{\bf p}|=\pi\sqrt{2}$ hence provides further evidence that disorder in the multifold fermion model dramatically rearranges the bulk conventional Weyl fermions that lie at free [unpinned] ${\bf k}$ in the crystalline limit, and likely drives a reduction of the surface-projected topological chiral charge at $p_{2}=p_{z}=0$.
In the strong-disorder regime [$\eta=0.5$ in (d,h)], $\bar{A}_{\text{surf}}(E,{\bf p})$ greatly differs from the moderate-disorder surface spectrum in (c,g), and no longer exhibits Fermi-arc-like spectral features.
Lastly, unlike in the crystalline limit [Fig.~\ref{fig:3F_figWL}(b)], $\bar{A}_{\text{surf}}(E,{\bf p})$ in the presence of even weak disorder [$\eta=0.1$ in (e)] no longer exhibits multiple distinct projected bulk gaps that are consecutively crossed by arc-like spectral features.}
\label{fig:3F_Arcs_amo}
\end{figure}

\begin{figure}[t]
\centering
\includegraphics[width=\linewidth]{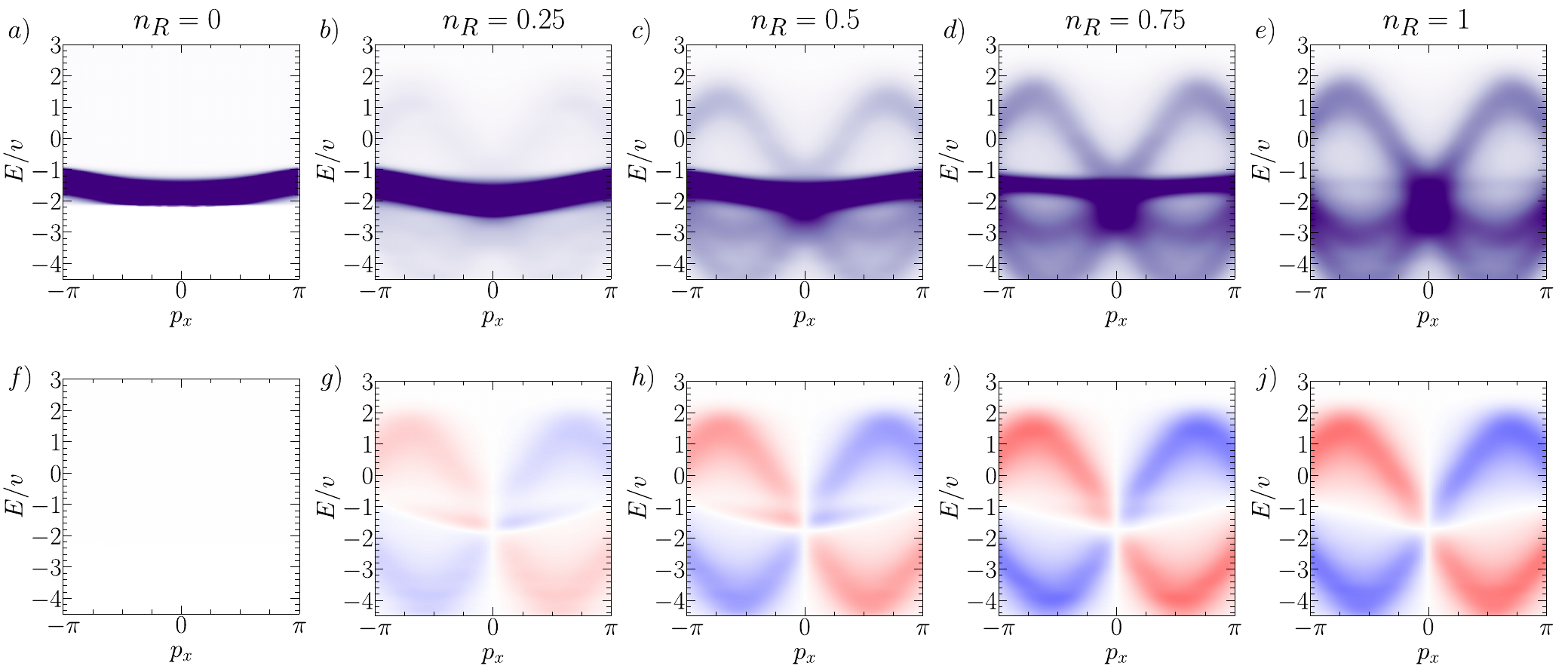}
\caption{Spectral function and orbital angular momentum texture of the disordered multifold fermion model with coexisting chiral and achiral domains.
To generate each panel in this figure, we place the non-crystalline chiral multifold model [Eq.~(\ref{eq:amorphous3FTmatrixFinalChirality})] with the parameters in Eq.~(\ref{eq:disordered3Fparams}) on a lattice with $N_{\mathrm{sites}}=15^3=3375$ and strong random nematic Gaussian structural and local frame disorder parameterized by the fixed standard deviation $\eta=0.5$, as detailed in Appendix~\ref{app:lattices}. 
We then generate the replica-averaged momentum-resolved Green's function $\bar{\mathcal{G}}(E,\mathbf{p})$ [Eq.~(\ref{eq:averageOneMomentumGreen})] by averaging the system over 50 disorder realizations [replicas] that -- unlike the previous analyses in this section -- each contain contiguous domains of either right-handed sites [$\chi_{\alpha}=1$] or \emph{achiral} sites [$\chi_{\alpha}=0$] with the respective concentrations $n_R=N_R/N_{\mathrm{sites}}$ and $n_{A}=1-n_{R}$.
(a-e) The disorder-averaged system spectral function $\bar{A}(E,{\bf p}) \propto \text{Im}\{\Tr[\bar{\mathcal{G}}(E,{\bf p})]\}$ [Eq.~(\ref{eq:SpecFunc})] for increasing $n_{R}$.  
Even for (b) a mostly achiral system [$n_{R}=0.25$], faint dispersive spectral features corresponding to the upper and lower bands of the disordered multifold fermion at ${\bf p}={\bf 0}$ [Fig.~\ref{fig:3F_bands_amo}] are visible on top of a large background signal of nondispersive trivial states at $E/v \approx -1.5$, and (c-e) become increasingly well-resolved for increasing $n_{R}$.
(f-j) The $\langle L^{x}(E,{\bf p})\rangle$ component of the orbital-angular-momentum- [OAM-] dependent spectral function vector [Eq.~\eqref{eq:OAMDOS}] for the systems in (a-e), respectively, plotted using a log-scale color map in which red is positive and blue is negative.
As shown below in Fig.~\ref{fig:3F_Spin_Pol}, the nondispersive trivial states exhibit a vanishing OAM polarization [$P_{\bf L}(E,{\bf p})\approx 0$ in Eq.~(\ref{eq:3FOAMPolarization}) for all values of $n_{R}$], whereas the dispersive multifold bands show a high degree of OAM polarization.  
$\langle L^{x}(E,{\bf p})\rangle$ in particular is only nonvanishing for the dispersive topological [upper and lower] multifold fermion bands, and may hence provide a signature of topological chirality in a strongly disordered system that can in principle be detected with OAM-sensitive experimental probes, such as circular dichroism angle-resolved photoemission spectroscopy [CD-ARPES]~\cite{OAMCDArpes1,OAMCDArpes2,OAMWeyl1,OAMWeyl2,OAMmultifold1,OAMmultifold2,OAMmultifold3,OAMmultifold4,OAMmultifold5,OAMnodalLine,OAMzxShenARPES3DTI}.}
\label{fig:3F_Spin_Bands}
\end{figure}

Lastly, we may contrast the $d_{A}=3$ non-crystalline multifold model symmetry action in Eq.~(\ref{eq:disorder3FKPSyms}) and the associated ALG small coreps with those of the $d_{A}=3$ nematic-disordered Kramers-Weyl model [Eq.~(\ref{eq:NumericsSecKWsymmetryAction})].
In both cases, the ${\bf k}\cdot {\bf p}$ Hamiltonians respect the symmetries of the chiral ALG $\tilde{G}_{\Gamma,3}$ in Eq.~(\ref{eq:3FnumericsALG}).
However, the Kramers-Weyl ${\bf k}\cdot {\bf p}$ Hamiltonian at ${\bf p}={\bf 0}$ [Eq.~(\ref{eq:KWdisorderKP})] exhibits a twofold [Kramers] degeneracy, whereas $\mathcal{H}_{\text{Eff}}({\bf p})$ at ${\bf p}={\bf 0}$ in Eq.~(\ref{eq:3FdisorderKP}) has four energy eigenvalues that subdivide into a singly degenerate trivial state at $E=3\tilde{m}'$ [which lies outside of the energy range of the spectral function in Fig.~\ref{fig:3F_bands_amo}] and a threefold degeneracy at $E=-\tilde{m}'$ that corresponds to the node of a non-crystalline chiral multifold fermion.
This can be understood by recognizing that the disordered Kramers-Weyl fermion in Eq.~(\ref{eq:KWdisorderKP}) transforms in a two-dimensional, double-valued [spinful] small corep of the double ALG $\tilde{G}_{\Gamma,3}$, whereas the topologically trivial single degeneracy and disordered chiral multifold fermion in Eq.~(\ref{eq:3FdisorderKP}) respectively transform in one- and three-dimensional, single-valued [spinless] small coreps of the single ALG $\tilde{G}_{\Gamma,3}$.
Importantly, this demonstrates that even in amorphous systems with the same average symmetry group, different local degrees of freedom and microscopic interactions can give rise to spectral features that transform in different \emph{irreducible coreps} of the average symmetry group, leading to the appearance of topologically distinct states~\cite{marsal_topological_2020}.

\paragraph*{\bf Disordered Fermi-Arc Surface States} -- $\ $ Having established that the bulk threefold [multifold] spectral feature at ${\bf p} = {\bf 0}$ in Fig.~\ref{fig:3F_bands_amo} is a non-crystalline spin-1 chiral multifold fermion, we will next explore disorder-driven shifts in spectral signatures of its bulk-boundary correspondence. To begin, we place the non-crystalline chiral multifold tight-binding model [Eq.~(\ref{eq:amorphous3FTmatrixFinalChirality})] on a lattice with increasing random nematic structural disorder [$d_{A}=3$ in Eq.~(\ref{eq:finiteDimAmorph})] parameterized by the standard deviation $\eta$, using the same structural chirality imbalance percentages and system parameters as the bulk calculations in Fig.~\ref{fig:DOS_3F}.
Unlike in our previous calculations, we now take the system to have periodic boundary conditions in the Cartesian $(\hat{x}-\hat{y})$- and $\hat{z}$-directions and open boundary conditions in the $(\hat{x}+\hat{y})$-direction. 
We next compute the $(\hat{x}+\hat{y})$-normal surface-projected spectral function $\bar{A}_{\text{surf}}(E,{\bf p})$ [Eq.~(\ref{eq:averageSpectrumSurface})] averaged over 50 disorder replicas [noting that the Cartesian $(\hat{x}+\hat{y})$-normal surface can no longer be designated the $(110)$-surface in a lattice-disordered system].
Each disorder replica in our system is initially constructed with $20^{3}=8000$ sites, which we then reduce by removing [``evaporating''] dangling [decoupled] surface atoms that represent numerical artifacts of the disorder implementation process [see the text preceding Eq.~(\ref{eq:RealSpaceGreenSlab})]. 
Additionally, because surface chirality domain walls can bind flat-band-like states that overwhelm and obscure spectral signatures of topological surface Fermi arcs~\cite{InternalChiralTheory,InternalChiralExp,SussmanJammedDiracWeyl}, then we only place chirality domain walls deep within the bulk of each disorder replica.

In Fig.~\ref{fig:3F_Arcs_amo}, we plot the disorder-averaged surface spectral function $\bar{A}_{\text{surf}}(E,{\bf p})$ of the non-crystalline multifold model for increasing disorder parameterized by $\eta$.
Specifically, in Fig.~\ref{fig:3F_Arcs_amo}(a-d), we plot $\bar{A}_{\text{surf}}(E,{\bf p})$ as a function of $p_{2}=(1/\sqrt{2})(p_{x}-p_{y})$ and $p_{z}$ for increasing $\eta$ at a fixed energy, and in Fig.~\ref{fig:3F_Arcs_amo}(e-h), we respectively plot $\bar{A}_{\text{surf}}(E,{\bf p})$ at the same $\eta$ as a function of energy on counterclockwise circular paths, parameterized by $\vartheta$, surrounding $p_{2}=p_{z}=0$.
In the weak-disorder regime [$\eta=0.1$ in Fig.~\ref{fig:3F_Arcs_amo}(a,e)], $\bar{A}_{\text{surf}}(E,{\bf p})$ continues to exhibit four clear -- but increasingly diffuse -- Fermi-arc surface states with the same connectivity and topological chirality [positive slopes] as the right-handed enantiomer of the crystalline chiral multifold model [Fig.~\ref{fig:3F_figWL}(a,b)], consistent with the average right-handedness of the disordered system.

However in the moderate-disorder regime [$\eta=0.15-0.2$ in Fig.~\ref{fig:3F_Arcs_amo}(b,c,f,g)], one time-reversed pair of Fermi arcs has disappeared from $\bar{A}_{\text{surf}}(E,{\bf p})$ [the upper-left and bottom-right arcs in Fig.~\ref{fig:3F_Arcs_amo}(a)].
To understand this, we first recall that in the crystalline limit -- and in the $\eta=0.1$ weak-disorder regime in Fig.~\ref{fig:3F_Arcs_amo}(a,e) -- the surface Fermi pocket at $p_{2}=p_{z}=0$ contains the superposed projections of a bulk $C=+2$ chiral multifold fermion at ${\bf p}={\bf 0}$ and a $C=+2$ cluster of conventional Weyl fermions at $M$ [$|{\bf p}|=\pi\sqrt{2}$, see Figs.~\ref{fig:3F_figBulk}(b) and~\ref{fig:3F_extraCrossings} and text surrounding Eq.~(\ref{eq:multifoldStructuralChirality})].
Hence for vanishing and weak disorder, the surface Fermi pocket at $p_{2}=p_{z}=0$ carries a net chiral charge of $C=4$, consistent with the four surface Fermi arcs visible in Fig.~\ref{fig:3F_Arcs_amo}(a,e).
Next, we previously observed in the constant-energy bulk spectral function [Fig.~\ref{fig:DOS_3F}] and OAM texture [Fig.~\ref{fig:OAMTexture3F}] that as the system enters the moderate-disorder regime, the bulk Fermi pockets in the momentum star of $M$ [see the text surrounding Fig.~\ref{fig:3F_figWL}] move away from $|{\bf p}|=\pi\sqrt{2}$, rather than merge into an isotropic [sphere- or ring-like] spectral feature with radius $|{\bf p}|=\pi\sqrt{2}$.
The coincident disappearance in $\eta$ of two $(\hat{x}+\hat{y})$-normal surface Fermi arcs and the bulk spectral features at $|{\bf p}|=\pi\sqrt{2}$ hence provides further evidence that disorder in the multifold fermion model dramatically rearranges the conventional Weyl fermions that lie at free [unpinned] ${\bf k}$ in the crystalline limit, including those forming the Weyl-point cluster at $M$.
Specifically, if the conventional Weyl fermions near $|{\bf p}|=\pi\sqrt{2}$ were displaced by disorder, this would in turn lead to a reduction in the $(\hat{x}+\hat{y})$-normal surface-projected topological chiral charge at $p_{2}=p_{z}=0$, consistent with the observed reduction from four to two well-resolved Fermi arcs in $\bar{A}_{\text{surf}}(E,{\bf p})$ under moderate disorder [$\eta=0.15-0.2$ in Fig.~\ref{fig:3F_Arcs_amo}(b,c,f,g)].

Finally, when the system enters the strong-disorder regime [$\eta=0.5$ in Fig.~\ref{fig:3F_Arcs_amo}(d,h)], the $(\hat{x}+\hat{y})$-normal surface spectrum greatly differs from $\bar{A}_{\text{surf}}(E,{\bf p})$ in the moderate-disorder regime, and no longer exhibits Fermi-arc-like spectral features.
We hence conclude that like the non-crystalline Kramers-Weyl and double-Weyl models [Figs.~\ref{fig:KWamorphousArcs}(d,h) and~\ref{fig:C2_arcs_amo}(d,h), respectively], the bulk multifold fermion semimetal state remains topological and gapless for strong disorder, but no longer exhibits topological Fermi arcs in the amorphous regime.  
This can be contrasted with the nonsymmorphic multifold model introduced in Ref.~\cite{chang2017large} to capture the electronic structure of B20 chiral cubic topological semimetals [see the text preceding Eq.~(\ref{eq:Amo3F})], in that the nonsymmorphic multifold model is instead trivialized by moderate Gaussian structural disorder~\cite{Franca2024}.
The absence of Fermi-arc surface states under strong disorder in the non-crystalline chiral multifold model introduced in this work [Eq.~(\ref{eq:amorphous3FTmatrixFinalChirality})] can be understood by recognizing that the multifold model with strong nematic lattice disorder is spectrally isotropic in 3D ${\bf p}$ space [Fig.~\ref{fig:DOS_3F}(c)], and specifically exhibits larger-$|{\bf p}|$ ring- [sphere-] like spectral features that surround the multifold fermion at ${\bf p}={\bf 0}$.
Similar to the strongly disordered Kramers-Weyl model in Fig.~\ref{fig:KWamorphousArcs}(d,h) and topologically chiral nodal-surface semimetals~\cite{KramersWeyl,WeylNodalSurfaceFollowupPRB}, the sphere-like spectral features in the strongly disordered multifold model ensure that there do not exist topological Fermi arcs for any choice of surface termination direction, due to the absence of topologically nontrivial projected bulk Fermi pockets.

We lastly note that in the crystalline limit, the $(\hat{x}+\hat{y})$-normal surface spectrum of the multifold model displays two projected bulk gaps [the energy gaps at $E/|v|=-1$ and $E/|v|=-2$ in Fig.~\ref{fig:3F_figWL}(b)], which are each spanned by four Fermi-arc surface states.
Conversely, in the presence of even weak disorder [$\eta=0.1$ in Fig.~\ref{fig:3F_Arcs_amo}(e)], $\bar{A}_{\text{surf}}(E,{\bf p})$ in the non-crystalline multifold model no longer exhibits multiple well-resolved projected bulk gaps that are consecutively crossed by arc-like spectral features.

\begin{figure}[t]
\centering
\includegraphics[width=\linewidth]{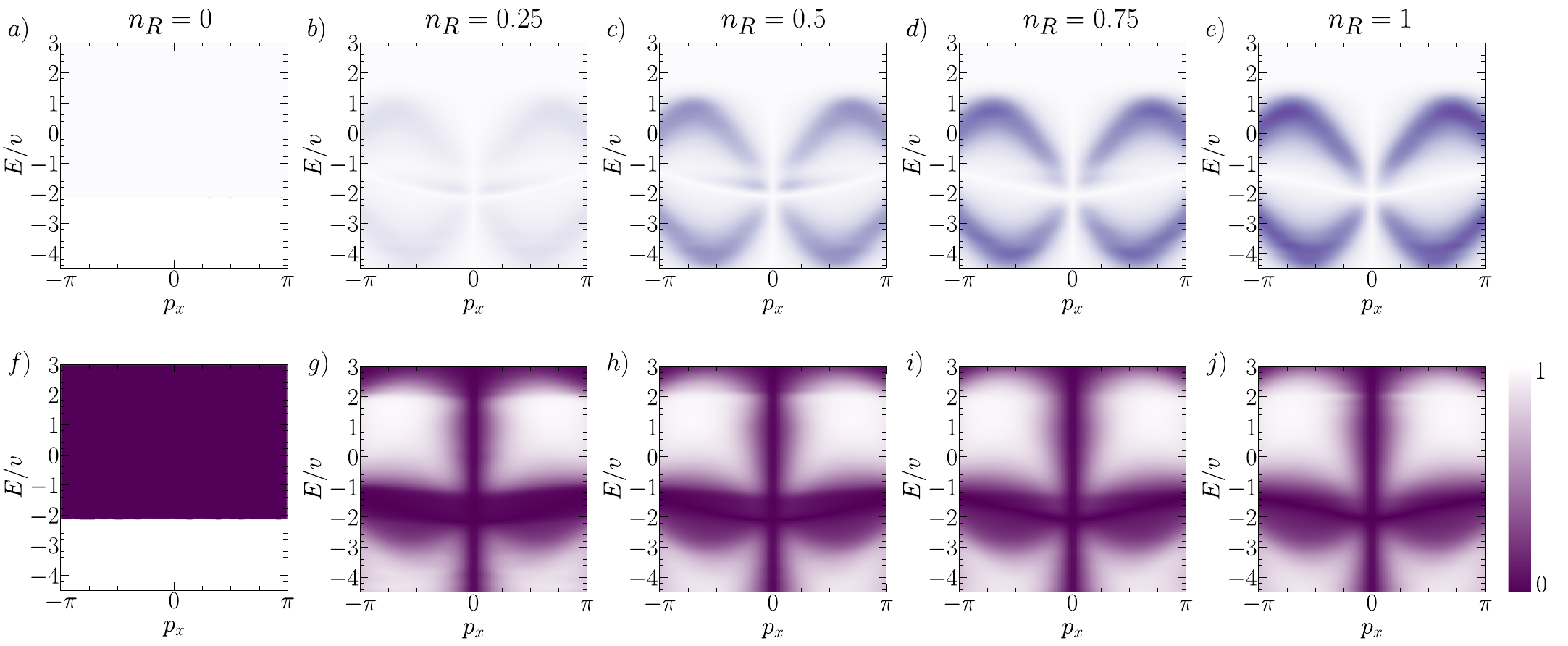}
\caption{Orbital angular momentum polarization of the disordered multifold fermion model with coexisting chiral and achiral domains.  
(a-e) Magnitude of the orbital-angular-momentum- [OAM-] dependent spectral function vector $|\langle {\bf L}(E,{\bf p})\rangle|$ [Eq.~(\ref{eq:OAMDOS})] for the strongly disordered chiral multifold model in Fig.~\ref{fig:3F_Spin_Bands} in which each replica contains contiguous domains of either right-handed or achiral sites with the respective concentrations $n_R=N_R/N_{\mathrm{sites}}$ and $n_{A}=1-n_{R}$, where $n_{R}$ is successively increased from (a) $n_R=0$ to (e) $n_R=1$.  
(f-j) OAM polarization $P_{\bf L}(E,{\bf p})$ [Eq.~(\ref{eq:3FOAMPolarization})] of the disordered multifold systems in (a-e), respectively. 
In the system spectral function [Fig.~\ref{fig:3F_Spin_Bands}(a-e)], dispersive topological states [the upper and lower bands of the disordered multifold fermion at ${\bf p}={\bf 0}$ in Fig.~\ref{fig:3F_bands_amo}] coexist with nondispersive trivial states near $E/v\approx -1.5$, where the relative spectral weights of the topological and trivial states respectively scale with $n_{R}$ and $n_{A}$.
In the disordered multifold model [Eq.~(\ref{eq:amorphous3FTmatrixFinalChirality})], only the dispersive topological multifold bands carry a nonvanishing OAM polarization, whereas the nondispersive trivial states exhibit vanishing $|\langle {\bf L}(E,{\bf p})\rangle|$ and $P_{\bf L}(E,{\bf p})$ for all $n_{R}$.}
\label{fig:3F_Spin_Pol}
\end{figure}

\paragraph*{\bf Experimental Signatures of Coexisting Chiral and Achiral Domains} -- $\ $ In Appendix~\ref{app:amorphousKramers}, we previously considered the physically motivated scenario~\cite{ChiralGlass1,ChiralGlass2,ChiralGlass3,ChiralGlassGiuliaSummary,ChiralGlassMolecules,AmorphousTeDFT,AmorphousChalcogenideNatCommH,ChiralPhaseChange1,ChiralPhaseChange2,MixedChiralPhotonicAlloy} of systems with strong structural disorder in which some -- or even most -- sites are structurally \emph{achiral} [$\chi_{\alpha}=0$ in Eq.~(\ref{eq:amorphous3FTmatrixFinalChirality})].
We specifically in Appendix~\ref{app:amorphousKramers} investigated the non-crystalline Kramers-Weyl model [Eq.~(\ref{eq:amorphousKWTmatrixFinalChirality})] with varying concentrations of chiral and achiral sites.
As detailed in the text surrounding Figs.~\ref{fig:KW_Spin_Bands} and~\ref{fig:KW_Spin_Pol}, we observed that even when most sites carry a vanishing local chirality, the system spectrum subdivides into faint, dispersive, highly spin-polarized topological bands and a large background signal of nondispersive trivial bands with negligible spin polarization. 
We will next show that when the disordered multifold model [Eq.~(\ref{eq:amorphous3FTmatrixFinalChirality})] is realized with a mixture of chiral and achiral domains, the spectrum analogously contains dispersive topological bands that instead exhibit a high degree of \emph{OAM polarization} [$P_{\bf L}(E,{\bf p})\approx 1$ in Eq.~(\ref{eq:3FOAMPolarization})].
To implement a mixed chiral-achiral system, we begin by placing the non-crystalline multifold model on a lattice with $N_{\mathrm{sites}}=15^3=3375$ and strong [$\eta=0.5$] random nematic Gaussian structural and local frame disorder [see Appendix~\ref{app:lattices}].
We then generate the replica-averaged momentum-resolved Green's function $\bar{\mathcal{G}}(E,\mathbf{p})$ [Eq.~(\ref{eq:averageOneMomentumGreen})] by averaging the system over 50 disorder realizations [replicas].
However, unlike the previous analyses in this section, each replica now contains contiguous domains of either right-handed sites [$\chi_{\alpha}=1$] or achiral sites [$\chi_{\alpha}=0$], whose concentrations are respectively given by $n_R=N_R/N_{\mathrm{sites}}$ and $n_{A}=1-n_{R}$.

In Fig.~\ref{fig:3F_Spin_Bands}(a-e), we plot the disorder-averaged spectral function $\bar{A}(E,{\bf p}) \propto \text{Im}\{\Tr[\bar{\mathcal{G}}(E,{\bf p})]\}$ [Eq.~(\ref{eq:SpecFunc})] of the mixed chiral-achiral system for increasing $n_{R}$.
We find that $\bar{A}(E,{\bf p})$ continues to exhibit dispersive spectral features that correspond to the upper and lower topological bands of the disordered multifold fermion at ${\bf p}={\bf 0}$ [Fig.~\ref{fig:3F_bands_amo}].
However, the dispersive topological bands in Fig.~\ref{fig:3F_Spin_Bands}(a-e) also coexist with nondispersive trivial states near $E/v \approx -1.5$, which originate from both the central flat band of the multifold fermion in locally structurally chiral regions with $\chi_{\alpha}=1$, and from achiral regions in which all three bands comprising the multifold fermion are nondispersive [\emph{i.e.} $\chi_{\alpha}=\chi_{\beta}=0$ such that the OAM coupling term proportional to $v$ vanishes in Eq.~(\ref{eq:amorphous3FTmatrixFinalChirality})].
We further observe in Fig.~\ref{fig:3F_Spin_Bands}(a-e) that the relative spectral weights of the dispersive topological and nondispersive trivial states respectively scale with $n_{R}$ and $n_{A}$.
Notably, even for a mostly achiral system [$n_{R}=0.25$ in Fig.~\ref{fig:3F_Spin_Bands}(b)], faint linearly dispersive bands are visible near ${\bf p}={\bf 0}$ on top of a large background signal of nondispersive trivial states, and become increasingly well-resolved for increasing $n_{R}$ [Fig.~\ref{fig:3F_Spin_Bands}(c-e)].

Importantly, we find that the dispersive topological multifold fermion bands are better revealed by comparing the overall energy spectrum to the OAM-dependent spectrum.
To demonstrate this, we show in Fig.~\ref{fig:3F_Spin_Bands}(f-j) the OAM textures of the mixed chiral-achiral disordered systems, focusing on the $\langle L^{x}(E,{\bf p})\rangle$ component of the OAM-dependent spectral function vector [Eq.~\eqref{eq:OAMDOS}]. 
While the central grouping of nondispersive states near $E/v \approx -1.5$ still carries some residual spectral weight in the $\langle L^{x}(E,{\bf p})\rangle$ OAM texture, it is much smaller than that of the dispersive topological bands, \emph{even for small $n_R$}. 
To rule out analysis artifacts due to our choice of the $\langle L^{x}(E,{\bf p})\rangle$ OAM-texture component, we also compute in Fig.~\ref{fig:3F_Spin_Pol}(a-e) the magnitude of the OAM-dependent spectral function vector $|\langle {\bf L}(E,{\bf p})\rangle|$ [Eq.~(\ref{eq:OAMDOS})]. 
As with $\langle L^{x}(E,{\bf p})\rangle$ in Fig.~\ref{fig:3F_Spin_Bands}(f-j), $|\langle {\bf L}(E,{\bf p})\rangle|$ in Fig.~\ref{fig:3F_Spin_Pol}(a-e) shows appreciable spectral weight on the two dispersive bands of the non-crystalline multifold fermion at ${\bf p}={\bf 0}$, and exhibits a much smaller weight on the central grouping of nondispersive states compared to the overall spectral function in Fig.~\ref{fig:3F_Spin_Bands}(a-e).

Finally, to properly account for the difference in spectral weight between the nondispersive trivial states and the dispersive topological multifold bands in Fig.~\ref{fig:3F_Spin_Bands}(a-e), we calculate the total OAM polarization $P_{\bf L}(E,{\bf p})$ [Eq.~(\ref{eq:3FOAMPolarization})] of the mixed-achiral system, which is shown in Fig.~\ref{fig:3F_Spin_Pol}(f-j).
Crucially, we see in Fig.~\ref{fig:3F_Spin_Pol}(f-j) that \emph{only} the dispersive topological bands of the multifold fermion at ${\bf p}={\bf 0}$ show a high degree of OAM polarization, whereas the nondispersive trivial states near $E/v \approx -1.5$ exhibit a vanishing OAM polarization [$P_{\bf L}(E,{\bf p})\approx 0$ in Eq.~(\ref{eq:3FOAMPolarization})], even for small values of $n_{R}$.
This pattern of faint -- but highly OAM-polarized -- dispersive topological bands on top of a large background signal of nondispersive trivial bands with vanishing OAM polarization can in principle be detected in OAM-sensitive experiments, such as CD-ARPES~\cite{OAMCDArpes1,OAMCDArpes2,OAMWeyl1,OAMWeyl2,OAMmultifold1,OAMmultifold2,OAMmultifold3,OAMmultifold4,OAMmultifold5,OAMnodalLine,OAMzxShenARPES3DTI}.

\begin{figure}[t]
\centering
\includegraphics[width=\linewidth]{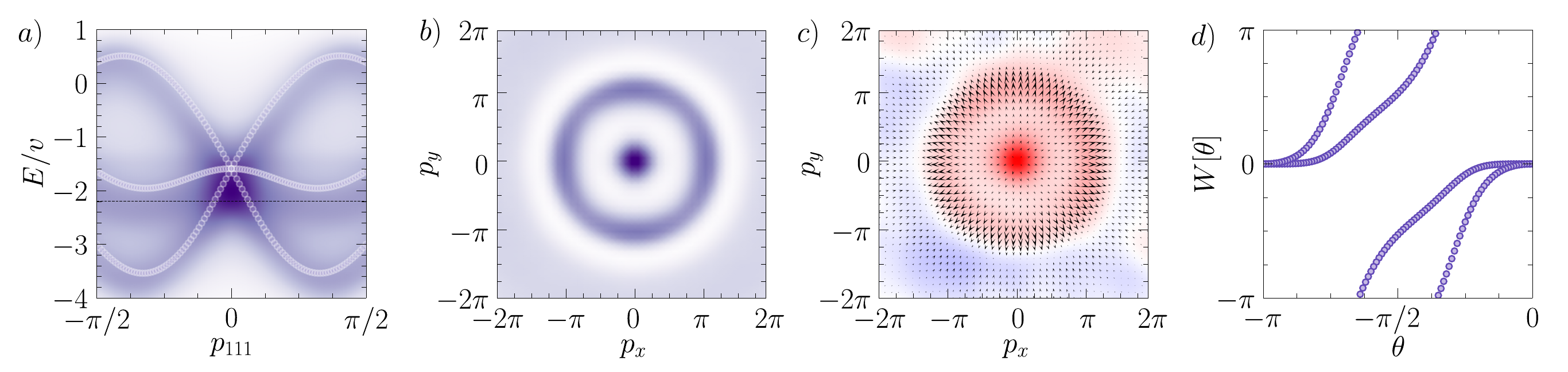}
\caption{Spin-1 chiral multifold fermions on random lattices.
(a-d) Bulk spectrum, spin texture, and topology of the non-crystalline chiral multifold model [Eq.~(\ref{eq:amorphous3FTmatrixFinalChirality})] on a lattice with randomly located sites in $d=d_A=3$ dimensions, random frame disorder parameterized by the standard deviation $\eta=0.5$, and chirality domains of unequal volume [Appendix~\ref{app:lattices}].
For all panels in this figure, the data were generated using Eq.~(\ref{eq:amorphous3FTmatrixFinalChirality}) with the tight-binding parameters in Eq.~(\ref{eq:disordered3Fparams}) implemented with a single domain in each replica of right-handed sites with $n_R=N_R/N_{\mathrm{sites}}=2/3$, and with the remaining volume in each replica containing a contiguous domain of left-handed sites with a corresponding concentration of $n_L=1-N_R/N_{\mathrm{sites}}=1/3$.
Each panel shows data generated by averaging over 50 randomly-generated replicas with $\sim 20^3$ sites each, as detailed in Appendix~\ref{app:PhysicalObservables}.
In (a), the bulk average spectral function $\bar{A}(E,{\bf p})$ [Eq.~(\ref{eq:SpecFunc})] exhibits an amorphous threefold [multifold] fermion at ${\bf p}={\bf 0}$. 
(b) $\bar{A}(E,{\bf p})$ at $E/v= -2.2$ [the dashed line in (a)], plotted at $p_{z}=0.1$ as a function of $p_{x}$ and $p_{y}$.
In addition to the multifold fermion at ${\bf p}={\bf 0}$, broad, ring- [sphere-] like spectral features appear in (b) at $|{\bf p}|=\pi/\bar{a}$ and $|{\bf p}|=\pi\sqrt{3}/\bar{a}$, where $\bar{a}=1$ is the average nearest-neighbor spacing.
As discussed in Figs.~\ref{fig:DOS_3F} and~\ref{fig:OAMTexture3F}, the ring-like features in (b) originate from both symmetry-enforced multifold fermions pinned to high-symmetry ${\bf k}$ [TRIM] points and conventional Weyl fermions at unpinned ${\bf k}$ in the crystalline limit [see Figs.~\ref{fig:3F_figBulk}(b) and~\ref{fig:3F_extraCrossings} and the text surrounding Eq.~(\ref{eq:multifoldStructuralChirality})], which are not only broadened and rotationally averaged, but also shifted in ${\bf p}$ by disorder.
(c) The orbital angular momentum [OAM] texture [Eq.~(\ref{eq:OAMDOS})] of the spectrum in (b).  
In (c), the random-lattice multifold fermion at ${\bf p}={\bf 0}$ exhibits a nearly perfect outward-pointing, monopole-like OAM texture like the Gaussian-disordered chiral multifold fermion at ${\bf p}={\bf 0}$ in Fig.~\ref{fig:OAMTexture3F}.
Also like in the Gaussian-disordered multifold model [Fig.~\ref{fig:OAMTexture3F}(b,c)], the ring-like features at $|{\bf p}|=\pi/\bar{a}$ and $|{\bf p}|=\pi\sqrt{3}/\bar{a}$ in (c) respectively exhibit outward- and inward-pointing OAM textures, and are hence more generic features of the non-crystalline multifold model with strong structural disorder [when realized with the tight-binding parameters in Eq.~(\ref{eq:disordered3Fparams})].
(d) The two-band [non-Abelian] amorphous Wilson loop spectrum of the non-crystalline multifold fermion at ${\bf p}={\bf 0}$ in (a), generated from an effective Hamiltonian [light purple circles in (a)] constructed with its reference energy cut $E_C/v = -1.8$ centered at the largest value of $\bar{A}(E,{\bf p})$ at ${\bf p}={\bf 0}$ [excluding the higher-energy states that arise from disordering the singly-degenerate trivial band at $E/|v| \approx 4.5$ in Fig.~\ref{fig:3F_figBulk}(b), see Appendix~\ref{app:EffectiveHamiltonian} for further details].
In (d), the two Wilson loop eigenvalues as a set wind twice in the positive direction, indicating that the lower two bands of the non-crystalline multifold fermion at ${\bf p}={\bf 0}$ together carry a chiral charge of $C=2$, consistent with the average right-handedness of the disordered system [$n_{R}>n_{L}$, see Fig.~\ref{fig:3F_WL}].}
\label{fig:3F_Unif}
\end{figure}

\paragraph*{\bf Random Lattices} -- $\ $ Lastly, one might be concerned that the topological and spectral properties of non-crystalline spin-1 chiral multifold fermions obtained in this section are specific to our use of Gaussian structural disorder, which admits a well-defined crystalline limit.
To show that this is not the case, we will conclude this section by computing the bulk energy spectrum, OAM texture, and ${\bf p}={\bf 0}$ amorphous Wilson loops of the non-crystalline chiral multifold model [Eq.~(\ref{eq:amorphous3FTmatrixFinalChirality})] on randomly generated lattices without well-defined [unique] crystalline limits.
To begin, we place the non-crystalline multifold model on a $d=d_{A}=3$ lattice [Eq.~(\ref{eq:finiteDimAmorph})] with randomly located sites in three dimensions, strong random frame disorder parameterized by the standard deviation $\eta=0.5$, and contiguous chirality domains of unequal volume with the respective concentrations of right-handed sites $n_R=N_R/N_{\mathrm{sites}}=2/3$ and left-handed sites $n_L=1-N_R/N_{\mathrm{sites}}=1/3$ [see Appendix~\ref{app:lattices}].
To simulate a large, self-averaging system, we further randomly generate 50 lattices [replicas] with $20^3 = 8000$ sites each, and then compute the replica-averaged momentum-resolved Green's function $\bar{\mathcal{G}}(E,{\bf p},{\bf p}')$ [Eq.~(\ref{eq:averageTWoMomentumGreen})].
As previously for the chiral multifold model with strong Gaussian disorder [Fig.~\ref{fig:3DGreen3F}(c)], we find that the off-diagonal-in-momentum elements $\bar{\mathcal{G}}(E,{\bf p},{\bf p}')$ vanish [on the average] as well for the non-crystalline multifold model on random lattices [Fig.~\ref{fig:3DGreen3F}(d)].
We may therefore again limit consideration to the diagonal-in-momentum replica-averaged momentum-resolved Green's function $\bar{\mathcal{G}}(E,\mathbf{p})$ [Eq.~(\ref{eq:averageOneMomentumGreen})], from which we below compute the spectral and topological properties of random-lattice chiral multifold fermions.

In Fig.~\ref{fig:3F_Unif}(a,b), we show the disorder-averaged, momentum-resolved spectral function $\bar{A}(E,{\bf p}) \propto \text{Im}\{\Tr[\bar{\mathcal{G}}(E,{\bf p})]\}$ [Eq.~(\ref{eq:SpecFunc})] of the random-lattice chiral multifold model respectively plotted as a function of energy and momentum [$p_{x}$] and at a fixed energy and $p_{z}$ as a function of the remaining two momenta $p_{x,y}$.
Like in the Gaussian-disordered lattice [Fig.~\ref{fig:3F_bands_amo}], $\bar{A}(E,{\bf p})$ in the random-lattice multifold model [Fig.~\ref{fig:3F_Unif}(a)] exhibits a threefold [multifold] nodal degeneracy at ${\bf p}={\bf 0}$ that to leading order consists of two linearly dispersing spectral features and a central nondispersing feature.
As previously for the random-lattice Kramers-Weyl [Fig.~\ref{fig:KW_Unif}(b)] and double-Weyl [Fig.~\ref{fig:C2_Unif}(b)] models, $\bar{A}(E,{\bf p})$ in the random-lattice multifold model also exhibits broadened, ring- [sphere-] like spectral features~\cite{spring_amorphous_2021,springMagneticAverageTI,corbae_evidence_2020,Ciocys2023}, which for the tight-binding parameters in Eq.~(\ref{eq:disordered3Fparams}) lie at $|{\bf p}|=\pi/\bar{a}$ and $|{\bf p}|=\pi\sqrt{3}/\bar{a}$, where $\bar{a}=1$ is the average nearest-neighbor spacing [Fig.~\ref{fig:3F_Unif}(b)].
The same ring-like spectral features at $|{\bf p}|=\pi$ and $|{\bf p}|=\pi\sqrt{3}$ also appear in the non-crystalline multifold model with Gaussian lattice disorder [Fig.~\ref{fig:DOS_3F}(b,c)], and originate from both symmetry-enforced multifold fermions pinned to high-symmetry ${\bf k}$ [TRIM] points and conventional Weyl fermions that lie at unpinned ${\bf k}$ in the crystalline limit [see Figs.~\ref{fig:3F_figBulk}(b) and~\ref{fig:3F_extraCrossings} and the text surrounding Eq.~(\ref{eq:multifoldStructuralChirality})], which become mixed by disorder through a more complicated process than simple broadening and rotational averaging. 
Specifically, unlike in the strongly disordered Kramers-Weyl [Figs.~\ref{fig:DOSKW}(c),~\ref{fig:SpinTextureKW}(c), and~\ref{fig:KW_Unif}(b,c)] and double-Weyl [Figs.~\ref{fig:DOSC2}(d) and~\ref{fig:C2_Unif}(b)] models, the radii $|{\bf p}|$ of the ring-like spectral features in the strongly disordered multifold model are not in one-to-one correspondence with the crystal momenta magnitudes $|{\bf k}|$ of the crystalline-limit bulk chiral fermions, indicating that the chiral fermions are also rearranged by lattice disorder.

To further analyze the random-lattice multifold spectral features in Fig.~\ref{fig:3F_Unif}(a,b), we next use $\bar{\mathcal{G}}(E,\mathbf{p})$ to compute the OAM texture [Eq.~(\ref{eq:OAMDOS})], which is plotted in Fig.~\ref{fig:3F_Unif}(c) at the same energy and $p_{z}$ values as the bulk spectral function $\bar{A}(E,{\bf p})$ in Fig.~\ref{fig:3F_Unif}(b).  
Like in the non-crystalline multifold model with Gaussian lattice disorder [Fig.~\ref{fig:OAMTexture3F}], the random-lattice multifold fermion at ${\bf p}={\bf 0}$ in Fig.~\ref{fig:3F_Unif}(c) exhibits a nearly perfect outward-pointing, monopole-like OAM texture.  
Also like in the Gaussian-disordered multifold model [Fig.~\ref{fig:OAMTexture3F}(b,c)], the ring-like spectral features at $|{\bf p}|=\pi$ and $|{\bf p}|=\pi\sqrt{3}$ in Fig.~\ref{fig:3F_Unif}(c) respectively exhibit outward- and inward-pointing OAM textures. 
The ring-like spectral features and their OAM textures are hence more generic features of the non-crystalline multifold model with strong structural disorder [though restricting to cases where the multifold model is realized with the tight-binding parameters in Eq.~(\ref{eq:disordered3Fparams}) and the same frame-disorder implementation and chirality imbalance percentages employed in Figs.~\ref{fig:DOS_3F},~\ref{fig:OAMTexture3F}, and~\ref{fig:3F_Unif}].
As discussed in the text surrounding Fig.~\ref{fig:OAMTexture3F}, the $|{\bf p}|=\pi\sqrt{3}$ radius and inward-pointing OAM texture of the outermost ring-like feature in Fig.~\ref{fig:3F_Unif}(b,c) is consistent with the possibility that the ring-like feature at $|{\bf p}|=\pi\sqrt{3}$ represents a many-particle disordered chiral multifold fermion with the opposite chiral charge as the multifold nodal degeneracy at ${\bf p}={\bf 0}$.
However, addressing this possibility would require a more careful treatment of tight-binding basis-state tails at small wavelengths [see the text surrounding Fig.~\ref{fig:dadtdf} and Eq.~(\ref{eq:TBbasisStates})], as well as formulating a Green's-function-based method for computing the many-particle chiral charges of disordered, sphere-like Fermi pockets at larger $|{\bf p}|$ than the ${\bf p}\approx {\bf 0}$ amorphous Wilson-loop method introduced in this work [Appendix~\ref{sec:WilsonBerry}].
We therefore leave a closer analysis of the ring-like spectral features in Fig.~\ref{fig:3F_Unif}(b,c) -- and their relationship to the Nielsen-Ninomiya chiral fermion doubling theorem~\cite{NielNino81a,NielNino81b} -- for future works.

Lastly, to precisely confirm that the threefold nodal degeneracy at ${\bf p}={\bf 0}$ in Fig.~\ref{fig:3F_Unif}(a) is indeed a topologically chiral multifold fermion, we compute the sphere Wilson loop spectrum of the random-lattice multifold model at ${\bf p}={\bf 0}$.  
To obtain the Wilson loop spectrum, we first construct an effective Hamiltonian $\mathcal{H}_{\mathrm{Eff}}({\bf p})=\mathcal{H}_{\mathrm{Eff}}(E_{C},{\bf p})$ [Eq.~(\ref{eq:AvgHEff}), light purple circles in Fig.~\ref{fig:3F_Unif}(a)] using a reference energy cut $E_{C}$ set to the energy of the largest spectral weight $\bar{A}(E,{\bf p})$ at ${\bf p}={\bf 0}$ [excluding the higher-energy states that arise from disordering the singly-degenerate trivial band at $E/|v| \approx 4.5$ in Fig.~\ref{fig:3F_figBulk}(b)].
As shown in Fig.~\ref{fig:H_eff_benchmark}, this choice of $E_{C}$ qualitatively maximizes the accuracy of $\mathcal{H}_{\mathrm{Eff}}({\bf p})$ in the vicinity of the threefold spectral feature at ${\bf p}={\bf 0}$.
We then use the lower two bands of $\mathcal{H}_{\mathrm{Eff}}({\bf p})$ in energy to compute the non-Abelian Wilson loop spectrum on a sphere surrounding the threefold degeneracy at ${\bf p}={\bf 0}$, as detailed in Appendix~\ref{sec:WilsonBerry}.
For the random-lattice multifold system with $n_{R} =2/3$ and $n_{L}=1/3$, the two Wilson loop eigenvalues as a set wind twice in the positive direction [Fig.~\ref{fig:3F_Unif}(d)], indicating that the nodal spectral feature at ${\bf p}={\bf 0}$ is a $C=2$ amorphous chiral multifold fermion.
This is consistent with our earlier Wilson loop analysis in Fig.~\ref{fig:3F_WL}, in which we demonstrated that the Gaussian-disordered multifold model hosts a non-crystalline spin-1 chiral multifold fermion at ${\bf p}={\bf 0}$ whose low-energy topological chirality is controlled by the average system chirality [\emph{i.e.} the ratio $n_{R}/n_{L}$].
Though not shown in Fig.~\ref{fig:3F_Unif}, we have numerically confirmed that the topological chiral charge of the random-lattice multifold fermion at ${\bf p}={\bf 0}$ is also controlled by the ratio $n_{R}/n_{L}$, and specifically also undergoes a topological phase transition between $C = \pm 2$ when $n_{R}/n_{L} \approx 1$.
Additionally, though not shown in Fig.~\ref{fig:3F_Unif}, we have further confirmed that the non-crystalline multifold model [Eq.~(\ref{eq:amorphous3FTmatrixFinalChirality})] on \emph{Mikado} lattices [see Fig.~\ref{appfig:structuraldisorder}(c) and Refs.~\cite{marsal_obstructed_2022,marsal_topological_2020}] also exhibits ${\bf p}={\bf 0}$ multifold fermions with nontrivial topological chiral charges that are controlled by the average system chirality.
Lastly, we have confirmed that the random-lattice multifold model in Fig.~\ref{fig:3F_Unif} continues to exhibit a spin-1 chiral multifold fermion at ${\bf p}={\bf 0}$ with a quantized $|C|=2$ non-Abelian Wilson loop winding number under the subsequent addition of weak Anderson [on-site chemical potential] disorder.

\end{appendix}

\bibliography{a_NN}
\end{document}